\RequirePackage{luatex85}
\documentclass[11pt,a4paper]{report}
\usepackage[twoside]{geometry}
\usepackage{fancyhdr}
\usepackage{afterpage}
\usepackage[compat=1.1.0]{tikz-feynman}
\usepackage[utf8]{inputenc}
\usepackage[polish, english]{babel}
\let\babellll\lll
\let\lll\relax
\usepackage[T1]{fontenc}
\usepackage{placeins}
\usepackage{mathtools}
\usepackage{hhline}
\usepackage{varwidth}
\usepackage{slashed}
\usepackage{cite}
\usepackage{wasysym}
\usepackage[nottoc]{tocbibind}
\usepackage{url}
\usepackage{rotating}
\usepackage[breaklinks]{hyperref}
\hypersetup{
    colorlinks = false,
    linkbordercolor = {white},
}
\usepackage{graphicx}
\usepackage{subcaption}
\usepackage{appendix}

\usepackage{afterpage}
\usepackage{mathbbol}
\usepackage{rotating}
\usepackage{multicol}
\usepackage{footnote}
\usepackage{footmisc}
\usepackage{breakurl}
 \usepackage{multirow}
\usepackage{amssymb}

\let\lll\babellll
\usepackage{amsmath}
\usepackage{amsfonts} 
\usepackage{afterpage}

\newcommand\blankpage{%
    \null
    \thispagestyle{empty}%
    \addtocounter{page}{-1}%
    \newpage}
 
\newcounter{savefootnote}
\newcounter{symfootnote}
\newcommand{\symfootnote}[1]{%
   \setcounter{savefootnote}{\value{footnote}}%
   \setcounter{footnote}{\value{symfootnote}}%
   \ifnum\value{footnote}>8\setcounter{footnote}{0}\fi%
   \let\oldthefootnote=\thefootnote%
   \renewcommand{\thefootnote}{\fnsymbol{footnote}}%
   \footnote{#1}%
   \let\thefootnote=\oldthefootnote%
   \setcounter{symfootnote}{\value{footnote}}%
   \setcounter{footnote}{\value{savefootnote}}%
}

\newenvironment{dedication}
    {\vspace{6ex}\begin{quotation}\begin{center}\begin{em}}
    {\par\end{em}\end{center}\end{quotation}}

\newcommand{\myparagraph}[1]{\paragraph{#1}\mbox{}\\}

\begin{document}
\graphicspath{ {./figures/} }

\newgeometry{
inner=3cm,
outer=3cm,
top=1in,
bottom=1in,
 }

\begin{titlepage}
\centering
\includegraphics[width=\textwidth]{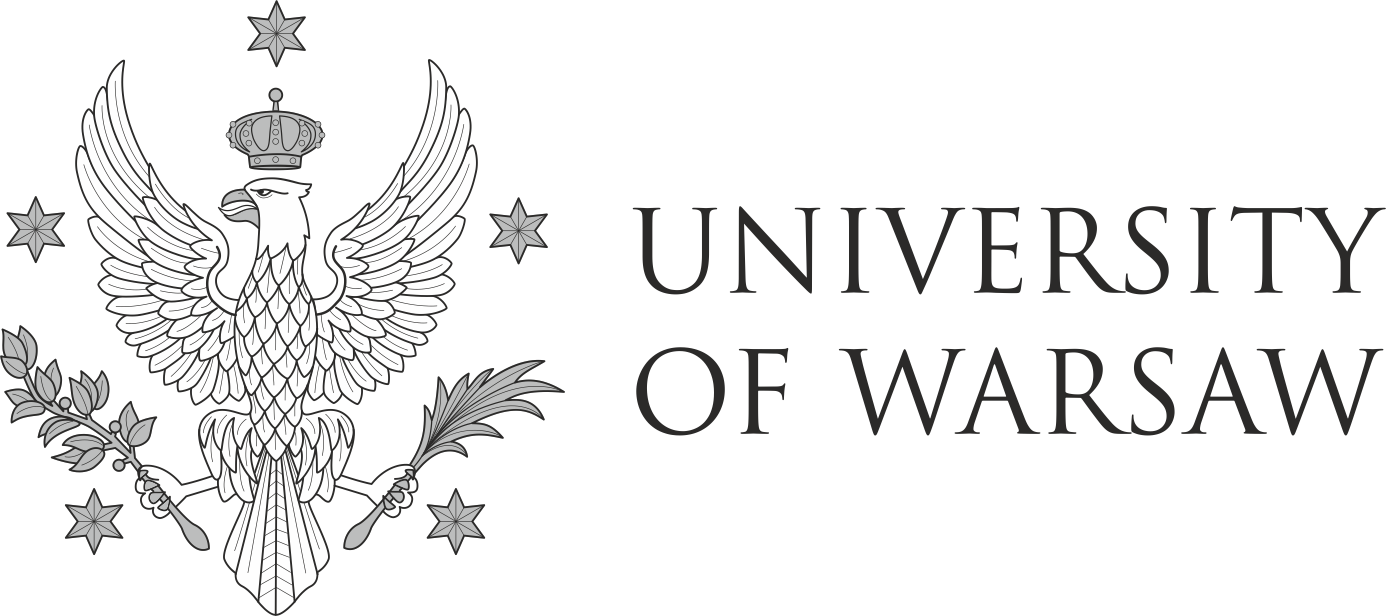}

{\vspace{2.5cm} \Huge\bfseries Prospects for detecting long-lived particles at the Large Hadron Collider \par}
\vspace{1.5cm}
{\huge\itshape Rafał Masełek\par}
\vfill
{\Large

	supervised by\par
	Kazuki Sakurai
}
	\vfill
	{\large  March 2023 \par}
	\afterpage{\blankpage}
\end{titlepage}

\newpage
\begin{dedication}
\thispagestyle{empty}
\vspace*{\fill}
\Large
Pamięci babci Stasi i dziadka Jana.
\end{dedication}
\newpage

\setcounter{page}{1}
\pagenumbering{Roman}

\begin{abstract}
\thispagestyle{plain}
This thesis summarises four years of research aiming at revealing the prospects for detection of long-lived particles at the LHC. It 
contains results of four projects, which have been published as independent articles.

In the first project, the possibility to utilise 
the MoEDAL detector to search for charged long-lived particles is considered for the first time. MoEDAL is a small effectively background-free and mostly 
passive detector located at the IP8 of the LHC. It can detect a particle with charge $Q$, if its decay length is of the order of $\mathcal{O}(1~\rm{m})$ and its velocity $\beta$ satisfies $\beta \lesssim 0.15 \cdot |Q/e|$. This makes MoEDAL complementary to
major general-purpose LHC experiments, i.e. ATLAS and CMS, which are 
sensitive only to particles with $\beta \gtrsim 0.5$. As a demonstration of the 
capabilities of the MoEDAL detector, a particular supersymmetric model is 
considered, in which a pair of gluinos ($\tilde g$) is produced, each of them 
decays to SM jets and a long-lived neutralino ($\tilde \chi_1^0$), which subsequently decays to a metastable stau $(\tilde\tau)$ and an off-shell tau. Schematically: 
$
pp \to \tilde g \tilde g \to \left( \tilde \chi_1^0 j j  \right) \left( \tilde \chi_1^0 j j  \right) \to 
\left( \tilde \tau_{1}  \tau^* jj \right)
\left( \tilde \tau_{1}  \tau^* jj \right)$.
For the considered scenario, it has been revealed that Run 3 MoEDAL ($L=30~\rm{fb}^{-1}$) can test regions in the gluino mass ($m_{\tilde g}$) vs. neutralino decay length ($c \tau_{\tilde \chi_1^0}$)
parameter plane, which are beyond the reach of ATLAS. This interesting result inspired a whole series of studies.

In the second 
project, various pair-produced long-lived supersymmetric particles are studied: gluino $\tilde g$, stop $\tilde t$, five light-flavour 
squarks $\tilde q=(\tilde u, \tilde d, \tilde c, \tilde s, \tilde b)$, wino- and higgsino-like charginos $\widetilde W, \tilde h$, and 
sleptons $\tilde l$. Moreover, doubly charged ($Q=\pm 2e$) scalars and fermions, transforming under $SU(2)_L$-singlet and 
triplet representations, are also considered. Except for $SU(2)_L$-triplet fermions, the MoEDAL detector is found to have 
worse sensitivity than ATLAS and CMS, mainly because of the lower amount of data available. However, the limits that MoEDAL can 
provide are complementary to the ones by ATLAS and CMS, due to completely different detector design and uncorrelated 
systematic uncertainties.

In the third project, a particular 1-loop radiative neutrino mass model is considered, which predicts 
the existence of long-lived scalar particles. Two versions of the model are studied, in the uncoloured version all BSM fields are 
singlets under $SU(3)_C$, while in the coloured version, they are promoted to colour-(anti)triplets. In the uncoloured version there 
are long-lived scalars $S^{\pm 2}$, $S^{\pm 3}$, and $S^{\pm 4}$, with charges $\pm 2 e$, $\pm 3e$, and $\pm 4e$, 
respectively. There are also triply charged fermions, $F^{\pm 3}$, however, they are short-lived. The coloured version of the 
model also contains long-lived scalars: $S^{\pm 4/3}$, $S^{\pm 7/3}$, and 
$S^{\pm 10/3}$, with charges $\pm 4/3 e$, $\pm 7/3e$, and $\pm 10/3e$, 
respectively. 
Possibility to detect long-lived particles in Run 3 and HL-LHC 
MoEDAL have been estimated and compared with the available limits from 
ATLAS and CMS collaborations. 
In the first part of the analysis, a model-independent detection reach of MoEDAL for multiply charged LLPs is 
obtained, where the lifetimes of BSM particles are treated as free parameters. 
It have been found that most of the parameter space accessible to Run 3 MoEDAL is 
already excluded by large general-purpose experiments. On the other hand, 
MoEDAL has a chance to detect multiply charged LLPs during the HL-LHC data-taking phase, but ATLAS is 
expected to provide better sensitivity.
The second part of the analysis targets the possibility to constrain parameters 
of the specific radiative neutrino mass model considered in the project. In particular, the impact of the dimensionless coupling $\lambda_5$, responsible for lepton number violation, is studied. It have been found that for $\lambda_5 \sim 10^{-5}$ MoEDAL achieves the highest sensitivity, which roughly corresponds to the longest lifetime of $S^{\pm 4}$ ($\tilde S^{\pm 10/3}$).

The fourth project aims at providing a comprehensive overview of possibilities 
to detect charged LLPs at the LHC. Four types of particles are studied: scalars 
and spin-1/2 fermions transforming under the $SU(3)_C$ gauge group as 
singlets or triplets. All particles are assumed to be $SU(2)_L$-singlets and 
have integer electric charges, $Q$, in the range $1 e \leq |Q| \leq 8e$.
Three different BSM search methods are considered: (i) large $dE/dx$ 
searches for detector-stable charged LLPs in ATLAS and CMS, (ii) searches for 
charged LLPs in MoEDAL, (iii) diphoton resonance searches by ATLAS and CMS, 
targeting diphoton decays of positronium/quarkonium-like bound states 
formed by the new BSM particles. The sensitivity of MoEDAL is estimated with 
the procedure established during the previous three projects, i.e. by varying 
the mass and lifetime of BSM particles and calculating the expected number of 
signal events for Run 3 $(L=30~\rm{fb}^{-1})$ and HL-LHC $(L=300~\rm{fb}^{-1})$. In the case of ATLAS and CMS searches, the most recent analyses are 
recast in order to obtain 95\% CL upper cross section limits, which are 
compared with theoretical calculations. A projection for Run 3 $(L=300~\rm{fb}^{-1})$ and HL-LHC $(L=3~\rm{ab}^{-1})$ is also made.

It has been found that large $dE/dx$ searches are most sensitive for smaller electric 
charges $|Q| \lesssim (3-4)e$, depending on the type of the BSM particle, while 
for larger charges, $|Q| \gtrsim (4-5)e$, diphoton resonance searches provide 
the strongest detection reach. The sensitivity of Run 3 MoEDAL is found to be 
in between the expected bounds of the aforementioned two types of ATLAS/CMS 
searches. However,  
for the intermediate charges, $3e \lesssim |Q| \lesssim 7e$,
MoEDAL 
might provide better sensitivity than major general-purpose experiments at the end of the HL-LHC phase, 
thanks to the background-free nature of MoEDAL's detector design.

The fourth project includes also an important discussion about the impact of 
photon-induced production processes on the correct interpretation of 
experimental results. It has been shown that for particles with $|Q| \gtrsim 4e$ 
photon fusion (and photon-gluon fusion for coloured particles) production 
cross section becomes comparable to that of the Drell-Yan process. This is an important result since 
the ATLAS and CMS collaborations did not take this effect into account, which 
resulted in an underestimation of experimental limits for multiply charged 
particles.
\end{abstract}
{
\begin{center}
\bfseries \large Szanse na detekcję cząstek długożyjących w Wielkim Zderzaczu Hadronów
\end{center}

\begin{center}
\bfseries \large Streszczenie
\end{center}

Niniejsza rozprawa doktorska podsumowuje cztery lata badań naukowych nastawionych na określenie szans na detekcję cząstek długożyjących w Wielkim Zderzaczu Hadronów. Rozprawa zawiera wyniki czterech projektów naukowych, które zostały opublikowane w formie osobnych artykułów.

Pierwszy projekt dotyczył nie zbadanej wcześniej możliwości wykorzystania detektora MoEDAL do poszukiwań naładowanych elektrycznie cząstek długożyjących. MoEDAL to 
mały pasywny detektor pozbawiony tła od Modelu Standardowego, który jest zlokalizowany w pobliżu IP8 przy Wielkim Zderzaczu Hadronów. MoEDAL jest w stanie wykryć cząstkę o ładunku 
elektrycznym $Q$, jeżeli jej czas życia jest na tyle długi, że jest w stanie przebyć odległość rzędu $\mathcal{O}(1~\rm{m})$, oraz jej szybkość $\beta$ spełnia warunek $\beta \lesssim 0.15 \cdot |Q/e|$. 
Dzięki temu poszukiwania nowej fizyki w doświadczeniu MoEDAL uzupełniają badania w eksperymentach ATLAS i CMS, które są czułe jedynie na cząstki o szybkościach $\beta \gtrsim 0.5$. W celu zademonstrowania możliwości detektora MoEDAL, rozważono pewien szczególny model supersymetryczny, w którym dwa gluina $(\tilde g)$ są produkowane, każde rozpada się 
natychmiastowo do 2 dżetów i długożyjącego nautralina $(\tilde \chi_1^0)$, które następnie rozpada się na długożyjący staon i wirtualny taon. Schematycznie:
$pp \to \tilde g \tilde g \to \left(\tilde \chi_1^0 j j \right) \left(\tilde \chi_1^0 j j \right) \to \left(\tilde \tau_1 \tau^* j j\right)  \left(\tilde \tau_1 \tau^* j j\right)$. Dla rozważanego modelu wykazano, iż MoEDAL w trakcie Run 3 $(L=30~\rm{fb}^{-1})$ jest w stanie zbadać wartości masy gluina $m_{\tilde g}$ i czasu życia neutralina $(\tau_{\tilde \chi_1^0})$, które są poza zasięgiem eksperymentu ATLAS. Ten interesujący wynik zainspirował całą serię badań.

W drugim projekcie rozważono róznego rodzaju długożyjące cząstki supersymetryczne: gluino $\tilde g$, stop $\tilde t$, 5 pozostałych skwarków $\tilde q = (\tilde u, \tilde d, \tilde c, \tilde s, \tilde b)$, naładowane wino $\widetilde W$ i higgsino $\tilde h$, oraz naładowane elektrycznie sleptony $\tilde l$. Dodatkowo zbadano podwójnie naładowane $(Q=\pm 2 e)$ cząstki skalarne i fermiony o spinie 1/2, transformujące się jak singlet lub triplet względem grupy cechowania $SU(2)_L$. Za wyjątkiem tripletu fermionów, MoEDAL przejawia niższą czułość na nowe cząstki niż ATLAS i CMS. Główną przyczyną jest mniejsza ilość danych zbieranych przez MoEDAL. Jednakże wyniki dostarczone przez doświadczenie MoEDAL są komplementarne względem rezultatu eksperymentów ATLAS i CMS, ze względu na całkowicie odmienną budowę detektora i nieskorelowane niepewności systematyczne.

W trzecim projekcie zbadano radiacyjny model generacji mas nautrin, który przewiduje istnienie naładowanych elektrycznie cząstek długożyjących. Rozważono dwie wersje modelu, 
w niekolorowej wersji wszystkie nowe cząstki są singletami względem grupy cechowania $SU(3)_C$, natomiast w wersji kolorowej zastępują je (anty)triplety. Długożyjące cząstki 
skalarne w niekolorowej wersji modelu to: $S^{\pm 2}$, $S^{\pm 3}$, oraz $S^{\pm 4}$, z ładunkami $\pm 2 e$, $\pm 3 e$, oraz $\pm 4e$, odpowiednio. Model przewiduje również istnienie potrójnie naładowanych fermioniów, $F^{\pm 3}$, jednakże cechują się one krótkim czasem życia. Kolorowa wersja modelu również zawiera długożyjące cząstki skalarne:
$S^{\pm 4/3}$, $S^{\pm 7/3}$, oraz $S^{\pm 10/3}$, z ładunkami $\pm 4/3 e$, $\pm 7/3 e$, oraz $\pm 10/3e$, odpowiednio. W projekcie zbadano szanse na detekcję cząstek 
długożyjących w trakcie Run 3 i HL-LHC w eksperymencie MoEDAL, które następnie porównano z dostępnymi wynikami z doświadczeń ATLAS i CMS. W pierwszej części analizy 
określono czułość detektora MoEDAL na wielokrotnie naładowane cząstki długożyjące w sposób niezależny od przyjętego modelu teoretycznego, tzn. traktując ich masy i czas życia 
jako wolne parametry. Odkryto, iż dla większości zakresu wartości parametrów, które MoEDAL może przetestować w trakcie Run 3, istnienie rozważanych cząstek zostało wykluczone przez duże eksperymenty przy LHC. Z drugiej strony, dla HL-LHC MoEDAL ma szansę odkryć wielokrotnie naładowane cząstki długożyjące, jednakże spodziewana czułość doświadczenia ATLAS jest większa. Druga część analizy była poświęcona możliwości zbadania parametrów konkretnego modelu generacji mas neutrin. W szczególności zbadano znaczenie bezwymiarowej stałej sprzężenia $\lambda_5$, odpowiedzialnej za złamanie zasady zachowania liczby leptonowej. Odkryto, iż dla wartości $\lambda_5 \sim 10^{-5}$, która z grubsza odpowiada najdłuższemu czasu życia $S^{\pm 4}$ $(\tilde S^{10/3})$,
MoEDAL osiąga największą czułość.

Celem czwartego projektu było przedstawienie wyczerpującego podsumowania perspektyw na detekcję cząstek długożyjących w Wielkim Zderzaczu Hadronów. Zbadano cztery rodzaje cząstek: skalary i fermiony o spinie 1/2 transformujące się względem grupy cechowania $SU(3)_C$ jako singlety lub triplety. Założono, iż wszystkie cząstki są singletami względem $SU(2)_L$ oraz mają całkowite ładunki elektryczne, $Q$, z zakresu $1e \leq |Q| \leq 8e$. Zbadano trzy różne metody poszukiwań nowej fizyki:
(i) poszukiwania stabilnych cząstek wysokojonizujących w eksperymentach ATLAS i CMS,
(ii) poszukiwania długożyjących cząstek naładowanych w doświadczeniu MoEDAL,
(iii) poszukiwania rezonansów rozpadających się na dwa fotony przez ATLASa i CMSa, które zastosowano dla układów związanych typu pozytronium/kwarkonium stworzonych przez nowe cząstki.
Czułość doświadczenia MoEDAL określono zgodnie z procedurą wypracowaną w trakcie poprzednich trzech projektów, tzn. traktując masy i czasy życia nowych cząstek jako wolne 
parametery i estymując liczbę rejestrowanych przypadków w trakcie przyszłych faz zbierania danych: Run 3 $(L=30~\rm{fb}^{-1})$ i HL-LHC $(L=300~\rm{fb}^{-1})$. W przypadku 
doświadczeń ATLAS i CMS, najnowsze analizy zostały zreinterpretowane w celu uzyskania górnych limitów na przekroje czynne, które zostały porównane z wynikami obliczeń 
teoretycznych. Uzyskano również spodziewane limity dla Run 3 $(L=300~\rm{fb}^{-1})$ i HL-LHC $(L=3~\rm{ab}^{-1})$.

W trakcie czwartego projektu odkryto, iż poszukiwania cząstek wysokojonizujących są najlepszą metodą na odkrycie cząstek o niskich ładunkach, $|Q|\lesssim (3-4)e$, gdzie 
dokładna wartość górnej granicy zależy od rodzaju cząstki. Z kolei dla większych ładunków, $|Q| \gtrsim (4-5)e$, poszukiwania rezonansów rozpadających się na dwa fotony są 
najczulsze. Czułość eksperymentu MoEDAL w trakcie trzeciej tury zbierania danych w LHC (Run 3) okazuje się być pomiędzy czułością wymienionych dwóch typów analiz. Jednakże dla średnich wartości ładunków elektrycznych, $3e \lesssim |Q| \lesssim 7e$, HL-LHC MoEDAL wykazuje się lepszą czułością niż duże eksperymenty takie jak ATLAS i CMS. Jest to spowodowane pomijalnie małym tłem od procesów Modelu Standardowego w MoEDALu.

Czwarty projekt rozwinął również ważną dyskusję na temat znaczenia procesów z fotonami w stanie początkowym na produkcję i właściwości cząstek wielokrotnie naładowanych w Wielkim Zderzaczu Hadronów. Wykazano, iż dla cząstek o ładunku $|Q| \gtrsim 4e$ rprzekrój czynny dla fuzji dwóch fotonów (i fuzji fotonu z gluonem w przypadku cząstek kolorowych) jest porównywalny z przekrojem czynnym dla procesu typu Drell-Yan. Jest to istotne spostrzeżenie, gdyż jak dotąd doświadczenia ATLAS i CMS nie brały tego efektu pod uwagę, co skutkowało niedoszacowaniem limitów na masy nowych cząstek.\par
}
\newpage\null\thispagestyle{plain}\newpage
{
	 \begin{center}%
    \bfseries\Large Acknowledgements (Podziękowania)

  \end{center}
  
\vspace{2em}
I would like to express gratitude towards my PhD supervisor Kazuki Sakurai, who taught me
how to be a diligent and dutiful scientist by his own example. 
I thank you for all your time and effort spent during endless discussions, and for your hard work side by side with me. I am forever indebted to you for the 
support and assistance you have shown, and for your tremendous patience that I have tested many times. I am grateful for your kindness and encouragement, which helped me to endure difficult moments in the past four years.
I acknowledge your advice and mentoring, and I am impressed by your creativity and enthusiasm for Physics.
I deeply hope that our collaboration and friendship may last and thrive.

Dziękuję z całego serca moim kochanym rodzicom za ich wsparcie w trakcie robienia doktoratu i pisania pracy. Dziękuję również całej mojej rodzinie za jej życzliwość i pomoc.

Na specjalne podziękowania zasługuje Pan Dionizy Kyc, który był moim nauczycielem w gimnazjum i jako pierwszy zainteresował mnie fizyką.

I appreciate and thank all my colleagues from the University of Warsaw for their support and collaboration. Special thanks to Fotis Koutroulis, Priyanka Lamba, Ayuki Kamada and Mohammad Altakach. Thanks to you, the four years of my PhD were a wonderful time.

I am grateful to John Ellis, Vasiliki Mitsou, Mihoko Nojiri and Stefan Pokorski for their support of my career. I hope we will continue to work together in the future.

I thank all my collaborators without whom I could not conduct the research presented in this thesis.

Many thanks to wonderful and passionate people in the MoEDAL collaboration led by James Pinfold, thanks to whom this thesis was possible.

I thank Wojciech Bryliński for the hospitality he has treated me with during my visits to CERN.

I thank my friends: Jacek, Marta, Radek, Wojtek, and Maciek, for listening to me talk endlessly about my PhD and the thesis.

I would like to thank the Polish National Science Centre, which provided me with the necessary means for my research, through the following grants:
\begin{itemize}
\item Sonata Bis 7
(2017/26/E/ST2/00135),
\item
Preludium 20 
(2021/41/N/ST2/00972),
\item
Beethoven 2 
(2016/23/G/ST2/04301),
\item
GRIEG 
(2019/34/H/ST2/00707).
\end{itemize}
I also acknowledge the Polish National Agency For Academic Exchange for awarding me the Bekker 2022 grant 
(BPN/BEK/2022/1/00253).

}


\begingroup
\makeatletter
  \def\@pnumwidth{2em}
 \def\@tocrmarg {3.5em} 
\makeatother
\let\cleardoublepage\clearpage
\tableofcontents
\endgroup

%

\restoregeometry
\fancyhead{}
\fancyhead[CE]{Prospects for detecting long-lived particles at the Large Hadron Collider}
\fancyfoot{}
\fancyfoot[LE,RO]{\thepage}
\fancyfoot[CO,CE]{Chapter \thechapter}
\fancyhead[CO]{Rafał Masełek}
\interfootnotelinepenalty=10000

\chapter{Particle Physics}\label{chap:one}
\pagenumbering{arabic}
\setcounter{page}{1}

\FloatBarrier
\section{Origins of Particle Physics}

\subsection*{Atoms}

The origins of Particle Physics can be traced back to antiquity, when ancient 
Greek philosophers tried to describe the surrounding world and observable 
phenomena. One of the major questions was what are the fundamental 
ingredients that matter is made of. Leucippus and his student Democritus of 
Abdera proposed that the physical universe is composed of fundamental 
indivisible components called \textit{atoms} (from Greek $\alpha \theta o \mu 
o \nu$, \textit{atomon}, meaning something indivisible), which differed in 
sizes, shapes and other properties. According to them, everything material and 
spiritual was created of an infinite number of types of eternal atoms, and nothing 
except atoms and void was fundamentally real. Another Greek philosopher, 
Epicurus of Samos, shared atomistic belief, but rejected the idea of infinite 
variety of atoms, claiming that there is only a finite number of them. Other 
philosophers had their own ideas, usually very different from our understanding 
of the physical reality. Although atomism was not the most popular philosophy 
in ancient and medieval times, it was not forgotten and inspired new 
generations of philosophers, alchemists and early scientists.

One of the greatest physicists ever, sir Isaac Newton, formulated the corpuscular theory of light. He claimed that the light is composed of tiny discrete parts, called \textit{corpuscles}. The authority of Newton caused a prevalence of his corpuscular theory for a hundred years until it was replaced by the wave theory of light based on works by Christiaan Huygens, Leonard Euler, Thomas Young and Augustin-Jean Fresnel. The theory by Newton was brought back to life in the 20th century by Albert Einstein's work on the photoelectric effect.

At the very beginning of the 19th century, British chemist John Dalton 
discovered that if two elements can form more than one compound, the ratios 
of their masses entering these compounds are small natural numbers. It lead 
him to formulate a theory of atoms, stating that matter is composed of 
indivisible parts that are identical for a single element, but they have different 
masses for distinct elements. Other chemists, like Amadeo Avogadro and 
Stanislao Canizzaro, developed Dalton's theory. In the middle of the 19th century, the 
atomic theory became widely accepted among chemists, but physicists were 
reluctant, many believed that chemical atoms have little to do with the 
fundamental physical reality. It changed after the success of statistical 
mechanics developed by James Clerk Maxwell, Ludwig Boltzmann, Rudolf 
Clausius and Josiah Willard Gibbs.

\subsection*{Quantum reality}

Soon after the physics community accepted the existence of eternal building blocks of matter, the belief was challenged by studies on radioactivity. 
In 1897 Joseph John Thomson studied the properties of cathode rays and concluded that rather than being a form 
of light, these rays were composed of negatively charged particles. Later these particles were called electrons.
Thomson made a bold hypothesis that atoms, as known to chemists at that time, were composed of a more fundamental form of 
charged matter, and cathode rays were a result of emitting such matter. The existence of electrons and the electric neutrality of 
atoms implied the existence of positively charged constituents. The open question was how the charged matter is distributed in an 
atom. J. J. Thomson suggested a plum pudding model, in which positively charged matter is continuous (like a pudding) and 
electrons are distributed within it (like plums). However, Ernest Rutherford showed that in reality atom is composed of a very 
heavy and positively charged nucleus surrounded by light electrons. It was natural to assume a planetary model, in which 
electrons move around the nucleus like planets around the Sun, but with gravity exchanged for electric force. However, the planetary 
model based on classical electromagnetism could not explain the stability of atoms, i.e. why the electron does not radiate its energy 
and fall to the nucleus. Another mystery was the origin of spectral lines observed in the emission spectrum of the hydrogen 
(Balmer series). To resolve these issues, Quantum Mechanics was necessary. 

The fatherhood of Quantum Theory is attributed to Max Planck, who theoretically explained the observed spectral 
density of the black-body radiation, by assuming that the energy can only be emitted in discrete packages of energy 
\begin{equation}\label{eq:photon-energy}
E= h \nu, 
\end{equation}
which were later called \textit{quanta}. $\nu$ in Eq. \eqref{eq:photon-energy} is the frequency of light, $E$ is the energy of a quanta (photon), and $h= 6.62607015 \times 10^{-34} \rm{J} \cdot {\rm Hz}^{-1}$ is a constant named after Planck.
At that time Planck, who was raised on classical electromagnetism, did not attribute a 
deep meaning to his interpretation, treating quanta as a kind of mathematical trick. A fresh point of view by the young Albert Einstein was 
needed. In his \textit{annum mirabelis} 1905 he published four groundbreaking papers on Brownian motion, the equivalence of mass 
and energy, special relativity and the photoelectric effect. The latter publication explained the emission of electrons from a metal plate 
irradiated by light, which depends on the light frequency rather than the intensity, as opposed to classical predictions. Einstein was able 
to theoretically describe the phenomenon by assuming that light is composed of energy quanta, described by the same 
relation as Planck had assumed. It was a revolutionary step because what had been considered just a mathematical trick to 
explain black body radiation was promoted to the fundamental description of electromagnetic radiation. As with any novel idea, 
Einstein's explanation was heavily attacked and criticised by the older generation of acknowledged scientists. History, however, has 
proven Einstein right.

In 1913, Danish physicist Niels Bohr postulated a new atomic model, in which he adopted quantum principles. In his model, 
electrons were orbiting the nucleus on circular orbits, but only a discrete set of orbits was allowed, each with an associated value of 
energy and angular momentum. When orbiting, electrons did not radiate and had a specific value of energy, however, they could 
change the orbit and radiate/absorb the energy difference between two orbits in a form of an energy quanta. The 
new atomic model was very successful. Not only it allowed to explain the observed hydrogen emission spectrum, but also predicted 
the existence of additional lines in the ultraviolet, which were observed by Theodore Lyman in 1914. 
In 1915 a German scientist, Arnold Sommerfeld, improved Bohr's model by introducing orbital quantum number $l$.
Despite all the evidence, the 
community was still reluctant to accept the quantum nature of electromagnetic 
radiation. The situation changed after the experiment of Stern and Gerlach, which 
proved the quantisation of spin \footnote{At that time spin was not known. Stern 
and Gerlach thought that their result proved the space quantisation of magnetic 
moments of atoms of silver. Now it is known that in a silver atom, all electrons 
but one form closed shells with a total magnetic moment equal to zero. Therefore, 
an atom's magnetic moment is given by the magnetic moment of a single 
electron.}, and scattering experiments conducted by Arthur H. Compton  
could only be explained by accepting the existence of photons. While most of 
the physicists were stunned by the apparent contradiction between wave and 
corpuscular nature of light, 
Luis de Broglie went even further, postulating that electrons should also exhibit 
wave-like behaviour with the wavelength $\lambda$ give by the relation $\lambda=h/p$. His hypothesis was confirmed by Clinton 
J. Davisson and L. Germer, who discovered the diffraction of electrons on 
crystals in 1927.

\subsection*{Quantum Mechanics}

The old quantum theory successfully described some aspects of the physics of the hydrogen atom, but its usability was highly limited, and eventually lead to the formulation 
of a new quantum theory, called \textit{Quantum Mechanics}, which was 
created in just few years by several young physicists. In 1925 Werner K. 
Heisenberg introduced the new quantum theory using matrix formulation. Soon 
after, Paul Dirac understood that Poisson brackets and commutators are closely 
related, which allowed him to connect Heisenberg's theory with classical 
mechanics. In the meantime, Wolfgang Pauli proposed the famous 
\textit{exclusion principle}, which prohibits two electrons with the same
quantum numbers to exist in a single atom. Ralph Kronig, and independently
George Uhlenbeck with Samuel Goudsmit, proposed that electrons possess 
intrinsic angular momentum -- \textit{spin}. In 1926 Erwin Schroedinger introduced Wave Mechanics and postulated his famous equation:
\begin{equation}\label{eq:schroedinger}
\left(-\frac{\hbar^2}{2m} \nabla^2  + V \left(\vec{r}, t \right) \right) \psi\left(\vec{r}, t \right)  = i \hbar \partial_t \psi\left(\vec{r}, t \right),
\end{equation}
where $\psi$ is the \textit{wave function}, a complex object describing the 
physical quantum particle, e.g. an electron, with mass $m$ and position $\vec{r}$, moving in a scalar potential $V \left(\vec{r}, t \right)$. $\hbar$ in Eq. \eqref{eq:schroedinger} is the \textit{reduced Planck constant} defined as $\hbar \equiv h/(2\pi)$.
Schroedinger was able to solve 
hydrogen atom and achieve very good agreement with experimental 
measurements.
The Wave Mechanics was 
accepted much more enthusiastically than Heisenberg's theory, because 
mathematically it resembled classical theories. Despite apparent differences, it 
was quickly proven that both theories are equivalent. The physical sense 
of the wave function was unclear, Schroedinger speculated that it represented
charge density of an electron. In the summer of 1926 Max Born proposed 
statistical interpretation of Quantum Mechanics, presently known as the 
\textit{Copenhagen interpretation}, in which $\left |\psi \left(\vec{r} \right)  \right|^2$ corresponds to the 
probability density of detecting a particle in position 
$\vec{r}$.  The new interpretation lead Heisenberg to formulate his famous 
\textit{uncertainty principle} --  there exist pairs of observables that cannot be 
simultaneously measured with arbitrary precision, e.g. for position and momentum 
of a particle moving in 1D:
\begin{equation}
\sigma_x \sigma_p\geq \frac{\hbar}{2},
\end{equation}
where $\sigma_\xi$ is the standard deviation of the operator $\hat \xi \in \{\hat x, \hat p \}$, defined as 
$\sigma_\xi \equiv \sqrt{\langle \xi^2 \rangle - \langle \xi \rangle^2}$, with $\langle \cdot \rangle$ representing the expectation value calculated for the probability density $\rho(\vec{r},t) = |\psi (\vec{r},t)|^2$.

Several attempts were made to construct a relativistic version of the quantum theory. Before postulating the Eq. \eqref{eq:schroedinger}, Erwin Schroedinger investigated the relativistic energy-momentum relation 
$E^2 = m^2c^4 +\vec{p}^2c^2$, 
where $c=299 792 458 m/s$ is the speed of light in the vacuum, $E$ is the energy of a particle with mass $m$ and momentum $\vec{p}$,
and arrived at the equation:
\begin{equation}
\left[  \hbar^2 \partial_\mu \partial^\mu + \left({m c}\right)^2 \right] \psi(x^\mu) = 0,
\end{equation} 
which was later rediscovered independently by Oskar Klein and Walter Gordon. The new equation was considered invalid because it lead to negative probability solutions. Therefore, Paul Dirac proposed a new relativistic equation:
\begin{equation}\label{eq:dirac}
\left(i \hbar \slashed{\partial} - mc \right) \psi(x^\mu),
\end{equation}
where the slashed derivative is defined by contraction of ordinary partial derivative with Dirac gamma matrices $ \slashed{\partial} \equiv \gamma^\mu \partial_\mu$, which will be explained in Eqs. \eqref{eq:gamma-matrices} and \eqref{eq:slashed}.
The equation \eqref{eq:dirac} led to solutions with negative energy.
Dirac understood that these solutions were correct and described antiparticles. Antiparticles have the same mass, but 
opposite quantum numbers with respect to ordinary particles.

Albert Einstein was initially enthusiastic about the new quantum theory, however, he quickly changed his mind after the 
probabilistic interpretation was proposed, and became one of the harshest critics of Quantum Mechanics. His famous quote:
\begin{quote}\centering
God does not play dice!
\end{quote}
started a vivid discussion with Niels Bohr about interpretations of quantum phenomena and the ``true'' nature of  
physical reality. Einstein did not appreciate the randomness of Quantum Mechanics, and he argued that there should exist  
classical ``hidden variables'' associated with particles, which would predetermine the outcome of measurements, contrary 
to the statistical interpretation of the wave function collapse.
After the formulation of the famous Bell inequalities in 1964, and their experimental check in 1972 by Stuart J. Freedman and John F. Clauser, it became evident that 
the physical reality was not described by a hidden variable theory but was consistent with the predictions of Quantum Mechanics.
However, the discussion about the ``true'' meaning of quantum 
phenomena is still ongoing and it is a popular topic among philosophers and some of the scientists.

\subsection*{Towards modern Particle Physics}

In 1927 Paul Dirac argued that Quantum Mechanics ought to be applied not only to the atom but also to the electromagnetic 
field, which is considered the birth of \textit{Quantum Electrodynamics}, the quantum field theory describing 
electromagnetic interactions. Dirac's theory was further developed and achieved good agreement with experiments. 
However, after World War II more precise measurements were conducted revealing discrepancy between Quantum 
Electrodynamics and experiments, which was correctly attributed to neglecting the electron self-interaction in 
theoretical calculations. Intense efforts by Richard Feynman, Julian Schwinger and Sin-Itiro Tomonaga allowed to introduce 
renormalisability concepts to Quantum Field Theory, and led to more precise calculations. In 1954, Chen-Ning Yang and 
Robert Mills developed gauge theories, which were necessary for the unification of electromagnetic and weak interactions by 
Sheldon Glashow, Abdus Salam and Steven Weinberg in the late 1960s. In their works, Weinberg and Salam incorporated 
the Higgs mechanism introduced in 1964 to Glashow's weak interaction. The electroweak theory became widely accepted 
after neutral weak currents were discovered at CERN in 1973, and further confirmed by discoveries of the $W^\pm$ and 
$Z^0$ bosons in 1983.

Understanding the strong interactions was possible thanks to the works of several generations of brilliant physicists. In 1934 
Hideki Yukawa proposed the existence of particles mediating interactions between nucleons. These particles, which we now 
call \textit{pions} $\pi$, were discovered in 1947 by Cecil Powell et al.  Many other new particles were observed in the 
following years, causing a lot of confusion in the community. Fortunately, in 1961 Murray Gell-Mann found a way to relate 
hadrons using the $SU(3)$ symmetry, and just three years later he proposed the existence of three quarks: 
\textit{up (u)}, \textit{down (d)} and \textit{strange (s)}.
The quark 
hypothesis allowed to simplify the picture of particle physics, but only after the discovery of the asymptotic freedom it
was possible to formulate a consistent theory called \textit{Quantum Chromodynamics}, in which $SU(3)_C$-gauged bosons, \textit{gluons}, mediate the strong interaction. 
In 1974 the fourth quark, 
\textit{charm (c)}, was discovered, and together with the strange quark formed the second family of quarks.
Around that time the term ``Standard Model'' was coined. 
In 1973 Makoto Kobayashi and Toshihide Maskawa proposed the existence of the third generation of quarks to explain the previously observed violation of the CP symmetry\footnote{CP symmetry is the combination of charge (C) and parity (P) 
symmetries. A process is symmetric with respect to the CP symmetry if exchanging all left-handed particles to the corresponding right-handed antiparticles (C symmetry) and inverting spatial coordinates (P symmetry) results in the same amplitudes. CP symmetry is obeyed by the electromagnetic interaction, 
but it is violated by the weak force.} in kaon decays. Their predictions were confirmed when the
\textit{bottom (b)} and \textit{top (t)} quarks were discovered in 1977 and 1995, respectively. 

Discoveries of muons, taus and different neutrino flavours allowed to group elementary fermions into three generations of matter. The last particle predicted by the Standard Model, the Higgs boson, was discovered at the \textit{Large Hadron Collider} at CERN in 2012.

When looking back at the history of Particle Physics, or science in general, one might think that it was a firm and 
straight march towards the progress and understanding. This couldn't be further from the truth. In reality, science is a 
process based on testing various ideas, which often leads to dead ends and forces a complete change in our assumptions. 
It is thoroughly described in an excellent book by Andrzej K. Wróblewski \cite{Wroblewski:book}, which was the main source for writing 
this chapter. The book contains many details about the events, circumstances and people connected to fundamental 
discoveries in Physics, and it provides a curious reader with numerous references to historically the most important 
research works.

\FloatBarrier
\section{Notation and conventions}\label{sec:notation}
\FloatBarrier

\noindent Before we discuss the mathematical aspects of the particle Physics theory, we need to agree on the notations we will be using.

\noindent From now on, we are using natural units in which $c=\hbar=\varepsilon_0 = 1$.
Spacetime indices are indicated with greek letters, usually from the middle of the alphabet, e.g. $\mu$, $\nu$. Contravariant 
four-vectors, e.g. positions and momenta, are defined with raised indices, and covariant vectors, e.g. derivatives, with lowered 
indices:

\begin{eqnarray}
x^\mu = (t; \vec x), \\
p^\mu = (E; \vec p), \\
\partial_\mu \equiv \frac{\partial}{\partial x^\mu} = ( \partial_t ; \vec \nabla ).
\end{eqnarray}

\noindent Raising and lowering of indices is done with the Minkowski metric in the ``West Coast'' convention:

\begin{equation}
g_{\mu\nu} = g^{\mu\nu} = \rm{diag}\left(+1, -1, -1, -1\right)
\end{equation}

\noindent Einstein summation convention is used, e.g. mass of the particle with four-momentum $p^\mu$ is given by:

\begin{equation}
m = p^2 = p_\mu p^\mu
\end{equation}

\noindent Pauli matrices are defined as:
\begin{equation}
\sigma_1 = \sigma_x =  
\begin{pmatrix}
    0 & 1 \\
    1 & 0
\end{pmatrix}
,~~~
\sigma_2 = \sigma_y =  
\begin{pmatrix}
    0 &-i  \\
    i & 0
\end{pmatrix}
,~~~
\sigma_3 = \sigma_z =  
\begin{pmatrix}
    1 & 0 \\
    0 & -1
\end{pmatrix}.
\end{equation}

\noindent We define the Pauli vector, Pauli four-vector, and its barred version:
\begin{equation}\label{eq:pauli-vector}
\vec{\sigma}  \equiv \left(  \sigma_x, \sigma_y, \sigma_z \right),~~~
\sigma^\mu \equiv  \left( \mathbb{1}_2; \vec{\sigma} \right),~~~
\bar{\sigma}^\mu \equiv \left( \mathbb{1}_2; -\vec{\sigma} \right).
\end{equation}

\noindent  Dirac matrices are defined in the Weyl representation:
\begin{equation}\label{eq:gamma-matrices}
\gamma^\mu =
\begin{pmatrix}
    0 & \sigma^\mu \\
   \bar{ \sigma}^\mu  & 0
\end{pmatrix},
~~~
\gamma^5 =
\begin{pmatrix}
    -\mathbb{1}_2 & 0 \\
   0  & \mathbb{1}_2
\end{pmatrix}.
\end{equation}

\noindent Gell-Mann matrices are given by:
\begin{equation}\label{eq:gell-mann}
\begin{split}
\lambda_1 &= \begin{pmatrix}
0 & 1 & 0 \\
1 & 0 & 0 \\
0 & 0 & 0
\end{pmatrix},
~~
\lambda_2 = \begin{pmatrix}
0 & -i & 0 \\
i & 0 & 0 \\
0 & 0 & 0
\end{pmatrix},
~~
\lambda_3 = \begin{pmatrix}
1 & 0 & 0 \\
0 & -1 & 0 \\
0 & 0 & 0
\end{pmatrix},
~~
\lambda_4 = \begin{pmatrix}
0 & 0 & 1 \\
0 & 0 & 0 \\
1 & 0 & 0
\end{pmatrix},
\\
\lambda_5 &= \begin{pmatrix}
0 & 0 & -i \\
0 & 0 & 0 \\
i & 0 & 0
\end{pmatrix},
~~
\lambda_6 = \begin{pmatrix}
0 & 0 & 0 \\
0 & 0 & 1 \\
0 & 1 & 0
\end{pmatrix},
~~
\lambda_7 = \begin{pmatrix}
0 & 0 & 0 \\
0 & 0 & -i \\
0 & i & 0
\end{pmatrix},
~~
\lambda_8 = 
\frac{1}{\sqrt{3}}
\begin{pmatrix}
1 & 0 & 0 \\
0 & 1 & 0 \\
0 & 0 & -2
\end{pmatrix}.
\end{split}
\end{equation}

A four-component Dirac spinor is an object transforming according to the reducible representation of the Lorentz group: $(\frac{1}{2}, 0) \oplus (0, \frac{1}{2}) $.
Therefore, it can be expressed using 2 anticommuting two-component Weyl spinors (bispinors).
The left-handed spinor $\psi_\alpha$ transforms according to  $(\frac{1}{2}, 0)$ representation, while the right-handed spinor
$ \psi^\dag_{\dot \alpha}$ transforms as $(0,\frac{1}{2})$.
Two distinct types of spinor indices $\alpha=1,2$, and $\dot \alpha = 1, 2$ are used for left and right-handed spinors, respectively, to indicate that they cannot be directly contracted to form Lorentz invariant quantities. 
The $(\frac{1}{2}, 0)$ and $(0, \frac{1}{2}) $ representations are related by the hermitian conjugation, hence if $\psi_\alpha$ is a left-handed spinor, then $(\psi_\alpha)^\dag$ transforms as a right-handed spinor. It allows us to describe all fermionic degrees of freedom using only left-handed spinors and hermitian conjugates:
\begin{equation}
\psi_{\dot \alpha}^\dag \equiv  (\psi_{ \alpha})^\dag .
\end{equation}
There are two additional irreducible representations, which are dual to  $(\frac{1}{2}, 0)$ and $(0, \frac{1}{2})$. We denote spinors transforming under these representations with raised indices, $\psi^\alpha$ and $\psi^{\dag \dot \alpha}$. We have:
\begin{equation}
\psi^{\dag \dot \alpha} \equiv  (\psi^{ \alpha})^\dag .
\end{equation}
Raising and lowering of the spinor indices is done using:
\begin{equation}\label{eq:weyl-epsilon}
\epsilon^{\alpha \beta} = \epsilon^{\dot \alpha \dot \beta} = -\epsilon_{\alpha \beta} =
 -\epsilon_{\dot \alpha \dot \beta} =
 i \sigma_2 = 
\begin{pmatrix}
0 & 1 \\
-1 & 0
\end{pmatrix},
\end{equation}
according to:
\begin{equation}
\psi_\alpha = \epsilon_{\alpha \beta} \psi^\beta,
~~~
\psi^\alpha = \epsilon^{\alpha \beta} \psi_\beta,
~~~
 \psi_{\dot \alpha }^\dag = \epsilon_{\dot {\alpha} \dot{ \beta}}  \psi^{\dag  \dot \beta},
~~~
 \psi^{\dag \dot{\alpha} }= \epsilon^{\dot \alpha \dot \beta} \psi_{ \dot \beta}^\dag.
\end{equation}
One has to remember that $\epsilon_{\alpha \beta}$ and $\epsilon_{\dot \alpha \dot \beta}$ in Eq. \eqref{eq:weyl-epsilon} are different objects acting on different spinors and spaces, however, the matrix form is sometimes 
useful for practical calculations.
To be more explicit:
\begin{equation}
\begin{split}
\psi_1 &= \epsilon_{11} \psi^1 + \epsilon_{12} \psi^2 = - \psi^2\\
\psi_2 &= \epsilon_{21} \psi^1 + \epsilon_{22} \psi^2 =  \psi^1
\end{split}
\end{equation}
and
\begin{equation}
\begin{split}
\psi^{\dot 1} &= \epsilon^{\dot 1 \dot 1} \psi_{\dot 1} + \epsilon^{\dot 1 \dot2} \psi_{\dot 2} =  \psi_{\dot 2}\\
\psi^{\dot 2} &= \epsilon^{\dot 2\dot 1} \psi_{\dot 1} + \epsilon^{\dot 2\dot 2} \psi_{\dot 2} =  -\psi_{\dot 1}
\end{split}
\end{equation}
Similarly to spacetime indices, when two spinor indices are contracted, they will be often omitted, e.g. for two left-handed spinors $\xi$ and $\chi$:
\begin{equation}\label{eq:spinor-contraction}
\xi \chi = \xi^\alpha \chi_\alpha =  \xi^\alpha \epsilon_{\alpha \beta} \chi^\beta =
- \chi^\beta \epsilon_{\alpha \beta}  \xi^\alpha = \chi^\beta \epsilon_{\beta \alpha}  \xi^\alpha
= \chi^\beta \xi_\beta = \chi \xi,
\end{equation}
where the minus sign in the third passage appeared because the components of a Weyl fermion are Grassman numbers and anticommute.
Note that in Eq. \eqref{eq:spinor-contraction} we were able to switch the order of contracted spinor fields without an additional minus sign on the right-hand side, because of the antisymmetric property of the $\epsilon_{\alpha \beta}$.

In order to build Lorentz invariant Lagrangians, one has to construct Lorentz tensors by contracting all spinor indices. For example, multiplying two left- or right-handed spinors results in a scalar:
\begin{align}
\xi \eta &\equiv \xi^\alpha \eta_\alpha \\
\xi^\dag \eta^\dag &\equiv \xi_{\dot \alpha}^\dag \eta^{\dag \dot \alpha}.
\end{align}
Lorentz vectors can be constructed with the help of Pauli vectors from Eq. \eqref{eq:pauli-vector} in the following way:
\begin{align}
\xi^\dag \bar{\sigma}^\mu \eta &\equiv \xi^\dag_{\dot \alpha} \bar{\sigma}^{\mu \dot \alpha \beta} \eta_\beta \\
\xi \sigma^\mu \eta^\dag &\equiv \xi^{\alpha} \sigma^\mu_{\alpha \dot \beta} \eta^{\dag \dot \beta}.
\end{align}
Higher rank tensors can be obtained by multiplying $\sigma^\mu$ and $\bar \sigma^\nu$, for example:
\begin{equation}
\xi \sigma^\mu \bar{\sigma}^\nu \eta \equiv 
\xi^{\alpha} \sigma^\mu_{\alpha \dot \beta} 
\bar{\sigma}^{\nu \dot \beta \gamma} \eta_\gamma.
\end{equation}

\noindent In the introduced notation a Dirac spinor $\psi$ can be written as:
\begin{equation}\label{eq:dirac-spinor-general}
\Psi_D =
\begin{pmatrix}
\xi_\alpha \\  \chi^{ \dag \dot \alpha}
\end{pmatrix},
\end{equation}
where $\xi$ and $\chi$ are independent left-handed bispinors.
The adjoint (barred) Dirac spinor is:
\begin{equation}
\bar \Psi_D = \Psi_D^\dag \cdot \gamma^0 =
\begin{pmatrix}
 (\xi_{ \alpha})^\dag & ( \chi^{\dag \dot \alpha})^\dag 
\end{pmatrix}
\begin{pmatrix}
0 & \mathbb{1}_2 \\
\mathbb{1}_2 & 0
\end{pmatrix}
= 
\begin{pmatrix}
( \chi^{\dag \dot \alpha})^\dag & (\xi_{ \alpha})^\dag
\end{pmatrix}=
\begin{pmatrix}
\chi^{ \alpha} &   \xi_{ \dot \alpha}^\dag
\end{pmatrix}.
\end{equation}
One can introduce chirality operators:
\begin{equation}
P_L \equiv \frac{1-\gamma^5}{2}, ~~~P_R \equiv \frac{1+\gamma^5}{2},
\end{equation}
such that
\begin{equation}
P_L \Psi_D = 
\begin{pmatrix}
\xi_\alpha \\ 0
\end{pmatrix},
~~~
P_R\Psi_D = 
\begin{pmatrix}
0 \\ \chi^{\dag  \dot \alpha}
\end{pmatrix}.
\end{equation}
\noindent We define the charge conjugation of a Dirac spinor $\psi$ as:
\begin{equation}\label{eq:charge-conj}
\Psi^C = -i\gamma_2 \Psi^*.
\end{equation}
Majorana fermion is a special case of a Dirac fermion, which satisfies $\Psi^C = \Psi$. It follows that:
\begin{equation}\label{eq:charge-conjugate}
\Psi^C = 
\begin{pmatrix}
0 & \epsilon_{\alpha \beta}  \\
 \epsilon^{\dot \alpha \dot \beta} & 0
\end{pmatrix}
\begin{pmatrix}
\xi^\dag_{\dot \beta} \\   \chi^{ \beta}
\end{pmatrix}=
\begin{pmatrix}
\chi_{ \alpha} \\   \xi^{\dag  \dot \alpha}
\end{pmatrix}.
\end{equation}
From Eq. \eqref{eq:charge-conjugate} it follows that for Majorana fermion $\xi = \chi$, and it can be expressed as:
\begin{equation}\label{eq:dirac-majorana}
\Psi_{\rm Majorana} = \begin{pmatrix}
\xi_{ \alpha} \\   \xi^{\dag  \dot \alpha}
\end{pmatrix}
=\begin{pmatrix}
\xi_{ \alpha} \\  \left(\epsilon^{\alpha \beta} \xi_{ \beta}\right)^\dag
\end{pmatrix}.
\end{equation}
 $\left(\epsilon^{\alpha \beta} \xi_{ \beta}\right)^\dag$ in Eq. \eqref{eq:dirac-majorana} is built out of a left-handed Weyl spinor 
 $\xi_\beta$, but transforms as a right-handed Weyl spinor. A general Dirac spinor, like the one in Eq. \eqref{eq:dirac-spinor-general}, is built out of two independent bispinors and has four complex degrees of freedom. Majorana fermion has only two 
 complex degrees of freedom associated with a single Weyl spinor.
This allows us to write a Lorentz-invariant mass term for particles for which only the left-handed (or equivalently only the right-handed) field exists, e.g. right-handed neutrinos in the seesaw model.

Sometimes we will be using a slashed notation introduced by Feynman. In this notation:
\begin{align}\label{eq:slashed}
\bar \psi \slashed{\partial} \psi  &\equiv \bar \psi \gamma^\mu \partial_\mu \psi, \\
\xi^\dag \slashed{\partial} \chi  &\equiv \xi^\dag_{\dot \alpha}  \bar{\sigma}^{\mu \dot \alpha \beta} \partial_\mu \chi_\beta, \\
\xi \slashed{\partial}  \chi^\dag  &\equiv \xi^\alpha \sigma^\mu_{\alpha \dot \beta} \partial_\mu  \chi^{\dag \dot \beta},
\end{align}
where $\psi$ is the Dirac spinor, $\xi$ and $\chi$ are left-handed Weyl spinors.

We close this section with a comment regarding another popular notation for Weyl spinors. The notation introduced so far is 
useful when doing the actual computations, because all fields are defined using only the left-handed bispinors. Thanks to this, is 
it trivial to deduce the transformation properties of all objects, if they have a dagger, they transform as right-handed spinors, 
otherwise, they are left-handed. However, there is another popular convention in which a subscript $L$ or $R$ is used to indicate the chirality. This notation allows us to trace the evolution of left and right-handed particles in the theory. The best way to understand the difference between these two notations is by studying a simple example. Let us consider a free Dirac field, $\Psi$, and its Lagrangian:
\begin{equation}\label{eq:free-Dirac-lagrangian} 
\mathcal{L} = i\bar \Psi \slashed{\partial} \Psi - m \bar \Psi \Psi
\end{equation}
The equation Eq. \eqref{eq:free-Dirac-lagrangian} can be rewritten using Weyl spinors. Let us introduce two left-handed bispinors $e$ and $\tilde e$, such that $\Psi^T = \begin{pmatrix}
e_\alpha & \tilde{e}^{\dag \dot \alpha }\end{pmatrix}$. Then Eq. \eqref{eq:free-Dirac-lagrangian} takes the form:
\begin{equation}\label{eq:Dirac-notation1}
\mathcal{L} = i e^\dag \bar{\sigma}^\mu \partial_\mu e + i \tilde e \sigma^\mu \partial_\mu \tilde e^\dag - m 
\left(
\tilde e  e + e^\dag \tilde e^\dag
\right),
\end{equation}
where contracted spinor indices were omitted.
In the second notation, let us introduce a left-handed bispinor $e_L$ and a right-handed Weyl spinor $e_R$, such that
$\Psi^T = \begin{pmatrix}
(e_L)_\alpha & {e_R}^{\dot \alpha }\end{pmatrix}$. Then Eq. \eqref{eq:free-Dirac-lagrangian} takes the form:
\begin{equation}\label{eq:Dirac-notation2}
\mathcal{L} = i e_L^\dag \bar{\sigma}^\mu \partial_\mu e_L + i  e_R^\dag \sigma^\mu \partial_\mu  e_R - m 
\left(
e_R^\dag  e_L + e_L^\dag  e_R
\right).
\end{equation}
By comparing Eqs. \eqref{eq:Dirac-notation1} and \eqref{eq:Dirac-notation2} we can notice that the Eq. \eqref{eq:Dirac-notation2} has a simpler form because all daggered fields appear on the left-hand sides of the expressions. Moreover, it is easy to see that for $m\to 0$, the chiral symmetry is restored, i.e. $e_L$ and $e_R$ do not mix. On the other hand, in Eq. \eqref{eq:Dirac-notation2} there is $e_R^\dag$, which transforms as a left-handed spinor despite the $R$ subscript, which may be confusing. Transformation properties of all fields in Eq. \eqref{eq:Dirac-notation1} are crystal clear. The daggered notation is popular in modern supersymmetry textbooks, while the subscript notation is frequently used in SM books. In this thesis we will use both notations, but always consistently within a given section.

\FloatBarrier
\section{Standard Model}
\FloatBarrier

The following section contains a brief survey of the Standard Model of Particle Physics with the description of the field content, 
mathematical formulation, and a discussion of its shortcomings. A reader who is unfamiliar with the theory is advised to familiarize 
oneself with any 
of the comprehensive textbooks, e.g. \cite{Mandl:textbook, Peskin:textbook, Schwartz:textbook, Pokorski:1987ed}.

\FloatBarrier
\subsection{Mathematical description}
\FloatBarrier

\textit{Standard Model of Particle Physics (SM)} is the current paradigm of Elementary Particle Physics. It is a quantum field theory 
gauged under $\rm SU(3)_C \times SU(2)_L \times U(1)_Y$ group. Associated charges of the gauge groups are called: colour, weak 
isospin and weak hypercharge, respectively. Each of the gauge groups have massless vector bosons associated with it: $G^{a\mu}$ 
$(a\in [1,8])$ for $\rm SU(3)_C$, $W^{a \mu}$ $(a\in [1,3])$ for $\rm SU(2)_L$, and $B^\mu$ for $\rm U(1)_Y$. The fermion sector of 
the theory consists of particles named \textit{quarks} that belong to triplet representation of $SU(3)_C$, and \textit{leptons} that 
are $\rm SU(3)_C$ singlets. Left-handed fermions are grouped into two $\rm SU(2)_L$ doublets, there is a quark doublet 
containing \textit{up} and \textit{down} quarks, and a lepton doublet containing charged and neutral leptons, where the latter one is 
called \textit{neutrino}. Right-handed quarks and charged leptons transform as singlets with respect to $\rm SU(2)_L$. Right-
handed neutrinos are not part of the SM, however, they can be easily introduced in order to explain neutrino masses (see Sec. \ref{sec:seesaw}). In the 
fermionic sector of the SM,  there are three generations of matter, each containing two quarks and two leptons (and their antiparticles). 
Standard Model contains also a complex scalar doublet $H$ named the \textbf{Higgs field}, which is a $\rm SU(2)_L$ doublet, and 
leads to the Higgs mechanism after electroweak symmetry is spontaneously broken: $SU(2)_L \times U(1)_Y \to U(1)_{EM}$. All 
Standard Model fields are listed in Tab. \ref{tab:sm_fields}.

\begin{table}[h]
\caption{\small List of the Standard Model fields and their representations. We follow the notation in which the chirality of the fields is indicated by a subscript $L$ or $R$ for the left and right-handed bispinors, respectively. Index $j$ for the fermionic fields enumerates generations of matter, i.e. $j=1,2,3$. The convention relating the hypercharge $Y$ and the 3rd component of the isospin $T_3$ with the electric charge $Q$ (in the units of elementary charge $e$) of fields is the following: $Q = Y+T_3$.
}
\centering
\begin{tabular}{c c c c c }
\multicolumn{5}{c}{\textbf{Spin 1 -- gauge fields}}                                                                                                                                                                                                                                 \\ 
\hline
\multicolumn{1}{c|}{\textbf{Symbol}} & \multicolumn{1}{c|}{\textbf{Associated charge}} & \multicolumn{1}{c|}{\textbf{Group}}         & \multicolumn{1}{c|}{\textbf{Coupling}}      & \textbf{Representation} \\ 
\hline
\multicolumn{1}{c|}{\textit{B}}      & \multicolumn{1}{c|}{weak hypercharge}           & \multicolumn{1}{c|}{$\rm U(1)_Y$}                  & \multicolumn{1}{c|}{$g_1$}                     & $(1,1,0) $                \\ 
\multicolumn{1}{c|}{\textit{W}}      & \multicolumn{1}{c|}{weak isospin}               & \multicolumn{1}{c|}{$\rm SU(2)_L$}                 & \multicolumn{1}{c|}{$g_2$}                     & $(1,3,0)  $               \\ 
\multicolumn{1}{c|}{\textit{G}}      & \multicolumn{1}{c|}{colour}                     & \multicolumn{1}{c|}{$\rm SU(3)_C$}                 & \multicolumn{1}{c|}{$g_3$}                     & $(8,1,0)$                 \\ 
\\
\multicolumn{5}{c}{\textbf{Spin $\frac{1}{2}$ -- fermions}}                                                                                                                                                                                                                                   \\ \hline
\multicolumn{1}{c|}{\textbf{Symbol}} & \multicolumn{1}{c|}{\textbf{Name}}              
& \multicolumn{1}{c|}{\textbf{\begin{tabular}[c]{@{}c@{}}Baryon\\ number\end{tabular}}}
& \multicolumn{1}{c|}{\textbf{\begin{tabular}[c]{@{}c@{}}Lepton\\ number\end{tabular}}} & \textbf{Representation} \\
 \hline
\multicolumn{1}{c|}{$Q^j=(u_L^j~d_L^j)$}          & \multicolumn{1}{c|}{left-handed quark (doublet)}          & \multicolumn{1}{c|}{$\frac{1}{3}$}                    & \multicolumn{1}{c|}{0}                      & $(3,2,\frac{1}{6}$)               \\ 
\multicolumn{1}{c|}{$u^j_R$}           & \multicolumn{1}{c|}{right-handed up quark}         & \multicolumn{1}{c|}{$\frac{1}{3}$}                    & \multicolumn{1}{c|}{0}                      & $(3,1,\frac{2}{3}$)               \\
\multicolumn{1}{c|}{$d^j_R$}           & \multicolumn{1}{c|}{right-handed down quark}         & \multicolumn{1}{c|}{$\frac{1}{3} $}                   & \multicolumn{1}{c|}{0}                      & $(3,1,-\frac{1}{3}$)              \\
\multicolumn{1}{c|}{$L^j=(\nu^j_L~e^j_L)$}           & \multicolumn{1}{c|}{left-handed lepton (doublet)}         & \multicolumn{1}{c|}{0}                      & \multicolumn{1}{c|}{1}                      & $(1,2,-\frac{1}{2}) $               \\ 
\multicolumn{1}{c|}{$e^j_R$}           & \multicolumn{1}{c|}{right-handed charged lepton}        & \multicolumn{1}{c|}{0}                      & \multicolumn{1}{c|}{1}                      & $(1,1,-1) $           \\
\\
\multicolumn{5}{c}{\textbf{Spin 0 -- scalar boson}}                                                                                                                                                                                                                                 \\ 
\hline
\multicolumn{1}{c|}{\textbf{Symbol}} & \multicolumn{1}{c|}{\textbf{Name}}              & \multicolumn{3}{c}{\textbf{Representation}}                                                                                                \\ 
\hline
\multicolumn{1}{c|}{\textit{H}}      & \multicolumn{1}{c|}{Higgs boson (doublet)}                & \multicolumn{3}{c}{$(1,2,\frac{1}{2})$}                                                                                                                \\ 
\end{tabular}
\label{tab:sm_fields}
\end{table}

{
The Lagrangian of the Standard Model can be schematically written in the following way:
\begin{equation}\label{eq:sm-lagr}
\mathcal{L} = \mathcal{L}_{\rm {GAUGE}} + \mathcal{L}_{\rm{MATTER}} + \mathcal{L}_{H}+ \mathcal{L}_{\rm{YUKAWA}}+ \mathcal{L}_{\theta}
\end{equation}

The first term in Eq. \eqref{eq:sm-lagr} describes the kinematics of the gauge fields, and is given by\footnote{When performing explicit calculations, one should also include gauge-fixing and ghost terms, but since they are not crucial to understand the physical consequences of the Standard Model, we restrain from discussing them.}:
\begin{equation}\label{eq:sm-gauge}
\mathcal{L}_{\rm{GAUGE}} = 
- \frac{1}{4} B_{\mu\nu} B^{\mu\nu} 
-\frac{1}{4} W_{\mu\nu}^a W^{a\mu\nu}  
-\frac{1}{4} G_{\mu\nu}^a G^{a\mu\nu}.
\end{equation}
$B$, $W$ and $G$  are gauge fields of $U(1)_Y$, $SU(2)_L$ and $SU(3)_C$ gauge groups, respectively.
Group indices $a$ in Eq. \eqref{eq:sm-gauge} are contracted and correspond to generators of the groups. $U(1)_Y$ has only one generator, hence the index $a$ is ommited.
Gauge fields enter the Eq. \eqref{eq:sm-gauge} in a form of field strength tensors, defined as:
\begin{equation}
F_{\mu\nu}^a = \partial_\mu A_\nu^a - \partial_\nu A_\mu^a - g f^{abc} A^b_\mu A^c_\nu,
\end{equation}
where $F_{\mu\nu}^a$ is the field strength tensor corresponding to the gauge field potential $A_\mu$ of a given Yang-Mills 
theory of $SU(N)$ gauge group, $g$ is the (running) coupling constant, and $f^{abc}$ are structure constants of the appropriate 
Lie algebra. In case of abelian groups, like $\rm U(1)$, we have $f^{abc} \equiv 0$.

The second term in Eq. \eqref{eq:sm-lagr} describes interactions between gauge bosons and fermions and is given by:
\begin{equation}\label{eq:sm-kin}
\begin{split}
\mathcal{L}_{\rm{MATTER}} =\sum_{j=1}^3 
&i  L^{j \dag}
 \left(  \slashed{\partial} - ig_1  Y_{L}  \slashed{B} - ig_2\tau^a \slashed{W}^a \right) L^j
+ i e_R^{j\dag}  \left( \slashed{\partial} - i g_1  Y_{e}  \slashed{B}  \right) e_R^j
+
\\
+
&i  Q^{j\dag}
 \left( \slashed{\partial} - ig_1  Y_{Q}  \slashed{B} - i g_2\tau_a \slashed{W}^a 
 - i g_3 \slashed{G}^a T^a  
 \right) 
Q^j
+
\\
+ &i u_R^{j\dag} \left(\slashed{\partial} - i g_1  Y_{u}  \slashed{B}  - i g_3 \slashed{G}^a T^a  \right) u_R^{j}
+ i d_R^{j\dag}  \left( \slashed{\partial} - i g_1  Y_{d}  \slashed{B}  - i g_3 \slashed{G}^a T^a  \right) d_R^{j}
.
\end{split}
\end{equation}
The summation in Eq. \eqref{eq:sm-kin} runs over three generations of 
fermions. $Y_{\chi}$ stands for the weak 
hypercharge associated with the fermion field $\chi$. $\tau^a = \frac{1}{2}\sigma^a$ are generators of the $\rm SU(2)_L$ group. 
Terms in brackets are covariant derivatives. Notice the absence of right-handed neutrinos and the fact that the right-handed charged 
fermions do not couple to the $W^{a\mu}$ gauge boson.
$T^a=\frac{1}{2}\lambda^a$ are generators of the $\rm SU(3)_C$ group, and $\lambda^a$ are Gell-Mann matrices defined in Eq. \eqref{eq:gell-mann}.

The next term in Eq. \eqref{eq:sm-lagr} stands for the Higgs sector. It consists of a standard kinetic term and the potential given by Eq. \eqref{eq:higgs-potential}:
{
\begin{equation}\label{eq:higgs-L}
\mathcal{L}_H =\left|\left( \partial_\mu - i g_1 Y_H B_\mu - ig_2 W^a_\mu \tau_a\right)H\right|^2 - V\left(H\right)
\end{equation}
}
\begin{equation}\label{eq:higgs-potential}
V\left(H\right) =  -\mu^2 H^\dag H + \lambda \left( H^\dag H \right)^2
\end{equation}

The fourth term in Eq. \eqref{eq:sm-lagr} stands for the Yukawa interaction, which is the interaction between fermions and the Higgs field. 
\begin{equation}
\mathcal{L}_{\rm{YUKAWA}} = 
- Y_{ij}^d  Q^{i\dag} H d^{j}_R
- Y_{ij}^u  Q^{i\dag} \left( i \sigma_2 H^*\right) u^{ j}_R 
- Y_{ij}^e  L^{i\dag} H e_R^{ j} + h.c.
\end{equation}
$Y_{ij}$ are \textit{Yukawa matrices (couplings)}, where indices $i$ and $j$ label three families of matter, and implicit summation convention is used. For each generation, 
there are 3 Yukawa matrices: for up-type quarks, for down-type quarks and charged leptons. After spontaneous symmetry 
breaking fermions acquire masses proportional to the Higgs vacuum expectation value (vev) and Yukawa couplings. Since in the 
original version of the SM neutrinos are massless, they do not have Yukawa terms.

The last term in Eq. \eqref{eq:sm-lagr} is called the ``$\theta$-term'', and is given by:
\begin{equation}\label{eq:theta-term}
\mathcal{L}_\theta = \theta_{\rm QCD}\frac{g_3^2}{32 \pi^2} \varepsilon^{\mu\nu\alpha\beta}
G^a_{\mu\nu} G^a_{\alpha \beta},
\end{equation}
where $ \varepsilon^{\mu\nu\alpha\beta}$ is totally antisymmetric Levi-Civita symbol, $G$ is the $SU(3)_C$ gauge field strength tensor, 
and $\theta_{\rm QCD}$ is a parameter named \textit{QCD vacuum angle}. QCD has a topologically non-trivial vacuum, and $\theta_{\rm QCD}$ defines the choice of the vacuum among an infinity of distinct and generally inequivalent vacua.
The term in Eq. \eqref{eq:theta-term} is different from previously discussed parts of the SM Lagrangian in Eq. \eqref{eq:sm-lagr}, 
because it can be shown that it is a total derivative, $\mathcal{L}_\theta = \partial_\mu K^\mu$,
hence it does not contribute in the 
perturbation expansion of the theory. However, $K$ is not gauge invariant and the operator in Eq. \eqref{eq:theta-term} has 
severe non-perturbative effects, most notably it leads to significant CP-violation and gives rise to the neutron electric dipole moment proportional to the \textit{strong CP phase} $\bar \theta \equiv \theta - \arg\det (Y_d Y_u)$\footnote{
The $\arg\det (Y_d Y_u)$ term arises due to diagonalising the mass matrix of quarks.}
. $\bar \theta$ can take any value between $0$ and $2\pi$, but the experimental bound is very tight $\bar \theta < 10^{-10}$. 
It is a mystery, called \textit{the strong CP problem}, why the value of $\bar \theta$ and the resulting CP violation is so small, despite the sizeable CP violation in the weak sector.
One approach to solve it is by introducing massless pseudo-Nambu-Goldstone bosons named \textit{axions}\cite{Peccei:1977hh}.

A crucial aspect of understanding the Standard Model of Particle Physics is the breaking of electroweak symmetry. $\rm SU(2)_L \times U(1)_Y$ is 
\textit{spontaneously broken} to electromagnetism $U(1)_{EM}$ by Higgs field acquiring non-zero vacuum expectation value. It is 
possible when the potential given by the Eq. \eqref{eq:higgs-potential} is characterised by {$\mu>0$} and $\lambda > 0$. 
Such potential has a shape of a ``Mexican hat'', see Fig. \ref{fig:higgs-pot}, with a local unstable maximum at $H^\dag H= 0$, 
and the minimum at: 
\begin{equation}\label{eq:higgs-vev}
 H^\dag H =  \frac{ \mu^2 }{2 \lambda}
\end{equation}

\begin{figure}[bht]
\center
\includegraphics[width=0.7\textwidth]{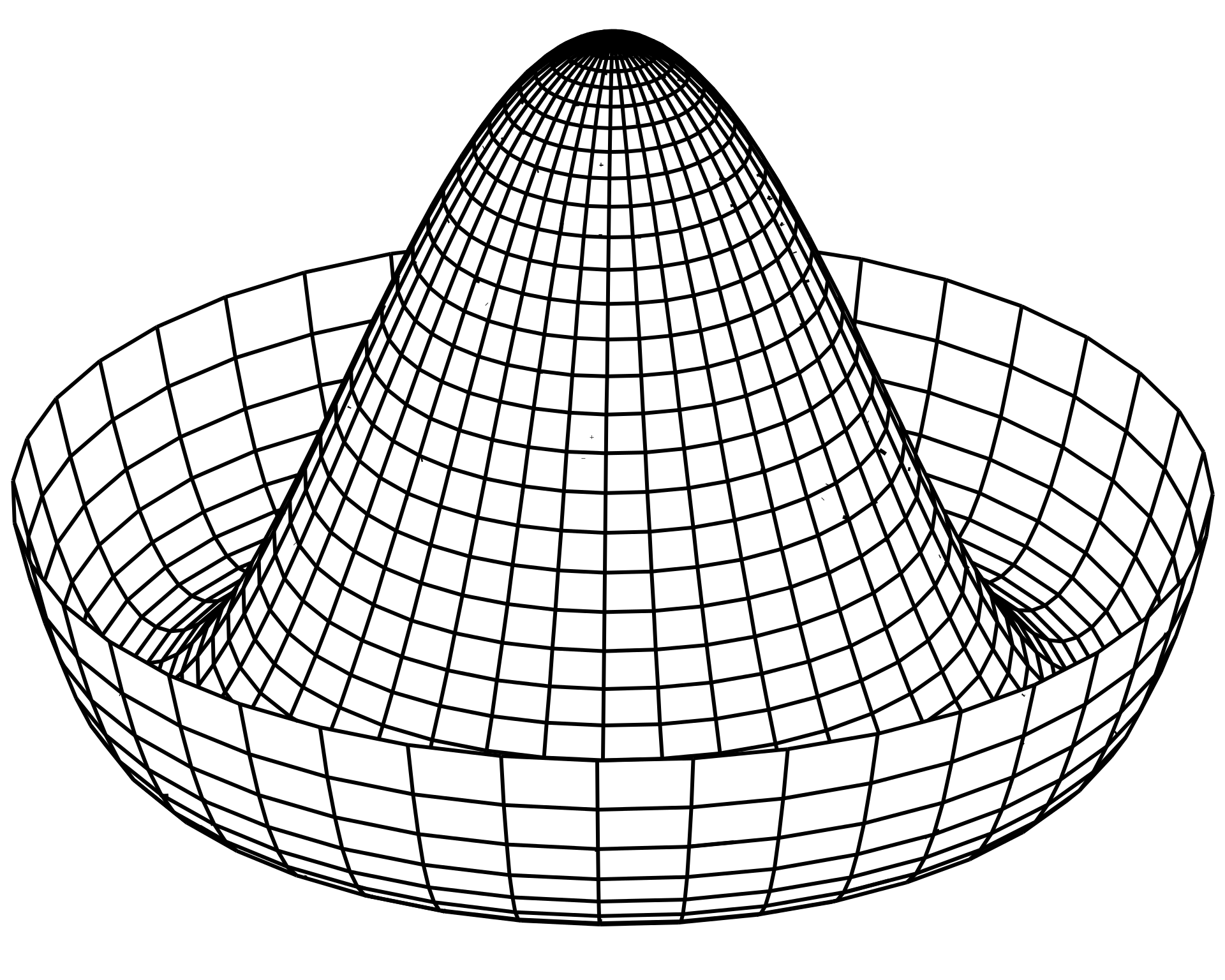}
\caption{\small A 3D visualisation of the Higgs potential in Eq. \eqref{eq:higgs-potential}. Image taken from \cite{higgs_pot}.}
\label{fig:higgs-pot}
\end{figure}

\noindent From Eq. \eqref{eq:higgs-vev} it follows that:
\begin{equation}\label{eq:higgs-vev2}
|H| =   \frac{ \mu }{\sqrt{2 \lambda}} \equiv \frac{v}{\sqrt{2}},
\end{equation}
where we have introduced $v \equiv = \mu/\sqrt{\lambda}$, which was measured to be $v \approx 246~\rm GeV$. We have
\begin{equation}
|H| = \sqrt{H^\dag H} = H_1 H_1^* +  H_2 H_2^* = \frac{v}{\sqrt{2}},
\end{equation}
for $H_i,~i\in \{1,2\}$ being the complex components of the Higgs doublet. The vacuum expectation value of the Higgs field should have a single real value satisfying Eq. \eqref{eq:higgs-vev2}. There are infinite ways to do it, each related by a $SU(2)_L \times U(1)_Y$ rotation. The typical choice is to select:
\begin{equation}\label{eq:higgs-vev3}
\langle H \rangle = \begin{pmatrix}
0 \\ \frac{v}{\sqrt{2}}
\end{pmatrix}.
\end{equation}
The potential in Eq. \eqref{eq:higgs-potential} is symmetric under $\rm SU(2)_L$, but the ground state in Eq. \eqref{eq:higgs-vev3} is not, hence we talk 
about symmetry breaking. It is spontaneous because at the level of Lagrangian the symmetry is intact, it is the 
vacuum expectation value that breaks it.

The complex scalar Higgs doublet has four degrees of freedom, but after the electroweak symmetry breaking (EWSB) three of 
them give mass to electroweak gauge bosons, and the remaining one can be associated with a single real scalar field $h$. It can 
be seen by expanding the Higgs field around vev: 

\begin{equation}\label{eq:higgs-pert}
H = 
\frac{1}{\sqrt{2}}
\exp{\left(2 i \frac{\pi_a \tau^a}{v} \right)}
\begin{pmatrix}
0 \\
v+h
\end{pmatrix},
\end{equation}
where $\tau^a = \frac{1}{2} \sigma^a$ are generators of the $\rm SU(2)_L$, and $h$ is a perturbation around the vev. 
It is convenient to work in the unitary gauge for which $\pi_a = 0$.

If we insert Eq. \eqref{eq:higgs-pert} into Eq. \eqref{eq:higgs-L}, after simple algebraic transformations we will find out mass terms for the physical Higgs field $h$ and gauge bosons, from which we can read their tree-level masses:
\begin{align}
m_h &=  v \sqrt{2 \lambda}, \label{eq:mh}\\
m_W &= g_2 \frac{v}{2},\label{eq:mw}\\
m_Z &=  \frac{v}{2} \sqrt{g_1^2 + g_2^2}. \label{eq:mz}
\end{align}
where $m_h$, $m_W$ and $m_Z$ are masses of Higgs, $W^\pm$ and $Z^0$ boson, respectively.

The massive W boson field is defined as:
\begin{equation}
W^{\pm,\mu} \equiv  \frac{1}{\sqrt{2}}\left( W_1^\mu \mp i W_2^\mu \right) ,
\end{equation}
while
the massive Z boson field is defined as:
\begin{equation}
Z^\mu \equiv \cos \theta_W W_3^\mu - \sin \theta_W B^\mu,
\end{equation}
where the \textit{weak mixing angle} is defined as:
\begin{equation}\label{eq:tanW}
\tan \theta_W = \frac{g_1}{g_2}.
\end{equation}
The remaining electromagnetic gauge boson, photon, is massless and defined as:
\begin{equation}
A^\mu\equiv \sin \theta_W W_3^\mu + \cos \theta_W B^\mu.
\end{equation}

By measuring the electron magnetic dipole moment $g_e$ one can determine the fine structure constant $\alpha$, or equivalently the elementary charge $e$, since $\alpha \equiv \frac{e^2}{4\pi}$. One finds that:
\begin{equation}
 e\approx 0.303.
 \end{equation}
Next, by a precise measurement of the muon's lifetime, one can estimate the value of the Fermi constant, $G_F$, which is related to the Higgs vacuum expectation value $v$ through the relation $G_F = (\sqrt{2} v^2)^{-1}$. This gives us:
\begin{equation}\label{eq:vev}
v \approx 246.22~\rm{GeV}.
\end{equation}
For the third parameter we can take the pole mass of the Z boson, which is measured to be:
\begin{equation}
m_Z \approx 91.19~\rm{GeV}.
\end{equation}
Knowing $m_Z$, $v$, and $\alpha$ evaluated at the scale of $m_Z$\footnote{The discussion in this paragraph is simplified and omits some important but technical details. A more detailed explanation can be found in one of the popular SM textbooks, e.g. in Sec. 31.1 of Ref. \cite{Schwartz:textbook}}, one can get $\sin \theta_W$ from the following equation:
\begin{equation}
\sin^2 \theta_W \left(1 - \sin^2 \theta_W \right) = \pi \frac{\alpha(m_Z)}{m_Z^2}v^2.
\end{equation}
Plugging the numbers and using $\sin\theta_W \approx 1 - \frac{m_W^2}{m_Z^2}$ to determine the correct root, one can find:
\begin{equation}\label{eq:sinW}
\sin^2 \theta_W \approx 0.234.
\end{equation}
Using the value of $sin^2\theta_W$ from Eq. \eqref{eq:sinW} one arrives at:
\begin{align}
g_1 &= \frac{e}{\sin\theta_W} \approx 0.63, \\
g_2 &=\frac{e}{\cos\theta_W} \approx 0.35.
\end{align}
The tree-level pole mass of the W boson might be calculated from Eq. \eqref{eq:mw}:
\begin{equation}\label{eq:mw-calc}
m_W \approx 77~\rm{GeV}.
\end{equation}
The predicted value in \ref{eq:mw-calc} is much below the measured one:
\begin{equation}
m_W^{\rm measured} \approx 80.4~\rm{GeV}.
\end{equation}
Even if we perform our calculations more precisely, we will get the mass of the W boson significantly lower than the measured value ($m_W \approx 79~\rm{GeV})$. The discrepancy can be resolved by including higher-order corrections and carefully renormalising.

The mass of the Higgs boson was measured to be:
\begin{equation}
m_h^{\rm measured}  = 125.25~\rm{GeV}.
\end{equation}
Using Eqs. \eqref{eq:mh} and \eqref{eq:vev}, one may estimate the $\lambda$ coupling from the Higgs potential in Eq. \eqref{eq:higgs-potential} to be:
\begin{equation}
\lambda \approx 0.13.
\end{equation}

The Higgs mechanism gives also rise to fermion masses. Consider Yukawa term for the down-type quark:
\begin{equation}\label{eq:fermion-mass}
-  Y_{ij}^d  (Q^i)^\dag H d_R^{ j}   \to - \frac{v}{\sqrt{2}} Y_{ij}^d  (d^i_L)^\dag  d_R^{ j}.
\end{equation}
Yukawa matrices are not diagonal, therefore one needs to perform a change of basis for quarks in order to read the physical 
masses. After going to the \textit{mass basis}, interactions between quarks and gauge bosons in Eq. \eqref{eq:sm-kin} will change. In 
the original \textit{flavour basis} there were no interactions between up- and down-type quarks, but after the basis change, an 
interaction involving the physical $W^\pm$ boson leads to flavour mixing:
\begin{equation}\label{eq:charged-curr}
\mathcal{L}_{\rm{MATTER}}^{\small \rm mass~basis} \supset \frac{e}{\sqrt{2} \sin{\theta_W}}
\left[
W^+_\mu  (u^i_L)^\dag 
\bar \sigma^\mu V^{ij} d^j_L
+
W^-_\mu  (d^i_L)^\dag 
\bar \sigma^\mu \left( V^\dag \right)^{ij}  u^j_L
\right].
\end{equation}
The matrix $V$ in \eqref{eq:charged-curr} is called \textit{Cabibbo-Kobayashi-Maskawa matrix (CKM)}, and is schematically given by:
\begin{equation}
V = 
\begin{pmatrix}
 V_{ud} & V_{us} & V_{ub} \\
  V_{cd} & V_{cs} & V_{cb} \\
   V_{td} & V_{ts} & V_{tb} 
   \end{pmatrix}
\end{equation}
The CKM is a complex unitary matrix with nine real degrees of freedom. If CKM was real, it would have only three degrees of 
freedom -- three rotation angles. One can therefore say that the real degrees of freedom of the CKM matrix are three angles $
\theta_{12},\theta_{23},\theta_{31}$ representing rotations in the three ``flavour planes'', and six phases. Residual $U(1)^6$ symmetry (see Ref. \cite{Schwartz:textbook})
can be utilised to set 5 out of 6 phases to zero, leaving only one phase $\delta$. With this convenient parametrisation the CKM matrix can be  written in the following form:
\begin{equation}\label{eq:sm-ckm-parametrisation}
V = \begin{pmatrix}
c_{12}c_{13}
& s_{12}c_{13}
& s_{13}e^{-i\delta} \\
-s_{12}c_{23} - c_{12}s_{23} s_{13} e^{i\delta}
& c_{12} c_{23} - s_{12} s_{23} s_{13} e^{i\delta} 
& s_{23}c_{13} \\
s_{12} s_{23} - c_{12} c_{23} s_{13} e^{i\delta} 
& -c_{12} s_{23} - s_{12} c_{23} s_{13} e^{i\delta} 
& c_{23} c_{13}
\end{pmatrix},
\end{equation}
with $s_{ij} \equiv \sin \theta_{ij}$ and $c_{ij} \equiv \cos \theta_{ij}$.
If there were only two generations of quarks, then the CKM matrix could be taken real, and there would be no CP violation. Historically, CP violation in the kaon system was observed when only $u, d, s$ quarks were known, and the third generation was predicted as necessary to explain this phenomenon.

In the lepton sector, if one takes into account that neutrinos are massive, an analogous matrix called \textit{Pontecorvo-Maki-Nakagawa-Sakata (PMNS) matrix} (see Eq. \eqref{eq:seesaw-pmns}) might be introduced, which explains flavour oscillations of the left-handed neutrinos.

All physical particles of the Standard Model are shown in Fig. \ref{fig:sm}, with their symbols, masses, electric charges and spins.

\begin{figure}[hbt]
\center
\includegraphics[width=1\textwidth]{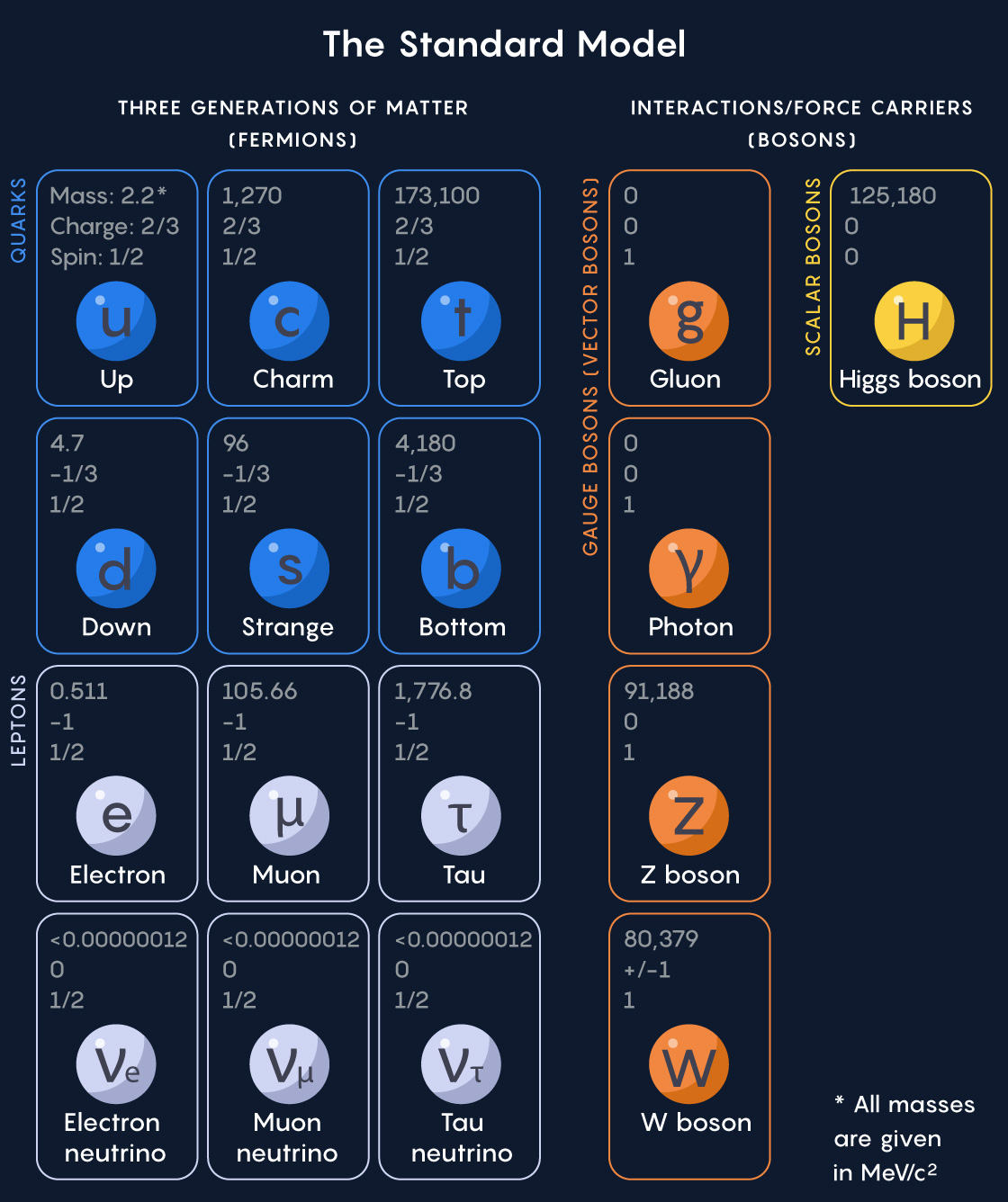}
\caption{\small Particle content of the Standard Model of Particle Physics. Based on image from \cite{smfig}.}
\label{fig:sm}
\end{figure}

\subsection{Why do we have to go beyond the Standard Model?}

Without any doubt, the Standard Model of Particle Physics is a beautiful and tremendously successful theory. It describes three out 
of four fundamental interactions, unifies electromagnetism with the weak force, provides a self-consistent explanation of the 
origin of masses for both fermions and gauge bosons, allows calculating cross sections and decay widths with unprecedented precision, 
and it allowed to predict the existence of many particles long before high-energy experiments were able to observe them. 
However, Standard Model has certain issues, which might be classified into two distinct groups, and various solutions to these 
problems have been proposed, which are generally referred to as \textit{Beyond the Standard Model (BSM)} Physics scenarios.

The first type of problems with the Standard Model is of a theoretical nature. There are four known fundamental interactions: 
electromagnetism, weak force, strong force and gravity. The latter is not included in the Standard Model and it is still described 
by Einstein's General Relativity. Hence, Standard Model is not a complete description of fundamental physics and a more 
sophisticated theory has to be found. It directly leads to another issue called \textit{the hierarchy problem}. The Standard Model Higgs particle is a scalar, and it acquires 
large loop corrections to its mass, scaling as $\mathcal{O}(\Lambda^2)$, where $\Lambda$ is the mass scale of the heaviest 
particle in the theory. For the Standard Model, the top quark is the heaviest with $m_t \approx 173$ GeV, however, it is reasonable to 
assume that there are heavier new particles somewhere between the electroweak scale $\Lambda_{EW} \sim 100\rm ~GeV$ and 
the Planck scale $\Lambda_{\rm Planck} \sim 10^{19}~ \rm GeV$. The existence of a such heavy particle would result in an elevation 
of the Higgs boson mass to a much higher value than the observed $m_h \approx 125~\rm GeV$ unless some protection 
mechanism is involved. One idea, discussed in Sect. \ref{sec:susy}, is supersymmetry \cite{Martin:1997ns, Terning:2006bq, Wess:1992cp}, which provides 
exact cancellation of quadratic divergences in all orders of the perturbative expansion. The hierarchy problem can also be resolved if 
the Higgs is not a fundamental scalar particle, for example in the Composite Higgs models \cite{Kaplan:1983fs, Dugan:1984hq}. Another, more 
exotic solution to the hierarchy problem is \textit{cosmological relaxation mechanism} \cite{Graham:2015cka}.
Furthermore, Standard Model is not a satisfactory theory for modern Cosmology, since it does not include an inflaton field, which 
is believed to be driving inflation, which is an exponential expansion of the early Universe. Inflation helps to solve several major 
issues of the Big Bang cosmology, e.g. flatness and horizon problems.
Finally, the critics of the Standard Model raise the \textit{inelegance} argument, claiming that there is a large number of a priori 
unconstrained parameters in the theory. It is true, Standard Model has 19 free parameters: masses of quarks (6), charged leptons 
(3) and the Higgs boson (1); vacuum expectation value (1); CKM mixing parameters and a CP violation phase (3+1); couplings of 
gauge groups (3); and the QCD vacuum angle (1). All SM parameters with their values are listed in Tab. \ref{tab:SMparams}.

The second group of issues comes from experimental observations that cannot be explained within the Standard Model. The most 
obvious discrepancy is the existence of massive neutrinos \cite{Heeger:2004mp}, which are absent in the original version of the 
theory, however, they are quite easy to be introduced, as explained in Sect. \ref{sec:neutrino-mass}.
Another problem is the need to explain the abundance of electrically neutral matter populating the Universe, as suggested by 
astronomical observations. This matter, called \textit{Dark Matter}, cannot be any of the Standard Model particles, hence new 
models are needed. There is a plethora of theoretical scenarios, e.g. WIMPs \cite{Pospelov:2007mp}, SIMPs 
\cite{Hochberg:2014dra}, axion-like particles \cite{Marsh:2015xka}, primordial Black Holes \cite{Carr:2016drx}. 
Finally, some experimental results are in tension with precise SM calculations. In 2021, the Muon g-2 experiment at 
the Fermilab published new results \cite{Muong-2:2021ojo} of a measurement of an anomalous magnetic moment of a muon, 
which lead to a 4.2$\sigma$ discrepancy\footnote{At the same time as the new experimental result appeared, the BMW lattice 
collaboration published SM predictions for $(g-2)_\mu$ acquired using lattice calculations, and their result, later confirmed by 
other lattice groups \cite{Borsanyi:2020mff}, corresponded to only $1.5\sigma$ discrepancy.}  between the global experimental average and the SM prediction based on data-driven 
approach \cite{Aoyama:2020ynm}. 
Another source of anomalies appears in the flavour sector. In the Standard Model, there is an accidental approximate symmetry, 
\textit{lepton flavour universality (LFU)}, 
broken only by the Yukawa interactions. The consequence of LFU is that electroweak gauge bosons couple to all three generations 
of leptons with the same strength. A deviation from this property is an indication of the BSM Physics, and can be tested by 
studying semileptonic decays of heavy hadrons, in which quark flavour is changed, e.g. $b \to c  l \nu_l$ via $B \to D^{(*)} l \nu_l
$, or 
$b \to s  l l$ through $B \to K^{(*)} l l$. Since flavour-changing processes of quarks are mediated only by the charged $W^\pm$ boson,  
transitions between quarks of the same type occur through tree-level diagrams, but between up-type and down-type quarks only 
through loop-suppressed diagrams.  
In the Standard Model, the dynamics of such processes are predicted with a high level of accuracy, and several anomalies with 
$2-3\sigma$ significance have been reported (c.f. \cite{Graverini:2018riw} and references therein) in the recent years
\footnote{A very recent result by the LHCb \cite{LHCb:2022zom} suggests that anomalies in $B$ decaying to kaons are not there.}. 
Finally, very recently a new measurement of $W^\pm$ boson mass was published \cite{CDF:2022hxs}, reporting a $7\sigma$ 
discrepancy with the SM prediction.

The Standard Model of Particle Physics is still the main tool physicists
use to describe the subatomic world, however, more and more cracks appear on its surface indicating deep problems hidden in 
the depths of the theory. The existence of BSM Physics is obvious, and its discovery is the biggest challenge the Particle 
Physics community will face in the upcoming years.

\begin{table}[htb]
\caption{\small Free parameters of the Standard Model of Particle Physics. Values of the parameters were taken from \cite{ParticleDataGroup:2022pth}. Parameter values with very low uncertainty compared to the central value, e.g. electron mass $m_e=0.510 998 950 00(15) MeV$, are presented in the table without the uncertainty for clarity. The choice of the free 
parameters is not unique, for example, one can exchange weak mixing angle and fine structure constant for $g_1$ and $g_2$ 
coupling constants.}\label{tab:SMparams}
\begin{tabular}{ccccc}
\multicolumn{5}{c}{\textbf{\Large Parameters of the Standard Model}}                                                                                                                                                                                                              \\ \hline
\multicolumn{1}{c|}{\textbf{\#}} & \multicolumn{1}{c|}{\textbf{Symbol}}    & \multicolumn{1}{c|}{\textbf{Description}}           & \multicolumn{1}{c|}{\textbf{\begin{tabular}[c]{@{}c@{}}Renormalization\\scheme\\ (point)\end{tabular}}} & \textbf{Value}                  \\ \hline
\multicolumn{1}{c|}{1}           & \multicolumn{1}{c|}{$m_e$}              & \multicolumn{1}{c|}{Electron mass}                  & \multicolumn{1}{c|}{}                                                                                  & $0.511~\rm MeV$                 \\ 
\multicolumn{1}{c|}{2}           & \multicolumn{1}{c|}{$m_\mu$}            & \multicolumn{1}{c|}{Muon mass}                      & \multicolumn{1}{c|}{}                                                                                  & $105.7~\rm MeV$                 \\ 
\multicolumn{1}{c|}{3}           & \multicolumn{1}{c|}{$m_\tau$}           & \multicolumn{1}{c|}{Tau mass}                       & \multicolumn{1}{c|}{}                                                                                  & $1776.86 \pm 0.12~\rm MeV$      \\
\multicolumn{1}{c|}{4}           & \multicolumn{1}{c|}{$m_u$}              & \multicolumn{1}{c|}{Up quark mass}                  & \multicolumn{1}{c|}{$\mu_{\overline {MS}}=2~\rm GeV$}                                                  & $2.16^{+0.49}_{-0.26}~\rm MeV$  \\ 
\multicolumn{1}{c|}{5}           & \multicolumn{1}{c|}{$m_d$}              & \multicolumn{1}{c|}{Down quark mass}                & \multicolumn{1}{c|}{$\mu_{\overline {MS}}=2~\rm GeV$}                                                  & $4.67^{+0.48}_{-0.17}~\rm MeV$  \\ 
\multicolumn{1}{c|}{6}           & \multicolumn{1}{c|}{$m_s$}              & \multicolumn{1}{c|}{Strange quark mass}             & \multicolumn{1}{c|}{$\mu_{\overline {MS}}=2~\rm GeV$}                                                  & $93.4^{+8.6}_{-3.4}~\rm MeV$    \\ 
\multicolumn{1}{c|}{7}           & \multicolumn{1}{c|}{$m_c$}              & \multicolumn{1}{c|}{Charm quark mass}               & \multicolumn{1}{c|}{$\mu_{\overline {MS}}=m_c$}                                                        & $1.27\pm 0.02~\rm GeV$          \\ 
\multicolumn{1}{c|}{8}           & \multicolumn{1}{c|}{$m_b$}              & \multicolumn{1}{c|}{Bottom quark mass}              & \multicolumn{1}{c|}{$\mu_{\overline {MS}}=m_b$}                                                        & $4.18^{+0.03}_{-0.02}~\rm GeV$  \\ 
\multicolumn{1}{c|}{9}           & \multicolumn{1}{c|}{$m_t$}              & \multicolumn{1}{c|}{Top quark mass}                 & \multicolumn{1}{c|}{direct measurement}                                                                & $172.69\pm 0.30~\rm GeV$        \\ 
\multicolumn{1}{c|}{10}          & \multicolumn{1}{c|}{$\sin \theta_{12}$} & \multicolumn{1}{c|}{CKM 12 mixing angle}            & \multicolumn{1}{c|}{}                                                                                  & $0.22500 \pm 0.00067$           \\ 
\multicolumn{1}{c|}{11}          & \multicolumn{1}{c|}{$\sin \theta_{23}$} & \multicolumn{1}{c|}{CKM 23 mixing angle}            & \multicolumn{1}{c|}{}                                                                                  & $0.04182^{+0.00085}_{-0.00074}$ \\
\multicolumn{1}{c|}{12}          & \multicolumn{1}{c|}{$\sin \theta_{13}$} & \multicolumn{1}{c|}{CKM 13 mixing angle}            & \multicolumn{1}{c|}{}                                                                                  & $0.00369 \pm 0.00011$           \\ 
\multicolumn{1}{c|}{13}          & \multicolumn{1}{c|}{$\delta$}           & \multicolumn{1}{c|}{CKM CP violation phase}         & \multicolumn{1}{c|}{}                                                                                  & $1.144 \pm 0.027$               \\ 
\multicolumn{1}{c|}{14}          & \multicolumn{1}{c|}{$\alpha^{-1}$}         & \multicolumn{1}{c|}{fine structure constant}            & \multicolumn{1}{c|}{low energy limit}                                                        &  137.03599                        \\ 
\multicolumn{1}{c|}{15}          & \multicolumn{1}{c|}{$\sin^2 \theta_W$}         & \multicolumn{1}{c|}{weak mixing angle}           & \multicolumn{1}{c|}{$on-shell$}                                                        &                                $0.22339 \pm 0.00010$
 \\ 
\multicolumn{1}{c|}{16}          & \multicolumn{1}{c|}{$\alpha_S$}         &
 \multicolumn{1}{c|}{strong coupling constant}           & \multicolumn{1}{c|}{$\mu_{\overline {MS}}=m_Z$}                                                        & $0.1179 \pm 0.0009$             \\ 
\multicolumn{1}{c|}{17}          & \multicolumn{1}{c|}{$\theta_{QCD}$}     & \multicolumn{1}{c|}{QCD vacuum angle}               & 
 \multicolumn{1}{c|}{}                                     &               $\leq 10^{-10}$                  \\ 
\multicolumn{1}{c|}{18}          & \multicolumn{1}{c|}{$v$}                & \multicolumn{1}{c|}{Higgs vev} & \multicolumn{1}{c|}{}                                                                                  & $246.22~\rm GeV$                   \\ 
\multicolumn{1}{c|}{19}          & \multicolumn{1}{c|}{$m_H$}              & \multicolumn{1}{c|}{Higgs mass}                     & \multicolumn{1}{c|}{}                                                                                  & $125.25 \pm 0.17~\rm GeV$       \\ 
\end{tabular}
\vspace{1em}
\end{table}

\section{Large Hadron Collider and BSM Physics searches}

\textit{Large Hadron Collider (LHC)} at the \textit{European Centre for Nuclear Research (CERN)} in the outskirts of Geneva is the 
biggest and most powerful particle accelerator built so far (see Fig. \ref{fig:lhc-plan}). It is a hadron collider of 27 kilometres 
length, capable of providing 
proton-proton, proton-ion and ion-ion collisions. It first started up on October 2008, and four years later led to the discovery 
of the Higgs boson \cite{ATLAS:2012yve, CMS:2012qbp}, the last missing piece of the Standard Model. In 2022 LHC entered its 
third data taking period, called \textit{Run 3}, after a substantial upgrade and maintaince. Currently, there are four major and 
five small experiments at the LHC:
\begin{itemize}
\item \textit{A Toroidal LHC Apparatus (ATLAS)} is the largest general-purpose experiment at the LHC,
\item \textit{Compact Muon Solenoid (CMS)} is the second major general-purpose experiment,
\item \textit{LHC-beauty (LHCb)} is a major experiment designed to study the physics of the bottom quark,
\item \textit{A Large Ion Collider Experiment (ALICE)} is the fourth major experiment devoted to Heavy Ion Physics research,
\item \textit{TOTal Elastic and diffractive cross section Measurement (TOTEM)} is a small experiment aiming at measuring total 
cross section, diffraction processes and elastic scattering,
\item \textit{LHC-forward (LHCf)} detector measures particles produced in the forward region in order to better understand the 
origin of the high-energy cosmic rays,
\item \textit{Monopole and Exotics Detector At the LHC (MoEDAL)} is a small and largely passive detector designed to search for 
magnetic monopoles, but in this thesis it is proven it can be also utilised for searches for long-lived charged BSM particles,
\item \textit{ForwArd Search ExpeRiment (FASER)} is a detector designed to search for new light and weakly coupled particles,
\item \textit{Scattering and Neutrino Detector (SND)} is an experiment to study collider neutrinos and Feebly Interacting Dark 
Matter Particles.
\end{itemize}

\begin{figure}[ht]
\includegraphics[width=\textwidth]{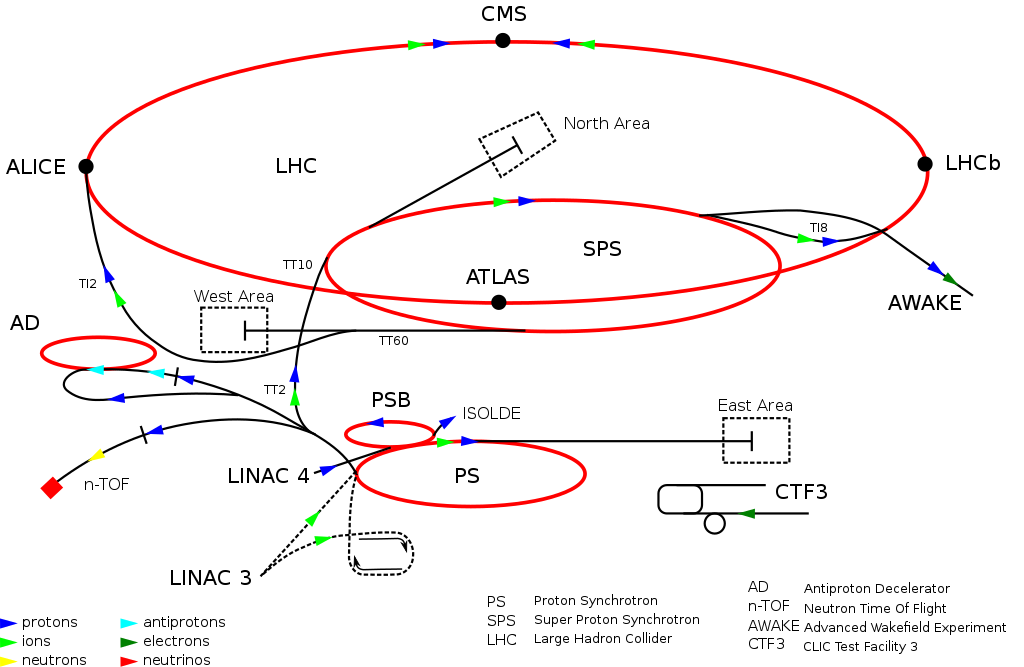}
\caption{\small A schematic drawing of the CERN accelerator complex.  Protons (ions) are first accelerated in the Linac 4 (Linac 3), then go to the Proton Synchrotron Booster, next they are further accelerated in the Proton Synchrotron and the Super Proton Synchrotron, and finally they are injected into the LHC ring, where two beams circulate the collider in opposite directions until they reach the desired energy, then collide in one of the 8 beam intersection points.
Image taken from \cite{cern_complex}.}
\label{fig:lhc-plan}
\end{figure}

For the research presented in this thesis only ATLAS, CMS and MoEDAL detectors are relevant and are discussed in detail in Chapter 3. These three experiments have a rich BSM Physics search programme. As a general purpose detectors, ATLAS and CMS are capable of searching for various supersymmetric particles : 
neutralinos \cite{ATLAS:2021yqv, ATLAS:2021moa, ATLAS:2021fbt, ATLAS:2021hza, CMS:2022sfi, CMS:2022vpy, CMS:2021edw, CMS:2021few, CMS:2021cox, CMS:2020bfa}, 
charginos  \cite{ATLAS:2021yqv, ATLAS:2022rme, ATLAS:2021moa, ATLAS:2022hbt, ATLAS:2022pib,CMS:2022sfi, CMS:2021edw, CMS:2021few, CMS:2021cox, CMS:2020bfa}, 
first two generation squarks \cite{ATLAS:2021twp, ATLAS:2020syg,CMS:2020bfa }, 
sbottoms and stops \cite{ATLAS:2021fbt, ATLAS:2021yij, ATLAS:2021pzz, ATLAS:2021jyv, ATLAS:2021hza, ATLAS:2020xzu, CMS:2021edw, CMS:2021eha, CMS:2021knz, CMS:2020bfa, CMS:2020pyk, CMS:2020cpy}, 
gluinos \cite{ATLAS:2021fbt, ATLAS:2021twp, ATLAS:2020syg, CMS:2022vpy, CMS:2020bfa, CMS:2020fia, CMS:2020cpy}
and 
sleptons \cite{ATLAS:2022hbt, ATLAS:2022pib, ATLAS:2020wjh, CMS:2022rqk, CMS:2020bfa}. 

So far, no supersymmetric particle has been discovered, which 
brings attention to other, more exotic models. One such theory is the leptoquark model \cite{ATLAS:2021jyv, ATLAS:2021yij, ATLAS:2021oiz, ATLAS:2020xov, CMS:2022nty, CMS:2021far}, predicting the existence of a coloured 
particle carrying a lepton number. Another candidate is the Composite Higgs model
\cite{ATLAS:2019fgd, ATLAS:2019zfh, CMS:2022yjm}, which treats the Higgs boson as a composite rather than an elementary particle, similar to QCD pion. There are various searches for models predicting the existence of a new heavy neutral gauge boson, the so-called
$Z'$ models \cite{ATLAS:2019fgd, CMS:2022zoc}. Furthermore, many simplified models of Dark Matter were also tested at the LHC
\cite{ATLAS:2021yij, ATLAS:2022ygn, ATLAS:2021gcn, ATLAS:2021kxv, CMS:2021far, CMS:2021rwb}.

The above list of searches and references is quite extensive, but it is not complete neither in terms of the models tested at the 
LHC, nor when it comes to multiple years of data taking. Despite all the efforts, there is still no indisputable evidence for a BSM 
particle discovery at the LHC, which calls for a reconsideration of the experimental methods and research directions. One of the 
ideas is to search for long-lived BSM particles, predicted by some New Physics theories. Such models have not 
been extensively tested so far and might be realised in Nature, hence they acquired more attention in recent years. 
In this thesis, prospects of detecting such particles are investigated, in the light of the Run 3 data-taking period at the LHC, and its 
anticipated upgrade to the High Luminosity phase (see Fig.  \ref{fig:lhc-plan}). Chapter \ref{chap:two} contains brief explanation 
of supersymmetry and neutrino mass models, which are theoretic scenarios relevant for the presented studies on long-lived 
particles. Chapter \ref{chap:three} describes detectors and experimental searches for the particles considered. Next, the final results  
of the PhD are presented and discussed in Chapter \ref{chap:four}. The last Chapter \ref{chap:five} contains the summary and  
conclusions.

\begin{sidewaysfigure}[ht]
\includegraphics[width=\textwidth]{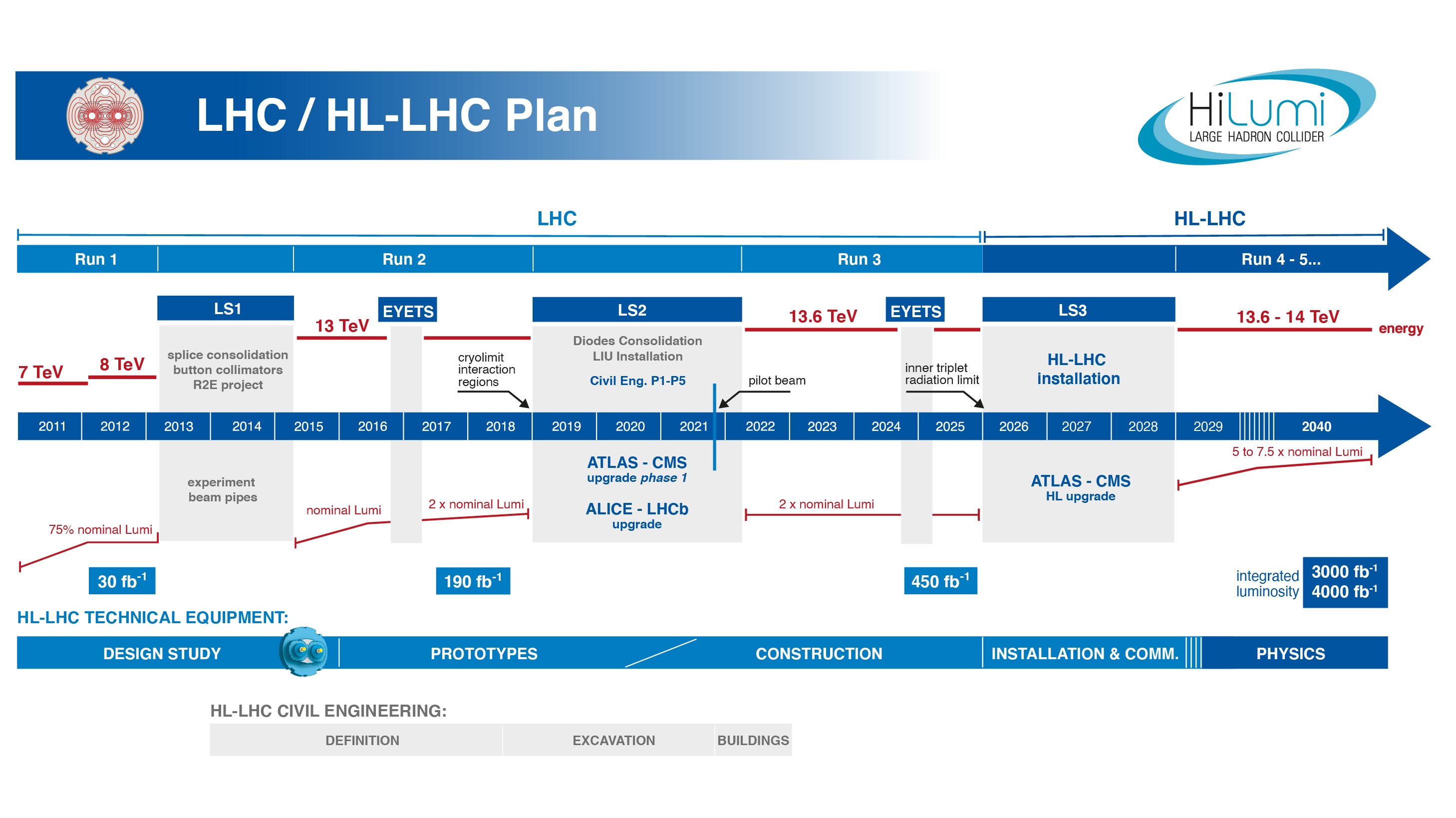}
\caption{\small LHC/HL-LHC upgrade plan from \cite{lhc_plan}.}
\label{fig:lhc-plan}
\end{sidewaysfigure} 
    \chapter{Beyond the Standard Model Physics}\label{chap:two}

\section{Supersymmetry}\label{sec:susy}

\subsection{Introduction}\label{sec:susy:intro}

The success of the Standard Model has proven that symmetry is the key to understanding physics at the subatomic level. The natural avenue in developing BSM field theories is therefore to introduce new symmetries in order to unify the description of previously distinct aspects of the theory. One of the most promising ideas so far is to propose the existence of a \textit{supersymmetry}, which connects bosons and fermion. Supersymmetric theories postulate the existence of operators $Q^a$ transforming bosons to fermions and vice versa:
\begin{equation}
Q^a | \mathrm{Boson} \rangle = | \mathrm{Fermion} \rangle, 
\;\;\;\;\;\;\;\;\;\;\;\;
Q^a | \mathrm{Fermion} \rangle = | \mathrm{Boson} \rangle.
\end{equation}
Operators $Q^a$ must be anti-commuting spinors carrying spin angular momentum $1/2$, hence their hermitian conjugates $\left(Q^a\right)^\dag$ are also symmetry operators. 
One can have multiple supersymmetries connecting bosons and fermions, each having its own pair of $Q, Q^\dag$ operators labelled with $a$. However, these theories are not 
phenomenologically interesting, hence from now on we will drop $a$ indices and discuss only theory with single pair of $Q, Q^\dag$ operators, the so-called \textit{N=1 supersymmetry}. 
Properties of supersymmetric operators in an interacting quantum field theory are severely limited by the Haag-Lopuszanski-Sohnius extension \cite{Haag:1974qh} of the Coleman-Mandula theorem \cite{Coleman:1967ad}. If we want to build a realistic supersymmetric extension of the Standard Model, then the symmetry operators form an algebra satisfying the following set of conditions:
\begin{align}
\{ Q_\alpha, Q_{\dot{\alpha}}^\dag \} &= 2 \sigma_{\alpha \dot{\alpha}}^\mu    P_\mu , \label{eq:susyalgebra1}\\
\{ Q_\alpha, Q_{\beta}^\dag \} &= 
\{ Q_{\dot{\alpha}}^\dag, Q_{\dot{\beta}}^\dag \} = 0 ,\\
\left[ Q_\alpha , P^\mu \right] &= 
\left[ Q_{\dot{\beta}}^\dag , P^\mu \right] = 0,
\end{align}
where $P^\mu$ is the momentum operator. 
Note that from Eq. \eqref{eq:susyalgebra1} it follows that the Hamiltonian operator is given by the sum of squares of supersymmetry generators:
\begin{equation}\label{eq:susy-hamiltonian}
H = P^0 = \frac{1}{4} \left(
Q_1 Q_1^\dag + Q_1^\dag Q_1 + Q_2 Q_2^\dag + Q_2^\dag Q_2
\right).
\end{equation}

The supersymmetry algebra has irreducible representations, called \textit{supermultiplets}, containing single-particle states of the supersymmetric theory. Each supermultiplet 
consists of fermionic and bosonic states, which are called \textit{superpartners} of each other. If two states $| \sigma \rangle$ and  $| \sigma' \rangle$ both belong to the same 
supermultiplet, then $| \sigma' \rangle$ can be represented as a linear combination of $Q$ and $Q^\dag$ acting on $| \sigma \rangle$, up to spacetime translations and rotations. 
The mass operator $P^2$ commutes with $Q$, $Q^\dag$, and all spacetime rotation and translation operators, hence superpartners have the same mass. If supersymmetry was the 
exact symmetry of nature, SM particles and their superpartners would be mass degenerate, and we would have already discovered supersymmetry a long time ago. The lack of such 
discoveries implies that supersymmetry, if real, must be broken. The exact mechanism of supersymmetry breaking is one of the major research problems, and its discussion is beyond 
the scope of this work.

It can be easily proven (c.f.\cite{Terning:2006bq}) that supermultiplets must contain the same number of bosonic and fermionic degrees of freedom. The simplest possibility is 
a supermultiplet containing a single Weyl fermion with two helicity states. In order to balance fermionic degrees of freedom, such supermultiplet may contain two real scalar fields or 
a single complex scalar field. This combination is called a \textit{chiral supermultiplet}.
Another possibility comes from considering a supermultiplet with a spin-1 vector boson. In 
order to be renormalizable, the vector boson must be massless, at least before the associated gauge symmetry is broken. Such multiplet contains two bosonic degrees of freedom that 
have to be balanced by a single Weyl fermion field \footnote{If one tried to use spin-3/2 field instead, the resulting theory would not be renormalizable.}. A supermultiplet consisting of 
a gauge boson and its fermionic superpartner (gaugino) is called \textit{gauge supermultiplet}. One of the key observations is that superpartners differ in spin by $1/2$.

Supersymmetry attracted tremendous interest because of its ability to solve the hierarchy problem.
The hierarchy problem is the sensitivity of low-energy physics, e.g. the mass 
spectrum of the SM, to a much higher energy scale of the BSM Physics. In order to 
understand it, consider a Yukawa interaction $\mathcal{L}_f = -\lambda_f h \bar f f$, between SM Higgs boson $h$ and BSM fermion $f$, with coupling $\lambda_f$. Correction to the self-energy of the Higgs scalar comes from the 
diagram in Fig. \ref{fig:higgs-corr-f}, and is given by \cite{Drees:1996ca}:
\begin{equation}\label{eq:higgs-corr-f}
\begin{split}
\pi^f_{hh}(0) &= 
- N(f) \int \frac{d^4 k}{(2\pi)^4} \rm{Tr} \left[
\left( i \frac{\lambda_f}{\sqrt{2}} \right) \frac{i}{\slashed{k} - m_f}
\left( i \frac{\lambda_f}{\sqrt{2}} \right) \frac{i}{\slashed{k} - m_f}
\right] =\\
&=-2 N(f) \lambda_f^2 \int \frac{d^4 k}{(2\pi)^4}
\left[ 
\frac{1}{k^2-m_f^2}+\frac{2 m_f^2}{(k^2-m^2_f)^2}
\right],
\end{split}
\end{equation} 
where $k$ in Eq. \eqref{eq:higgs-corr-f} stands for the momentum of a fermion in the 
loop, $m_f=\lambda_f v/\sqrt{2}$ is the mass of the new fermion, $N(f)$ is a multiplicity factor (e.g. 
$N(f)=3$ for 
quarks). The first term in Eq. \eqref{eq:higgs-corr-f} is quadratically divergent, 
hence $\pi^f_{hh}(0) \propto \Lambda_{\rm UV}^2$, where $\Lambda_{\rm UV}$ is the energy scale of a New Physics. At least one physical scale 
of a BSM Physics is known, the Planck scale, if we take $ \Lambda_{\rm UV} \sim M_{\rm Pl}$ then the correction to the Higgs mass will be many orders of 
magnitude greater than the physical SM Higgs mass, $m_h \approx 125~\rm GeV$. The reason behind this problem is the lack of a SM mechanism protecting 
the mass of an elementary scalar, unlike fermions, which masses are 
protected by the chiral symmetry, or photon masslessness protected to all 
orders of perturbative expansion by the exact $U(1)$ gauge invariance of QED. One could 
simply renormalize the quadratic divergence in Eq. \eqref{eq:higgs-corr-f}, but 
a finite correction $N(f)m_f^2\lambda_f^2/8\pi$ would persist. This correction 
is proportional to $m_f^2$, which is expected to be of the order of $\Lambda_{\rm UV}^2$, hence very large.

\begin{figure}[!tbhp]
\centering
\includegraphics[width=0.49\textwidth]{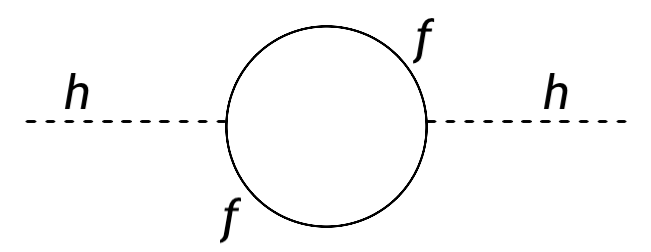}
\caption{\small One loop correction to the Higgs self-energy resulting from Yukawa interaction.}
\label{fig:higgs-corr-f}
\end{figure}

\begin{figure}[!bthp]
\centering
  \begin{subfigure}[b]{0.49\textwidth}
    \includegraphics[width=\textwidth]{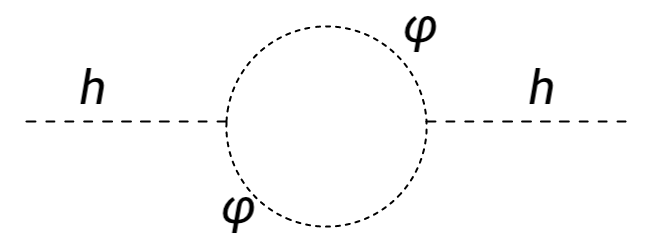}
    \label{fig:higgs-corr-s1}
  \end{subfigure}
  \hfill
  \begin{subfigure}[b]{0.49\textwidth}
    \includegraphics[width=\textwidth]{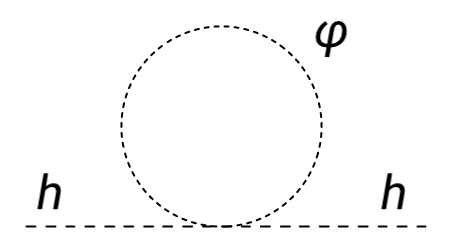}
    \label{fig:higgs-corr-s2}
  \end{subfigure}
  \caption{\small Contributions to the Higgs self-energy from loops involving new scalar particles. $\phi$ stands for either $\phi_L$ or $\phi_R$.}
    \label{fig:higgs-corr-s}
\end{figure}

In supersymmetric theories the hierarchy problem is solved by additional contributions to $\pi_{hh}$. Let us introduce two new complex scalars $\phi_L,\phi_R$, coupled to the Higgs field in the following way:
\begin{equation}\label{eq:higgs-corr-s-lagr}
\mathcal{L} = \frac{1}{2} \lambda_\phi h^2 \left( |\phi_L|^2 + |\phi_R|^2 \right)
+ v \lambda_\phi h \left( |\phi_L|^2 + |\phi_R|^2 \right)
+ 
\left(
\frac{\lambda_f}{\sqrt{2}} A_\phi h \phi_L \phi_R^* + h.c.
\right),
\end{equation}
where $v$ is the vacuum expectation value of the SM Higgs. 
Feynman diagrams representing interactions in Eq. \eqref{eq:higgs-corr-s-lagr} are depicted in Fig. \ref{fig:higgs-corr-s}.
The second term in 
Eq. \eqref{eq:higgs-corr-s-lagr} originates from the breaking of the $SU(2)_L\times U(1)_Y$ and its coefficients are related to those of the first term. The 
coeffcient of the last term in Eq. \eqref{eq:higgs-corr-s-lagr} is arbitrary, but conventionally $\lambda_f$ is factored out. If we assume $N(\phi_L)=N(\phi_R)$, then the Lagrangian in Eq. \eqref{eq:higgs-corr-s-lagr} contributes to the 
self-energy in the following way:
\begin{equation}\label{eq:higgs-corr-s}
\begin{split}
\pi_{hh}^\phi(0) &= - \lambda_\phi N(\phi) \int \frac{d^4 k}{(2\pi)^4}
\left[
\frac{1}{k^2-m_{\phi_L}^2}
+
\frac{1}{k^2-m_{\phi_R}^2}
\right] \\
&+
(\lambda_\phi v)^2 N(\phi)  
\int \frac{d^4 k}{(2\pi)^4}
\left[
\frac{1}{(k^2-m_{\phi_L}^2)^2}
+
\frac{1}{(k^2-m_{\phi_R}^2)^2}
\right] \\
&+ 
|\lambda_f A_\phi|^2 N(\phi) 
\int \frac{d^4 k}{(2\pi)^4}
\frac{1}{(k^2-m_{\phi_L}^2)(k^2-m_{\phi_R}^2)}.
\end{split}
\end{equation}
The first line in Eq. \eqref{eq:higgs-corr-s} contains the quadratically divergent term. Comparison with Eq. \eqref{eq:higgs-corr-f} leads to the conclusion that quadratic divergences can be avoided if:
\begin{equation}\label{eq:higgs-corr-condition}
\begin{split}
N(\phi_L) &= N(\phi_R) = N(f), \\
\lambda_\phi &= - \lambda_f^2.\\
\end{split}
\end{equation}
By resorting to the $\overline{\rm MS}$ regularisation with the renormalisation scale $\mu$, and assuming for simplicity $m_{\phi_L} = m_{\phi_R}$ one can arrive \cite{Drees:1996ca} at:
\begin{equation}\label{eg:higgs-corr-full}
\begin{split}
\pi_{hh}^{f+\phi} (0) =
 i \frac{\lambda_f^2 N(f)}{16\pi^2} &
\left[
 -  2m_f^2 \left( 1 - \ln\frac{m_f^2}{\mu^2} \right)
+ 4m_f^2 \ln \frac{m_f^2}{\mu^2}  + \right.\\
&+2 m^2_\phi \left( 1 - \ln\frac{m^2_\phi}{\mu^2} \right)
-4 m_\phi^2 \ln \frac{m_\phi^2}{\mu^2} +\\
&- |A_\phi|^2 \left. \ln \frac{m_\phi^2}{\mu^2}
\right].
\end{split}
\end{equation}
From Eq. \eqref{eg:higgs-corr-full} it follows that the correction can be completely cancelled if:
\begin{equation}\label{eq:higgs-corr-vanishing}
\begin{split}
m_\phi &= m_f\\
A_f &= 0.
\end{split}
\end{equation}

If one could guarantee the fulfilment of the conditions in Eqs. \eqref{eq:higgs-corr-condition} and \eqref{eq:higgs-corr-vanishing}, then there would be a theoretically consistent and elegant mechanism for solving the hierarchy problem. This is precisely what supersymmetry does, and 
it will be elaborated in Sec. \ref{sec:susy:formalism}. Before moving on, it is interesting to consider a small relaxation of requirements for vanishing corrections to the Higgs self-energy, i.e. let us take $m^2_\phi = m_f^2 + \delta^2$, with $\delta, |A_\phi| \ll m_f$, so that
$\ln (m_\phi^2/\mu^2) \approx \ln (m_f^2/\mu^2) + \delta^2/m_f^2$. One can show \cite{Drees:1996ca} that the correction in Eq. \eqref{eg:higgs-corr-full} becomes:
\begin{equation}\label{eq:higgs-corr-pert}
\pi_{hh}^{f+\phi} (0) \approx -i \frac{\lambda_f^2 N(f)}{16\pi^2} \left[ 4\delta^2 + \left(2\delta^2 + |A_\phi|^2 \right) \ln \frac{m_f^2}{\mu^2} \right] + \mathcal{O} \left( \delta^4, A^2_\phi\delta^2 \right).
\end{equation}
From Eq. \eqref{eq:higgs-corr-pert} we see that even if $m_f$ is very large, the quantum correction will remain moderate, as long as the mass splitting parameter $\delta$ and trilinear coupling $A_\phi$ are small compared to the New Physics scale.

The reason why supersymmetry allows us to solve the hierarchy problem might be 
also understood by studying the chiral symmetry in the SM. Chiral symmetry allows to rotate phases of the left ($\psi_L$) and right-handed ($\psi_R$) fermion fields independently:
\begin{align}
\psi_L \to e^{i \theta_L} \psi_L, ~\psi_R \to \psi_R,\\
\psi_R \to e^{i \theta_R} \psi_R,  ~\psi_L \to \psi_L,\\
\end{align}
where $\theta_L$ and $\theta_R$ are independent parameters corresponding to the chiral symmetry group $U(1)_L \times U(1)_R$. In the case of $N$ flavours, the appropriate symmetry group is $U(N)_L \times U(N)_R$.
As can be seen 
from Eq. \eqref{eq:Dirac-notation2}, 
chiral symmetry is broken by the mass term for electrons, and similarly for other SM fermions.
The mass of the electron (or other SM fermion) is a technically natural parameter in the sense of t'Hooft, i.e.
when $m_e \to 0$ then the chiral symmetry is restored. This protects the mass of 
the SM fermions, i.e. the radiative correction is at most logarithmically 
divergent. 
On the other hand, there is no such mechanism for SM scalars, which results in the hierarchy problem for the 
Higgs boson. 
However, supersymmetry imposed $m_f=m_\phi$ for fermion $f$ 
and scalar $s$ belonging to the same supermultiplet. In this way, the mass of  
scalars in supersymmetric extensions of the SM is protected via ``inherited'' 
chiral symmetry.

However, as already mentioned before, supersymmetry has to be broken. If we 
allow for soft supersymmetry breaking through a parameter $M$ with a 
positive mass dimension, then the quantum correction to the scalar mass 
squared, $m_\phi^2$, has to be proportional to $M^2$ to ensure that 
the correction vanishes for $M\to 0$. But this implies that the correction cannot 
be proportional to a positive power of the UV cutoff $\Lambda$, because of the dimensional analysis. 
Therefore, the scalar mass is protected up to $M$, despite the soft supersymmetry breaking. This result holds for all orders of the perturbation theory.

\FloatBarrier
\subsection{Mathematical description}\label{sec:susy:formalism}
\FloatBarrier

In this section, we will provide a very brief overview of the mathematical structure of $N=1$ supersymmetric theories. A more detailed and comprehensive introduction can be found in one of the available textbooks, e.g. \cite{Martin:1997ns, Terning:2006bq, Wess:1992cp}.

Supersymmetric theories can be conveniently studied in terms of 
\textit{superfields} and \textit{superalgebras}. However, introducing this 
new formalism is not necessary to 
understand the phenomenological consequences of the MSSM (described in Sec. \ref{sec:mssm}), and it is 
beyond the scope of this thesis. Therefore we will use standard QFT 
description to familiarise the reader with the 
theory.

The simplest SUSY theory, called \textit{free Wess-Zumino model}\cite{Wess:1974tw}, is constructed out of non-interacting chiral supermultiplets.
Each chiral supermultiplet contains a complex scalar field $\phi_i$ and two-
component Weyl spinor $\psi_i$. The Lagrangian density describing 
free chiral field propagation is given by:
\begin{equation}\label{eq:susy-chiral-free}
\mathcal{L}_{\rm chiral}^{\rm free} = 
\partial_\mu \phi_i \partial^\mu \phi^{*i} 
+
i \psi^{\dag i} \bar{\sigma}^\mu \partial_\mu \psi_i 
+ F_i^*F^i,
\end{equation}
with the off-shell SUSY transformation of the form:
\begin{align}\label{eq:susy-chiral-offshell}
\delta \phi_j &= \epsilon \psi_j , 
&
 \delta \psi_j^* &= \epsilon^\dag \psi^{\dag j},
 \\ 
\delta \psi_{j \alpha} &= -i \left( \sigma^\mu \epsilon^\dag \right)_\alpha \partial_\mu \phi_j + \epsilon_\alpha F_j, 
&
\delta \psi^{\dag j}_{\dot \alpha} &= i \left( \epsilon \sigma^\mu \right)_{\dot \alpha} \partial_\mu \phi^{*j} + \epsilon_{\dot \alpha}^\dag F^{*j}, 
\label{eq:susy-chiral-offshell-2}
\\
\delta F_j &= -\epsilon^\dag \bar \sigma^\mu \partial_\mu \psi_j, 
&
\delta F^{*j} &= i \partial_\mu \psi^{\dag j} \bar \sigma^\mu \epsilon,
\label{eq:susy-chiral-offshell-3}
\end{align}
where $\epsilon^\alpha$ is a constant, infinitesimal, anticommuting two-component Weyl object parametrising SUSY 
transformation. $F$ is a the \textit{auxiliary} field of mass dimension 2. This is not a physical field, but rather a mathematical trick allowing 
us to write Lagrangian density that is invariant under off-shell SUSY transformation. Otherwise, invariance of the free chiral lagrangian 
under the SUSY transformation would hold only after applying the equation of motion for $\phi_i$ and $\psi_i$ fields. 

The most general set of renormalizable interactions for $\phi_i$, $\psi_i$, $F_i$ fields is of the form:
\begin{equation}\label{eq:susy-chiral-int}
\mathcal{L}_{\rm int} = - \frac{1}{2} W^{jk} \psi_j \psi_k + W^j F_j + h.c.
\end{equation}
with $W^{jk}$ ($W^j$) being a linear (quadratic) function of scalar fields and their conjugates. 

By adding Eq. \eqref{eq:susy-chiral-int} to Eq.  \eqref{eq:susy-chiral-free}, working out equations of motion for auxiliary fields 
and integrating them out\footnote{One has to remember to update transformation properties of the spinor field in 
Eq. \eqref{eq:susy-chiral-offshell-2}, by plugging in the solution of EOM for the auxiliary fields.}, one arrives at:
\begin{equation}\label{eq:susy-chiral}
\mathcal{L}_{\rm chiral} = 
\partial_\mu \phi_i \partial^\mu \phi^{*i} 
+
i \psi^{\dag i} \bar{\sigma}^\mu \partial_\mu \psi_i 
- \frac{1}{2} \left( W^{ij} \psi_i \psi_j + W^*_{ij} \psi^{\dag i} \psi^{\dag j} \right)
- W^i W_i^*.
\end{equation}

\noindent It is possible and convenient to relate $W^{jk}$ and $W^j$:
\begin{align} \label{eq:Wi}
W^i &\equiv \frac{\delta W}{\delta \phi_i}, \\ \label{eq:Wij}
W^{ij} &\equiv \frac{\delta^2 W}{\delta \phi_i \delta \phi_j}.
\end{align}
$W$ is called \textit{superpotential}, and it is a holomorphic function of scalar fields $\phi_i$ treated as complex variables. It can 
be shown that for a theory like MSSM\footnote{In general, it is also possible to include a term that is linear in scalar fields: $W \supset L^i \phi_i$.}, superpotential has the form:
\begin{equation}\label{eq:superpotential}
W = \frac{1}{2} M^{ij} \phi_i \phi_j + \frac{1}{6} y^{ijk} \phi_i \phi_j \phi_k.
\end{equation}
$M^{ij}$ in Eq.\eqref{eq:superpotential} is a symmetric mass matrix for the fermion fields, while $y^{ijk}$ is a Yukawa coupling of 
two fermions $\psi_i \psi_j$ and a scalar $\phi_k$, which has to be totally symmetric under the change of any pair of indices. If 
we plug in Eq. \eqref{eq:superpotential} to Eq. \eqref{eq:susy-chiral}, we obtain the Lagrangian density for the \textit{interacting 
Wess-Zumino model}:
\begin{equation}\label{eq:susy:lagr}
\begin{split}
\mathcal{L}_{\rm chiral} &= 
\partial^\mu \phi^{*i} \partial_\mu \phi_i
+ i \psi^{\dag i} \overline{\sigma} ^\mu \partial_\mu \psi_i +\\
&- \frac{1}{2} M^{ij} \psi_i \psi_j
- \frac{1}{2} M^*_{ij} \psi^\dag_i \psi^\dag_j
- V\left( \phi, \phi^* \right)+\\
&-\frac{1}{2} y^{ijk} \phi_i \psi_j \psi_k 
-\frac{1}{2} y^*_{ijk} \phi^{*i} \psi^{\dag j} \psi^{\dag k}
\end{split}
\end{equation}
The scalar potential $V(\phi, \phi^*)$ in Eq. \eqref{eq:susy:lagr} is given by:
\begin{equation}\label{eq:susy-scalarpot}
\begin{split}
&V(\phi, \phi^*) = W^j W^*_j = F_j F^{*j} = M_{ik}^* M^{kj} \phi^{*i} \phi_j +\\
&+ \frac{1}{2} M^{in} y_{jkn}^* \phi_i \phi^{*j} \phi ^{*k}
+\frac{1}{2} M^*_{in} y^{jkn} \phi^{*i} \phi_j \phi_k
+ \frac{1}{4} y^{ijn} y^*_{kln} \phi_i \phi_j \phi^{*k} \phi^{*l}
\end{split}
\end{equation}

\noindent There are three important features of the interacting Wess-Zumino model worth pointing out:
\begin{enumerate}
\item Interactions between components of chiral multiplets are given by superpotential, which is just a holomorphic function of 
scalar fields.
\item Scalar potential is quadratic in $F$, hence it satisfies $V(\phi_i, \phi_i^*) \geq 0$.
\item By looking at the linearised equations of motion arising from Eq. \eqref{eq:susy-chiral}, one can find that scalar and 
fermionic counterparts of the chiral supermultiplet have the same mass. Therefore, supersymmetry must be broken, otherwise we 
would have already discovered it.
\end{enumerate}

Gauge supermultiplets contain gauge boson fields $A^a_\mu$ and fermionic superpartners $\lambda^a_\alpha$ (gauginos), 
where the index $a$ runs over the adjoint representation of the gauge group (e.g. $a=1,2,...8$ for $SU(3)_C$).
The Lagrangian density for a free gauge supermultiplet is given by:

\begin{equation}\label{eq:susy-gauge-free}
\mathcal{L}_{\rm gauge}^{\rm free} =  - \frac{1}{4} F^a_{\mu\nu} F^{\mu \nu a} + i \lambda^{\dag a} \bar{\sigma}^\mu
D_\mu \lambda^a + \frac{1}{2}D^a D^a,
\end{equation}
with the field strength tensor given by:
\begin{equation}
F_{\mu\nu}^a = \partial_\mu A_\nu^a - \partial_\nu A_\mu^a - g f^{abc} A^b_\mu A^c_\nu,
\end{equation}
and covariant derivative:
\begin{equation}
D_\mu \lambda^a = \partial_\mu \lambda^a - gf^{abc} A^b_\mu \lambda^c.
\end{equation}

The field $D^a$ is the auxiliary boson field with mass dimension 2. Supersymmetry transformation laws are given by:
\begin{align}\label{eq:susy-gauge-offshell}
\delta A^a_\mu &= - \frac{1}{\sqrt{2}} \left[ \epsilon^\dag \bar \sigma_\mu \lambda^a + \lambda^{\dag a} \bar \sigma_\mu \epsilon \right], \\
\delta \lambda_\alpha^a &= - \frac{i}{2\sqrt{2}} \left( \sigma^\mu \bar \sigma^\nu \epsilon \right)_\alpha F^a_{\mu\nu} + \frac{1}{\sqrt{2}} \epsilon_\alpha D^a, \\
\delta \lambda^{\dag a}_{\dot \alpha} &= \frac{i}{2\sqrt{2}} ( \epsilon^\dag \bar \sigma^\nu \sigma^\mu ) _{\dot \alpha} F^a_{\mu\nu} + \frac{1}{\sqrt{2}} \epsilon^\dag_{\dot \alpha} D^a, \\
\delta D^a &= \frac{-i}{\sqrt{2}} \left[ \epsilon^\dag \bar \sigma^\mu D_\mu \lambda^a - D_\mu \lambda^{\dag a} \bar\sigma^\mu \epsilon \right].
\end{align}

In a full theory we have both chiral and gauge supermultiplets, and interactions between them. The naive way of constructing such 
a theory is to simply promote ordinary derivatives in Eqs. \eqref{eq:susy:lagr}, \eqref{eq:susy-chiral-offshell-2} and \eqref{eq:susy-chiral-offshell-3} to covariant derivatives:

\begin{align}
D_\mu  \phi_j &= \partial_\mu \phi_j + i g A_\mu^a T^a \phi_j ,
\\
D_\mu  \phi^{*j} &= \partial_\mu \phi^{*j} - i g A_\mu^a T^a \phi^{*j},
\\
D_\mu \psi_j &= \partial_\mu \psi_j + i g A^a_\mu T^a \psi_j,
\end{align}
with $T^a$ being generators of the relevant gauge group and $g$ the corresponding coupling constant. However, in a theory with 
chiral and gauge supermultiplets, there are 
three additional renormalizable interactions that can be written:
\begin{equation}\label{eq:susy-3inter}
(\phi^* T^a \psi)\lambda^a, 
~~~~
\lambda^{\dag a} ( \psi^\dag T^a \phi),
~~~~
(\phi^* T^a \phi) D^a
\end{equation}
in fact all three terms with appropriate coefficients related to gauge coupling constants are necessary for SUSY invariance. The complete Lagrangian of a general SUSY gauge theory is:

\begin{equation}\label{eq:susy-complete}
\mathcal{L} = \mathcal{L}_{\rm gauge} +  \mathcal{L}_{\rm chiral} - 
\sqrt{2}g \left[
(\phi^* T^a \psi)\lambda^a + \lambda^{\dag a} ( \psi^\dag T^a \phi)
\right]
+g (\phi^* T^a \phi) D^a,
\end{equation}
where $\mathcal{L}_{\rm gauge}$ is given by Eq. \eqref{eq:susy-gauge-free} while $\mathcal{L}_{\rm chiral}$ is the Lagrangian 
density in Eq. \eqref{eq:susy:lagr}, but with all ordinary derivatives promoted to covariant derivatives. Transformation properties of chiral fields are generalisations of Eqs. \eqref{eq:susy-chiral-offshell}, \eqref{eq:susy-chiral-offshell-2} and \eqref{eq:susy-chiral-offshell-3}
\footnote{In this section we are using the so-called \textit{Wess-Zumino gauge}.}:

\begin{align}\label{eq:susycomplete-offshell}
\delta \phi_j &= \epsilon \psi_j , 
 \\
\delta \psi_{j \alpha} &= -i ( \sigma^\mu \epsilon^\dag )_\alpha D_\mu \phi_j + \epsilon_\alpha F_j, 
\\
\delta F_j &= -\epsilon^\dag \bar \sigma^\mu D_\mu \psi_j +\sqrt{2} g (T^a \phi)_j \epsilon^\dag \lambda^{\dag a}. 
\end{align}

Scalar potential of the theory has two terms, \textit{F-term} and \textit{D-term}:
\begin{equation}\label{eq:susy-scalarpot}
V\left( \phi, \phi^* \right) = F^{*j} F_j + \frac{1}{2} D^a D^a = W_j^* W^j + \frac{1}{2}g^2 \left( \phi^* T^a \phi \right)^2 \geq 0.
\end{equation}
The supersymmetric vacuum state, $|0 \rangle$, is defined as $Q_{1,2} | 0 \rangle = 0$. From Eq. \eqref{eq:susy-hamiltonian} it follows that for unbroken supersymmetry $H | 0 \rangle = 0$. 
If space-time effects and fermion condensates can be neglected, then we have 
$\langle 0 | H | 0 \rangle = \langle 0 | V | 0 \rangle$, where $V$ is given by Eq. \eqref{eq:susy-scalarpot}. 
Therefore, the vanishing of all $D^a$ and $F_j$ is equivalent to supersymmetry conservation in the vacuum state.
However, we know that for the vacuum state we live in, the 
supersymmetry is broken\footnote{Another possibility is that our vacuum state is a metastable supersymmetry-
breaking state and the true ground state preserves supersymmetry. The requirement is that the lifetime of the metastable state is 
of the order of the age of the Universe or greater.}, hence it should be $ \langle 0 | V | 0 \rangle >0$ for symmetry to be 
spontaneously broken in the ground state. Supersymmetry breaking is a vast topic and detailed discussion is beyond the scope of this thesis, therefore curious readers can refer to the specialistic literature, e.q. \cite{Martin:1997ns, Terning:2006bq, Wess:1992cp}.

\FloatBarrier
\subsection{Minimal Supersymmetric Standard Model}\label{sec:mssm}
\FloatBarrier

\textit{Minimal Supersymmetric Standard Model 
(MSSM)} is the simplest extension of the Standard Model incorporating $N=1$ global 
supersymmetry.
One could think that in order to build a supersymmetric extension of the Standard Model, SM particles can be groupped together into supermultiplets, e.g. neutrino and Higgs boson. It 
is known that this approach does not work, and all superpartners of the SM fields are completely new BSM particles. SM fermions, both quarks and leptons, are put into chiral
supermultiplets, so they have scalar superpartners. Names of these superpartners are created by appending a prefix "s-" to names of SM fermions, e.g. electron $\to$ \textbf{s}electron,
 top quark $\to$ \textbf{s}top. Names like \textit{left-handed smuon} are used, but one has to remember that ``left-handed'' corresponds to the property 
of the SM superpartner, scalar particles do not have ``handedness''. SM gauge bosons are put into gauge supermultiplets together with their fermionic superpartners. Names of 
these superpartners are created by adding a suffix -ino, e.g. gluon $\to$ gluino. After electroweak symmetry breaking, winos and binos mix, and the mass eigenstates are called zino and 
photino. The situation is little more complicated for the Higgs boson. Obviously, it has to be a part of a chiral supermultiplet, but one can prove that a single Higgs supermultiplet is not 
enough. There have to be two Higgs supermultiplets in order to avoid gauge anomaly in the electroweak gauge symmetry. Another reason is that the structure of supersymmetric 
theory requires the presence of two Higgs supermultiplets in order to give mass to all quarks. One supermultiplet, labelled $H_u$, has Yukawa interactions with up-type quarks $(u, 
c,t)$, while the other, named $H_d$, couples to down-type quarks $(d,c,b)$. The spin-zero components of the Higgs supermultiplets are $(H_u^+, H_u^0)$ for $H_u$, and 
$(H_d^0, H_d^+)$ for $H_d$. The physical boson with $m\approx 125~\rm GeV$ is a linear combination of $H^0_u$ and $H^0_d$. Spin-1/2 components of Higgs supermultiplets are called \textit{higgsinos}.

The general convention is to label supersymmetric partners of SM particles using the same symbols, but appended with a tilde on 
top, e.g. $e_R \to \tilde{e}_R$. Chiral supermultiplets of the MSSM are listed in Tab. \ref{tab:susy:chiral} while gauge supermultiplets are in 
Tab. \ref{tab:susy:gauge}. 

In Tab. \ref{tab:susy:chiral} we follow the same naming scheme for the SM 
fields as in Tab. \ref{tab:sm_fields}. 
We apply the usual \cite{Martin:1997ns, Terning:2006bq} convention that all chiral supermultiplets are defined in terms 
of the left-handed Weyl spinors only, hence conjugates of the right-handed 
quarks and leptons (and their superpartners) appear in Tab. 
\ref{tab:susy:chiral}.
The index $j$ in Tab. \ref{tab:susy:chiral} labels generations, e.g. $u^j_L = (u_L ,c_L, t_L)$. $L$ and $R$ in the names of scalars, 
i.e. $ \tilde e_R^{*}$, inform us about the chirality of their SM superpartners.
Symbols provided in the first column of Tab. 
\ref{tab:susy:chiral} represent the whole supermultiplet, i.e. $Q$ stands for a $SU(2)_L$-doublet chiral supermultiplet containing
$\tilde u_L, u_L$ ($T_3=1/2$) and $\tilde d_L, d_L$ ($T_3=-1/2$).
A bar over a symbol, i.e. $\bar u^j$, is a part of the name of the supermultiplet and does not correspond to any sort of conjugation.

\FloatBarrier
\begin{table}[htb]
\centering
\caption{\small Chiral supermultiplets in the MSSM. The spin-0 fields are complex scalars, and the spin-1/2 fields 
are left-handed two-component Weyl fermions. We use the same notation for SM fields as in Tab. \ref{tab:sm_fields}, i.e. we 
define both left- and right-handed spinors and label them with subscripts $L$ and $R$, respectively. Barred symbols without a 
subscript for up quark $(\bar u)$, down quark $(\bar d)$, and fermions $(\bar e)$, represent right-handed supermultiplets. Index $j$ labels generations of 
matter, i.e. $j=1,2,3$.
}
\begin{tabular}{cc|c|c|c}
\multicolumn{2}{c|}{\textbf{Names}}                                            & \textbf{spin 0}                             & \textbf{spin 1/2}                           & {$\mathbf{ SU(3)_C,SU(2)_L,U(1)_Y }$}        \\ 
\hline
\multicolumn{1}{c}{\multirow{3}{*}{\textbf{squarks, quarks}}}   & $Q^j$       & $\tilde Q \equiv \left( \tilde u_L^j~~\tilde d_L^j \right)$ & $\left( u_L^j~~d_L^j \right)$               & $\left( 3,2,\frac{1}{6} \right)$         \\ 
\multicolumn{1}{c}{}                                            & $\bar{u}^j$ & $\tilde u_R^{j*} $                         & $u_R^{ j \dag}$                              & $\left( \bar{3}, 1, -\frac{2}{3}\right)$ \\ 
\multicolumn{1}{c}{}                                            & $\bar{d}^j$ & $\tilde d_R^{j*}$                          & $d_R^{ j \dag}$                              & $\left( \bar{3}, 1, \frac{1}{3}\right)$  \\
 \hline
\multicolumn{1}{c}{\multirow{2}{*}{\textbf{sleptons, leptons}}} & $L^j$         & $\tilde L \equiv \left( \tilde \nu_L^j ~~ \tilde e_L^j \right)$  & $\left( \nu_L^j ~~ e_L^j \right)$                & $\left( 1, 2, -\frac{1}{2}\right)$       \\ 
\multicolumn{1}{c}{}                                            & $ \bar{e}^j$ & $ \tilde e_R^{j*} $           & $e_R^{ j \dag}$                              & $\left( 1, 1, 1 \right)$                 \\ 
\hline
\multicolumn{1}{c}{\multirow{2}{*}{\textbf{Higgs, higgsinos}}}  & $H_u$       & $\left( H_u^+~~H_u^0 \right)$               & $\left( \tilde H_u^+~~\tilde H_u^0 \right)$ & $\left( 1, 2, +\frac{1}{2} \right)$      \\ 
\multicolumn{1}{c}{}                                            & $H_d$       & $\left( H_d^0~~H_d^- \right)$               & $\left( \tilde H_d^0~~\tilde H_d^- \right)$ & $\left(1, 2, -\frac{1}{2} \right)$       \\ 
\end{tabular}\label{tab:susy:chiral}
\end{table}

\begin{table}[htb]
\centering
\caption{\small Gauge supermultiplets in the MSSM.}
\begin{tabular}{c|c|c|c}
\textbf{Names}          & \textbf{spin 1/2}                   & \textbf{spin 1}       & $\mathbf{ SU(3)_C, SU(2)_L, U(1)_Y}$ \\ \hline
gluino, gluon  & $\tilde g$                 & g            & $(8, 1, 0)$                \\ 
wino, W bosons & $\left. \widetilde W^\pm~~\widetilde W^0 \right.$ & $W^\pm~~W^0$ & $\left(1,3,0\right)$                  \\ 
bino, B boson  & $\widetilde B^0$               & $B^0$        & $(1,1,0)$                  \\
\end{tabular}\label{tab:susy:gauge}
\end{table}

The superpotential of the MSSM is\footnote{It is now known that neutrinos have masses, so SM and consequently MSSM should be amended to include couplings between neutrinos and Higgs. However, the exact origin of neutrino masses is still unknown, hence we do not consider such terms.}:
\begin{equation}\label{eq:susy-mssm-superpotential}
W_{\rm MSSM} = \bar u \mathbf{Y_u} Q H_u - \bar d \mathbf{Y_d} Q H_d - \bar e \mathbf{Y_e} L H_d + \mu H_u H_d,
\end{equation}
where all contracted indices are omitted and summation over the three generations of matter is assumed.
There are four additional renormalizable terms that can be appended to superpotential in Eq. \eqref{eq:susy-mssm-superpotential}:
\begin{align}
W_{\Delta {\rm L} =1} &= \frac{1}{2} \lambda^{ijk} L_i L_j \bar{e}_k 
+{ \lambda'}^{ijk}L_i Q_j \bar d_k 
+ {\mu'}^iL_i H_u, \label{eq:susy-lepton-violation}
 \\
W_{\Delta {\rm B} =1} &= \frac{1}{2} {\lambda''}^{ijk} \bar u_i \bar d_j \bar d_k, \label{eq:susy-baryon-violation}
\end{align}
where $\lambda^{ijk}$ and ${\lambda''}^{ijk} $ are antisymmetric under the exchange of family indices $i$ and $j$. Terms in Eq. 
\eqref{eq:susy-lepton-violation} violate lepton number, while the term in Eq. \eqref{eq:susy-baryon-violation} violates baryon 
number. Simultaneous violation of both lepton and baryon numbers would result in a rapid proton decay, e.g. $p \to e^+\pi^0$ 
depicted in Fig. \ref{fig:proton-decay}, 
which is clearly in contradiction to the observed reality. In order to forbid these terms, a new symmetry is imposed, 
called \textit{R-parity}. It is defined for each particle as:
\begin{equation}
P_R \equiv (-1)^{3(B-L)+2s},
\end{equation}
where $B$, $L$ and $s$ are the baryon number, lepton number and spin of the particle, respectively. All SM particles and Higgs 
bosons are \textit{R-parity even} ($P_R=1$), while squarks, sleptons gauginos and higgsinos are \textit{R-parity odd} ($P_R=-1$).
MSSM is defined to conserve R-parity, which has important phenomenological consequences:
\begin{enumerate}
\item The lightest supersymmetric particle (LSP) with $P_R=-1$ must be absolutely stable. If it is neutral, then it constitutes a natural candidate for Dark Matter.
\item Each particle different from the LSP must eventually decay to a state containing an odd number of LSPs.
\item In collider experiments SUSY particles are produced in pairs.
\end{enumerate}

\begin{figure}[htb]
\centering
\includegraphics[width=0.8\textwidth]{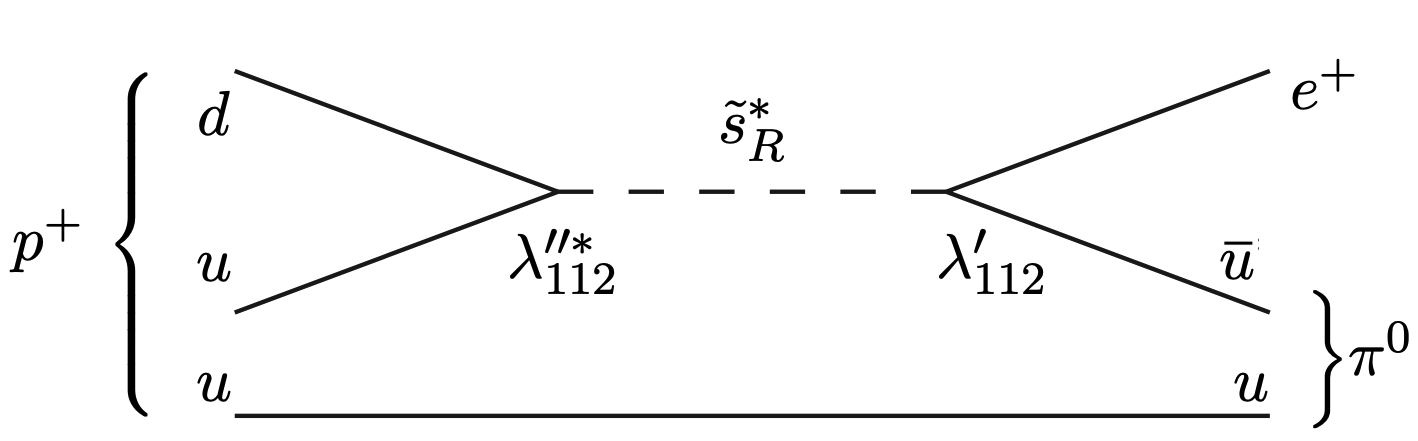}
\caption{\small Proton decay via an RPV-violating process described by operators in Eqs. \eqref{eq:susy-lepton-violation} and 
\eqref{eq:susy-baryon-violation}. In this process, proton decays to neutral pion $\pi^0$ and positron $e^+$, violating both lepton and baryon number conservation.}
\label{fig:proton-decay}
\end{figure}

As was already mentioned, Supersymmetry has to be broken, because we have not observed any sparticles yet. In order to 
maintain small corrections to the Higgs boson mass, it should be broken softly through the following terms:
\begin{equation}
\begin{split}
\label{eq:susy-mssm-soft}
\mathcal{L}^{\rm MSSM}_{\rm soft} =
& -\frac{1}{2}\left( M_3 \tilde g \tilde g + M_2 \widetilde W \widetilde W + M_1 \widetilde B \widetilde B \right) + h.c.\\
&-\left(\tilde u_R^* \mathbf{A_u} \tilde Q H_u - \tilde d_R^*\mathbf{A_d} \tilde Q H_d - \tilde e_R^*\mathbf{A_e} \tilde L H_d \right) + h.c.\\
&-\tilde Q^* \mathbf{m_Q^2} \tilde Q 
- \tilde L^* \mathbf{m_L^2} \tilde L
- \tilde u_R^* \mathbf{m_{\bar u}^2} \tilde u_R -  \tilde d_R^* \mathbf{m_{\bar d}^2} \tilde d_R - \tilde e_R^*\mathbf{m_{\bar e}^2} \tilde e_R\\
&-m^2_{H_u} H_u^* H_u - m^2_{H_d} H_d^* H_d - \left( b H_u H_d+h.c. \right),
\end{split}
\end{equation}
where all the fields in the 2nd, 3rd and 4th line of Eq. \eqref{eq:susy-mssm-soft} are scalars.
In Eq. \eqref{eq:susy-mssm-soft}, $M_1$, $M_2$ and $M_3$ are soft SUSY breaking mass terms of bino, wino and gluino, 
respectively. The second line of the equation contains trilinear scalar couplings, each of which is a complex $3\times 3$ matrix. The third 
line of Eq. \eqref{eq:susy-mssm-soft} contains mass terms for fermions. Finally, the last line of the equation contains soft SUSY-breaking contributions to the Higgs potential. It is expected that all soft supersymmetry breaking parameters should be related to 
the same energy scale $m_{\rm soft}$.

In the case of gluino, it is a colour octet fermion that cannot mix with anything, hence its physical mass is given by the soft SUSY 
breaking mass $|M_3|$. For sleptons and squarks situation is much more complicated, in principle one has to 
diagonalise $6\times 6$ matrix. However, the discussion can be simplified if one considers only the third generation and neglects 
the mixing between various families. Nevertheless, even in this simplified case, there are contributions to $\tilde t_L$ and $\tilde 
t_R$ that cannot be neglected. The mass terms for stops are:
\begin{equation}
\mathcal{L}_{\rm stop} = -
\begin{pmatrix}
\tilde{t}^*_L  & \tilde{t}^*_R 
\end{pmatrix}
\mathbf{m^2_{\tilde t}}
\begin{pmatrix}
\tilde{t}_L  \\ \tilde{t}_R 
\end{pmatrix},
\end{equation}
with the mass matrix given by:
\begin{equation}
\mathbf{m^2_{\tilde t}}\label{eq:susy-mssm-mix-stop}
=
\begin{pmatrix}
(\mathbf{m^2_Q})_{33} + m_t^2+\delta_u
&
v\left[ (\mathbf{A_u})_{33} s_\beta - \mu y_t c_\beta\right]\\
v\left[ (\mathbf{A_u})_{33} s_\beta - \mu y_t c_\beta\right]
&
(m^2_{\bar{u}})_{33} + m_t^2+\delta_{\bar u}
\end{pmatrix},
\end{equation}
with 
\begin{equation}
\delta_f= \left( T_f^3 - Q_f s^2_W \right)  M_Z^2 \cos 2\beta
\end{equation}
$s_W$ being sine if the weak mixing angle, and
\begin{align}
\langle H_u^0 \rangle &= \frac{v_u}{\sqrt{2}}, \\
\langle H_d^0 \rangle &= \frac{v_d}{\sqrt{2}}, \\
s_\beta &\equiv \sin \beta \equiv \frac{v_u }{v}, \\
c_\beta &\equiv \cos \beta \equiv \frac{ v_d }{v}, \\
v^2 &=  v_u^2 +  v_d^2 \approx (246 {\rm ~GeV})^2.
\end{align}

The mixing matrix in Eq. \eqref{eq:susy-mssm-mix-stop} can be diagonalised to give mass eigenstates $\tilde t_1$ and $\tilde 
t_2$, with $m^2_{\tilde t_1} < m_{\tilde{t}_2}^2$ by convention. Similarly, for sbottom and stau there are $\tilde b_1$, $\tilde 
b_2$, $\tilde \tau_1$ and $\tilde \tau_2$ mass eigenstates.

Charged winos and higgsinos can mix to form mass eigenstates named \textit{charginos}. 
{
In the basis 
$\psi^T = (
\widetilde W^+  \tilde H^+_u  \widetilde W^-  \tilde H_d^- )$, the mass terms of charginos are:
\begin{equation}
\mathcal{L}_{\rm chargino} = -\frac{1}{2} \psi^T \mathbf{M_{\tilde C}}\psi + h.c.
\end{equation}
with the mass matrix
\begin{equation}\label{eq:susy-mssm-chargino-mass}
\mathbf{M_{\tilde C}} = 
\begin{pmatrix}
0 & 0 &M_2 &  \sqrt{2} c_\beta M_W\\
0 & 0 &\sqrt{2} s_\beta M_W  & \mu\\
M_2 & \sqrt{2} s_\beta M_W  & 0 & 0\\
\sqrt{2} c_\beta M_W & \mu & 0 & 0
\end{pmatrix}
\end{equation}
}
The mass matrix in Eq. \eqref{eq:susy-mssm-chargino-mass} can be diagonalised, resulting in two mass eigenstates with masses:
\begin{equation}
\begin{split}
m^2_{\tilde C_1 \tilde C_2} = \frac{1}{2} \left[\left(
|M_2|^2 + |\mu|^2 + 2M_W^2\right)
\mp
\sqrt{
\left( |M_2|^2 + |\mu|^2 +  2M_W^2 \right)^2
-4 | \mu M_2 - M_W^2 \sin 2\beta |^2
}
\right]
\end{split}
\end{equation}
If we assume $||\mu| \pm M_2| \gg M_W$, charginos become approximately wino and higgsino with masses $|M_2|$ and $|\mu|$, respectively.

Neutral higgsinos and gauginos can mix in a similar manner to charginos, but in this case, there are four contributing fields: $\psi^0 = (
\widetilde B,  \widetilde W^0,  \tilde H_d^0,  \tilde H^0_u )$. The mass terms for \textit{neutralinos} are:
\begin{equation}
\mathcal{L}_{\rm neutralino} =  - \frac{1}{2} (\psi^0)^T \mathbf{M_{\widetilde N}} \psi^0 + h.c.
\end{equation}
with the $4 \times 4$ matrix $\mathbf{M_{\widetilde N}}$ given by:

\begin{equation}\label{eq:susy-mssm-neutrinomass}
\mathbf{M_{\widetilde N}} = \begin{pmatrix}
M_1 & 0 & -c_\beta s_W M_Z & s_\beta s_W M_Z \\
0 & M_2 & c_\beta c_W  M_Z & - s_\beta c_W M_Z \\
-c_\beta s_W M_Z  & c_\beta c_W  M_Z & 0 & -\mu \\
s_\beta s_W M_Z   & - s_\beta c_W M_Z  & -\mu & 0
\end{pmatrix}
\end{equation}
The matrix in Eq. \eqref{eq:susy-mssm-neutrinomass} can be diagonalised, resulting in four neutralino mass eigenstates. If $M_Z 
\ll |\mu \pm M_1|, |\mu\pm M_2|$ then neutralinos are approximately $\widetilde B, \widetilde W^0, (\tilde H^0_u \pm \tilde H_d^0 )/
\sqrt{2}$ with masses: $|M_1|, |M_2|, |\mu|, |\mu|$, respectively. If the lightest neutralino is the LSP, it can be a valid Dark Matter candidate.

MSSM is the simplest extension of the Standard Model, but other more complicated scenarios are also being studied. A natural 
extension of the MSSM is to allow for the R-parity violation by one or more of the terms in Eqs. \eqref{eq:susy-lepton-violation} 
and \eqref{eq:susy-baryon-violation}. Such scenarios are called \textit{R-parity violating (RPV) SUSY} \cite{Evans:2012bf}, and have to be carefully 
constructed in order to evade experimental constraints on lepton/baryon number violating processes. Another possibility is to 
consider local supersymmetry, called \textit{supergravity}\cite{Wess:1992cp}, in which an attempt to unify supersymmetry with gravity is made. In supergravity theories, the LSP is often \textit{gravitino} -- a spin-3/2 superpartner of the graviton. 

\FloatBarrier
\subsection{Long-lived sparticles}\label{sec:susy-llps}
\FloatBarrier

Supersymmetric scenarios can predict the existence of particles with long lifetimes. One reason is the existence of additional symmetries, e.g. R-parity in the MSSM causes the LSP to be absolutely stable. Long-lived particles might also emerge due to phase-space suppression or compressed mass spectrum.

\textbf{Sleptons} $\tilde l$ might be long-lived as next-to-the-lightest SUSY particles (NLSP) decaying to gravitino $(\tilde G )$ or a 
neutralino $\tilde \chi_1^0$ LSP. In gauge-mediated SUSY breaking (GMBS) 
models, decays of the $\tilde l$  NLPS to gravitino might be suppressed by 
the ``weak'' gravitational interaction \cite{Hamaguchi:2006vu}, resulting in partial compatibility with 
constraints on the Dark Matter abundance in super-weakly interacting massive 
particle models \cite{Feng:2015wqa}.
In other scenarios, e.g. the co-annihilation region in the MSSM, the 
lighter stau $\tilde \tau_1$ is a natural candidate for NLSP 
\cite{Jittoh:2005pq, Kaneko:2008re, Konishi:2013gda}. If the 
mass splitting between $\tilde \tau_1$ and $\tilde \chi_1^0$ is less 
than the mass of a tau, then $\tilde \tau_1$ will be long-lived.

\textbf{Gluino} cannot 
be directly detected, because it rapidly hadronises into colour-neutral 
hadrons, called \textit{R-hadrons}. Such states appear for example in 
Split SUSY \cite{Arkani-Hamed:2004ymt, Giudice:2004tc} due to 
extremely heavy squarks, which significantly suppress decays of $\tilde 
g$ to $\tilde q$ and quarks \cite{Arkani-Hamed:2004ymt, Arkani-Hamed:2004zhs}. The colour charge of gluino may be neutralised either by a gluon, leading to the formation of a neutral hadron, or by a colour-octet quark pair. 

\textbf{Squarks} might also form charged or neutral R-hadrons, for example in R-parity 
violating SUSY \cite{Evans:2012bf} and Gravitino Dark Matter model 
\cite{Diaz-Cruz:2007ewo}. 

\textbf{Charginos}, being a mixture of Wino $\widetilde W$ and Higgsino $\tilde h$, 
could be long-lived as LSPs in RPV models with small RPV couplings 
\cite{Bomark:2014rra} or as NLSPs in scenarios with gravitino LSP 
\cite{Kribs:2008hq}. Long lifetime might also result from small mass splitting between 
chargino and neutralino LSP $\tilde \chi_1^0$, e.g. in the focus-point region of 
the mSUGRA parameter space \cite{Gladyshev:2008ag} or in anomaly-mediated 
supersymmetry breaking (AMSB) scenarios 
\cite{Giudice:1998xp, Randall:1998uk}.

\textbf{Neutralinos} in the MSSM are usually LSP, 
hence absolutely stable due to the R-parity conservation. Such particles constitute a 
natural candidate for the Dark Matter  \cite{Jungman:1995df}. Other 
scenarios are also possible, e.g. RPV models with $\tilde \chi_1^0$ LSP and 
small RPV couplings \cite{Barbier:2004ez}. 

\section{Neutrino mass models}\label{sec:neutrino-mass}

The first indication that neutrinos are massive appeared in the 1960s 
when the Homestake experiment measured the flux of the solar (electron) neutrinos and reported a number which was much 
lower than nuclear theory predictions \cite{Davis:1968cp}. The discrepancy, called \textit{solar neutrino problem}, led to 
thorough
reconsideration of solar models and experimental methods. Eventually, it was understood that the result was correct and the 
discrepancy appeared because neutrinos were more complex than it had been anticipated. The reason behind the flux of electron 
neutrinos from the Sun being smaller than expected was that at least two neutrinos were massive and their mass matrix was not 
diagonal in the flavour eigenbasis. Neutrinos produced in the Sun and detected on Earth were pure flavour states, but a mixture of 
different mass eigenstates propagating with different velocities. In recent years, neutrino physics became a large and active 
field of research. In this section two most popular theoretical mechanisms leading to the origin of tiny neutrino masses are 
presented: seesaw and radiative models.

\subsection{Seesaw}\label{sec:seesaw}

When the Standard Model was proposed, neutrinos were assumed to be left-handed and massless. Now we know that they are 
massive, hence it is natural to consider right-handed neutrinos $\nu_R$. If they exist, their masses can be generated after the spontaneous electroweak 
symmetry breaking from interactions of the form $Y^\nu_{ij}{L}^{\dag i} (i\sigma_2 H^*) \nu_R^{ j}$. Since $ L$ and $H$ doublets have the 
same weak and hypercharge quantum numbers, the right-handed neutrino $\nu_R$ must be uncharged under both forces. Such particles are called \textit{sterile neutrinos}. 

The most general renormalizable mass terms for leptons are:
\begin{equation}\label{eq:seesaw-mass-general}
\mathcal{L}_{\rm mass} = 
-Y^e_{ij}  L^{i\dag} H e_R^{ j} 
- Y^\nu_{ij} L^{i\dag} (i\sigma_2 H^*) \nu_R^{ j} 
-  \frac{i}{2} M_{ij} \nu_R^\dag \sigma_2 \nu_R^{* j} + h.c,
\end{equation}
The second term in  \eqref{eq:seesaw-mass-general} corresponds to the ordinary Dirac mass, similarly to charged lepton masses.
The last term in Eq. \eqref{eq:seesaw-mass-general} describes the Majorana mass term for 
neutrinos. If right-handed neutrinos carry any conserved quantum numbers, e.g. lepton number, then the Majorana mass term is forbidden. 
One of the important consequences of the Majorana mass term is the explicit violation of the lepton number.

\noindent In order to investigate the masses of neutrinos, let us construct two Majorana fermions out of left- ($\nu_L$) and right-handed ($\nu_R$) neutrino fields:
\begin{align}
\psi_L &= \begin{pmatrix}
\nu_L \\
i \sigma_2 \nu_L^*
\end{pmatrix},
\\
\psi_R &= \begin{pmatrix}
i \sigma_2 \nu_R^* \\
\nu_R
\end{pmatrix},
\end{align}
which allows us to write neutrino mass terms in a concise form:
\begin{equation}\label{eq:seesaw-mass-matrix}
\mathcal{L}_{\rm mass} = -m_D \bar \psi_L \psi_R -\frac{M}{2} \bar \psi_R \psi_R = 
- \frac{1}{2}
\begin{pmatrix}
\bar \psi_L & \bar \psi_R
\end{pmatrix}
\begin{pmatrix}
0 & m_D \\
m_D & M
\end{pmatrix}
\begin{pmatrix}
\psi_L \\ \psi_R
\end{pmatrix},
\end{equation}
where $m_D \equiv vY^\nu/\sqrt{2}$ is the Dirac mass, and $M$ is the Majorana mass from Eq. \eqref{eq:seesaw-mass-general}.
Diagonalisation of
the mass matrix in Eq. \eqref{eq:seesaw-mass-matrix} will mix $\psi_L$ and $\psi_R$, and yield
two mass eigenstates, which we label $\nu_1$ and $\nu_2$.
The corresponding masses are:
\begin{equation}\label{eq:seesaw}
m_{1,2} = \sqrt{m_D^2+\frac{1}{4}M^2} \mp\frac{1}{2} M
\end{equation}
If the Majorana mass $M$ is much bigger than the Dirac mass $m_D$, then one of the neutrino mass eigenstates is very heavy, while 
the other is very light:
\begin{align}
m_1 &\approx \frac{1}{2}M \left( 1 + 2\frac{m_D^2}{M^2} \right) - \frac{1}{2} M =  \frac{m_D^2}{M} \ll m_D=\frac{(v Y^\nu )}{\sqrt{2}},  \\
m_2 &\approx \frac{1}{2}M \left( 1 + 2\frac{m_D^2}{M^2} \right) + \frac{1}{2} M =  M + \frac{m_D^2}{M} \approx M.
\end{align}

This mechanism is called \textit{seesaw},
and provides a theoretically aesthetic explanation of small masses of the observed neutrinos.

Neutrino masses might be studied in general using an effective dimension-5 
operator, called \textit{Weinberg operator}  \cite{Weinberg:1979sa}, which violates the lepton number conservation:
\begin{equation}\label{eq:seesaw-weinberg}
\mathcal{L}_5 = \frac{1}{\Lambda} (L H) (LH) 
\xrightarrow[]{\rm EWSB}
 \mathcal{L}_5 \ni -i \frac{ v^2}{2 \Lambda} \nu_L^T \sigma_2 \nu_L,
\end{equation}
where $\Lambda$ is the UV cutoff scale, $H$ and $L$ are, respectively, SM Higgs doublet and left-handed lepton doublet. After 
electroweak symmetry is spontaneously broken, the Weinberg operator in Eq. \eqref{eq:seesaw-weinberg} generates Majorana 
mass term for left-handed neutrinos. A crucial aspect of the operator is that it permits only a restricted number of UV completions. 
If we consider a minimal extension of the Standard Model with only one additional multiplet, then there are only three possibilities 
\cite{Ma:1998dn} allowing to obtain the Weinberg operator at the tree level: \textit{type I}, \textit{type II}, \textit{type III} seesaw models.

\textbf{Minimal type I seesaw}  \cite{Minkowski:1977sc, Yanagida:1979as, Gell-Mann:1979vob, Yanagida:1980xy, Glashow:1979nm, Mohapatra:1979ia} 
is a model in which SM is augmented with a right-handed neutrino $\nu_R$, transforming as a singlet $(1,1,0)$ under the SM 
gauge group. This right-handed neutrino has a Majorana mass term and interacts with a single generation of SM leptons through 
Yukawa interaction. If Yukawa coupling is of the order of $1$, Majorana mass has to be of the order of 
$\sim10^{14}~\rm GeV$ to
provide mass for the SM neutrinos of order $0.1~\rm eV$.

\textbf{Minimal type II seesaw} \cite{Schechter:1980gr, Konetschny:1977bn, Cheng:1980qt, Lazarides:1980nt, Mohapatra:1980yp} is a 
model, in which an additional Higgs field $\Delta$ with mass $M_\Delta$ is introduced in a triplet representation of $SU(2)_L$. 
The field transforms as $(1,3,1)$ under the SM gauge group. Light neutrino masses originate from Majorana mass terms for the 
left-handed neutrino fields, and are given by $m_{\nu_L} \approx Y v_{\Delta}$, with $v_{\Delta}$ being the vacuum expectation 
value of the neutral component of the Higgs triplet, and
$Y$ being a Yukawa coupling. After electroweak symmetry is spontaneously broken, SM Higgs doublet mixes with the new triplet 
via a dimensionful parameter $\mu$, leading to a relation between the mass of the neutrino and Higgs triplet:
$m_{\nu_L} \sim Y \frac{\mu v_0^2}{M^2_\Delta}$. An important aspect of type II seesaw models is that they do not require 
right-handed neutrinos. Another, phenomenologically interesting property is the existence of doubly charged constituents of the 
Higgs triplet.

\textbf{Minimal type III seesaw} \cite{Foot:1988aq} is similar to type I and type II models. In this scenario a new fermionic triplet in 
adjoint representation is introduced, which transforms as $(1,3,0)$ under the $SU(3)_C\times SU(2)_L \times U(1)$ gauge group. 
The neutrino mass matrix has the same form as in type I but features charged heavy leptons. The new physics scale from Eq. 
\eqref{eq:seesaw-weinberg} is replaced by the mass of new leptons.

While there are three basic types of seesaw scenarios, the mechanism giving mass to SM neutrinos might be more complicated, 
hence a multiplicity of models were introduced, e.g. \textit{Inverse} \cite{Mohapatra:1986aw, Mohapatra:1986bd, Bernabeu:1987gr,Gavela:2009cd } and \textit{Linear} \cite{Akhmedov:1995ip, Akhmedov:1995vm} seesaw models of type I and III, 
and hybrid models combining two or more seesaw mechanisms. Discussion of seesaw models in this thesis is limited to the 
necessary basics, curious readers are encouraged to see one of the many reviews on the topic, e.g. Deppisch et al.\cite{Deppisch:2015qwa} and Cai et 
al. \cite{Cai:2017mow}.

The non-zero masses of the SM neutrinos result in severe phenomenological consequences, regardless of the exact mechanism 
leading to their emergence. In the flavour basis, neutrino couplings to the W boson are diagonal, but masses are not:

\begin{equation}
\mathcal{L}_{\nu W} =
-\frac{g_2}{\sqrt{2}} 
\left(
 e_L^\dag \slashed W \nu_{L e} +
 \mu_L^\dag  \slashed W \nu_{L \mu} +
 \tau_L^\dag \slashed W \nu_{L \tau} +
h.c.
\right).
\end{equation}
Neutrino mass eigenstates $\nu_{L 1}, \nu_{L 2}, \nu_{L 3}$ are related to flavour eigenstates by a unitary transformation:
\begin{equation}
\mathcal{L}_{\nu W} = - \frac{g_2}{\sqrt{2}} U^{ij} 
\left(
 e_{L i}^\dag \slashed W \nu_{L j} + h.c.
\right),
\end{equation}
where $\nu_{L\tau} = U^{\tau 1}\nu_{L 1} + U^{\tau 2}\nu_{L 2} +U^{\tau 3}\nu_{L 3}$, and analogously for other flavours. Unitary 
matrix $U$ is called \textit{Pontecorvo-Maki-Nakagawa-Sakata (PMNS) matrix}. It can be written as:
\begin{equation}\label{eq:seesaw-pmns}
U = \begin{pmatrix}
c_{12} c_{13} 
& s_{12} c_{13} 
& s_{13} e^{-i\delta} \\
-s_{12}c_{23} -c_{12}s_{23}s_{13}e^{i\delta} 
& c_{12}c_{23} -s_{12}s_{23}s_{13}e^{i\delta} 
& s_{23}c_{13} \\
s_{12}s_{23} -c_{12}c_{23}s_{13}e^{i\delta} 
& -c_{12}s_{23} -s_{12}c_{23}s_{13}e^{i\delta} 
&c_{23}c_{13}
\end{pmatrix}
\begin{pmatrix}
1 & &\\
& e^{i\frac{\alpha_{12}}{2}} & \\
& &  e^{i\frac{\alpha_{31}}{2}}
\end{pmatrix}.
\end{equation}
In Eq. \eqref{eq:seesaw-pmns} a similar parametrisation as in the Eq. \eqref{eq:sm-ckm-parametrisation} is used, but the angles are different 
from the CKM mixing angles. The phase $\delta$ is called the \textit{Dirac phase}. If neutrinos are Dirac fermions, this is the only 
non-zero phase. If there are Majorana mass terms contributing to neutrino masses then there are two additional phases $
\alpha_{12}$ and $\alpha_{31}$. 

Neutrino oscillation is a phenomenon in which neutrino changes its flavour over time. One can consider an experiment on the 
Earth designed to detect electron neutrinos emitted from the Sun. Pure electron neutrinos are produced, which are linear 
combinations of the mass eigenstates. Each eigenstate travels with a different speed, leading to an effective rotation in the flavour 
basis, hence the number of electron neutrinos detected on the Earth will not match the number of neutrinos produced in the Sun.

The hierarchy of the neutrino masses is currently unknown, we only know that at least two out of three mass eigenvalues are non-zero. Situation when $m_3>m_2>m_1$ 
is called the \textit{normal hierarchy}, while $m_2>m_1>m_3$ is the \textit{inverted hierarchy}.

\FloatBarrier
\subsection{Radiative mass generation}\label{sec:radiative-mass-gen}
\FloatBarrier

The Weinberg operator in Eq. \eqref{eq:seesaw-weinberg} is the lowest order and the simplest way to achieve small neutrino 
masses using only SM particles and gauge interactions. However, if one considers its realisations beyond the tree level, neutrino 
Majorana masses might be generated radiatively, with loop suppression factors explaining their smallness. \textit{Radiative neutrino mass generation} models require the absence of the tree-level contributions to the neutrino mass, which can be achieved 
either by the absence of the necessary particles, e.g. right-handed neutrino singlet, or by additional symmetries imposed on the 
theory. The one-loop radiative neutrino mass generation models were introduced by Zee \cite{Zee:1980ai} and Hall and Suzuki 
\cite{Hall:1983id}. Two-loop versions were developed by Cheng and Li \cite{Cheng:1980qt}, Zee \cite{Zee:1985id}, and Babu 
\cite{Babu:1988ki}, while three-loop models were proposed by Krauss et al.\cite{Krauss:2002px}. In this thesis, we will consider 
and describe a recent one-loop model introduced by Hirsch et al. \cite{R:2020odv}. In the model, the SM is augmented with two 
additional scalar fields, $S_1$ and $S_3$, which are singlet and triplet under $SU(2)_L$, respectively. In addition, three pairs of 
vector-like spin-1/2 fermions are introduced, $(F_i \bar F_i)(i=1,2,3)$, in $SU(2)L$ doublet representations. All new fields and 
their quantum numbers are listed in Tab. \ref{tab:hirsch}.

\begin{table}[]
\caption{\small New fields and their quantum numbers introduced in the one-loop radiative neutrino mass generation model from \cite{R:2020odv}. The lepton number conservation is broken in the model, hence the lepton number assignment in this table should be understood as valid in the limit $\lambda_5 \to 0$.}
\centering
\begin{tabular}{c|c|c|c|c}
                       & \textbf{$\mathbf {S_1}$} & \textbf{$\mathbf{ S_3}$} & \textbf{$\mathbf {F_i}$} & \textbf{$\mathbf{\bar{F}_i}$} \\ \hline
\textbf{Spin}          & 0              & 0              & $\frac{1}{2}$  & $\frac{1}{2}$       \\
\textbf{$\mathbf{ SU(3)_C}$}     & 1              & 1              & 1              & 1                   \\
\textbf{$\mathbf {SU(2)_L}$}     & 1              & 3              & 2              & 2                   \\
\textbf{$\mathbf {U(1)_Y}$}      & 2              & 3              & $\frac{5}{2}$  & $-\frac{5}{2}$      \\
\textbf{Lepton number} & -2             & -4             & -3             & 3                   \\ 
\end{tabular}\label{tab:hirsch}
\end{table}

The Lagrangian density of the new fields has the following form:

\begin{equation}\label{eq:neutrino-lagrangian}
\begin{split}
\mathcal{L}_{\rm BSM} &= \mathcal{L}_{\rm KIN}  -\left[ (h_{ee})_{ij} (e_R^{ i})^C (e_R^{ j})^C S_1^\dag +
(h_F)_{ij} L_i F_j S_1^\dag
+ (h_{\bar F})_{ij} L_i \bar F_j S_3 + h.c. \right] +\\
&+ \lambda_2 |H|^2 |S_1|^2 + \lambda_{3a} |H|^2 |S_3|^2 + \lambda_{3b} |H S_3|^2 +\\
&- \left[ \lambda_5 H H S_1 S_3^\dag + h.c. \right] +\\
&+ \lambda_4 |S_1|^4 + \lambda_{6a} |S_3^\dag S_3 |^2 +  \lambda_{6b} |S_3 S_3 S_3^\dag S_3^\dag | 
+ \lambda_7 |S_1|^2 |S_3|^2,
\end{split}
\end{equation}
where Dirac mass terms for the new fields were included in $\mathcal{L}_{\rm KIN}$, and 
$(e_R^i)^C \equiv (e_R^i)^T\sigma_2$ stands for the charge 
conjugated right-handed SM charged lepton of generation $i$.

The term proportional to $\lambda_5$ 
breaks the lepton number conservation, therefore in the limit $\lambda_5 \to 0$, BSM fields have definite lepton number 
assignments, as listed in Tab. \ref{tab:hirsch}. Note that in all new operators BSM fields appear an even number of times, except for the term with the 
$h_{ee}$ coupling. One can therefore introduce BSM parity and assign odd (even) charges to BSM (SM) fields. In the limit $h_{ee}\to 
0$, the parity is exact symmetry and the lightest BSM particle is absolutely stable, therefore $h_{ee}$ controls the decay lifetime of the 
lightest BSM particle. $h_{ee}=0$ is ruled out by the cosmology, since it would lead to a charged stable relic.

A non-zero value of $\lambda_5$ results in a violation of the lepton number, hence one can take $\lambda_5$ to be small and 
the theory will be technically natural in the sense of t'Hooft. Similarly, the absence of $(h_F)_{ij}$ or  $(h_{\bar{F}})_{ij}$ leads to 
the conservation of the lepton number, and if both couplings are simultaneously zero, then BSM fermion number symmetry, i.e.
$F_i \to e^{i\theta} F_i, \bar F_i \to e^{-i\theta} \bar F_i$, is restored. Therefore, a model in which both $(h_F)_{ij}$ and  $(h_{\bar{F}})_{ij}$ are small is technically natural and radiatively stable.

In the discussed model, neutrino masses emerge radiatively from the one-loop diagram shown in Fig. \ref{fig:neutrino-gen}. The Weinberg operator takes the form:
\begin{equation}
\frac{\lambda_5 N_C}{32 \pi^2 \Lambda^2} \left[ h^T_{\bar F} m_F h_F + h_F^T m_F h_{\bar F} \right]_{ij} \cdot L_i H L_j H,
\end{equation}
where $N_C$ is the number of colours of particles in the loop ($N_C=1$ for now, but later a coloured version of the model will be discussed) and $\Lambda$ is the mass scale of BSM fields. If we assume that 
all heavy masses are of the same order of magnitude, and that $h_F$ and $h_{\bar F}$ are diagonal, one can roughly estimate 
neutrino masses by:
\begin{equation}\label{eq:neutrino-mass}
m_\nu \sim 0.05 \cdot N_C \cdot \left( \frac{\lambda_5}{10^{-6}} \right) \left( \frac{h_F h_{\bar{F}}}{10^{-4}}\right) 
\left( \frac{ 1~\rm TeV}{\Lambda} \right) \rm eV.
\end{equation}

\begin{figure}[htb]
\centering
\includegraphics[width=0.9\textwidth]{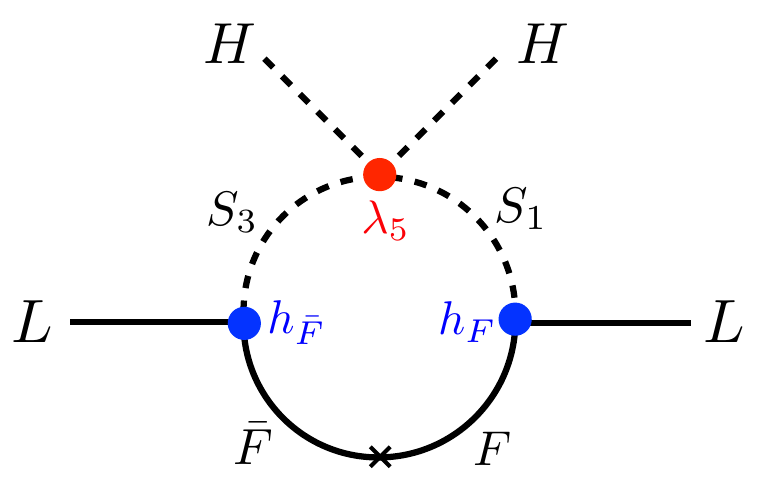}
\caption{\small Neutrino mass generation diagram for a model from \cite{R:2020odv}.}
\label{fig:neutrino-gen}
\end{figure}

New BSM fields have large $SU(2)_L$ and $U(1)_Y$ charges, resulting in mass eigenstates with high electric charges, e.g. scalar 
triplet consists of doubly $S^{+2}$, triply $S^{+3}$ and quadruply $S^{+4}$ charged particles. Doubly charged scalar singlet 
$S_1$ mixes with the doubly charged component of the triplet, leading to the following mass matrix:
\begin{equation}
\begin{pmatrix}
m_{S_1}^2 - \frac{1}{2} \lambda_2 v^2 & \frac{1}{2}\lambda_5 v^2 \\
 \frac{1}{2}\lambda_5 v^2 & m_{S_3}^2 - \frac{1}{2} \lambda_{3a} v^2,
\end{pmatrix}
\end{equation}
where $v$ denotes the SM Higgs vacuum expectation value. Interestingly, if 
\begin{equation}\label{eq:neutrino-assumption}
m_{S_3} \ll m_{S_1}, m_F,
\end{equation}
then multicharged triplet 
states might be long-lived. From now on, such hierarchy of masses will be assumed. Moreover, it will be also assumed that 
$\lambda_i v^2 \ll m^2_{S_3}$, for $i=2, (3a), (3b), 5$, resulting in members of the scalar triplet being nearly mass degenerate. 
For $m_{S_3} \ll m_{S_1}$, the lighter $S^{+2}$ and the heavier  $S_1^{+2}$ doubly charged mass eigenstates are approximately 
pure singlet and pure triplet, respectively:

\begin{align}
m_{S_1^{+2}}^2 &\approx m_{S_1}^2 - \frac{1}{2} \lambda_2 v^2,\\
m_{S^{+2}}^2 &\approx m_{S_3}^2 - \frac{1}{2} \lambda_{3a} v^2.
\end{align}

The mass degeneracy of the triplet components is lifted after the electroweak symmetry breaking, resulting in:
\begin{align}
m^2_{S^{+3}} &= m^2_{S_3} - \frac{1}{2} \left( \lambda_{3a} + \frac{1}{2} \lambda_{3b} \right) v^2,\\
m^2_{S^{+4}} &= m^2_{S_3} - \frac{1}{2} \left( \lambda_{3a} + \lambda_{3b} \right) v^2.
\end{align}

When the assumption in Eq. \eqref{eq:neutrino-assumption} holds, triplet mass eigenstates $S^{+4},S^{+3}, S^{2+}$ are lighter 
than any other BSM particles. However, their decays to SM particles are phase-space suppressed, because of their large electric 
charge and the necessity to produce a high multiplicity of SM particles in the final state.

The radiative neutrino mass generation model considered in this section can be easily modified to obtain a coloured 
version. The first step is to promote the BSM fields to colour (anti-)triplets. Next, the interaction leading to breaking of the BSM 
parity and decay of the $S_1$ to SM particles has to be modified accordingly:
\begin{equation}\label{eq:neutrino-colour-change}
(h_{ee})_{ij} (e_R^i)^C (e^j_R)^C S_1^\dag \to (h_{ed})_{ij} (d_R^i)^C (e^j_R)^C \tilde S_1^\dag,
\end{equation}
where we have introduced a notation in which coloured versions of the BSM fields are indicated with a tilde on top. The Eq. 
\eqref{eq:neutrino-colour-change} implies that $\tilde S_1$ and $S_1$ must differ in the hypercharge by $-2/3$. Moreover, by assuming 
that the breaking of the lepton number in the model happens only through the $\lambda_5$ term, not by the operator in Eq. 
\eqref{eq:neutrino-colour-change}, the coloured singlet scalar $\tilde S_1$ must be assigned $L=-1$. Lepton number assignment of other fields must be modified accordingly. Quantum numbers of fields in the coloured model are listed in Tab. \ref{tab:hirsch-color}.

\begin{table}[]
\caption{\small New fields and their quantum numbers introduced in the coloured version of the one-loop radiative neutrino mass generation model from \cite{R:2020odv}.}
\centering
\begin{tabular}{c|c|c|c|c}
                       & \textbf{$\mathbf {\tilde S_1}$} & \textbf{$\mathbf{ \tilde S_3}$} & \textbf{$\mathbf {\tilde F_i}$} & \textbf{$\mathbf{\tilde{\bar{F}}_i}$} \\ \hline
\textbf{Spin}          & 0              & 0              & $\frac{1}{2}$  & $\frac{1}{2}$       \\
\textbf{$\mathbf{ SU(3)_C}$}     & $\bar 3$              & $\bar 3$              & $3$              & $\bar 3$                   \\
\textbf{$\mathbf {SU(2)_L}$}     & 1              & 3              & 2              & 2                   \\
\textbf{$\mathbf {U(1)_Y}$}      & $\frac{4}{3}$              & $\frac{7}{3}$              & $\frac{11}{6}$  & $-\frac{11}{6}$      \\
\textbf{Lepton number} & -1             & -3             & -2             & 2                   \\ 
\end{tabular}\label{tab:hirsch-color}
\end{table}

The Lagrangian of the coloured model is the same as in Eq. \eqref{eq:neutrino-lagrangian}, with the replacement of fields with 
their coloured versions and one term modified
according to Eq. \eqref{eq:neutrino-colour-change}. The mechanism and formulas related to neutrino mass generation remain 
unchanged, however, one has to put $N_C=3$ in Eq. \eqref{eq:neutrino-mass}. The shift of the hypercharge implies a shift in 
electric charge by $-2/3$, hence the coloured triplet mass eigenstates have charges $Z=4/3$, $7/3$ and $10/3$. 


    \chapter{Searching for long-lived BSM particles}\label{chap:three}
\section{ATLAS and CMS}

\subsection{Detector set up}\label{sec:atlascms-detector}
\textit{A Toroidal LHC Apparatus (ATLAS)} is a major general-purpose experiment 
located in an underground cavern around the IP1 of 
the LHC \cite{ATLAS:2008xda}. The detector has a forward-background-symmetric cylindrical set-up covering 
nearly $4\pi$ in solid angle. ATLAS is composed of an inner 
detector (ID) tracking system to measure trajectories of charged particles, surrounded by 
a thin superconducting solenoid providing a 2 T field, followed by a calorimeter system 
allowing to measure the energy of particles interacting electromagnetically or 
hadronically,
and a muon spectrometer (MS) inside toroidal magnets in order to provide additional 
tracking information for muons. The schematic picture of the detector is depicted in Fig. \ref{fig:atlas-det}.

The inner tracking detector is organised in three concentric regions covering 
$|\eta|<2.5$, where $\eta$ is the pseudorapidity of the particle defined as 
$\eta \equiv - \ln \left(\tan \frac{\theta}{2} \right)$, with $\theta$ being the angle between the particle's three-momentum and positive direction of the beam axis.
The outermost layer (TRT) is built out of 4-mm-diameter cylindrical drift tubes 
and covers $|\eta|<2$ in the radial region 60-100 cm. The middle layer, covering the 
radial region from 30 cm to 60 cm, is made of silicon microstrip detectors (SCT) and covers $|\eta|<2.5$. 
The 
innermost detector (IBL) at radial distances from 3.4 cm to 13 cm from the LHC beam 
line is composed of silicon pixel detectors. Compared to TRT and SCT, the innermost 
detector is characterised by reduced thickness, smaller pixels, faster electronics,
and provides a charge measurement with lower resolution and dynamic range.

The calorimeter system covers $|\eta|<4.9$ and consists of the electromagnetic 
calorimeter (ECAL) and hadronic calorimeter (HCAL) parts. The ECAL system within 
$|\eta|<3.2$ is composed of barrel and endcap lead/liquid-argon (LAr) calorimeters 
characterised by a high granularity. In addition, a thin LAr presampler at $|\eta|<1.8$ 
is installed to correct for the energy loss in the material upstream of the 
calorimeters. HCAL system is made of steel/scintillator-tile calorimeters distributed 
into a central barrel and 2 extended-barrel cylindrical structures within $|\eta|<1.7$, 
and 2 copper/LAr endcap hadronic calorimeters.

The muon spectrometer consists of separate trigger and high-precision tracking 
chambers installed to measure the deflection of muons in a magnetic field. 
The MS 
trigger system, designed to select events with muon candidates, 
covers $|\eta|<2.4$, with thin-gap chambers in the endcap region and 
resistive-plate chambers in the barrel.
Precision chambers cover $|\eta|<2.7$ with 3 
layers of monitored drift tubes augmented with cathode-strip chambers in the forward region.

The amount of data produced in the LHC is enormous\footnote{1 Petabyte (PB) = 1000 Terabytes (TB).}, about 1PB/s. Therefore, a system called \textit{trigger} is needed, which selects and registers only the most interesting events.  
The ATLAS detector relies on a two-level trigger system to select events for the 
analysis. The first-level (L1) trigger implemented in the hardware allows for reduction of the 
event rate from $\sim 1~\rm{GHz}$ down to 100 kHz, by utilising a subset of detector information.
Data selected by the L1 trigger is passed to a software-based high-level trigger (HLT), which allows to 
reduce the accepted event rate even further, typically to 1 kHz.

\textit{Compact Muon Solenoid (CMS)} is another major general-purpose experiment
located at the IP5 of the LHC \cite{CMS:2008xjf}. The key feature of its design is a 
superconducting solenoid with a diameter of $6~\rm m$, providing a 3.8 T 
magnetic field, and enabling the reconstruction of charged particles' 
trajectories as 
they traverse the tracking system. CMS tracker, electromagnetic (ECAL) and 
hadronic (HCAL) calorimeter are all located within the volume encapsulated 
by the solenoid. The muon detection system comprises gas-ionisation chambers embedded in the steel flux-return yoke located outside of the solenoid. The schematic drawing of the CMS detector is shown in Fig. \ref{fig:cms-det}.

The CMS tracking system consists of two main parts: the smaller inner pixel detector covering the radial region from
4 to 15 cm,
and the larger silicon strip detector located from 25 to 110 cm from the beam line. The CMS pixel detector consists of three concentric cylindrical barrels and four 
fan-blade disks closing the barrel ends. The pixel system provides efficient three-pixel coverage in $|\eta|<2.2$ and efficient 
two-pixel coverage in $|\eta|<2.5$. The CMS silicon strip consists of three main subsystems. The Tracker Inner Barrel and Disks (TIB/TID) extend in radius to 55 cm and are composed of four barrels augmented with three disks at each end. The TIB/TID is enclosed within the Tracker Outer Barrel (TOB), which has an outer radius of 116 cm and is composed of 6 barrel layers. The TOB covers the 
$|z| < 118~\rm cm$ region, beyond which there are Tracker End Caps (TEC). The TEC instruments the region $124 < |z|<280~\rm cm$ and $22.0 < r < 113.5~\rm cm$. All TEC comprise of nine disks, each with up to seven rings of radial strip detectors \cite{CMS:2010vmp}.

The calorimetric system of the CMS detector consists of electromagnetic (ECAL) and hadronic (HCAL) calorimeters. The 
electromagnetic calorimeter is built out of lead tungstate crystals, covering $|\eta|<1.48$ in the barrel region, and 
$1.48 < |\eta| < 3.0$ in two endcap regions. The HCAL consists of four subdetectors, a barrel detector providing the coverage in 
pseudorapidity $|\eta| < 1.3$, two endcap subdetectors covering $1.3 < |\eta| < 3.0$, two forward detectors covering 
$2.8 < |\eta| < 5.0$, and a 
detector located outside of the solenoid and covering $|\eta| < 1.3$.

The CMS muon system (MS) consists of three types of gas-ionisation chambers: resistive-plate chambers (RPCs), drift tube chambers (DTs), and cathode strip 
chambers (CSCs). 
RPCs are double-gap chambers, providing timing information for the muon trigger.
DTs are divided into drift cells, allowing experimentalists to determine the position of a 
muon by measuring the drift time to an anode wire of a cell with a shaped electric field.
CSCs function as standard multi-wire proportional counters, but add a segmented cathode strip readout.
The coverage in pseudorapidity of DTs, CSCs and RPCs is $|\eta|<1.2$, $0.9<|\eta|<2.4$ and $|\eta|<1.9$, respectively. The 
placement of chambers is organised to maximise the coverage and provides some overlap where it is possible. The chambers in 
both barrel and the endcap layers are grouped into four ``muon stations'', separated by the steel absorber of the magnet return yoke. In the barrel, RPC and DT stations are situated in 5 wheels along the beam direction.

CMS experiment relies on a two-tiered trigger system, similar to ATLAS. The first level trigger (L1) is based on custom hardware processors. It 
utilises information from the calorimeters and muon detectors to reduce the event rate from $1~\rm{GHz}$ down to 100 kHz. The second level of the system, 
called the high-level trigger (HLT), comprises a farm of processors executing a specific version of a full event reconstruction, 
and allows to lower the event rate down to 1 kHz.

Although ATLAS and CMS experiments differ in size, placement, parameters and choice of the subdetectors used, they have much in common, in particular:
\begin{itemize}
\item Both experiments exhibit a rotational symmetry in the azimuthal plane, thanks to subdetector placement in concentric cylinders around the beam line.
\item The experiments have large coverage in $|\eta|$, however, their sensitivity in forward and backward regions is typically worse than in the barrel.
\item ATLAS and CMS subdetectors are grouped into three main components: the inner tracking detector, electromagnetic and hadronic calorimeters, and outermost muon system.
\item Data is collected in a continuous manner, which requires fast electronics and two-level trigger system.
\item Thanks to their size and many subdetectors, ATLAS and CMS can test a variety of interesting theoretical scenarios, both SM and New Physics. Interesting results published by one of the collaborations are often verified by the measurements in the other detector.
\item Both experiments operate in similar conditions with respect to the beam properties, e.g. with the same centre-of-mass energy and luminosity.
\item Large luminosity and continuous data taking result in the so-called \textit{pileup}. Pileup is a pollution of the final state of the 
event caused by multiple proton collisions within a single bunch-crossing. Removing pileup is a crucial experimental challenge in 
face of the planned HL-LHC upgrade.
\end{itemize}
The similarities between ATLAS and CMS listed above justify treating both experiments interchangeably in some of the phenomenological discussions. 

\begin{figure}[!thbp]
  \begin{subfigure}[b]{0.49\textwidth}
    \includegraphics[width=\textwidth]{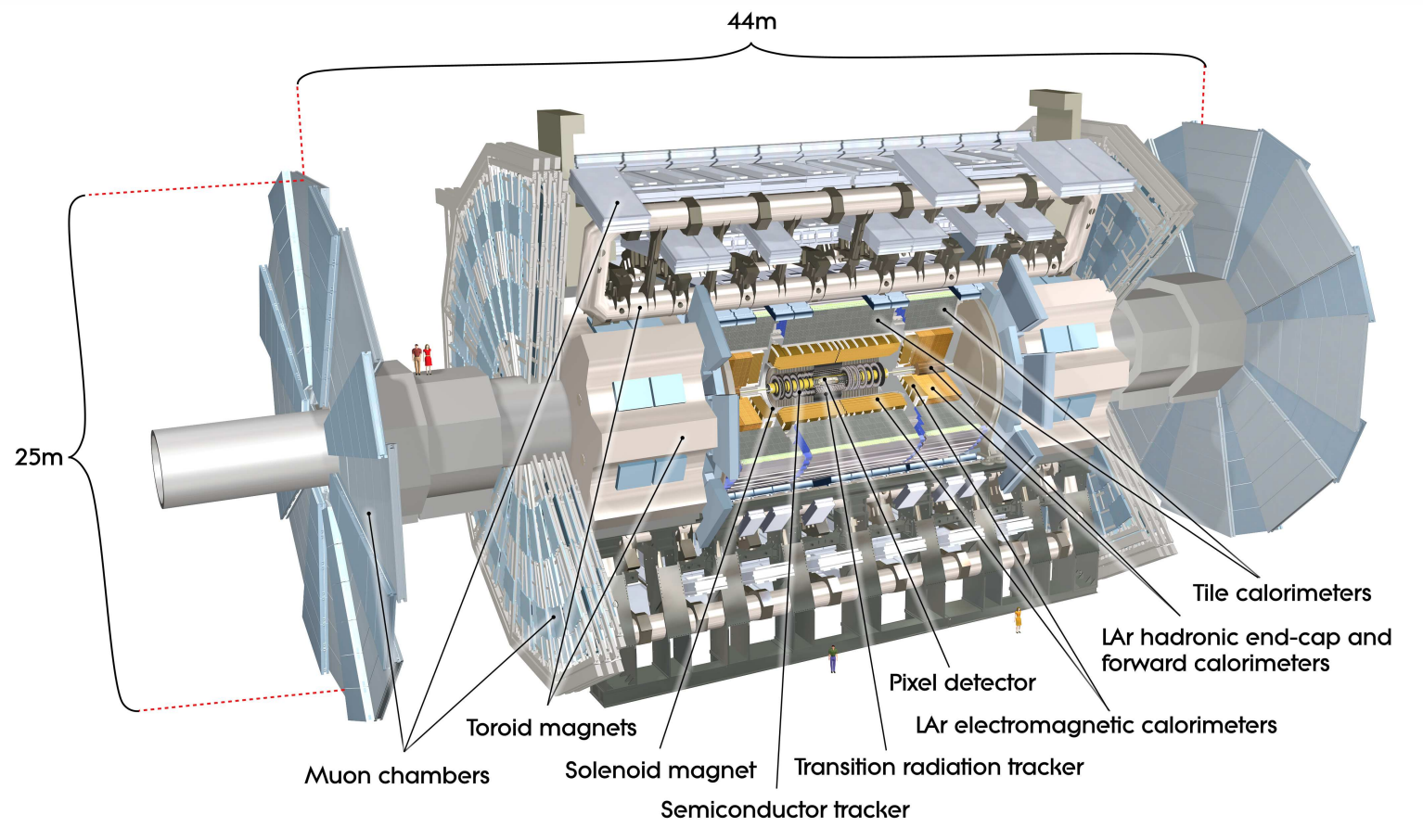}
    \caption{\small ATLAS detector}
    \label{fig:atlas-det}
  \end{subfigure}
  \hfill
  \begin{subfigure}[b]{0.49\textwidth}
    \includegraphics[width=\textwidth]{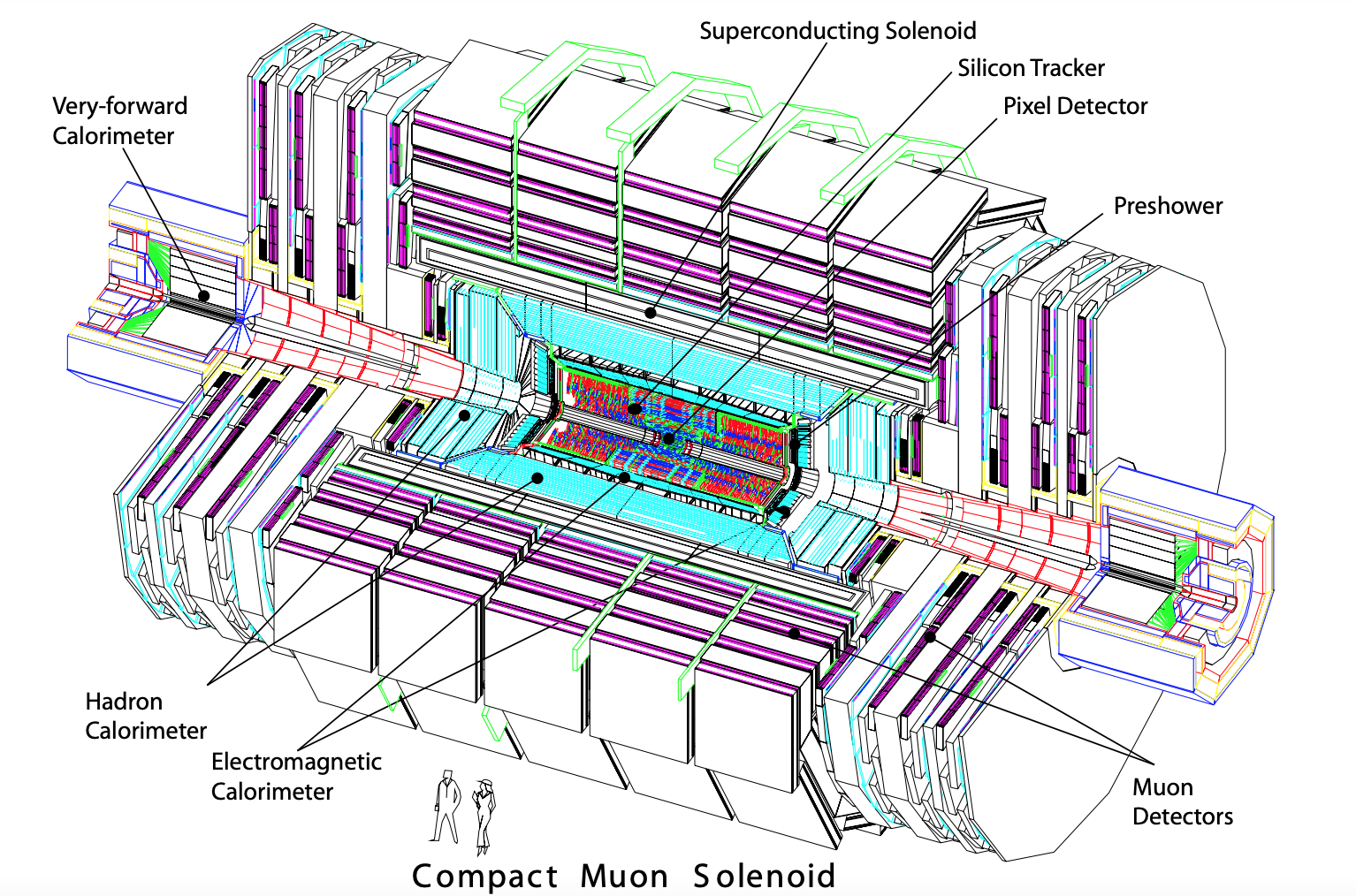}
    \caption{\small CMS detector}
    \label{fig:cms-det}
  \end{subfigure}
  \caption{\small Schematic drawings of ATLAS\cite{ATLAS:2008xda} (left) and CMS\cite{CMS:2008xjf} (right) detectors. Both detectors exhibit a very similar design, all subdetectors are arranged in concentric cylinders, the inner-most layer is responsible for tracking, the middle layer consists of electromagnetic and hadronic calorimeters, and the outermost layer of detectors comprises the muon system. Images taken from \cite{ATLAS:2008xda, CMS:2008xjf}. }
\end{figure}

\subsection{Searches for LLPs}\label{sec:atlascms-searches}

Experimental searches for BSM long-lived particles at the ATLAS and CMS experiments are usually driven by characteristic 
detector signatures rather than {the} theory. In this section, we discuss the most important search channels for LLPs, schematically depicted in Fig. \ref{fig:llp-overview}.

\begin{figure}[!tbh]
\includegraphics[width=\textwidth]{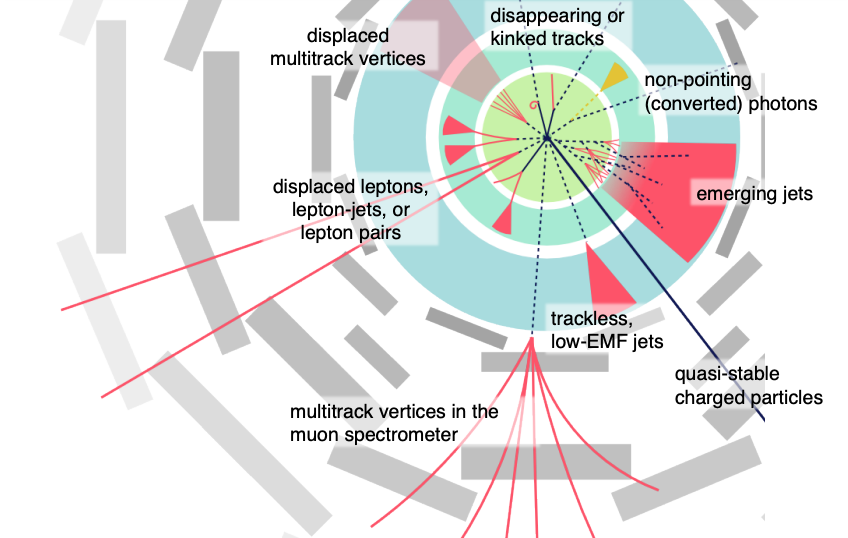}
\caption{\small Overview of different signatures of long-lived particles at the ATLAS and CMS detectors. Image taken from \cite{Alimena:2019zri}, based on a graphic from \cite{llp_overview}. }
\label{fig:llp-overview}
\end{figure}

\myparagraph{Disappearing or kinked tracks}
Trajectories of charged long-lived particles can be reconstructed in the tracker. 
If a charged LLP decays inside the inner detector into a pair of new particles, one of which is electrically charged and the other neutral, then the decay
can be observed as a kinked or disappearing 
track signature.
The track is called ``disappearing'' if it ends before reaching the outer layers of the tracker. If a track changes its 
direction at some point in the tracker, creating a characteristic kink shape ``>'', then the track is said to be ``kinked''.
The neutral decay product is invisible to the tracker, but the charged particle could be measured, if it is hard 
enough, resulting in a kinked track, which appears because of the four-momentum carried away by the neutral particle. 
If the charged decay product is soft, it will not be properly reconstructed by the tracker, and the signature will be of a 
disappearing track. In the latter situation, there will be also no sign of the charged particle in the electromagnetic calorimeter.

Searches for disappearing and kinked tracks have been carried out by both ATLAS \cite{ATLAS:2022rme, ATLAS:2017oal, ATLAS:2012jp} and CMS \cite{CMS:2020atg, CMS:2018rea}.

\myparagraph{Displaced multitrack vertices}
Most of the charged particle tracks originate from the primary vertex, which is 
the reconstructed point of the collision. However, often events contain tracks 
emerging from vertices shifted away from the collision point, such vertices are 
called ``displaced''. When a neutral LLP decays into coloured particles, the experimental signature is a displaced vertex with multiple outgoing tracks.

ATLAS has searched for displaced vertices emerging from decays of dark photons to quarks in \cite{ATLAS:2022izj}, 
Higgs decaying to hidden sector bosons \cite{ATLAS:2021jig}, and long-lived gluinos \cite{ATLAS:2017tny}. 
CMS has conducted analyses looking for 
long-lived sparticles \cite{CMS:2021tkn, CMS:2020iwv, CMS:2019dqq, CMS:2019qjk, CMS:2018qxv, CMS:2018tuo, CMS:2017poa}
and other LLPs decaying to SM quarks \cite{CMS:2021yhb, CMS:2020iwv}.

\myparagraph{Displaced leptons, lepton-jets, or lepton-pairs}
A neutral LLP might decay inside the inner detector into leptons, resulting in leptonic tracks or lepton jets with a displaced vertex. 

ATLAS conducted searches for the displaced leptons emerging from dark photon decays \cite{ATLAS:2022izj}, long-lived heavy 
neutral lepton decaying to a pair of charged SM leptons \cite{ATLAS:2022atq, ATLAS:2019kpx, ATLAS:2018rjc}, sleptons decaying to SM leptons 
and gravitinos \cite{ATLAS:2020wjh}, LLPs decaying into hadrons and at least one muon \cite{ATLAS:2020xyo}, neutralino 
decaying to opposite sign leptons in an RPV scenario \cite{ATLAS:2019fwx}.
CMS has performed searches for LLPs decaying to muons \cite{CMS:2022qej, CMS:2021sch}, and both muons and electrons \cite{CMS:2021kdm}.

\myparagraph{Trackless, low-EMF jets}
It is possible that a heavy neutral long-lived particle will decay into two 
collimated coloured particles, resulting in jets. However, if the lifetime of the 
LLP is long enough, it may decay after traversing the inner detector and part of the 
ECAL, leading to the observation of jets in the 
hadronic calorimeter, but with only a small fraction of energy deposited in the electromagnetic calorimeter, and no preceding track.

ATLAS conducted searches for neutral LLPs \cite{ATLAS:2022zhj, ATLAS:2019qrr, ATLAS:2018niw} decaying hadronically into displaced jets in the 
hadronic calorimeter. In addition, the collaboration searched for stopped R-
hadrons \cite{ATLAS:2021mdj}, decaying inside the calorimeter significantly 
later than the bunch crossing in which they were produced.
CMS has performed searches for supersymmetry \cite{CMS:2022wjc} using 
neural networks to analyse trackless and out-of-time jet information.
Moreover, CMS collaboration has also searched for exotic
very long-lived BSM particles stopped in the calorimeter \cite{CMS:2017kku}, which decay out-of-time with respect to the presence of proton bunches in the detector.

\myparagraph{Multitrack vertices in the muon spectrometer}
A neutral LLP with decay length of the order of $\mathcal{O}(1~\rm{m})$
can decay inside the 
muon spectrometer, leaving no trace of its presence in the tracker or calorimeters. 
Such decays can be registered by ATLAS, 
which is able to reconstruct tracks inside the muon spectrometer.
The reconstructed tracks can be used to determine the displaced vertices in the MS and identify individual jets. 

Results of such searches were published by the ATLAS collaboration for $\sqrt{s}=13{\rm~TeV}$, $L=139\rm fb^{-1}$ \cite{ATLAS:2022gbw} and
$\sqrt{s}=13{\rm~TeV}$, $L=36.1\rm fb^{-1}$ \cite{ATLAS:2018tup}.
A study targeting a pair of neutral LLPs where one of them decays inside the ID and the other in the MS was also conducted \cite{ATLAS:2019jcm}. CMS has published a novel analysis targeting long-lived particles decaying in the endcap muon detectors \cite{CMS:2021juv}.

\myparagraph{Heavy stable charged particles}
Heavy stable particles (HSCPs) do not decay in the detector volume. If they are charged, 
they traverse the entire detector like muons but leaving a different signature. A 
typical method of searching for charged LLPs is measuring their ionisation 
energy loss $dE/dx$ in the tracker.

ATLAS has conducted several searches for highly-ionising particles in the context of the supersymmetry 
\cite{ATLAS:2022pib, ATLAS:2019gqq, ATLAS:2018lob}, and an analysis targeting mulicharged LLPs \cite{ATLAS:2018imb}. Similar studies were conducted by the CMS experiment \cite{CMS:2011arq, CMS:2012wcg, CMS:2013czn, CMS:2016kce}.

\myparagraph{Emerging jets}
One of the unresolved questions in modern Particle Physics is the nature of 
Dark Matter. 
Some of the theoretical ideas try to explain it by assuming the existence of a dark sector with a strong QCD-like interaction.
Such sector 
could have dark quarks and dark baryons, where the lightest of the latter 
would be stable, in analogy to the proton, and constitute a neutral Dark Matter 
particle. Models of this type can predict the existence of a portal between the dark 
sector and QCD at the TeV scale, resulting in phenomenologically attractive 
scenarios,
in which dark hadrons decay to SM particles observable in the LHC experiments.
A single dark parton, when showering and hadronising, would result in the creation of multiple 
dark hadrons, which would then decay into coloured SM particles. 
Coloured SM particles would hadronise and appear in the detector as jets.
If the decay length of the 
dark parton is of the order of a few centimetres, then the experimental signature 
would be the emergence of one or more jets inside the inner detector, with many displaced vertices and jet evolution different from the ordinary SM jets. The emerging jets signature has been investigated by the CMS collaboration \cite{CMS:2018bvr}.

\myparagraph{Non-pointing photons}
The non-pointing photon signature arises in a BSM scenario in which a neutral long-lived particle decays, and as a result, a photon 
is produced with a different time of arrival and angle in ECAL, with respect to photons produced in the primary vertex. An example 
scenario leading to such a signature is the GMSB model, in which neutralino is a long-lived NLSP, and decays to photon and 
gravitino. 
Since gravitino is invisible to the detector, the signature of the process is a delayed photon coming from a different direction than the primary vertex, without any charged particle's track associated.
Such a scenario has been tested by the CMS experiment \cite{CMS:2019zxa}.

\vspace{1em}
\noindent For more information about searches for long-lived BSM particles at the ATLAS and CMS experiments, please refer to Ref. \cite{Alimena:2019zri} for a comprehensive overview, or to Ref. \cite{Knapen:2022afb} for a pedagogical introduction to the topic.

\section{MoEDAL}\label{sec:moedal}

\subsection{Overview}\label{sec:moedal-overview}

\textit{Monopole and Exotics Detector at the LHC (MoEDAL) } \cite{MoEDAL:2009jwa} is a small experiment located approximately 2 meters away from the IP8, outside the LHCb \textit{VErtex LOcator (VELO)} detector. MoEDAL was envisioned primarily to search for magnetic monopoles (MMs), hence its design 
relies on unconventional passive detection methodologies, making
it completely different from ATLAS and CMS experiments. 
The four main subsystems of the MoEDAL detector are: 
\textit{Nuclear Track Detectors (NTDs)},
\textit{Magnetic Monopole Trackers (MMTs)},
 \textit{TimePix (TPX)},
 and \textit{MoEDAL Apparatus for Penetrating Particles (MAPP)}. A 3D visualisation of the MoEDAL detector is shown in Fig. \ref{fig:moedal-detector}.

\begin{figure}[tbh]
\includegraphics[width=\textwidth]{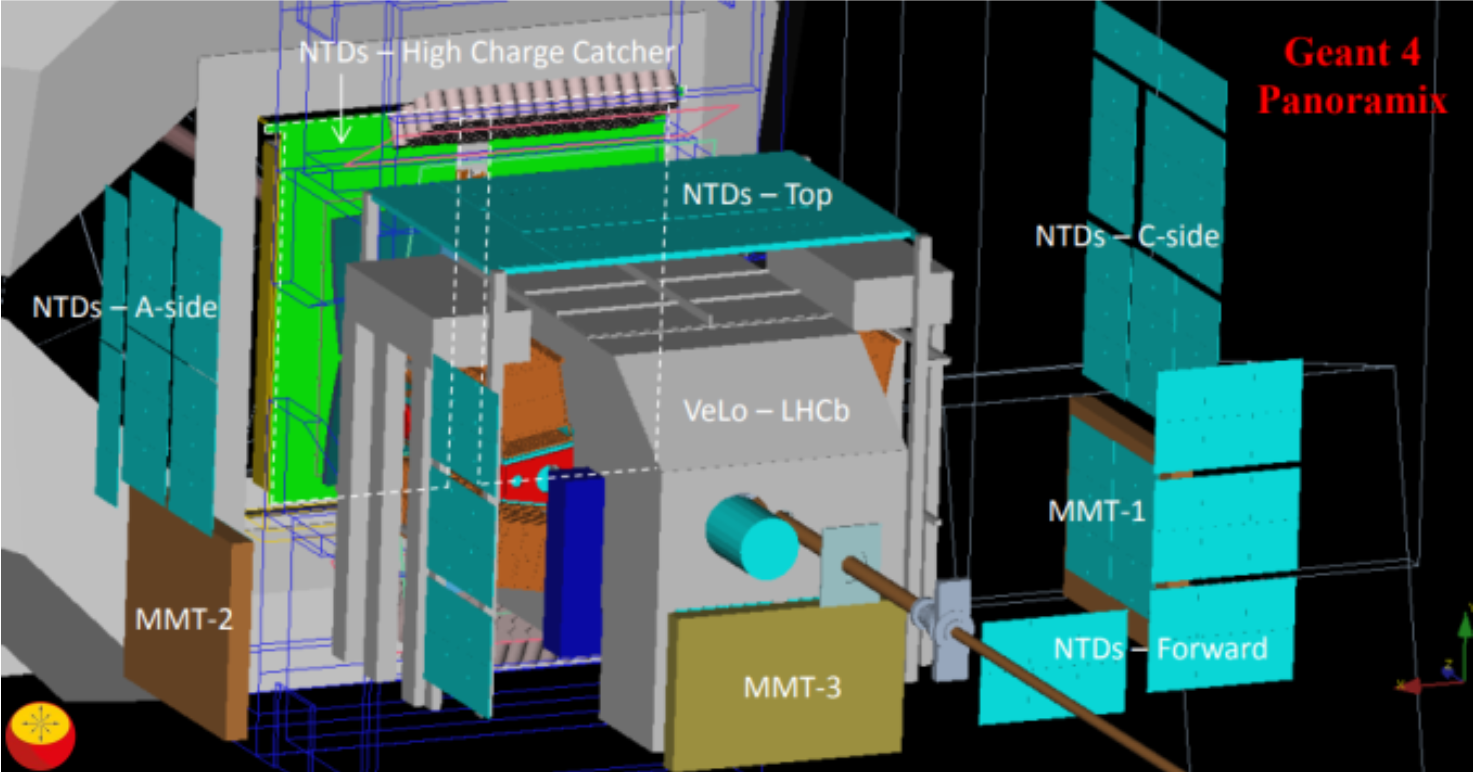}
\centering
\caption{\small A 3D visualisation of the MoEDAL detector located outside of the LHCb VELO. NTDs and MMTs are displayed, but the TimePix pixel array and MAPP detectors are not shown.
Image from \cite{Staelens:2021efs}.}
\label{fig:moedal-detector}
\end{figure}

\begin{figure}[!tbh]
\centering
\includegraphics[width=\textwidth]{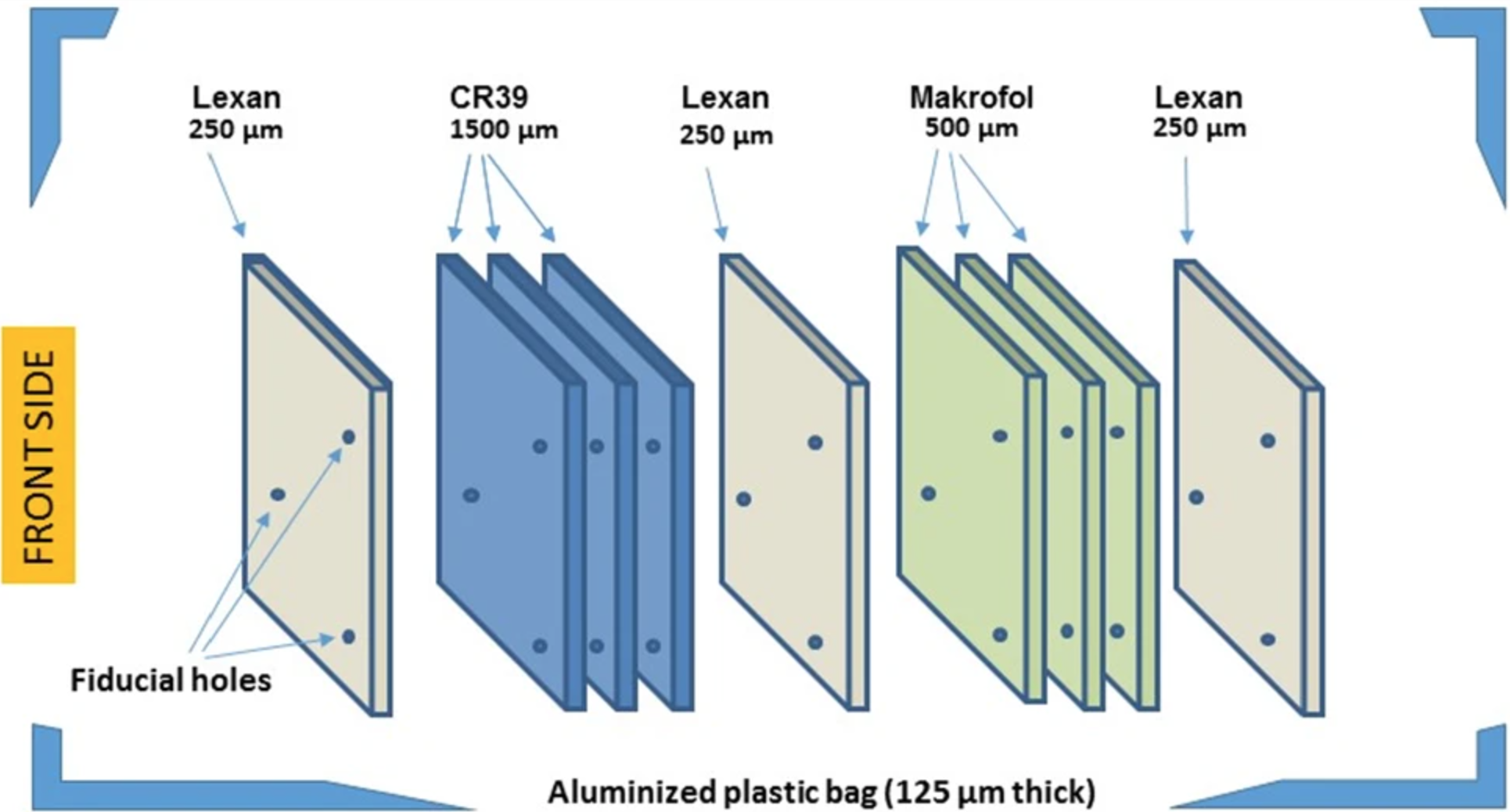}
\caption{\small Nuclear Track Detectors in MoEDAL experiments consists of $25 \times 25 ~\rm cm^2$ CR39™, Makrofol®, and Lexan™ layers enveloped in a thick aluminium bag.
Image from \cite{MoEDAL:2021mpi}.}
\label{fig:moedal-layers}
\end{figure}

The main sub-detector of the MoEDAL is a large array of passive NTDs covering an area of $\sim 100~\rm m$. Each NTD consists of three layers of CR39™ plastic (each $\sim 1.5~\rm mm$ thick)
and three layers of 
Makrofol® polymer (each $\sim 0.5~\rm mm$ thick),
and three Lexan™ layers (each $\sim 0.25~\rm mm$ thick): between CR39™ and Makrofol® stacks, in the front, and in the back of the NTD panel. In addition, panels have a $125~\rm \mu m$ thick aluminium shielding, as shown 
in Fig. \ref{fig:moedal-layers}. 
NTD panels have $25\times 25~\rm cm^2$ dimensions in width and length, and only a few millimetres in depth.
The six plastic layers of CR39™ and Makrofol® enable tracking of \textit{Highly Ionising Particles (HIPs)}, which can be magnetic monopoles, dyons\footnote{Dyon is a hypothetical
particle, which possesses both magnetic and electric charges, first proposed by Schwinger in \cite{Schwinger:1969ib}. }, ions or electrically charged elementary particles. When a HIP traverses the Nuclear Track Detector, it 
damages the plastic material at the level of polymeric bounds in the cylindrical region around its trajectory, resulting in the creation of a so-called \textit{latent track}. The latent track 
can be revealed by putting NTD panels into an etching solution. The array of NTDs at MoEDAL is exposed to the beam for a year, then it is dismantled and transported to the 
laboratory in Bologna, where the etching process is conducted. Each panel is disassembled, and plastic {sheets} are placed in the etching solution, leading to the dissolution of the outer 
layer of the plastic material. The material damaged by the traversing HIP dissolves faster than the unexposed plastic, leading to the formation of conical etch pits on both sides of the 
plastic layer, as depicted in Fig. \ref{fig:moedal-etching}. 
If the etching process is long enough, etch-pits formed on both sides of the polymer layer will eventually join and a hole in 
the detector will appear, along the trajectory of {a heavy-ionising} particle. These holes can be found using optical scanners. By analysing the shape and size of etch-pits, one 
can deduce the BSM particle's charge and incidence angle.
A signature of a BSM particle would be six aligned holes in all layers of plastic, pointing back to the 
interaction vertex.

\begin{figure}[!htb]
\centering
\includegraphics[width=\textwidth]{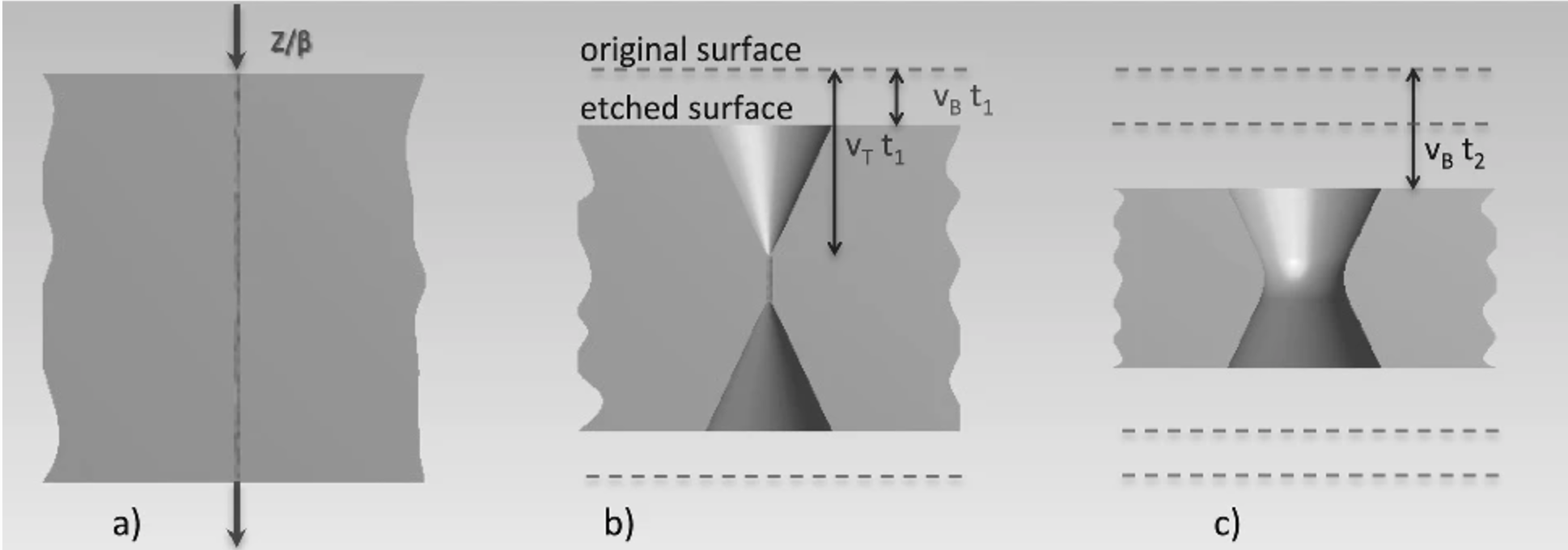}
\caption{\small Illustration of the etching technique used by the MoEDAL experiment to reveal tracks of HIP particles: \textbf{a)} a highly-ionising particle traverses the plastic layer leading to the formation of a latent track; \textbf{b)} conical etch-pits develop during the etching process; \textbf{c)} etch-pits join and form a hole in the detector after a prolonged etching.
Image from \cite{MoEDAL:2021mpi}.}
\label{fig:moedal-etching}
\end{figure}

The incidence angle of a particle is crucial for the acceptance of the detector \cite{Patrizii:2010jla} because only the normal component of the particle's 
velocity contributes to the formation of an etch-pit. If the perpendicular component is too small, then the unavoidable removal of 
outer layers of the polymer during the etching will hide any signature emerging from energy deposition by HIP.
Another critical characteristic of NTD panels is that they are sensitive only to particles which are slow (or highly charged), which can be expressed within a rough approximation by the formula:
\begin{equation}\label{eq:moedal-charge-vel}
\frac{|Q/e|}{\beta} \gtrsim 5,
\end{equation}
where $Q$ is the electric charge of a particle and $e$ is the elementary charge.
This has to be compared with ATLAS and CMS, which 
are sensitive only to fast particles with $\beta > 0.5$, due to active readout and 25 ns time separation between consecutive bunch crossings.
This property, together with a very good resolution of NTD panels 
corresponding to precision of alignment of etch-pits of the order of $\sim 10 ~\mu \rm m$, allows to greatly reduce the 
possible SM background coming from neutron recoils and spallation products. Typically these backgrounds only travel $\sim 10 ~\mu \rm m$, if they manage to penetrate the whole plastic layer, which is very unlikely, they would produce an etch-pit signature of a 
stopping particle, whereas the signal goes straight without much change of an ionisation. In order to produce a false signal 
signature, several of such unlikely events would have to occur in the same direction within the $\sim 10 ~\mu \rm m$ resolution.  
It is so improbable that despite multiple years of exposure, MoEDAL was not able to find a set of etch-pits satisfying signal 
selection criteria, despite its sensitivity confirmed by calibration with heavy ions. Taking into account that the detector is placed 
2 m away from the interaction point and it is sensitive only to slowly moving particles, \textbf{the conclusion is that MoEDAL is 
effectively free from the Standard Model background, and any observation of the signal would come from the New 
Physics.} 

\noindent There is however a caveat to this statement, while no SM process can render the signature of a BSM signal, it is technically difficult 
to select etch-pits which may correspond to {the traversing} HIP. In Fig. \ref{fig:moedal-foil-unexposed} one can see etched foil with 
several calibration pits created by heavy ions. This plastic foil has not been exposed in MoEDAL, hence the reference pits are 
clearly visible. In Fig. \ref{fig:moedal-foil-exposed} one can see the same foil (slightly shifted in the picture), which has been 
exposed in MoEDAL and etched again. Both images depict the same calibration holes, but the analysis of etched plastic in Fig. \ref{fig:moedal-foil-exposed} is far more challenging because of {the high} multiplicity of shallow etch-pits. These pits arise because of 
various particles deposing energy in the polymer, e.g. neutrons, spallation products, alphas, etc. Although they cannot mimic the 
signal, they introduce a large noise, which makes the analysis of NTD panels a difficult and time-consuming process.

\begin{figure}[!t]
  \begin{subfigure}[b]{0.49\textwidth}
    \includegraphics[width=\textwidth]{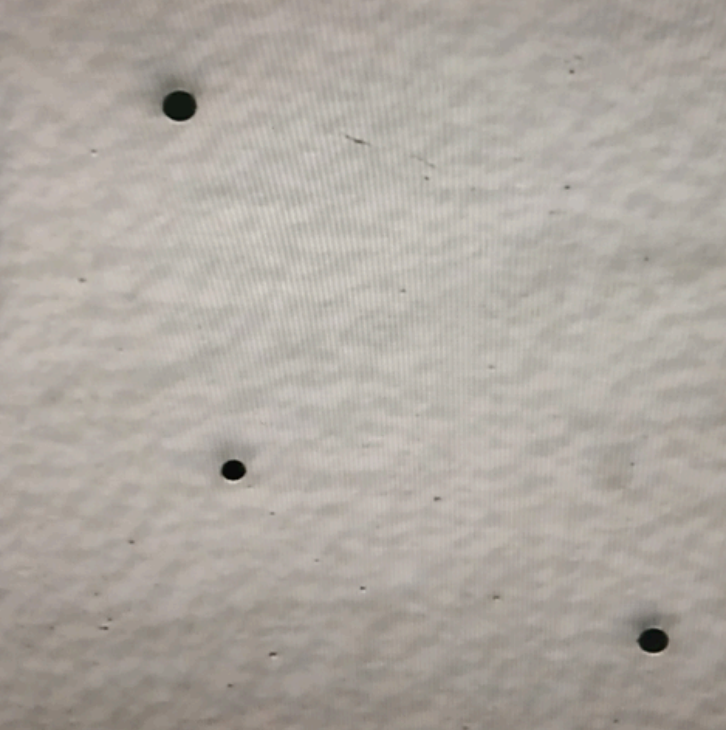}
    \caption{\small Etched NTD plastic foil unexposed in MoEDAL.}
    \label{fig:moedal-foil-unexposed}
  \end{subfigure}
  \hfill
  \begin{subfigure}[b]{0.49\textwidth}
    \includegraphics[width=\textwidth]{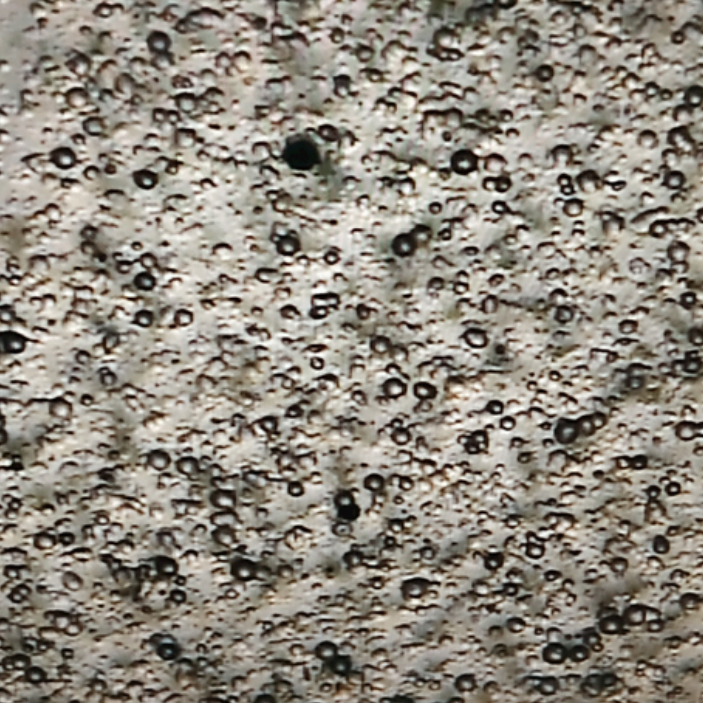}
    \caption{\small Etched NTD plastic foil previously exposed in MoEDAL.}
    \label{fig:moedal-foil-exposed}
  \end{subfigure}
  \caption{\small Etched polymer foil with reference holes produced by heavy ions. The picture on the left depicts a plastic foil, which has not been exposed in MoEDAL, hence the reference holes are clearly visible and easily recognised. The same foil after exposure in MoEDAL is depicted on the right (note that the images are slightly shifted). Multiple shallow etch-pits are visible, making it much harder to analyse and search for the signal. Image from \cite{laurapatrizii}.}
\end{figure}

Other passive subdetectors being a part of the MoEDAL experiment are 
\textit{Magnetic Monopole Trappers (MMTs)}, which allow for capturing magnetic monopoles. 
Any magnetic monopole produced in the $pp$ collision would lose its energy when traversing aluminium-based MMT detectors, 
and may eventually stop. Due to the large magnetic charge of a monopole, it is characterised by a strong magnetic dipole moment, 
which may lead to a strong binding with the nuclei of the aluminium MMT. Trapped monopoles might be revealed by passing 
MMTs through a \textit{superconducting quantum interference device (SQUID)} magnetometer, allowing to indirectly detect and 
measure magnetic charges. Additionally, MMTs can be also utilised to search for heavy quasi-stable charged particles, which are revealed by the detection of their decay products.

The third subdetector system of the MoEDAL experiment is the \textit{TimePix (TPX)} pixel array. 
It is an active system, the purpose of which is to 
monitor radiative background from the cavern.
It can simultaneously establish the time-of-arrival and time-over-threshold of any 
pixel data obtained. It is also capable of measuring the energy deposited in the detector 
material.
TimePix sensors are located in several regions of the MoEDAL detector in order to provide information about a local 
radiation field.

For the Run 3 data-taking period the MoEDAL collaboration {introduced} a new subdetector, called \textit{MoEDAL Apparatus 
for Penetrating Particles (MAPP)} \cite{Staelens:2021efs}, {deployed} in the UA83 tunnel situated 100 meters away from the IP8, at an angle of 7 degrees 
with respect to the beam axis. MAPP is an active state-of-the-art scintillator detector, designed to search for \textit{Feebly 
Interacting Particles}, e.g. \textit{Milli-Charged Particles} possessing electric charges up to $\sim 0.001 e$. MAPP is a major 
development of the MoEDAL experiment allowing it to extend its physics programme, hence the collaboration has recently started using the combined name ``MoEDAL-MAPP''.

There are important differences between MoEDAL and {general-purpose} experiments like ATLAS and CMS, which are worth pointing out explicitly:
\begin{itemize}
\item ATLAS and CMS can conduct a variety of searches, but MoEDAL is sensitive only to highly-ionising particles with decay length $\gtrsim 2~\rm m$.
\item Because of the requirements of the LHCb experiment, MoEDAL operates at approximately 10 times lower luminosity than ATLAS and CMS. It is one of the main limitations of the experiment.
\item 
Large experiments are sensitive to relativistic particles, while MoEDAL is sensitive to particles with $|Q|/\beta \gtrsim 5$, which for singly charged particles corresponds to $\beta \lesssim 0.2$.
\item Unlike ATLAS and CMS, MoEDAL is a mostly passive experiment, which does not suffer from pile-up and does not require a trigger for data selection. Instead, MoEDAL aggregates all data over a long period of exposure.
\item MoEDAL is effectively free from the SM background, which allows the collaboration to claim the discovery of a New Physics after observing very few signal events, in principle even one could be enough. 
\item MoEDAL delivers a permanent record of BSM particles in a form of particle tracks in etched plastic sheets and trapped monopoles in MMTs.
\end{itemize}

\myparagraph{Experimental results}
MoEDAL experiment carried out searches for magnetic monopoles using the 
prototype \cite{MoEDAL:2016jlb}, 
forward \cite{MoEDAL:2016lxh, MoEDAL:2017vhz},
and full MMT arrays \cite{MoEDAL:2019ort}. It also conducted searches for \textit{Highly Electrically Charged Objects} \cite{MoEDAL:2021mpi} for Run 1 data for 
$\sqrt{s}=8~\rm TeV$. 
MoEDAL was the first experiment to perform searches for: 
1) spin-1 magnetic monopoles \cite{Baines:2018ltl};
2) MMs produced in heavy-ion collisions through the Schwinger Mechanism \cite{MoEDAL:2021vix, Gould:2019myj};
3) dyons \cite{MoEDAL:2020pyb}, particles possessing both electric and magnetic charges.
In addition, MoEDAL published the only LHC's search for magnetic monopole production through photon fusion \cite{MoEDAL:2019ort}.

\subsection{Detector simulation}\label{sec:moedal-simulation}

In order to conduct the research presented in this thesis, an efficient and flexible method to simulate MoEDAL's detector was needed. 
For ATLAS and CMS there exists publically available software allowing for simplified detector simulation, e.g. {\tt Delphes} \cite{Ovyn:2009tx, deFavereau:2013fsa}.
However, the only existing Monte Carlo simulation code for 
MoEDAL involved full detector simulation using {\tt Geant4} \cite{GEANT4:2002zbu}, which is very precise, but time and resource-consuming. Phenomenological studies often do not require precise simulation of all detector effects, hence in the course of the PhD, a simplified simulation framework for MoEDAL was developed using a {\tt Python3} programming language.

The framework accepts \textit{Les Houches Event (LHE)} files, which contain information about the particles produced in a collision, and their subsequent decays. Such event files can be created with one of the popular Monte Carlo event generators. The research presented in this 
thesis was based on the {\tt MadGraph5} event generator \cite{Alwall:2011uj}, which contains implementations of the SM and MSSM, and allows to easily introduce new models, which can be created with the help of the {\tt FeynRules} \cite{Alloul:2013bka} package for {\tt Mathematica}. From each event, the final state particles of interest are recognised and their four-momenta are kept. 
Next, the detector geometry is loaded from a file, which has been used by the MoEDAL collaboration to perform {\tt Geant4} simulation for Run 2. 
The only subdetector system relevant to the presented research is the NTD array. 
The placement and dimensions of each of the NTD panels are read from the file, see Fig. \ref{fig:moedal-geometry}, and used to infer if particle trajectories cross the detector volume. Several simplifying assumptions are made here. First, it is assumed that the particle will travel from the interaction point to the MoEDAL detector unperturbed, without any interaction with the intermediate medium leading to a significant change of their momentum. The possibility of decay is included in a later stage of analysis by assigning appropriate weights to particles. Moreover, the depth of NTD panels is ignored, since it is 
three orders of magnitude smaller than their length and width. 
Particles with trajectories that cross an NTD panel are subject to velocity and incidence angle constraints, while the other particles are rejected.

\begin{figure}[!t]
\centering
\includegraphics[width=\textwidth]{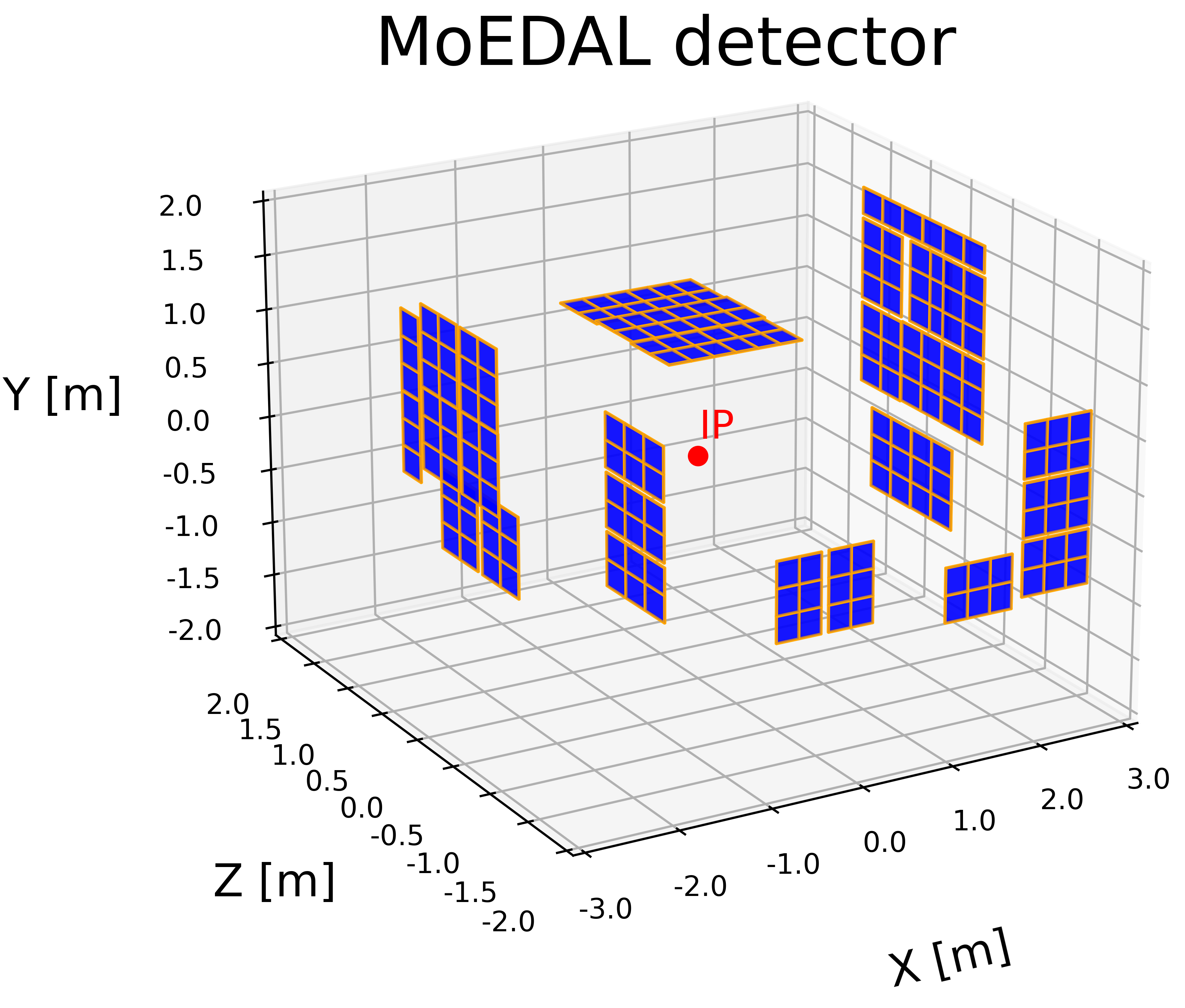}
\caption{\small Placement of NTD detectors for Run 2 MoEDAL used in the research presented in this thesis. Compare it with the full 3D visualisation in Fig. \ref{fig:moedal-detector}.}
\label{fig:moedal-geometry}
\end{figure}



Taking into account velocity and incidence angle constraints leads to the formula for partial efficiency $\epsilon_i$:
\begin{equation}\label{eq:moedal-epsilon-i}
\epsilon_i = P_{\rm NTD} \left( \vec{p}_i, \tau_i \right) \cdot \Theta \left( \delta_{\rm max}\left( \beta_i, Q \right) - \delta_i \right),
\end{equation} 
where index $i$ enumerates particles in the event; $ P_{\rm NTD} \left( \vec{p}_i, \tau_i \right)$ stands for the probability of reaching the NTD detector; $\Theta$ is the Heaviside function and corresponds to detector's acceptance; $\delta_i$ is the incidence angle of the ith particle, 
and $\delta_{\rm max}\left( \beta_i, Q \right)$ is the maximum angle of a particle with a given speed and charge, for which the track can be effectively revealed in the process of 
etching.
$P_{\rm NTD}$ in \eqref{eq:moedal-epsilon-i} is a function of the particle's three-momentum, $\vec p_i$, and lifetime, $\tau_i$. Its exact form will be provided later because it depends on the considered theoretic scenario, it is different for particles produced in the primary vertex and for particles emerging from a sequence of decays.
In Fig. \ref{fig:moedal-charge-vel}, the maximum allowed incidence angle as a function of the particle's speed is shown for singly and doubly charged particles. 
In Fig. \ref{fig:moedal-charge-vel} one can see that for $\beta \in [0.15, 0.20]$, the particle has to hit the NTD panel almost perpendicularly in order to be detected, therefore the
rough approximate constraint given in Eq. \eqref{eq:moedal-charge-vel} can be made more precise without much loss of efficiency\footnote{
The purpose of this discussion is to provide readers with a perspective allowing them to understand the change in the methodology between the first (see Sec. \ref{sec:paper1}) and second (see Sec. \ref{sec:paper2}) project.
Initially, $|Q|/\beta \gtrsim 5$ was provided by MoEDAL collaboration as a rough estimate of the sensitivity of MoEDAL NTDs, however, a more detailed investigation revealed that for a realistic description, one should rather take 
 $|Q|/\beta \gtrsim 6.67$ and include the incidence angle constraint.
}
 by taking $|Q|/\beta \gtrsim 6.67$, or equivalently:
\begin{equation}\label{eq:moedal-charge-vel2}
\beta \lesssim 0.15 \cdot |Q|.
\end{equation}
One can even consider relying only on Eq. \eqref{eq:moedal-charge-vel2} instead of using the function in Fig. \ref{fig:moedal-charge-vel}. Such approximation is valid if NTD panels directly face the interaction point resulting in small incidence angles of incoming long-lived particles, which is the case for Run 3 but not for Run 2. This issue will be discussed further in Chapter 4.

Within the approximation in Eq. \eqref{eq:moedal-charge-vel2}, it follows that in order to be detectable at MoEDAL, a singly charged particle has to have velocity $\beta < 0.15$, for doubly charged particle
($Q=\pm 2e$) it is $\beta < 0.30$, and so forth. If the charge of the particle is $|Q| \geq 6 \frac{2}{3}$ then (within the considered approximation) even ultrarelativistic particles can 
be measured in MoEDAL.

\begin{figure}[!t]
\centering
\includegraphics[width=\textwidth]{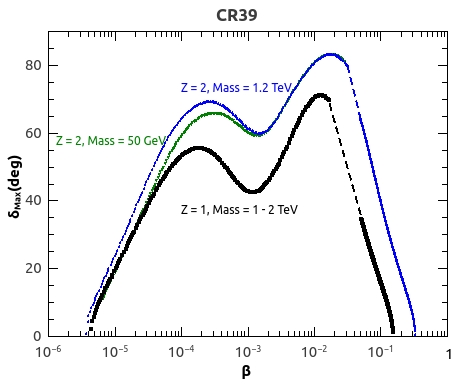}
\caption{\small Acceptance of CR39™ for singly and doubly charged particles. The plot shows the maximum allowed incidence angle $\delta_{\rm max}$ as a function of the particle's speed. If the incidence angle of a particle is larger than the maximum allowed value, the particle track will not be revealed in the etching process.}
\label{fig:moedal-charge-vel}
\end{figure}

The mean efficiency of the NTD detector array is calculated by summing all particles in an event, and averaging over the Monte Carlo sample, using the following formula:
\begin{equation}\label{eq:moedal-eps}
\epsilon = 
\left\langle \sum_{i=1}^N \epsilon_i \right\rangle_{\rm MC} = 
\left\langle \sum_{i=1}^N P_{\rm NTD} \left( \vec{p}_i \right) \cdot \Theta \left( \delta_{\rm max}\left( \beta_i, Q \right) - \delta_i \right) \right\rangle_{\rm MC}.
\end{equation}
In a typical New Physics scenario BSM particles are produced in pairs, hence $N=2$.

The quantity of interest for the MoEDAL experiment is the expected number of BSM particles that can be detected. It is estimated using the following equation:
\begin{equation}\label{eq:moedal-n}
N_{\rm sig} = \sigma\left(m\right) \times L \times \epsilon,
\end{equation}
with $ \sigma\left(m\right)$ being the production cross section for BSM particles with mass $m$, $L$ being the integrated luminosity at the end of data taking period at MoEDAL, 
and $\epsilon$ is given by Eq. \eqref{eq:moedal-eps}. The integrated luminosity for MoEDAL is $20~fb^{-1}$, $30~fb^{-1}$, and $300~fb^{-1}$ for Run 2, Run 3 and HL-LHC, 
respectively. 
Since MoEDAL is effectively background free, even a very small number of detected signal events,
$N_{\rm sig} \sim \mathcal{O}(1)$, might mark a discovery of the BSM Physics.

    \chapter{Research results}\label{chap:four}
The research results presented in the thesis were published in four articles:
\begin{enumerate}
\item 
\textbf{Prospects for discovering supersymmetric long-lived particles with MoEDAL}\\
D. Felea, 
J. Mamuzic, 
\underline{R. Masełek}, 
N.E. Mavromatos, 
V.A. Mitsou, 
J.L. Pinfold, 
R. Ruiz de Austr,
K. Sakurai,
A. Santra,
O. Vives\\
e-Print: 2001.05980 [hep-ph]\\
DOI: 10.1140/epjc/s10052-020-7994-7\\
Published in: Eur.Phys.J.C 80 (2020) 5, 431\\
Ref. \cite{Felea:2020cvf}
\item
\textbf{Prospects of searches for long-lived charged particles with MoEDAL}\\
B.S. Acharya,
A. De Roeck,
J. Ellis,
D.K. Ghosh,
\underline{R. Masełek},
G. Panizzo,
J.L. Pinfold,
K. Sakurai,
A. Shaa,
A. Wall\\
e-Print: 2004.11305 [hep-ph]\\
DOI: 10.1140/epjc/s10052-020-8093-5\\
Published in: Eur.Phys.J.C 80 (2020) 6, 572\\
Ref. \cite{Acharya:2020uwc}
\item 
\textbf{Detecting long-lived multi-charged particles in neutrino mass models with MoEDAL}\\
M. Hirsch,
\underline{R. Masełek},
K. Sakurai\\
e-Print: 2103.05644 [hep-ph]\\
DOI: 10.1140/epjc/s10052-021-09507-9 , 10.1140/epjc/s10052-022-10668-4 (erratum)\\
Published in: Eur.Phys.J.C 81 (2021) 8, 697, Eur.Phys.J.C 82 (2022) 8, 774 (erratum) \cite{Hirsch:2021wge}
\item
\textbf{Discovery prospects for long-lived multiply charged particles at the LHC}\\
M.M. Altakach,
P. Lamba, 
\underline{R. Masełek}, 
V.A. Mitsou,
K. Sakurai\\
e-Print: 2204.03667 [hep-ph]\\
DOI: 10.1140/epjc/s10052-022-10805-z\\
Published in: Eur.Phys.J.C 82 (2022) 9, 848\\
Ref. \cite{Altakach:2022hgn}
\end{enumerate}
The ordering of the articles is chronological so that the reader can trace the development of understanding and progress in the 
research topic. The general goal of these studies is to provide a comprehensive overview of prospects for the detection of heavy 
charged long-lived particles at the LHC, including for the first time predictions for the MoEDAL experiment. In the description of 
each of the projects, the ``most recent'' experimental constraints are discussed, which has to be understood as the most up-to-date at the moment of writing the article.

\noindent For other scientific articles and conference proceedings published by the author of this thesis in the course of his PhD, please refer to \cite{MoEDAL:2021mpi, Chakraborti:2022vds, Maselek:2022nie, Maselek:2022cjb, Lara:2022new, Acharya:2022nik, Feng:2022inv}.

\section{Prospects for discovering supersymmetric long-lived particles with MoEDAL}\label{sec:paper1}

\subsection{Introduction}\label{sec:paper1-sec1}
Supersymmetry described in Sec. \ref{sec:susy}, is a well-motivated and 
attractive candidate for the BSM Physics theory. Multiple searches of SUSY have 
been conducted at the LHC, and the lack of discoveries set strong constraints on 
the simplest models, e.g. $N=1$ supergravity and MSSM. Nevertheless, there 
are compelling arguments that supersymmetry might be discovered in the near 
future \cite{Ellis:2019ujs} since there are still unexplored regions in the parameter space.

In this project, we discuss the potential of the MoEDAL experiment to discover supersymmetry 
by presenting a case study,  which demonstrates a complementarity of the 
MoEDAL detector to ATLAS and CMS searches. We consider a specific SUSY 
model predicting heavy long-lived charged particles and we reveal the relevant 
parameter space in terms of lifetimes and masses for which MoEDAL can detect signal from BSM particles.

\subsection{Direct production of metastable sparticles at the LHC}\label{sec:paper1-sec2}

In this project we discuss the kinematics of long-lived particles produced in $13~\rm TeV~pp$ collisions, focusing on their velocity $\beta$, which is crucial for 
MoEDAL's acceptance. We use {\tt MadGraph5} \cite{Alwall:2011uj} and {\tt 
Pythia8} \cite{Bierlich:2022pfr} for Monte Carlo simulation. The velocity distribution of pair-produced right-handed staus $\tilde \tau_R$ is depicted in Fig. \ref{fig:paper1-stau-beta}.
One can 
see that the fraction of particles with velocity $\beta \lesssim 0.2$, for which NTD 
panels deployed in MoEDAL are able to detect BSM particles, 
grows with the stau mass $m_{\tilde \tau_R}$. Therefore, scenarios with
$m_{\tilde \tau_R} \gtrsim \mathcal{O}(1~\rm TeV)$
are favoured.
However, the production cross section for heavy sleptons is small, 
as can be seen in Fig. \ref{fig:paper1-stau-xs}, hence the possibility to detect directly-produced 
$\tilde \tau_R$ in an NTD array is very small.
\begin{figure}[!thb]
\centering
\includegraphics[width=0.86\textwidth]{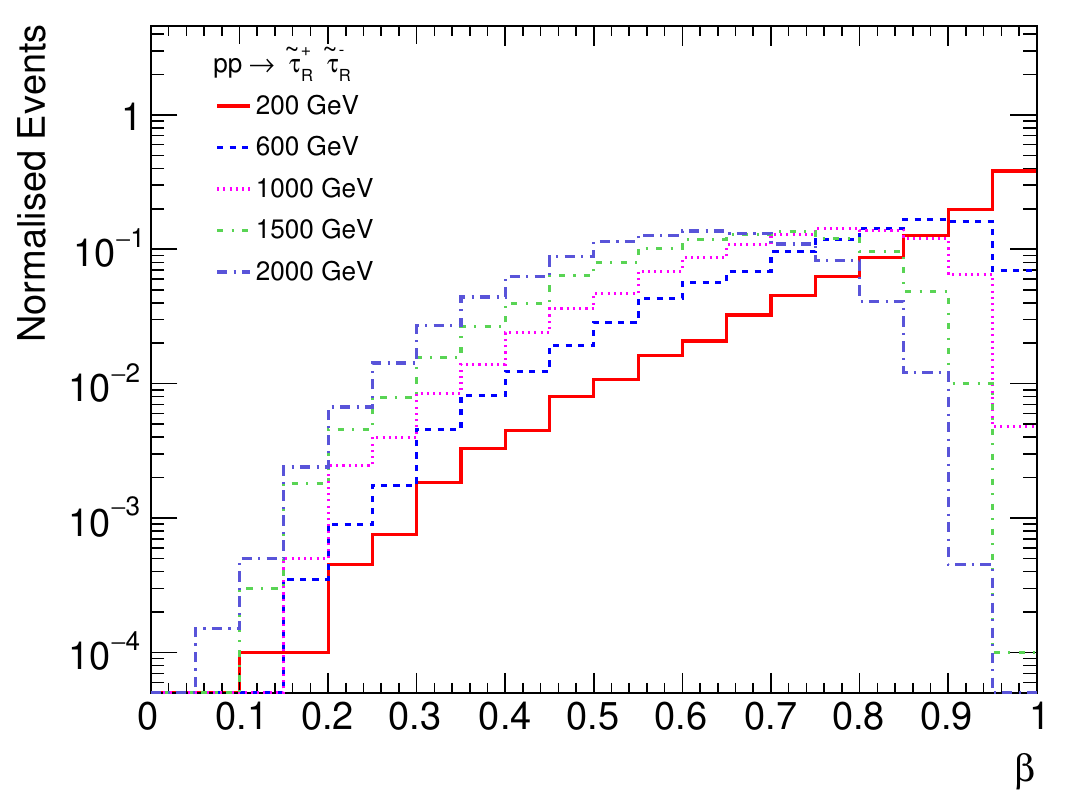}
\caption{\small Distribution of stau $\tilde \tau_R$ velocity for 
pair production at
$13~\rm TeV~pp$ collisions. Different histograms correspond to various masses of stau, from $200~\rm GeV$ to $2~\rm TeV$.}
\label{fig:paper1-stau-beta}
\end{figure}

\begin{figure}[!b]
\centering
\includegraphics[width=0.86\textwidth]{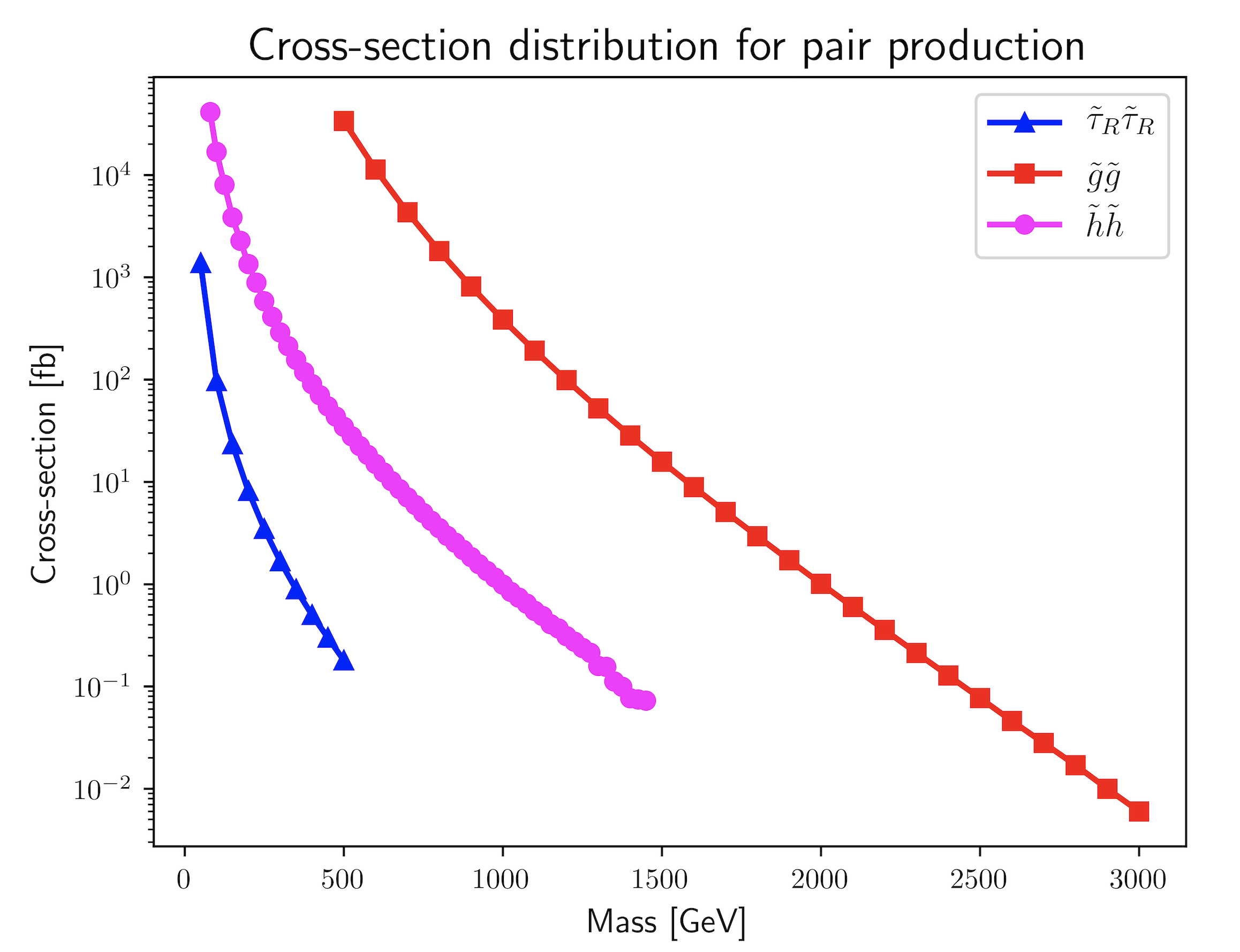}
\caption{\small The cross section for pair production at $13~\rm TeV$ LHC of 
staus (blue), gluinos (red), and higgsinos (magenta). Cross sections of staus and higgsinos were calculated at $\rm{NLO} + \rm{NLL}$ level, while gluinos with $\rm{NLLO_{approx}} + \rm{NNLL}$ precision \cite{susy_xs}.
}
\label{fig:paper1-stau-xs}
\end{figure}

Direct pair production of higgsinos $(\tilde \chi_1^0 \tilde \chi_1^\pm,\tilde \chi_2^0 \tilde \chi_1^\pm )$ and gluinos $(\tilde g \tilde g)$ is studied 
besides staus $(\tilde \tau_R^+ \tilde \tau_R^-)$. Comparison of velocity 
distributions shown in Fig. \ref{fig:paper1-velocity-comp} reveals that fermions (higgsinos and 
gluinos) are slower than bosons (staus), leading to larger ionisation loss of 
the former and higher chances for detection in MoEDAL. The reason behind 
this is that the main production process is an s-channel exchange of a spin-1 
gauge boson $(\gamma^*/Z^*)$ with quark-antiquark $q \bar{q}$ intial 
states. Helicity conservation in the initial vertex leads to the transverse 
polarisation of the gauge boson, hence the final state must have a non-zero 
total angular momentum. The production of scalar (spinless) staus $\tilde \tau$ suffers from the p-wave suppression, i.e. the production cross section 
vanishes as $\beta_{\tilde \tau} \to 0$ because of the conservation of 
angular momentum. In the case of fermions, there is no suppression, because 
they have non-zero spin.

In Fig. \ref{fig:paper1-stau-xs} the pair production cross section, taken from \cite{susy_xs}, is shown for right-handed staus, gluinos and higgsinos. The 
latter includes all production modes:
$ \tilde \chi_1^0 \tilde \chi_1^+ + 
 \tilde \chi_1^0 \tilde \chi_1^- + 
  \tilde \chi_2^0 \tilde \chi_1^+ + 
   \tilde \chi_2^0 \tilde \chi_1^- $, where the gauginos are assumed to be 
mass degenerate. The pair-production cross section for sleptons and 
higgsinos is calculated at the next-to-leading order (NLO) plus the next-to-leading logarithmic (NLL) level. For gluinos, the precision is the approximate 
next to NLO $\rm (NNLO_{\rm approx})$ plus next to NLL (NNLL).
Out of the sparticles considered, gluinos exhibit the highest production cross 
section and do not suffer from the p-wave suppression, hence their direct pair 
production constitutes the most promising scenario for MoEDAL.

\subsection{MoEDAL sensitivity to staus}\label{sec:paper1-sec3}
A preliminary study \cite{Sakurai:2019bac} assessing the sensitivity of MoEDAL in comparison with CMS revealed that MoEDAL 
experiment can be 
complementary to ATLAS/CMS despite the lower luminosity available at the IP8.
This study used a simplistic description of MoEDAL NTDs and CMS efficiencies for 
\textit{Heavy (detector-)Stable Charged Particles (HSCPs)} from 
\cite{CMS:2015lsu}, which were used to re-interpret the CMS analysis\cite{CMS:2013czn} in specific supersymmetric models for $\sqrt{s}=7~\rm TeV$ and $\sqrt{s}=8~\rm TeV$.

From Fig. \ref{fig:paper1-velocity-comp} one can see that even for gluinos the fraction of particles with $\beta \lesssim 0.2$, i.e. 
detectable in MoEDAL NTD array detector, is only $1\%$. Taking into account the lower luminosity available in the IP8, ATLAS and CMS 
are expected to be more sensitive than MoEDAL in most scenarios. Therefore, we are interested in particular SUSY models, for which ATLAS and CMS lose sensitivity to long-lived charged particles, while MoEDAL maintains it. 

\myparagraph{Model description}
A crucial aspect of ATLAS and CMS searches for charged detector-stable particles is the presence of a common 
requirement for multiple hits in the pixel detector, imposed in order to provide a good track reconstruction of a 
charged particle. However, a presence of a neutral particle in the decay cascade is likely to violate this selection criterium, leading 
to lower acceptance and sensitivity loss to the particular model. The effect becomes more severe if the intermediate neutral particle is 
long-lived.

This leads us to consider a model, in which a pair of gluinos is produced, $p p \to \tilde g \tilde g$, followed by a prompt decay 
of gluino to a long-lived neutralino and two quarks, $\tilde g \to \tilde \chi_1^0 q \bar q$. The neutralino is assumed to decay to 
a metastable stau and an off-shell tau, $\tilde \chi_1^0 \to \tilde \tau_1 \tau^*$,
after traversing a distance of $\mathcal{O}(1~\rm{m})$. 
The long lifetime of neutralino is a consequence of the assumed small mass splitting between the neutralino and stau. Metastability of stau, on the other hand, can be attributed to weak RPV couplings or gravitino LSP.
The full decay chain can be written as:
\begin{equation}\label{eq:paper1-model}
pp \to \tilde g \tilde g \to \left( j j \tilde \chi_{1,\rm{LL}}^0  \right) \left( jj \tilde \chi_{1,\rm{LL}}^0 \right) \to 
\left( jj\tilde \tau_{1,\rm{LL}}  \tau^*  \right)
\left( jj\tilde \tau_{1,\rm{LL}}  \tau^* \right),
\end{equation}
where $j$ represents a jet emerging from (anti)quark produced in the decay of gluino, $\tilde \tau_1$ is the lighter of stau mass 
eigenstates, and subscript ``LL'' indicates that the particle is long-lived.
The lifetime of the neutralino depends on its mass difference $\delta m$ with stau as $\propto (\delta m)^6$ in 3-body decay \cite{Khoze:2017ixx, Kaneko:2008re}.
Fig. \ref{fig:paper1-lifetime} shows the plot taken from \cite{Khoze:2017ixx}, which depicts the lifetime of an annihilation partner $\eta$ as a function of the mass splitting, $\Delta M$, between $\eta$ and Dark Matter particle $X$. Several simplified models were considered in \cite{Khoze:2017ixx}, in which the annihilation partner decays to a Dark Matter particle and an off-shell stau, $\eta \to X \tau^*$. Different models are depicted in the plot using distinct colours of the curves. As can be seen, all curves are very similar, which means that the lifetime of the annihilation partner is not very sensitive to the details of the model, and the result can be applied to our supersymmetric scenario, by treating $\tilde \chi_1^0$ as $\eta$.
One can see from Fig. \ref{fig:paper1-lifetime} that
for $\delta_m \sim 1~\rm GeV$, which is less than the mass of tau ($m_\tau \approx 1.77~\rm{GeV}$), neutralino has a lifetime $\sim 10^{-5}~\rm s$, which corresponds to decay lengths of $\sim \mathcal{O}(100~\rm{m})$. 
If the metastable stau 
is LSP, it can eventually decay to tau and other SM particles via very small RPV couplings. In a supergravity scenario, it could be an 
NLSP decaying via gravitational interaction to tau $\tau$ and gravitino $\tilde G$. Since 
we are interested in the interaction of stau with the MoEDAL detector,
the exact 
decay mode is irrelevant.

\begin{figure}[!t]
\centering
\includegraphics[width=0.81\textwidth]{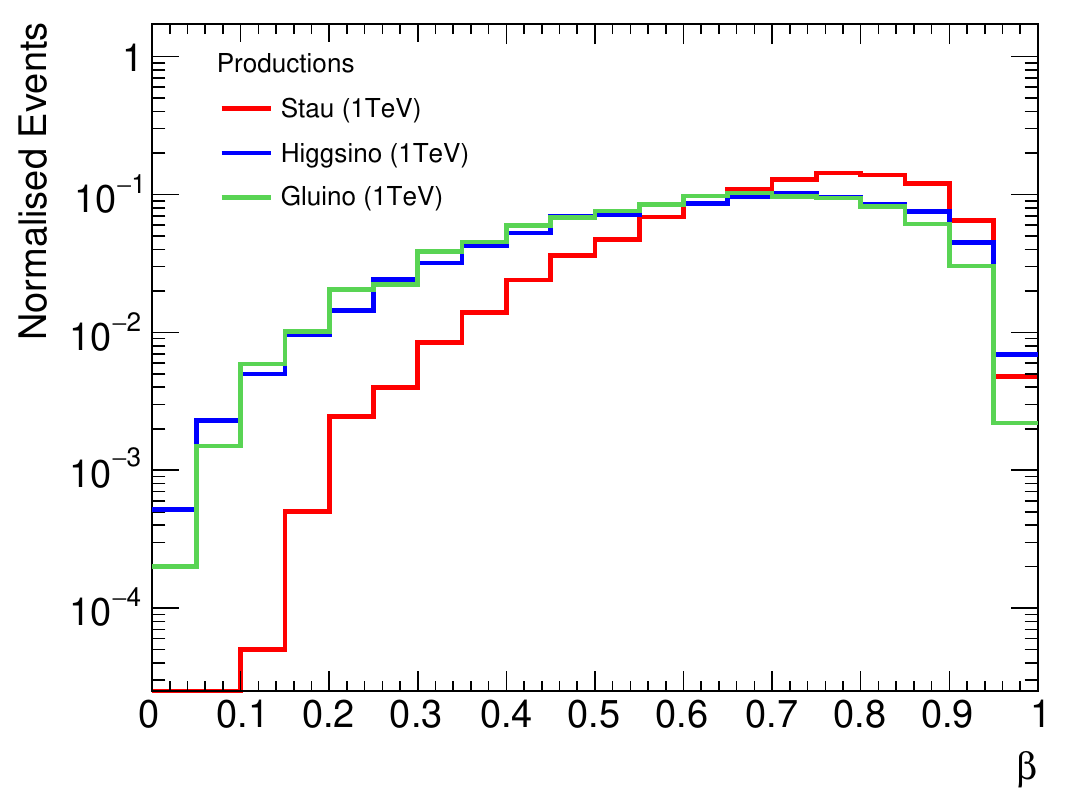}
\caption{\small Comparison of velocity distributions of staus (red), higgsinos (blue), and gluinos (green). Pair production in $13~\rm TeV~pp$ collisions was simulated. All particles have $m=1~\rm TeV$.
}
\label{fig:paper1-velocity-comp}
\end{figure}

\begin{figure}[!b]
\centering
\includegraphics[width=0.81\textwidth]{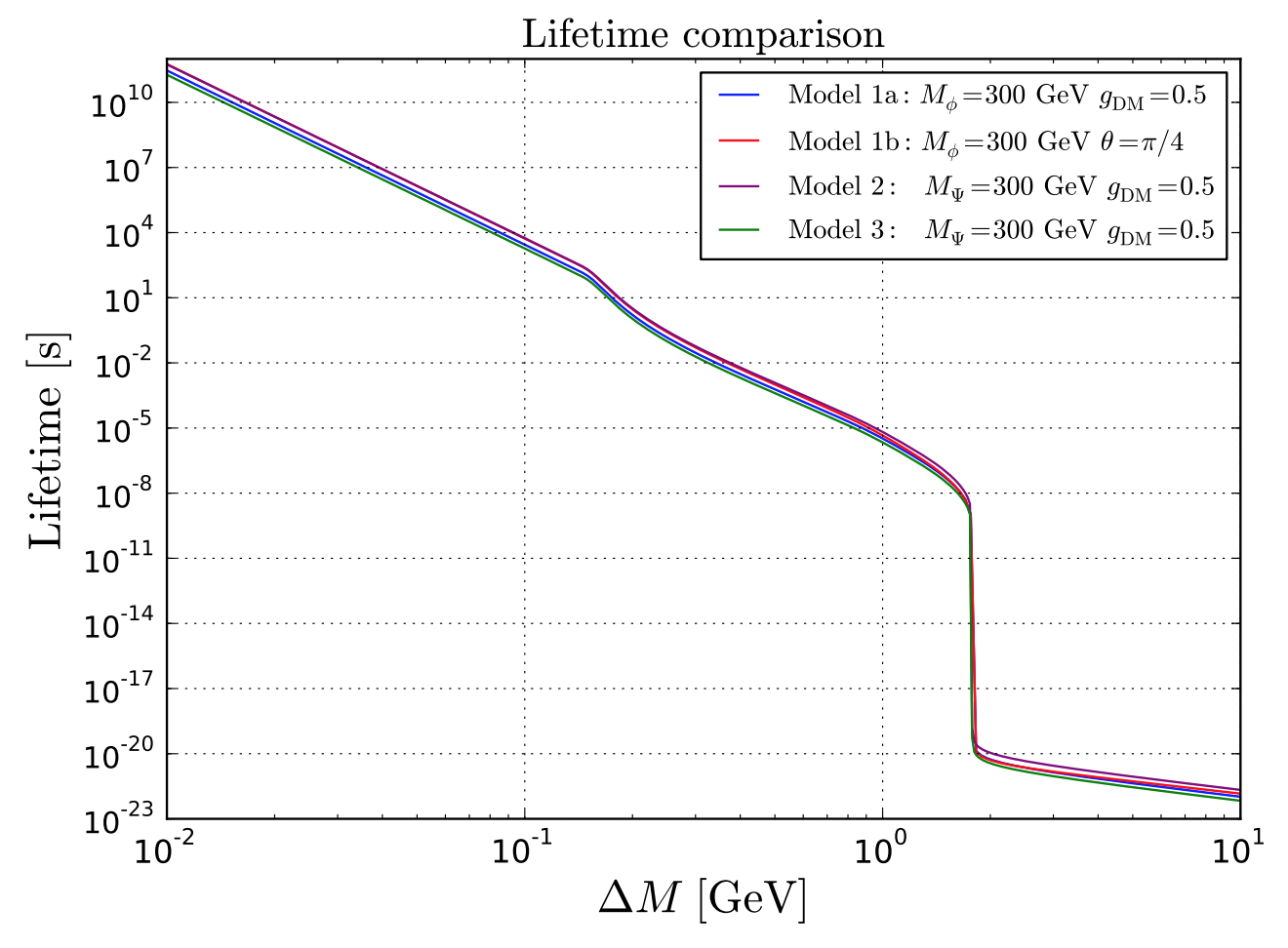}
\caption{\small Plot taken from \cite{Khoze:2017ixx}, showing the lifetime of an annihilation partner $\eta$ as a function of the mass splitting, $\Delta M$, between $\eta$ and the Dark Matter particle $X$. Several simplified models were considered in \cite{Khoze:2017ixx}, in which the annihilation partner decays to a Dark Matter particle and an off-shell stau, $\eta \to X \tau^*$. Different models are depicted in the plot using distinct colours of the curves. As can be seen, all curves are very similar, which means that the lifetime of the annihilation partner is not very sensitive to the details of the model, and the result 
of the calculations can be applied to our supersymmetric scenario, by treating $\tilde \chi_1^0$ as $\eta$.
}
\label{fig:paper1-lifetime}
\end{figure}

\myparagraph{ATLAS analysis recasting and other constraints}
At the time of the realisation of the project, the most recent LHC analysis searching for long-lived charged particles was by ATLAS \cite{ATLAS:2019gqq}, relying on $36.1~\rm{fb}^{-1}$ $13~\rm TeV$ data from Run 2. The latest CMS analysis \cite{CMS:2016kce} used only 
$2.5~\rm{fb}^{-1}$ of $13~\rm TeV$ data, therefore only the ATLAS analysis was considered. As argued in Sec. \ref{sec:atlascms-detector}, ATLAS and 
CMS detectors are very similar, hence the obtained result is applicable also to the CMS experiment.

If neutralinos in the decay cascade in Eq. \eqref{eq:paper1-model} have decay lengths of the order of $\mathcal{O}(1~\rm{m})$, 
multiple hits in the pixel detector cannot be expected, because neutral particles traverse it without interacting. In particular, the ATLAS analysis\cite{ATLAS:2019gqq} requires seven hits in the pixel detector. In order to satisfy this requirement, neutralino has to decay before reaching the pixel detector, which can be expressed in terms of the probability:
\begin{equation}\label{eq:paper1-pixel-prob}
P_{\rm pixel} = 1- \exp \left( - \frac{L_{\rm pixel}}{\beta \gamma c \tau_{\tilde \chi_1^0} \sin \theta } \right),
\end{equation}
where $\gamma \equiv (1 -\beta^2)^{-1/2}$ with $\beta$ being the velocity of neutralino, $\theta \in [0, \pi/2]$ is the angle 
between the $\tilde \chi_1^0$ momentum and beam axis, and $L_{\rm pixel}/\sin\theta$ is the distance between the pixel 
detector and interaction point. $L_{\rm pixel}=34~\rm mm$ is the minimal distance between the IP and the first layer of the pixel 
detector (at $\theta=\pi/2$). From Eq. \eqref{eq:paper1-pixel-prob} it is evident that $P_{\rm pixel} \ll 1$ for $c \tau_{\tilde \chi_1^0} \gg L_{\rm pixel}$.

In recasting the ATLAS analysis we follow the instructions provided in the 
{\tt HEPData} record \cite{hepdata} of \cite{ATLAS:2019gqq}, where detailed information 
about technical aspects of the analysis is provided, i.e. efficiency maps for 
signal reconstruction and trigger efficiencies. The limit is 
obtained by multiplying a carefully recast signal efficiency by $P_{\rm pixel}$, and it is
given in terms of the gluino mass $m_{\tilde g}$ and neutralino decay length $c \tau_{\tilde \chi_1^0}$.

\myparagraph{MoEDAL detector geometry and response}
Sensitivity of the MoEDAL detector to the considered SUSY scenario was 
estimated without resorting to time and resource-consuming {\tt Geant4} 
simulation. Instead, a completely new fast simulation framework was created, 
as described in Sec. \ref{sec:moedal-simulation}. As a starting point, the Run 
2 geometry, depicted in Fig. \ref{fig:moedal-geometry}, was considered, for 
which the geometrical acceptance of NTD array is $\sim 20\%$. In order for 
the model to be testable in MoEDAL, neutralino must decay into a charged 
stau before reaching the detector. However, it has to be long-lived in order to 
evade ATLAS constraint, therefore a mass splitting less than $m_\tau = 1.777~\rm GeV$ is assumed. As a consequence, $\tilde \tau_1$ and $\tilde \chi_1^0$ travel in almost the same direction with approximately equal 
momenta. For a neutralino with momentum  $\vec{p}_{\tilde \chi_1^0}$ and lifetime $\tau_{\tilde \chi_1^0}$, the probability of stau hitting an NTD panel is given by:
\begin{equation}\label{eq:paper1-pntd}
P_{\rm NTD}
\left( 
{\vec p}_{\tilde \chi_1^0}, \tau_{\tilde \chi_1^0}
 \right)  =
 \omega \left( \vec p_{\tilde \chi_1^0} \right) 
\left[ 
1 - \exp 
\left(
\frac{L_{\rm NTD}( \vec p_{\tilde \chi_1^0})}
{\beta \gamma c \tau_{\tilde \chi_1^0}}
\right)
\right],
\end{equation}
where 
${L_{\rm NTD}( \vec p_{\tilde \chi_1^0})}$ is the distance from IP to NTD panel along the trajectory of the particle\footnote{Since the momentum of neutralino and stau is almost the same, we do not differentiate between their velocities nor trajectories}, $\beta$ is particle's velocity, $\gamma$ is a 
Lorentz factor, and
$ \omega \left( \vec p_{\tilde \chi_1^0} \right)$ is 1 if there is an NTD panel in the direction of neutralino's three-momentum $ \vec p_{\tilde \chi_1^0}$, and 0 otherwise.

The efficiency of the detector is given by Eq. \eqref{eq:moedal-eps} and depends not only on
$P_{\rm NTD}$, but also on the incidence angle, i.e. for a given 
value of $\beta$ there exists a maximum incidence angle $\delta_{max}$ following the plot in Fig. \ref{fig:moedal-charge-vel}. A Monte Carlo 
event simulation was used to obtain the distribution of the incidence angle of 
staus in the considered model, as depicted in Fig. \ref{fig:paper1-incangle}. Parameters of the simulation were: $m_{\tilde g} = 1.2 ~ \rm TeV$, $m_{\tilde g} - m_{\tilde \chi_1^0} = 30 ~\rm GeV$ and $m_{\tilde \chi_1^0} - m_{\tilde \tau_1} = 1~\rm GeV$. One can see from Fig. \ref{fig:paper1-incangle} that about 50\% of produced staus 
have incidence angle smaller than 25 degrees, which requires $\beta \lesssim (0.08 \div 0.15)$ to be detected.

\begin{figure}[!htb]
\includegraphics[width=\textwidth]{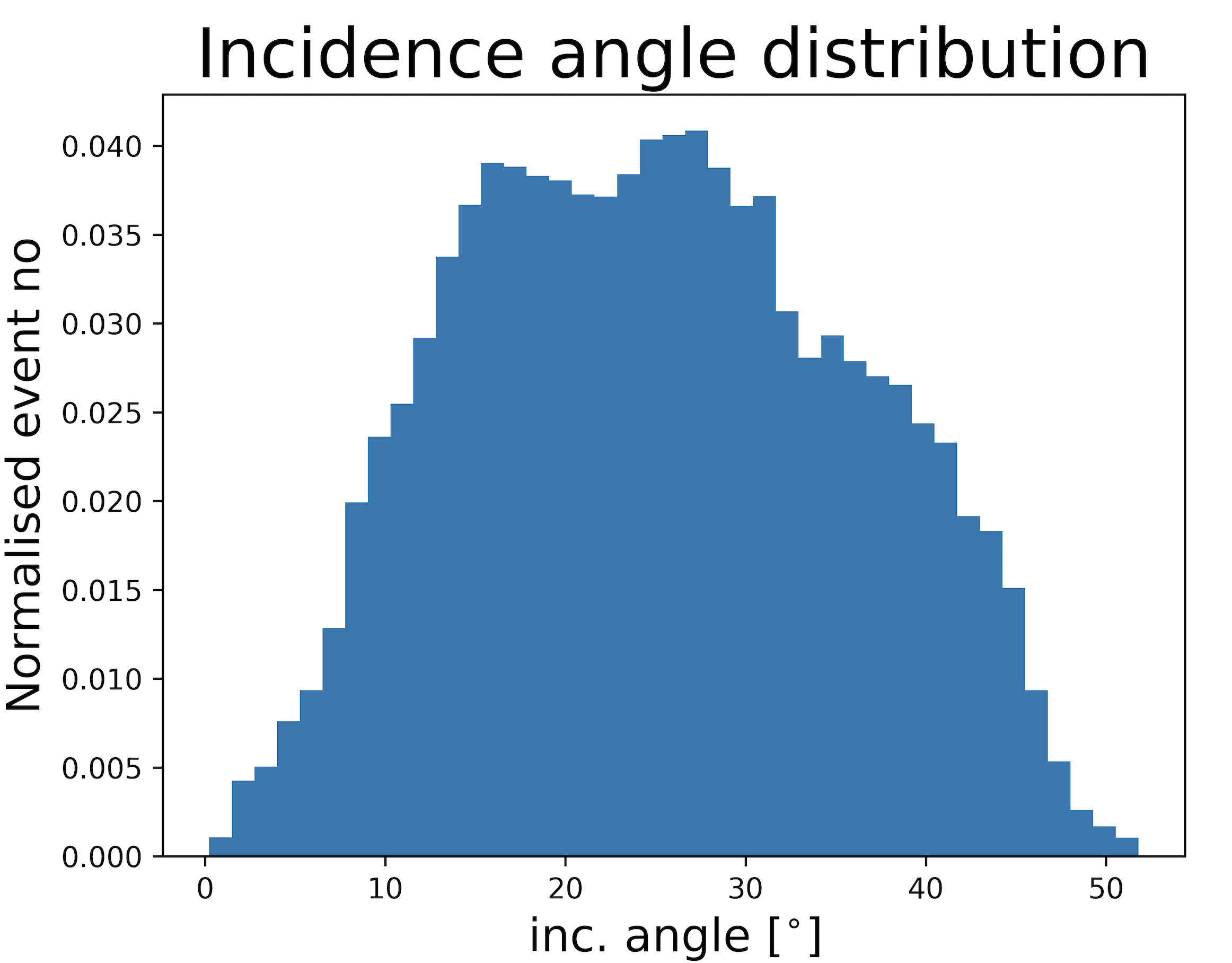}
\caption{\small The distribution of incidence angle of staus hitting NTD panels assuming Run 2 geometry.}
\label{fig:paper1-incangle}
\end{figure}

Incidence angle constraint is a major factor limiting the sensitivity of the 
MoEDAL detector, therefore a configuration of NTD panels minimising the 
angle is desired. The rearrangement of NTD panels in such a 
way that they face the interaction point is expected to enhance MoEDAL's 
sensitivity with respect to the Run 2 geometry. Therefore, in our study we 
additionally consider an ``ideal'' geometry set-up corresponding to NTD panels located 
spherically around the interaction point, for which the incidence angle 
constraint is negligible, i.e. $\delta \approx 0$.
However, the requirement $\beta < 0.15 \cdot |Q/e|$ still holds.
We do not change the number, size, or distance from IP8 of NTD panels. The realistic Run 3 detector placement is expected to 
be somewhere between Run 2 and ideal geometries.

\myparagraph{Analysis and results}
We follow the procedure described in Sec. \ref{sec:moedal-simulation} and estimate the expected number of BSM signal events in MoEDAL using the Eq. 
\eqref{eq:moedal-n}. Since MoEDAL is effectively a background-free 
experiment, we consider two small values: $N_{\rm sig}=1$ and $N_{\rm sig}=2$, and for each of them (and for two NTD geometries) we plot sensitivity 
contours in Fig. \ref{fig:paper1-result}. 
Contour lines correspond to detecting exactly 1 or 2 signal events, while
coloured areas enclosed by them correspond to 
detecting even larger number of BSM particles, i.e. the light-blue region 
depicts the parameter range for which
MoEDAL with an ``ideal'' NTD geometry is expected to detect $N_{\rm sig} \geq 
1$ staus, with $N_{\rm sig}= 1$ at the edge (contour line). The integrated luminosity at the end of Run 3 for MoEDAL is expected to be $L=30~\rm{fb}^{-1}$, and this value is used when calculating the event rate.
\begin{figure}[!htb]
\includegraphics[width=\textwidth]{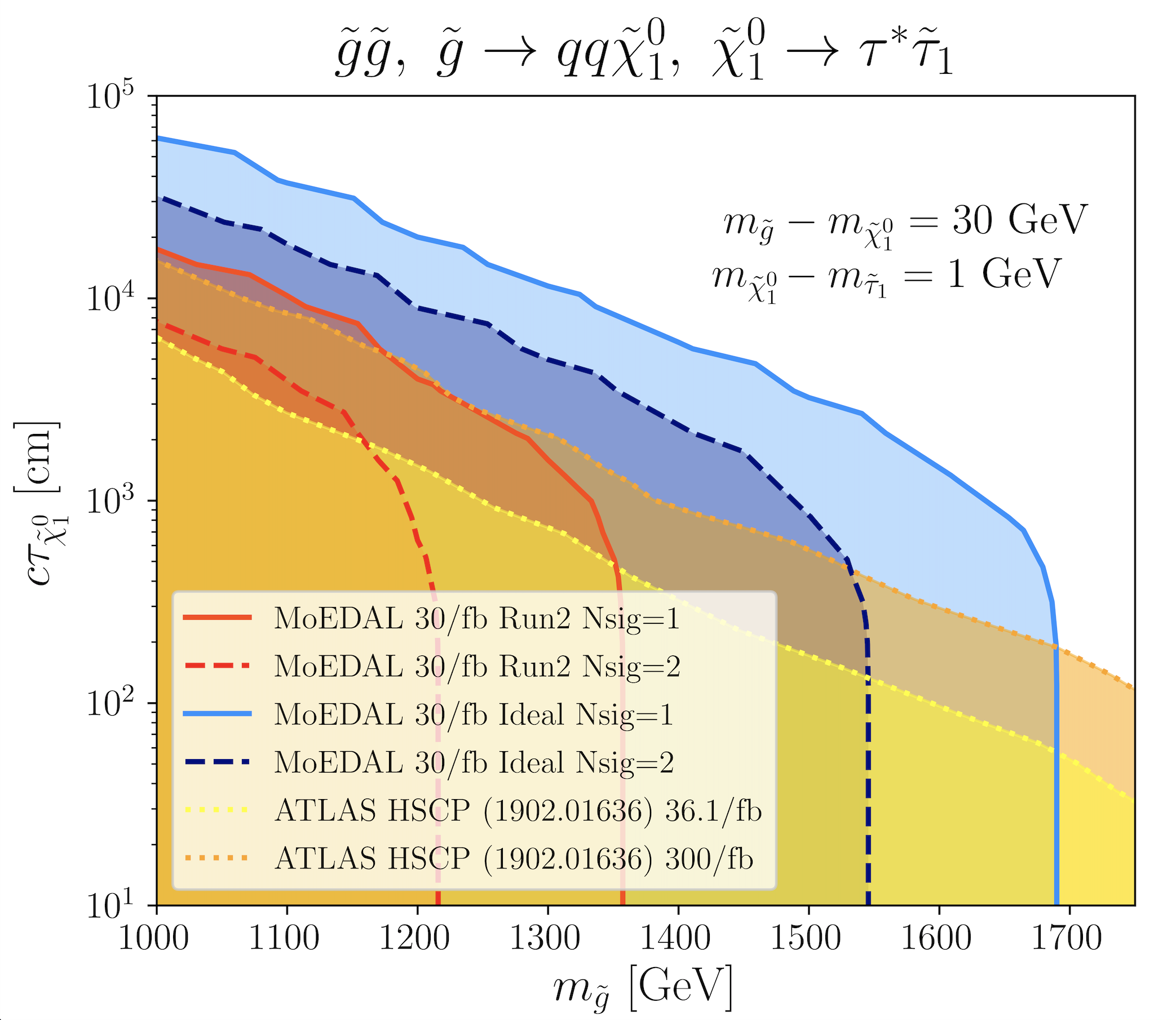}
\caption{\small The sensitivity of the MoEDAL detector to the considered SUSY scenario, presented in the $m_{\tilde g}$ vs. $c
\tau_{\tilde \chi_1^0}$ parameter plane. Solid contour lines correspond to $N_{\rm sig}=1$, while dashed contour lines to  $N_{\rm sig}=2$. In the 
shaded area enclosed by the contour, the expected number of detected signal events is equal to or larger than at the contour line. Red 
contours correspond to Run 2 NTD geometry, while blue contours are obtained for an ``ideal'' geometry, for which all NTD panels 
face the interaction point. The latest experimental constraint coming from the ATLAS search \cite{ATLAS:2019gqq} for Heavy (detector-)Stable Charged Particles (HSCPs) for $L=36.1~\rm{fb}^{-1}$ $13~\rm TeV$ is recast and plotted as a yellow contour with a dotted line. In 
addition, the exclusion contour by ATLAS is scaled up to $L=300~\rm fb^{-1}$ and plotted in orange with a dotted contour line.
For MoEDAL $L=30~\rm fb^{-1}$ at the end of Run 3 is assumed.
}
\label{fig:paper1-result}
\end{figure}
In the same figure we superimpose the current experimental limit obtained 
by recasting the ATLAS analysis \cite{ATLAS:2019gqq}. In addition, we plot the expected  
ATLAS limit for the Run 3 luminosity, $L=300~\rm fb^{-1}$, obtained by 
assuming that signal and background scale in the same way. We do not 
consider here any future upgrades of the ATLAS detector, nor improvements 
to the analysis, e.g. relaxing the requirement for seven hits in the pixel 
detector, which would likely allow ATLAS to outcompete MoEDAL in the whole parameter plane.

One can see from Fig. \ref{fig:paper1-result} that MoEDAL is able to investigate currently unconstrained parameter region with 
$m_{\tilde g} < 1.3~\rm TeV$ and $c \tau_{\tilde \chi_1^0} > 500~\rm cm$. Its expected detection reach is comparable to 
sensitivity of the ATLAS search for heavy charged LLPs if Run 2 geometry set-up is considered, however for the ``ideal'' NTD 
geometry MoEDAL is able to outperform ATLAS search, even if the ATLAS limit is rescaled to match integrated luminosity at the end of Run 3 
($L=300~\rm fb^{-1}$).

The detection reach of the MoEDAL detector in Fig. \ref{fig:paper1-result} follows a different trend than the ATLAS constraint. MoEDAL 
is capable of testing neutralinos with very long lifetimes due to the larger distance from the IP, however, it 
loses sensitivity when gluinos are heavy, because of a small production cross section. ATLAS can overcome the latter limitation
thanks to higher luminosity. We would like to stress out that the possibility to test the same parameter space of a BSM scenario using two experiments with totally different design philosophies is highly useful. In case of observing a positive signal in one of 
the detectors, the other can confirm it and provide additional information on the phenomenology. In case of the lack of 
BSM signal, two experiments can provide stronger constraints thanks to the different characteristics of their exclusion and 
uncorrelated systematic uncertainties.

Finally, we would like to comment on possible constraints on the considered model coming from prompt gluino searches in the 
jets-plus-MET channel. Recently,  stringent limits were placed on the mass of gluino by ATLAS \cite{ATLAS:2019vcq} (1100 GeV) and CMS \cite{CMS:2019ybf} 
(1300 GeV). The simplified SUSY model assumed decays of gluino to stable neutralino $(\tilde g \to q \bar q \tilde \chi_1^0)$ 
with a compressed mass spectrum $m_{\tilde g} - m_{\tilde \chi_1^0} \lesssim 50~\rm GeV$. Contrary to this model, in the SUSY 
scenario considered in this project, neutralino decays to metastable stau, hence the limits from \cite{ATLAS:2019vcq,CMS:2019ybf} cannot be applied directly. 
It is because the presence of a stau decaying inside the detector volume would affect the trigger efficiency and estimation of the 
missing transverse momentum. Although the estimation of this effect is complicated and beyond the scope of this project, one 
should bear in mind that the region of the parameter space with $m_{\tilde g} < 1200~\rm GeV$ might already be excluded.

\subsection{Conclusions and outlook}

A feasibility study on the detection of long-lived charged sparticles at the MoEDAL experiment was conducted, and   
complementarity with large general-purpose experiments, e.g. ATLAS, was demonstrated. The most favourable BSM scenarios for 
MoEDAL were characterised by a production of heavy (hence slow) fermions with a large production cross section (via strong 
interaction).

MoEDAL is sensitive only to slowly moving particles, with velocity $\beta \lesssim 0.2$, in opposite to ATLAS and CMS, which are 
able to detect particles with relativistic speeds. A key factor limiting the sensitivity of MoEDAL to simple BSM scenarios is the low 
luminosity available at the IP8. Nevertheless, the results of this project strongly suggest the utility of the MoEDAL detector to test more 
complex topologies, e.g. scenarios with a decay chain involving intermediate production of neutral LLPs. For such a model, MoEDAL
is able to cover a part of the parameter space, which is currently unconstrained by ATLAS and CMS, but could be probed by the 
general-purpose experiments if they manage to relax some of their selection criteria. 

Even for the models that can be tested by both MoEDAL and ATLAS/CMS, there is an added value of MoEDAL, because due to its 
passive detector design, MoEDAL provides coverage with totally uncorrelated systematics. Moreover, if the BSM signal is observed, 
the other experiment can confirm it, and provide additional information about the properties of new particles. For example,  
the determination of the velocity and mass of BSM particles in ATLAS/CMS relies on the assumption that the particle has a unit charge 
$Q=\pm 1e$, while NTDs in MoEDAL provide information about the energy and charge of the highly ionising particle. Velocity, on the 
other hand, can only be constrained by the maximum allowed value, which depends on the charge and incidence angle.

We would like to mention that in the published version of the project \cite{Felea:2020cvf}, a connection between the study 
and observed anomaly in the ANITA experiment \cite{ANITA:2016vrp, ANITA:2018sgj} was made. A possible explanation for the anomaly could be provided by the 
considered SUSY scenario, however, it was later demonstrated that the ANITA anomaly was due to experimental issues and did not 
correspond to BSM Physics. Therefore, the discussion of the solution to the anomaly is not included in the thesis, a curious reader 
is recommended to investigate the published article \cite{Felea:2020cvf}.

Finally, we would like to point out that more effort is needed to explore realistic SUSY scenarios in which 
decay sequences, similar to the one considered in this study, are embedded in a full theory and arise naturally.
We have concentrated on interactions of metastable sleptons with MoEDAL's NTD array, however, charginos and R-hadrons could also be detected in MoEDAL, and we leave their examination for the next study. 

\section{Prospects of searches for long-lived charged particles with MoEDAL}\label{sec:paper2}

\subsection{Introduction}\label{sec:paper2-intro}
In Sec. \ref{sec:paper1} it was shown that MoEDAL can conduct 
tests of New Physics scenarios
complementary to ATLAS and CMS, and in some cases it can investigate 
regions of the parameter space unprobed by the two major LHC experiments. 
In this project we continue our previous work and aim at providing a 
comparison between MoEDAL and ATLAS/CMS for a variety of BSM models. We 
begin by considering the production of charged R-hadrons induced by gluinos, 
stops or light-flavour squarks. Next, we study the production of charginos and sleptons, 
and prospects for their detection at the LHC. In both cases we do not make any 
specific assumptions about the models in which the particles appear, we parametrise their lifetime instead.

Another BSM Physics scenario, already introduced in Sec. \ref{sec:seesaw}, which potentially leads to the production of long-lived charged fermions is a Type 
III seesaw model \cite{Foot:1988aq, Bajc:2006ia, Arhrib:2009mz}, with at least two new $SU(2)_L$-triplet fermion fields 
$(\Sigma)$ with $Y=0$. In this model the neutrino mass is given by $m_\nu \approx Y_\nu^2v^2/m_\Sigma$, where $Y_\nu$ is the Yukawa coupling for 
neutrinos, $v\approx 246~\rm GeV$  is the Higgs vev, and $m_\Sigma$ is the 
mass of the new fermion. 
Typically, radiative corrections generate a mass 
splitting $m_{\Sigma^\pm} - m_{\Sigma^0} > m_{\pi^\pm}$ \cite{Jana:2020qzn}, making 
$\Sigma^\pm$ short-lived and
the 
decay $\Sigma^\pm \to \Sigma^0 + \pi^\pm$ kinematically allowed.
Experimental signature 
of such decays would be a disappearing track, because $\Sigma^0$ is neutral, 
and $\pi^\pm$ is very soft. Nevertheless, if there exists additional contribution 
to $\Sigma^{\pm,0}$ masses resulting in smaller mass splitting, the two-body 
decay of $\Sigma^\pm$ would not be allowed. Instead, a three-body decay 
would occur, which is phase space suppressed, and might
result in a signature of a charged detector-stable particle.

Doubly charged long-lived scalars can appear in many New Physics scenarios. One example is a type II seesaw model, in which 
SM is augmented with a complex $SU(2)_L$ triplet of scalar fields with hypercharge $Y=2$ \cite{Schechter:1980gr,Cheng:1980qt,Melfo:2011nx, BhupalDev:2013xol, Ghosh:2017pxl}. Spin-1/2 particles might 
also be doubly charged, e.g. a doubly charged Higgsino in supersymmetric L-R models 
\cite{Kuchimanchi:1993jg, Babu:2008ep, Frank:2014kma}. Another possibility is to simply 
add new doubly charged particles on top of the SM field content \cite{Delgado:2011iz, Alloul:2013raa}, write the possible interaction terms, and investigate 
phenomenological consequences.
In this project, we consider doubly charged scalars and spin-1/2 fermions that may be either singlets or triplets with respect to 
$SU(2)_L$ gauge symmetry group.

We note that parameters of L-R symmetric models, as well as type II and III seesaw models, are constrained based on theoretical 
considerations and low-energy experimental data \cite{Melfo:2011nx, BhupalDev:2013xol, Ghosh:2017pxl, BhupalDev:2018tox}. Since the goal of the project is to assess the sensitivity of the MoEDAL 
experiment in a model-independent way, we restrain from further discussing these indirect constraints.

\subsection{Previous LHC searches}\label{sec:paper2-searches}

Various types of searches for long-lived particles were conducted by both ATLAS and CMS, and have been already discussed in 
Sec. \ref{sec:atlascms-searches}. General-purpose experiments group their searches into two broad classes: 1) Heavy Stable 
Charged Particles (HSCPs), which do not decay within the detector volume; 2) decaying LLPs. Since we are interested in comparing 
the sensitivity of ATLAS and CMS to MoEDAL, we consider only the former class of searches, which relies on measurements of the 
anomalous ionisation loss $dE/dx$ and a very long time of flight (ToF) to the Muon System. Most of these searches were interpreted
by ATLAS and CMS in the context of supersymmetry.

The ATLAS collaboration used $36.1 ~\rm{fb}^{-1}$ of data collected for $13~\rm TeV$ $pp$ collisions,
and searched for HSCPs, but no significant signal was observed. It allowed setting $95\%$ CL upper-limits on the pair-production cross 
sections of long-lived R-hadrons, staus and charginos. These limits can be translated into lower mass limits on SUSY 
particles: gluino 2000 GeV, sbottom 1250 GeV, stop 1340 GeV, stau 430 GeV and chargino 1090 GeV \cite{ATLAS:2019gqq}\footnote{
A shorter-lived coloured sparticle can decay within the detector volume after hadronisation, e.g. a gluino hadron 
might decay to a pair of quarks and the lightest neutralino, $\tilde \chi_1^0$. Using the same data set ATLAS collaboration imposed
weaker lower mass limits of $1290-2060~\rm GeV$ on the gluino, assuming the aforementioned decay scheme and $m_{\tilde \chi_1^0}=100~\rm GeV$ \cite{ATLAS:2018lob}.
}.

ATLAS also performed a dedicated search \cite{ATLAS:2018imb} for the anomalous ionisation signal caused by the pair production of 
$SU(2)_L$ and $SU(3)_C$-singlet multiply charged ($|Q|=Z \cdot e$, $2 \leq Z \leq7$) HSCPs with spin-1/2, in the mass range $m\in [50,1400]~\rm GeV$. The data set consisted of $36.1~\rm{fb}^{-1}$ recorded $pp$ collisions at the center of mass energy of 13 
TeV. No excess signal events were observed, which led to setting a $95\%$ CL upper limits on the Drell-Yan pair production cross 
section as a function of the charge magnitude of the lepton-like HSCP. These upper cross section limits can be recast to lower mass limits for spin-1/2 fermions, ranging from 980 GeV for $Z=2$ to 1220 GeV for $Z=7$. 

CMS experiment searched for $SU(2)_L$ and $SU(3)_C$ singlet HSCPs, with single $|Q|=1e$, multiple $|Q|>1e$, and fractional $|Q|<1 e$ charge. The analysis \cite{CMS:2013czn} considered only spin-1/2 particles produced via the Drell-Yan process. The study was conducted 
using two datasets: $5~\rm{fb}^{-1}$ and  $18.8~\rm{fb}^{-1}$,
with center of mass $\sqrt{s}=7~\rm TeV$ and $\sqrt{s}=8~\rm TeV$ energies, respectively. The lower mass bounds obtained are: 480 GeV for $|Q|=2/3e$, 574 GeV for $|Q|=1e$, 685 GeV for $|Q|=2e$, 796 GeV for $|Q|=5e$, 781 GeV for $|Q|=6e$, 757 GeV for $|Q|=7e$, and 715 GeV for $|Q|=8e$.

The CMS collaboration used 13 TeV $2.5~\rm{fb}^{-1}$ data set
to set $95\%$ CL lower mass bounds on SUSY particles \cite{CMS:2016kce}: gluino (1610 GeV), stop (1040 GeV) and stau (490 
GeV). In addition, 550 GeV and 680 GeV mass limits were set on
singly and doubly charged lepton-like fermions, respectively.
\footnote{
CMS performed also a search \cite{CMS:2020atg} for an anomalous signal arising from disappearing tracks using 13 TeV data collected during Run 2. The 95\% CL limits provided were 
$m_{\tilde \chi_1^\pm} = 884 ~(474)~\rm GeV$
for purely wino LSP in the AMSB model with $\tau_{\tilde \chi_1^\pm} = 3~(0.2)~\rm ns$, with $L=140~\rm{fb}^{-1}$. Results for purely higgsino neutralino were
$m_{\tilde \chi_1^\pm} = 750 ~(175)~\rm GeV$
for $\tau_{\tilde \chi_1^\pm} = 3~(0.05)~\rm ns$, with $L=101~\rm{fb}^{-1}$..
A similar analysis \cite{CMS:2017ybg} was constructed by the CMS in order to constrain $SU(2)_L$-triplet charged fermion in the type III seesaw model. $35.9~\rm{fb}^{-1}$ of 13 TeV LHC data was used, and the lower mass bound was estimated to be $m_{\Sigma^\pm}>840~\rm GeV$.  
}

\subsection{Analysis framework}\label{sec:paper2-analysis}

In this project we consider pair production of new, singly or doubly charged 
particles at the LHC, $pp \to Y \bar Y+X$, where $\bar Y$ is the antiparticle of 
a new BSM particle $Y$, and $X$ stands for soft particles originating from the 
beam remnants or initial QCD radiation. More specifically, the particle $Y$ can 
be a SUSY particle, i.e. an R-hadron formed by gluino $\tilde g$, stop $\tilde t$, or light-flavour squark $\tilde q \in \{\tilde u, \tilde d, \tilde c, \tilde s, \tilde b\}$; or 
it can also be a long-lived chargino $\chi_1^\pm$ or a charged slepton $\tilde 
l^\pm$. In addition, we consider scenarios in which $Y$ is a doubly charged, 
$SU(2)_L$-singlet or triplet, scalar $S^{++}$ or spin-1/2 fermion $f^{++}$.

As already discussed in Sec. \ref{sec:moedal}, the MoEDAL NTD array is sensitive 
only to charged particles, hence coloured sparticles must hadronise to charged
R-hadrons in order to be detected. Since the probability $\kappa$ to form a 
charged R-hadron is not well understood \cite{ATLAS:2019duq}, we vary this parameter over 
the range $\kappa \in [0.5, 0.7]$. Distance from the interaction point to the 
MoEDAL NTD array is $\sim 2 ~ \rm m$, which corresponds to lifetimes of the 
order of $\mathcal{O}(1 ~\rm{m/c})$. For shorter lifetimes, $\tau$,
the detection probability will be exponentially suppressed. In this study, similarly 
to the project described in Sec. \ref{sec:paper1}, we treat the lifetime $\tau$ of the new 
BSM particle $Y$ as a free parameter, in order to achieve a higher level of model independence. The results will be presented in 2D planes, corresponding to 
$m_Y$ vs. $c \tau_Y$ parameter space.

As discussed previously in Sec. \ref{sec:moedal}, the incidence angle constraint is one of the key factors 
limiting the sensitivity of the MoEDAL detector. When the MoEDAL collaboration 
got acquainted with the results presented in Sec. \ref{sec:paper1-sec3}, a decision was made to 
reorganise the deployment of NTD panels in order for them to face the 
interaction point and minimise the incidence angle of incoming particles. 
Therefore, from now on only the ``ideal'' geometry set-up, introduced in Sec. \ref{sec:paper1-sec3}, is considered, for which 
the incidence angle condition in Eq. \eqref{eq:moedal-epsilon-i} can be 
approximated with:
\begin{equation}\label{eq:paper2-epsilon-i}
\Theta \left( \delta_{\rm max}\left( \beta_i, Q \right) - \delta_i \right) \to
\Theta \left( \beta_{\rm max}(Q) - \beta_i \right),
\end{equation}
where $\beta_{\rm max}(Q) = 0.15 \cdot |Q/e|$ was introduced in Eq. \eqref{eq:moedal-charge-vel2}. 
One can then replace the Eq. \eqref{eq:moedal-eps} with the following formula:
\begin{equation}\label{eq:paper2-moedal-eps}
\epsilon = 
\left\langle \sum_{i=1}^N \epsilon_i \right\rangle_{\rm{MC}} = 
\left\langle \sum_{i=1}^N P_{\rm NTD} \left( \vec{p}_i, \tau\right) \cdot \Theta \left( \beta_{\rm max}(Q) - \beta_i \right) \right\rangle_{\rm {MC}},
\end{equation}
where $N=2$ because we consider pair production in this study. The probability $ P_{\rm NTD}$ in Eq. \eqref{eq:paper2-moedal-eps} is given by:
\begin{equation}\label{eq:paper2-pntd}
P_{\rm NTD}(\vec p_Y , \tau_Y) = 
\omega \left( \vec p_Y
\right)
\exp 
\left(
\frac{L_{\rm NTD}( \vec p_{Y})}
{\beta \gamma c \tau_{Y}}
\right),
\end{equation}
where $\vec p_{Y}$ and $\tau_{Y}$ are the three-momentum and lifetime of the 
considered BSM particle $Y$, $\beta$ is the particle's velocity expressed in the units of the speed of light in the vacuum $c$, $\gamma \equiv (1-\beta^2)^{-\frac{1}{2}}$ is the Lorentz factor, and $L_{\rm NTD}( \vec p_{Y})$ is the distance from the IP8 to the point in which particle Y hit an NTD panel. 
$\omega \left( \vec p_Y \right)$ in Eq. \eqref{eq:paper2-pntd} is the geometrical acceptance factor, equal to 1 if there is an NTD panel in the direction of  $\vec p_{Y}$, and 0 otherwise.
Note that the probability in Eq. \eqref{eq:paper2-pntd} is different for the expression in Eq. \eqref{eq:paper1-pntd}.
It is because in the first project described in Sec. \ref{sec:paper1}, we considered a long-lived neutralino decaying to a metastable stau, where the latter could be detected by MoEDAL. Therefore, we required the neutralino to decay before reaching the MoEDAL detector, which is expressed in Eq. \eqref{eq:paper1-pntd}. In this project, on the other hand, we are interested in charged particles produced directly in the $pp$ collisions (or shortly afterwards in the case of R-hadrons), which have to live long enough ($\tau_Y \gtrsim \mathcal{O}(1~\frac{\rm{m}}{c}) $) to reach the MoEDAL detector. Hence the Eq. \eqref{eq:paper2-pntd} is proportional to the expression for exponential decay.

In order to evaluate the Monte Carlo average in Eq. \eqref{eq:paper2-moedal-eps} we used MC event generator {\tt MadGraph5\_aMC v-2.6.7} \cite{Alwall:2011uj} and 
implemented doubly charged particles with {\tt FeynRules 2} \cite{Alloul:2013bka}. The fast analysis framework described in Sec. \ref{sec:moedal-simulation} was properly adjusted to the needs of the project, and used to derive 
MoEDAL detection reach as a function of the mass and decay length for each of the BSM particles.

\subsection{Singly charged particles}\label{sec:paper2-susy}
\myparagraph{Motivation for candidates}
A general overview of long-lived SUSY particles was given in Sec. \ref{sec:susy-llps}. Sparticles which can be detected in MoEDAL are: gluino, stop, other squarks, charginos, and charged sleptons.

The gluino cannot be directly detected, because it rapidly hadronises into colour-neutral R-hadrons. If the colour charge of the gluino 
is neutralised by a gluon, the resulting R-hadron is electrically neutral and cannot be detected by MoEDAL. The other possibility is 
the formation of colour-neutral bound states involving gluino and two colour-octet quarks. By assuming that R-hadrons formed 
with the first-generation quarks are the lightest, we arrive at four possible bound states:
$\tilde u \bar u,~\tilde g d \bar d, ~\tilde g u \bar d, ~\tilde g d \bar u$\footnote{Here we use the bar notation to indicate antiquarks.}. 
The former two states are electrically neutral and cannot be detected in the MoEDAL experiment,
hence we will discuss the prospects for detection of the latter two charged R-hadrons. In the succeeding sections, we will introduce a parameter $\kappa$ to represent the fraction of charged states among all of the produced R-hadrons, and study MoEDAL's sensitivity to charged R-hadrons for different values of $\kappa$.

In the case of squarks, we distinguish between two scenarios. In the first model, stop is the lightest squark forming a long-lived 
R-hadron, while the other five flavours are much heavier. In the second scenario the situation is reversed, stop is very heavy, while 
the other five squarks are nearly mass degenerate and form long-lived R-hadrons. In both cases we assume that heavy squarks 
decay to the lightest one on a time scale $ \ll \mathcal{O}(1~\rm{m/c})$. Similarly to the gluino, the lightest squark is assumed to 
form bound states with the first generation of SM quarks: $\tilde u \bar u,~\tilde d \bar d,~\tilde u \bar d,~\tilde d \bar u$. The first two bound states are electrically neutral, while the other two are charged. Similarly to gluino R-hadrons, we will introduce a parameter $\kappa$ to represent the fraction of charged states among all of the produced R-hadrons, and present prospects for detection of charged R-hadrons in MoEDAL for different values of $\kappa$.

A long-lived chargino, being a mixture of wino $\widetilde W$ and higgsino $\tilde h$, can also be a long-lived particle. In this project 
we consider the case when one of the components dominates the production. Decays to the lightest state (wino-like or higgsino-
like) is again assumed to occur on a short time scale: $ \ll \mathcal{O}(1~\rm{m/c})$. We do not consider neutralinos, since they 
are chargeless and cannot be directly detected at the MoEDAL detector, therefore chargino is assumed to be lighter than neutralino\footnote{Typically, charged wino is heavier than the neutral one by $>m_{\pi^\pm}$ thanks to the radiative corrections. However, this mass hierarchy can be reversed by tuning the mixing in the neutralino and chargino sectors, which is the case considered in this study.}.

Finally, we admit a possibility to detect long-lived charged sleptons $\tilde l^\pm$ in MoEDAL. A long-lived slepton might be a superpartner of a left- or right-handed SM 
lepton and both cases are considered. 
We refer to the charged slepton as stau $\tilde \tau_{L/R}$, since in many SUSY scenarios staus are lighter than selectrons and 
smuons, however, our results apply to all flavours.
Unlike the project described in Sec. \ref{sec:paper1}, we consider a direct production of 
staus in $pp$ collisions.

\myparagraph{Run 3 projections for supersymmetric particles}
We begin the discussion of MoEDAL's expected sensitivity to long-lived singly charged sparticles at the end of Run 3, by 
showing in Fig. \ref{fig:paper2-susy-xs} pair production cross sections for various SUSY particles. All values were taken from the
{\tt LHC SUSY Cross Section Working Group} \cite{susy_xs}, except for $\tilde \tau_{L/R}$, for which {\tt Resummino 2.0.1} was 
used \cite{Fuks:2013vua}. The precision of cross section calculations for the coloured sparticles, $(\tilde g, ~\tilde t,~\tilde q)$, 
is at the approximate next-to-leading order ($\rm NNLO_{approx}$) precision level with the resummation of soft gluon emission at the next-to-next-to-leading-logarithmic (NNLL) accuracy \cite{Beenakker:2016lwe}. The accuracy for the weakly interacting sparticles, $(\widetilde W, ~\tilde h,~\tilde \tau_{L/R})$, is at the $NLO+NLL$ level.

\begin{figure}[tbh]
\includegraphics[width=\textwidth]{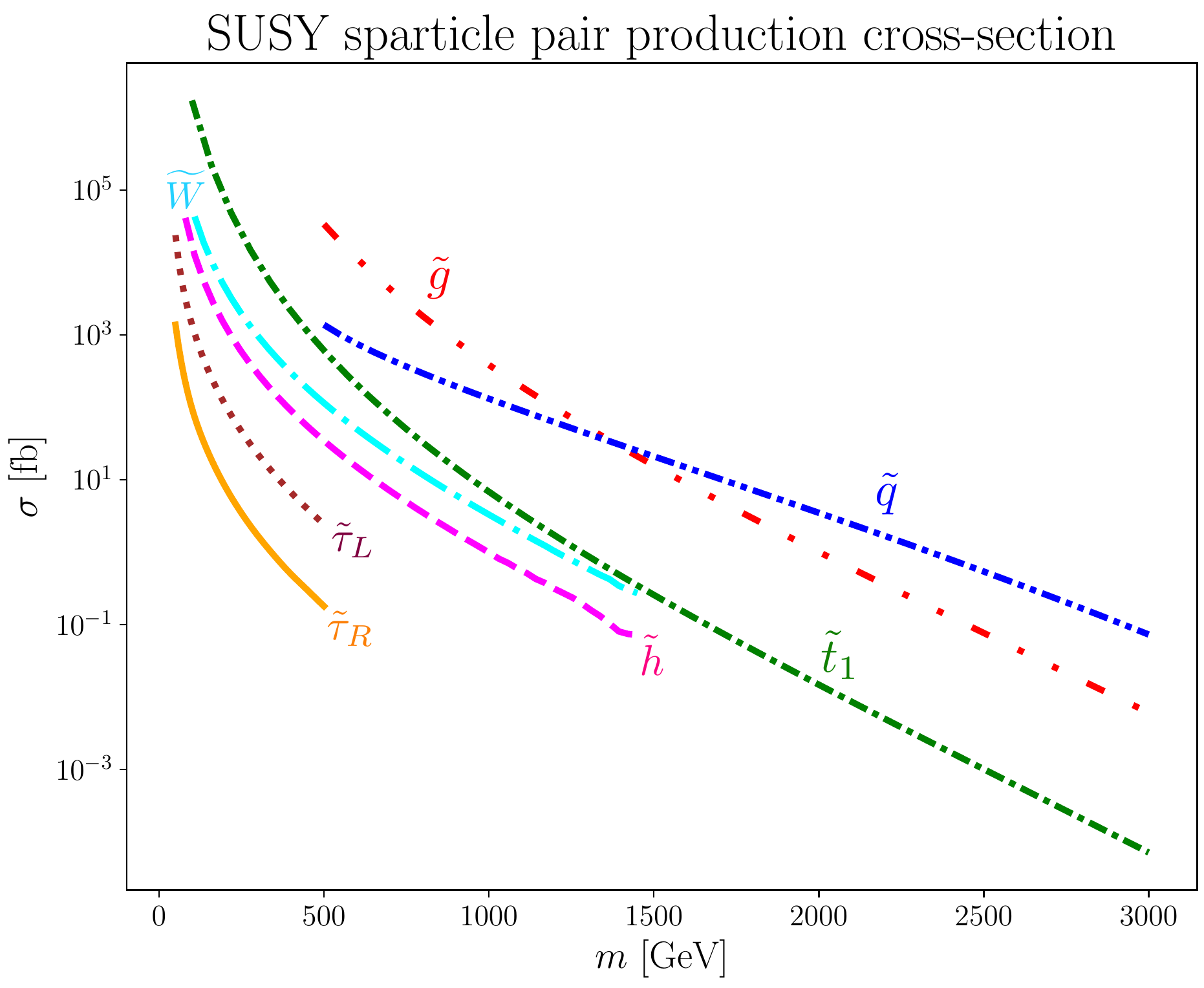}
\caption{\small Pair production cross section for various sparticles as a function of their mass. For the coloured sparticles, 
$(\tilde g, ~\tilde t,~\tilde q)$, cross sections were calculated at the $\rm NNLO_{approx}+NNLL$ level\cite{susy_xs, Beenakker:2016lwe}, while for the weakly interacting particles, $(\widetilde W, ~\tilde h,~\tilde \tau_{L/R})$, the accuracy was at the $\rm NLO+NLL$ level \cite{Beenakker:2016lwe, Fuks:2013vua}.
}
\label{fig:paper2-susy-xs}
\end{figure}

The cross section curve for $\tilde q$ in Fig. \ref{fig:paper2-susy-xs} is the sum of contributions to both left- and right-handed 
squarks of the first 5 flavours, calculated with the assumption $m_{\tilde g} = 3~\rm TeV$. For the weakly interacting sparticles,
cross sections are summed over all triplet (doublet) components of $\widetilde W$ ($\tilde h$ and $\tilde \tau_{L}$). To be more 
explicit, we include $\widetilde W^\pm \widetilde W^0$ and $\widetilde W^+ \widetilde W^-$ production modes for winos,
$\tilde h^+ \tilde h^-$, $\tilde h^0_{1,2} \tilde h^\pm$ and $\tilde h^0_{1} \tilde h^0_2$ production for higgsinos, and
$\tilde \tau_L \tilde \tau_L$, $\tilde \nu _\tau \tilde \tau^\pm_L$ and $\tilde \nu_\tau \tilde \nu_\tau$ for producing left-handed slepton, with the assumption that the heavier components of the multiplets decay to lighter long-lived charged 
sleptons \footnote{Phenomenological studies of such scenarios with a long-lived $\tilde \tau$ were described in \cite{Jittoh:2005pq, Kaneko:2008re, Konishi:2013gda}.}.

One can see in Fig. \ref{fig:paper2-susy-xs} that coloured sparticles have the largest cross section, while charged sleptons 
the smallest. For the cross section to be above $1~\rm pb$, sparticles need to be lighter than (approximately):
1 TeV $(\tilde g)$, 700 GeV  $(\tilde q,~\tilde t)$, 400 GeV  $(\widetilde W,~\tilde h)$, and 200 GeV $(\tilde \tau_{L/R})$.

The velocity distribution of BSM particles heavily affects the sensitivity of MoEDAL NTDs, therefore $\beta$ distributions for various SUSY particles are depicted in Fig. \ref{fig:paper2-susy-betas}. The following representative masses were taken to produce the plot in Fig. \ref{fig:paper2-susy-betas}: 
$\tilde g$: 1010 GeV, $\tilde q$: 920 GeV, 
$\tilde q$: 720 GeV, $\widetilde W$: 300 GeV, $\tilde \tau_{L}$: 80 GeV. The velocity distribution of $\tilde \tau_R$ (not shown) is similar to the distribution of $\tilde \tau_{L}$.

As evident from Fig. \ref{fig:paper2-susy-betas}, coloured sparticles have, in general, much lower velocities than charginos and 
sleptons. One reason is that the considered masses of coloured sparticles are larger than winos, higgsinos or sleptons. Another 
effect is that the production of coloured sparticles might be dominated by a t-channel fusion of two gluons, while the poduction of 
weakly interacting sparticles is mainly via the s-channel Drell-Yan process initiated by the quark-antiquark initial state.
We also observe that charginos are typically slower than sleptons. The reason is that the s-channel is mediated by the spin-1 gauge boson $(\gamma^*, ~Z^*)$, and the pair production rate for scalar particles vanishes in the $\beta \to 0$ limit because of the p-wave suppression, which is not the case for fermions.

\begin{figure}[tbh]
\includegraphics[width=\textwidth]{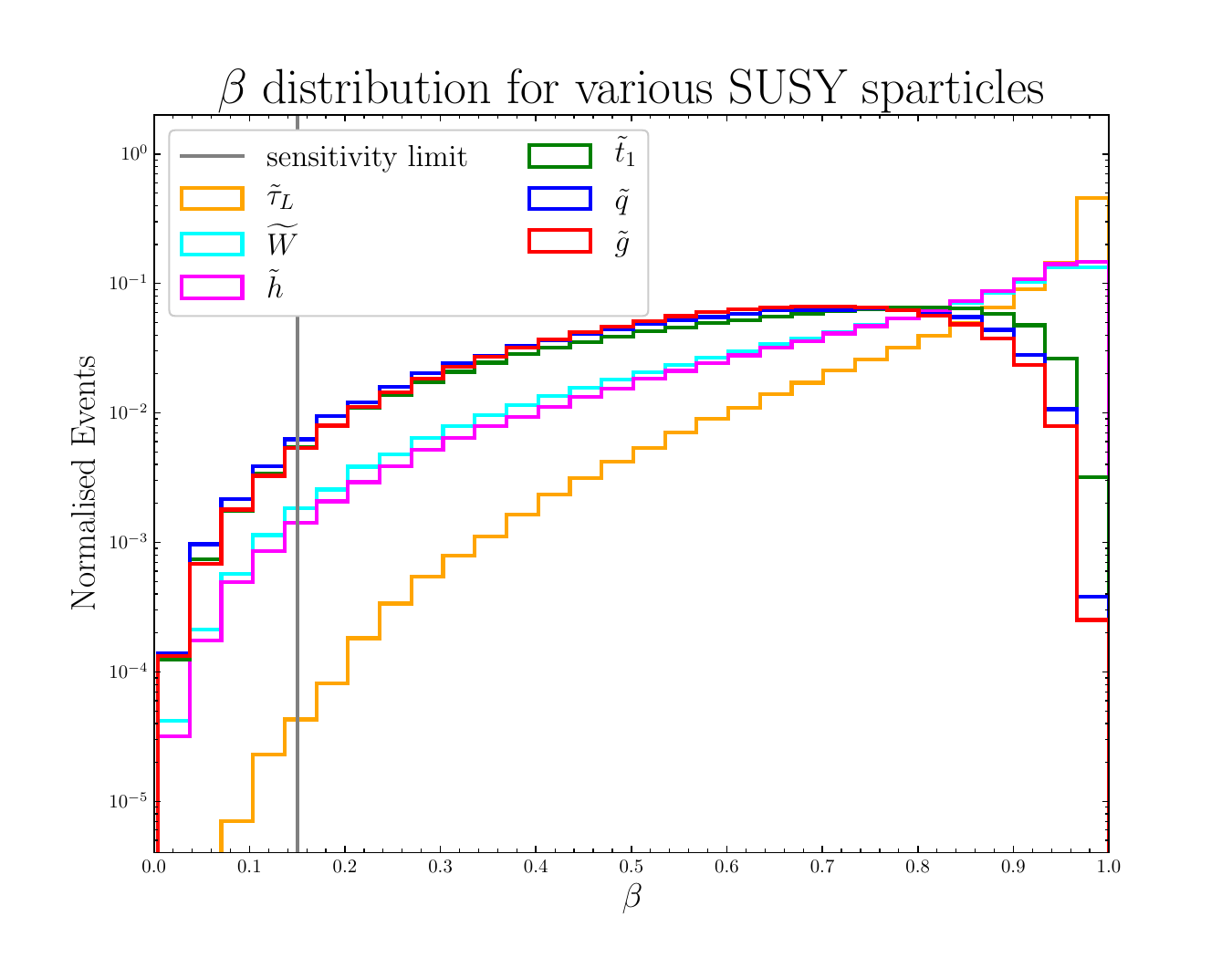}
\caption{\small Velocity distributions for various SUSY particles. The vertical grey line corresponds to the MoEDAL sensitivity limit
for singly charged particles, i.e. only particles with $\beta < 0.15$ can be registered by NTDs.
The following representative masses were taken to produce the plot: 
$\tilde g$: 1010 GeV, $\tilde q$: 920 GeV, 
$\tilde q$: 720 GeV, $\widetilde W$: 300 GeV, $\tilde \tau_{L}$: 80 GeV.
}
\label{fig:paper2-susy-betas}
\end{figure}

The vertical grey line in Fig. \ref{fig:paper2-susy-betas}
corresponds to the MoEDAL sensitivity limit
for singly charged particles, i.e. only particles with $\beta < 0.15$ can be registered by NTDs. We can see that the largest fraction 
of detectable particles is for coloured sparticles, intermediate for charginos, and the lowest for sleptons.

We now discuss the expected sensitivity of the MoEDAL detector to long-lived charged SUSY particles in Run 3 LHC, with the 
integrated luminosity $L=30~\rm{fb}^{-1}$. The contours in Fig. \ref{fig:paper2-susy-res} enclose the range of masses and decay 
lengths that MoEDAL can test. Solid lines correspond to $N_{\rm sig}=1$, while dashed represent $N_{\rm sig}=2$.

The plot in Fig. \ref{fig:paper2-susy-res-gluino} shows the expected MoEDAL's sensitivity for gluino R-hadrons, with red (blue) 
contours corresponding to $\kappa=0.5$ ($\kappa=0.7$) fraction of charged bound states produced. In the more conservative 
case $\kappa=0.5$, MoEDAL is expected to observe 1 (2) signal event(s) for $m_{\tilde g} \approx$ 1530 (1400) GeV, in the limit 
of a very long lifetime ($c\tau > 100~\rm m$). Under the more optimistic assumption $\kappa=0.7$, the maximum gluino 
mass that MoEDAL can probe is $\sim$1600 (1470) GeV if 1 (2) signal event(s) are demanded.

The sensitivity to the lighter stop, $\tilde t_1$, is plotted in Fig. 
\ref{fig:paper2-susy-res-stop}. The mass of the long-lived stop can be probed by MoEDAL up 
to $\sim$870 GeV ($N_{\rm sig}=1$) and $\sim$780 GeV  ($N_{\rm sig}=2$) for $\kappa=0.5$, while for $\kappa=0.7$ the limits are
$\sim$920 GeV ($N_{\rm sig}=1$) and $\sim$830 GeV  ($N_{\rm sig}=2$)

The sensitivity to squarks is shown in Fig. \ref{fig:paper2-susy-res-squark}. Since the production cross section depends on 
$m_{\tilde g}$, we show contours for $m_{\tilde g}=2~\rm TeV$ and $m_{\tilde g}=3~\rm TeV$. In the most optimistic case, i.e. 
for $\kappa = 0.7$, $m_{\tilde g}=2~\rm TeV$ and $c \tau > 100~\rm m$, MoEDAL can probe squark mass up to $\sim$1920 (1700) 
GeV for $N_{\rm sig}=1$ (2). On the other hand for $\kappa = 0.5$ and $m_{\tilde g}=3~\rm TeV$ the mass reach is 
$\sim$1670 (1450) GeV for $N_{\rm sig}=1$ (2).

We now turn to discuss the prospects for the detection of weakly interacting 
sparticles. MoEDAL sensitivity for charged winos (blue) and higgsinos (red) is 
shown in Fig. \ref{fig:paper2-susy-res-chargino}. MoEDAL can probe the wino 
mass up to $\sim$670 (570) GeV for $N_{\rm sig}=1$ (2). In the case of higgsinos, the 
mass reach is $\sim$530 (430) GeV for $N_{\rm sig}=1$ (2). Higher sensitivity to 
winos comes from the larger production cross section caused by the bigger 
$SU(2)_L$ charge.

Finally, we discuss the possibility to discover sleptons with MoEDAL. Run 3 sensitivities for $\tilde \tau_R$ (blue) and  $\tilde \tau_L$ (red) are shown in
Fig. \ref{fig:paper2-susy-res-stau}. The limits are weak, it is $\sim$61 (58) GeV for the metastable $\tilde \tau_L$ with $N_{\rm sig}=1$ (2), and $\sim$56 GeV for 
 $\tilde \tau_R$, corresponding to $N_{\rm sig}=1$ (there is no 
 $N_{\rm sig}=2$ contour for $\tilde \tau_R$). All masses are below the LEP 
 limit \cite{lep_limit} and already excluded. The reason why MoEDAL is insensitive to 
 sleptons is the s-channel production mode with a heavy spin-1 gauge
 boson propagator, resulting in velocity suppression of the production cross section and 
 unfavourable, from the perspective of MoEDAL, distribution of slepton velocities. 

\begin{figure}[!tbp]
\centering
  \begin{subfigure}[t]{0.49\textwidth}
    \includegraphics[width=\textwidth]{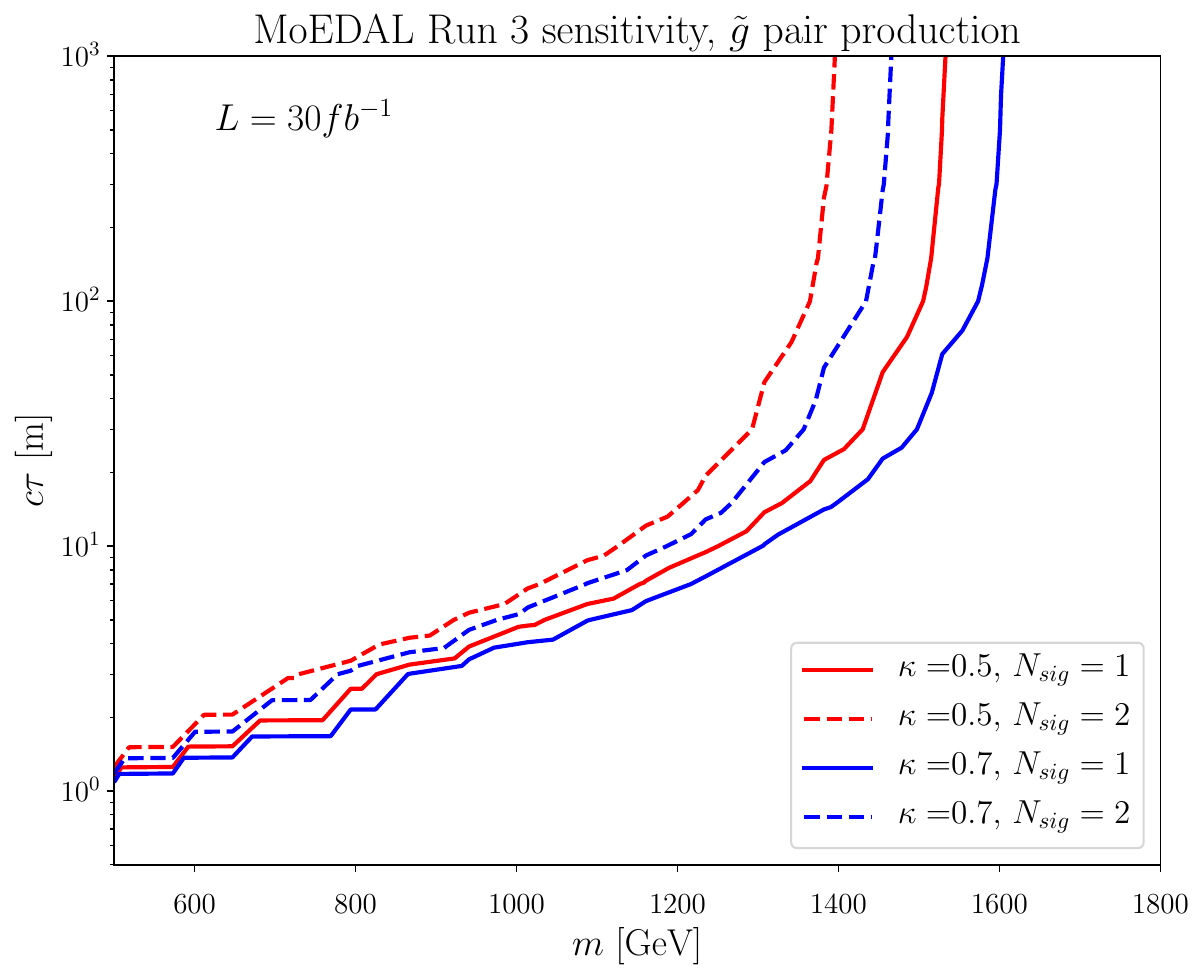}
    \caption{\small gluino R-hadrons}
    \label{fig:paper2-susy-res-gluino}
  \end{subfigure}
    \hfill
   \begin{subfigure}[t]{0.49\textwidth}
    \includegraphics[width=\textwidth]{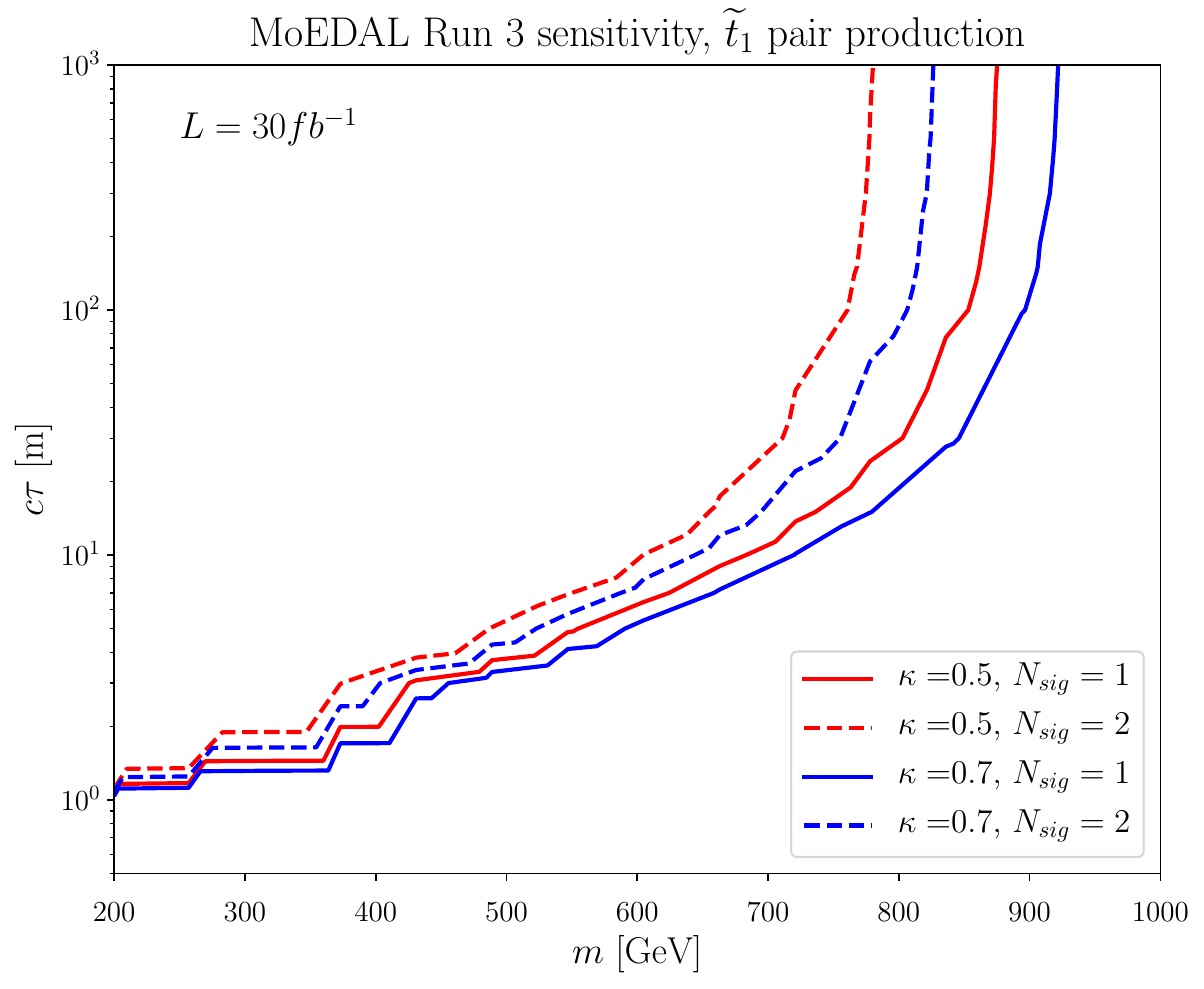}
    \caption{\small stop R-hadrons}
    \label{fig:paper2-susy-res-stop}
  \end{subfigure}
    \hfill
   \begin{subfigure}[t]{0.49\textwidth}
    \includegraphics[width=\textwidth]{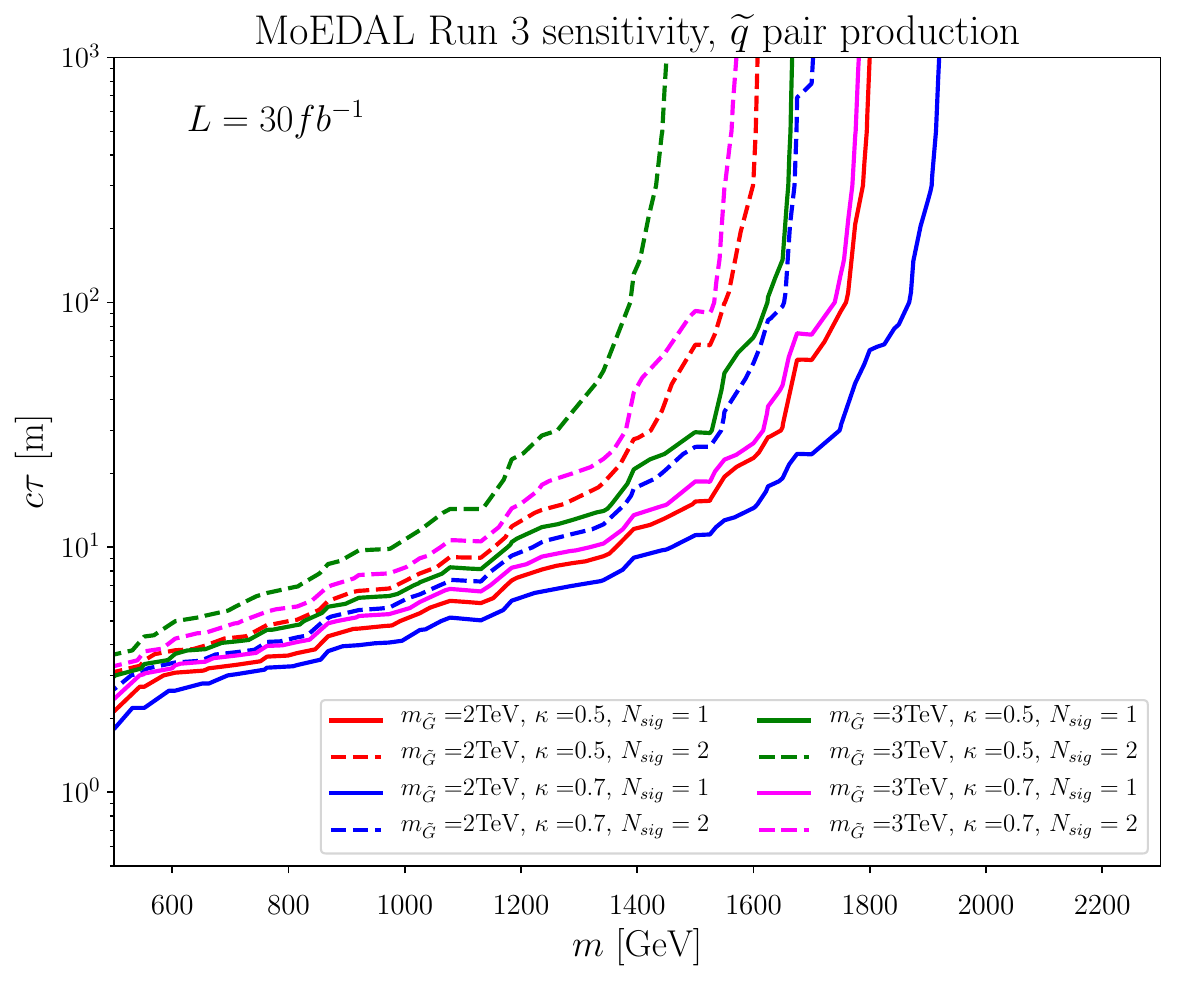}
    \caption{\small squark R-hadrons}
    \label{fig:paper2-susy-res-squark}
  \end{subfigure}
  \hfill
  \begin{subfigure}[t]{0.49\textwidth}
    \includegraphics[width=\textwidth]{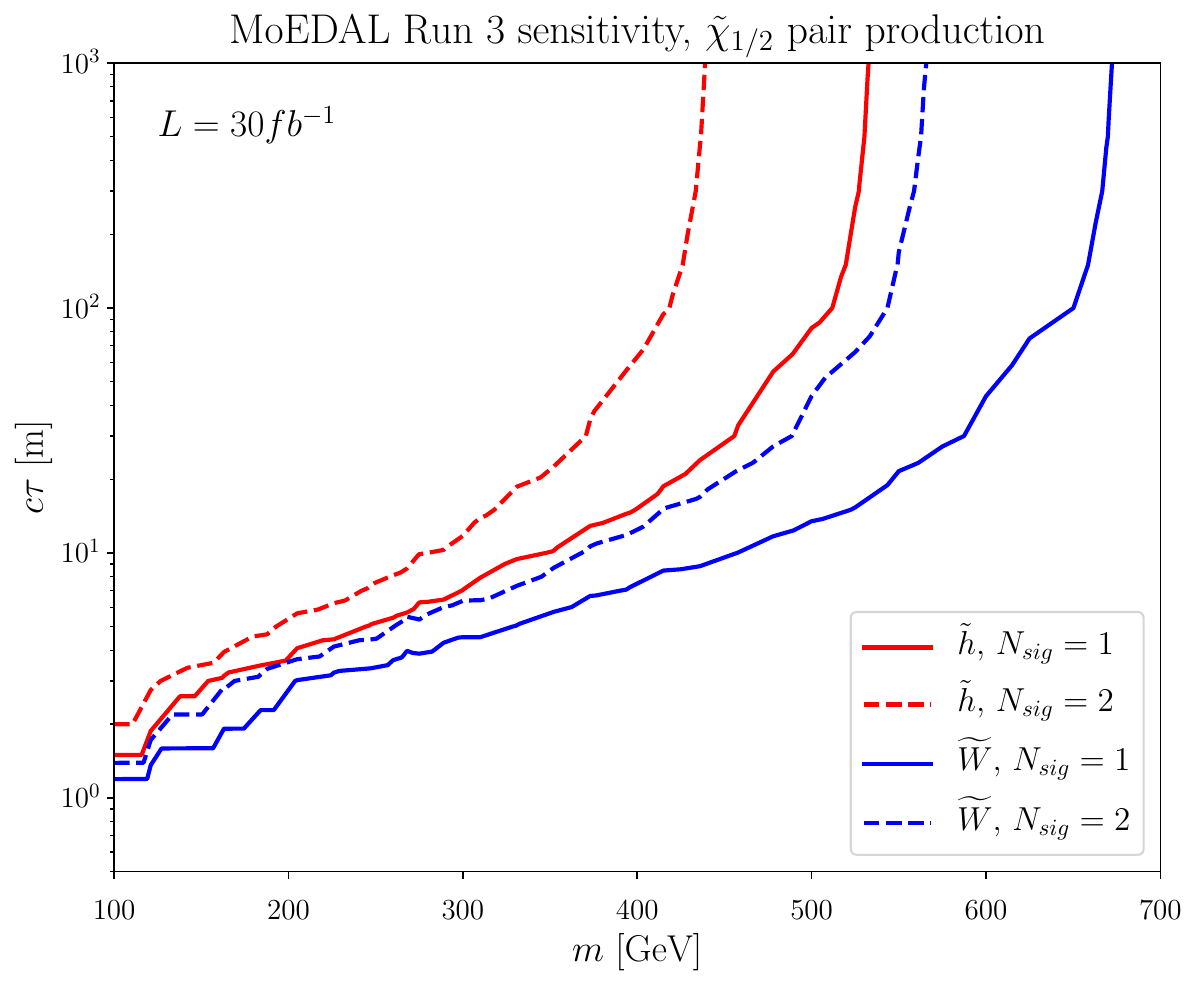}
    \caption{\small charginos}
    \label{fig:paper2-susy-res-chargino}
  \end{subfigure}
   \begin{subfigure}[t]{0.49\textwidth}
    \includegraphics[width=\textwidth]{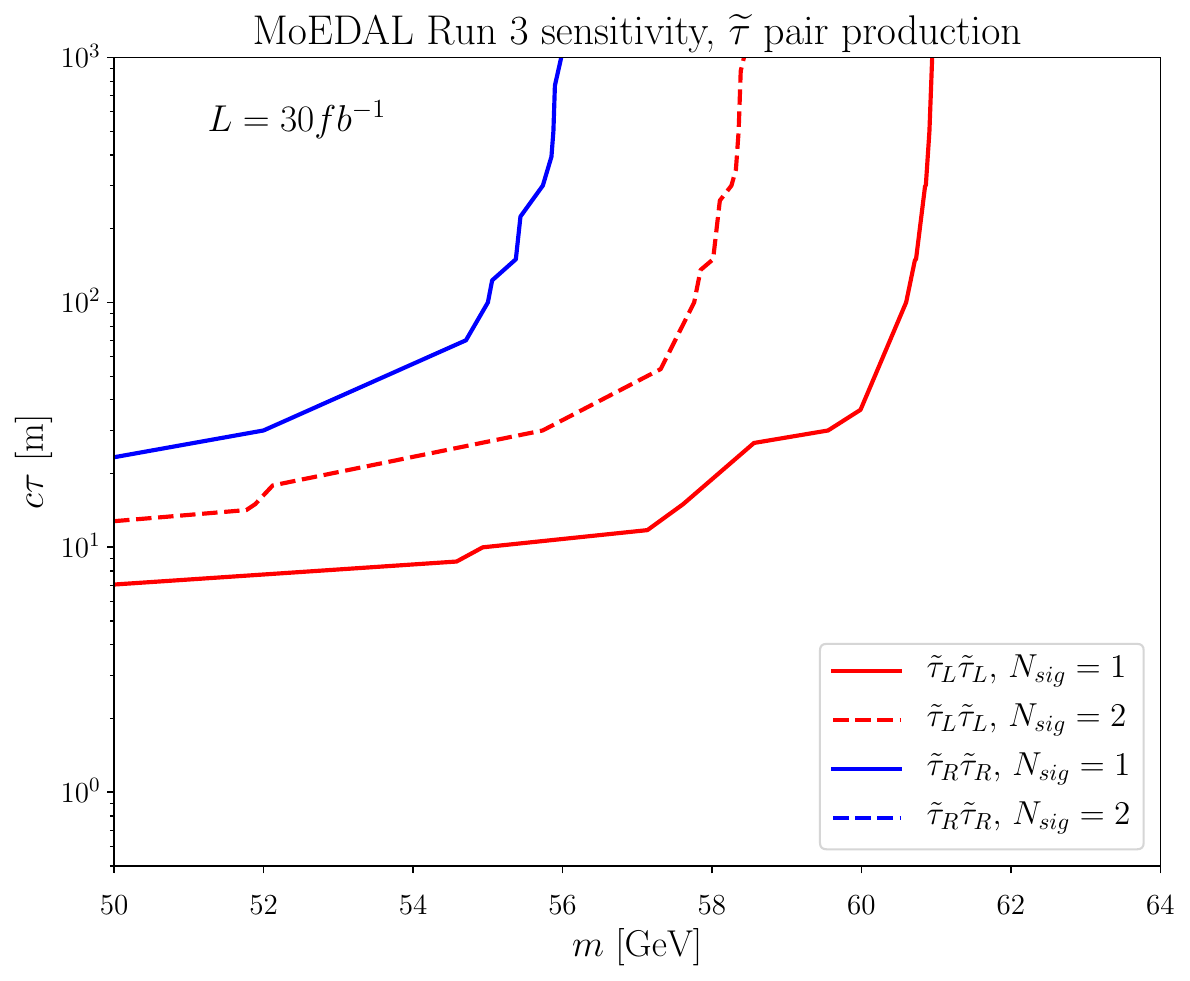}
    \caption{\small staus}
    \label{fig:paper2-susy-res-stau}
  \end{subfigure}
    \caption{\small The expected sensitivity of the MoEDAL detector, at the end of Run 3 ($L=30~\rm{fb}^{-1}$), to various types of SUSY particles.}
    \label{fig:paper2-susy-res}
\end{figure}

\myparagraph{Comparison with the existing searches}
We now compare the expected sensitivity of MoEDAL to supersymmetric 
particles with the existing constraints by ATLAS and CMS experiments. Multiple 
searches for long-lived particles are described in Sec. \ref{sec:atlascms-searches}, however, in most of them the assumed lifetimes are much shorter 
than in the scenarios considered in this project. Searches for Heavy Stable 
Charged Particles (HSCPs) are an exception since they assume that particles live long 
enough to be considered detector-stable. They rely on the ionisation energy loss 
and time of flight measurements and are less model dependent than other searches. 
Therefore, we focus on the bounds provided by the HSCP searches.

In Tab. \ref{tab:paper2-susy} we summarise the expected sensitivity of MoEDAL 
to supersymmetric particles at Run 3 for $\kappa=0.7$ and $N_{\rm sig}=1$. 
We compare these results with constraints from the latest ATLAS HSCP analysis \cite{ATLAS:2019gqq}
with $L=36.1~\rm{fb}^{-1}$. The ATLAS collaboration interpreted their results 
for long-lived stau, wino, stop and gluino, and derived the lower 95\% CL mass 
limits, which are shown in the third column in Tab. \ref{tab:paper2-susy}. At 
the time of conducting the project, the most recent CMS analysis\cite{CMS:2016kce} was 
based on 
a smaller set of data $L=2.5~\rm{fb}^{-1}$ and provided weaker limits on 
gluino, stop and stau. Lower mass limits from this analysis are presented in the 
right-most column in Tab. \ref{tab:paper2-susy}.

The limits on long-lived $\tilde g$, $\tilde t$ and $\widetilde W$ are nominally 
stronger than the expected detection reach of the MoEDAL detector. However, 
the ATLAS selection cuts might lead to gaps in the full phase space, which can 
be avoided by MoEDAL. In Tab. 1 of \cite{ATLAS:2019gqq}, ATLAS defined 5 signal regions, all of 
which incorporated single (isolated) high-momentum muons and $E_T^{\rm miss}$ triggers varying between 70 and 110 GeV. Similarly, the CMS analysis \cite{CMS:2016kce} 
assumed $E_T^{\rm miss}>170$ GeV. These requirements might not be satisfied in some of the long-lived sparticle scenarios, e.g. there would be no
$E_T^{\rm miss}$ signature in models with weak R-parity violation.

We also note that in \cite{ATLAS:2019gqq} ATLAS collaboration did not interpret their 
results for higginos nor light-flavour squarks, therefore we recast the cross 
section upper limits for sbottoms and winos. Our approach assumes that the 
detection efficiencies for HSCP analysis would be similar for sbottom and light-flavour squarks, and similarly for winos and higgsinos. Based on this 
assumption we obtain lower mass limits 1170 (2310) GeV for higgsinos 
(squarks), as written in double parentheses in Tab. \ref{tab:paper2-susy}.

Comparing the expected sensitivity of MoEDAL and current experimental 
constraints by ATLAS and CMS we note two points. Firstly, MoEDAL is sensitive 
only to particles with $Z/\beta \lesssim 7$, which reduced its ability to detect 
singly charged particles. Secondly, the amount of data which MoEDAL is expected to 
accumulate is already about 5 times smaller than the one already 
collected by ATLAS and CMS. It is because MoEDAL is located at the IP8 together 
with the LHCb detector, which requires a restricted instantaneous luminosity in order to suppress pile-up. 

As a final remark, we comment on the scenario with an $SU(2)_L$-triplet fermion$(\Sigma^\pm, \Sigma^0)$ in type III seesaw, 
already mentioned in Sec. \ref{sec:paper2-intro}. In general, the mass degeneracy between the charged and neutral fermions is 
uplifted by the radiative corrections:
$m_{\Sigma^\pm} - m_{\Sigma^0} > m_{\pi^\pm}$, resulting in a relatively short lifetime of $\Sigma^\pm$. However, if there are 
additional contributions to the mass splitting, such that $m_{\Sigma^\pm} - m_{\Sigma^0} < m_{\pi^\pm}$, the lifetime of $\Sigma^\pm$ might be long enough to allow for its detection at the MoEDAL detector. In the case when
$m_{\Sigma^0} > m_{\Sigma^\pm} + m_{\pi^\pm}$, $\Sigma^0$ decays promptly to $\Sigma^\pm$ 
and the mass reach of MoEDAL would be $\sim 670~\rm{GeV}$, similarly to Winos studied in this section.
For 
$m_{\Sigma^0} + m_{\pi^\pm} > m_{\Sigma^\pm} > m_{\Sigma^0}$, $\Sigma^\pm$ undergoes a three-body decay to $\Sigma^0$ with lifetime
$\mathcal{O}(1~\rm{s})$. In this case, $\Sigma^0$ does not contribute to the signal and the signal yield is reduced by 
approximately 1/2 because the dominant production mode is $pp\to\Sigma^\pm \Sigma^0$. We expect that the mass reach in 
such a scenario would be similar to the result for higgsinos.

\begin{table}[!t]
\centering
\caption{Comparison of the expected MoEDAL sensitivity at the end of Run 3 ($L=30~\rm{fb}^{-1},~\kappa=0.7,~N_{\rm sig}=1$) with the latest ATLAS and CMS constraints.}
\begin{tabular}{c|c|c|c}
              & MoEDAL & (ATLAS)  & (CMS)  \\ \hline
$\tilde g$    & 1600   & (2000)   & (1500) \\
$\tilde q$    & 1920   & ((2310)) & -      \\
$\tilde t$    & 920    & (1350)   & (1000) \\
$\widetilde W$    & 670    & (1090)   & -      \\
$\tilde h$    & 530    & ((1170)) & -      \\
$\tilde \tau$ & 61     & (430)    & (230) 
\end{tabular}\label{tab:paper2-susy}
\end{table}

\subsection{Doubly charged particles}\label{sec:paper2-doubly}
\myparagraph{Motivation for candidates}
Scalar particles with an electric charge twice the charge of a proton/electron can arise in many BSM scenarios. A typical example, 
already mentioned in Sec. \ref{sec:paper2-intro}, is the type II seesaw model \cite{Schechter:1980gr,Cheng:1980qt,Melfo:2011nx, BhupalDev:2013xol, Ghosh:2017pxl}, in which the SM is augmented with a complex $SU(2)_L$-triplet of scalar fields. Other BSM scenarios in which doubly charged scalars appear are: Left-Right symmetric model \cite{Pati:1974yy, Mohapatra:1974hk, Senjanovic:1975rk},
Georgi-Machacek model \cite{Georgi:1985nv, Chanowitz:1985ug, Gunion:1989ci, Ismail:2020zoz}, 3-3-1 model \cite{CiezaMontalvo:2006zt, Alves:2011kc} and Little Higgs model \cite{Arkani-Hamed:2002iiv}. Supersymmetric versions of these models 
might lead to doubly charged spin-1/2 higgsinos, e.g. the supersymmetric version of the Left-Right 
symmetric model based on $SU(3)_C \times SU(2)_L \times SU(2)_R \times U(1)_{B-L}$ gauge group. Doubly charged scalars can be also introduced in simplified models \cite{Delgado:2011iz, Alloul:2013raa}.

The type II seesaw model, already introduced in Sec. \ref{sec:seesaw}, contains $SU(2)_L$-triplet scalar multiplet $\Delta$ 
interacting with leptons and W bosons. The interaction strength is regulated by the vacuum expectation value, $v_T$, of the neutral 
component of the $\Delta$. Majorana masses of neutrinos are given by $M_\nu = \sqrt{2} Y_\nu v_T$, with $Y_\nu$ being the 
neutrino Yukawa coupling. Collider limits on the mass of the doubly charged scalar, $H^{\pm\pm}$, depend on the value of $v_T$.
For small $v_T \leq 10^{-4}~\rm{GeV}$ (corresponding to large $Y_\nu$), 
the like-sign dilepton (LSD) channel can be effectively used to search for a doubly charged Higgs.
The latest limit was set by the ATLAS collaboration \cite{ATLAS:2017xqs}, 
and for a doubly charged Higgs boson coupled to left-handed leptons $H_L^{\pm\pm}$, it varies between 770 and 870 GeV, depending on the branching ratio for a $H_L^{\pm\pm} \to e^\pm e^\pm$ decay, with an assumption that $\rm{Br}(H_L^{\pm\pm} \to l^\pm l^\pm)=100\%~(l=e,\mu)$. For $\rm{Br}(H_L^{\pm\pm} \to l^\pm l^\pm)=10\%$, the limit is $450~\rm{GeV}$.

On the other hand, if we assume a small Yukawa and large vev $v_T \geq 10^{-4}~\rm GeV$, then the LSD decay mode is highly 
suppressed, while several other competing modes open up if kinematically accessible, e.g. (i) pair of heavy bosons $W^\pm W^\pm$, 
(ii) $W^\pm h^\pm$ and (iii) $H^\pm H^\pm$. $W^\pm$ and $H^\pm$ decay subsequently to jets and leptons, resulting in a 
variety of rather complicated final states, and weak limits on the mass of a doubly charged scalar.
ATLAS considered production of a pair of doubly charged scalars, each decaying to two W bosons with 100\% 
branching ratio, for $L=36.1~{fb}^{-1}$ 13 TeV LHC data. No signal excess was observed, allowing to exclude $m_{H^{\pm\pm}}$ 
between 200 and 220 GeV are 95\% CL \cite{ATLAS:2018ceg}.

In the Left-Right (LR) symmetric model there are two scalar multiplets $\Delta_L$ and $\Delta_R$, corresponding to
$SU(2)_L$ and  $SU(2)_R$ triplet representations, respectively. Each of them has an associated doubly charged component 
$H^{\pm\pm}_{L,R}$. 
Since the scalars are gauged under different $SU(2)$ groups, their couplings to fermions and gauge 
bosons are distinct \cite{ATLAS:2017xqs,Huitu:1996su}. 
The ATLAS collaboration investigated the LSD channel looking for doubly charged scalars in $L=36.1~{\rm fb}^{-1}$ 13 TeV LHC 
data \cite{ATLAS:2017xqs}. No signal excess over the SM background was observed, allowing to set 95\% CL limits on $m_{H^{\pm\pm}_{L}}$ and 
$m_{H^{\pm\pm}_{R}}$.
The limit for $m_{H^{\pm\pm}_{L}}$ is the same as for type II seesaw scalar.
In the case of the ${H^{\pm\pm}_{R}}$,
with $\rm{Br}(H_R^{\pm\pm} \to l^\pm l^\pm)=100\%$ the limit varies between 660 and 760 GeV, depending on the branching ratio for $H_R^{\pm\pm} \to e^\pm e^\pm$. For $\rm{Br}(H_R^{\pm\pm} \to l^\pm l^\pm)=10\%$ the corresponding limit is $320~\rm{GeV}$.

The 95\% CL lower limit on a mass of a doubly charged scalar in the Georgi-Machacek model is 220 GeV. 
It was obtained 
by the ATLAS collaboration using $L=36.1~{\rm fb}^{-1}$ 13 TeV LHC data \cite{ATLAS:2018ceg}.

We would like to stress that in all the searches described above, the doubly charged scalar is assumed to decay promptly ($c\tau < 10~\rm \mu m$), hence they are complementary to the search for long-lived charged particles at the MoEDAL detector. There 
exist values of $v_T$ and $m_{H^{\pm\pm}}$ in the type II seesaw model for which the doubly charged scalar $H^{\pm\pm}$ can 
have a significantly longer lifetime, resulting in an HSCP-like signature. Such signatures were studied by both ATLAS and CMS and 
will be discussed later.

\myparagraph{Run 3 projections for doubly charged particles}
In the project, we estimate the sensitivity of MoEDAL to doubly charged scalars and spin-1/2 fermions, which are either $SU(2)_L$ 
singlets or triplets with quantum number assignments $(SU(2)_L, U(1)_Y)=(\mathbf{1}, 2)$ and  $(SU(2)_L, U(1)_Y)=(\mathbf{3}, 1)$, respectively. All these particles are assumed to be $SU(3)_C$ singlets\footnote{In this project we do not consider heavy coloured particles hadronising into doubly charged bound states. Such particles are interesting from the MoEDAL point of view, because their cross section is expected to be large, and velocity distribution might be favourable. Such particles will be discussed in the next two projects.}.

The pair production cross section for the four considered types of doubly charged particles is depicted in Fig. \ref{fig:paper2-doubly-xs}. For particles 
with the same spin, the triplet cross section is about an order of magnitude larger than for the singlet. The difference comes from 
including all neutral $pp \to Y^0Y^0,~Y^+Y^-,~Y^{++}Y^{--} + \ldots$ and charged
$pp \to Y^0Y^\pm,~Y^\pm Y^{\pm\pm}+\ldots$ triplet production modes, with the assumption that the heavier components 
decay to nearly-degenerate lightest doubly charged particle. In that way, all production modes lead to a final state with two doubly 
charged particles. Neutral production modes are produced via s-channel with $\gamma/Z$ exchange, while charged ones with s-
channel involving $W^\pm$. In the $SU(2)_L$-singlet case, there is only one neutral production mode, i.e. $pp \to Y^{++} Y^{--}+\ldots$.

\begin{figure}[!t]
\centering
\includegraphics[width=1\textwidth]{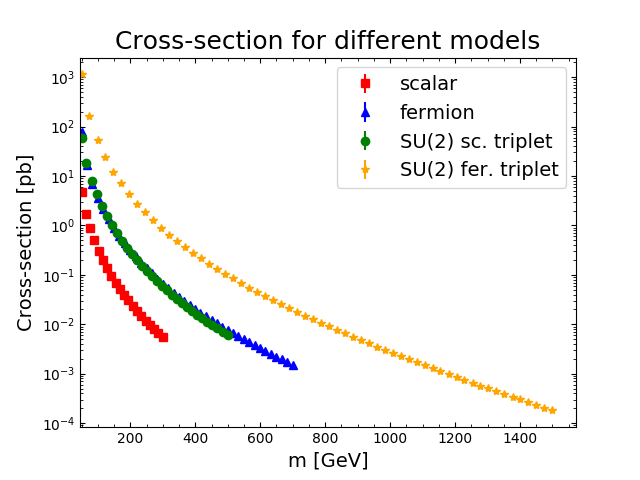}
\caption{\small Leading order cross section for pair production of various doubly charged particles at 13 TeV LHC.
}
\label{fig:paper2-doubly-xs}
\end{figure}

\begin{figure}[!tbh]
\centering
  \begin{subfigure}[t]{0.485\textwidth}
    \includegraphics[width=\textwidth]{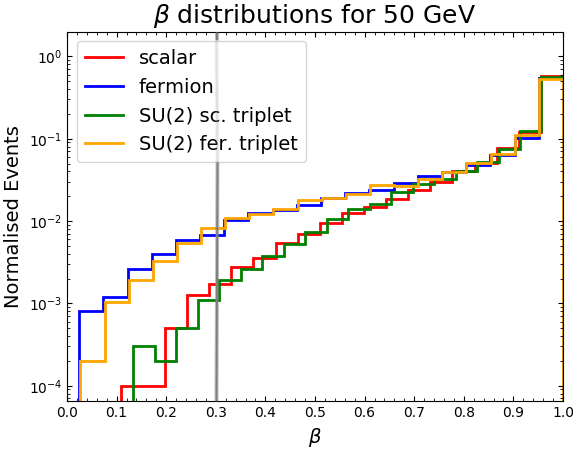}
    \label{fig:paper2-doubly-beta-50}
  \end{subfigure}
  \hfill
   \begin{subfigure}[t]{0.49\textwidth}
    \includegraphics[width=\textwidth]{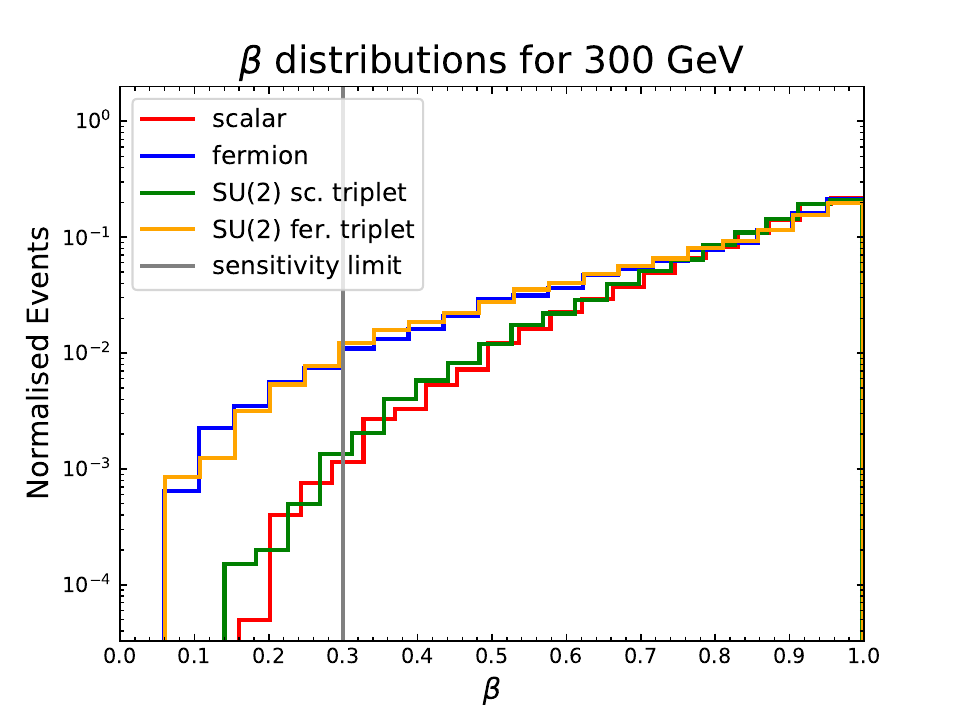}
    \label{fig:paper2-doubly-beta-300}
\end{subfigure}
\caption{The velocity distributions of various doubly charged particles $Y$, with $m_Y=50~\rm GeV$ (left) and $m_Y=300~\rm GeV$ (right). The vertical grey line corresponds to the MoEDAL's sensitivity limit, i.e. the requirement that $\beta < 0.3$.}
\label{fig:paper2-doubly-beta}
\end{figure}

Another observation is that the cross section in Fig. \ref{fig:paper2-doubly-xs} is an order of magnitude larger for fermions than scalars with the same 
quantum numbers. 
It is because a Dirac fermion has twice as many degrees of 
freedom as a complex scalar. Another reason,
already discussed earlier, is that scalars suffer from threshold velocity suppression, i.e. $\sigma \to 0$ as $\beta \to 0$, which is absent for fermions.

One important difference between sparticles in Sec. \ref{sec:paper2-susy} and doubly charged particles is the change of the sensitivity limit on velocity, below 
which MoEDAL can detect BSM particles. For singly charged particles, the limit is $\beta<0.15$, but for doubly charged particles it 
is $\beta<0.30$, which significantly enhances the sensitivity of MoEDAL. Fig. \ref{fig:paper2-doubly-beta} depicts normalised velocity distributions for various pair-produced
doubly charged particles, where the plot on the left (right) is for $m_Y=50$ (300) GeV. The vertical grey lines in Fig. \ref{fig:paper2-doubly-beta} 
correspond to MoEDAL's sensitivity limit for doubly charged particles. Comparing both plots in Fig. \ref{fig:paper2-doubly-beta} we observe that they are 
more mass-dependent in the high $\beta$ region, and mildly change for the low $\beta$. Nonetheless, the difference is still sizeable, for $m_Y= 300$ GeV about twice as many events fall into the $\beta<0.3$ region than for $m_Y= 50$ GeV.

Finally, in Fig. \ref{fig:paper2-doubly-result} we present the expected sensitivity of Run 3 MoEDAL to the four considered types of doubly 
charged particles: scalar singlet (red), fermion singlet (green), scalar triplet (blue) and fermion triplet (magenta). Solid 
contours correspond to $N_{\rm sig} = 1$, while dashed contours represent  $N_{\rm sig} = 2$. As expected, MoEDAL 
is most sensitive to fermion $SU(2)_L$-triplet, due to its large cross section and favourable velocity distribution. The 
detection reach in the $c\tau > 100~\rm m$ limit is 1130 (990) GeV for  $N_{\rm sig} =1$ (2). For the singlet fermion, 
MoEDAL can probe masses up to 650 (540) GeV for $N_{\rm sig} =1$ (2). Sensitivity to scalars is much lower than 
fermions due to lower production cross section and unfavourable velocity distribution. The MoEDAL mass reach for the 
scalar triplet is about 340 (280) GeV, and for the singlet, it is 160 (130) GeV, for $N_{\rm sig} =1$ (2) (if $c\tau > 100~
\rm m$).

\begin{figure}[tbh]
\includegraphics[width=\textwidth]{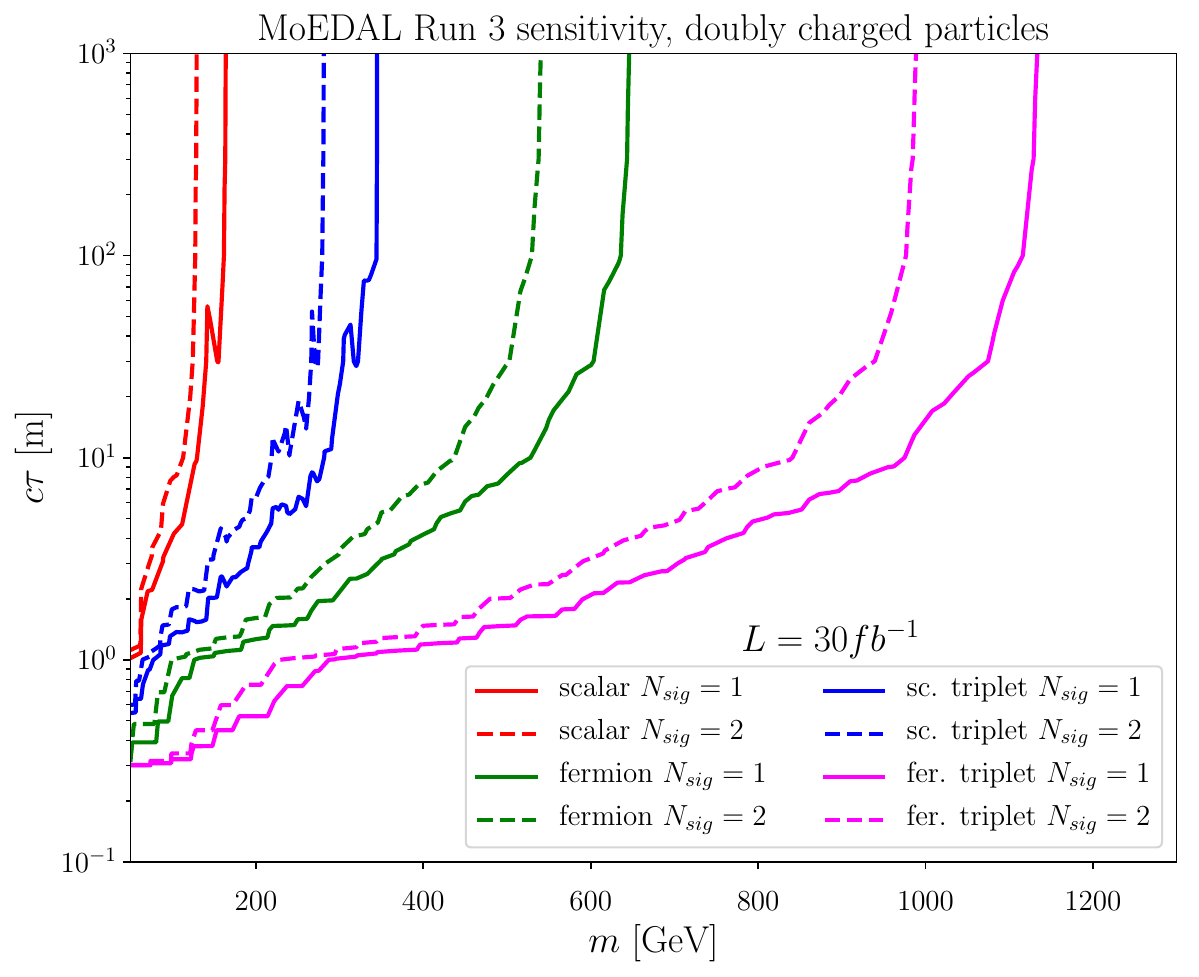}
\caption{
\small The expected sensitivity of the MoEDAL detector at the end of Run 3 to doubly charged singlet and triplet spin-0 and spin-1/2 particles. The integrated luminosity is assumed to be $L=30~\rm{fb}^{-1}$.
}
\label{fig:paper2-doubly-result}
\end{figure}

\myparagraph{Comparison with the existing searches}
We compare prospects for the detection of doubly charged particles at the Run 3 MoEDAL with the current constraints 
from ATLAS and CMS. As discussed in Sec. \ref{sec:paper2-susy}, the most relevant constraints come from HSCP searches, 
which are based on the ionisation energy loss and time of flight measurements, but also incorporate some 
additional trigger requirements. 
ATLAS analysis \cite{ATLAS:2019gqq} for $L=36.1~\rm{fb}^{-1}$ 13 TeV data utilises the largest data set, but the collaboration did 
not interpret its results for doubly charged particles. 

The CMS collaboration, on the other hand, published its result \cite{CMS:2016kce}
for smaller data set ($L=2.5~\rm{fb}^{-1}$), but interpreted it for long-lived doubly charged spin-1/2 $SU(2)_L$-
singlet. The mass bound provided in \cite{CMS:2016kce} is 680 GeV, which is 30 GeV higher than the MoEDAL mass reach for Run 3, 
as shown in Tab. \ref{tab:paper2-doubly}. We would like to stress that CMS analysis relies on extra assumptions, e.g. $E_T^{\rm miss} > 170 ~\rm{GeV}$, making its lower mass bound more model-dependent than the expected MoEDAL sensitivity.

Unfortunately, the CMS collaboration did not interpret their result in \cite{CMS:2016kce} for other types of doubly charged particles, 
hence we are forced to recast it. In order to derive the approximate mass bounds, we assume that detector efficiencies 
are not very sensitive to the type of particle, only to its electric charge. The mass bounds obtained within this assumption for long-lived doubly charged scalars and fermion triplet are in double parentheses in Tab. \ref{tab:paper2-doubly}.

The recast CMS sensitivity to scalar singlet (triplet) is estimated to be 320 (590) GeV, which is to be compared with 
the expected sensitivity of the MoEDAL detector of 160 (340) GeV (for $L=30~\rm{fb}^{-1}$,  $N_{\rm sig}=1$ and $c\tau>100~\rm m$). In the case of the fermion triplet, the estimated CMS bound is 900 GeV, to be compared with 
1130 GeV that can be probed at Run 3 MoEDAL. All lower mass bounds are summarised in Tab. \ref{tab:paper2-doubly}.

\begin{table}[!t]
\centering
\caption{ \small Comparison of the expected MoEDAL sensitivity ($N_{\rm sig} = 1$) to doubly charged particles
at Run 3, and
the current ($95\%$ CL) mass bounds 
for a long-lived fermion-singlet doubly charged particle 
obtained from the CMS HSCP search 
with $L = 2.5$ fb$^{-1}$~\cite{CMS:2016kce} (in parentheses). 
Additionally, we provide lower mass bound for fermion triplet and scalars obtained by re-interpreting  
\cite{CMS:2016kce} (in double parentheses).
}
  \begin{tabular}{c|c|c} 
& MoEDAL   & (CMS)  \\ 
\hline
Scalar singlet  & {160} & 
{((320))}\\ 
\hline
Fermion singlet  & {650} &
{(680)}\\
\hline
Scalar triplet  & {340} & 
{((590))} \\
\hline
Fermion triplet  & {1130} & {((900))}  \\
  \end{tabular}
\label{tab:paper2-doubly}
\end{table}

\subsection{Conclusions}\label{sec:paper2-conclusions}
In this project, we have estimated the expected sensitivity of the MoEDAL detector to various types of long-lived 
charged particles, and compared it to current lower mass bounds provided by ATLAS and CMS experiments.
First, various types of singly charged supersymmetric particles were considered. Next, long-lived doubly 
charged scalars and fermions, inspired by type II and type III seesaw models, were investigated. We emphasise that the 
MoEDAL search for long-lived BSM particles is completely model-independent and effectively free from the SM 
background, contrary to ATLAS and CMS analyses, 
which always rely on a set of triggers and selection cuts to reduce the SM background. However, the 
sensitivity of MoEDAL is severely limited due to two factors. The first reason is that MoEDAL is sensitive only to slowly 
moving particles, i.e. $\beta<$0.15 (0.30) for $|Q|=1e$ ($|Q|=2e$). The second reason is the low luminosity available to 
MoEDAL, due to requirements of the LHCb experiment located at the same interaction point. At the end of Run 3 
MoEDAL is expected to collect an order of magnitude fewer data than ATLAS and CMS.

The best sensitivity of MoEDAL to supersymmetric particles is for strongly interacting sparticles forming long-lived  R-hadrons, namely gluinos, light-flavour quarks and stops. Sensitivity to charged winos and higgsinos is also interesting, 
but there are no prospects to detect directly produced long-lived staus because the lower mass bound is below the 
model-independent limit established by LEP. Nonetheless, we regard the MoEDAL model-independent limits to be 
complementary to searches by ATLAS and CMS, and particularly interesting in the scenario with weakly-broken R-parity, in which there is no $E_T^{\rm miss}$.

Out of all the particles considered in this project, doubly charged $SU(2)_L$-triplet fermions are the most favourable 
for MoEDAL. Since they are doubly charged, the velocity threshold is relaxed to $\beta_{\rm th} = 0.30$. Unlike scalar 
particles, the production cross section of fermions does not vanish in the $\beta \to 0$ limit, which results in slower, 
thus detectable in MoEDAL, particles produced. Finally, the production cross section for $SU(2)_L$-triplet is enhanced 
with respect to the singlet. Our analysis suggests that MoEDAL can provide better limits than CMS, however, one has to 
remember that the CMS result is based on a small data set and the CMS lower mass bound was obtained using our 
approximate recasting procedure.

It is evident that at the end of Run 3, ATLAS and CMS will be able to provide stronger constraints on all of the 
considered doubly charged particles. However, once more we would like to emphasise that the limits of MoEDAL are 
complementary and free of typical auxiliary signal assumptions.

\section{Detecting long-lived multi-charged particles in neutrino mass models with MoEDAL}\label{sec:paper3}

\subsection{Introduction}

In the previous two projects, the prospects for the detection of supersymmetric and doubly charged particles at Run 3 MoEDAL 
were discussed. Now we turn to study radiative neutrino mass models introduced in \cite{R:2020odv} and
described in Sec. \ref{sec:radiative-mass-gen}. These two models (coloured and uncoloured) are particularly 
interesting to MoEDAL because they predict the existence of long-lived multiply charged particles, for which 
MoEDAL's sensitivity might be significantly enhanced. They are also interesting from a theoretical point of view. 
Many of the 1-loop models introduce additional discrete symmetry in orded to remove unwanted tree-level 
contributions to neutrino masses and provide a Dark Matter candidate. A prototype model of this class is the scotogenic model \cite{Ma:2006km}.
On contrary, the models introduced in \cite{R:2020odv} do not require imposing ad-hoc discrete symmetries, the leading order 
contributions to the mass of the neutrino are automatically at the 1-loop level.

In this project, we consider both coloured and uncoloured versions of the model described in Sec. \ref{sec:radiative-mass-gen}. We perform a scan over the model's parameters in order to determine for which values MoEDAL can test the 
considered scenario. Moreover, we present the sensitivity of MoEDAL to multiply charged particles in a general case, 
where the mass and decay length of the new particle are free parameters. We also demonstrate the significance of the photon 
fusion process to the production of particles with large electric charges, which allows us to update lower mass bounds from Sec. \ref{sec:paper2-doubly}.

\subsection{Neutrino mass models and multiply charged long-lived particles}\label{sec:paper3-model}

One of the consequences of the radiative neutrino mass model described in Sec. \ref{sec:radiative-mass-gen} is the existence of long-lived multiply charged scalars. In this section, we will study the phenomenology of these particles. 

Let us begin with the basic (uncoloured) version of the model, in which there is a scalar $SU(2)_L$-singlet $S_1$ with $Y=2$, a scalar 
$SU(2)_L$-triplet $S_3$ with $Y=1$, and three generations of fermionic $SU(2)_L$-doublets $F_i$ ($\bar{F}_i$) with $Y=+5/2$ ($Y=+5/2$), as listed in Tab. \ref{tab:hirsch}.
If $m_{S_3} \ll m_{F}, m_{S_1}$, then the components of the scalar $SU(2)_L$-triplet might be long-lived. These components are 
$S^{+2}$, $S^{+3}$, and $S^{+4}$, with charges $2e$, $3e$, and $4e$, respectively. We would like to investigate the decays of these particles and their dependence on the model's parameters in order to understand what controls their lifetime. The relevant interactions from Eq. \eqref{eq:neutrino-lagrangian} are:
\begin{equation}\label{eq:paper3-neutrino-lagr}
\mathcal{L}_{\rm BSM} \ni  \mathcal{L}_{\rm KIN}
-(h_{ee})_{ij} (e_R^{ i})^C (e_R^{ j})^C S_1^\dag -
(h_F)_{ij} L_i F_j S_1^\dag
- (h_{\bar F})_{ij} L_i \bar F_j S_3 
-  \lambda_5 H H S_1 S_3^\dag + h.c.
\end{equation}
The first term in Eq. \eqref{eq:paper3-neutrino-lagr} is just the ordinary kinetic term $\mathcal{L}_{\rm KIN}$ containing interactions between the SM gauge bosons and BSM fields.
The second term in Eq. \eqref{eq:paper3-neutrino-lagr} is responsible for the decay of the lightest BSM particle to SM leptons via $S_1$. The next two terms are Yukawa-type interactions connecting SM lepton doublet with BSM scalars and fermions. The last term in Eq. \eqref{eq:paper3-neutrino-lagr} allows for mixing between the singlet $S_1$ and triplet $S_3$ scalars. The dimensionless coupling $\lambda_5$ is connected to the violation of the lepton number conservation, i.e. when $\lambda_5 \to 0$ the lepton number is conserved. The last three terms in the Eq. \eqref{eq:paper3-neutrino-lagr} contribute to the neutrino mass generation loop diagram shown in Fig. \ref{fig:neutrino-gen}, hence the product $\lambda_5 \times h_F h_{\bar F}$ is constrained from above by the neutrino data (see Eq. \eqref{eq:neutrino-mass}), which implies that for phenomenologically viable parameter values $\lambda_5 \propto  (h_F h_{\bar F})^{-1}$.

Let us now discuss possible decay modes of long-lived scalars.
Decays of $S^{+4}$ within the triplet states, e.g. $S^{+4}\to l^+ \nu S^{+3}$ through an off-shell W boson are subdominant, because of 
the chiral suppression and proportionality to the final state lepton.  
Therefore, the main decay modes for the quadruply charged state are $S^{+4}\to 4l^+$ and $S^{+4} \to W^+ W^+ l l$, as depicted in Fig. 
\ref{fig:neutrino-s4decay}. The approximate formulas for partial decay widths, if we assume $v^2 \ll m_{S_3} \ll m_{S_1}, m_F$, are given by \cite{R:2020odv}:

\begin{align}\label{eq:neutrino-p4to4ldecay}
\Gamma \left(S^{+4}\to l^+_\alpha l^+_\beta l^+_\gamma l^+_\delta \right) &\sim
\frac{ | ( h_{ee} )_{\alpha \beta} |^2 }{192  (4\pi)^5} \left| \left\langle \frac{h^{\gamma\delta}_{F \bar F}}{m_F} \right\rangle \right|^2 \frac{m^7_{S_3}}{m^4_{S_1}}, \\
\Gamma \left( S^{+4} \to W^+ W^+ l^+_\alpha l_\beta^+ \right) &\sim
\frac{ |g_2^2  ( h_{ee} )_{\alpha \beta} |^2 }{48  (4\pi)^5} \left( \frac{\lambda_5 v^2}{m^2_{S_1}} \right)^2 \frac{m^5_{S_3}}{m_W^4}, \label{eq:neutrino-p4to2l2wdecay}
\end{align}
with 
\begin{equation}
 \left| \left\langle \frac{h^{\gamma\delta}_{F \bar F}}{m_F} \right\rangle \right| \equiv 
 \left| \sum_k \frac{ h_F^{\gamma k } h_{\bar F}^{\delta k} + h_F^{\delta k } h_{\bar F}^{\gamma k} }{m_{F_k}} \right|.
\end{equation}

\begin{figure}
\centering
\includegraphics[width=\textwidth]{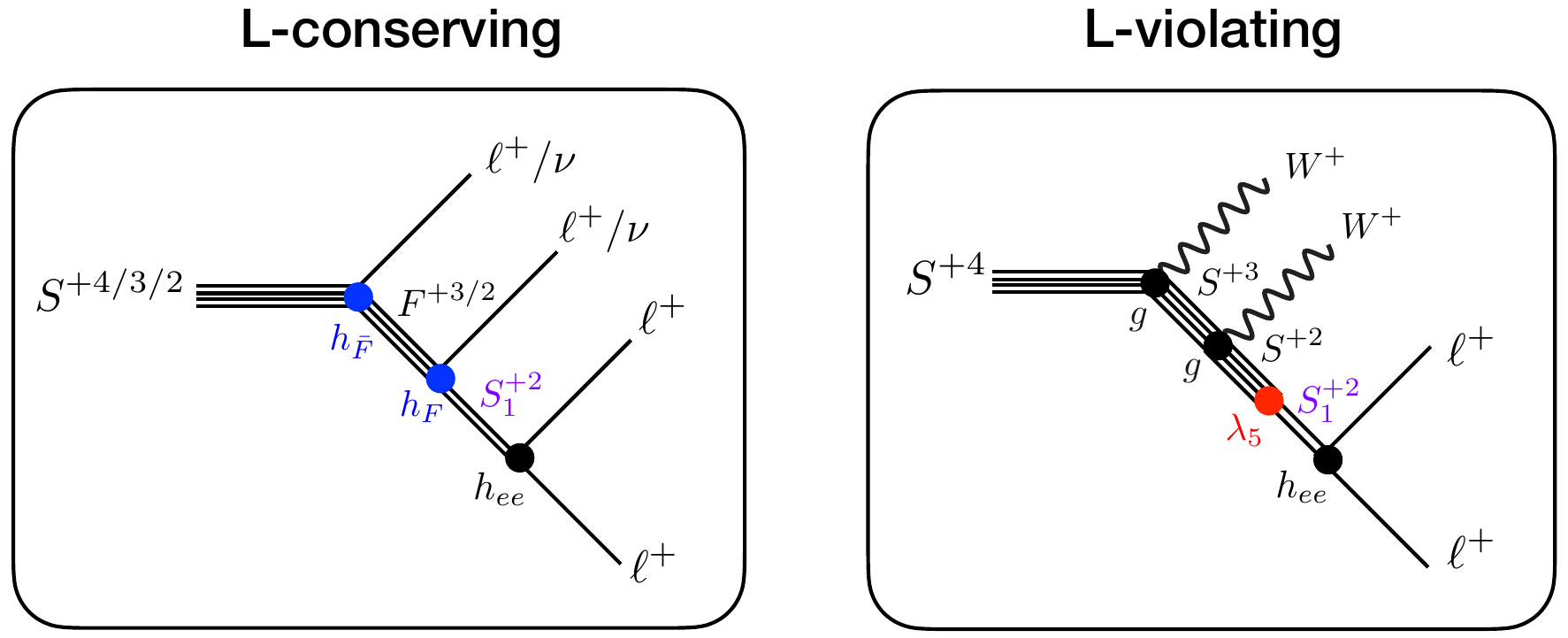}
\caption{\small Diagrams for possible decays of $S^{+4}$, $S^{+3}$ and $S^{+2}$.
The same particle can decay into $L=-4$ final state (left) and $L=-2$ (right), hence its lepton number is not fixed unless $\lambda_5$ or $h_F \times h_{\bar F}$ is non-zero.
}
\label{fig:neutrino-s4decay}
\end{figure}  

It is worth noting that the four-lepton decay width in Eq. \eqref{eq:neutrino-p4to4ldecay} is proportional to $h_F h_{\bar F}$,
while the decay width in Eq. \eqref{eq:neutrino-p4to2l2wdecay} is proportional to $\lambda_5$. A correct explanation of 
neutrino masses requires the product of all these couplings to be small, which follows from the Eq. \eqref{eq:neutrino-mass}.

The two relevant decay modes for the triply charged scalar are $S^{+3} \to l^+ l^+ l^+ \nu$ and
$S^{+3} \to W^+ l^+ l^+$. As for the $S^{+4}$, the first decay mode obeys the conservation of the lepton number, while the second violates it. Partial decay rates are (approximately):

\begin{align}\label{eq:neutrino-p3to3ldecay}
\Gamma \left(S^{+3}\to l^+_\alpha l^+_\beta l^+_\gamma \nu^+_\delta \right) &\sim
\frac{ | ( h_{ee} )_{\alpha \beta} |^2 }{192  (4\pi)^5} \left| \left\langle \frac{h^{\gamma\delta}_{F \bar F}}{m_F} \right\rangle \right|^2 \frac{m^7_{S_3}}{m^4_{S_1}}, \\
\Gamma \left( S^{+3} \to  W^+ l^+_\alpha l_\beta^+ \right) &\sim
\frac{ |g_2^2  ( h_{ee} )_{\alpha \beta} |^2 }{16  (4\pi)^3} \left( \frac{\lambda_5 v^2}{m^2_{S_1}} \right)^2 \frac{m^3_{S_3}}{m_W^2}. \label{eq:neutrino-p3to2l1wdecay}
\end{align}
Note that Eq. \eqref{eq:neutrino-p3to3ldecay} is identical to Eq. \eqref{eq:neutrino-p4to4ldecay}, because the two decays are 
related by the $SU(2)_L$ symmetry.

Decay modes for the doubly charged component of the scalar triplet are $S^{+2} \to \nu \nu l^+ l^+$ and $S^{+2} \to l^+l^+$, and the partial decay widths are:
\begin{align}\label{eq:neutrino-p2to2l2nudecay}
\Gamma \left(S^{+2}\to l^+_\alpha l^+_\beta \nu^+_\gamma \nu^+_\delta \right) &\sim
\frac{ | ( h_{ee} )_{\alpha \beta} |^2 }{192  (4\pi)^5} \left| \left\langle \frac{h^{\gamma\delta}_{F \bar F}}{m_F} \right\rangle \right|^2 \frac{m^7_{S_3}}{m^4_{S_1}}, \\
\Gamma \left( S^{+2} \to  l^+_\alpha l_\beta^+ \right) &\sim
\frac{ |  ( h_{ee} )_{\alpha \beta} |^2 }{4\pi} \left( \frac{\lambda_5 v^2}{m^2_{S_1}} \right)^2 {m_{S_3}}. \label{eq:neutrino-p2to2ldecay}
\end{align}
Again, the approximate formula for the 4-body decay in Eq. \eqref{eq:neutrino-p2to2l2nudecay} is the same as Eqs. 
\eqref{eq:neutrino-p4to4ldecay} and \eqref{eq:neutrino-p3to3ldecay}, because of the $SU(2)_L$ symmetry. This decay is 
subdominant in the case of the doubly charged scalar unless the $\lambda_5$ parameter is very small.

In the case of the coloured version of the model,
formulas for masses of new particles and their decay widths remain the same in their form, with the following replacements: 
$S^{+2} \to S^{+4/3}$,
$S^{+3} \to S^{+7/3}$,
$S^{+4} \to S^{+10/3}$,
$(h_{ee})_{\alpha \beta} \to (h_{ed})_{\alpha\beta}$, and one charged lepton in each of the considered final states is exchanged for the down antiquark $\bar d$, e.g.
$\Gamma \left( S^{+2} \to  l^+_\alpha l_\beta^+ \right)  \to \Gamma \left( S^{+4/3} \to  \bar{d}_\alpha l_\beta^+ \right) $.

\subsection{Model implementation and signal estimation}

In order to perform simulations, the studied model is first implemented in {\tt SARAH} \cite{Staub:2013tta, Staub:2012pb}. The output is 
plugged to {\tt SPheno} \cite{Porod:2003um, Porod:2011nf} in order to calculate mass spectra, mixing matrices and scalar two-body decay rates and cross sections. Next, this information is fed to {\tt MadGraph5} \cite{Alwall:2011uj} for event generation and 
numerical evaluation of cross sections and decay rates. For the cross section calculations, we use the
\\{\tt LUXqed17\_plus\_PDF4LHC15\_nnlo\_100} PDF, which is based on the work of Manohar et al. \cite{Manohar:2016nzj, Manohar:2017eqh}. It combines 
QCD partons from {\tt PDF4LHC15} \cite{Butterworth:2015oua} with an improved estimation of the photon density in the proton.

The expected signal estimation for the MoEDAL detector is conducted following the procedure introduced in
Sec. \ref{sec:moedal-simulation} with the modifications described in Sec. \ref{sec:paper2-analysis}, i.e. we assume an 
``ideal'' NTD array geometry and use Eqs. \eqref{eq:paper2-moedal-eps} and \eqref{eq:paper2-pntd}, with $\beta_{\rm max} = 0.15 \cdot |Q/e|$, to estimate the expected number of signal events $N_{\rm sig}$.

\subsection{Numerical analysis for the colour-singlet model}\label{sec:paper3-uncoloured}
Among the particles of the uncoloured model, the $SU(2)_L$-triplet scalar, $S_3$, may be long-lived if it is lighter than 
other exotic states: $SU(2)_L$-singlet scalar, $S_1$, and vector-like $SU(2)_L$-triplet fermions $F$ and $\bar F$. We 
assume here that the model parameters are tuned in a way to explain the neutrino masses. In this section, we first 
investigate general collider properties of multiply charged scalars, derive model-independent MoEDAL detection reach, 
and identify what range of parameters of the model can be probed by MoEDAL by the end of the Run 3 data-taking 
period. 

\begin{figure}[!t]
\includegraphics[width=\textwidth]{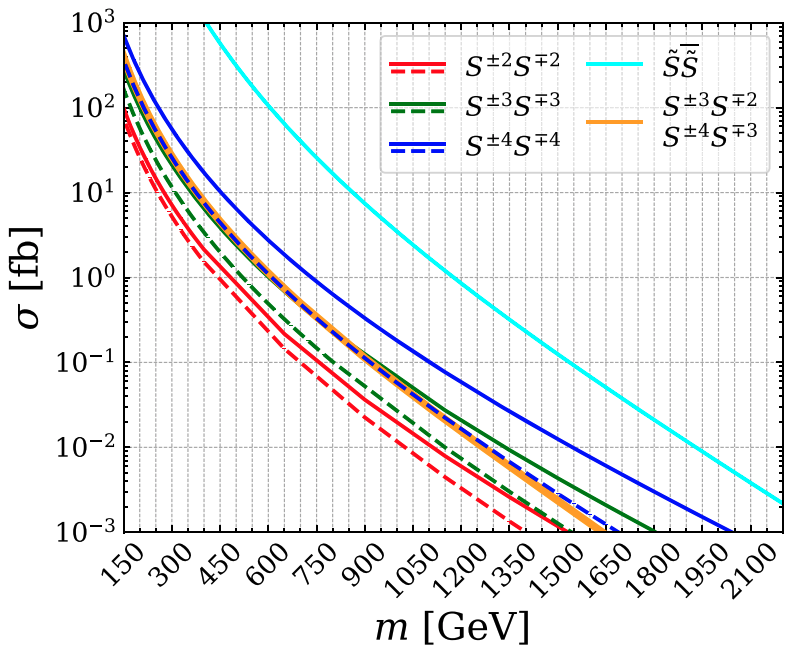}
\caption{
\small The leading order cross section for various modes of particle production in the coloured and uncoloured 
models introduced in Sec. \ref{sec:radiative-mass-gen}. The dashed curves correspond to the cross section 
\textbf{without} the photon fusion process. The orange curve corresponds to the associated production:
 $pp\to S^{\mp 2e} S^{\pm 3e}$ (or $S^{\mp 3e} S^{\pm 4e}$ since the values are the same), mediated by the s-channel W boson. The cyan curve depicts the production cross section for a pair of coloured scalar particles,
i.e. $pp \to \tilde S^{\pm 4/3}\tilde S^{\mp 4/3}$, $pp \to \tilde S^{\pm 7/3}\tilde S^{\mp 7/3}$, and
 $pp \to \tilde S^{\pm 10/3}\tilde S^{\mp 10/3}$.
}
\label{fig:paper3-xs}
\end{figure}

\myparagraph{MoEDAL sensitivity to multiply charged scalars}
The $S_3$ and its conjugate field have three almost mass degenerate eigenstates with charges: $\pm 2e$, $\pm 3e$, 
$\pm 4e$, denoted as: $S^{\pm 2e}$, $S^{\pm 3e}$, $S^{\pm 4e}$, respectively. All these particles might be long-lived and contribute to the signal observed in MoEDAL. There are three production mechanisms: 
(i) Drell-Yan process with s-channel exchange of a gauge boson $\gamma^*/Z^*$
(ii) photon fusion
(iii) associated productions:  $S^{\mp 2e} S^{\pm 3e}$ and $S^{\mp 3e} S^{\pm 4e}$, via W boson exchange.

In Fig. \ref{fig:paper3-xs} we plot the LO pair production cross section for $S^{\pm 2e}$ (red) $S^{\pm 3e}$ (green) and 
$S^{\pm 4e}$ (blue). In order to visualise the impact of photon fusion production, two sets of curves are plotted. 
Solid curves represent the total cross section, which includes both Drell-Yan and photon fusion production, while the 
dashed curves correspond to only the Drell-Yan component. One can see from Fig. \ref{fig:paper3-xs} that photon fusion 
production is especially important for particles with higher electric charges. For example, if the mass of new scalar is 
$m_{S_3}\approx 300~ \rm GeV$, then photon fusion enhances the production cross section by $\sim 30$\% for 
$S_3^{\pm 2}$, but for  $S_3^{\pm 4}$ its contribution is over 50\%. The reason behind this is that the contribution from 
the s-channel photon exchange is proportional to $Q^2$, while in the fusion process it is $Q^4$. Photon fusion also 
affects the velocity distribution of the produced scalar particles. As already discussed several times before, scalars 
produced via s-channel spin-1 gauge boson exchange suffer from the velocity suppression, i.e. $\sigma \to 0$ when $\beta \to 0$. However, there is no such effect for scalars produced via the photon fusion process. This explains why 
photon fusion contribution to the production cross section in Fig. \ref{fig:paper3-xs} becomes more significant for heavier 
particles. When scalars are heavy, the energetic cost of high velocity is larger, so the Drell-Yan production is 
disfavoured and the related cross section is suppressed by the parton distribution function.

\begin{figure}[!b]
\includegraphics[width=\textwidth]{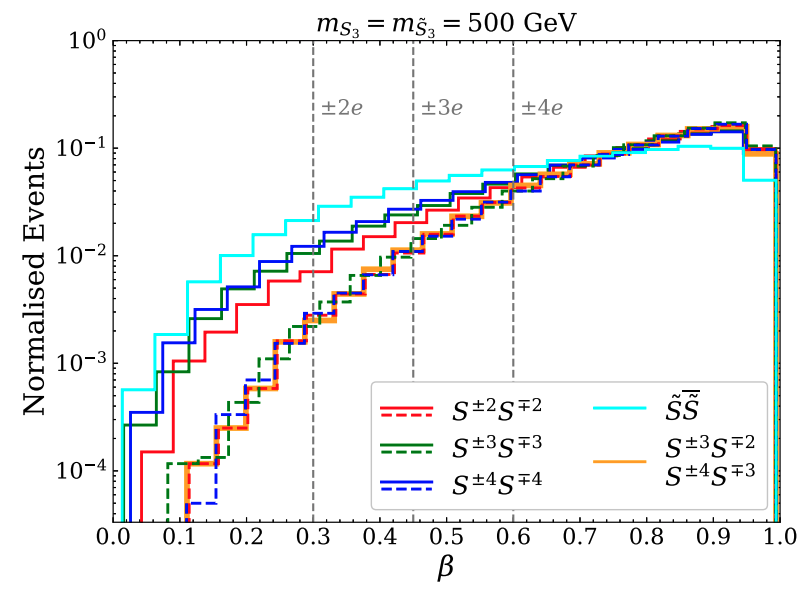}
\caption{
\small Velocity distributions for particles in various production modes. The dashed curves correspond to production
\textbf{without} the photon fusion process included. The orange curve corresponds to the associated production:
 $pp\to S^{\mp 2e} S^{\pm 3e}$ (or $S^{\mp 3e} S^{\pm 4e}$ since the values are the same), mediated by the s-channel W boson. The cyan curve depicts the production cross section for a pair of coloured scalar particles,
i.e. $pp \to \tilde S^{\pm 4/3}\tilde S^{\mp 4/3}$, $pp \to \tilde S^{\pm 7/3}\tilde S^{\mp 7/3}$, and
 $pp \to \tilde S^{\pm 10/3}\tilde S^{\mp 10/3}$. The masses of all particles were 500 GeV. Vertical grey lines depict MoEDAL NTD velocity thresholds for $|Q|\in \{2e, 3e, 4e\}$. Only particles moving below the appropriate velocity threshold can be detected.
}
\label{fig:paper3-beta}
\end{figure}

The associated production $pp\to S^{\mp 2e} S^{\pm 3e}$ (or $S^{\mp 3e} S^{\pm 4e}$ since the leading order values 
are the same) cross section is also plotted in Fig. \ref{fig:paper3-xs}, in orange. Its size is comparable to the cross 
section for the pair production of $S^{\pm3}$ and gives non-negligible effects when studying the MoEDAL's capability 
to test the parameters of the model. Finally, in Fig. \ref{fig:paper3-xs} we superimpose the production cross section for a pair of coloured scalars (cyan) predicted by the coloured version of the model. It will be discussed in the next section.

In Fig. \ref{fig:paper3-beta} we plot velocity distributions for various production modes, using the same colouring scheme as in Fig. 
\ref{fig:paper3-xs} and dashed lines to indicate production \textit{without} photon fusion. The masses of all particles in the plot are set to 500 GeV.
One can immediately notice 
from Fig. \ref{fig:paper3-beta} that including the photon fusion production significantly enhances the fraction of particles with low 
velocities, which is fortuitous for MoEDAL, because NTD detectors are only sensitive to slowly moving particles.
This effect reflects the lack of the p-wave suppression for scalars produced from two initial photons. The high 
sensitivity of the photon fusion to magnitude of the electric charge ($Q^4$) is also visible in Fig. \ref{fig:paper3-beta}, where the fraction of slowly moving particles is larger for 
particles with a higher magnitude of the electric charge.

As already explained, MoEDAL NTDs are sensitive only to slowly moving particles with $\beta < 0.15\cdot Z$, where 
$Z \equiv |Q|/e$. Vertical grey lines in Fig. \ref{fig:paper3-beta}, depict velocity thresholds for $Z=2,3,4$. One can 
see in Fig. \ref{fig:paper3-beta} that there is a twofold enhancement of MoEDAL's acceptance for particles with large 
electric charge, because of the higher velocity threshold, and more favourable velocity distribution with a larger fraction 
of slow particles.

Fig. \ref{fig:paper3-model1-res} depicts the sensitivity of the MoEDAL detector to scalars with $Z=2,3,4$ and triply charged fermions 
$F^{\pm 3} $. The plots show regions of the $m$ vs. $c\tau$ parameter plane, for which the expected number of 
events detected by MoEDAL, $N_{\rm sig}$, exceeds 1 (solid), 2 (dashed) and 3 (dotted), with the integrated luminosity 
of $30~\rm{fb}^{-1}$ (left) and $300~\rm{fb}^{-1}$ (right). These luminosities are predicted to be delivered at the end 
of the Run 3 and HL-LHC phases, respectively. In this part of the study, we consider only particle-antiparticle pair 
production, e.g. $pp \to S^{+2} S^{-2}$, which depends solely on the mass and lifetime of the particle treated here as 
free parameters. Therefore, the results in Fig. \ref{fig:paper3-model1-res} are model-independent.

The top row of Fig. \ref{fig:paper3-model1-res} depicts MoEDAL's sensitivity to $S^{\pm 2}$. Since MoEDAL is effectively free from 
the SM background, $N_{\rm sig}=3$ roughly corresponds to the 95\% CL exclusion\footnote{In the previous two projects we provided results only for $N_{\rm sig}=1,2$, because we estimated that $N_{\rm sig}=2$ should roughly correspond to 95\% CL limit, however, more precise computation revealed that it should be $N_{\rm sig}=3$.} when observing no signal event. The mass 
reach for $N_{\rm sig}=1$ (3) with $c\tau > 100~\rm m$ is $m_{S^{\pm 2}}\approx 290~(190)~\rm GeV$ in Run 3 ($30~\rm{fb}^{-1}$). This can be compared with Tab. \ref{tab:paper2-doubly}, where $m_{S^{\pm 2}}\approx 160~\rm GeV$ for $N_{\rm sig}=1$, 
where the photon fusion is not taken into account. For the HL-LHC ($300~\rm{fb}^{-1}$) the expected mass reach is 
extended to  $m_{S^{\pm 2}}\approx 600~(400)~\rm GeV$ for $N_{\rm sig}=1$ (3).

Second and third rows in Fig.  \ref{fig:paper3-model1-res} show the expected sensitivity to triply ($S^{\pm 3}$) and quadruply 
($S^{\pm 4}$) charged scalars, respectively. One can see that the detection reach is significantly larger than for  $S^{\pm 2}$. 
MoEDAL can expect to probe $S^{\pm 3}$ up to 610 (430) GeV for Run 3, and 1100 (850) GeV for the HL-LHC, with $N_{\rm sig}=1$ (3). In the case of the quadruply charged scalar, lower mass bounds are even higher, and extend to  960 (700) GeV for Run 
3, and 1430 (1200) GeV for HL-LHC, with $N_{\rm sig}=1$ (3) and $c\tau > 100~\rm m$.

The neutrino mass model considered in this project has a triply charged fermion, $F^{\pm 3}$, which is expected to be short-lived. 
Nonetheless, in the bottom row of Fig. \ref{fig:paper3-model1-res} we present the model-independent sensitivity of the MoEDAL 
detector to this kind of particle. In the large lifetime region, $c\tau > 100~\rm m$, Run 3 MoEDAL can probe masses of such fermions 
up to 1030 (800) GeV for $N_{\rm sig}=1$ (3). For HL-LHC the reach is extended to 1550 (1300) GeV for $N_{\rm sig}=1$ (3). 
MoEDAL is more sensitive to triply charged fermion than to scalar field with the same charge. The reason is the following. Dirac fermions have twice as many 
degrees of freedom as complex scalars and their Drell-Yan production does not suffer from the p-wave suppression, hence the 
production cross section, and consequently the signal event rate in MoEDAL, is enhanced.

One can see in Fig. \ref{fig:paper3-model1-res} that the sensitivity of MoEDAL does not change for $c\tau > 100~\rm m$. It is a 
general feature of the detector set-up, and applies to all kinds of particles. The signal in MoEDAL comes from energy 
deposition in all layers of an NTD panel. In order to produce this signature, a particle has to live long enough to reach and traverse 
the detector. The probability for this to happen approaches asymptotically one for very long lifetimes. It corresponds to the situation in 
which almost all particles live long enough to reach the NTD array. Increasing the lifetime even further has little benefit 
for the MoEDAL's sensitivity, which is now determined by the cross section and velocity distribution, as can be seen from Eqs. \eqref{eq:moedal-eps} and \eqref{eq:moedal-n}.

\begin{figure}[!p]
\fancypagestyle{plain}
\centering
  \vspace{-0.5em}
  \begin{subfigure}[t]{0.44\textwidth}
    \includegraphics[width=\textwidth]{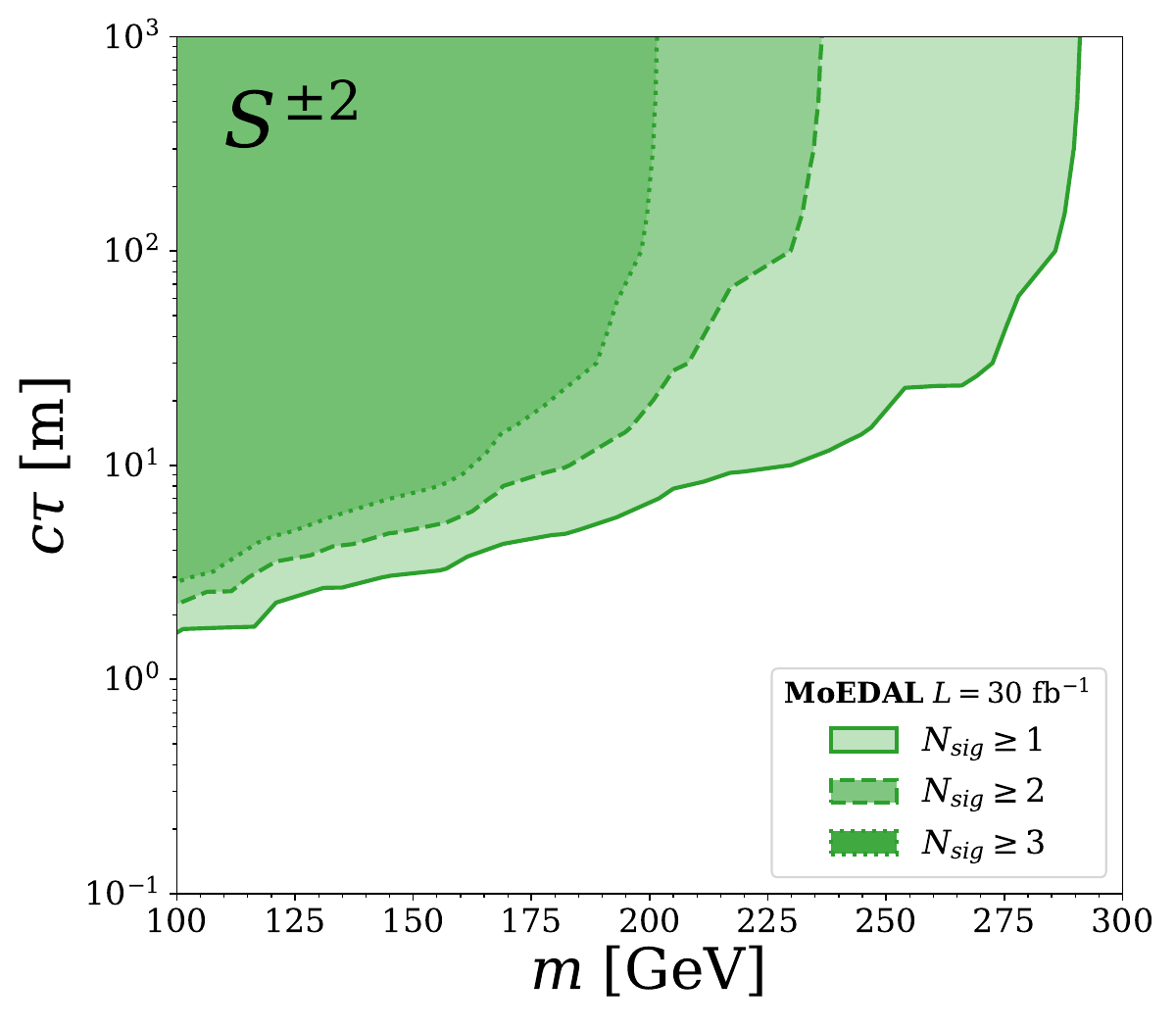}
  \end{subfigure}
    \hfill
   \begin{subfigure}[t]{0.44\textwidth}
    \includegraphics[width=\textwidth]{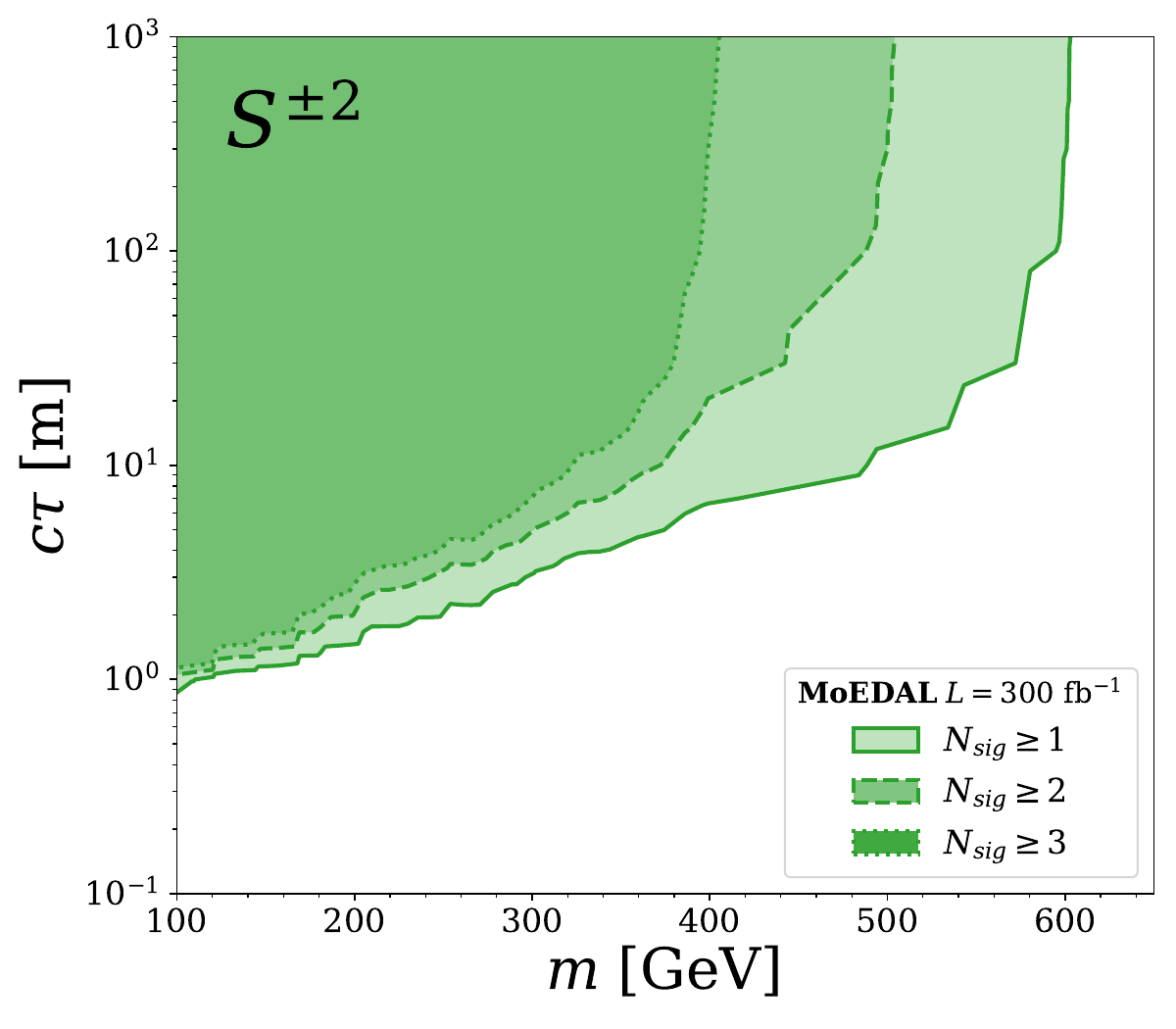}
  \end{subfigure}
    \hfill
   \begin{subfigure}[t]{0.44\textwidth}
    \includegraphics[width=\textwidth]{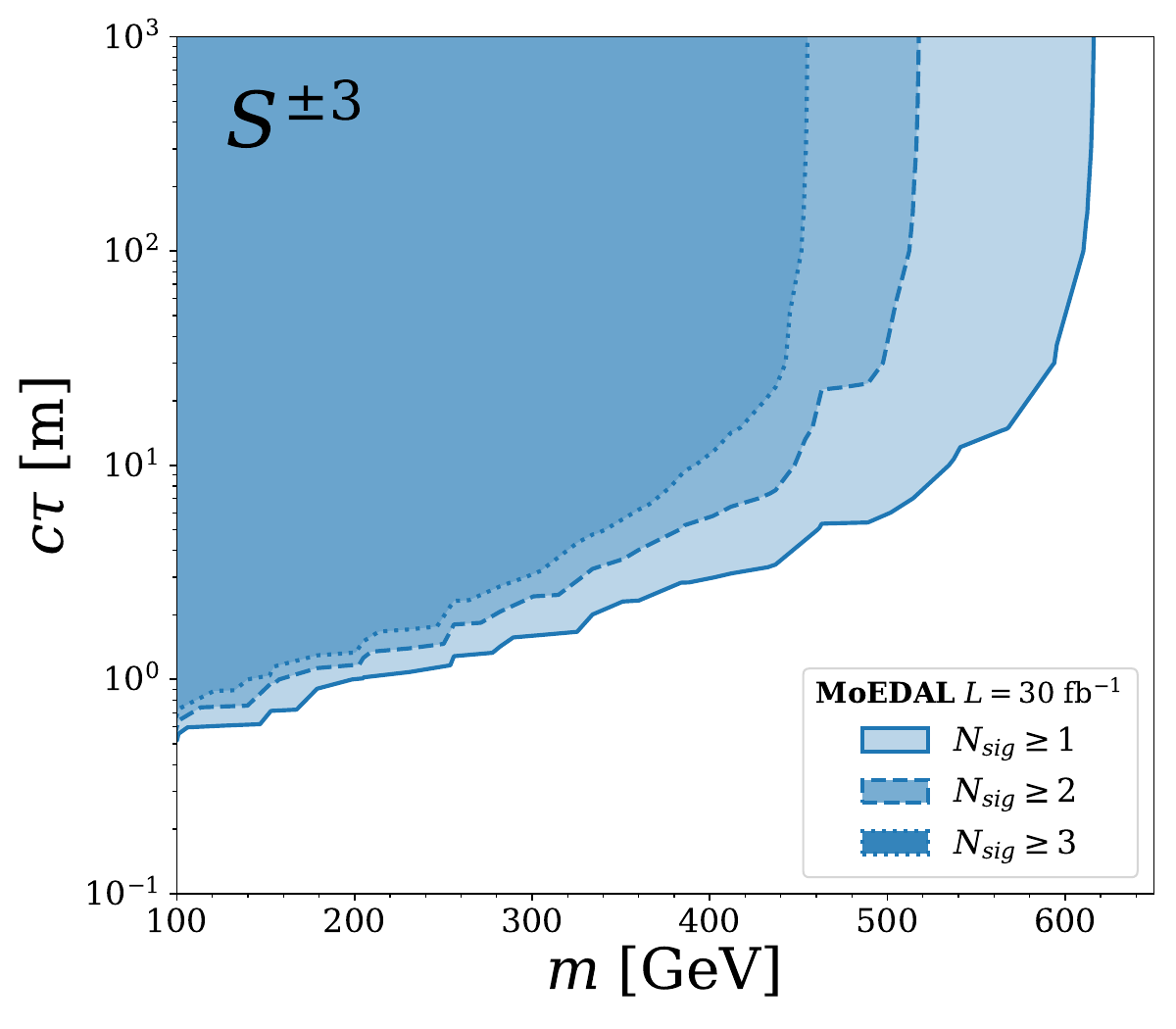}
  \end{subfigure}
  \hfill
  \begin{subfigure}[t]{0.44\textwidth}
    \includegraphics[width=\textwidth]{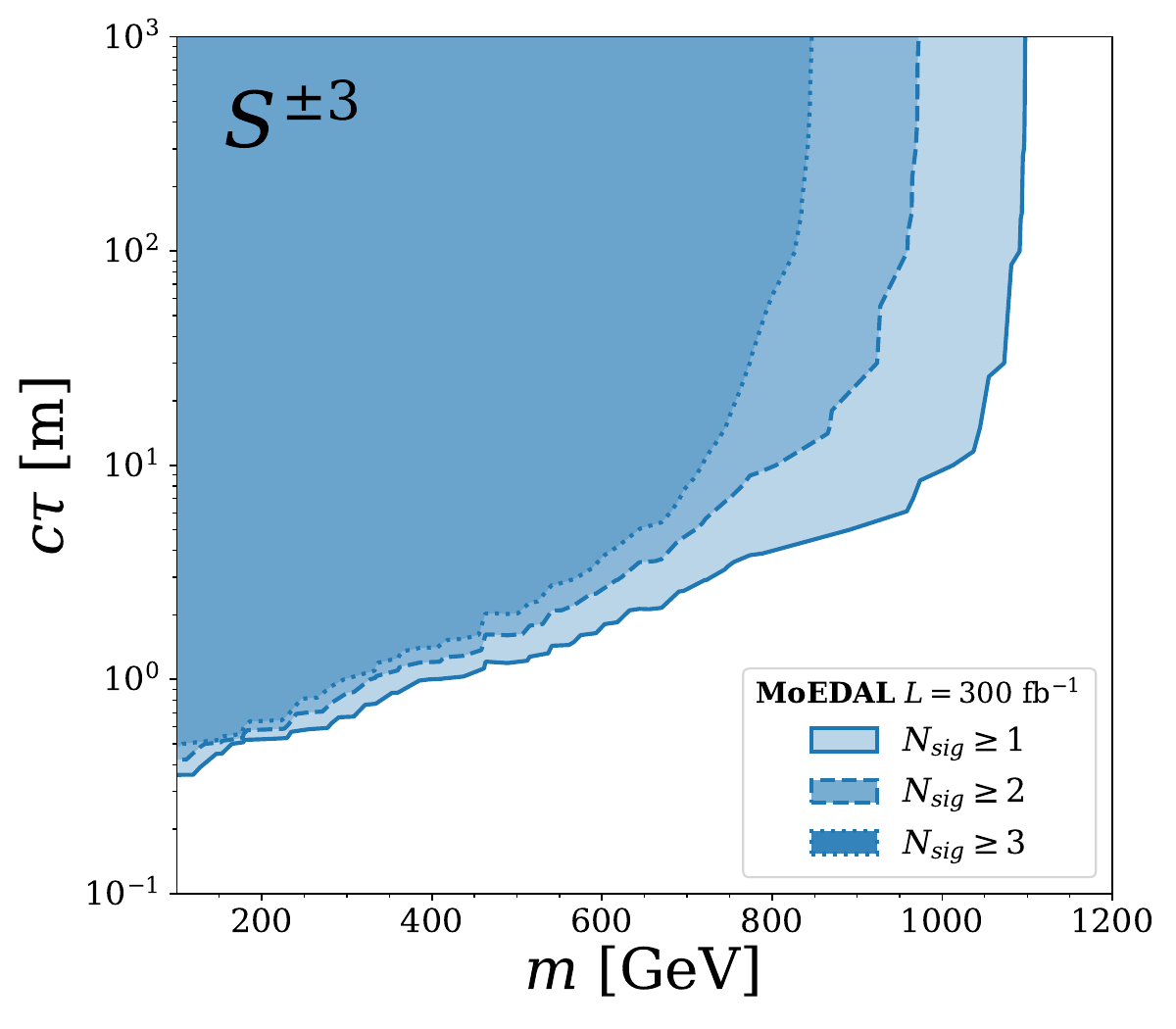}
  \end{subfigure}
    \hfill
   \begin{subfigure}[t]{0.44\textwidth}
    \includegraphics[width=\textwidth]{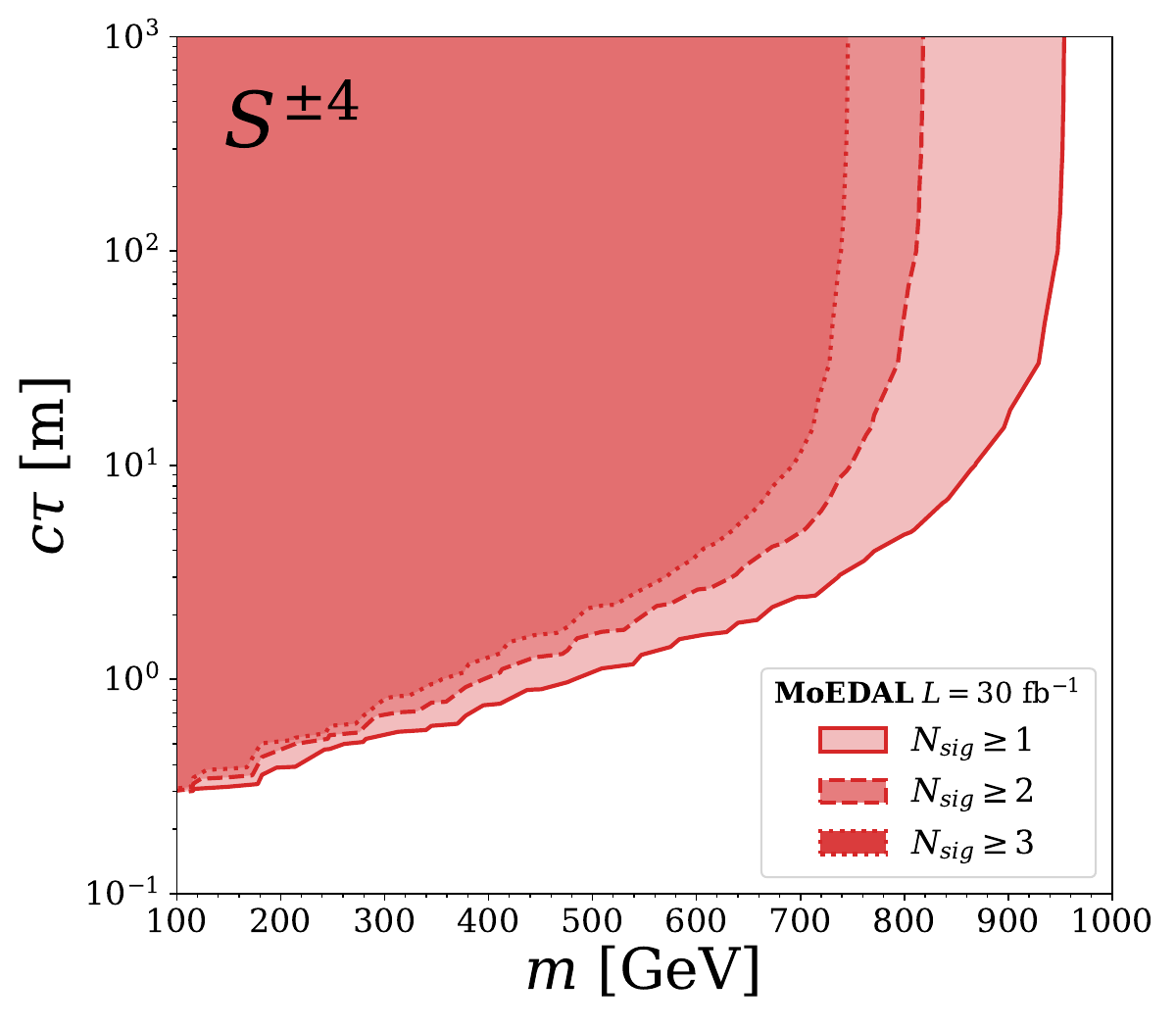}
  \end{subfigure}
  \hfill
     \begin{subfigure}[t]{0.44\textwidth}
    \includegraphics[width=\textwidth]{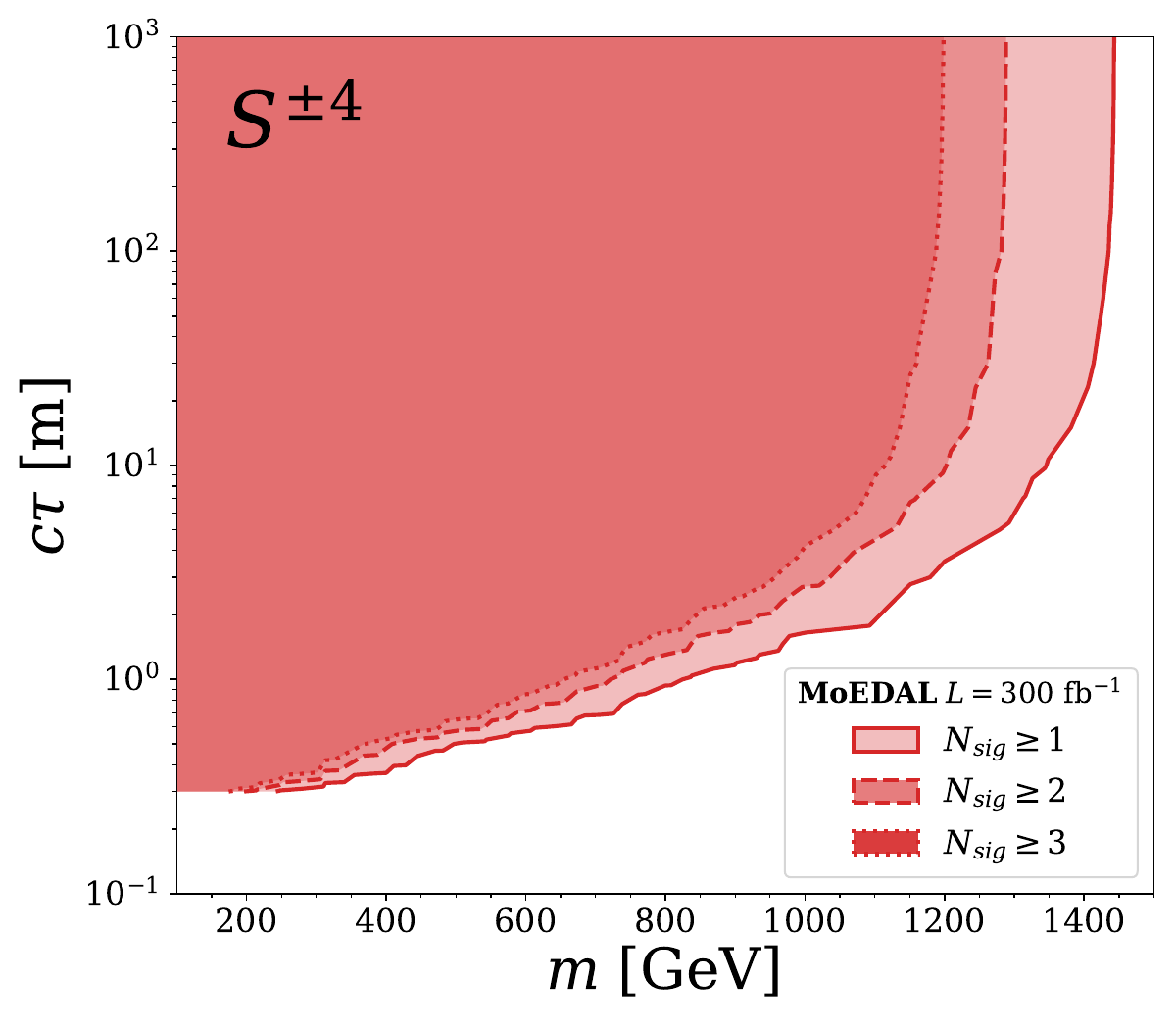}
  \end{subfigure}
    \hfill
       \begin{subfigure}[t]{0.44\textwidth}
    \includegraphics[width=\textwidth]{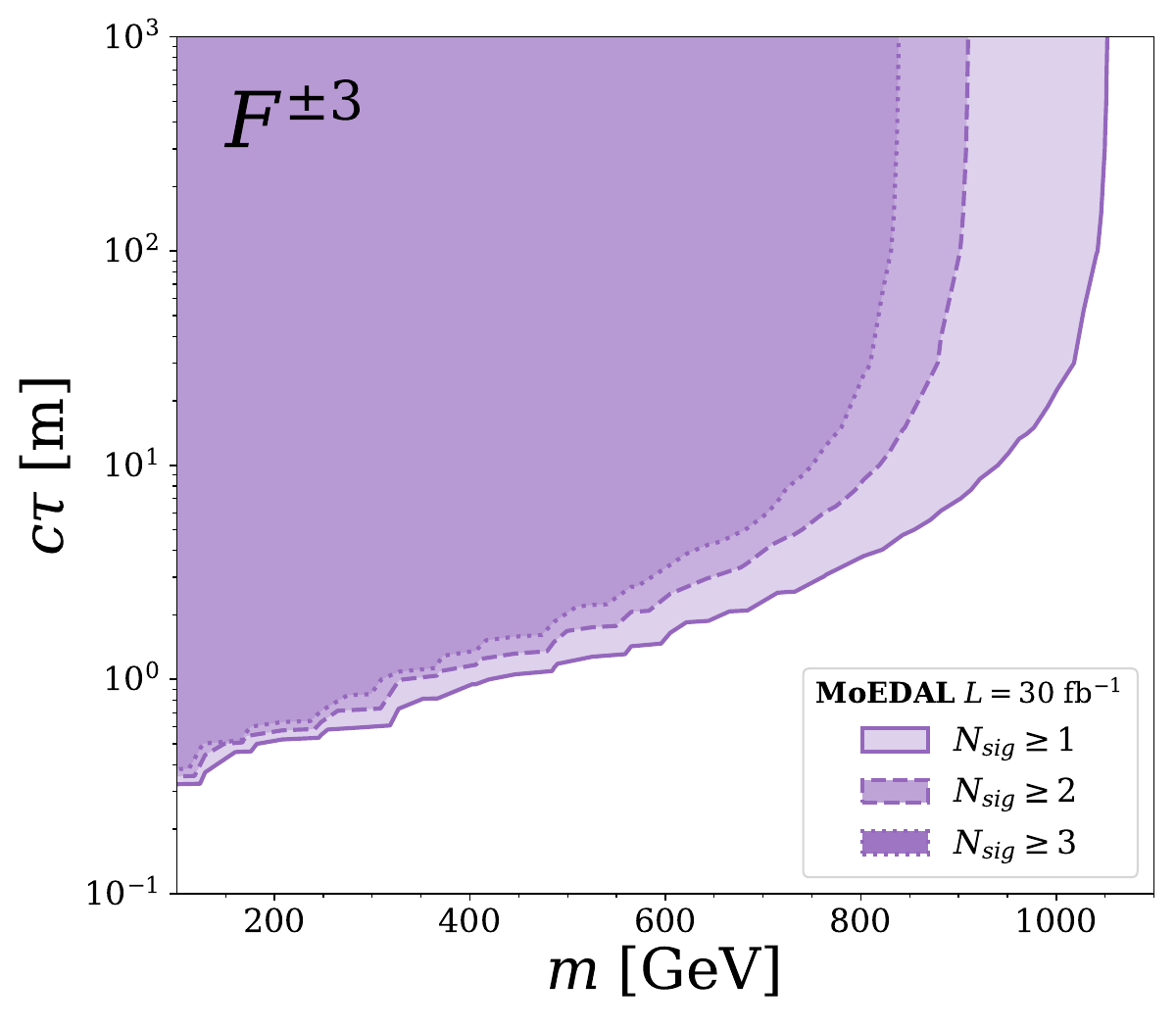}
  \end{subfigure}
  \hfill
     \begin{subfigure}[t]{0.44\textwidth}
    \includegraphics[width=\textwidth]{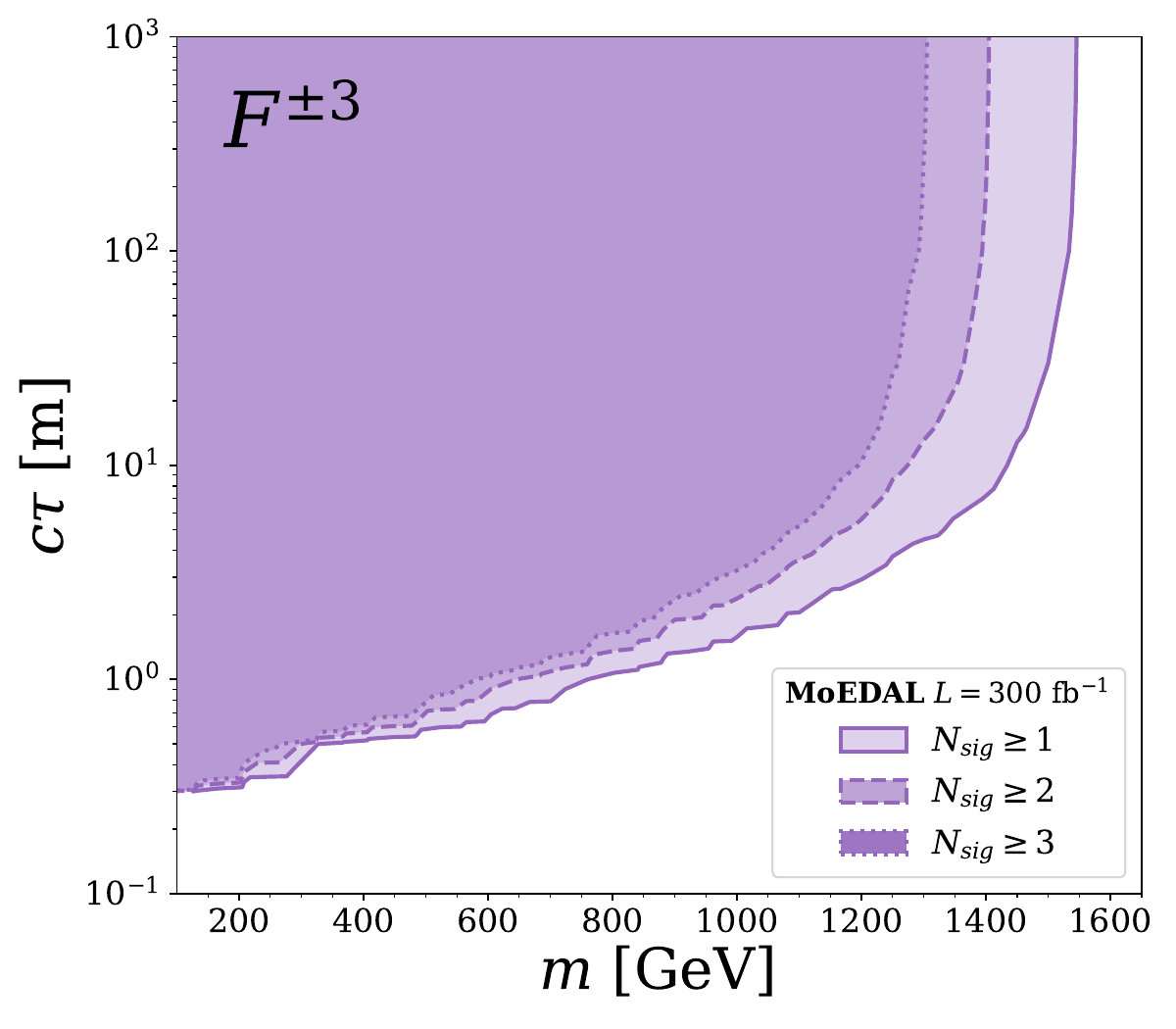}
  \end{subfigure}
  \vspace{-0.5em}
    \caption{\small The model-independent detection reach of Run 3 (left, $L=30~{\rm fb}^{-1}$) and HL-LHC (right, $L=300~{\rm fb}^{-1}$) MoEDAL detector for colour-singlet multiply charged particles. Solid, dashed and dotted contours 
    correspond to $N_{\rm sig}=1$, $2$, and $3$, respectively.}
    \label{fig:paper3-model1-res}
\end{figure}

In Tab. \ref{tab:paper3-uncoloured} MoEDAL mass reach for multiply charged colour-singlet particles is summarised and compared to the latest mass 
bounds from ATLAS and CMS (if available). The current experimental limits are taken (estimated) from the ATLAS analysis \cite{ATLAS:2018imb} 
relying on $L=36~\rm{fb}^{-1}$ data. In this analysis, the collaboration interpreted its result only for spin-1/2 particles, therefore 
we estimated the lower mass bounds of scalar particles by naively imposing the cross section bounds from \cite{ATLAS:2018imb}, assuming that 
the ATLAS efficiencies are not very sensitive to spins. Numbers obtained in such a way are written in double parentheses in Tab. \ref{tab:paper3-uncoloured} and have to be treated with caution. To our knowledge, the Run 3 projection is available only for fermions \cite{Jager:2018ecz}, and 
estimated to be $\sim$1500 GeV for $F^{\pm 3}$. The numbers outside (inside) the brackets in the last two columns correspond to 
MoEDAL's mass reach with $N_{\rm sig} \geq$3 (1), for Run 3 and HL-LHC.

As evident from the comparison in Tab. \ref{tab:paper3-uncoloured}, there is not much hope to observe particles predicted by the studied model in the 
Run 3 MoEDAL, since the majority of accessible mass ranges is already excluded by ATLAS, except for $S^{\pm 4}$, 
for which there is a small $\sim 40$ GeV window. The situation is more optimistic for the HL-LHC, for which MoEDAL can probe 
larger masses, which are currently unconstrained.

\begin{table}[t!]
\centering
\caption{\small Comparison of the most recent \cite{ATLAS:2018imb, Jager:2018ecz} lower mass bounds on multiply charged particles from ATLAS for 
36 fb$^{-1}$ (column 1) and  300 fb$^{-1}$ (column 2), with the detection reach by MoEDAL for 30 fb$^{-1}$ (column 3) and 300 fb$^{-1}$ (column 4). Values in double parentheses were obtained by recasting the ATLAS search \cite{ATLAS:2018imb}. MoEDAL reach outside (inside) the parentheses corresponds to $N_{\rm sig}= 3$ (1). All masses are in GeV units.}
\begin{tabular}{c|c|c|c|c}
           & current HSCP bound & HSCP (Run-3) & MoEDAL (Run-3) & MoEDAL (HL-LHC) \\
           & 36 fb$^{-1}$ \cite{ATLAS:2018imb} & 300 fb$^{-1}$ \cite{Jager:2018ecz} & 30 fb$^{-1}$ & 300 fb$^{-1}$ \\        
\hline
$S^{\pm2}$ & ((650))  & -- & 190 (290) & 400 (600)   \\
\hline
$S^{\pm3}$ & ((780))  & -- & 430 (610) & 850 (1100)   \\
\hline
$S^{\pm4}$ & ((920))  & -- & 700 (960) & 1200 (1430)   \\
\hline
$F^{\pm3}$ & 1130   & 1500  & 800 (1030) & 1300 (1550)   \\
\end{tabular}
\label{tab:paper3-uncoloured}
\end{table}

\myparagraph{Interpretation of the results for the uncoloured version of the model}

So far we have treated multiply charged particles in a model-independent way, parametrising their mass and decay length.
Now we will estimate the capability of the MoEDAL experiment to test the parameters of the neutrino mass generation model described in Sec. \ref{sec:radiative-mass-gen}.
The Lagrangian of the studied
model is given by Eq. \eqref{eq:neutrino-lagrangian}.
There are many parameters, but phenomenologically interesting are $(h_{ee})_{ij}$, $\lambda_5$ and the 
product $h_F h_{\bar F}$. For simplicity, we will assume from now on that only the 1-1 component of $(h_{ee})_{ij}$ is non-zero, 
but extensions to other cases are simple and will be discussed later. 
The L-violating decays:
$S^{+4} \to W^+ W^+ l^+ l^+$, $S^{+3} \to W^+ l^+ l^+$, and $S^{+2} \to l^+ l^+$, are controlled by 
$h_{ee}=(h_{ee})_{11}$ and $\lambda_5$ couplings, as evident from Eqs. \eqref{eq:neutrino-p4to2l2wdecay}, \eqref{eq:neutrino-p3to2l1wdecay} and \eqref{eq:neutrino-p2to2ldecay}.
The decay rates of the L-conserving processes:
$S^{+4} \to 4l^+$, $S^{+3} \to \nu l^+l^+l^+$ and $S^{+2} \to \nu \nu l^+ l^+$,
depend on $h_{ee}=(h_{ee})_{11}$ and  $h_F h_{\bar F}$, as can be seen from Eqs. \eqref{eq:neutrino-p4to4ldecay}, \eqref{eq:neutrino-p3to3ldecay}, and \eqref{eq:neutrino-p2to2l2nudecay}.
L-violating processes depend on the masses of scalar singlet, $m_{S_1}$, and triplet, $m_{S_3}$, while L-conserving decays depend also on $m_{F_i}~(i=1,2,3)$. 
Particles in the model can be long-lived if $m_{S_3} < m_{S_1}, m_{F_i}$, hence we assume this hierarchy and set $m_{F_i} = 3~\rm TeV$ for simplicity.
 
The values of dimensionless couplings and masses are important for the determination of the neutrino masses. The approximate 
 formula is provided in Eq. \eqref{eq:neutrino-mass}, however, in our analysis we do not use it. Instead, we do a parameter fit for the neutrino data 
 and choose the value of  $h_F h_{\bar F}$ for a given $\lambda_5$ and masses. In Fig. \ref{fig:paper3-model1-lifetime} we show decay lengths of $S^{\pm 2}$ (red dotted), $S^{\pm 3}$ (green dashed), and $S^{\pm 4}$ (blue solid) as a function of $\lambda_5$, with $h_F h_{\bar F}$ from the fit. Other 
 parameters are set to $h_{ee}=0.1$, $m_{S_3} = 750~\rm GeV$, and $m_{S_1} = 3$ (10) TeV for the left (right) plot in Fig. \ref{fig:paper3-model1-lifetime}.
Lifetimes for other values of $(h_{ee})_{ij}$ can be obtained by simple rescaling, since they are proportional to $(\sum_{ij}|(h_{ee})_{ij}|^2)^{-1}$.

\begin{figure}[!tbp]
\centering
  \begin{subfigure}[t]{0.49\textwidth}
    \includegraphics[width=\textwidth]{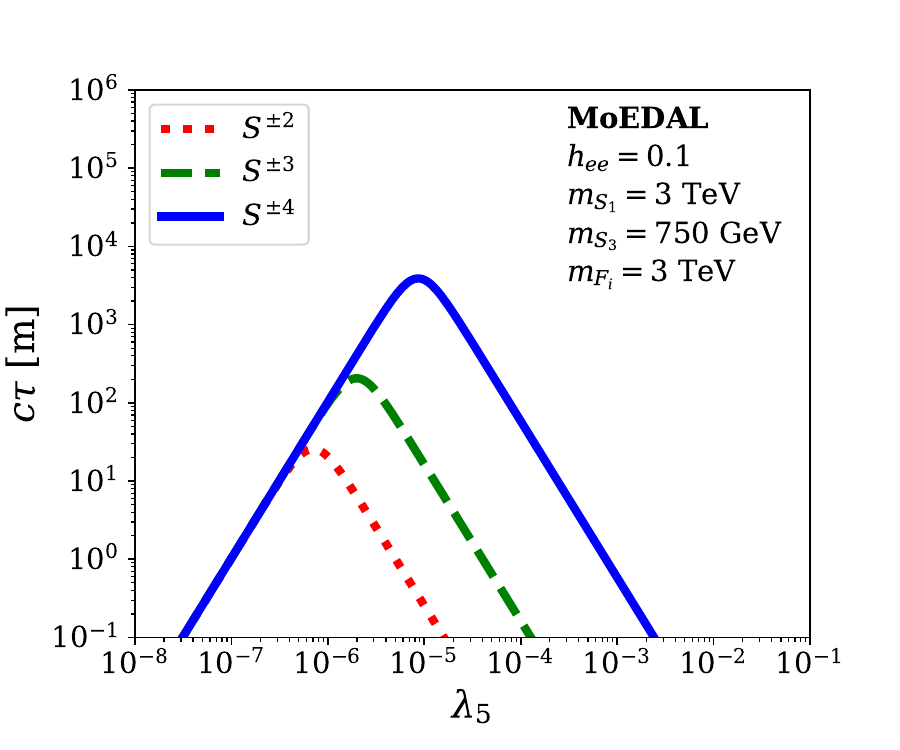}
  \end{subfigure}
    \hfill
   \begin{subfigure}[t]{0.49\textwidth}
    \includegraphics[width=\textwidth]{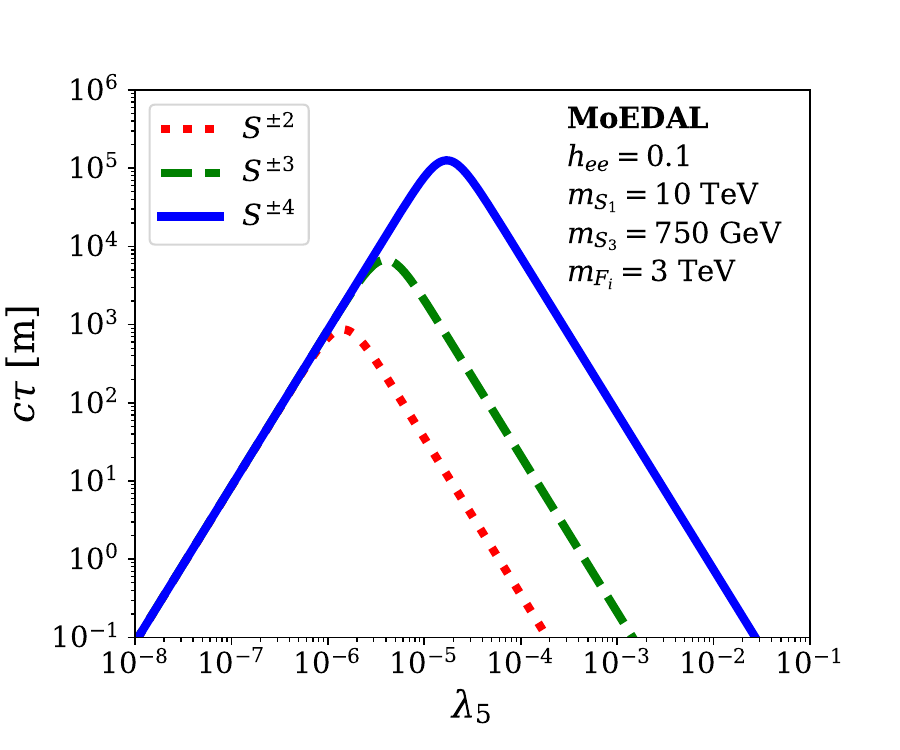}
  \end{subfigure}
    \caption{\small Decay lengths of $S^{\pm 4}$ (blue solid), $S^{\pm 3}$ (green dashed) and $S^{\pm 2}$ (red dotted) as a function of $\lambda_5$. Other 
 parameters are set to $h_{ee}=0.1$, $m_{S_3} = 750~\rm GeV$, and $m_{S_1} = 3$ (10) TeV for the left (right) plot.
 For the given set of mass and $\lambda_5$ parameters,  $h_F h_{\bar F}$  is obtained from fitting the neutrino data.
 }
    \label{fig:paper3-model1-lifetime}
\end{figure}

One can see from the plots in Fig. \ref{fig:paper3-model1-lifetime} that for $\lambda_5 \gtrsim 10^{-4}~\rm$, lifetimes of the particles decrease with the growth of $\lambda_5$. This is because in this region the L-violating decay processes, proportional to $\lambda_5$, are dominant. On the other hand, when $\lambda_5 \lesssim 10^{-6} ~\rm $, decrease of $\lambda_5$ implies 
shorter lifetimes. The reason is that in this region the L-conserving decays are dominant, and their decay rates depend on $h_F$ 
and $h_{\bar F}$, the product of which is inversely related to $\lambda_5$ through the neutrino mass constraints (see Eq. 
\eqref{eq:neutrino-mass}). In the region where L-conserving decays dominate, lifetimes of $S^{\pm 2}$, $S^{\pm 3}$, and 
$S^{\pm 4}$ coincide. This is expected, since these decays are related by the $SU(2)_L$ symmetry, and the decay rates formulas 
are identical, when small effects of the electroweak symmetry breaking are neglected, as discussed in Sec. \ref{sec:paper3-model}. Lifetimes of the considered particles peak for the intermediate values of $\lambda_5$, at which L-conserving and L-violating decay processes are comparabale. It happens at $\lambda_5 \sim 10^{-5}$, $3\times 10^{-6}$, and  $10^{-6}$ for 
$S^{\pm 4}, S^{\pm 3}, S^{\pm 2}$, respectively.

In Fig. \ref{fig:paper3-model1-reach} we present the sensitivity of the MoEDAL NTD detector for the considered radiative neutrino mass models, in the 
$m_{S_3}$ vs. $\lambda_5$ parameter plane. The plots on the left are for Run 3 MoEDAL with $L=30~\rm{fb}^{-1}$, while the 
plots on the right are for HL-LHC with $L=300~\rm{fb}^{-1}$. Plots in the upper row of Fig. \ref{fig:paper3-model1-reach} assume $m_{S_1}=3~\rm{TeV}$, 
while in the lower row $m_{S_1}=10~\rm{TeV}$. Other parameters, common to all plots, are $m_{F_i}=3~\rm{TeV}$ and $h_{ee}=0.1$. In Fig. \ref{fig:paper3-model1-reach} all production modes are considered, including the associated production: $pp\to S^{\mp 2e} S^{\pm 3e}$ 
and $S^{\mp 3e} S^{\pm 4e}$. The region shaded in grey, around $\lambda_5 \sim 10^{-9}$, is non-perturbative, because in that 
region $\mathrm{max}_{ij}\left(|(h_F)_{ij}|, |(h_{\bar F})_{ij}|\right) \geq 2$.

\begin{figure}[!bth]
\centering
  \begin{subfigure}[t]{0.49\textwidth}
    \includegraphics[width=\textwidth]{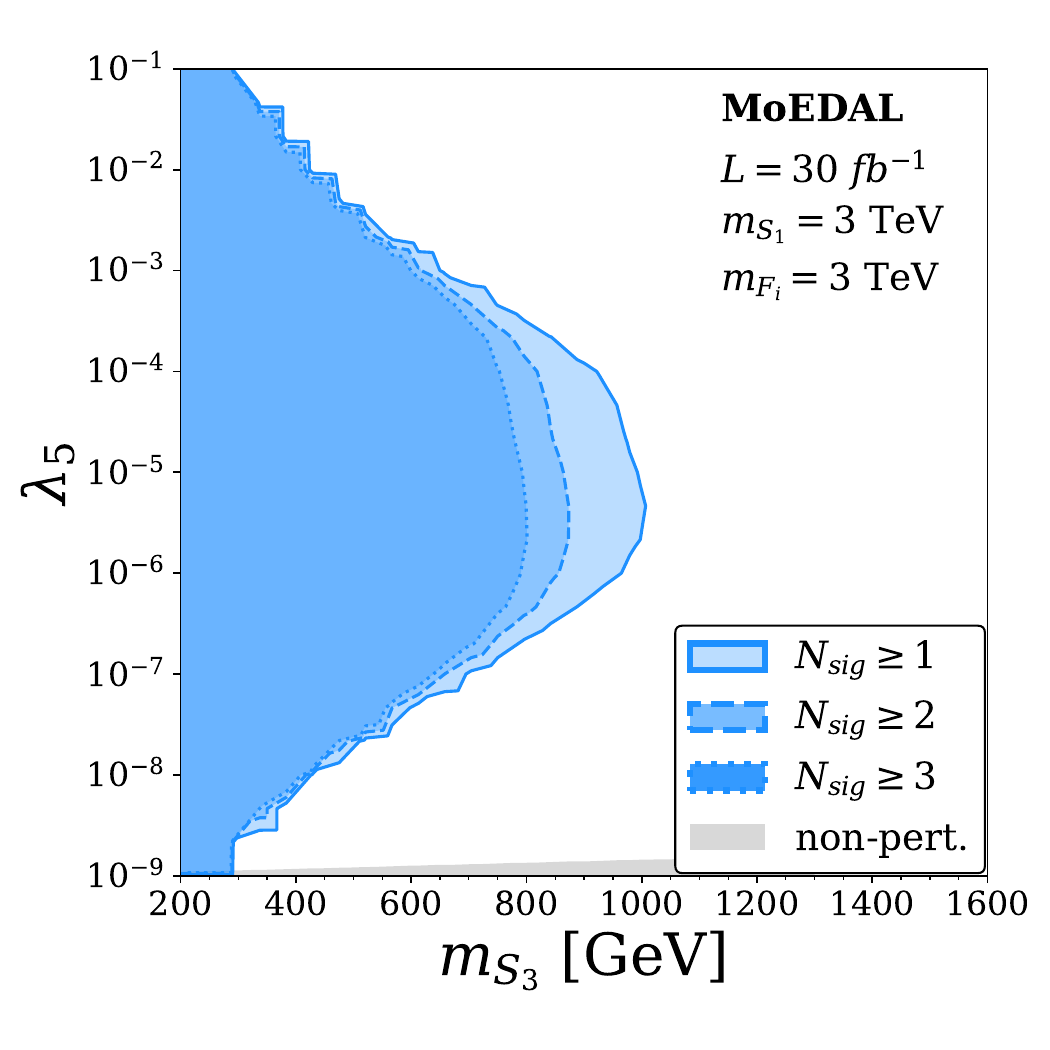}
  \end{subfigure}
    \hfill
   \begin{subfigure}[t]{0.49\textwidth}
    \includegraphics[width=\textwidth]{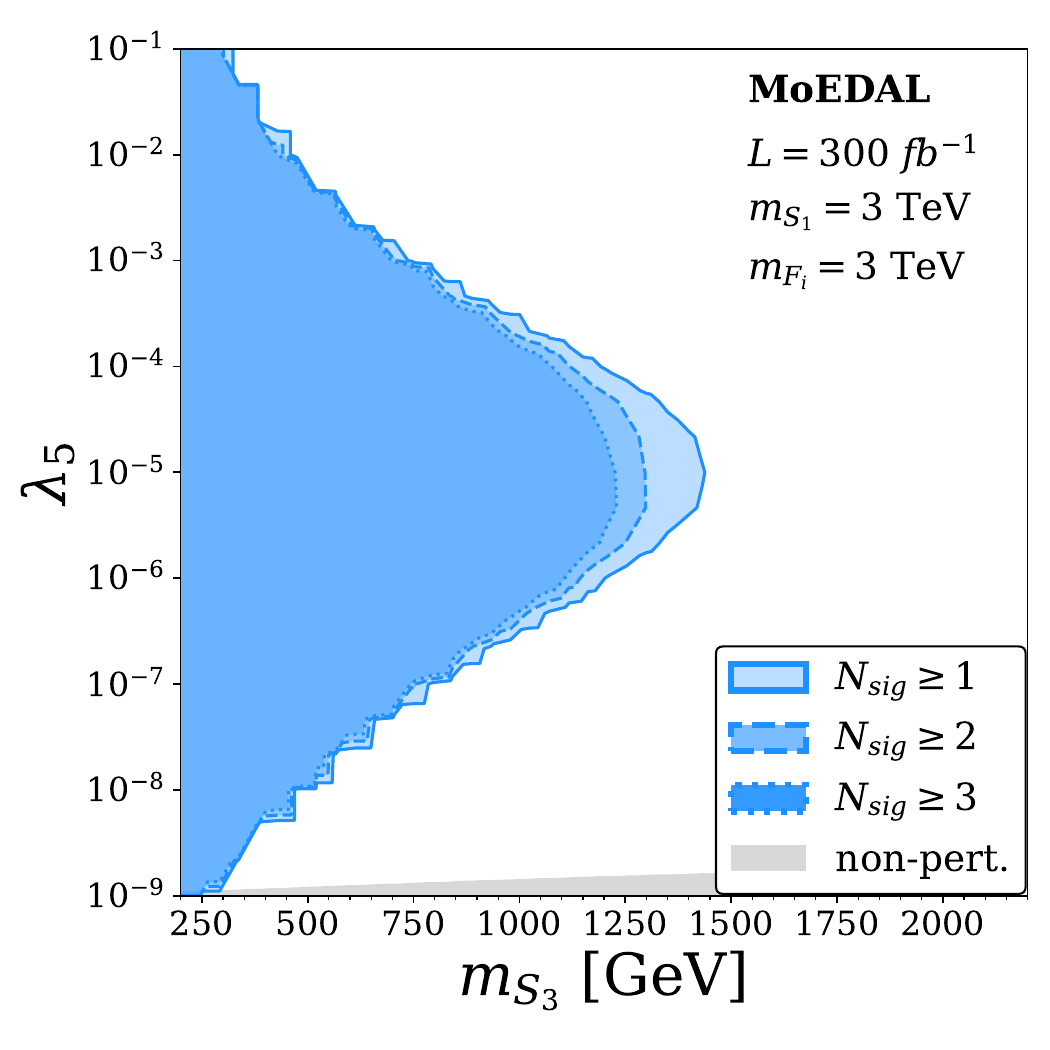}
  \end{subfigure}
  \begin{subfigure}[t]{0.49\textwidth}
    \includegraphics[width=\textwidth]{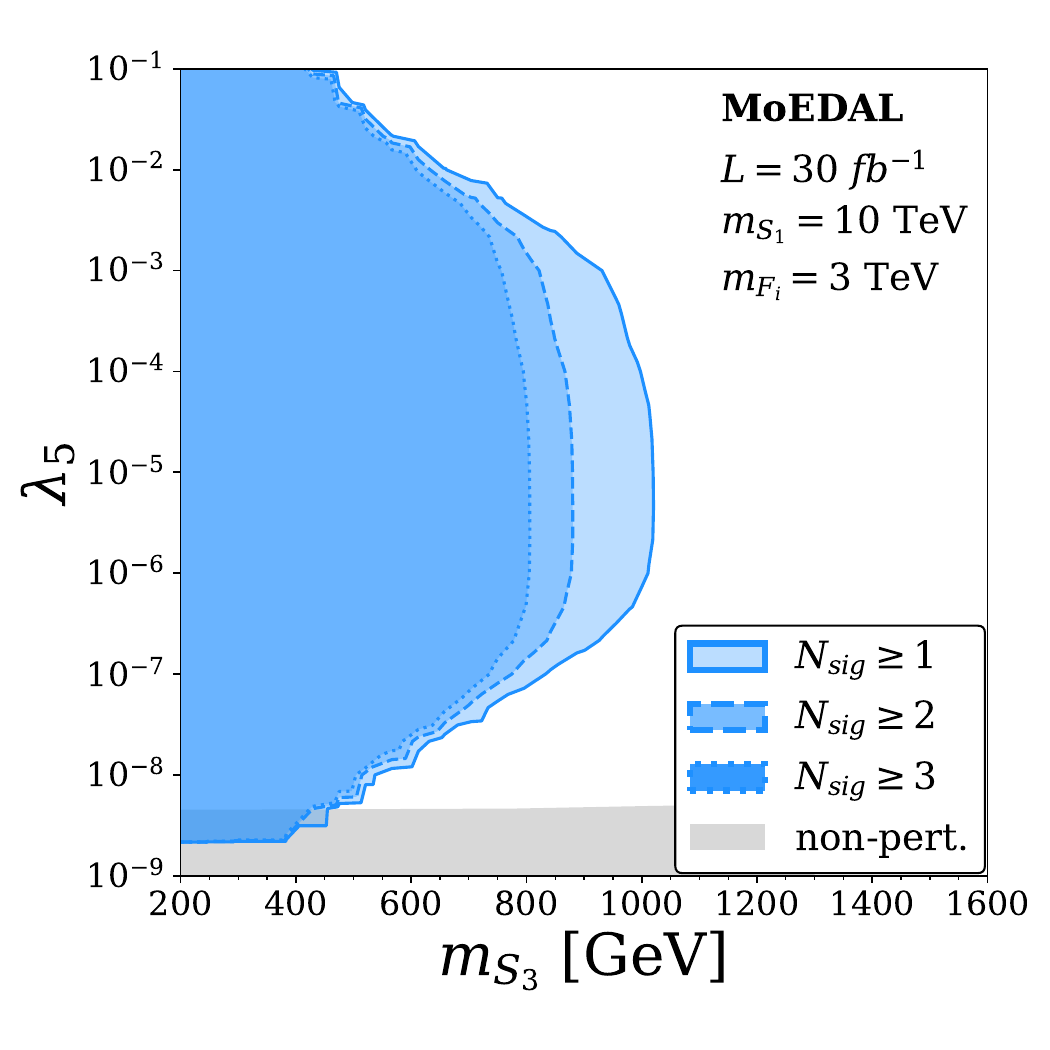}
  \end{subfigure}
    \hfill
   \begin{subfigure}[t]{0.49\textwidth}
    \includegraphics[width=\textwidth]{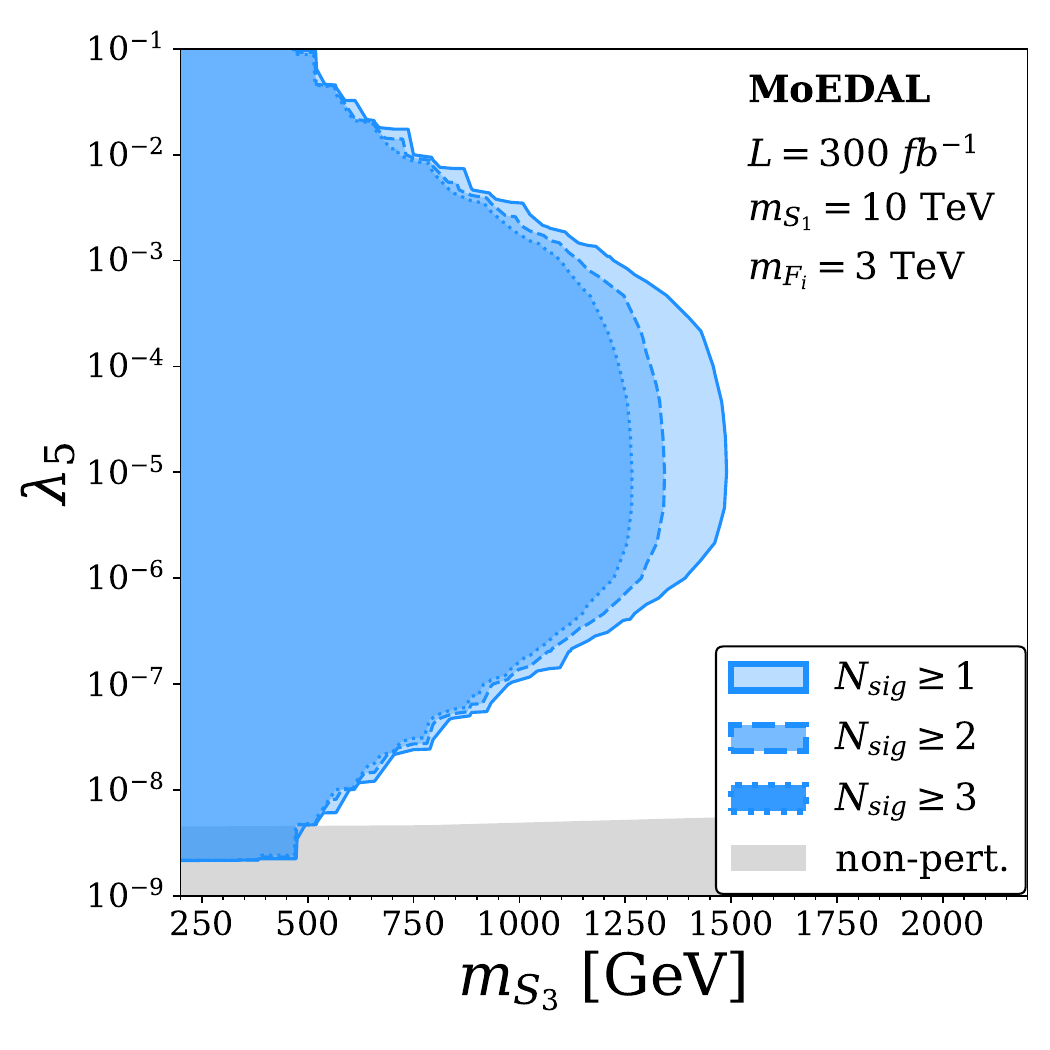}
  \end{subfigure}
    \caption{\small MoEDAL's sensitivity to the uncoloured version of the studied neutrino mass model on the $(m_{S_3},\lambda_5)$ parameter plane. Regions enclosed by solid, dashed and dotted contour lines correspond to $N_{\rm sig}\geq$1, 2, and 3 signal event observations, respectively. The region with $N_{\rm sig}\geq 3$ will be excluded at 95\% CL if MoEDAL observes no events.
    The left and right panels correspond to Run 3 ($L=30~\rm{fb}^{-1}$) and HL-LHC ($L=300~\rm{fb}^{-1}$). 
    The upper (lower) two panels assume  $m_{S_1}=3$ (10) $\rm{TeV}$. The $h_F$ and $h_{\bar F}$ are fitted to neutrino data.
    The region shaded in grey, around $\lambda_5 \sim 10^{-9}$, is non-perturbative, because in that 
region $\mathrm{max}_{ij}\left(|(h_F)_{ij}|, |(h_{\bar F})_{ij}|\right) \geq 2$.
    Other parameters, common to all plots, are $m_{F_i}=3~\rm{TeV}$ and $h_{ee}=0.1$.
 }
    \label{fig:paper3-model1-reach}
\end{figure}

In the upper left plot in Fig. \ref{fig:paper3-model1-reach}, corresponding to $m_{S_1}=3$ TeV and $L=30~\rm{fb}^{-1}$, we 
can see that MoEDAL can probe $m_{S_3}$ up to 1000 (800) GeV with $N_{\rm sig}=1$ (3) for $\lambda_5 \sim 10^{-5}$, at 
which the lifetime of quadruply charged scalar is maximised. If one considers $\lambda_5 \sim 10^{-3}$ or  $\lambda_5 \sim 10^{-7}$, then the mass reach is lowered to $\sim 600~\rm{GeV}$ for both $N_{\rm sig}=1$ and $N_{\rm sig}=3$. In 
the HL-LHC case, the sensitivity is improved to 1450 (1250) GeV with $N_{\rm sig}=1$ (3) for $\lambda_5 \sim 10^{-5}$. 
For other values of the $\lambda_5$ coupling, the sensitivity degrades, e.g. for $\lambda_5\sim 10^{-3}$ the maximum 
reach is $\sim 750$ GeV, similarly for $\lambda_5\sim 10^{-7}$.
The bottom plots in Fig. \ref{fig:paper3-model1-reach}, which correspond to $m_{S_3}=10~\rm{TeV}$, tend to be less sharply 
peaked around $\lambda_5\sim 10^{-5}$, which signalises smaller sensitivity to variation of the $\lambda_5$ coupling. 
The reason behind this is that decay rates get extra suppression due to larger $m_{S_1}$, and $c\tau$ comes down to the $\sim 10~\rm{m}$ range only for $\lambda_5 \lesssim 10^{-7}$ or $\lambda_5 \gtrsim 10^{-3}$, as can be seen in the 
right panel of Fig. \ref{fig:paper3-model1-lifetime}. For $\lambda_5 \in (10^{-4}, 10^{-6})$ and $m_{S_1}=10~\rm{TeV}$, Run 3 MoEDAL can probe the model 
up to $m_{S_3}\sim 1000$ (800) GeV for $N_{\rm sig}=1$ (3). In the HL-LHC case, the mass reach is extended to 1500 (1250) 
GeV for $N_{\rm sig}=1$ (3). In all plots in Fig. \ref{fig:paper3-model1-reach}, the maximum reach for $m_{S_3}$ roughly 
corresponds to the maximum lifetime of $S^{\pm 4}$.

So far we have assumed $h_{ee} = 0.1$. In Fig. \ref{fig:paper3-model1-hee} we 
show the impact of $h_{ee}$ on the MoEDAL's sensitivity. Distinct colours in the 
plots correspond to $N_{\rm sig}=1$ contours for different values of the $h_{ee}$ 
coupling. We set $m_{F_i} = 3~\rm{TeV}$ and $m_{S_1}=3$ (10) TeV 
for the left (right) plot in Fig. \ref{fig:paper3-model1-hee}. Both L-conserving 
and L-violating decays are proportional to $h_{ee}$ at the amplitude level, and 
the lifetimes of multiply charged particles in the $S_3$ triplet are inversely 
proportional to $|h_{ee}|^2$. 
In the left panel of Fig. \ref{fig:paper3-model1-hee}, for which $m_{S_1}=3~\rm{TeV}$, one can observe that for smaller values of $h_{ee}$, the maximum mass reach is less dependent on the $\lambda_5$ coupling. It is because a smaller value of $h_{ee}$ results in decay suppression.
For the right plot in Fig. \ref{fig:paper3-model1-hee}, corresponding to 
$m_{S_1}=10~\rm{TeV}$, the effect is even stronger, because the decay modes 
receive an additional suppression due to very large $m_{S_1}$.

\begin{figure}[!tbp]
\centering
  \begin{subfigure}[t]{0.49\textwidth}
    \includegraphics[width=\textwidth]{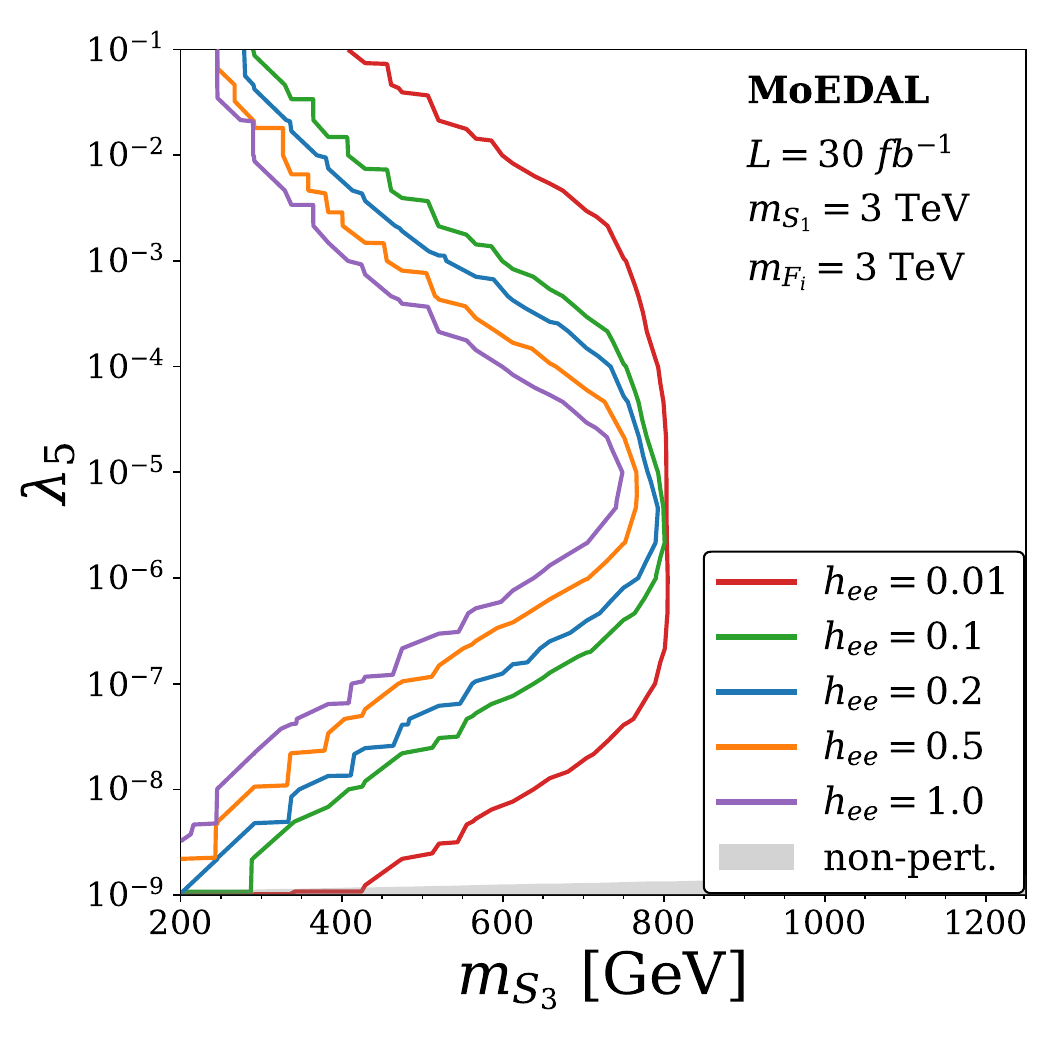}
  \end{subfigure}
    \hfill
   \begin{subfigure}[t]{0.49\textwidth}
    \includegraphics[width=\textwidth]{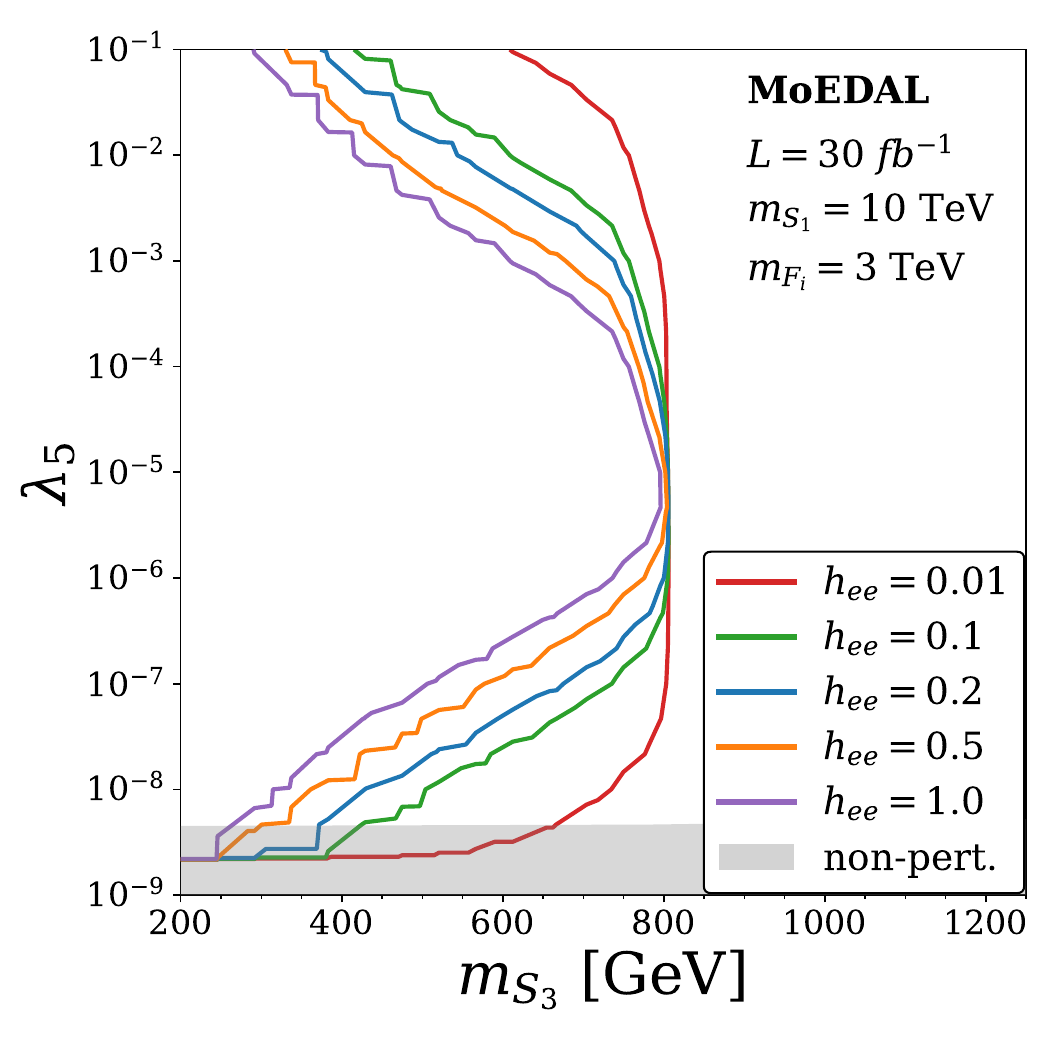}
  \end{subfigure}
    \caption{\small $N_{\rm sig}=1$ contours for different values of the $h_{ee}$ coupling, ranging from 0.01 to 1.0. The other parameters are assumed to take the following values:$m_{F_i}=3~\rm{TeV},~(i=1,2,3)$, 
    $L=30~\rm{fb}^{-1}$, and $m_{S_1}=3$ (10) TeV in the left (right) plot.
 }
    \label{fig:paper3-model1-hee}
\end{figure}

\subsection{Numerical analysis for the coloured model}

In this section, we describe the coloured version of the radiative neutrino mass 
model introduced in Sec. \ref{sec:radiative-mass-gen}. The coloured version is 
obtained by promoting the fields of the basic model from $SU(3)_C$-singlets 
to (anti)triplets, $S \to \tilde S$ and $F \to \tilde F$ according to Tab.  \ref{tab:hirsch-color}. Another modification is the BSM parity breaking term:
$(h_{ee})_{ij} (e_R^i )^C (e_R^j)^C S_1^\dag \to (h_{ed})_{ij} (d_R^i)^C (e_R^j)^C S_1^\dag$. This is necessary for gauge invariance and making the lightest of 
the new particles unstable. The relevant decays for the coloured model are:
\begin{equation}
\begin{array}{lcll}
& & L~{\rm conserving}, & L~{\rm violating} \\
\tilde S^{+10/3} & \to & \ell^+ \ell^+ \ell^+ \bar d, & W^+ W^+ \ell^+ \bar d \\
\tilde S^{+7/3} & \to & \nu \ell^+ \ell^+ \bar d, & W^+ \ell^+ \bar d  \\
\tilde S^{+4/3} & \to & \nu \nu \ell^+ \bar d, & \ell^+ \bar d 
\end{array}
\label{eq:paper3-decays}
\end{equation}
We assign $L=3$ for $\tilde S_3$, and call the first (second) column of decay modes in Eq. 
\eqref{eq:paper3-decays} ``L-conserving'' (``L-violating''). With the replacements
$l^+ \to \bar{d}$ and $h_{ee} \to h_{ed}$ the decay rate formulas remain the 
same as for the uncoloured version of the model.

There is however one complication.
One can (for example) consider a process
$\tilde S^{+10/3} \to l^+l^+l^+\bar{d}$, which is a coloured version of the decay depicted in the left panel of Fig. \ref{fig:neutrino-s4decay}. 
It is possible to move the d-antiquark ($\bar d$) from the final state to the initial state, and study a hadron 
$(\tilde S^{+10/3} d)_{\mathrm{had}}$ composed of a d-quark and a BSM scalar. This hadron can undergo a three-body decay: 
$(\tilde S^{+10/3} d)_{\mathrm{had}} \to l^+ l^+l^+$, however, the contribution from this process is subdominant. For example, if we set $m_{\tilde S_3}=600~\rm{GeV}$, we can estimate \cite{Hirsch:2021wge}:
\begin{equation}
\frac{\Gamma 
\left[ 
(\tilde S^{+10/3} d)_{\mathrm{had}} \to l^+ l^+ l^+ 
\right]
}
{\Gamma \left[
\tilde S^{+10/3} \to l^+ l^+ l^+ \bar d
\right]}
\sim
\frac{P_3 f^2_\pi}{P_4 m^2_{\tilde S_3}}
\sim 10^{-5},
\end{equation}
with $f_\pi$ being the pion decay constant, and $P_n \equiv [4\cdot (4\pi)^{2n-3} \cdot (n-1)! \cdot (n-2)!]^{-1}$ being the n-body phase-space factor. Since the rate of the three-body decay of a hadron state is several orders of magnitude smaller than the decay rate for $\tilde S^{+10/3} \to l^+l^+l^+\bar{d}$, we do not include the former process in our analysis.

Long-lived particles in the coloured model are members of the $SU(2)_L$-triplet $\tilde S_3$: $\tilde S_3^{\pm 10/3}$, $\tilde S_3^{\pm 7/3}$, and $\tilde S_3^{\pm 4/3}$. These fields are analogous to $\tilde S^{\pm 2}$, $\tilde S^{\pm 3}$, and $\tilde S^{\pm 4}$, but with electric charges reduced by $2/3e$ and charged under $SU(3)_C$.

The dominant production processes are particle-antiparticle pair creations via 
QCD interactions with the $gg$ or $q\bar q$ initial states. Contributions from 
electroweak processes are negligible, hence cross sections for pair production 
of different components of the $SU(3)_C$-triplet are almost the same. This 
cross section is depicted in Fig. \ref{fig:paper3-xs} by a cyan curve. One can 
see from Fig. \ref{fig:paper3-xs} that the production cross section for coloured 
scalars is almost two orders of magnitude larger than for $pp \to S^{+4}S^{-4}$ 
at $m\sim 400~\rm{GeV}$. The difference is biggest for small masses because 
the parton distribution function of gluons is enhanced for smaller values of the 
parton energy fraction, x. 

Fig. \ref{fig:paper3-beta} contains the velocity distribution of coloured scalars 
(cyan). As can be seen, coloured particles are characterised by smaller velocities 
than uncoloured ones. The reason is that, unlike the Drell-Yann production of 
the electroweak particles, gluon fusion does not suffer from the p-wave 
suppression. What is more, as mentioned above, gluon PDF is significantly 
enhanced for smaller energy fraction, x, favouring near-threshold production 
with low velocities.

Multi-charged long-lived coloured particles will hadronise, and the charges of the colour-singlet states will be shifted with respect 
to the original BSM particles. Precise simulation of this process is complicated and beyond the scope of this work, therefore we 
resort to a crude hadronisation model in order to investigate the impact of the charge shifting on MoEDAL's sensitivity. In our 
hadronisation model, $\tilde S = (\tilde S^{+10/3}, \tilde S^{+7/3}, \tilde S^{+4/3})$ hadronises into mesonic state with the 
probability $k$, and into a baryonic state with the probability $1-k$. We assume that colour-singlet states will be formed with the 
first generation of SM quarks and the probability to bound to $u$ quark is the same as for the $d$ quark. The two possible mesonic states are:
\begin{equation}\label{eq:paper3-meson}
\begin{split}
& \mbox{Spin-1/2 mesons: probability $k$~~~~~~~}  \\
 &\tilde S + u_{L/R}  ~~~~ (+ 2/3)  \\
 &\tilde S + d_{L/R}  ~~~~ (- 1/3) \,, 
 \end{split}
\end{equation}
where the charge shift is provided in the brackets. For baryons, we assume four spin-0 and two spin-1 states:
\begin{equation}\label{eq:paper3-spin0}
\begin{split}
 &\mbox{Spin-0 baryons: probability $\frac{2}{3}(1-k)$}  \\
 &\tilde S + \bar u_L \bar u_R ~~~~ (- 4/3) \\
 &\tilde S + \bar d_L \bar d_R ~~~~  (+ 2/3) \\
 &\tilde S + \bar u_L \bar d_R ~~~~ (- 1/3)  \\
 &\tilde S + \bar d_L \bar u_R ~~~~ (- 1/3)  \\
 \end{split}
\end{equation}
\begin{equation}\label{eq:paper3-spin1}
\begin{split}
& \mbox{Spin-1 baryons: probability $\frac{1}{3}(1-k)$}  \\
 &\tilde S + \bar u_L \bar d_L ~~~~ (- 1/3)   \\
 &\tilde S + \bar u_R \bar d_R ~~~~ (- 1/3)  \,,
 \end{split}
\end{equation}

In order to roughly estimate the size of the uncertainty associated with our crude hadronisation model, we vary the $k$ parameter 
from $0.3$ to $0.7$. We stress that we use our hadronisation model only for approximate estimation of the MoEDAL's 
detection reach for the considered BSM scenario. More rigorous treatment of the hadronisation effects might be necessary to fully 
understand the sensitivity of the MoEDAL detector.

The expected sensitivity of the MoEDAL detector to multiply charged colour particles is presented in Fig. \ref{fig:paper3-model2-res}. The results can be 
regarded as model-independent because masses and decay lengths of the new particles are treated as free parameters. Analysis 
related directly to parameters of the coloured model from Sec. \ref{sec:radiative-mass-gen}. will be presented later. The panel on 
the left (right) of Fig. \ref{fig:paper3-model2-res} corresponds to $L=30~\rm{fb}^{-1}$ $(L=300~\rm{fb}^{-1})$ integrated luminosity expected at 
the end of the Run 3 (HL-LHC) data-taking phase. 
$N_{\rm sig} = $1, 2, and 3, correspond to solid, dashed, and dotted contour lines, respectively.
Sensitivities to all particles are calculated for three different values of the $k$ parameter, $k \in \{0.3, 0.5, 0.7\}$, in order to parametrise the uncertainty of our crude hadronisation model. One can immediately observe, that 
the dependence of the maximum mass reach (for $c\tau > 100~\rm{m}$) on $k$ is relatively small. The difference between $k=0.3$ and 
$k=0.7$ mass reach is only $\sim 50~\rm{GeV}$ for $\tilde S^{\pm 4/3}$. For $\tilde S^{\pm 10/3}$ the effect is even smaller, 
approximately 30 GeV. This is because the change of the signal acceptance by varying $\beta_{\rm th}$ is milder for larger 
charges, as can be seen in Fig. \ref{fig:paper3-beta}. Another observation is that the detection reach of MoEDAL increases for 
larger values of $k$. It can be easily understood 
by looking at the Eqs. \eqref{eq:paper3-meson}, \eqref{eq:paper3-spin0} and 
\eqref{eq:paper3-spin1}.
Greater $k$ corresponds to a higher probability of forming a meson state, half of which increases the 
absolute value of the electric charge and thus the signal acceptance of the MoEDAL detector. It also implies a lower probability of forming a baryon state, which in  
most cases have an electric charge lowered. 

\begin{figure}[!tbp]
\centering
  \begin{subfigure}[t]{0.49\textwidth}
    \includegraphics[width=\textwidth]{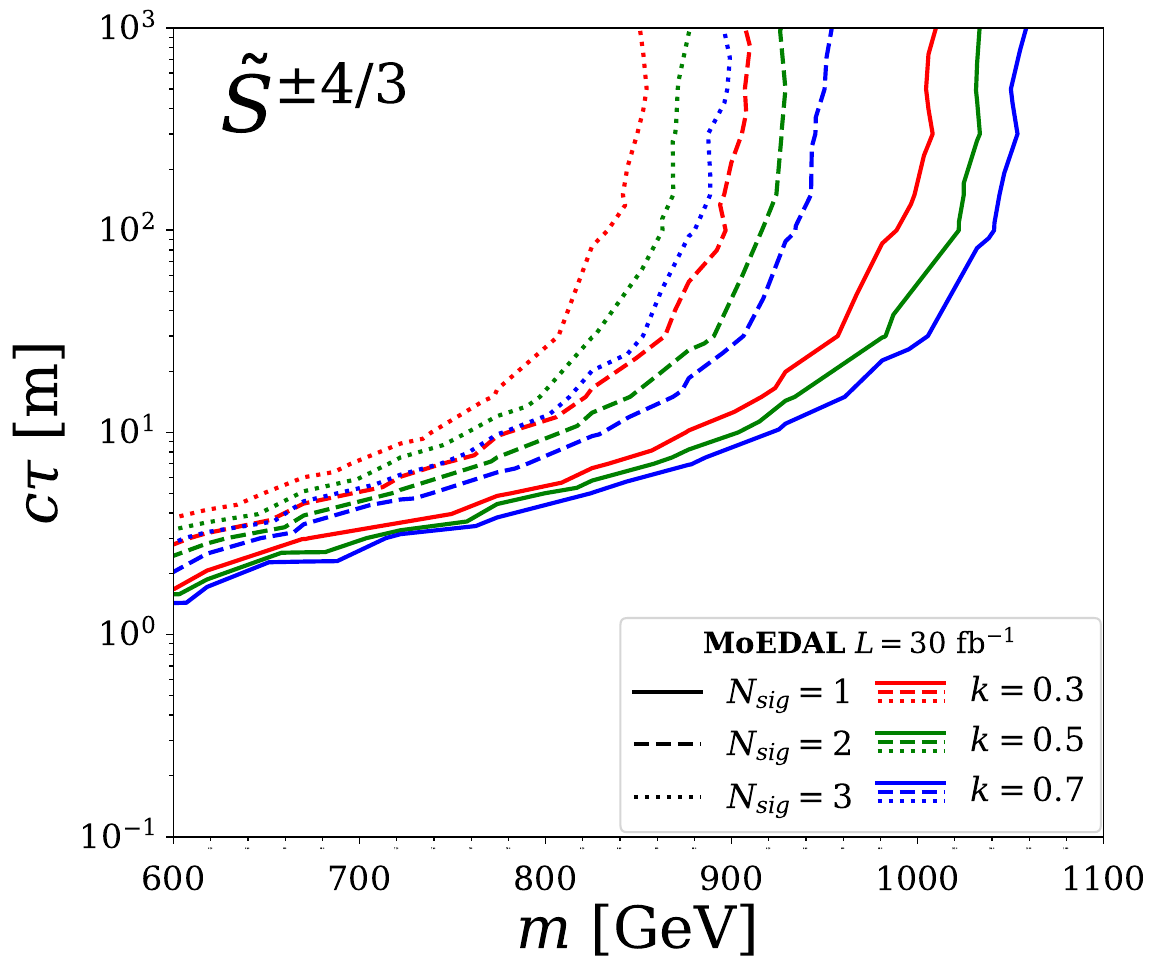}
  \end{subfigure}
    \hfill
   \begin{subfigure}[t]{0.49\textwidth}
    \includegraphics[width=\textwidth]{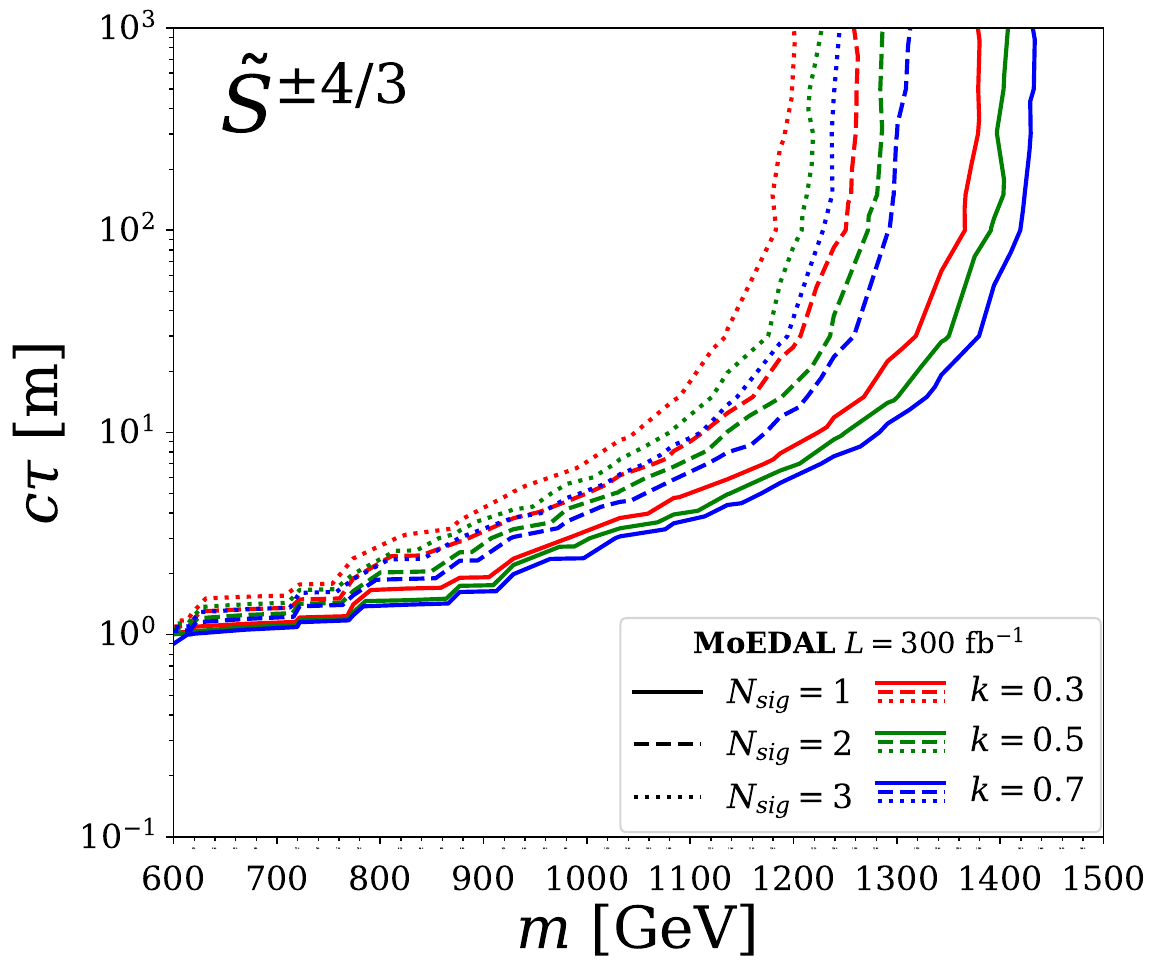}
  \end{subfigure}
    \hfill
   \begin{subfigure}[t]{0.49\textwidth}
    \includegraphics[width=\textwidth]{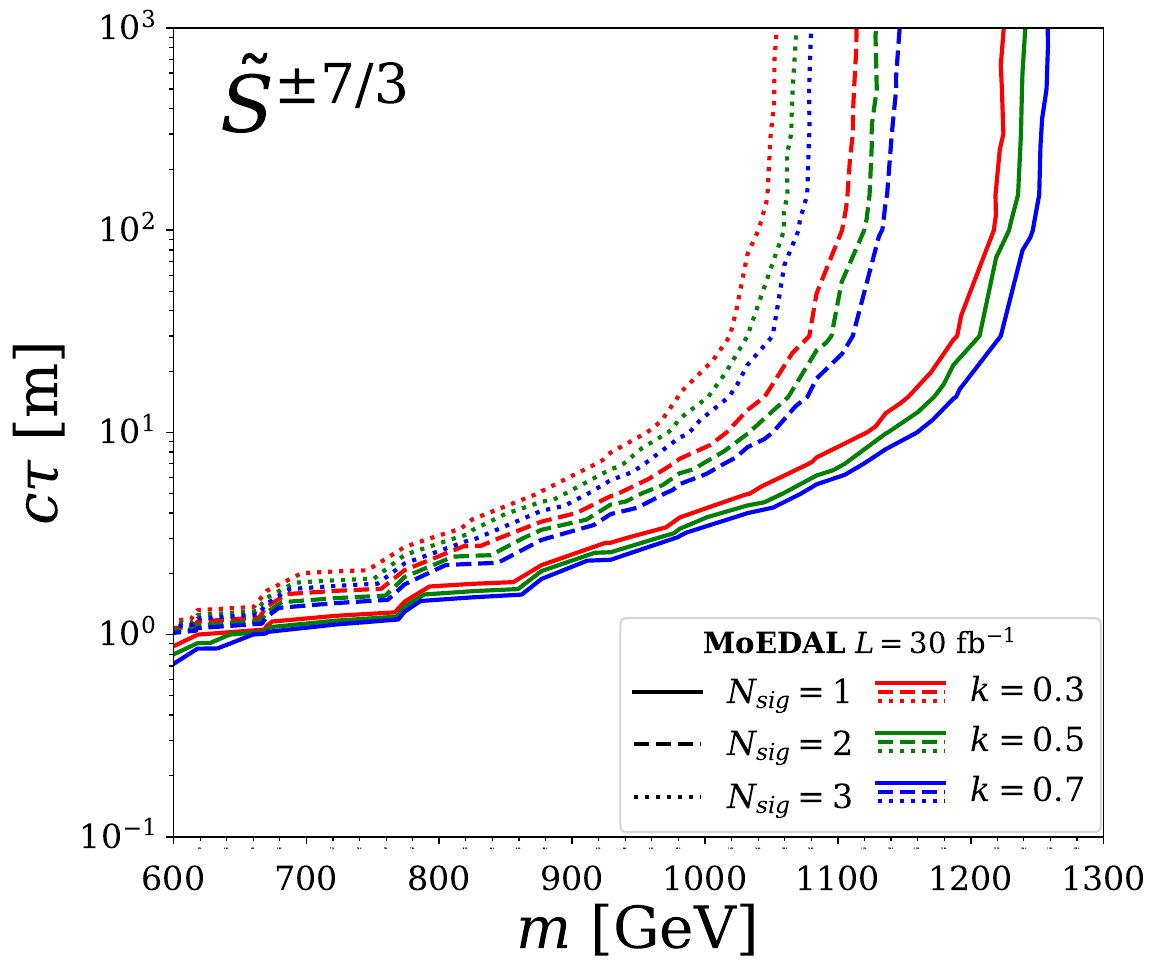}
  \end{subfigure}
  \hfill
  \begin{subfigure}[t]{0.49\textwidth}
    \includegraphics[width=\textwidth]{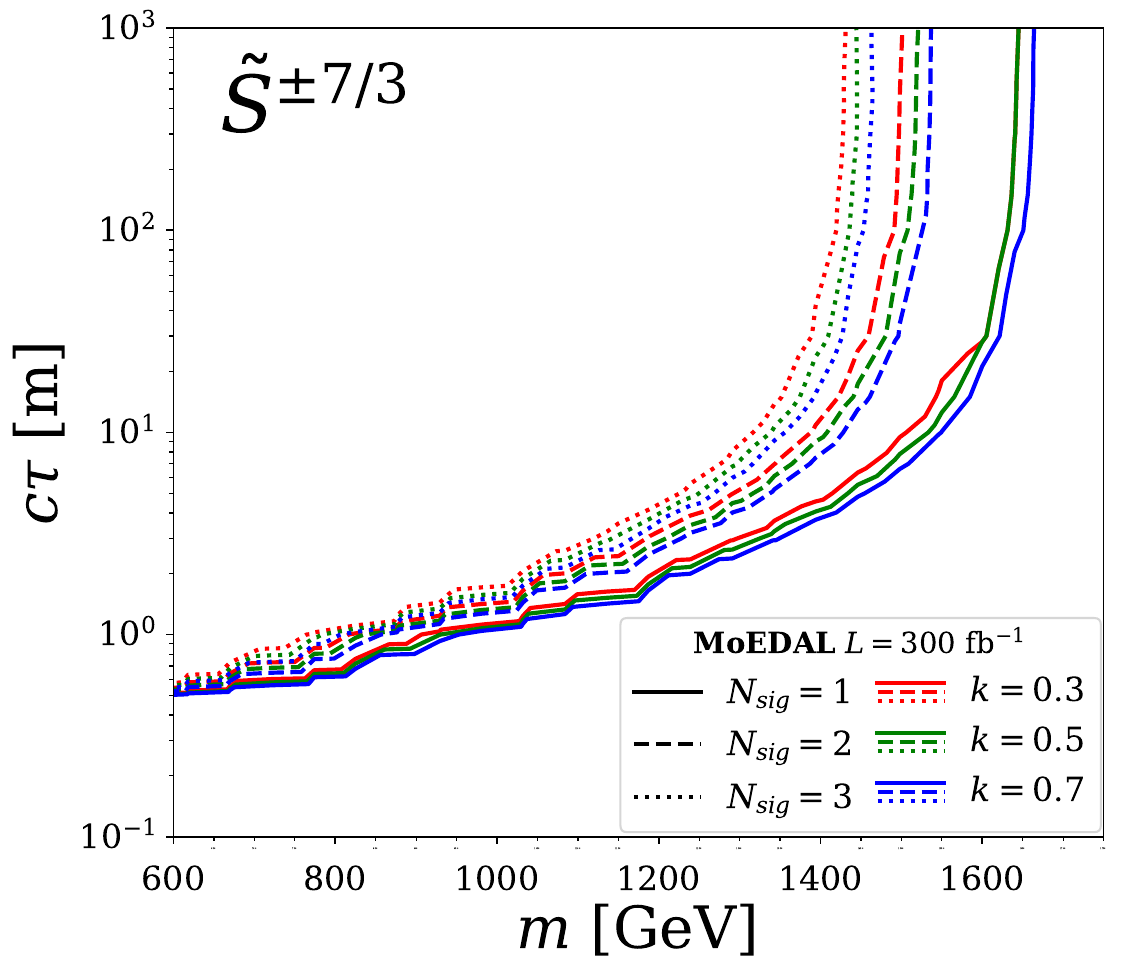}
  \end{subfigure}
    \hfill
   \begin{subfigure}[t]{0.49\textwidth}
    \includegraphics[width=\textwidth]{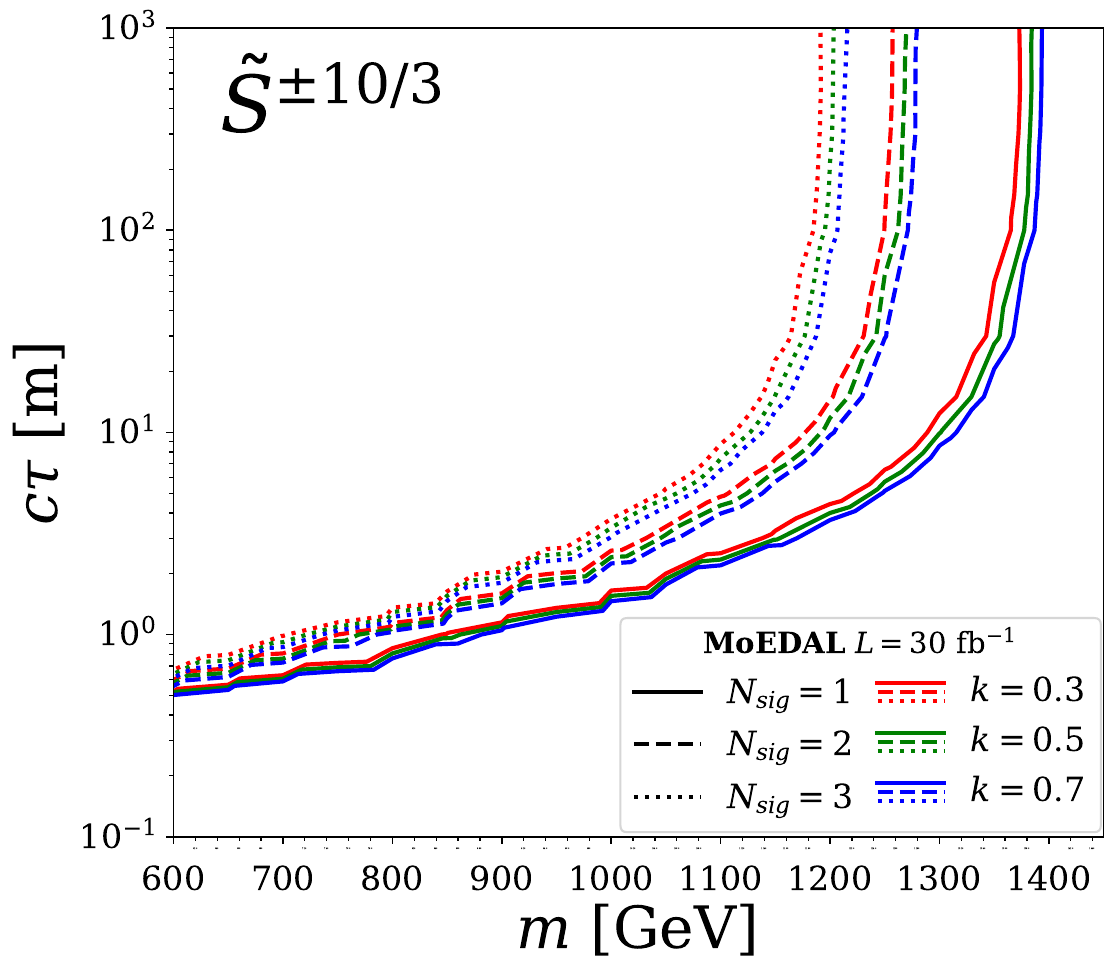}
  \end{subfigure}
  \hfill
     \begin{subfigure}[t]{0.49\textwidth}
    \includegraphics[width=\textwidth]{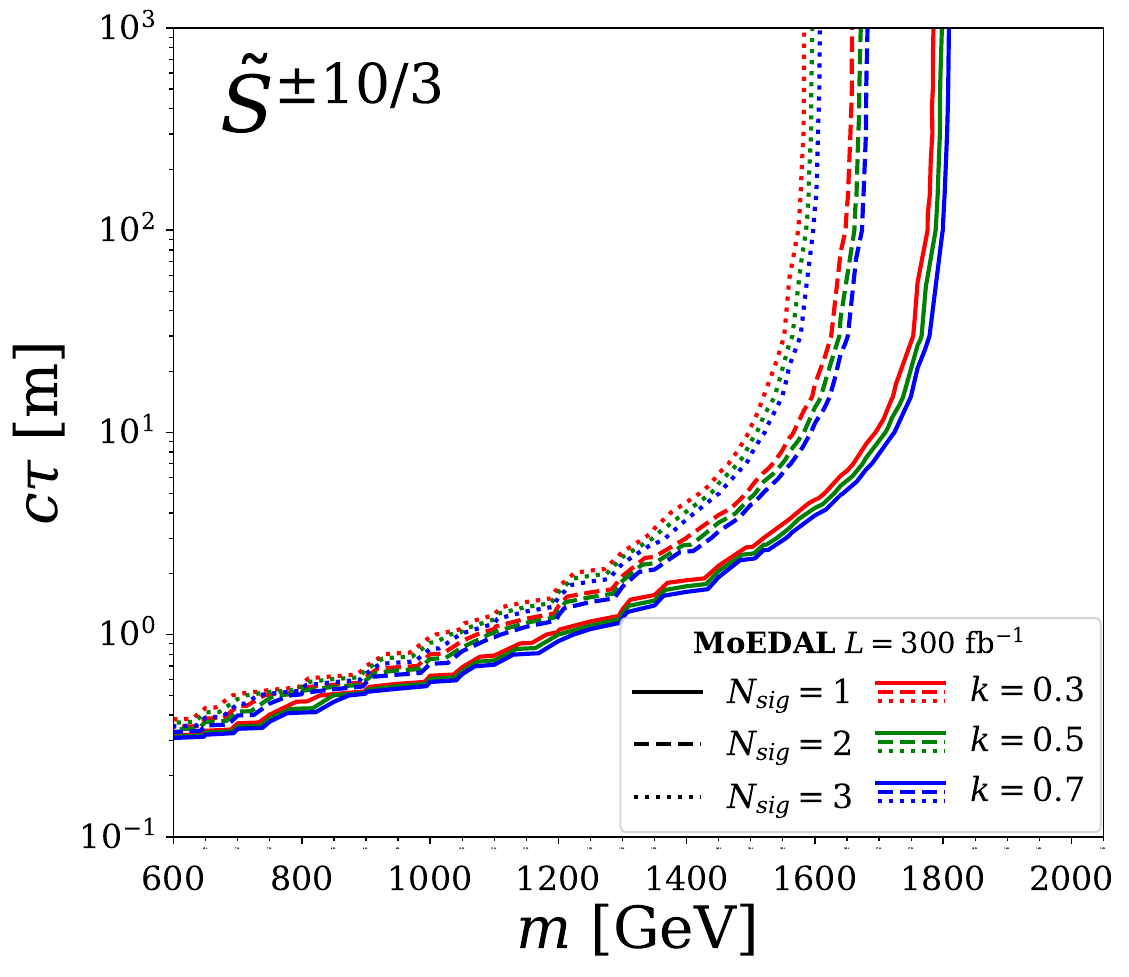}
  \end{subfigure}
    \caption{\small The model-independent detection reach of the MoEDAL detector. Results for the multiply charged colour-triplet particles are presented in the $m$ vs. $c\tau$ parameter plane . Solid, dashed and dotted contours correspond to $N_{\rm sig}=1$, $2$, and $3$, respectively. Red, green and blue colours represent $k=0.3$, $k=0.5$,  and $k=0.7$, respectively.
    The integrated luminosity is $30~\rm{fb}^{-1}$ (left) and $300~\rm{fb}^{-1}$ (right).}
    \label{fig:paper3-model2-res}
\end{figure}

The top row of Fig. \ref{fig:paper3-model2-res} presents results for $\tilde S^{\pm 4/3}$. For $N_{\rm sig}=1$ (3) Run 3 MoEDAL ($L=30~\rm{fb}^{-1}$) can probe mass 
of $\tilde S^{\pm 4/3}$ up to $~1050$ (880) GeV. MoEDAL for HL-LHC ($L=300~\rm{fb}^{-1}$) can 
improve this result up to $\sim 1400$ (1250) GeV with $N_{\rm sig}=1$ (3). The middle row of Fig. \ref{fig:paper3-model2-res} depicts MoEDAL's sensitivity to $\tilde S^{\pm 7/3}$.
One can see that for $N_{\rm sig}=1$ (3) Run 3 MoEDAL can test mass 
of $\tilde S^{\pm 7/3}$ up to $\sim 1250$ (1080) GeV. In the case of the HL-LHC, it can 
improve this result up to $\sim 1650$ (1450) GeV with $N_{\rm sig}=1$ (3).
Finally, the prospects for detection of $\tilde S^{\pm 10/3}$ are shown in the bottom row. The maximum $m_{\tilde S^{\pm 10/3}}$ 
reach of Run 3 MoEDAL is $\sim 1400$ (1200) GeV for $N_{\rm sig}=1$ (3). HL-LHC allows to raise these values up to $\sim 1800$ (1600) GeV with
$N_{\rm sig}=1$ (3).

The sensitivity of the MoEDAL detector to long-lived multiply charged coloured scalars is summarised in Tab. \ref{tab:sum_col}. In the same 
table prospects for BSM detection in MoEDAL are compared with the latest HSCPs searches by ATLAS and CMS (if available). All 
masses in Tab. \ref{tab:sum_col} are in GeV units. The numbers in double parentheses in the first column represent the estimated mass bounds 
obtained in \cite{Jager:2018ecz} by rescaling the 8 TeV CMS result \cite{CMS:2013czn} to 13 TeV with $L=36~\rm{fb}^{-1}$.
The second column presents the
  projected mass reach for Run 3 (300 fb$^{-1}$) obtained in
  \cite{Jager:2018ecz}. The numbers outside (inside) the brackets in
  the third and fourth columns represent MoEDAL's mass reaches with
  $N_{\rm sig} \geq 3$ (1) assuming $L = 30$ (Run 3) and 300 (HL-LHC)
  fb$^{-1}$, respectively.
  By comparing the values in Tab. \ref{tab:sum_col} one can see that the current HSCP bounds are rather strong and masses testable in Run 3 
  MoEDAL are already excluded. However, in the case of $\tilde S^{\pm 7/3}$ and $\tilde S^{\pm 10/3}$ HL-LHC MoEDAL is capable 
  of testing currently unconstrained mass regions, but even then its overall sensitivity is expected to be lower or comparable to that of 
  ATLAS/CMS.

\begin{table}[t!]
\centering
\caption{\small Model-independent mass reaches (in GeV) of
  the multiply charged particles in the coloured model by MoEDAL (2nd and 3rd
  columns).  In the first column, the numbers in the double parentheses
  correspond to the estimated mass bounds obtained in \cite{Jager:2018ecz} by
  rescaling the 7 and 8 TeV CMS result \cite{CMS:2013czn} to 13
  TeV with $L = 36$ fb$^{-1}$. The second column presents the
  projected mass reach for Run 3 (300 fb$^{-1}$) obtained in
  \cite{Jager:2018ecz}. The numbers outside (inside) the brackets in
  the third and fourth columns represent MoEDAL's mass reaches with
  $N_{\rm sig} \geq 3$ (1) assuming $L = 30$ (Run 3) and 300 (HL-LHC)
  fb$^{-1}$, respectively.}
  
\begin{tabular}{c|c|c|c|c}
           & current HSCP bound & HSCP (Run-3) & MoEDAL (Run-3) & MoEDAL (HL-LHC) \\
           & 36 fb$^{-1}$ \cite{Jager:2018ecz} & 300 fb$^{-1}$ \cite{Jager:2018ecz} & 30 fb$^{-1}$ & 300 fb$^{-1}$ \\        
\hline
$\tilde S^{\pm 4/3}$ & ((1450))  & 1700 & 880 (1050) & 1250 (1400)   \\
\hline
$\tilde S^{\pm 7/3}$ & ((1480))  & 1730 & 1080 (1250) & 1450 (1650)   \\
\hline
$\tilde S^{\pm 10/3}$ & ((1510))  & 1790 & 1200 (1400) & 1600 (1800)   \\
\end{tabular}

\label{tab:sum_col}
\end{table}

\myparagraph{Interpretation of the results for the coloured version of the model}
We now move on to discussing the testability of the coloured model's parameters in MoEDAL. Our goal is to reveal the parameter 
region that can be probed by Run 3 (30 fb$^{-1}$) and HL-LHC (300 fb$^{-1}$) MoEDAL within the subspace of parameters for 
which neutrino masses and mixing angles can be explained.

We begin by showing in Fig. \ref{fig:paper3-model2-lifetime} decay lengths of multiply charged coloured scalar particles as a 
function of the $\lambda_5$ parameter. $h_F$ and $h_{\bar F}$ are fitted from the neutrino data. We fix $m_{\tilde S_3} = 750~\rm{GeV}$, $m_{\tilde F_i} = 3~\rm{TeV}$ and $h_{ed}=0.1$. The plot on the left (right) in Fig. \ref{fig:paper3-model2-lifetime} 
corresponds to $m_{\tilde S_1}=3$ (10) TeV. 
Black horizontal lines in Fig. \ref{fig:paper3-model2-lifetime} correspond to $c\tau\sim 1~\rm{m}$, which is a typical decay length for particles measurable in MoEDAL.
The result in Fig. \ref{fig:paper3-model2-lifetime} is very similar to the one for 
the uncoloured model in Fig. \ref{fig:paper3-model1-lifetime}, which is natural since decay rate formulas in both models are identical, 
and contribution to neutrino masses is also the same up to a colour factor $N_C$. As a consequence, decay lengths in Fig. 
\ref{fig:paper3-model2-lifetime} are about $\sim 2$ times larger than in Fig. \ref{fig:paper3-model1-lifetime}. As mentioned in Sec. 
\ref{sec:paper3-uncoloured}, for large $\lambda_5 \gg 10^{-5}$ the L-violating decay modes, which are proportional to $|\lambda_5|^2$, dominate. On the other hand, in the low  $\lambda_5 \ll 10^{-6}$ region the L-conserving decays dominate 
because their decay rate is proportional to $h_F h_{\bar F}$, which is inversely proportional to $\lambda_5$ due to neutrino mass 
constraints. As a result, the particles predicted in the studied neutrino mass model are long-lived for intermediate values of $\lambda_5$, i.e. $10^{-7} \lesssim \lambda_5 \lesssim 10^{-3}$ for $m_{\tilde S_1}=3~\rm{TeV}$, and 
$5\cdot 10^{-8} \lesssim \lambda_5 \lesssim 10^{-2}$  for $m_{\tilde S_1}=10~\rm{TeV}$.

\begin{figure}[!tbp]
\centering
  \begin{subfigure}[t]{0.49\textwidth}
    \includegraphics[width=\textwidth]{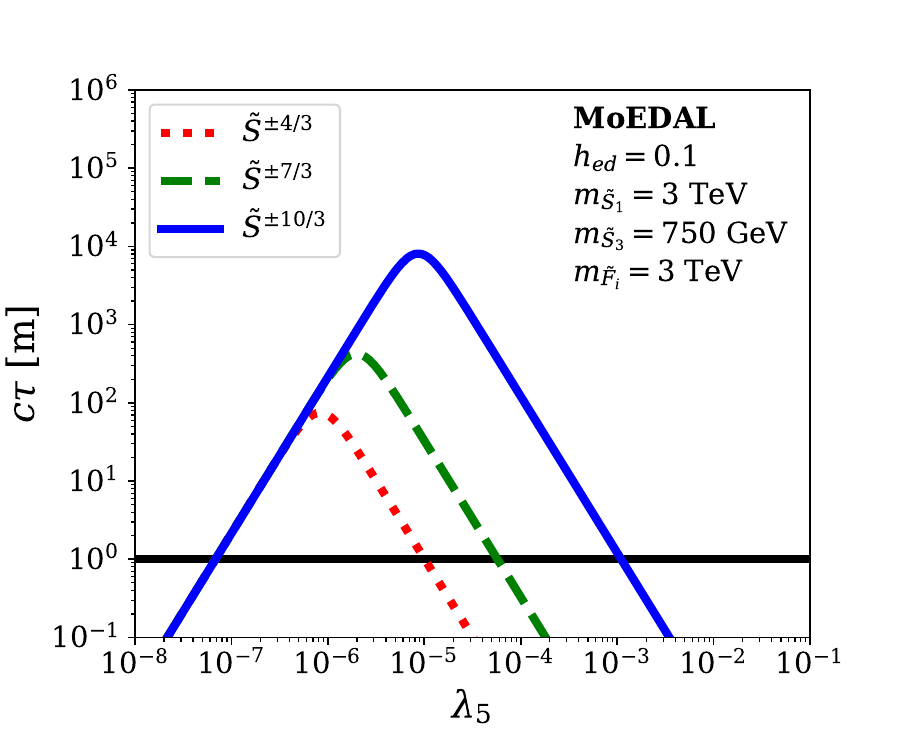}
  \end{subfigure}
    \hfill
   \begin{subfigure}[t]{0.49\textwidth}
    \includegraphics[width=\textwidth]{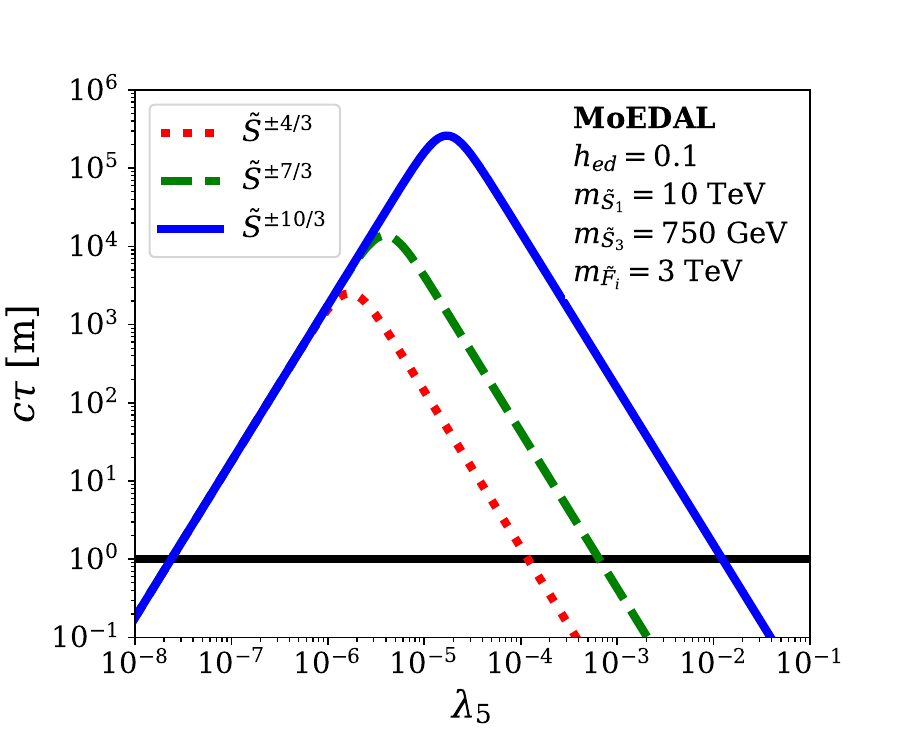}
  \end{subfigure}
    \caption{\small Decay lengths of $S^{\pm 10/3}$ (blue solid), $S^{\pm 7/3}$ (green dashed) and $S^{\pm 4/3}$ (red dotted) as a function of $\lambda_5$. Other 
 parameters are set to $h_{ed}=0.1$, $m_{S_3} = 750~\rm GeV$, and $m_{S_1} = 3$ (10) TeV for the left (right) plot.
 For the given set of mass and $\lambda_5$ parameters,  $h_F h_{\bar F}$  is obtained from fitting the neutrino data.
 The thick black horizontal line corresponds to $c\tau = 1~\rm{m}$, which is a typical distance for the MoEDAL detector. 
 }
    \label{fig:paper3-model2-lifetime}
\end{figure}

Fig. \ref{fig:paper3-model1-reach2} shows the sensitivity of MoEDAL to the coloured version of the studied model in a $m_{\tilde S_3}$ vs. $\lambda_5$ 
parameter plane, with $h_F h_{\bar{F}}$ fitted to neutrino data.
The top (bottom) plots in Fig. \ref{fig:paper3-model1-reach2} correspond to $m_{\tilde S_1}=3$ (10) TeV, and left (right) panels are for $L=30~\rm{fb}^{-1}$ ($L=300~\rm{fb}^{-1}$) integrated luminosity. Other parameters are fixed to $m_{\tilde F_i}=3~\rm{TeV}$ and $h_{ed}=0.1$ 
for all plots. Solid, dashed and dotted contours in Fig. \ref{fig:paper3-model1-reach2} correspond to $N_{\rm sig}=1$, 2 and 3, respectively. Similarly to Fig. \ref{fig:paper3-model2-res}, we vary the value of the parameter $k$ to reveal the uncertainty of our crude hadronisation model. 
Red, green and blue curves in Fig. \ref{fig:paper3-model1-reach2} represent $k=0.3$, $k=0.5$, and $k=0.7$, respectively. The grey region for small values of
$\lambda_5$ is non-perturbative, because of a large coupling $\mathrm{max}_{ij}\left(|(h_F)_{ij}|, |(h_{\bar F})_{ij}|\right) \geq 2$, and
cannot be trusted. By looking at Fig. \ref{fig:paper3-model1-reach2} we can see that varying $k$ parameter from 0.3 to 0.7 changes the maximum mass reach 
only slightly, about $\sim 30~\rm{GeV}$, which leads to a conclusion that our analysis is quite robust despite a crude model of 
hadronisation. The highest mass of the $SU(2)_L$ triplet, $m_{\tilde S_3}$, which can be probed by MoEDAL roughly agrees with the 
mass reach for $m_{\tilde S^{\pm 10/3}}$, though for $m_{\tilde S_3}$ it is slightly higher due to the effect of other particles.

\begin{figure}[!htb]
\centering
  \begin{subfigure}[t]{0.49\textwidth}
    \includegraphics[width=\textwidth]{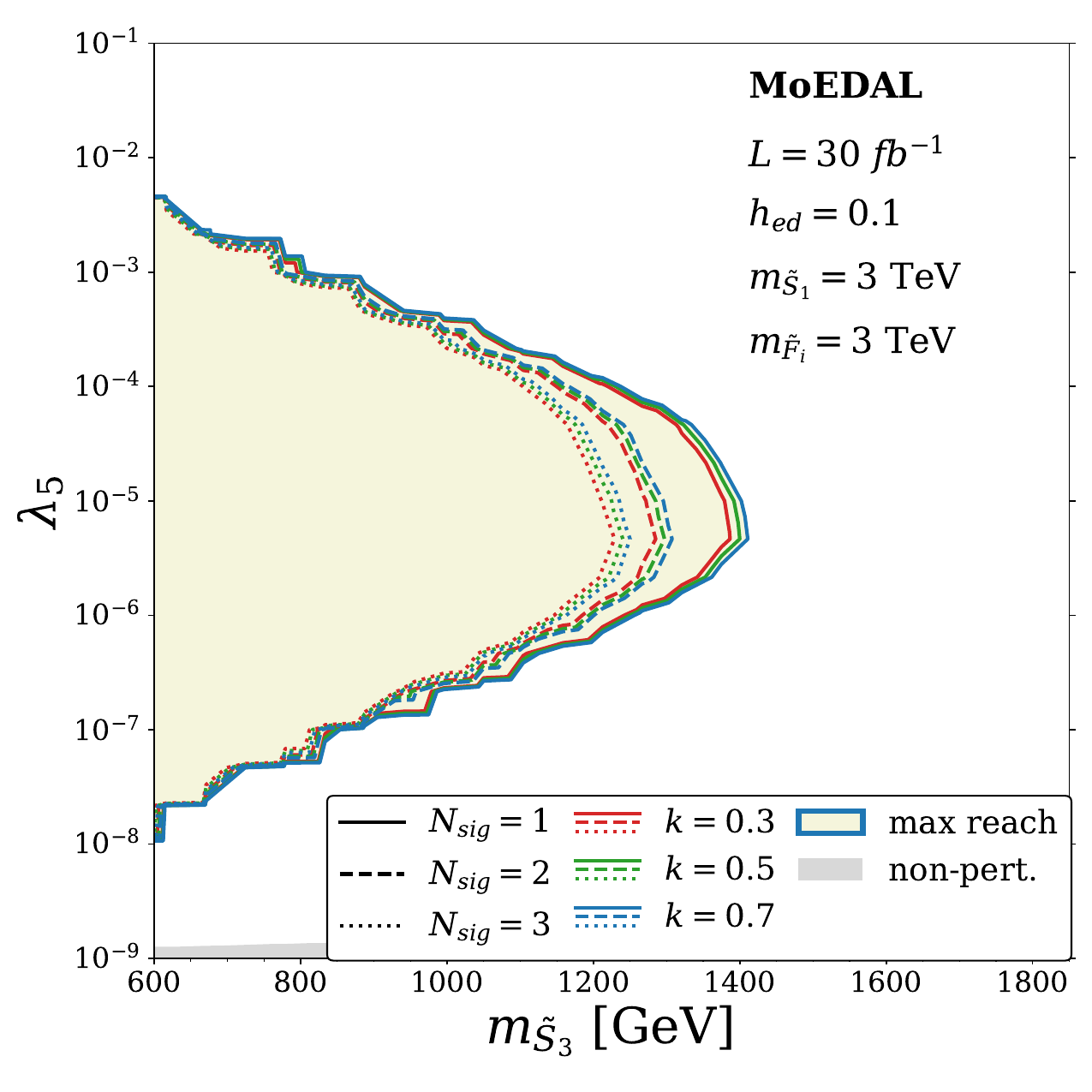}
  \end{subfigure}
    \hfill
   \begin{subfigure}[t]{0.49\textwidth}
    \includegraphics[width=\textwidth]{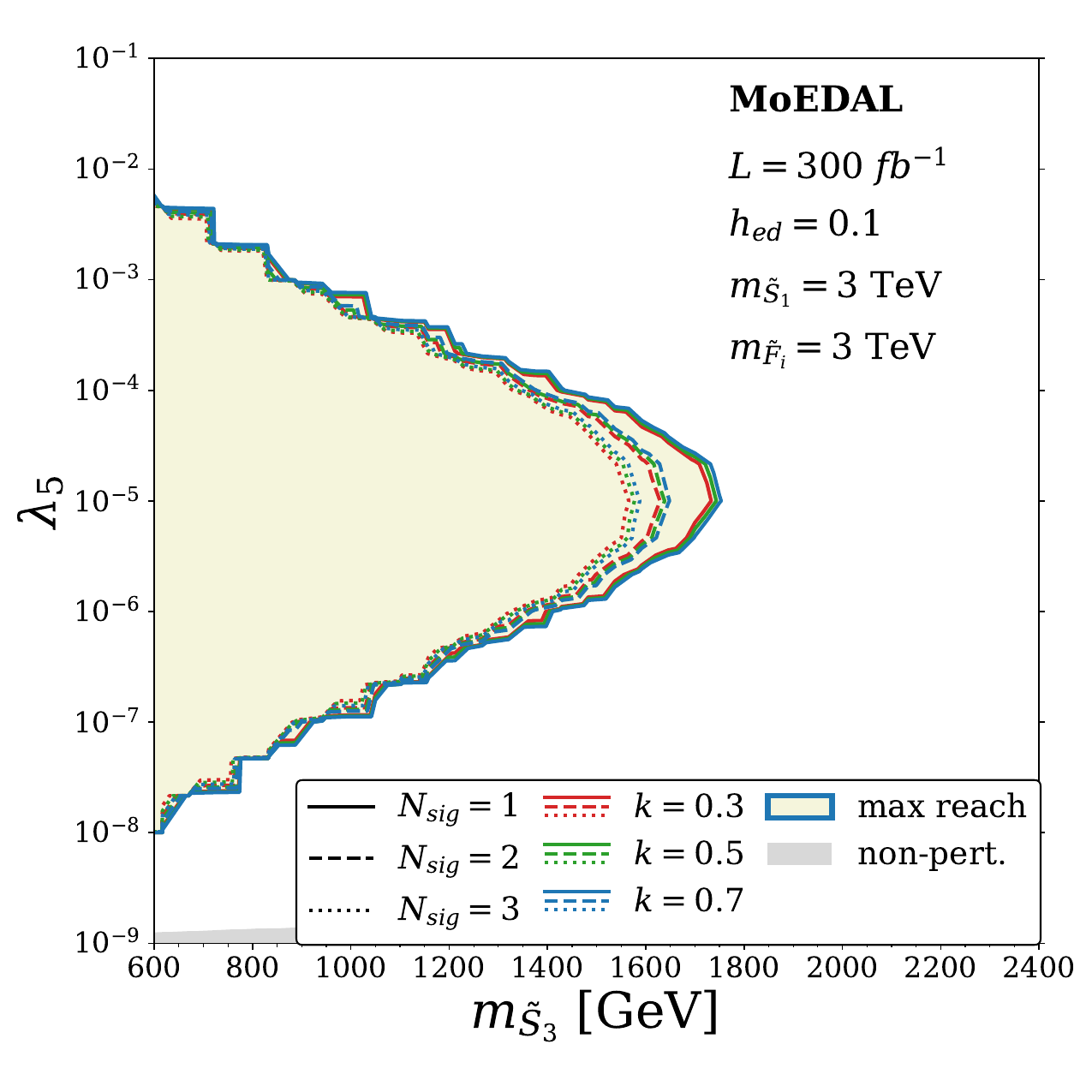}
  \end{subfigure}
  \begin{subfigure}[t]{0.49\textwidth}
    \includegraphics[width=\textwidth]{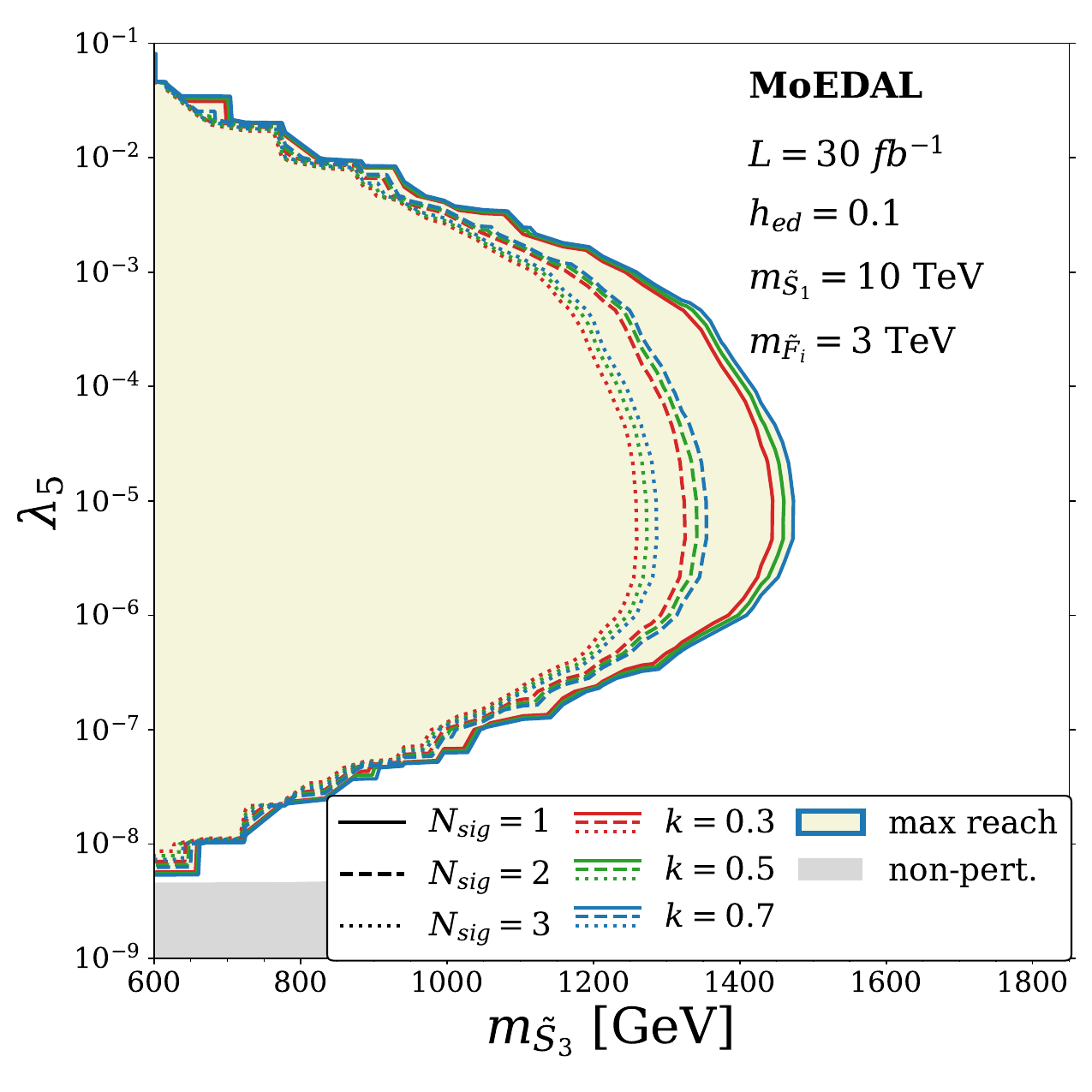}
  \end{subfigure}
    \hfill
   \begin{subfigure}[t]{0.49\textwidth}
    \includegraphics[width=\textwidth]{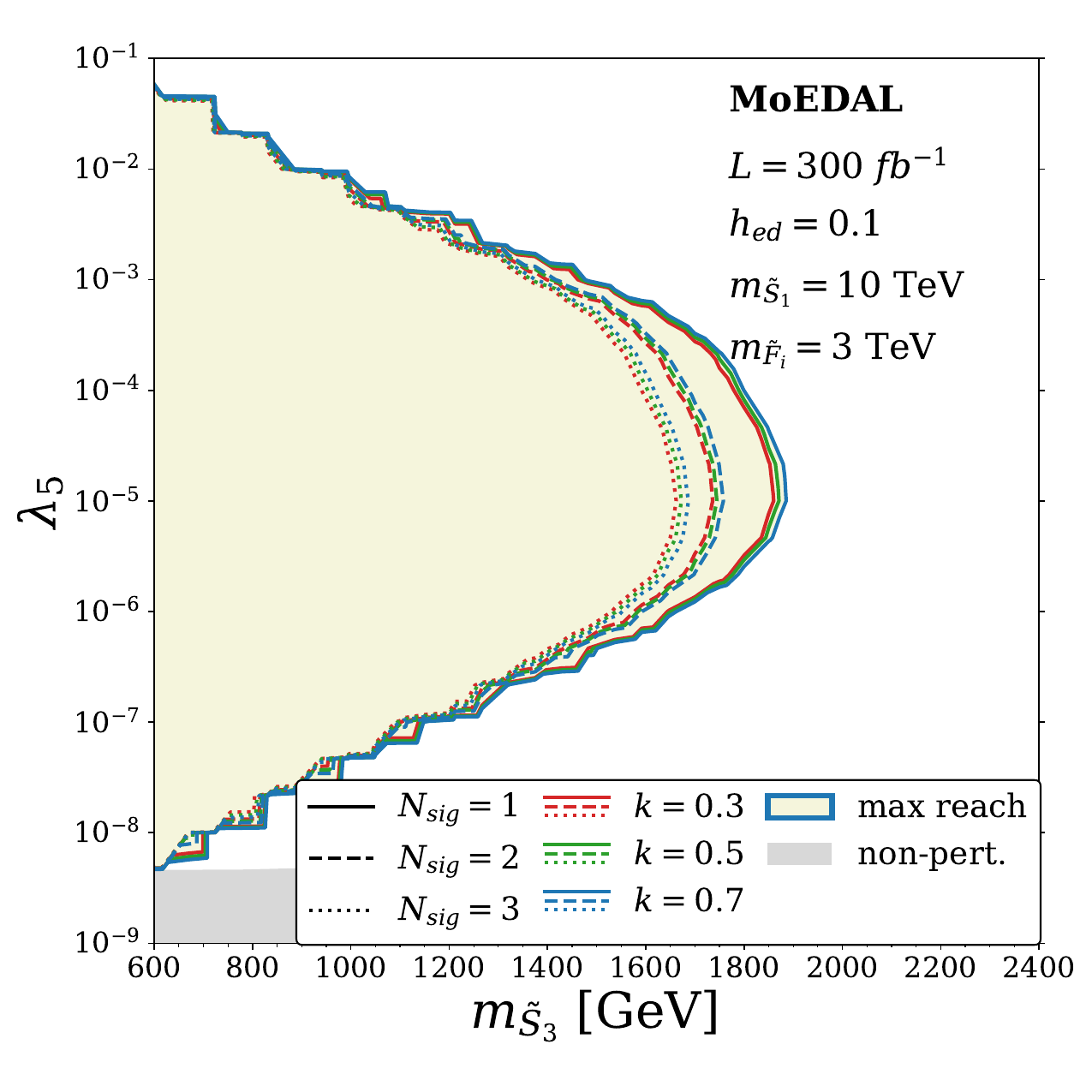}
  \end{subfigure}
    \caption{\small Sensitivity of the MoEDAL detector to the coloured version of the neutrino mass model studied in this project. 
    Results are presented in $m_{\tilde S_3}$ vs. $\lambda_5$ parameter plane, with $h_F h_{\bar{F}}$ fitted to neutrino data.
     The top (bottom) panels correspond to $m_{\tilde S_1}=3$ (10) TeV, while other parameters are set to $m_{\tilde F_i}=3~\rm{TeV}$ and $h_{ed}=0.1$ for all plots.
    Plots on the left (right) correspond to $L=30~\rm{fb}^{-1}$ ($L=300~\rm{fb}^{-1}$) integrated luminosity. Solid, dashed and dotted contours represent $N_{\rm sig}=1$, 2 and 3, respectively. 
    Red, green and blue curves correspond to $k=0.3$, $k=0.5$, and $k=0.7$, respectively. The grey region for small values of
$\lambda_5$ is non-perturbative, because of a large coupling $\mathrm{max}_{ij}\left(|(h_F)_{ij}|, |(h_{\bar F})_{ij}|\right) \geq 2$, and 
cannot be trusted.
    }
    \label{fig:paper3-model1-reach2}
\end{figure}

The top two panels in Fig. \ref{fig:paper3-model1-reach2} correspond to $m_{\tilde S_1} = 3~\rm{TeV}$. One can see from the 
left (right) panel that the maximum mass reach for the Run 3 (HL-LHC) MoEDAL with $L=30~\rm{fb}^{-1}$ ($L=300~\rm{fb}^{-1}$) is $m_{\tilde S_3}=1400$ (1800) GeV, for $N_{\rm sig}=1$ and $\lambda_5 \sim 10^{-5}$. This value of $\lambda_5$ roughly 
corresponds to the longest lifetime of the $\tilde S^{\pm 10/3}$, explaining why the MoEDAL's sensitivity is enhanced. For $\lambda_5 \sim 10^{-3}$ or $10^{-7}$ MoEDAL is expected to observe one or more signal events only if the mass of $\tilde S_3$ does not 
exceed $\sim 800$ (1000) GeV for $L=30$ (300) fb$^{-1}$. Results of the two bottom panels in Fig. \ref{fig:paper3-model1-reach2} correspond to $m_{\tilde S_1} = 10~\rm{TeV}$. One can see that for the higher mass of $m_{\tilde S_1} $, the range of $\lambda_5$ values giving the maximum mass reach is broader, and the reach is slightly higher. The latter is caused by the enhancement of 
lifetimes of multiply charged particles in the $\tilde S_3$ triplet when $m_{\tilde S_1}$ becomes large, due to the presence of an off-shell scalar singlet particle
$\tilde S_1^*$ in decays of the triplet (see Fig. \ref{fig:neutrino-s4decay}). The largest reach for $m_{\tilde S_3} $ is obtained around $\lambda_5 \sim 10^{-5}$ and it is $\sim 1500$ (1900) GeV with $L=30$ (300) fb$^{-1}$ and $N_{\rm sig}=1$. For values of the $\lambda_5$ coupling much larger or much lower than $10^{-5}$ the detection reach is degraded as can be seen in Fig. \ref{fig:paper3-model1-reach2}.

We finish our discussion of the MoEDAL's sensitivity to the coloured version of the studied neutrino mass model by investigating 
the dependence of the maximum reach for $m_{\tilde S_3}$ with $N_{\rm sig}=1$, when the value of the $h_{ed}$ coupling is varied 
between 0.01 and 1.0. The results are depicted in the $(m_{\tilde S_3}, \lambda_5)$ parameter plane, which is shown
in Fig. \ref{fig:paper3-model2-hee}, where the left and right panels correspond to $m_{\tilde S_1}=3~\rm{TeV}$ and $m_{\tilde S_1}=10~\rm{TeV}$, 
respectively. We assume Run 3 integrated luminosity $L=30~\rm{fb}^{-1}$ and we set other parameters to $k=0.5$ and $m_{\tilde F_i}=3~\rm{TeV}$. As can be seen in Fig \ref{fig:paper3-model2-hee}, when the value of the $h_{ed}$ coupling is smaller, MoEDAL is sensitive to a 
broader range of $\lambda_5$ values. It is a similar behaviour as in Fig. \ref{fig:paper3-model1-hee}, caused by the suppression 
of decay rates, which are proportional to $|h_{ed}|^2$ (see for example Eq. \eqref{eq:neutrino-p4to4ldecay}). 
Nonetheless, the sensitivity of MoEDAL to $h_{ed}$ in Fig. \ref{fig:paper3-model2-hee} is stronger than for $h_{ee}$
\ref{fig:paper3-model1-hee}, which reveals that MoEDAL is more sensitive to the lifetimes of particles in the coloured model.

\begin{figure}[!htb]
\centering
  \begin{subfigure}[t]{0.49\textwidth}
    \includegraphics[width=\textwidth]{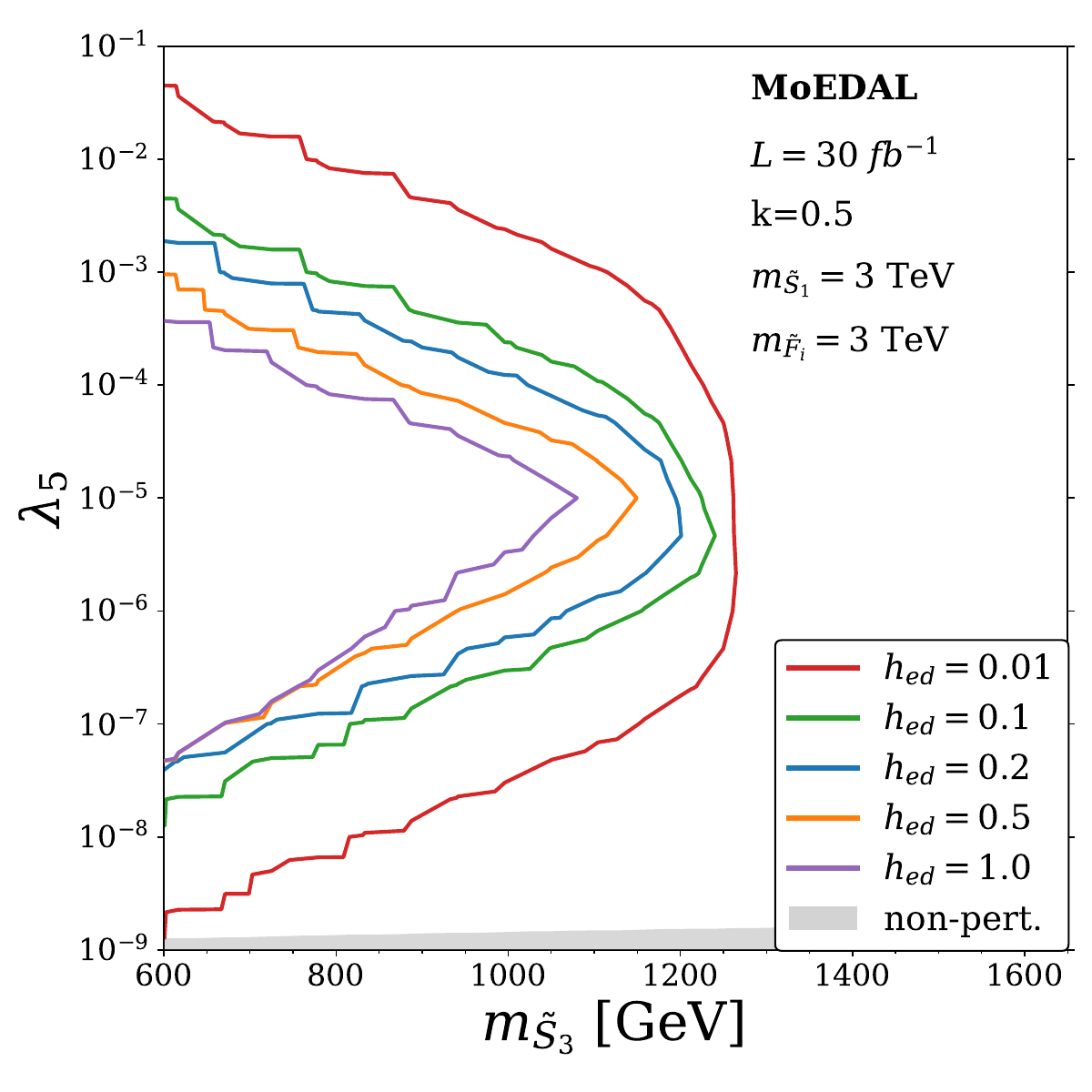}
  \end{subfigure}
    \hfill
   \begin{subfigure}[t]{0.49\textwidth}
    \includegraphics[width=\textwidth]{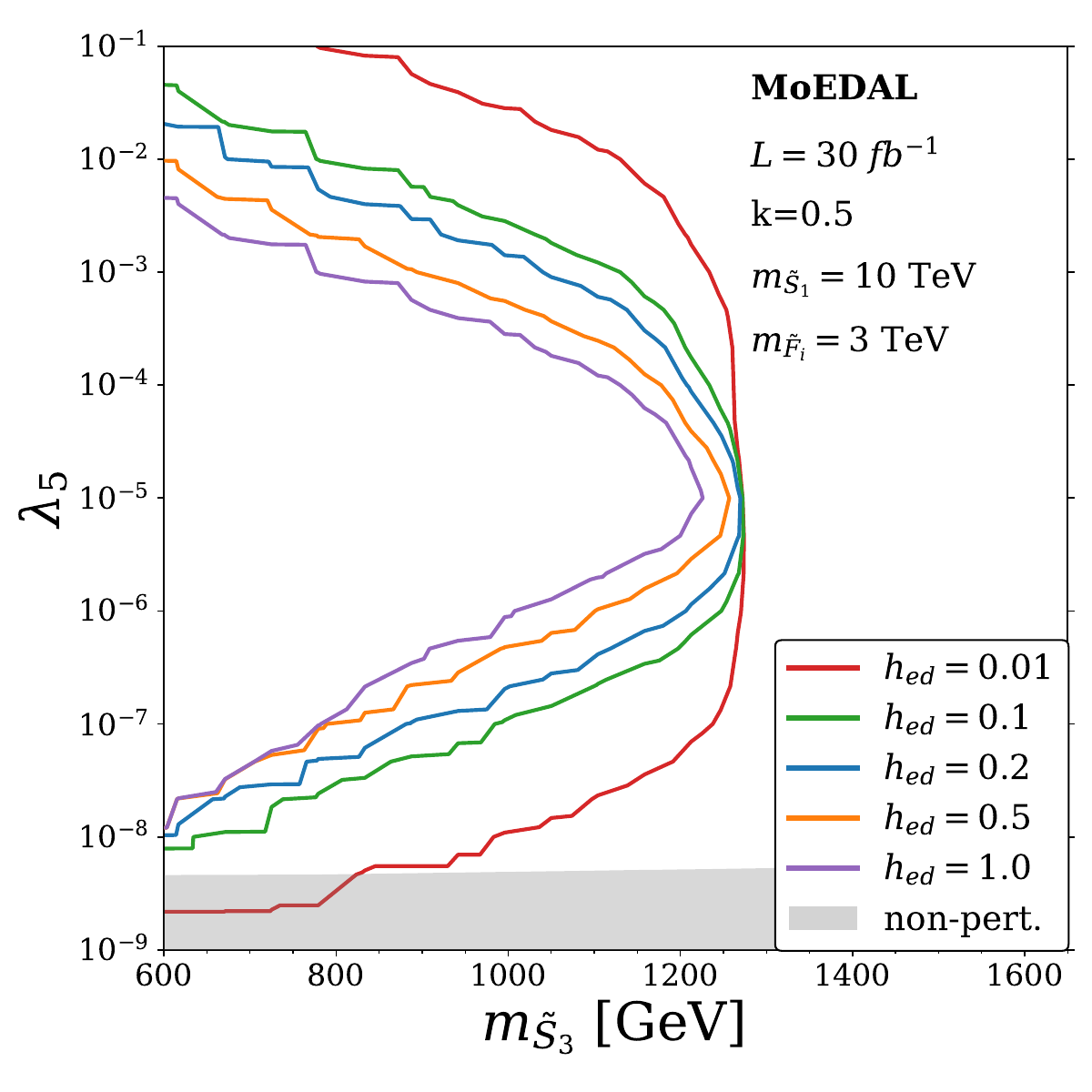}
  \end{subfigure}
    \caption{\small $N_{\rm sig}=1$ contours for different values of the $h_{ed}$ coupling, ranging from 0.01 to 1.0. 
    Other parameters were $m_{F_i}=3~\rm{TeV},~(i=1,2,3)$, 
    $L=30~\rm{fb}^{-1}$, and $m_{S_1}=3$ (10) TeV for the left (right) plot.
 }
    \label{fig:paper3-model2-hee}
\end{figure}

\subsection{Conclusions}

In this project, we investigated the prospects for detection of multiply charged long-lived particles with the MoEDAL detector for 
Run 3 and HL-LHC phases of the Large Hadron Collider. The BSM particles were inspired by two 1-loop radiative neutrino mass models, introduced in Sec. \ref{sec:radiative-mass-gen}. In the first model, all BSM fields are singlets under $SU(3)_C$ 
gauge group. In this model, there are three multiply charged long-lived scalars: $S^{\pm 2}$, $S^{\pm 3}$, $S^{\pm 4}$, and 
a short-lived triply charged fermion $F^{\pm 3}$. BSM fields in the second model are colour (anti)triplets, and the long-lived 
scalars predicted in this scenario are: $S^{\pm 4/3}$, $S^{\pm 7/3}$, $S^{\pm 10/3}$. In both models the lifetimes of multiply charged 
particles depend on $\lambda_5$, $h_F$, $h_{\bar F}$ couplings, as well as the BSM parity breaking coupling $h_{ee}$ ($h_{ed}$) 
for the uncoloured (coloured) version of the model. What is more, $\lambda_5$ and 
the product  $h_F \cdot h_{\bar F}$ are related through constraints on the 
neutrino masses and mixing angles.

We demonstrated that multiply charged LLPs are favourable from the point of 
view of the MoEDAL detector. Firstly, the ionisation power increases with the 
magnitude of the electric charge, allowing to relax the velocity threshold below 
which MoEDAL can detect BSM particles. Secondly, a large electric charge 
enhances the photon fusion production mode, which becomes as important as 
the Drell-Yann process and leads to a higher total production cross section than 
for the singly charged particles. The lower significance of the Drell-Yan production 
mode has another benefit for MoEDAL because scalar production via the Drell-Yan process is an s-channel 
exchange of a gauge boson and suffers from the p-wave suppression, i.e. $\sigma\to 0$ when $\beta \to 0$. Photon fusion, on the other hand, is free 
from this effect and allows for the production of slow particles, which improves 
MoEDAL's detection efficiency.

Our study consiseds of both model-independent and model-specific parts. In the 
former, we considered particles from the considered neutrino mass model, but we 
simply parametrised their masses and decay lengths without relating them to 
neutrino constraints nor parameters of the model's Lagrangian. This approach 
allowed us to derive MoEDAL's sensitivity to multiply charged particles in the $(m, c\tau)$ plane, which could be compared with any theoretic scenario 
predicting such particles. This approach led to results summarised in Fig. 
\ref{fig:paper3-model1-reach} and Tab. \ref{tab:paper3-uncoloured} for the 
uncoloured version of the model, and in Fig. \ref{fig:paper3-model1-reach2} 
and Tab. \ref{tab:sum_col} for the model with particles charged under $SU(3)_C$. In the optimistic case, i.e. $c\tau > 100~\rm{m}$ and $N_{\rm sig} \geq 1$, 
MoEDAL could search for $S^{\pm 2}$,  $S^{\pm 3}$, and $S^{\pm 4}$, up to 290 
(600) GeV, 610 (1100) GeV and 960 (1430) for Run 3 (HL-LHC) with the 
integrated luminosity of $L=30~\rm{fb}^{-1}$ ($L=300~\rm{fb}^{-1}$), 
respectively. These values ought to be compared with the constraints from the 
HSCP searches, which are 650 GeV, 780 GeV and 920 GeV for
$S^{\pm 2}$,  $S^{\pm 3}$, and $S^{\pm 4}$, respectively.
Even though the most of the parameter space accessible for Run 3 
MoEDAL has already been excluded by the HSCP searches, for the HL-LHC phase MoEDAL 
can test masses of multiply charged BSM scalars that are still unprobed. In the 
case of the coloured version of the studied neutrino mass model, for very long 
lifetimes $c\tau > 100~\rm{m}$ with $N_{\rm sig} \geq 1$ MoEDAL can search 
for $\tilde S^{\pm 4/3}$,  $\tilde S^{\pm 7/3}$, and $\tilde S^{\pm 10/3}$, up 
to 1050 (1400) GeV, 1250 (1650) GeV, and 1400 (1800) GeV for Run 3 (HL-LHC) with the 
integrated luminosity of $L=30~\rm{fb}^{-1}$ ($L=300~\rm{fb}^{-1}$), 
respectively. Comparing these values to estimated LHC bounds: 
1450 GeV, 1480 GeV and 1510 GeV for $\tilde S^{\pm 4/3}$,  $\tilde S^{\pm 7/3}$, and $\tilde S^{\pm 10/3}$, respectively, we arrive to conclusion that there is little 
hope for discovering multiply charged coloured scalars in Run 3 MoEDAL, but 
for the HL-LHC phase MoEDAL can investigate part of the unconstrained 
parameter space. Nonetheless, we would like to emphasise that results 
provided by MoEDAL are complementary to those by ATLAS and CMS, because 
MoEDAL has a completely different design and detection principle with respect to the 
major LHC experiments.

In the part of the study that was model-specific, we investigated the subspace of the model's
parameters satisfying current constraints on neutrino masses and mixing 
angles. We revealed that for $\lambda_5 \sim 10^{-5}$, which roughly 
corresponded to the longest lifetime of $S^{\pm 4}$ (and $\tilde S^{\pm 10/3}$), the sensitvity of MoEDAL was the largest. The maximum range of scalar triplet 
masses $m_{S_3}$ ($m_{\tilde S_3}$) that MoEDAL could probe was found to be 
approximately the same as for $m_{S^{\pm 4}}$ ($m_{\tilde S^{\pm 10/3}}$) found 
in the model-independent part of the study.

As a final remark, we point out that our analysis relies on a simplified model of 
NTD acceptance, which depends on the velocity and charge of the BSM particle. 
In reality, the incidence angle of the particle also plays a role, however, this 
effect is expected to be negligible since the MoEDAL collaboration decided to 
slightly rearrange the deployment of NTD panels in order to minimise the 
incidence angle.

\section{Discovery prospects for long-lived multiply charged particles at the LHC}\label{sec:paper4}

\subsection{Introduction}

In this project, we investigate in a model-independent way the prospects 
for the detection of multiply charged long-lived particles at the LHC. We study 
BSM particles with spin 0 and 1/2, singlet or triplet under $SU(3)_C$,
with electric charge $Q$ in the range $1 \leq |Q/e| \leq 8$, where $e$ is 
the elementary charge. We consider the production of particle-antiparticle pairs 
propagating through the detector, as well as the formation of positronium/quarkonium-like bound states decaying to two photons. In this study we are interested in particles with lifetimes long enough to be 
considered as detector-stable by ATLAS and CMS, i.e. $c\tau\gtrsim 1~\rm{m}$. We estimate the lower 
mass bounds on new BSM particles that ATLAS, CMS and MoEDAL 
experiments can provide at the end of the Run 3 and HL-LHC 
phases. 

CMS and ATLAS collaborations have searched for multiply charged LLPs \cite{ATLAS:2015hau, ATLAS:2018imb, CMS:2013czn, CMS:2016kce} with electric charge $Q$ in the range $2 \leq |Q/e| \leq 7$. Their 
analyses are based on highly ionising tracks with anomalously large energy 
loss, $dE/dx$, which is a very characteristic signature of highly charged 
LLPs. Up to this point, ATLAS and CMS searches for multiply charged LLPs  
are limited to colourless fermions and include only Drell-Yan production, 
which for a particle with a large electric charge is subdominant with respect 
to the photon fusion production, as demonstrated in Sec. \ref{sec:paper3}.
In this project we recast a search for the large $dE/dx$ \cite{CMS:2016kce} in order to 
derive limits for various multiply charged LLPs with spin 0 and 1/2.
Our study includes photon-induced production modes and takes into 
account two important experimental effects. The first one is the 
underestimation of the transverse momentum, $p_{\rm T}$, of charged tracks 
for multiply charged particles. The reason behind this effect is that the 
transverse momentum is estimated based on the curvature of the track 
$(r \propto Q/p_{\rm T})$ with the assumption $Q=\pm 1 e$, which is true for 
SM and SUSY, but not for the BSM scenario we are interested in. As a 
result, multiply charged particles are assigned lower $p_{\rm T}$ than they really 
have, and part of the signal is removed due to $p_{\rm T}$ cuts. Another effect 
is the rejection of slowly moving particles. Highly charged particles loose 
kinetic energy by multiple electromagnetic interactions inside the detector's 
volume, and may arrive in the muon system after 
the next bunch crossing occurs. Such events are rejected resulting in the 
loss of sensitivity.

Another interesting possibility is the formation of bound states, $\mathcal{B}$, 
by pair-produced BSM particles. These bound states are typically short-lived and decay into SM particles. Out of all possible decay channels, $\mathcal{B} \to \gamma \gamma$ is the most promising, where 
New Physics can be discovered by observing a significant bump in the 
diphoton mass distribution. In this study, we evaluate the event rate of the
$pp \to \mathcal{B} \to \gamma \gamma$ process for different types of 
charged LLPs, we obtain current limits, and estimate the expected sensitivities at Run 3 
and HL-LHC.  

It was demonstrated in the previous three projects that the MoEDAL detector 
may be used to search for electrically charged long-lived particles. 
These projects concentrated on $|Q| \leq 4e$, and led to a 
conclusion that ATLAS/CMS searches for large $dE/dx$ are generally more 
sensitive than MoEDAL, except for some special cases, e.g. 
supersymmetric model in Sec. \ref{sec:paper1}. One of the reasons is 
much lower $(\times 10)$ integrated luminosity available to MoEDAL. 
Additionally, MoEDAL is restricted only to slowly moving particles, with the 
threshold depending on particle's charge, $\beta < 0.15 \cdot |Q/e|$, 
while particles at the LHC are often produced with relativistic velocities, 
thus inaccessible to MoEDAL.

The result might, however, be different for particles with $|Q| \gtrsim 4 e$. For the MoEDAL NTD array, the efficiency of the BSM signal increases with 
the magnitude of the electric charge, while for large $dE/dx$ searches it 
eventually decreases due to the aforementioned difficulties: $p_{\rm T}$
underestimation and late arrival to the muon system. 

In this project we 
calculate, for the first time, MoEDAL's sensitivity to highly charged ($|Q| > 4 e$) LLPs, and compare it to the limits obtained by recasting ATLAS and 
CMS analyses. Moreover, in order to provide a comprehensive overview of 
the prospects for detecting LLPs at the LHC, we estimate lower mass 
bounds by ATLAS, CMS and MoEDAL for long-lived spin-0 and spin-1/2 particles with $|Q| \leq 4 e$, 
including production modes with photons in the initial state. 
The results of this project are 
obtained for Run 3 and HL-LHC, where in the latter case we perform a ``naive scaling'' of luminosity, assuming no significant change in detector 
set-up. This assumption is certainly not satisfied but allows us to provide 
a rough estimate of what progress in the LLP field we can hope for.

\subsection{Model selection}
In this project we consider four types of electrically charged LLPs:
(spin 0 and 1/2)$\times$(colour-singlet and triplet). 
We use the simplified model approach, which allows us to derive results characterised by a high level of model independence. 
Each type of particle has its own simplified model, which is built by introducing a BSM 
scalar field, $\phi$, or a Dirac fermion field, $\psi$. We postulate that 
these fields are singlets under the $SU(2)_L$ gauge group, and
we assign to them hypercharges $Y=Q/e$, where $Q$ is the electric charge of a given particle and $e$ is the elementary charge.

We augment the SM lagrangian, $\mathcal{L}_{\rm SM}$, with these new fields:
\begin{equation}\label{eq:paper4-lagr-scalar}
\mathcal{L} = \mathcal{L}_{\rm SM} + |D_\mu \phi|^2 -m^2 |\phi|^2+ \ldots,
\end{equation}
for scalar particles and
\begin{equation}\label{eq:paper4-lagr-fermion}
\mathcal{L} = \mathcal{L}_{\rm SM} + i \bar{\psi} \slashed{D} \psi -m \bar \psi \psi +\ldots
\end{equation}
for fermions with $\slashed D = \gamma^\mu D_\mu$. When $SU(3)_C$-
singlets are considered, the covariant derivative is $D_\mu \partial_\mu - i g_Y Q B_\mu$, while for $SU(3)_C$-triplets it is $D_\mu \partial_\mu - i g_Y Q B_\mu -i g_C T^a G^a_\mu$, where $B_\mu$ and $G^a_\mu$ are
$U(1)_Y$ and $SU(3)_C$ gauge fields, respectively, and $T^a$ stands for 
$SU(3)_C$ generators. After the electroweak symmetry is spontaneously 
broken, interactions with the $B_\mu$ field should be rewritten in terms of the Z-boson, $Z_\mu$, and photon, $A_\mu$, fields:
\begin{equation}
g_Y Q B_\mu \to e Q A_\mu - e Q \tan \theta_W Z_\mu,
\end{equation}
with $\theta_W$ being the weak mixing angle.

We use the $\phi^{+Q}$ $(\psi^{+Q})$ symbol for scalar (fermionic) particle 
with electric charge $+Q$, and $\phi^{-Q}$ $(\psi^{-Q})$ for the 
corresponding antiparticle. It is useful to introduce $\xi^{+Q}$ and $\xi^{- Q}$ to symbolise particle and antiparticle without specifying the spin. ``$\ldots$'' in Eqs. \eqref{eq:paper4-lagr-scalar} and \eqref{eq:paper4-lagr-fermion} indicate presence of additional interactions leading to decay of 
the considered BSM particle. Their presence is required due to 
cosmology, but we do not specify their form nor details of the decays in 
order to remain as model-independent as possible. Instead, when studying 
the sensitivity of MoEDAL, we treat the lifetime of $\xi^{\pm Q}$ as a free 
parameter, while for ATLAS and CMS we assume that the BSM particles are 
detector-stable, i.e. they typically traverse the entire detector volume without 
decaying.

The descibed models are implemented in {\tt MadGraph5} \cite{Alwall:2011uj} using {\tt FeynRules} \cite{Alloul:2013bka}. {\tt MadGraph5} is used to calculate tree-level 
production cross sections and generate Monte Carlo events.

\subsection{Open production mode}
The long-lived BSM particles described in the previous section might be produced 
in pairs and traverse the detector volume without decaying. This production 
process will from now on be referred to as the \textit{open production mode}. 
The production processes in this case are similar to other scenarios with exotic 
particles, e.g. magnetic monopoles \cite{Baines:2018ltl}.

\myparagraph{Colour-singlet particles}

Colour-singlet BSM particles can be produced via (i) Drell-Yan with $q \bar q$ 
initial state and s-channel exchange of $\gamma^*/Z^*$ boson; (ii) photon fusion 
with $\gamma \gamma$ initial state and t-channel interaction. Additionally, 
scalar particles have a 4-point interaction with $\gamma \gamma$ initial state. 
Tree-level diagrams for the production of colour-singlet particles are shown in Fig. \ref{fig:paper4-diagrams}.
As can be deduced from the diagrams in Fig. \ref{fig:paper4-diagrams}, the production rate for the Drell-Yan process initiated by 
$q \bar q$ is proportional to $Q^2$, while for the processes initiated by $\gamma \gamma$, it is proportional to $Q^4$. The 
latter type of process is typically suppressed by the PDF, however, it can be significant for larger charges \footnote{One might argue that 
the processes with a Z boson in the initial state should also be taken into account. However, we have checked that this 
contribution is small $\lesssim 15\%$ compared to photon fusion. For further information see Appendix A in \cite{Altakach:2022hgn}.}.

\begin{figure}[htb]
\centering
\includegraphics[width=0.36\textwidth]{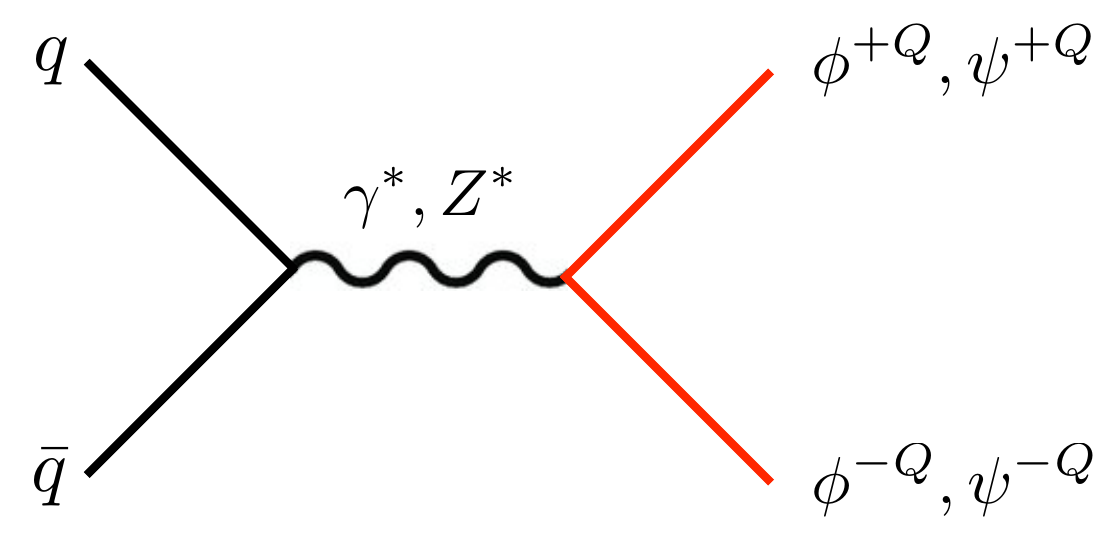}
\hspace{0.2cm}
\includegraphics[width=0.3\textwidth]{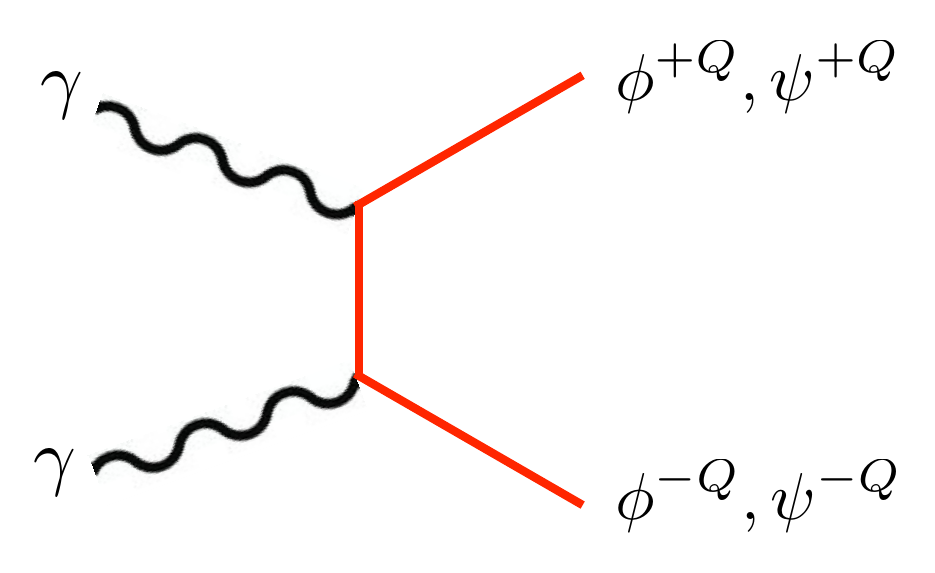}
\hspace{0.1cm}
\includegraphics[width=0.26\textwidth]{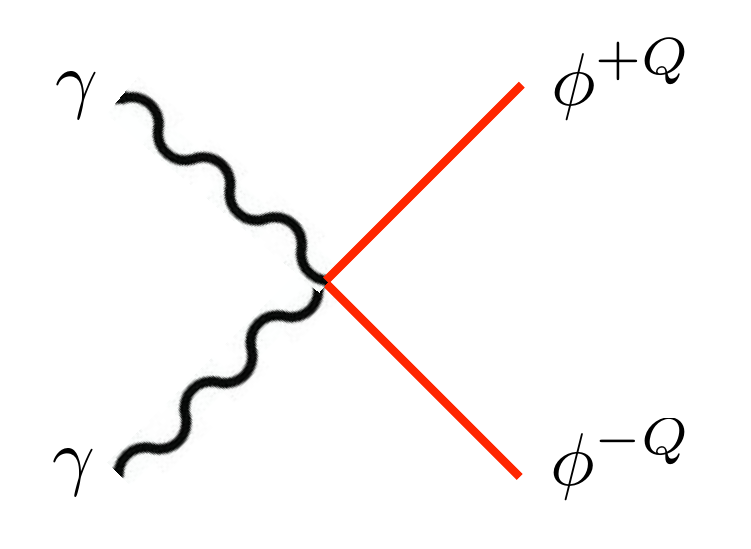}
\caption{\label{fig:paper4-diagrams} \small Feynman diagrams for colour-singlet open production modes.}
\end{figure}

Leading order cross sections for the open production mode are depicted 
in Fig. \ref{fig:paper4-xs_cless}, where the panel on the left (right) corresponds to colour-singlet 
scalars (fermions). Cross sections are calculated with {\tt MadGraph5} \cite{Alwall:2011uj} 
using the \\{\tt LUXqed17\_plus\_PDF4LHC15\_nnlo\_100} PDF \cite{Manohar:2016nzj,Manohar:2017eqh}, which provides 
an improved calculation of the photon PDF. One can see in Fig. \ref{fig:paper4-xs_cless} that for 
$m=100~\rm{GeV}$ the production cross section spans over two orders of 
magnitude from $|Q|=1e$ to $|Q|=8e$. The impact of the magnitude of the electric charge on the production cross section is 
more evident for scalars and heavy masses because for these cases the photon fusion production channel is relatively more 
important than the Drell-Yan process. For example, for scalars with $m=2~\rm{TeV}$ and $Q=8e$, the production cross section 
is about three orders of magnitude larger than for $Q=1e$. We expect that for $|Q| \gg 1e$, the electromagnetic higher order 
corrections are considerable, however, NLO cross sections for multiply charged particles have not been well studied and including 
these effects is beyond the scope of this work.

\begin{figure}[tbh]
\centering
\includegraphics[width=0.49\textwidth]{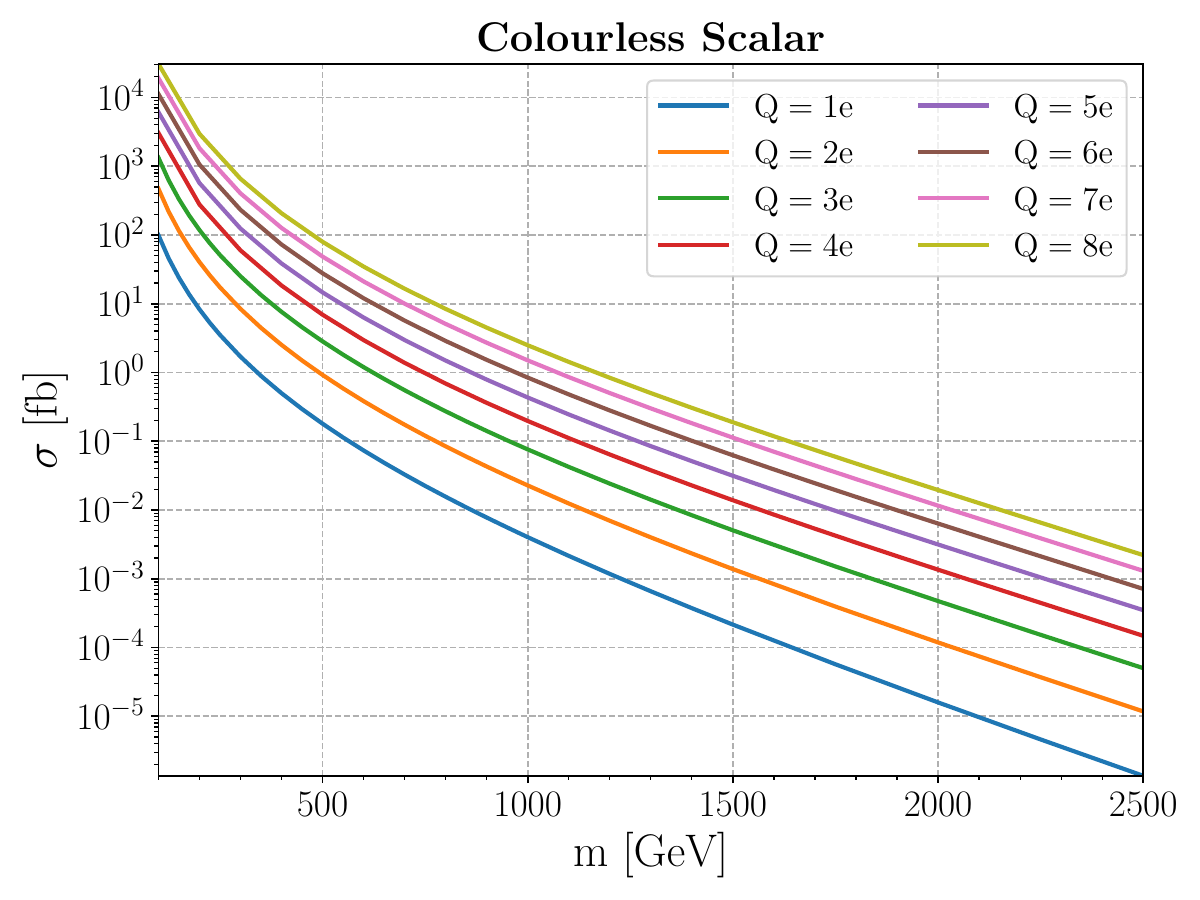}
\includegraphics[width=0.49\textwidth]{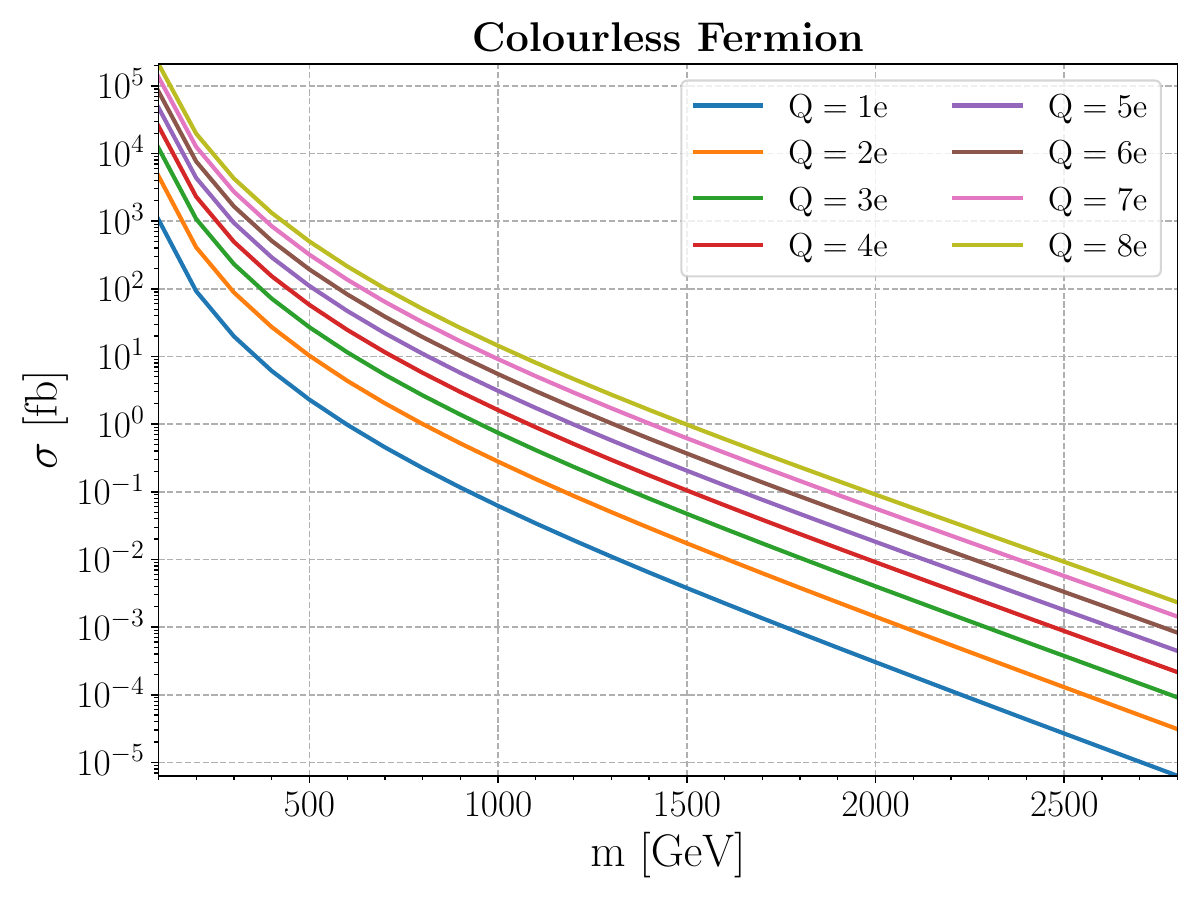}
\caption{\label{fig:paper4-xs_cless}  \small Leading order cross section for pair-production production of scalar (left) and 
fermion (right) particles. Curves with different colours correspond to particles with different magnitudes of the electric charge, $Q$, from $\pm 1e$ to $\pm 8e$.
}
\end{figure}

In order to better understand the impact of different production processes, in Fig. \ref{fig:paper4-frac} we plot the relative contribution of 
photon fusion and Drell-Yan processes to the total cross section of colour-singlet scalars (left) and fermions (right), assuming $m=1~\rm{TeV}$. One can see that for singly charged particles, $|Q|=1e$, the photon fusion is subdominant because of the 
suppression of the photon PDF. However, for larger charges photon fusion becomes relatively more important since $\hat \sigma_{\gamma \gamma}/\hat\sigma_{q \bar q} \propto Q^2$, and becomes dominant for $|Q/e| > 3$ (5) for scalars (fermions). 
The relative impact of the photon fusion grows more rapidly for scalars than fermions, due to the p-wave suppression in the 
Drell-Yan channel. Drell-Yan process for pair production is an s-channel exchange of a spin-1 gauge boson $\gamma/Z$ from the 
initial $q \bar q$ state. In order for a non-zero cross section, the final state configuration is required to have a non-zero angular 
momentum. In the case of scalars, it leads to velocity suppression of the production cross section. This effect is absent for scalars 
produced via photon fusion and spin-1/2 fermions. 

\begin{figure}[htb]
\centering
\includegraphics[width=0.49\textwidth]{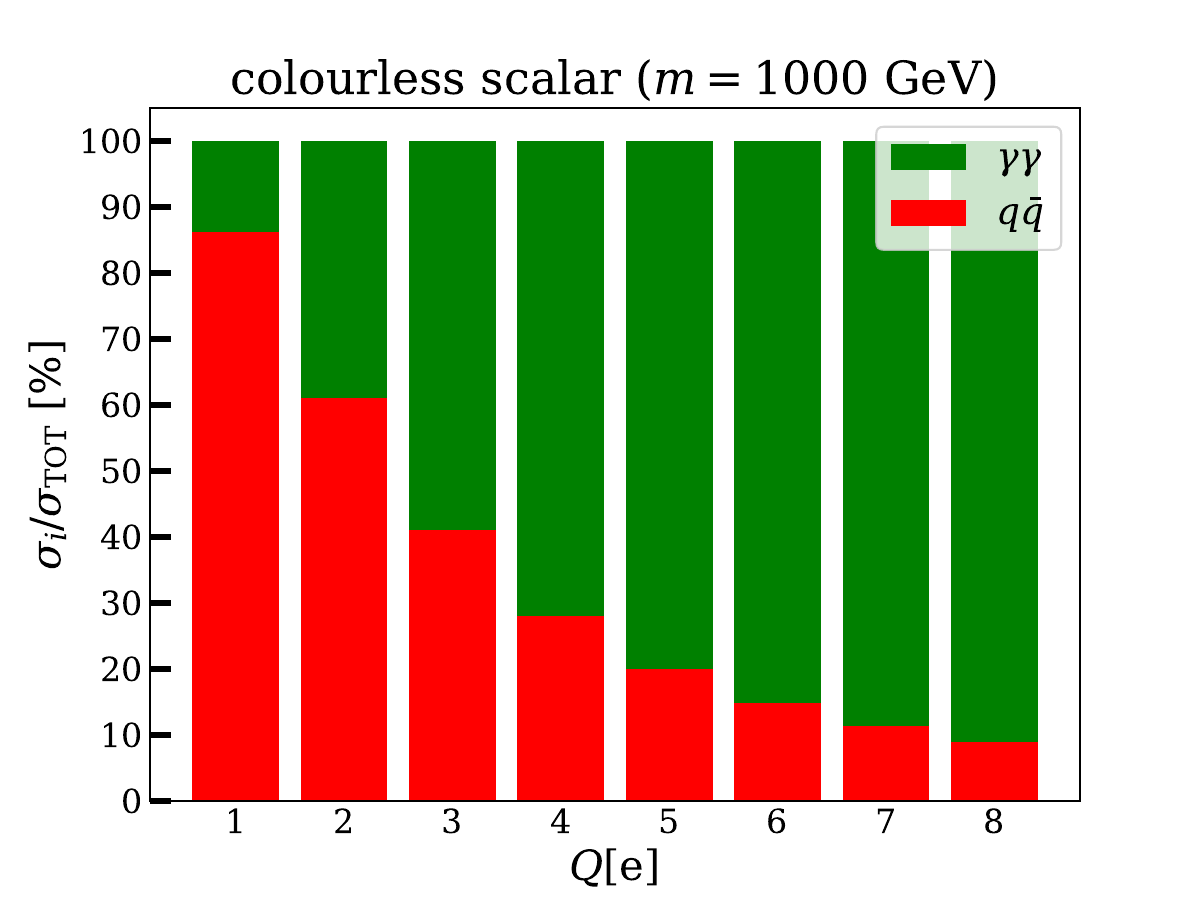}
\includegraphics[width=0.49\textwidth]{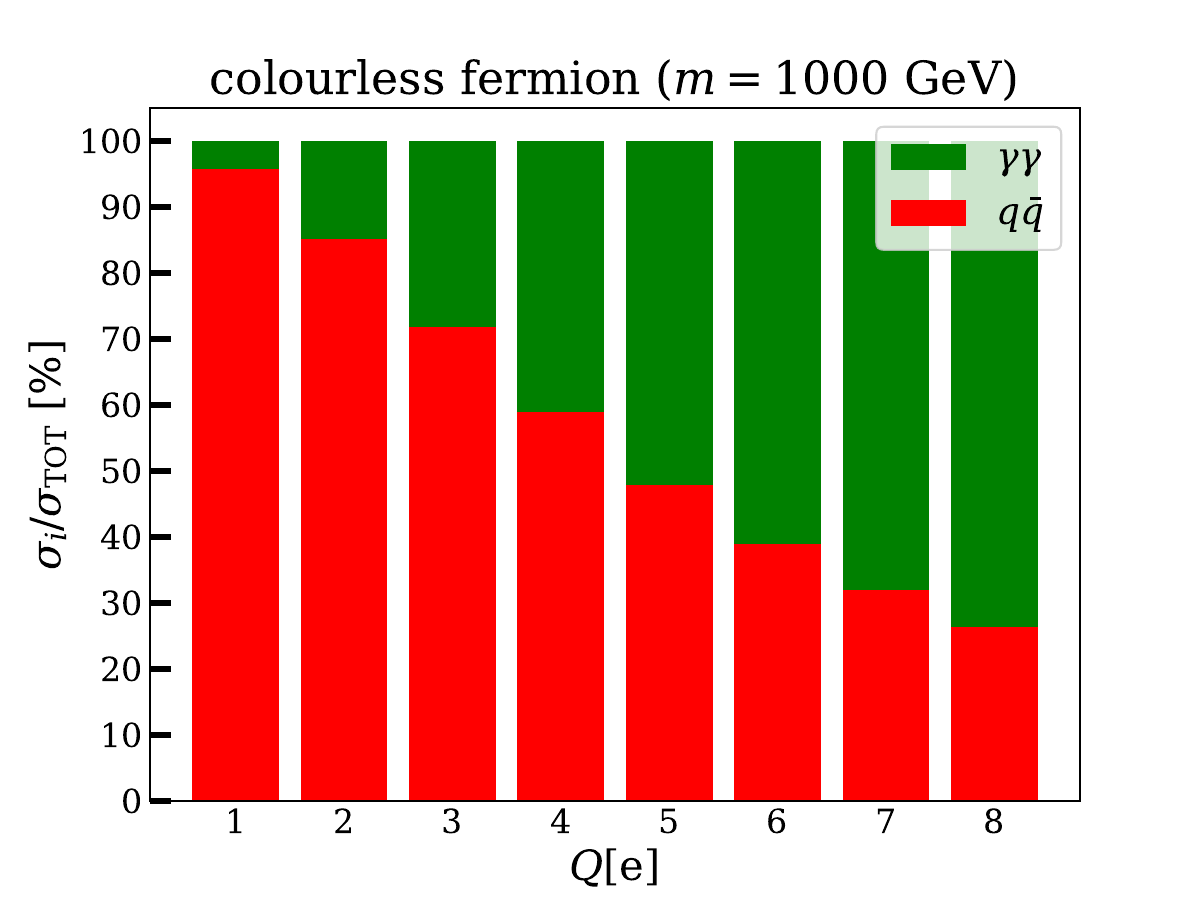}
\caption{\small\label{fig:paper4-frac}
Relative contribution to the total production cross section of multiply charged scalars (left) and fermions (right) coming from $q \bar q$ (red) and $\gamma \gamma$ (green) initial states. All particles were assumed to have $m=1~\rm{TeV}$.
}
\end{figure}

\myparagraph{Colour-triplet particles and a hadronisation model}
In the case of colour-triplet particles, the open production mode receives contributions from the following set of processes:
(i) Drel-Yan process with a $q \bar q$ initial 
state and an s-channel exchange of a spin-1 boson $\gamma^*/Z^*/g^*$; 
(ii) gluon fusion; 
(iii) photon-gluon fusion with the
mixed QED and QCD interaction via t-channel; 
(iv) photon fusion with the t-channel QED interaction and $\gamma \gamma$ in the initial state. 
In the case of scalar particles, there is also a 4-point QED vertex with $\gamma \gamma$ in the initial state.

In Fig. \ref{fig:paper4-xs_c} we depict the leading order open mode production ($pp \to \xi^+ \xi^-$) cross section for colour-triplet scalars 
(left) and fermions (right), as a function of mass of the produced BSM particles. By comparing Fig. \ref{fig:paper4-xs_c} with the analogous plot for 
colour-singlet particles in Fig. \ref{fig:paper4-xs_cless}, we observe that the impact of the electric charge magnitude $|Q|$ on the 
cross section is weaker, particularly for small masses. The reason behind this is that production processes with the QCD 
interactions, i.e. DY with gluon exchange and $gg/\gamma g$ fusions, constitute a large fraction of the total production cross 
section for all values of $Q$, as visible from Fig. \ref{fig:paper4-cfrac}. The dependence of these processes on the value of $Q$ is 
weak or none. The strong interaction is relatively more important for smaller masses, because of the gluon PDF enhancement for 
small momentum fraction, $x$. As already mentioned, electromagnetic higher-order corrections for multiply charged particles are 
not available to date. In order to remain consistent, we use the leading-order cross section calculations also for the QCD 
processes.

If the lifetime of $SU(3)_C$-triplet particles is sufficiently long, they will hadronise into colour-neutral states by drawing quark-antiquark pairs from the vacuum. The electric charge of a colour-neutral hadron state, $\tilde Q$, will be shifted with respect to 
the charge $Q$ of the involved $SU(3)_C$-triplet component by $\Delta Q$ due to extra quarks: $\tilde Q = Q + \Delta Q$. The 
shift of the electric charge is an important effect resulting in a change in detector's sensitivity. We 
estimate it by introducing a simple hadronisation model, similar to the one considered in Sec. \ref{sec:paper3}. The model has only one free parameter, $k$, which is the probability to form a mesonic 
final state. Among the possible mesonic (baryonic) states, the probability is distributed equally. Possible 
colour-singlet hadronic states are listed in Tab. \ref{tab:paper4-had_scalar} and \ref{tab:paper4-had_fermion} for scalar ($\phi^{+Q}$) and fermion ($\psi^{+Q}$) particles, respectively, together with the probability of forming a given state and the 
corresponding charge shift $\Delta Q$. We assume that hadrons will be formed with the first generation of SM quarks only.
When estimating the sensitivity of MoEDAL, ATLAS and CMS, the free parameter $k$ is varied 
between 0.3 and 0.7 in order to estimate the uncertainty of the hadronisation model.

\begin{figure}[!t]
\centering
\includegraphics[width=0.49\textwidth]{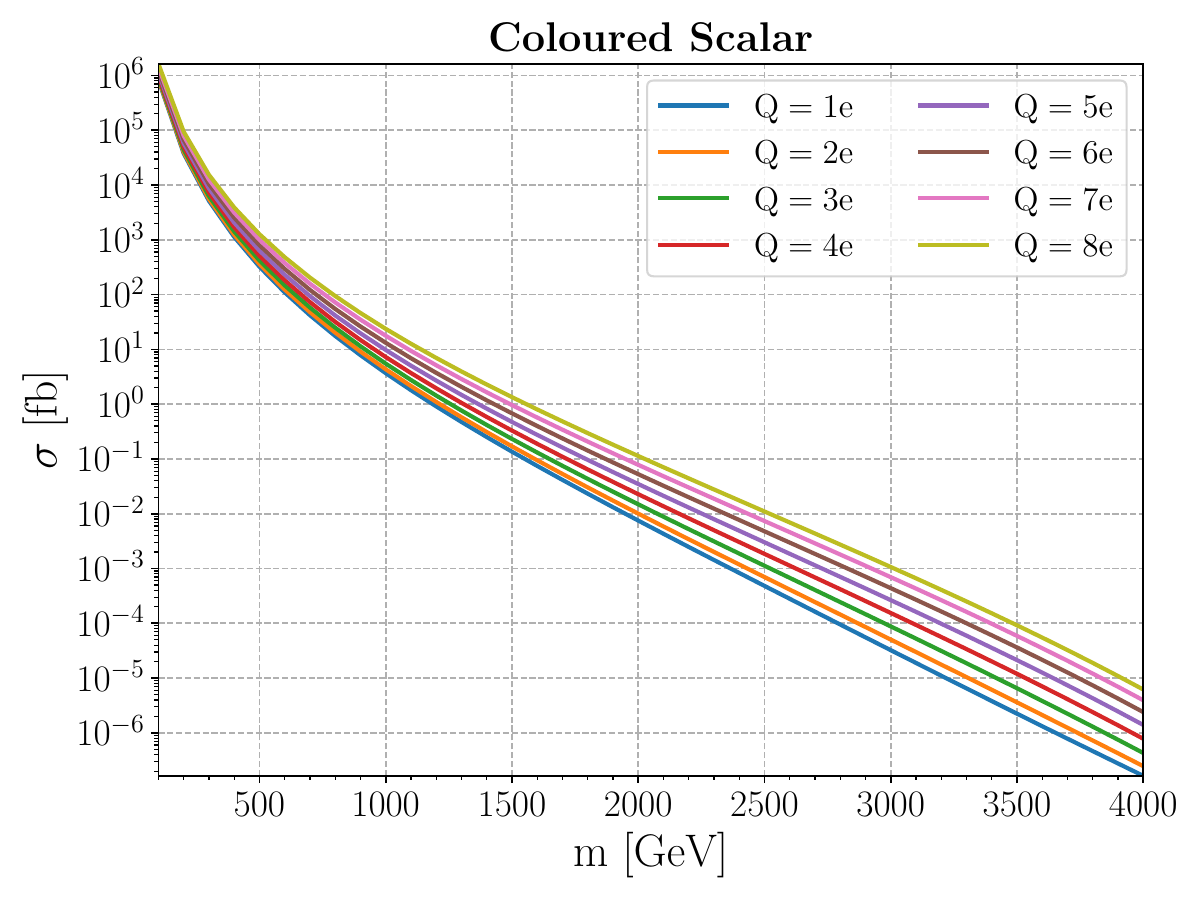}
\includegraphics[width=0.49\textwidth]{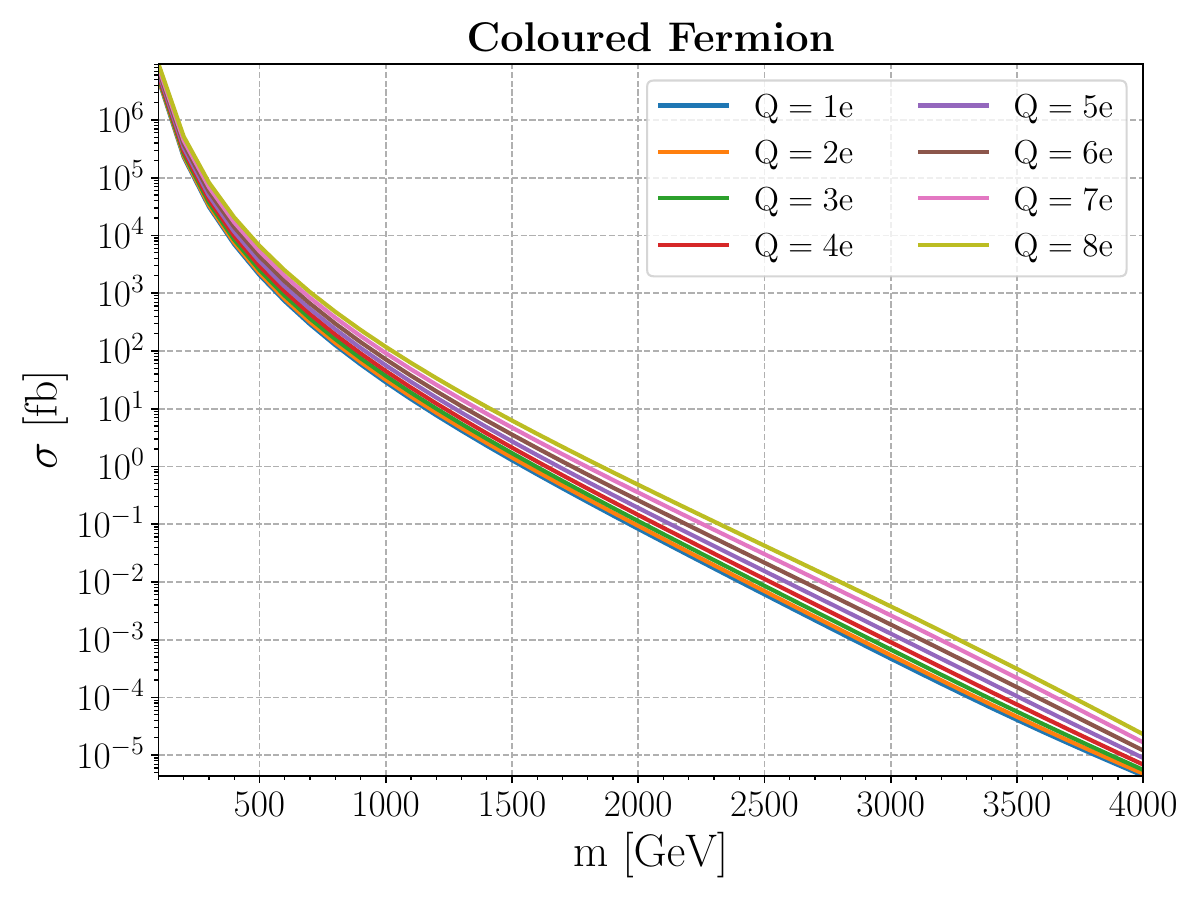}
\caption{\label{fig:paper4-xs_c}  
\small LO cross section for open mode production ($pp \to \xi^+ \xi^-$)  of colour-triplet scalars (left) and fermions (right).
Curves with different colours correspond to particles with different magnitudes of the electric charge, $Q$, 
from $\pm 1e$ to $\pm 8e$.}
\end{figure}

\begin{figure}[!t]
\centering
\includegraphics[width=0.49\textwidth]{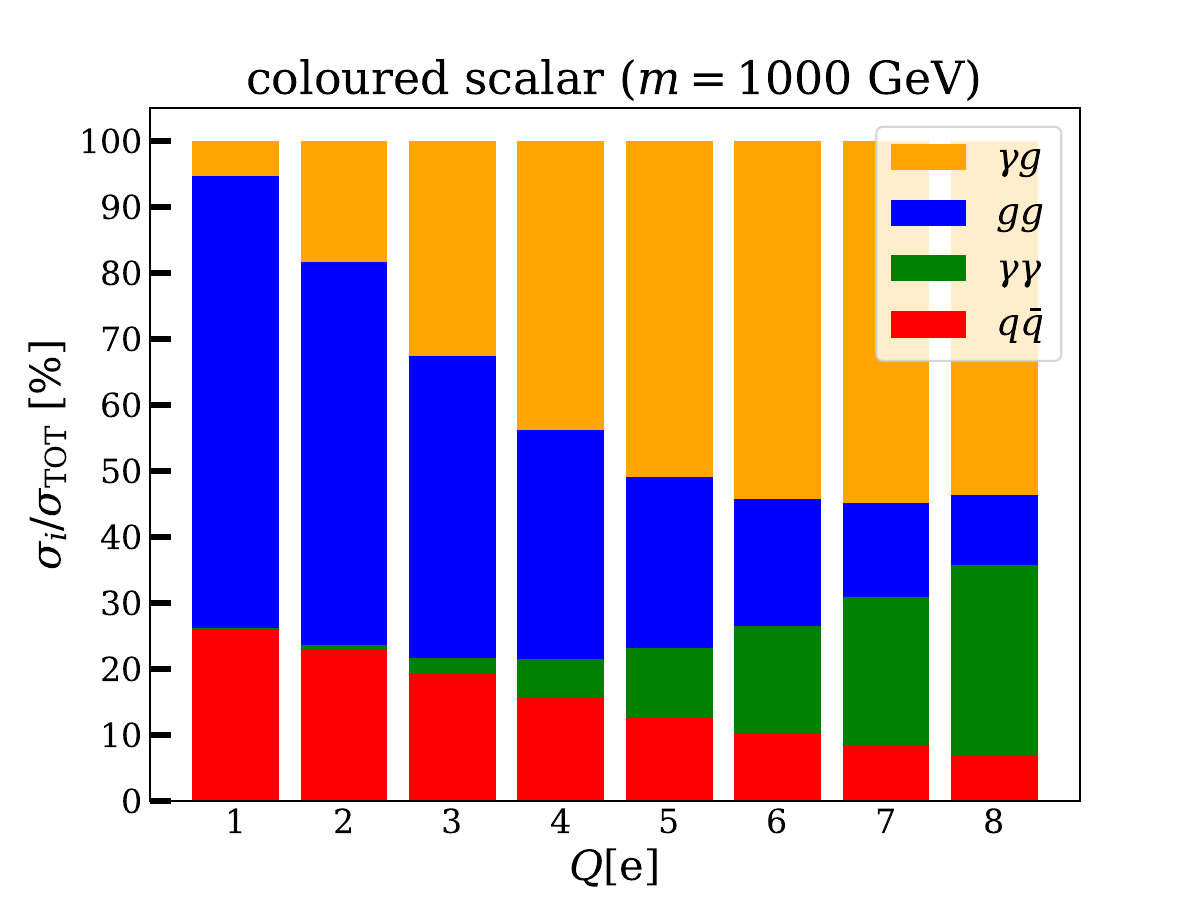}
\includegraphics[width=0.49\textwidth]{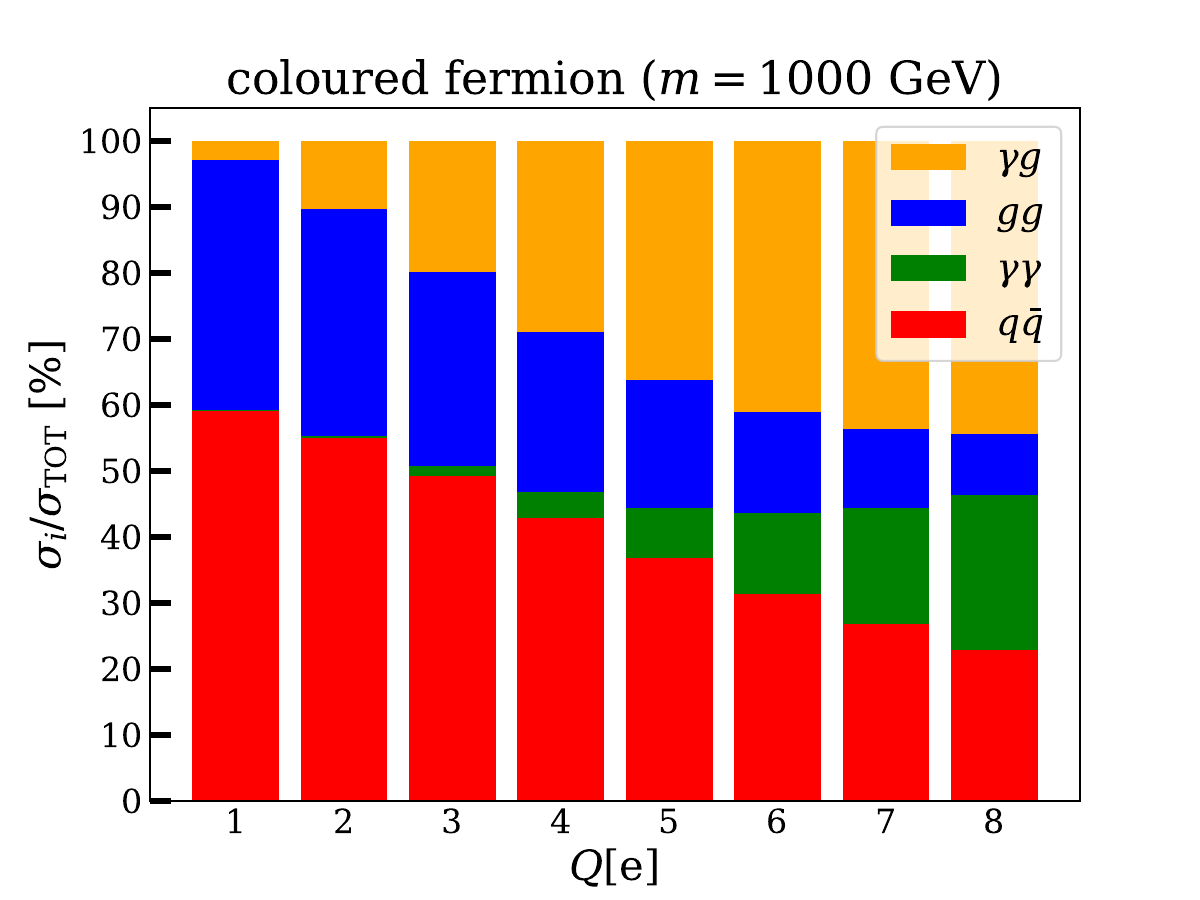}
\caption{\small\label{fig:paper4-cfrac} Relative contributions of different initial states, 
$q \bar q$ (red),
$\gamma \gamma$ (green),
$gg$ (yellow),
$\gamma g$ (blue),
to the total production cross section of $SU(3)_C$-triplet scalars (left)
and fermions (right), assuming $ m= 1$ TeV.} 
\end{figure}

\begin{table}[!h]
\small
\centering
\caption{\small The hadronisation model for colour-triplet scalars ($\phi^{+Q}$) explained. Possible spin-1/2, spin-0 and spin-1 colour-singlet states are listed in the left, centre and right tables, respectively. Each state has a probability of creation and charge shift assigned to it. The only free parameter of the model, $k$, being the probability to form a mesonic state, is varied between 0.3 and 0.7 when estimating the sensitivity of the LHC experiments to coloured LLPs.
}
\label{tab:paper4-had_scalar}
\begin{tabular}{c|c|c}

\multicolumn{3}{c}{spin 1/2 mesons}\\
\hline \hline
state & $\Delta Q/e$ & $p$ \\
\hline
$\phi^{+Q} + \bar u_{L/R}$ & $- \frac{2}{3}$ & $\frac{k}{2}$ \\
$\phi^{+Q} + \bar d_{L/R}$  & $+ \frac{1}{3}$ & $\frac{k}{2}$ \\ 

\end{tabular}
\hspace{2mm}
\begin{tabular}{c|c|c}

\multicolumn{3}{c}{spin 0 baryons}\\
\hline \hline
state & $\Delta Q/e$ & $p$ \\
\hline
$\phi^{+Q} + u_L u_R$ & $+ \frac{4}{3}$ & $\frac{1-k}{6}$ \\
$\phi^{+Q} + d_L d_R$  & $- \frac{2}{3}$ & $\frac{1-k}{6}$ \\ 
$\phi^{+Q} + u_L d_R$  & $+ \frac{1}{3}$ & $\frac{1-k}{6}$ \\ 
$\phi^{+Q} + d_L u_R$  & $+ \frac{1}{3}$ & $\frac{1-k}{6}$ \\ 

\end{tabular}
\hspace{2mm}
\begin{tabular}{c|c|c}

\multicolumn{3}{c}{\small spin 1 baryons}\\
\hline \hline
state & $\Delta Q/e$ & $p$ \\
\hline
$\phi^{+Q} + u_L d_L$ &$+ \frac{1}{3}$ & $\frac{1-k}{6}$ \\
$\phi^{+Q} + u_R d_R$  & $+ \frac{1}{3}$ & $\frac{1-k}{6}$ \\ 
\end{tabular}
\end{table}

\begin{table}[!h]
\small
\centering
\caption{\small The hadronisation model for colour-triplet fermions ($\psi^{+Q}$) explained. Possible spin-1/2, spin-0 and spin-1 colour-singlet states are listed in the left, centre and right tables, respectively. Each state has a probability of creation and charge shift assigned to it. The only free parameter of the model, $k$, being the probability to form a mesonic state, is varied between 0.3 and 0.7 when estimating the sensitivity of the LHC experiments to coloured LLPs.
}
\label{tab:paper4-had_fermion}
\begin{tabular}{c|c|c}
\multicolumn{3}{c}{\small spin 0 mesons}\\
\hline\hline
state & $\Delta Q/e$ & $p$ \\
\hline
$\psi^{+Q} + \bar u_{L}$  & $- \frac{2}{3}$ & $\frac{k}{4}$ \\
$\psi^{+Q} + \bar d_{L}$  & $+ \frac{1}{3}$ & $\frac{k}{4}$ \\ 

\end{tabular}
\hspace{3mm}
\begin{tabular}{c|c|c}
\multicolumn{3}{c}{\small spin 1 mesons}\\
\hline\hline
state & $\Delta Q/e$ & $p$ \\
\hline
$\psi^{+Q} + \bar u_{R}$  & $- \frac{2}{3}$ & $\frac{k}{4}$ \\
$\psi^{+Q} + \bar d_{R}$  & $+ \frac{1}{3}$ & $\frac{k}{4}$ \\ 

\end{tabular}
\hspace{3mm}
\begin{tabular}{c|c|c}
\multicolumn{3}{c}{\small spin $\frac{1}{2}$ baryons}\\
\hline\hline
state & $\Delta Q/e$ & $p$ \\
\hline
$\psi^{+Q} + u_L u_R$ & $+ \frac{4}{3}$ & $\frac{1-k}{5}$ \\
$\psi^{+Q} + d_L d_R$  & $- \frac{2}{3}$ & $\frac{1-k}{5}$ \\ 
$\psi^{+Q} + u_L d_R$  & $+ \frac{1}{3}$ & $\frac{1-k}{5}$ \\ 
$\psi^{+Q} + d_L u_R$  & $+ \frac{1}{3}$ & $\frac{1-k}{5}$ \\ 
$\psi^{+Q} + u_L d_L$  & $+ \frac{1}{3}$ & $\frac{1-k}{5}$ \\ 
\end{tabular}
\end{table}

\subsection{Detecting open production at ATLAS and CMS}

ATLAS and CMS have conducted searches for charged long-lived particles, out of which the most relevant 
for our study are analyses targeting the large $dE/dx$ channel, which is characterised by anomalously 
high deposition of energy in the inner tracker. ATLAS searched for highly ionising particles in \cite{ATLAS:2015hau}, 
where the full Run 1 data (8 TeV, $L=20.3~\rm{fb}^{-1}$) was analysed, and the results were interpreted 
for spin-1/2 fermions with $2 \leq |Q/e| \leq 6$. The search was updated for $L=36.1~\rm{fb}^{-1}$ 13 
TeV data \cite{ATLAS:2018imb} and interpreted for long-lived fermions with $2 \leq |Q/e| \leq 7$. CMS collaboration 
studied a combined dataset of 7 and 8 TeV $pp$ collisions with integrated luminosity of 5.0 and 18.8 fb$^{-1}$, respectively \cite{CMS:2013czn}. The most recent CMS result was obtained using only a small $L=2.5~\rm{fb}^{-1}$ 13 TeV data set \cite{CMS:2016kce}. The results of the CMS analyses were interpreted for supersymmetric 
particles and charged fermions with $|Q|<2e$.

For particles with $|Q| \geq 2 e$, the results of the aforementioned ATLAS and CMS studies were 
interpreted only to colourless spin-1/2 fermions. What is more, these searches included only Drell-Yan 
production channel, which has been revealed in the previous section to be subdominant for multiply 
charged particles. In Ref. \cite{Jager:2018ecz}, results of the CMS analysis \cite{CMS:2013czn} were recast for 
coloured scalars and fermions, moreover, vector boson fusion processes ($\gamma \gamma$, $gg$, $\gamma g$) were included. In this project, we follow the approach in \cite{Jager:2018ecz} and extend it 
by including colourless scalar particles and estimation of projected sensitivities for 13 TeV Run 3 and HL-LHC. 

Our recasting procedure follows the approach in \cite{Jager:2018ecz} and relies on the CMS analyses 
\cite{CMS:2013czn, CMS:2016kce}. It is based on the \textit{Tracker+Time of Flight (TOF)} selection. 
Signal regions are constructed with several kinematic cuts, e.g. $p_{\rm T}$ and $\eta$ of charged tracks, 
along with more subtle conditions, e.g. requirement on the number of hits in the pixel detector and the 
uncertainty of the $1/\beta$ measurement. The former class of data selections is sensitive to particle's 
spin and production mechanism, and its efficiency might be predicted with Monte Carlo simulation. The 
latter class of selections, on the other hand, depends highly on the mass and charge of the BSM 
particle. The signal efficiency, $\epsilon$, is obtained in a two-step procedure. First, the efficiency of the 
online (muon trigger) selection, $\epsilon_{\rm on}$, is estimated using MC simulations. Then, the offline 
selection efficiency, $\epsilon_{\rm off}$, is based on the cut-flow tables provided in 
\cite{Veeraraghavan:2013rqa}. $\epsilon_{\rm off}$ is calculated as a product of relative efficiencies of 
selection cuts with respect to the previous ones in the cut-flow table:
$\epsilon_{\rm off}(m, Q) = \prod_i \epsilon_i(m, Q)$ with $\epsilon_i \equiv$(number of events passing
all cuts up to $i$)/(number of events passing all cuts up to $i-1$)\footnote{The relative efficiency of the first 
offline cut, $\epsilon_1$, is defined with respect to the online selection.}. The relative efficiencies of the kinematical cuts are estimated with Monte Carlo simulations, while for the non-kinematical selection cuts 
the efficiencies are taken directly from \cite{Veeraraghavan:2013rqa}. This is justified because the non-kinematical selections are not very sensitive neither to the spin of the BSM particle, nor the production 
mechanism, nor the collision energy. The total efficiency is given by $\epsilon=\epsilon_{\rm on}\cdot \epsilon_{\rm off}$.

Several important effects need to be taken into account when estimating detector efficiencies. First of all, 
the ATLAS and CMS experiments underestimate transverse momentum, $p_{\rm T}$, for multiply charged 
particles. The reason is that $p_{\rm T}$ is inferred from the curvature of a track in a magnetic field, which is 
proportional to $p_{\rm T}/|Q|$, assuming $|Q|=1e$. For a particle with charge $Q=Ze$, its reconstructed 
momentum will be $Z$ times smaller than the real value, resulting in worse sensitivity to higher 
charges, e.g. one of the online selection cuts demands the measured $p_{\rm T}$ to be larger than 50 GeV 
\cite{CMS:2016kce}, which can be expressed in terms of the true transverse momentum,  $p_{\rm T}^{\rm true}$, as $p_{\rm T}^{\rm true} > 50~\rm{GeV} \times |Q/e|$. In the extreme, when $|Q|=8e$, the condition 
is $p_{\rm T}^{\rm true} > 400~\rm{GeV}$, which significantly diminishes the sensitivity of the analysis to BSM 
particles with large electric charges.

The large $dE/dx$ analysis relies on the muon trigger with an implicit assumption regarding the time of 
flight of the highly ionising particles. The tracks in the inner pixel detector are expected to match the 
signal in the muon system within the same bunch crossing. However, heavy highly ionising particles 
lose a significant fraction of their kinetic energy via electromagnetic interactions in the detector volume. 
As a consequence, they slow down and are registered in the muon system after the subsequent bunch 
crossing. Such events are rejected. In our analysis we follow the approach described in \cite{Jager:2018ecz} in order to include this effect. We denote by $x_{\rm trigger}$ a minimum distance that a 
particle has to travel in order to be detected at the MS, as a function of pseudorapidity $\eta$. Next, we 
calculate for each simulated particle the time of flight, $t_{\rm TOF}$, needed to travel to the distance 
$x_{\rm trigger}$. The exact procedure is provided in the appendix of \cite{Jager:2018ecz}. It is based on 
the information about the detector's geometry, materials used, and the Bethe-Bloch formula \cite{ParticleDataGroup:2020ssz}. 
Consecutive bunch crossings at the LHC occur every 25 ns, therefore we require $t_{\rm TOF} - x_{\rm trigger}/c < 25~\rm{ns}$ to satisfy the muon trigger selection. Charged particles with $\eta < 1.6$ can 
be detected using the \textit{resisitve plate chamber (RPC) muon trigger}, which has a relaxed threshold 
of 50 ns. We would like to note that the energy loss in the Bethe-Bloch formula depends quadratically on 
the electric charge, $\propto Q^2$, therefore we expect the effect to be more relevant for higher charges.

To sum up, the muon trigger (online) event selection requires at least one candidate particle satisfying:
\begin{equation}
p_{\rm T} \,>\, 50 \cdot |Q/e| ~{\rm GeV} \,, 
~~~
|\eta|  \,<\,  2.1 \,,
\nonumber
\end{equation}
\begin{equation}
t_{\rm TOF} - \frac{x_{\rm trigger}}{c}
\, < \,
\left\{
\begin{array}{ll}
50 ~{\rm ns} & \cdots~ |\eta| < 1.6  \\
25 ~{\rm ns} & \cdots~ 1.6 \le |\eta| < 2.1
\end{array}
\right.
\end{equation}
Next, we use Monte Carlo simulation to estimate and replace relative efficiencies of the kinematical cuts in the offline selection:
\begin{eqnarray}
p_{\rm T} &>& 65 \cdot |Q/e| ~{\rm GeV} \,, \nonumber \\
\frac{1}{\beta_{\rm MS}} & \equiv &  \frac{c \cdot t_{\rm TOF}}{x_{\rm trigger}} 
~ > ~ 1.25\,,
\end{eqnarray}
with ${\beta_{\rm MS}}$ being the velocity measured in the MS. As explained 
earlier, for relative efficiencies of the non-kinematical cuts we simply take the 
relevant values from cut-flow tables in \cite{Veeraraghavan:2013rqa}. Values 
in these tables are provided only up to $m=1~\rm{TeV}$, however, all non-kinematical cuts reach plateaus for $m\lesssim 1~\rm{TeV}$. Therefore, if the mass of the BSM particle is larger than 1 TeV, we assume its relative efficiency is approximately the same as for $m=1~\rm{TeV}$.

The total signal efficiency, $\epsilon =\epsilon_{\rm on} \cdot \epsilon_{\rm off}$, is shown in Fig. \ref{fig:paper4-eff} as a function of electric charge $Q$. The left 
(right) panels in Fig. \ref{fig:paper4-eff} correspond to scalar (fermionic) particles, and the 
upper (lower) panels are for colour-singlets (triplets). Distinct colours of the 
curves represent LLPs with different masses. A general tendency can be 
observed in Fig. \ref{fig:paper4-eff}. For a fixed mass, increasing the electric charge from 1$e$ to 2$e$ results in higher efficiency, however, for large charges the 
efficiency diminishes. The overall peak is around $2e \lesssim |Q| \lesssim 3e$. The reason behind the initial growth of the efficiency comes from larger 
energy loss in the pixel detector, leading to a distinctive $dE/dx$ signature. 
However, when the charge is large, i.e. $|Q| \gtrsim 4e$, the $p_{\rm T}$ cut ($p_{\rm T} > 65\times |Q/e|~\rm{GeV}$) requires particles to have large velocities, 
which prevents them from satisfying the $1/\beta_{\rm MS} > 1.25$ 
selection. For a fixed momentum and bigger mass, the velocity is lower, 
which explains why the signal efficiency in Fig. \ref{fig:paper4-eff}, as a function of the electric charge, 
decreases less for larger masses.
On the other hand, for particles other than colourless scalar, at $m=500~\rm{GeV}$ the highest 
efficiency is for $Q=\pm 1 e$. This effect is caused by the reduction of the efficiency of the charge track isolation criterium for heavier particles. In the case of colourless scalars, the typical velocity is larger than for other types of particles. It causes the loss of signal efficiency due to $1/\beta_{MS} > 1.25$ selection.

\begin{figure}[t!]
    \centering
    \includegraphics[width=0.49\textwidth]{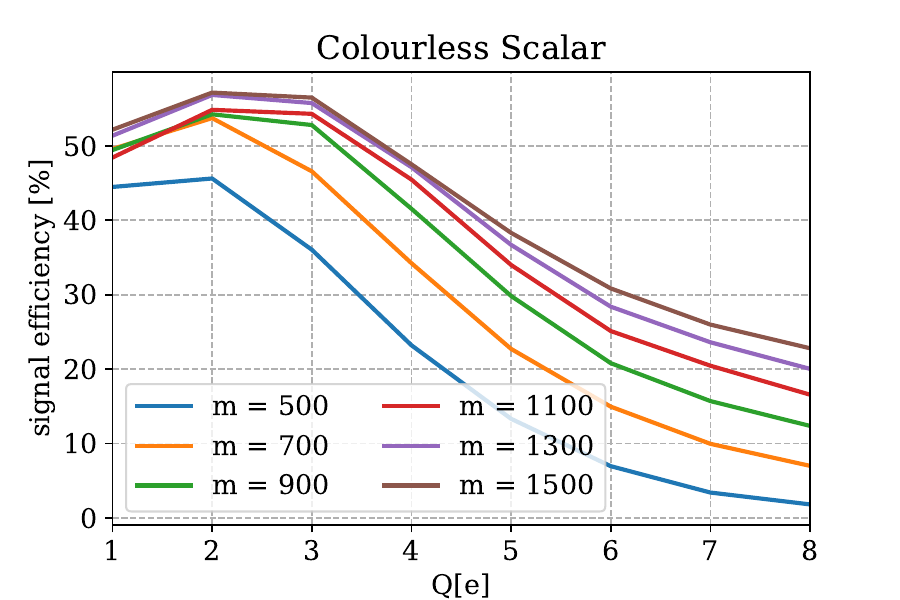}
    \includegraphics[width=0.49\textwidth]{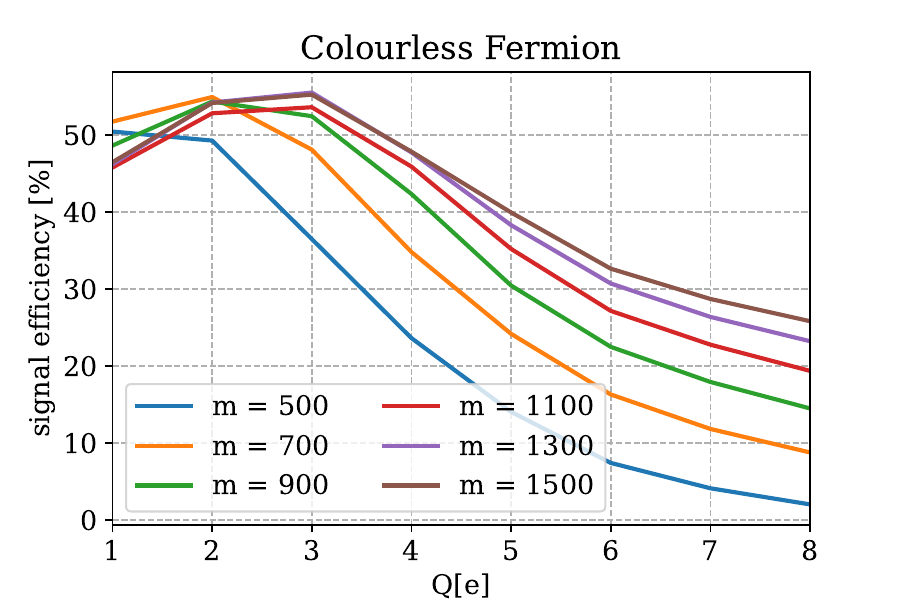}
    \includegraphics[width=0.49\textwidth]{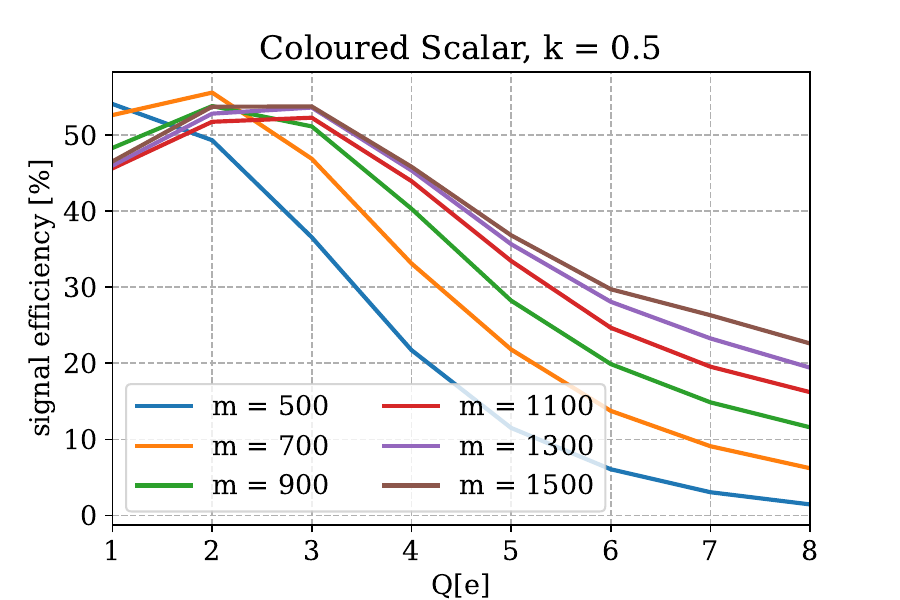}
    \includegraphics[width=0.49\textwidth]{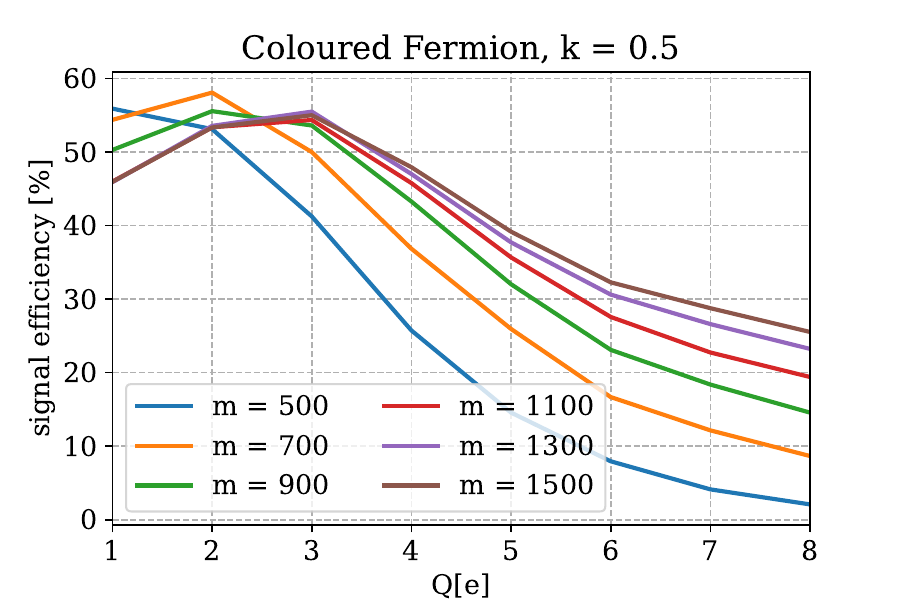}
    \caption{\small Signal efficiency as a function of BSM particle's charge $Q$, for scalar (left) and fermionic (right) particles. The panels in the bottom (top) correspond to colour-triplet (singlet) particles. Curves with different colours represent different masses.}
    \label{fig:paper4-eff}
\end{figure}

The effective cross section, $\sigma_{\rm eff} \equiv \sigma \cdot \epsilon_{\rm eff}$, is shown in Fig. 
\ref{fig:paper4-xs_eff}. The left (right) panels correspond to scalar (fermionic) particles, bottom (top) 
plots are for colour-triplets (singlets). As evident from Fig. \ref{fig:paper4-xs_eff}, the effective cross 
section of colourless particles is significantly enhanced for larger electric charges. This is because in the 
case of colourless particles, the photon fusion production process is crucial, and its production rate increases with $Q^4$.

When it comes to $SU(3)_C$-triplet particles, we can see in Fig. \ref{fig:paper4-xs_eff} that their effective 
cross section also grows for larger electric charges, but the effect is much milder than for colour-singlet 
particles. The reason is that photon fusion is never as significant as other production modes (cf. Fig. 
\ref{fig:paper4-cfrac}), which either do not depend on electric charge (gluon-gluon fusion and gluon-
mediated Drell-Yan) or they depend only quadratically (photon-gluon fusion and DY mediated by 
electroweak gauge bosons). What is interesting, for $m \lesssim 800~\rm{GeV}$ the effective cross 
section \textit{drops} as a function of the electric charge. This is because for larger charges ($|Q| \gtrsim 3 e$) the signal efficiency drops, and this effect is stronger for lighter particles, as can be seen in Fig. 
\ref{fig:paper4-eff}.

The effective cross sections in Fig. \ref{fig:paper4-xs_eff} are superimposed with the (projected) model-independent upper limits ($\sigma_{\rm eff}^{\rm UL}$). 
The latest 13 TeV $L=2.5~\rm{fb}^{-1}$ CMS analysis \cite{CMS:2016kce} was re-interpreted in order to
include photon-induced production modes for all four types of the considered particles.
The obtained limits are shown in Fig. \ref{fig:paper4-xs_eff} as green dashed lines.
Projections for Run 3 ($L=300~\rm{fb}^{-1}$) and HL-LHC ($L=3000~\rm{fb}^{-1}$) were also made, and are shown in
Fig. \ref{fig:paper4-xs_eff} as grey and black lines, respectively.
The sensitive (excluded) mass region in the future (present) 
corresponds to
the condition $\sigma_{\rm eff} \geq \sigma_{\rm eff}^{\rm UL}$

\begin{figure}[t!]
    \centering
    \includegraphics[width=0.49\textwidth]{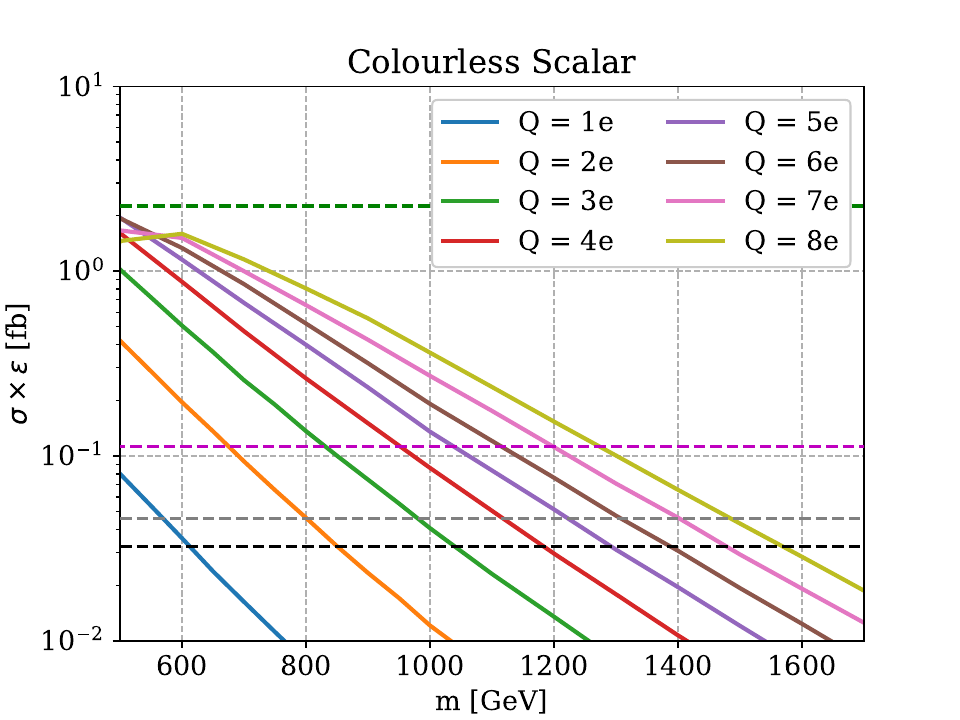}
    \includegraphics[width=0.49\textwidth]{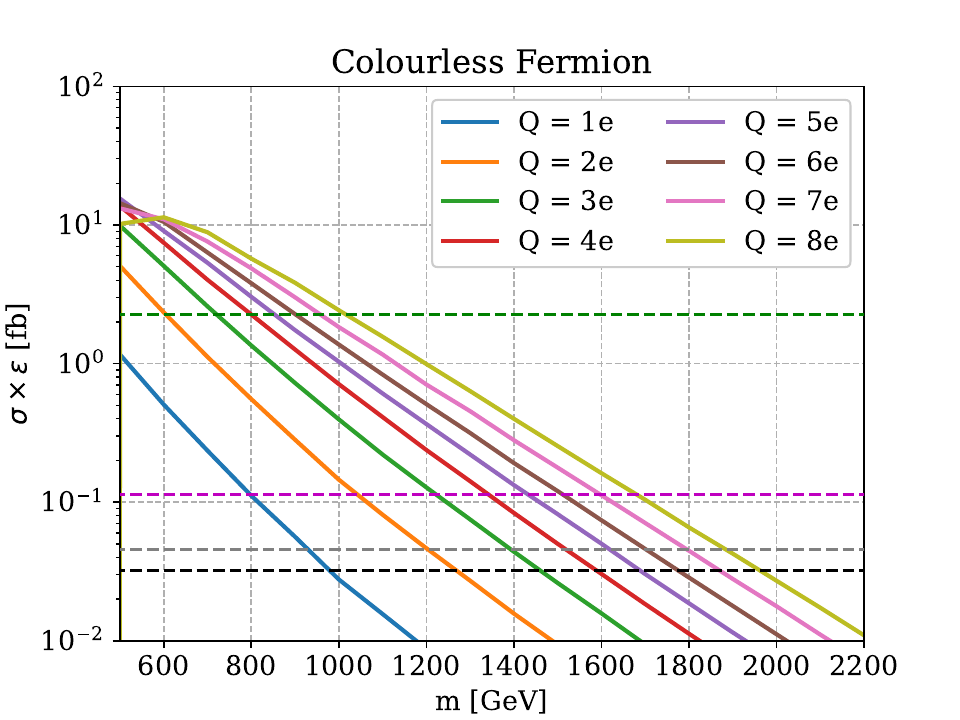}
    \includegraphics[width=0.49\textwidth]{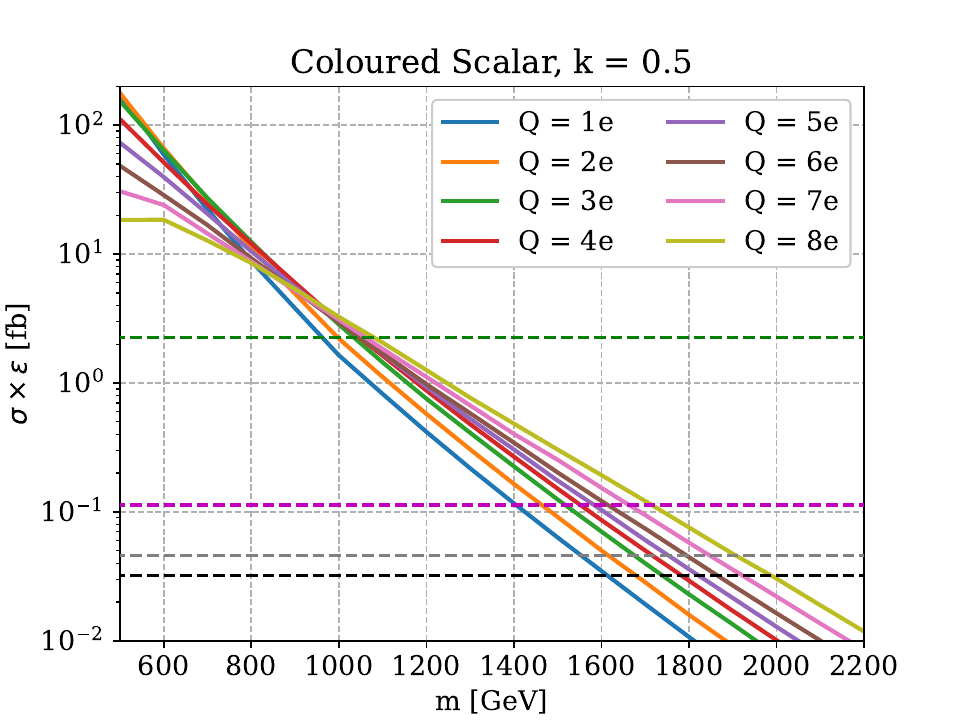}
    \includegraphics[width=0.49\textwidth]{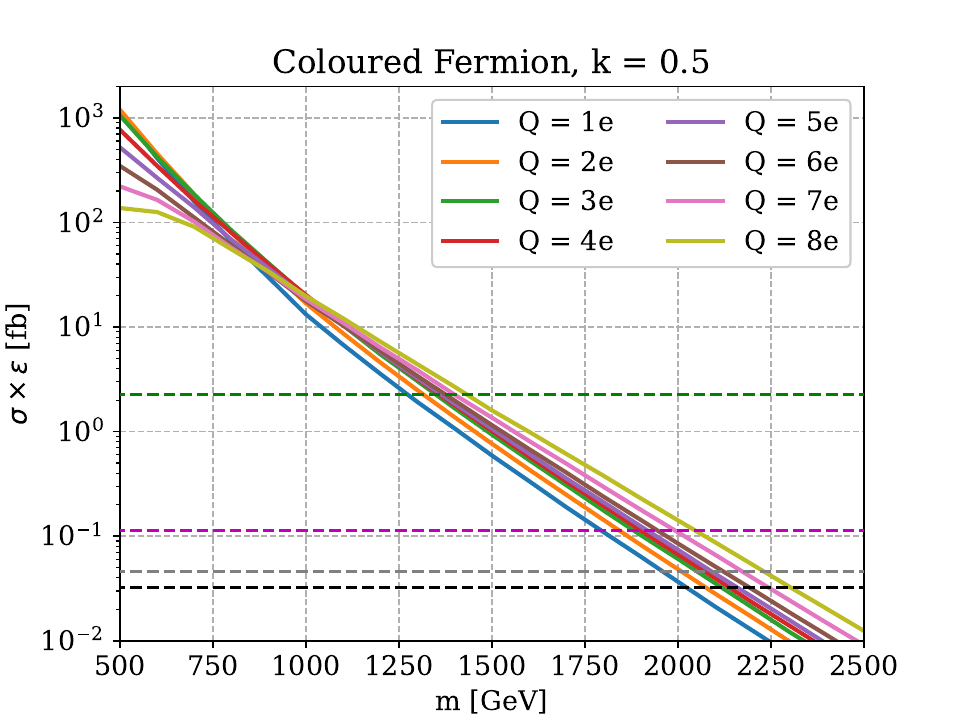}
    \caption{\small Effective cross section, $\sigma_{\rm eff} \equiv \sigma \cdot \epsilon_{\rm eff}$, for colourless scalars (top left), colourless fermions (top right), coloured scalars (bottom left), and coloured fermions (bottom right). Distinct colours of solid lines correspond to the effective cross section of particles with different electric charges. The dashed green, magenta, grey and black lines correspond to upper bounds for 2.5 fb$^{-1}$, 138 fb$^{-1}$, 300 fb$^{-1}$, and 3000 fb$^{-1}$, respectively. Limits were obtained by re-interpreting the CMS analysis \cite{CMS:2016kce}.}
    \label{fig:paper4-xs_eff}
\end{figure}

\subsection{Detecting open production in MoEDAL}

A characteristic property of the MoEDAL NTD array is that its sensitivity grows with the magnitude of the 
electric charge of BSM particles. It is a different behaviour to large $dE/dx$ searches, for which the signal 
efficiency peaks around $Q=(2-3)e$, then declines for bigger charges. This property is a huge 
advantage of the MoEDAL detector and might allow it to provide a comparable sensitivity to ATLAS and 
CMS, despite approximately 10 times lower integrated luminosity. Therefore, in our analysis we include 
the MoEDAL detector and use the simulation framework developed in the course of the previous 
projects (see Sec. \ref{sec:moedal-simulation}) to derive prospects for the detection of long-lived multi-charged particles.

We use a similar approach to analysis as in the project described in Sec. \ref{sec:paper3}. To be more 
precise, we aim at estimating the expected number of signal events, $N_{\rm sig}$, to be observed at 
MoEDAL in Run 3 ($L=30~\rm{fb}^{-1}$) and HL-LHC ($L=300~\rm{fb}^{-1}$) data taking phases. The 
expression for $N_{\rm sig}$ is given in Eq. \eqref{eq:moedal-n}. 

\begin{figure}[t!]
    \centering
    \includegraphics[width=0.47\textwidth]{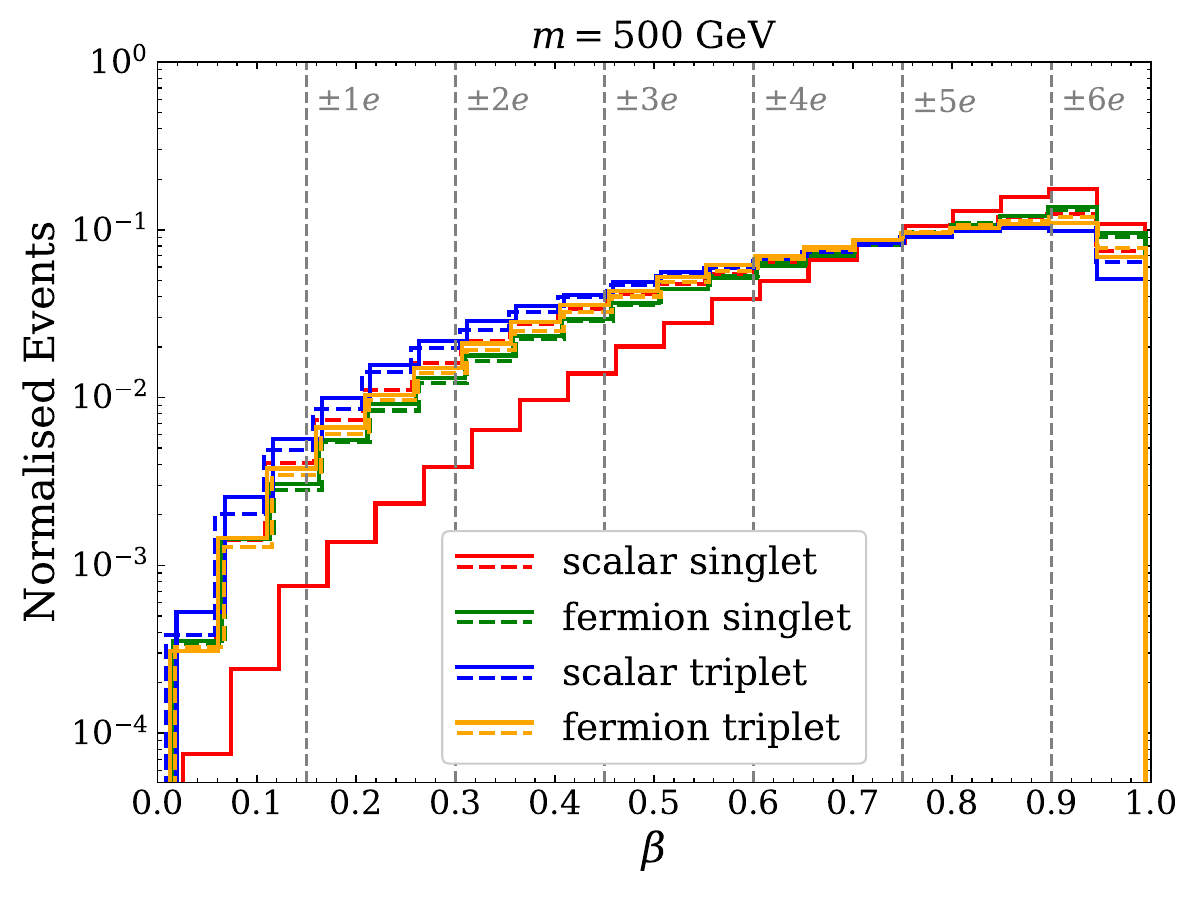}
    \hspace{3mm}
    \includegraphics[width=0.47\textwidth]{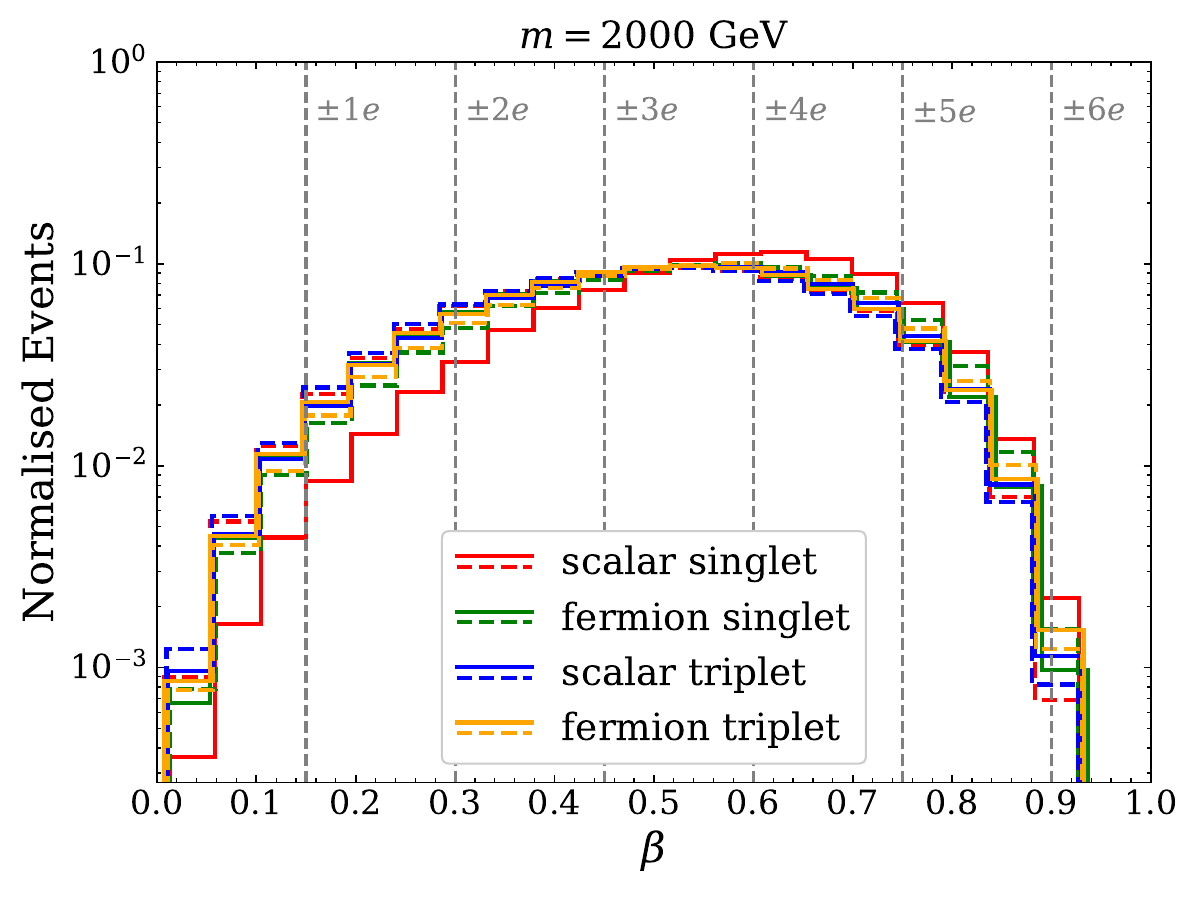}
    \caption{\small
    Velocity distributions of multiply charged long-lived particles with mass 500 GeV (left) and 2 TeV (right). Solid (dashed) histogram lines correspond to particles with $Q=1e$ ($8e$).
    Vertical dashed grey lines correspond to MoEDAL velocity thresholds for different magnitudes of the electric charge. 
    }
    \label{fig:paper4-beta}
\end{figure}

Several comments regarding the MoEDAL's efficiency calculation. In general, the efficiencies of NTDs 
depend on the velocity and incidence angle of BSM particles, however, the MoEDAL collaboration decided to rearrange the 
placement of detectors in order for NTD panels to directly face the IP8. Therefore, incidence angles are expected to be small, and the 
averaged detector efficiency can be approximated by Eq. \eqref{eq:paper2-moedal-eps} with $P_{\rm NTD}$ given by Eq. \eqref{eq:paper2-pntd}. In this approximation, the velocity threshold 
constraint, $\beta < \beta_{\rm th} = 0.15 \cdot |Q/e|$, becomes relaxed for particles with $|Q/e| = (0.15)^{-1} = 6\frac{2}{3}$. Since $\beta \leq 1$ by definition, MoEDAL can detect particles with any 
speed if their electric charge is $|Q| \geq  6\frac{2}{3}e$. Moreover, we assume that the interactions of 
particles moving through the LHCb VELO detector do not alter their four-momentum significantly, which 
is a fair approximation for the considered range of electric charges. For very highly charged particles, $|Q/e| \gg 8$, ionisation loss becomes so large that more precise treatment is needed to properly estimate  
the detector's efficiency.

To illustrate the impact of the velocity threshold on the signal acceptance, we plot in Fig. \ref{fig:paper4-beta} 
velocity distributions for all four considered types of charged LLPs: colourless scalars (red), colourless 
fermions (green), coloured scalars (blue), and coloured fermions (yellow). 
Solid and dashed histograms correspond to $Q=1e$ and $Q=8e$, respectively. 
The panel on the left (right) in Fig. \ref{fig:paper4-beta} is made for $m=500~\rm{GeV}$ ($m=200~\rm{GeV}$). By 
comparing both panels one can observe that the velocities of lighter particles are generally larger. Moreover, 
colour-singlet scalars with $|Q|=1e$ are characterised by the highest velocities, notably larger than 
other particles. This is because singly charged colour-singlet scalar particles are produced via s-channel 
Drell-Yan exchange of a spin-1 gauge boson. Such a process suffers from the p-wave suppression, i.e. $\sigma \to 0$ when $\beta \to 0$, as discussed previously. As can be seen from Fig. 
\ref{fig:paper4-frac}, for $Q=\pm 8e$ colourless scalar Drell-Yan production is subdominant, and 
indeed in Fig. \ref{fig:paper4-beta} we do not observe the effect of the p-wave suppression.
The grey vertical lines in Fig. \ref{fig:paper4-beta} represent velocity thresholds for different magnitudes 
of the electric charge, according to $\beta_{\rm th} = 0.15 \cdot |Q/e|$. One can see that for $Q=\pm1e$ only a small fraction of the produced particles satisfy the velocity condition, while for $|Q|= (5-6)e$ 
the majority of particles can be detected in MoEDAL.

As argued in Sec. \ref{sec:moedal}, the MoEDAL detector is effectively free from the SM background, hence it 
makes sense to consider low thresholds on the expected number of observed signal events. In our study, 
we focus on $N_{\rm sig}=1$, $2$, and $3$. In Sec. \ref{sec:paper4-appendix} we present sensitivity limits to all studied BSM particles in $(m, c\tau)$ parameter planes, where both $m$ and $c\tau$ are free parameters. These results are 
obtained for Run 3 and HL-LHC MoEDAL with $L=30~\rm{fb}^{-1}$ and $L=300~\rm{fb}^{-1}$, 
respectively.

\subsection{Closed production and diphoton signature}

\myparagraph{Bound state and diphoton decay mode}
If pair-produced BSM particles of mass $m$ have large electric charges, they can form a positronium-like bound state, $\mathcal{B}$, with mass $M_\mathcal{B} \approx 2~\rm{ m}$, as shown in Fig. \ref{fig:bound_diagram}. 
From now on, we 
refer this case as \textit{closed production mode}.

\begin{figure}[htb!]
\centering
\includegraphics[width=0.5\textwidth]{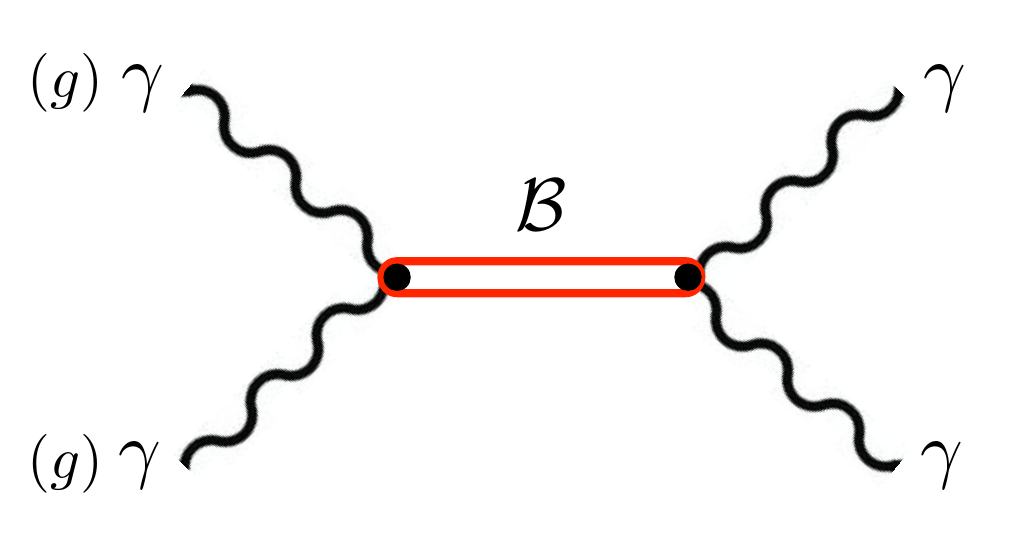}
\caption{\small
Diagrammatic representation of a bound state formation
and decay into the diphoton channel,
$pp \to {\cal B} \to \gamma \gamma$.}
\label{fig:bound_diagram}
\end{figure} 

The spin of the bound state $\mathcal{B}$ depends on the initial partons. For instance, if the initial state is $q\bar q$ (or $gg$ for $SU(3)_C$-triplet particles $\xi^{+Q}$), the s-channel exchange of a gauge boson $\gamma^*/Z^*$ (or $g^*$) allows $\mathcal{B}$ to have spin 1. Initial states with two gauge bosons result in the formation of $\cal B$ with spin 0 or 2, via t-channel or 4-point interaction diagrams. 

The spin of the bound state regulates its decay. Spin 0 and 2 bound states decay to $ZZ$, $\gamma Z$ and $\gamma \gamma$ 
final states, and additionally for coloured particles into $gg$. Bound states with spin 1, on the other hand, can decay into $W^+W^-$ or any SM fermion particle-antiparticle pair through the s-channel $\gamma^*/Z^*/g^*$ exchange. Amongst all final 
states, the diphoton channel ($\cal B \to \gamma \gamma$) has the best experimental sensitivity, therefore we focus on this 
search channel in order to derive constraints on the closed production mode \cite{Jager:2018ecz}.

In the case of coloured BSM particles, the bound state formed by a pair $\xi^{+Q} \xi^{-Q}$ might be a colour singlet or octet, since $\mathbf{3} \otimes \mathbf{\bar 3} = \mathbf{1} \oplus \mathbf{8}$. However, the octet configuration in the QCD does not lead 
to attractive force, which makes the production rate subdominant with respect to the production of colour-singlet bound states. 
Moreover, the octet state cannot decay to $\gamma \gamma$ final state, which we are interested in, hence in this study we will 
consider the formation of colour-singlet states only. 

The system consisting of $\xi^{+Q}\xi^{-Q}$ can be described near the threshold, in a fair approximation, by the non-relativistic 
Quantum Mechanics. The calculations of energy eigenstates for the considered bound state $\cal B$ are analogous to solving the 
hydrogen atom. Labelling the displacement from the position of $\xi^{+Q}$ to the position of $\xi^{-Q}$ as $\vec r ~(r=|\vec r |)$, we 
can write the static potential between the two BSM particles as:
\begin{equation}
V \left( \vec r \right) = - \frac{C \alpha_S(r^{-1})+Q^2 \alpha}{r},
\end{equation}
with $\alpha_S(r^{-1})$ being the running strong coupling at the scale $r^{-1}$, $\alpha$ being the electromagnetic fine-structure 
constant, $C=(C_1+C_2-C_{\cal B})/2$, $C_{\cal B}$ and $C_{1/2}$ being the quadratic $SU(3)_C$ Casimirs for the bound state 
and two constituent particles, respectively ($C=4/3$ (0) for the colour-triplet (singlet)).
We introduce the reduced mass: $\mu \equiv (   m_{\xi^{+Q}} \cdot m_{\xi^{-Q}}         )/(m_{\xi^{+Q}} + m_{\xi^{-Q}}    ) = m/2$ ($m \equiv  m_{\xi^{+Q}} = m_{\xi^{-Q}} $); which allows us to write the Schroedinger equation \eqref{eq:schroedinger} in the same form 
as for the hydrogen atom. The relevant energy eigenstates and wave functions, $\Psi_{nlm}(\vec r)$, can be found in analogy to 
this textbook calculation. 

The decay rate for the considered bound state is proportional to the probability of finding the two particles in the same point in space, $|\Psi_{nlm}(0)|^2$, which is non-vanishing only for the s-wave. The relevant wave function ($\Psi_n \equiv \Psi_{n00}$) at $\vec r = \vec 0$ is given by:
\begin{equation}\label{eq:paper4-wave-function}
\Psi_n(0) = \frac{1}{\sqrt{\pi}} (n r_b)^{-3/2},~r_b^{-1} = (C \bar \alpha_S + Q^2 \alpha) \mu.
\end{equation}
The $r_b$ in Eq. \eqref{eq:paper4-wave-function} stands for the Bohr radius of this bound system, $\bar \alpha_s \equiv \alpha_S (r^{-1}_{\rm rms})$ for a ground state $r_{\rm rms} = \sqrt {3} r_b$. The probability of finding two particles at the same 
point is therefore expressed by:
\begin{equation}\label{eq:paper4-prob}
|\Psi_n(0) |^2 = \frac{(C \bar\alpha_S+Q^2 \alpha)^3m^3}{8 \pi n^3}.
\end{equation}
From Eq. \eqref{eq:paper4-prob} we see that radial excitations are suppressed by $n^{-3}$, therefore we include only the ground 
state contribution with $n=1$.

The partial decay rates to $gg$ and $\gamma \gamma$ are given by:
\begin{align}
\Gamma_{\mathcal{B} \to \gamma \gamma} &= \frac{8 \pi Q^4 \alpha^2}{M^2_{\mathcal{B}}} n_C n_f |\Psi(0)|^2, \\
\Gamma_{\mathcal{B} \to gg} &= \frac{4 \pi C \alpha_S^2}{M^2_{\mathcal{B}}} n_f |\Psi(0)|^2,
\end{align}
with $n_C$ being the number of colours (1 for colour-singlet and 3 for colour triplet), and $n_f$ being the dimension of the 
Lorentz representation (1 for scalars and 2 for fermions). In the colour-singlet case, the $gg$ final state is unavailable, hence we 
set $\Gamma_{gg} =0$.

The partonic cross section for the $\cal B$ production is given by:
\begin{equation}\label{eq:paper4-sigmahat}
\hat \sigma_{ab \to \cal B}(\hat s) =
c_{ab} \cdot 
\frac
{2\pi(2J_{\cal B}+1)D_{\mathcal{B}}}
{D_a D_b}
\cdot
\frac{\Gamma_{\mathcal{B} \to {a b} }}{M_{\mathcal{B}}}
\cdot 
2 \pi \delta(\hat s - M^2_{\mathcal{B}}),
\end{equation}
with the partonic centre of mass energy denoted as $\hat s$, $c_{ab}$ being a symmetry factor $(c_{ab} = 1+\delta_{ab})$, $D_p$ 
being the dimension of the colour representation of particle $p$, and $J_{\cal B}$ labelling the spin of the bound state $\cal B$.

In order to get the hadronic production cross section, $\sigma_{pp\to\mathcal{B}}$, one needs to convolve the partonic cross section, $\hat \sigma_{ab \to \cal B}( s \tau)$, with the luminosity function $d L_{ab}/d\tau$:
\begin{equation}
\sigma_{pp\to\cal B} = \sum^{\gamma\gamma, gg}_{ab} \int_0^1 d\tau \frac{d L_{ab}(\tau)}{d\tau}\hat \sigma_{ab \to \cal B}(\hat s \tau),
\end{equation}
with the collision energy denoted as $ s$ and $\tau \equiv \hat s /s$. The integration is conducted using the {\tt ManeParse} 
\cite{Clark:2016jgm} package for Mathematica, with {\tt LUXqed17\_plus\_PDF4LHC15\_nnlo\_100} PDF set \cite{Manohar:2016nzj,Manohar:2017eqh}.

The signal rate for a production of a bound state that decays to the diphoton channel can be obtained using the narrow-width approximation:
\begin{equation}
\sigma_{pp\to\cal B \to \gamma\gamma} = \sigma_{pp\to \mathcal{B}} \cdot \rm{BR}_{\mathcal{B} \to \gamma \gamma},
\end{equation}
with
\begin{equation}
\rm{BR}_{\mathcal{B} \to \gamma \gamma} = \frac{\Gamma_{\mathcal{B} \to \gamma \gamma}}
{\Gamma_{\mathcal{B} \to \gamma \gamma} + \Gamma_{\mathcal{B} \to Z \gamma} +\Gamma_{\mathcal{B} \to ZZ} +\Gamma_{\mathcal{B} \to g g }}
\end{equation}
\begin{equation}
\Gamma_{\mathcal{B} \to \gamma Z} = \Gamma_{\mathcal{B} \to \gamma \gamma} \cdot 2 \tan^2 \theta_W \left( 1 -\frac{m^2_Z}{M^2_{\mathcal{B}}}\right),
\end{equation}
and
\begin{equation}
\Gamma_{\mathcal{B} \to ZZ} = \Gamma_{\mathcal{B} \to \gamma \gamma} \cdot \tan^4 \theta_W \sqrt{1-\frac{4m^2_Z}{M_{\cal B}}}.
\end{equation}

\myparagraph{Constraints from diphoton resonance searches}

The signal rate in the $\mathcal{B} \to \gamma \gamma$ channel calculated for the considered BSM particles should be compared with 
the latest experimental constraints. The most recent CMS result \cite{CMS:2018dqv} corresponds to 13 TeV $35.9~\rm{fb}^{-1}$ data set, while 
the latest ATLAS analysis \cite{ATLAS:2021uiz} used 13 TeV $139~\rm{fb}^{-1}$ data. Because the ATLAS analysis \cite{ATLAS:2021uiz} is based on a larger data set, 
we use it to derive the current limit on the bound state production and we estimate the projected sensitivities for Run 3 ($L=300~\rm{fb}^{-1}$) and HL-LHC ($L=3~\rm{ab}^{-1}$).

In Ref. \cite{ATLAS:2021uiz}, ATLAS collaboration analysed the diphoton channel and did not observe any significant signal excess over the SM 
background. The results were interpreted for spin-0 and spin-2 particles for several different values of decay widths. The upper 
limits on the production cross section times the branching ratio to two photons were obtained, as a function of the resonance 
mass. In our simple BSM model, we have $\Gamma_{\cal B}/M_{\cal B} \lesssim 2\%$ for the relevant range of $M_{\cal B}$ and $Q$ parameters, therefore it is justified to use the upper limit provided by ATLAS, which relies on the narrow width approximation.

We present the diphoton signal cross section, $\sigma_{pp\to\cal B \to \gamma\gamma}$, as a function of the bound state mass ($M_{\cal B} = 2 m$), in Fig. \ref{fig:paper4-xs_diphoton}. The panels on the left(right)-hand side correspond to scalar $\phi^{\pm Q}$ (fermionic $\psi^{\pm Q}$ ) particles. Panels in the upper (lower) row in Fig. \ref{fig:paper4-xs_diphoton} are for colour-singlet (triplet) particles. The 95\% CL upper 
limit provided by ATLAS \cite{ATLAS:2021uiz} for 13 TeV 139 fb$^{-1}$ dataset is depicted in Fig. \ref{fig:paper4-xs_diphoton} as a red dashed line. The ATLAS study \cite{ATLAS:2021uiz}
provided results only up to $M_{\cal B} = 2.5~\rm{TeV}$. In order to derive the upper limit for heavier states, we use the same 
value as for $M_{\cal B} = 2.5~\rm{TeV}$. This approach is justified because in the $M_{\cal B} > 2.5~\rm{TeV}$ region ATLAS 
observed no events and the SM background is below one event.

\begin{figure}[htb]
\centering
\begin{subfigure}[t]{.48\linewidth}
\includegraphics[width=1.\textwidth]{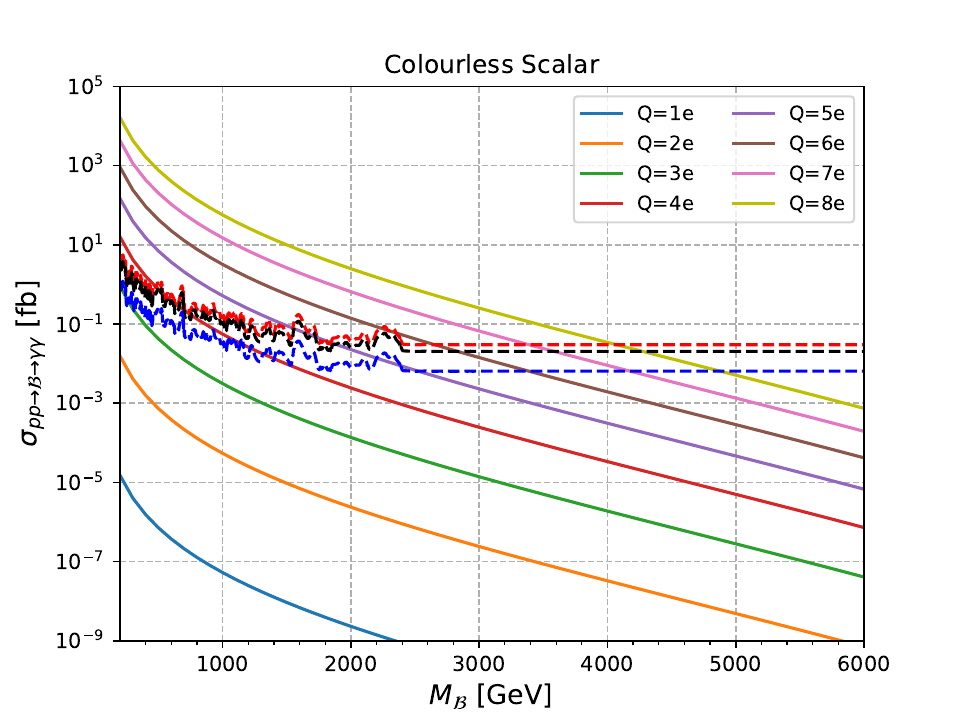}
\end{subfigure}
\begin{subfigure}[t]{.48\linewidth}
\includegraphics[width=1.\textwidth]{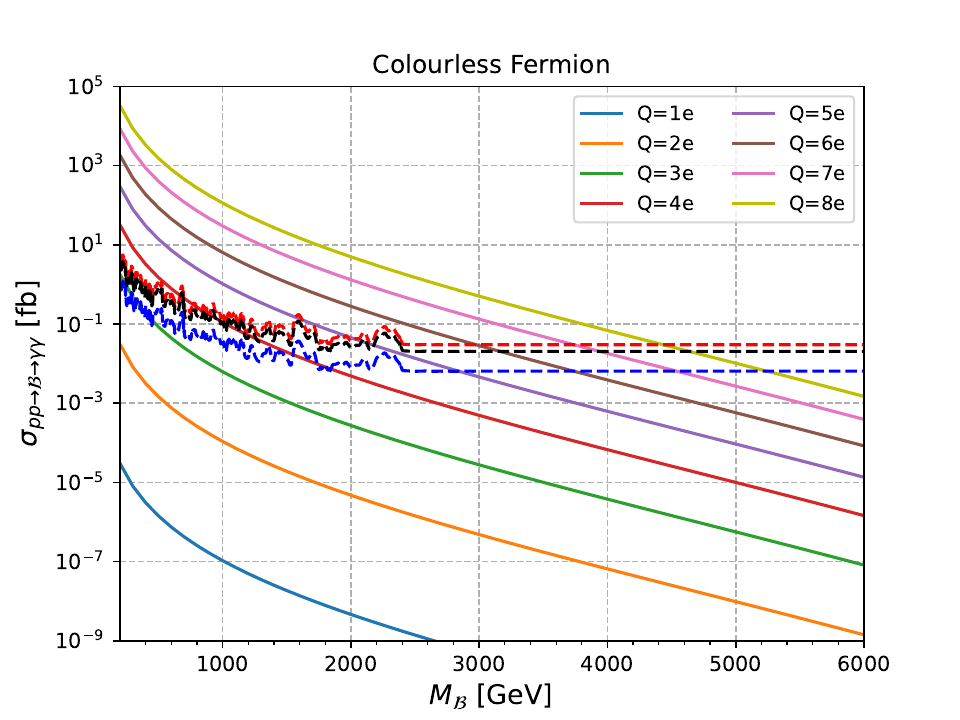}
\end{subfigure}
\begin{subfigure}[t]{.48\linewidth}
\includegraphics[width=1.\textwidth]{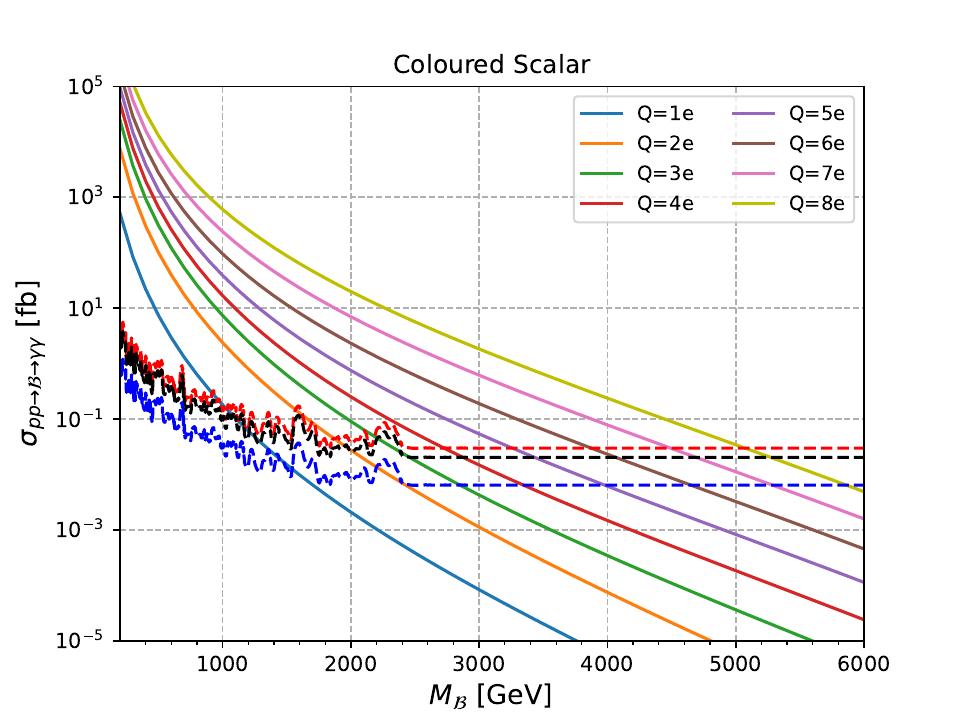}
\end{subfigure}
\begin{subfigure}[t]{.48\linewidth}
\includegraphics[width=1.\textwidth]{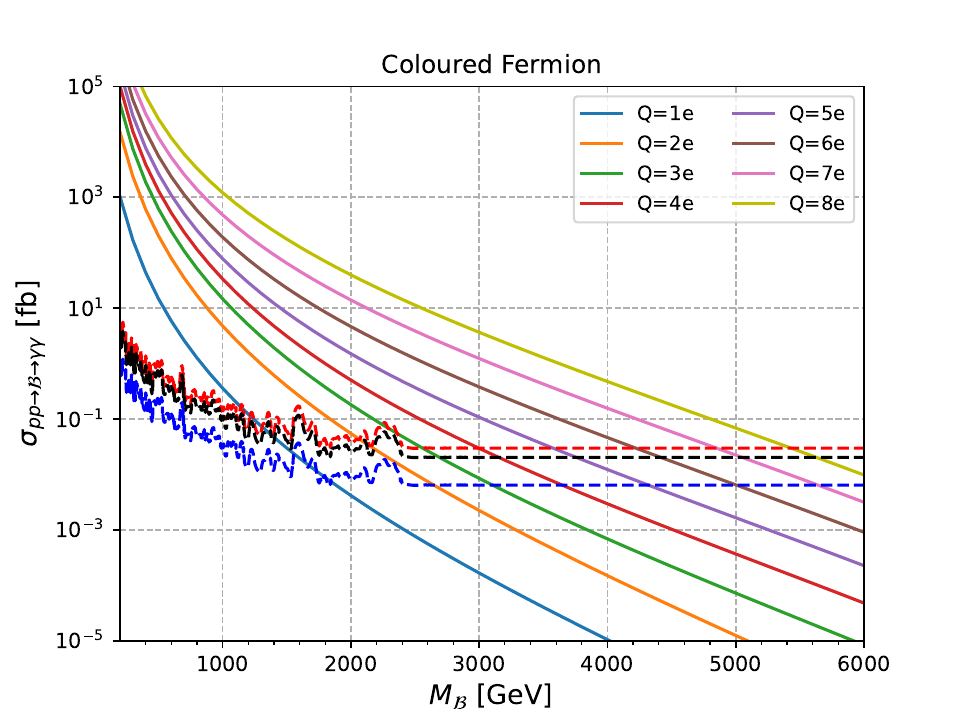}
\end{subfigure}
\caption{\small Diphoton resonant production cross section for colourless (top row) and coloured (bottom row) particles, with spin-0 (left column) and spin-1/2 (right column). The cross section is presented as a function of the mass of the formed bound state, $M_{\cal B} \approx 2 m$. Dashed lines represent exclusion (sensitivity) upper limits for 139 (red), 300 (black), and 3000 (blue) fb$^{-1}$ ATLAS, based on \cite{ATLAS:2021uiz}.
\label{fig:paper4-xs_diphoton}  }
\end{figure}

In order to estimate the projected sensitivity of ATLAS for Run 3 ($L_{\rm Run3}=300~\rm{fb}^{-1}$) and HL-LHC ($L_{\rm HL}=3~\rm{ab}^{-1}$), we 
rescale the current ($L_0=139~\rm{fb}^{-1}$) upper limit $\sigma_0^{\rm UL}$ in the following manner:
\begin{align}
\sigma_{\rm Run3}^{\rm UL} &= \sigma_0^{\rm UL} \sqrt{\frac{L_0}{L_{\rm Run3}}} \\
\sigma_{\rm HL}^{\rm UL} &= \sigma_0^{\rm UL} \sqrt{\frac{L_0}{L_{\rm HL}}}.
\end{align}
We assume that the signal/background efficiencies and systematic uncertainties will not change drastically. This approach is 
conservative because improving the efficiencies or reducing systematics would result in better sensitivity. The projected upper 
cross section limits are shown in Fig. \ref{fig:paper4-xs_diphoton}, where black and blue dashed curves correspond to Run 3 and 
HL-LHC, respectively.

\subsection{Results}

In the following section, we summarise calculations for the open and closed production modes of long-lived BSM particles and 
compare them with the sensitivity of ATLAS, CMS and MoEDAL experiments. For the major general-purpose LHC experiments, we 
recast large $dE/dx$ and diphoton resonance searches for Run 3 ($L=300~\rm{fb}^{-1}$) and HL-LHC ($L=3~\rm{ab}^{-1}$), and 
obtain 95\% CL limits. For MoEDAL, we present the expected number of observed BSM signal events, $N_{\rm sig}=1,2,3$, for Run 
3 ($L=30~\rm{fb}^{-1}$) and HL-LHC ($L=300~\rm{fb}^{-1}$).

We begin by presenting in Fig. \ref{fig:paper4-res_curr} the current 95\% CL mass limits as a function of $Q$. The upper (lower) 
panels in Fig. \ref{fig:paper4-res_curr} correspond to colour-singlet (triplet) particles. The left and right panels in Fig. \ref{fig:paper4-res_curr} are for scalars and fermions, respectively. The parameter region in Fig. \ref{fig:paper4-res_curr} shaded in red and 
enclosed between two red curves is excluded by the most recent CMS search \cite{CMS:2016kce} for large $dE/dx$ with 13 TeV $2.5~\rm{fb}^{-1}$ data. The analysis was re-interpreted in order to provide mass limits on all four types of studied particles and to include initial 
states with photons. The lower boundary of the exclusion region originates from the fact that for large $|Q|$ the $p_{\rm T}$ selection 
cut ($p_{\rm T} > 65 \cdot |Q/e|$ GeV) makes it hard for light particles to satisfy the $1/\beta_{\rm MS} > 1.25$ condition. However, one 
can observe that the parameter regions below the lower boundaries of the large $dE/dx$ limits in Fig. \ref{fig:paper4-res_curr} 
are mostly excluded by the ATLAS diphoton resonance search \cite{ATLAS:2021uiz}, depicted in blue. Moreover, we believe that the remaining 
unconstrained low masses are already excluded by the previous large $dE/dx$ analysis \cite{CMS:2013czn}, based on 7 and 8 TeV data, in 
which the $p_{\rm T}$ cut is weaker. For the colour-singlet scalars no exclusion is found for $|Q|\geq 6e$, due to $1/\beta_{\rm MS}>1.25$ and generally large production velocity. The latest large $dE/dx$ search by ATLAS \cite{ATLAS:2018imb} with 36.1 fb$^{-1}$ of 13 TeV 
data reported results only for colourless fermions with $2e \leq |Q| < 7e$. We include this limit in the top right panel of Fig. 
\ref{fig:paper4-res_curr}. This limit is based on the leading-order cross section calculations without including the photon fusion 
production. The amount of data used in the ATLAS analysis is larger than in the CMS study \cite{CMS:2016kce}, however, the signal efficiency for the former analysis is unavailable and therefore we cannot re-interpret it.

\begin{figure}[!tb]
    \centering
    \includegraphics[width=0.49\textwidth]{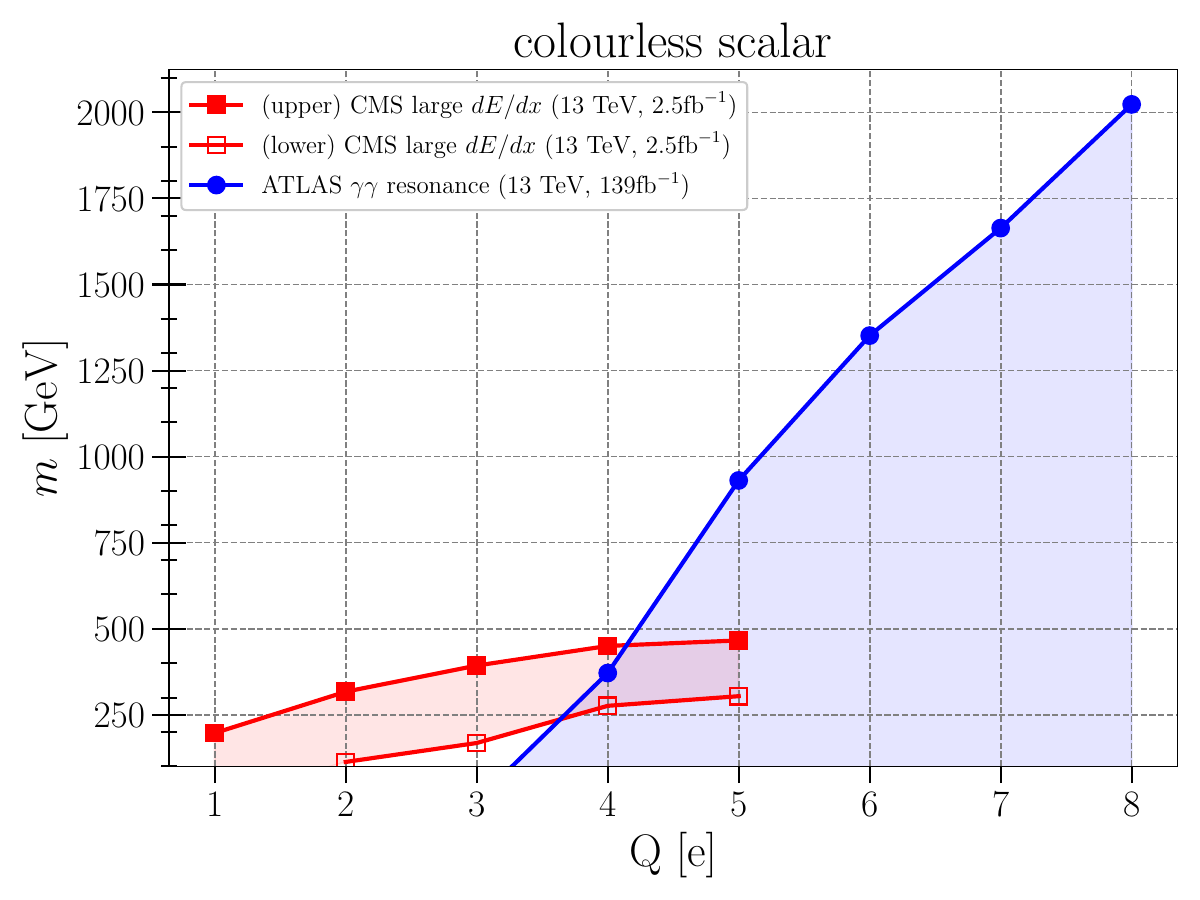}
    \includegraphics[width=0.49\textwidth]{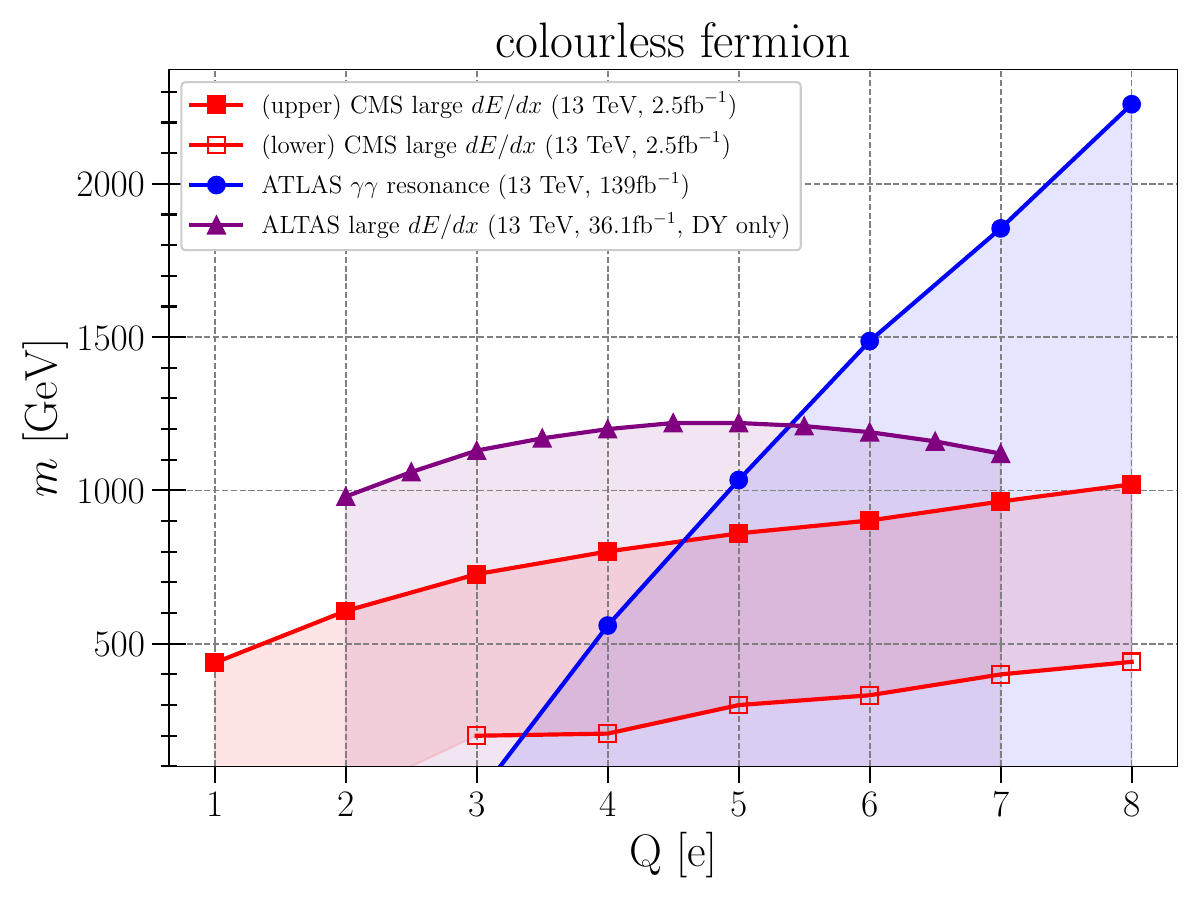}
    \includegraphics[width=0.49\textwidth]{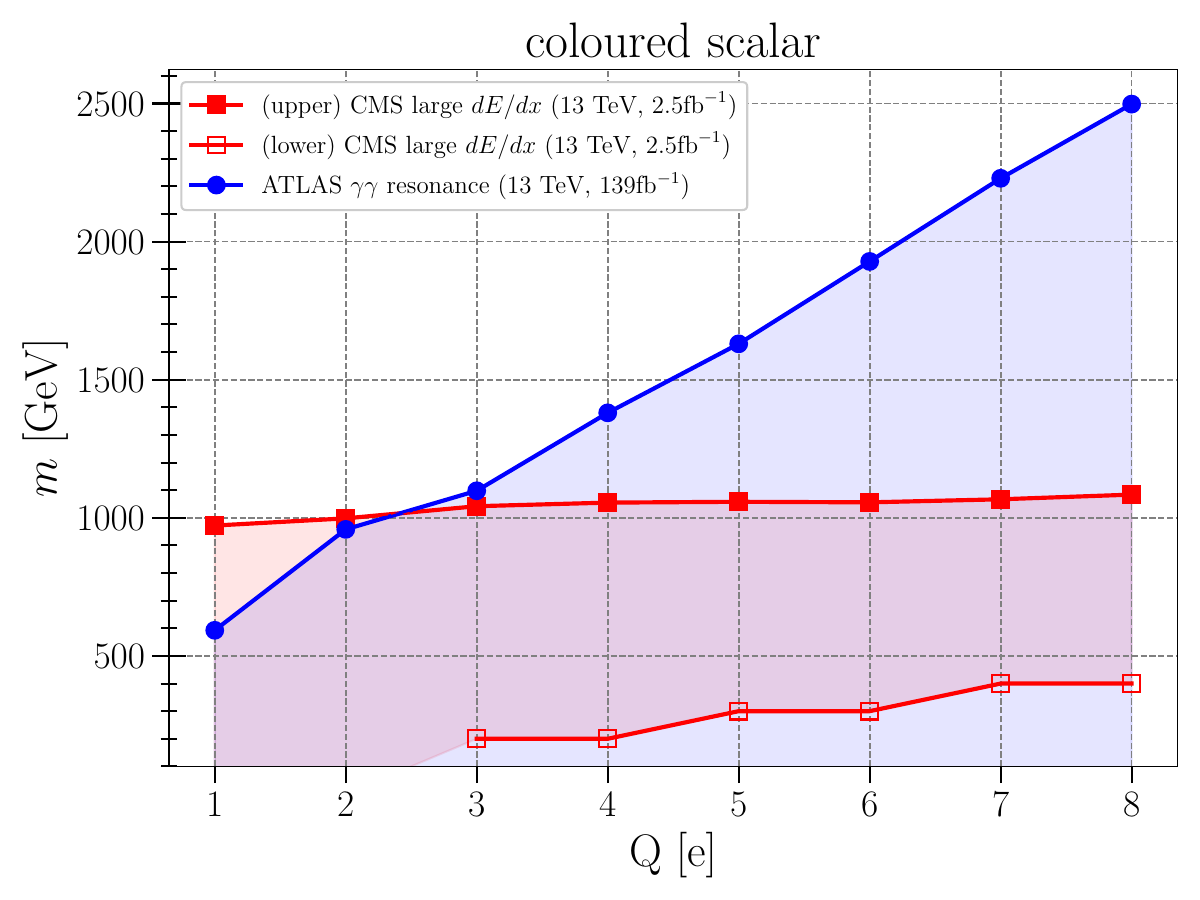}
    \includegraphics[width=0.49\textwidth]{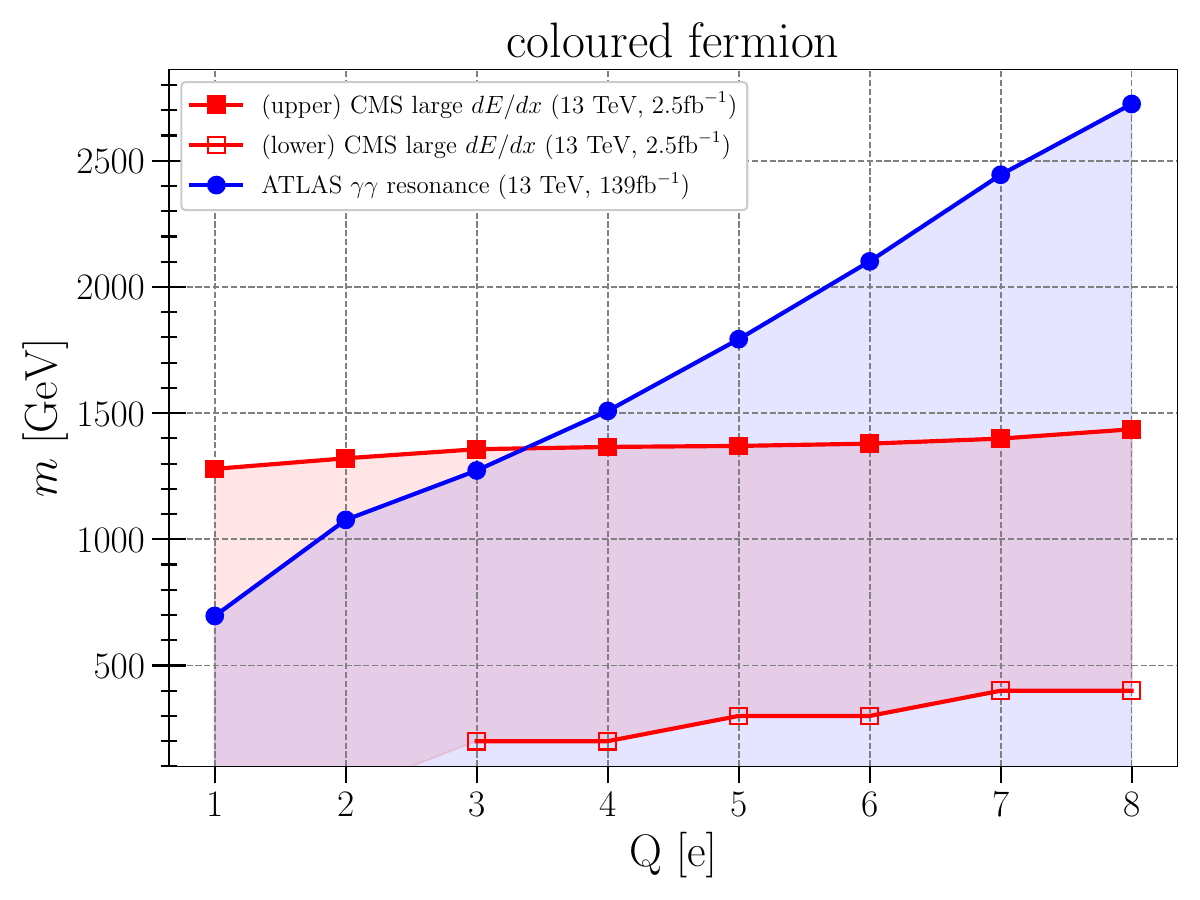}
    \caption{ \small
    Current lower limits on masses of multiply charged particles: colourless scalars (top left), colourless fermions (top right), coloured scalars (bottom left), and coloured fermions (bottom right). The limits were recast to include production processes with initial state photons. The region shaded in red is excluded by the large $dE/dx$ CMS search \cite{CMS:2016kce}, based on 13 TeV $2.5~\rm{fb}^{-1}$ data set. For colourless fermions with $2 e \leq |Q| \leq 7e$ there is an additional constrain from the ATLAS search \cite{ATLAS:2018imb} for large $dE/dx$ based on 13 TeV 36.1 fb$^{-1}$ data. The region shaded in blue is excluded by the diphoton resonance search by ATLAS\cite{ATLAS:2021uiz}, based on 13 TeV 139 fb$^{-1}$ data set.
    }
    \label{fig:paper4-res_curr}
\end{figure}

The mass limits from recasting the 13 TeV $L=139~\rm{fb}^{-1}$ ATLAS analysis \cite{ATLAS:2021uiz}, which searched for diphoton resonances, 
are shown in Fig. \ref{fig:paper4-res_curr} as blue curves. The corresponding regions shaded in blue correspond to excluded values of mass and 
electric charge parameters. From Fig. \ref{fig:paper4-res_curr} we can see that for particles with small electric charges ($Q \lesssim 2\rm{-}4 e$) the diphoton resonance search provides weaker constraints than the search for large $dE/dx$. The limit 
becomes rapidly much stronger for larger charges, and it exceeds the large $dE/dx$ limit for $Q \sim 2\rm{-}4 e$, where the 
exact magnitude of charge depends on the type of the BSM particle. This behaviour can be understood by looking at Eqs. 
\eqref{eq:paper4-prob}-\eqref{eq:paper4-sigmahat}, from which it follows that the production cross section for the bound state 
from $\gamma\gamma$ initial state grows as $Q^{10}$. The limits from large $dE/dx$ searches, on the other hand, increase only 
mildly with the growth of the electric charge. This is because of the two opposite effects: the open production mode cross section 
grows with the charge (cf. Fig. \ref{fig:paper4-xs_cless} and Fig. \ref{fig:paper4-xs_c}), but the signal efficiency drops, as depicted in Fig. \ref{fig:paper4-eff}, due to 
underestimation of particle's transverse momentum and the velocity loss through electromagnetic interactions with the detector material.

In Fig. \ref{fig:paper4-res_run3} we present the projected mass limits for Run 3. The red and blue curves in Fig. \ref{fig:paper4-res_run3} correspond to 95\% CL limits 
from searches for the large $dE/dx$ and resonance decaying to two photons, respectively. The green curves with diamond, 
triangle and circle markers represent the sensitivity of MoEDAL, and correspond to $N_{\rm sig}=3,2,1$, respectively. In Fig. \ref{fig:paper4-res_run3} 
we do not show the lower boundary from the large $dE/dx$ search, because the parameter region below this boundary is tested 
by open channel MoEDAL and ATLAS diphoton analyses, as well as by other searches done with Run 1 and Run 2 data. For coloured 
particles, we vary the free parameter of the hadronisation model, $k$, from 0.3 to 0.7. This parameter changes the distribution of 
charge of the final state hadrons and in principle might alter the results. We see this effect in MoEDAL searching for particles with 
$|Q|\lesssim 3e$, which is depicted in the lower panels of Fig. \ref{fig:paper4-res_run3} by the width of the green curves. However, for bigger charge 
magnitudes, and for large $dE/dx$ searches, the effect is so small that it cannot be seen in the plots. This proves that the 
uncertainty from using our crude hadronisation model is small.

\begin{figure}[!tb]
    \centering
    \includegraphics[width=0.49\textwidth]{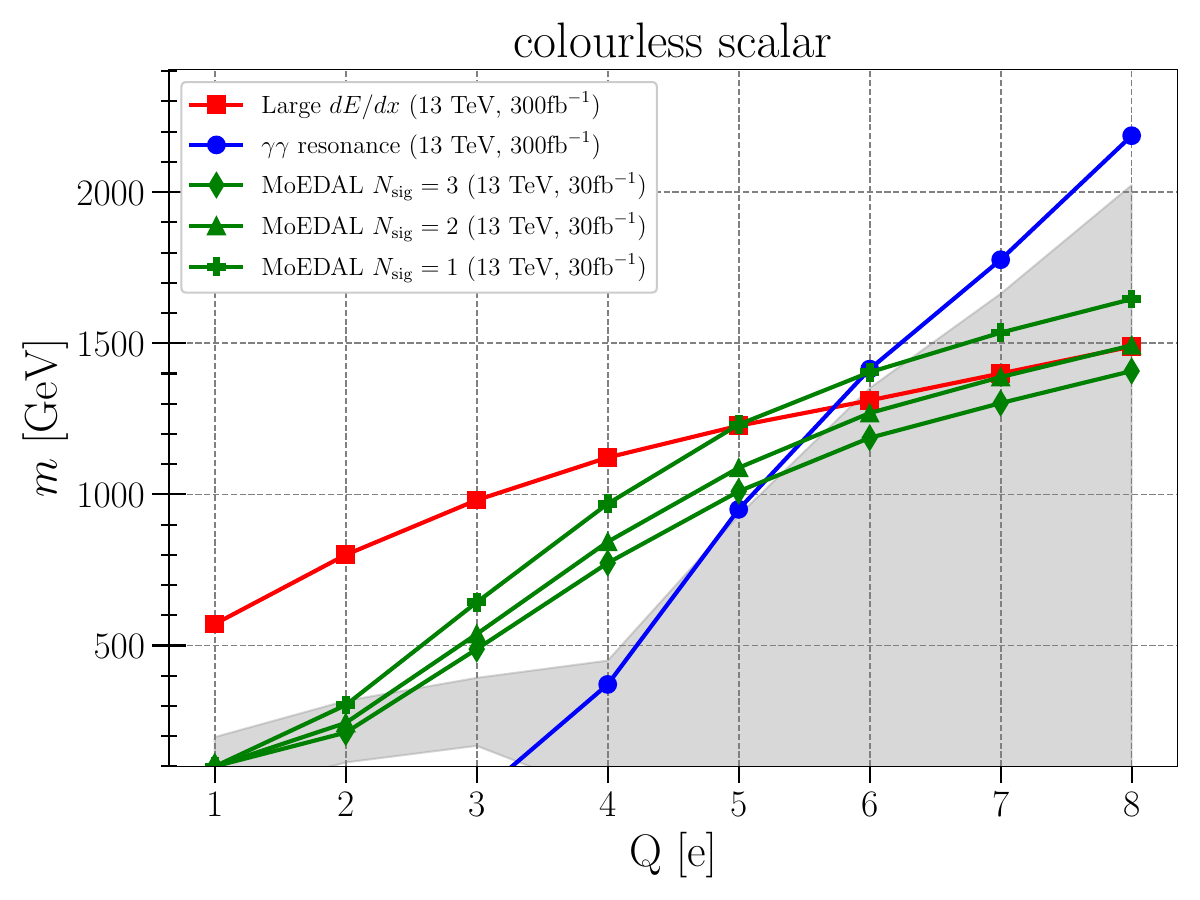}
    \includegraphics[width=0.49\textwidth]{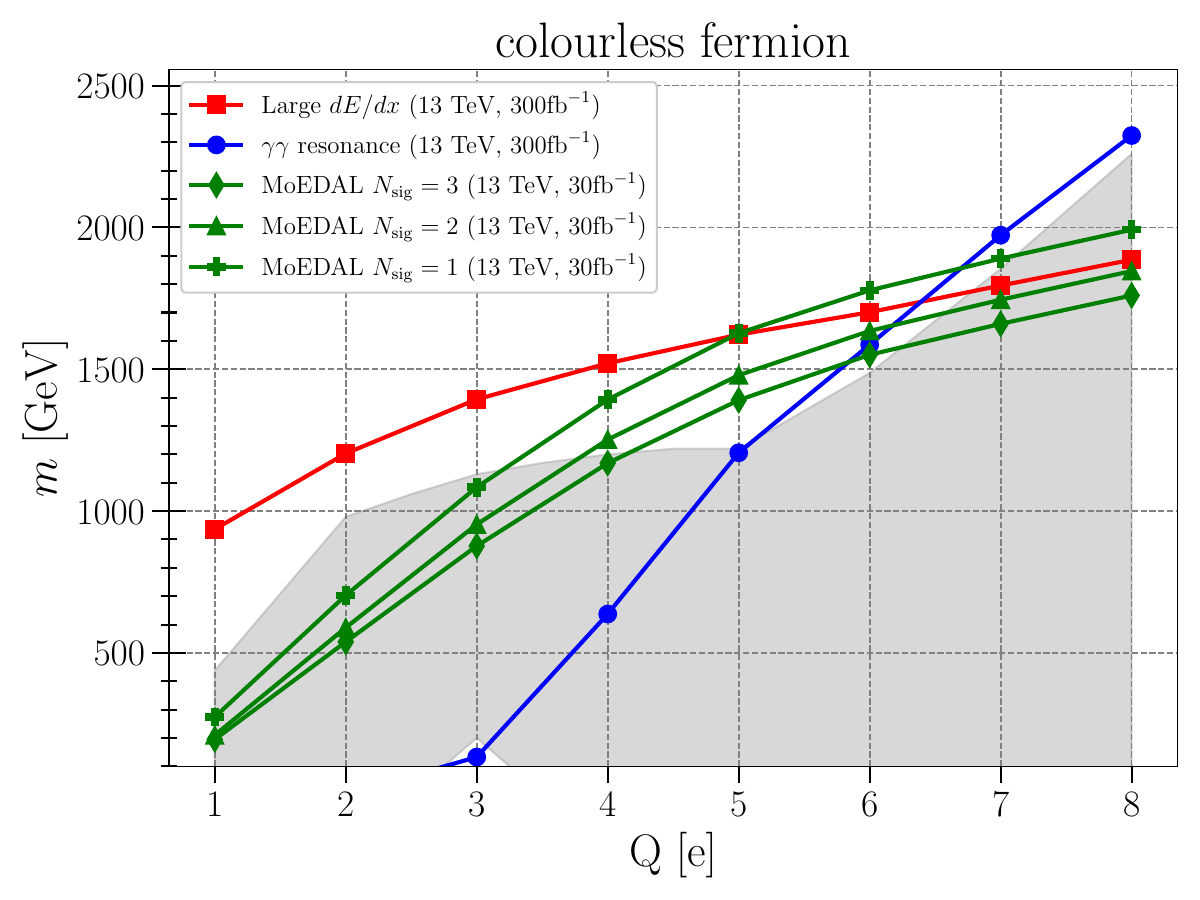}
    \includegraphics[width=0.49\textwidth]{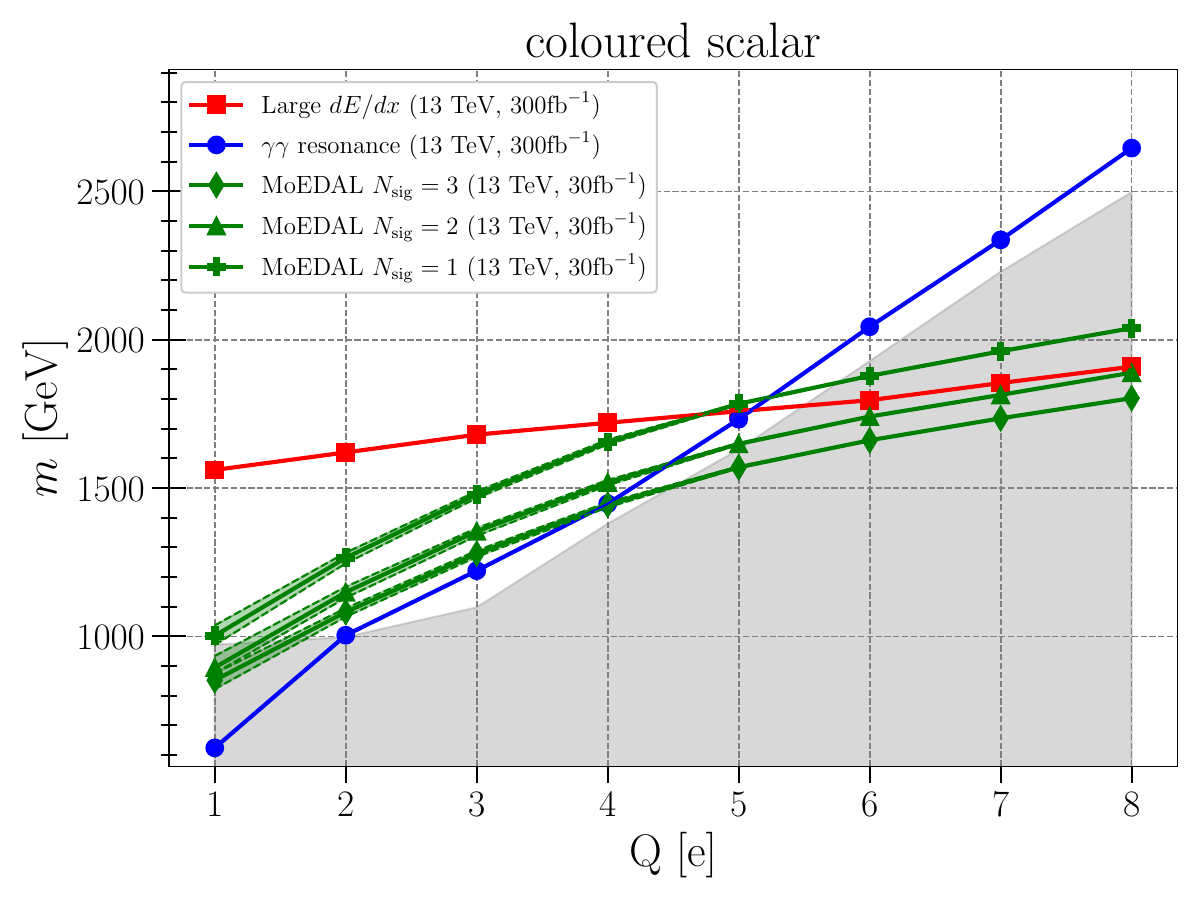}
    \includegraphics[width=0.49\textwidth]{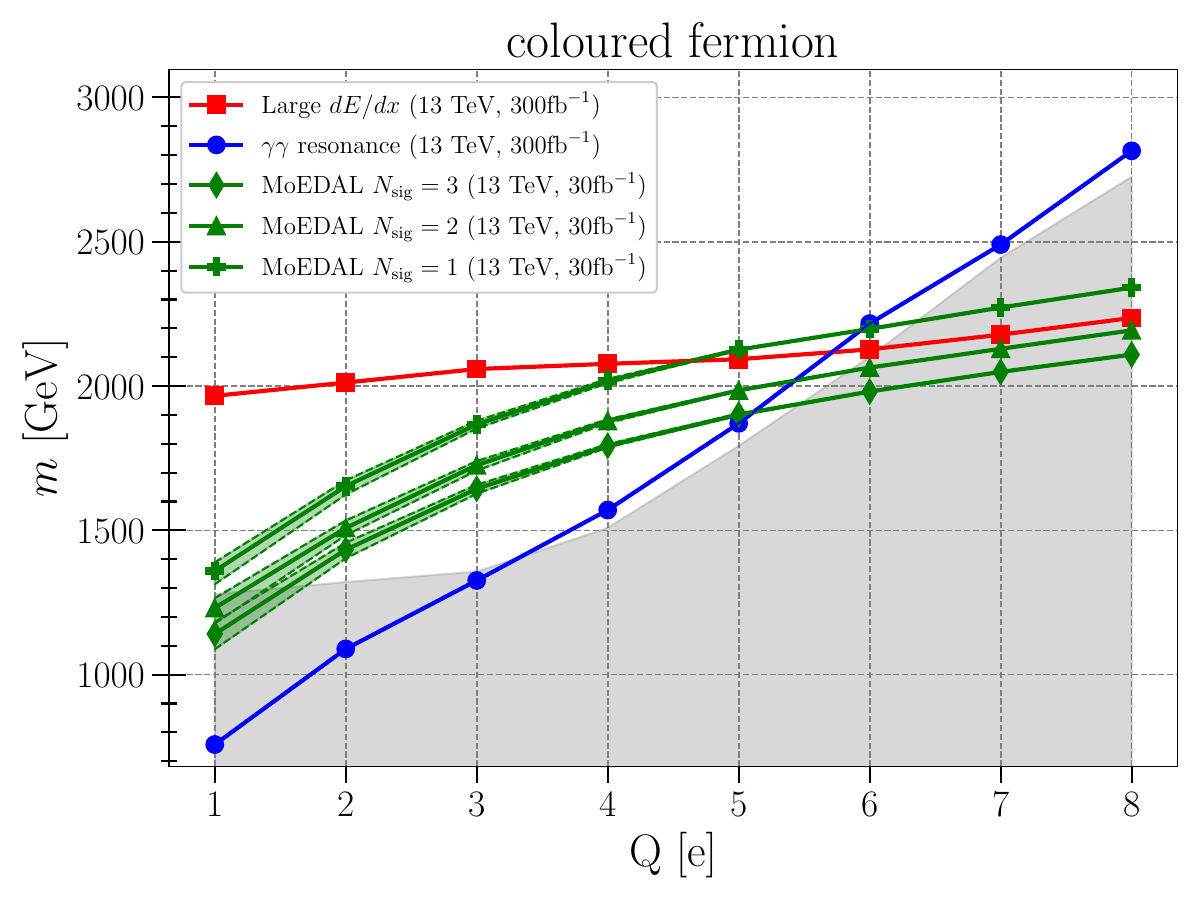}
    \caption{\small 
    Sensitivities of the Run 3 LHC experiments to multiply charged long-lived BSM particles: colourless scalars (top left), colourless fermions (top right), coloured scalars (bottom left), and coloured fermions (bottom right). The integrated luminosity is assumed to be $30~\rm{fb}^{-1}$ for MoEDAL, and $300~\rm{fb}^{-1}$ for ATLAS and CMS. The region shaded in grey colour is already excluded (cf. Fig. \ref{fig:paper4-res_curr}). Red, blue and green curves correspond to limits provided by large $dE/dx$, diphoton resonance, and MoEDAL searches, respectively. The uncertainty of the hadronisation model for colour-triplet particles is depicted as the variable width of the curves. This effect is visible only for MoEDAL and particles with $Q \leq 3e$.
    }
    \label{fig:paper4-res_run3}
\end{figure}

When recasting the CMS search for the large $dE/dx$ and scaling the luminosity from $2.5~\rm{fb}^{-1}$ to $300~\rm{fb}^{-1}$, 
we follow the steps in \cite{Allanach:2011wi, Sakurai:2011pt} to obtain the 95\% CL upper limit on the detected signal. We assume 
that the systematic uncertainty stays at the same level, $\sim 66\%$, relative to the total background. In this regard our limits are 
conservative. The grey-shaded regions in Fig. \ref{fig:paper4-res_run3} show currently constrained parameters. One can see in 
Fig. \ref{fig:paper4-res_run3} that the Run 3 improvement from the diphoton resonance search is modest, it is because the data 
used in the ATLAS search \cite{ATLAS:2021uiz} is already quite large, $L=139~\rm{fb}^{-1}$. On the other hand, the recast 
limits from large $dE/dx$ search \cite{CMS:2016kce} are based on $L=2.5~\rm{fb}^{-1}$ data set, and scaling up to Run 3 
luminosity, $L=300~\rm{fb}^{-1}$, allows to derive much stronger constraints. For colour-singlet fermions with $2 e\leq |Q| \leq 7 e$, there is another limit 
from the ATLAS search for large $dE/dx$ \cite{ATLAS:2018imb}, based on 13 TeV 36.1 fb$^{-1}$ data, hence the change of the 
limits when 
considering Run 3 luminosity is smaller for these particles.
As already stated before, open mode searches for large $dE/dx$ provide predominant constraints for smaller charges, while 
diphoton resonance analysis for the closed production mode effectively constrains particles with high electric charges. For Run 3, 
both types of searches provide a similar discovery potential at $|Q| \sim (5-6) e$. 
The lower mass bounds, which MoEDAL is expected to provide at the end of Run 3, typically lie between the open and closed 
channel limits by ATLAS and CMS. For $Q \sim (4-6)e$ with $N_{\rm sig}=1$, MoEDAL can provide a comparable sensitivity to 
large $dE/dx$ and diphoton resonance searches. We would like to emphasise that the limits provided by MoEDAL are 
complementary to ATLAS and CMS, because of the completely different detector design and uncorrelated systematic uncertainties.

\begin{figure}[!tb]
    \centering
    \includegraphics[width=0.49\textwidth]{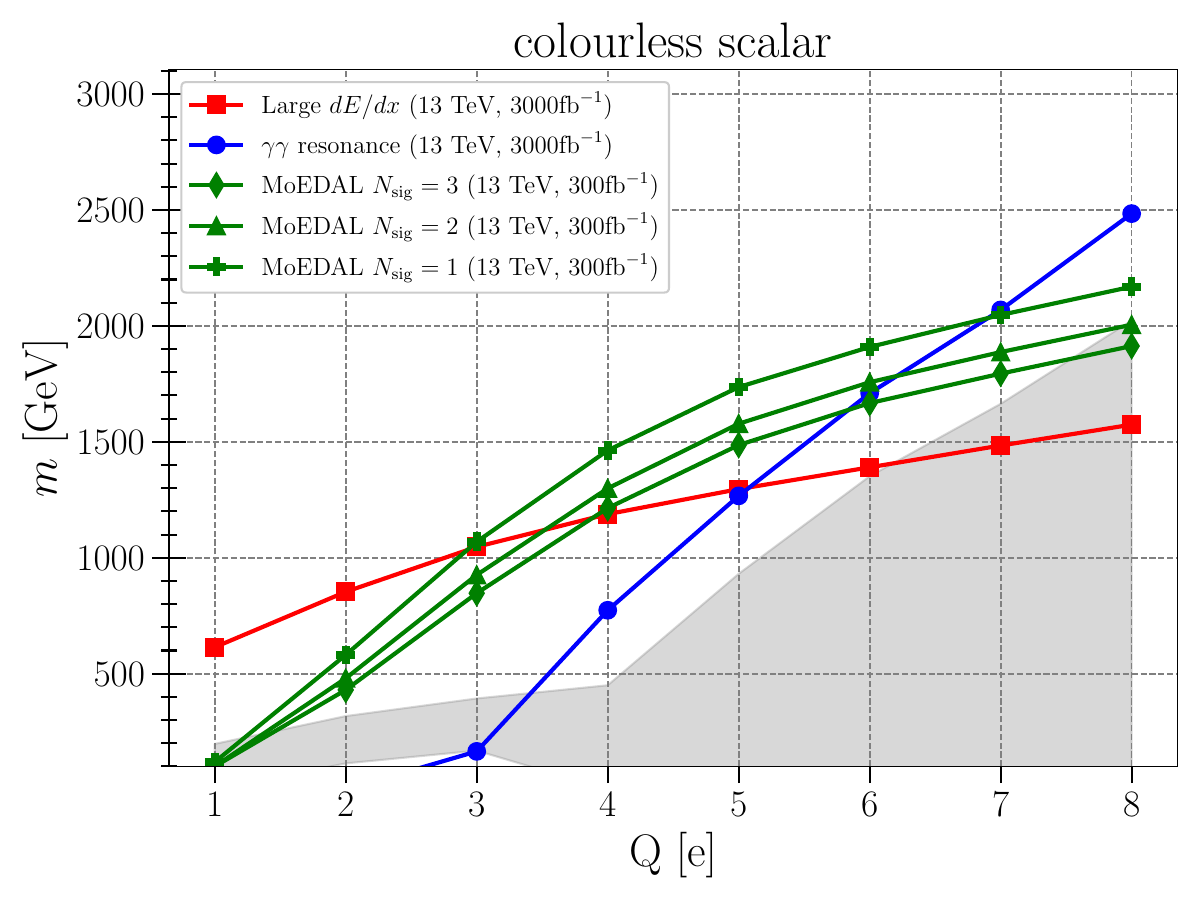}
    \includegraphics[width=0.49\textwidth]{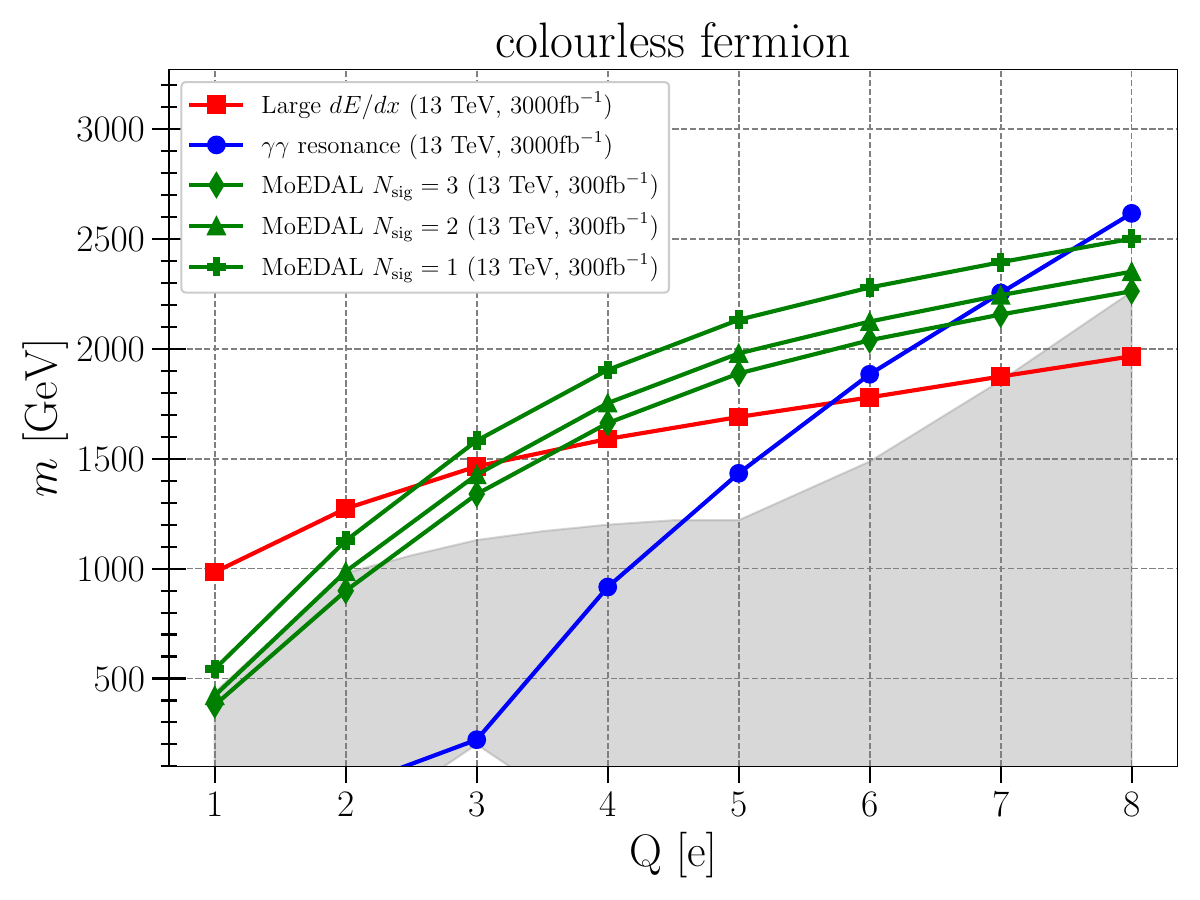}
    \includegraphics[width=0.49\textwidth]{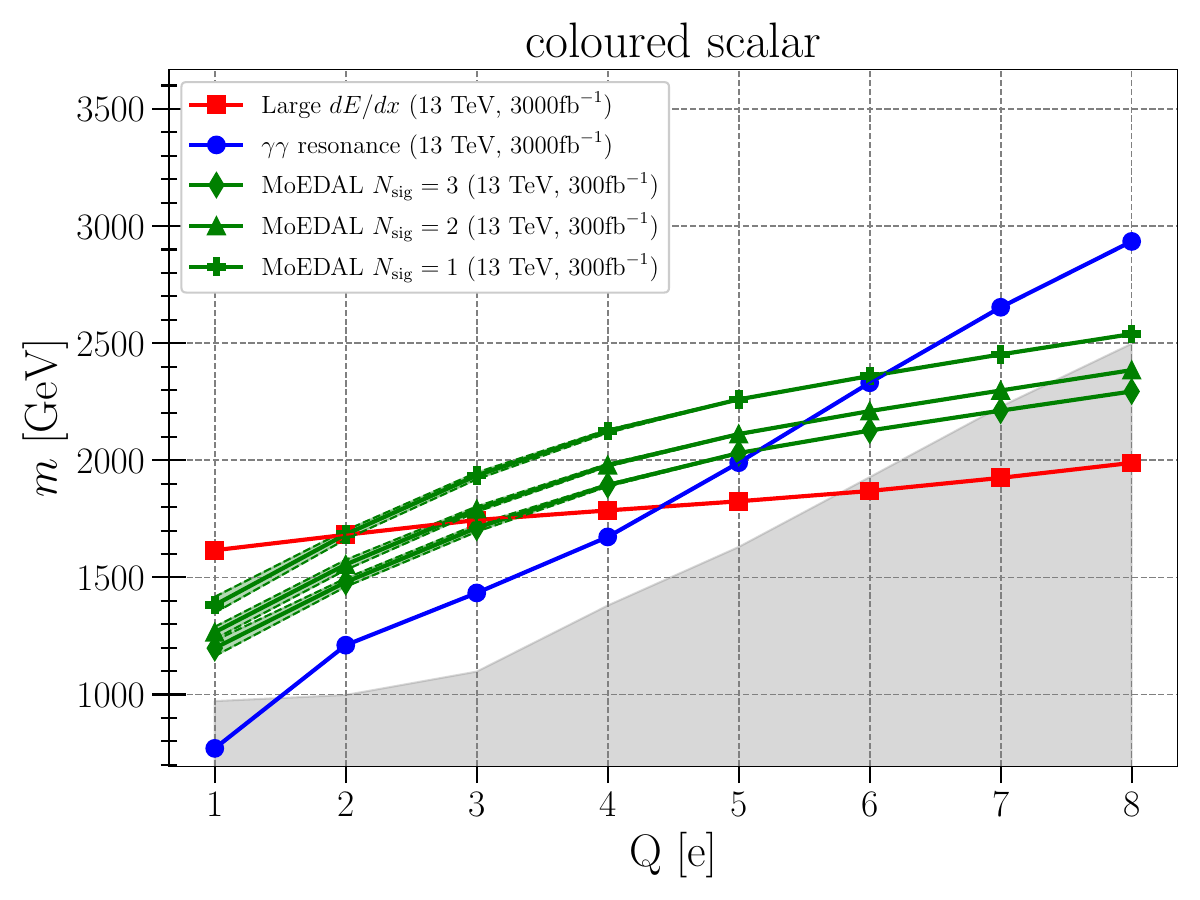}
    \includegraphics[width=0.49\textwidth]{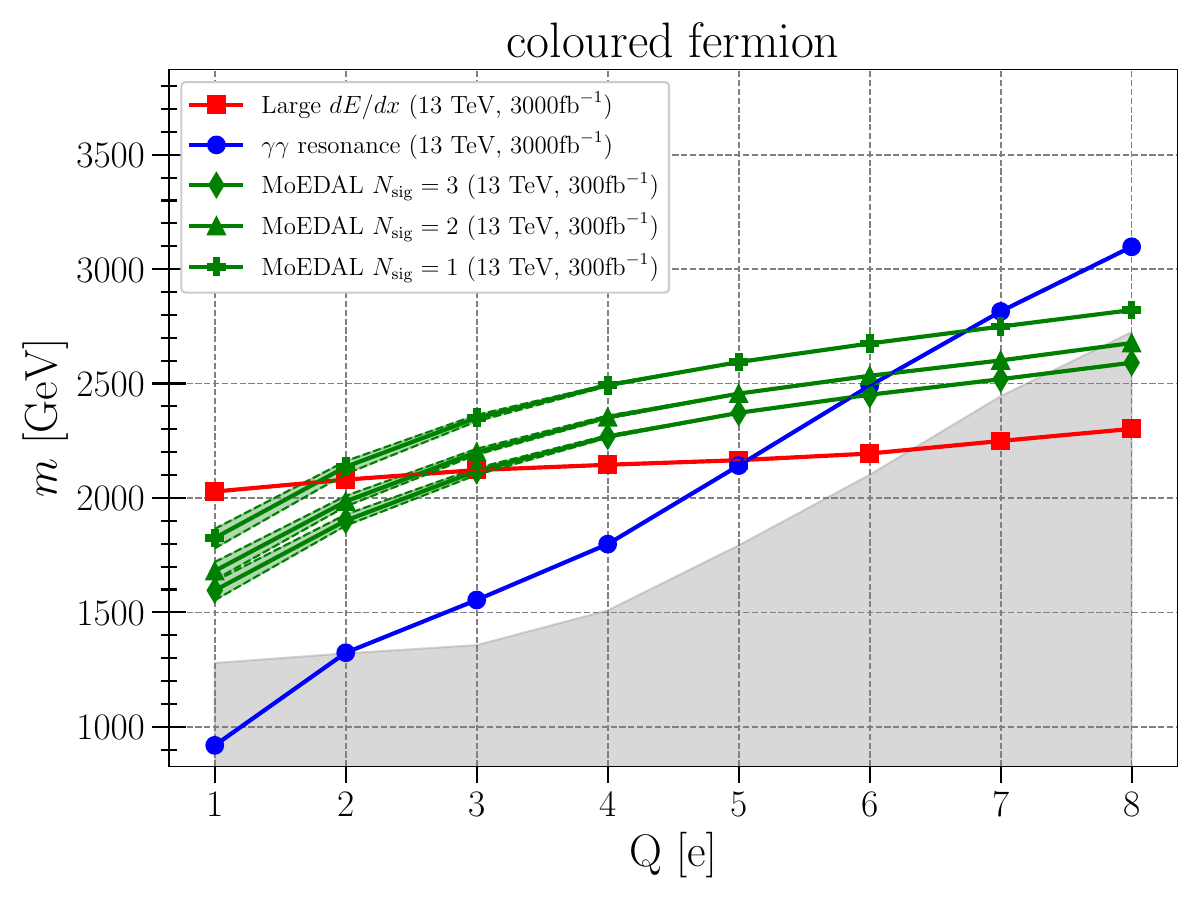}
    \caption{\small
  Sensitivities of the HL-LHC experiments for multiply charged long-lived BSM particles: colourless scalars (top left), colourless fermions (top right), coloured scalars (bottom left), and coloured fermions (bottom right). The integrated luminosity is assumed to be $300~\rm{fb}^{-1}$ for MoEDAL, and $3~\rm{ab}^{-1}$ for ATLAS and CMS. The region shaded in grey colour is already excluded (cf. Fig. \ref{fig:paper4-res_curr}). Red, blue and green curves correspond to limits provided by large $dE/dx$, diphoton resonance, and MoEDAL searches, respectively. The uncertainty of the hadronisation model for colour-triplet particles is depicted as the variable width of the curves. This effect is visible only for MoEDAL and particles with $Q \leq 3e$.
    }
    \label{fig:paper4-res_hl}
\end{figure}

In Fig. \ref{fig:paper4-res_hl} we present the expected sensitivities at the HL-LHC, where we use the same graphical conventions as in Fig. 
\ref{fig:paper4-res_run3}. From Fig. \ref{fig:paper4-res_hl} one can see that the improvement in lower mass bounds from large $dE/dx$ search is 
very mild, when compared to Run 3 estimation in Fig. \ref{fig:paper4-res_run3}. The event selection used was the same as in 
\cite{CMS:2016kce}, where the background contribution to the relevant signal region was estimated by the CMS collaboration to be 0.06 with $L=2.5~\rm{fb}^{-1}$. In the case of the HL-LHC with $L=3~\rm{ab}^{-1}$, the same contribution is expected to be 72 events, 
which demands a large number of signal events for exclusion. We expect that the large $dE/dx$ could provide better limits if 
optimised for the HL-LHC, however, it is beyond the scope of this work. When MoEDAL's sensitivity is compared to large $dE/dx$ 
and diphoton resonance searches, one can observe a significant improvement for HL-LHC, with respect to Run 3. The main reason 
is that the SM background for MoEDAL is negligibly small. While for ATLAS and CMS searches both the background and signal event rates increase with 
the luminosity, for MoEDAL only the signal increases, while the background remains zero. This creates a possibility for MoEDAL to compete with major experiments, and provide superior mass limits for LLPs with $3e \lesssim |Q| \lesssim 7e$.

\subsection{Conclusions}

In this project, we have studied prospects for the detection of long-lived particles with electric charges in the range $1e \leq |Q| \leq 8e$ at the LHC. We compared Run 3 and HL-LHC sensitivities of three distinct types of searches: (i) anomalous track searches (i.e. 
large $dE/dx$ analyses) at ATLAS and CMS, (ii) searches for charged LLPs in MoEDAL, (iii) diphoton resonance searches at ATLAS 
and CMS. The former two types of searches target the open production mode, $pp \to \xi^{+Q}\xi^{-Q}$, for which pair produced 
long-lived particles are directly detected, while the diphoton resonance searches consider the production of a bound state decaying 
into two photons: $pp \to \cal B \to \gamma \gamma$. 

We have proven the relevance of the photon fusion on the open mode production cross section for highly charged particles, which 
was not considered by the ATLAS and CMS large $dE/dx$ analyses. Moreover, we have also investigated the impact of the photon 
fusion production on the MoEDAL's sensitivity, which is twofold: (i) it significantly increases the total production cross section for 
particles with $Q \gtrsim 4e$, (ii) it alters the velocity distribution of produced particles, especially colourless scalars, such that 
more particles can be detected in MoEDAL.

To estimate the current and projected bounds from large $dE/dx$ searches, we recast the CMS analysis \cite{CMS:2016kce} for various types of 
charged long-lived BSM particles, including the photonic initial states. We investigated the effect of underestimation of the 
transverse momentum for multiply charged particles, as well as velocity loss due to (strong) electromagnetic interaction with the 
detector medium. Both effects become stronger for larger charges and lead to reduced efficiency, therefore the large $dE/dx$ 
searches were found to be the most powerful for smaller $Q$.

The sensitivity of the diphoton resonance search, on the other hand, expands rapidly for larger charges, since the event rate of $pp \to \cal B \to \gamma \gamma$ grows as $Q^{10}$.

We revealed that the sensitivity of MoEDAL is typically between those of diphoton resonance studies and anomalous track 
searches. Run 3 MoEDAL might be competitive to ATLAS and CMS for $|Q|\sim(5-6)e$. The advantage of MoEDAL over ATLAS and 
CMS comes from the lack of the SM background and becomes evident for the HL-LHC data-taking phase, for which MoEDAL can 
provide superior sensitivity for $3e \lesssim |Q|\lesssim 7e$ than the two general-purpose experiments.

Although the derived mass bounds rely on the leading-order cross section calculations, our conclusion is qualitatively robust.
However, one has to remember that our estimates of the projected 
sensitivities of the large $dE/dx$ and diphoton resonance searches rely on event selection designed for Run 2 and 
conservative assumptions on the systematic uncertainties.

\subsection{Appendix}\label{sec:paper4-appendix}

In this section we provide a model-independent mass reach ($N_{\rm sig} \ge 1$, $2$ and $3$)
of the MoEDAL experiment
for the long-lived multiply charged particles.
The results are obtained both for Run 3 ($L = 30$ fb$^{-1}$) 
and HL-LHC ($L = 300$ fb$^{-1}$), and presented
in the mass versus lifetime planes ($m$, $c \tau$).
Figs.~\ref{fig:lim_sHighQ},
\ref{fig:lim_fHighQ},
\ref{fig:lim_csHighQ} and
\ref{fig:lim_cfHighQ}
correspond to colour-singlet scalars,
colour-singlet fermions, 
colour-triplet scalars 
and colour-triplet fermions, respectively.

\begin{figure}[!b]
\centering
      \includegraphics[width=0.4\textwidth]{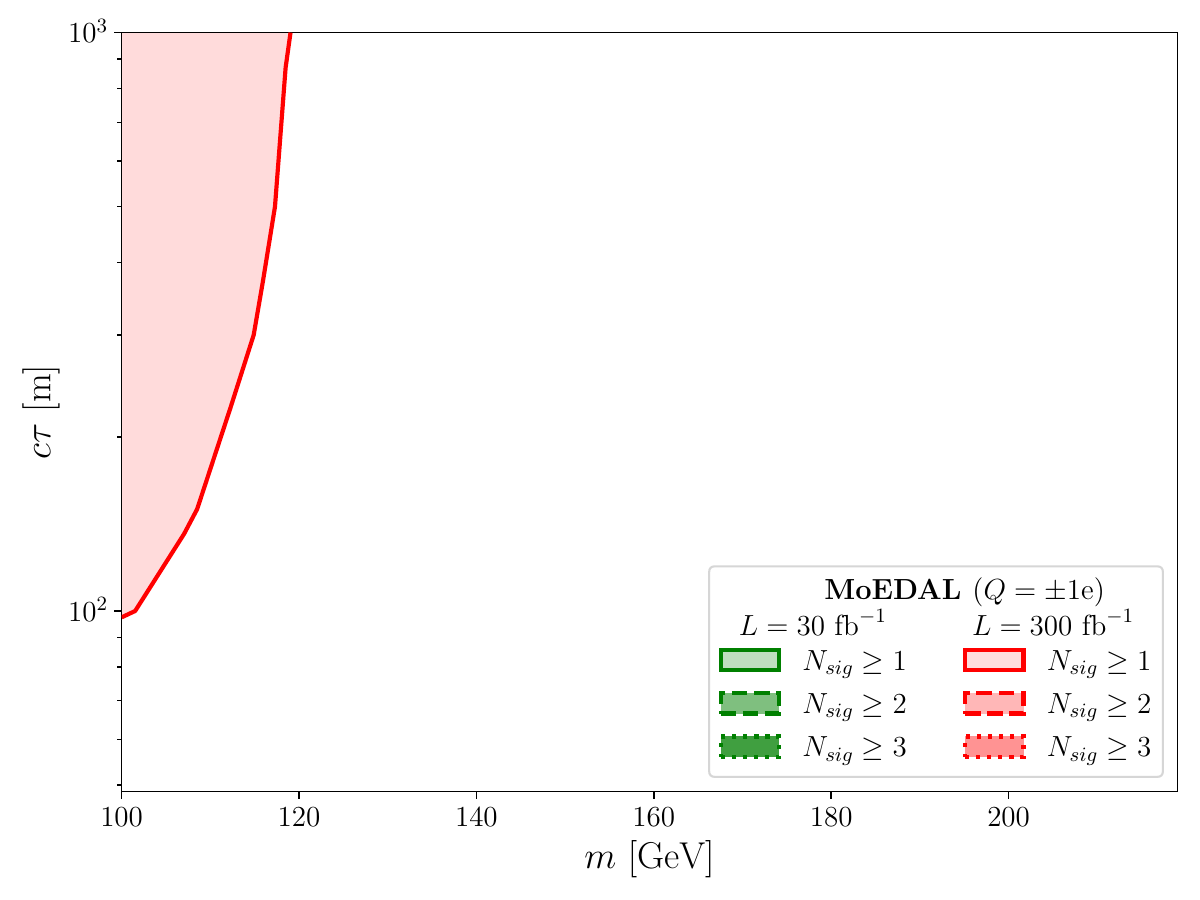} \hspace{5mm}
      \includegraphics[width=0.4\textwidth]{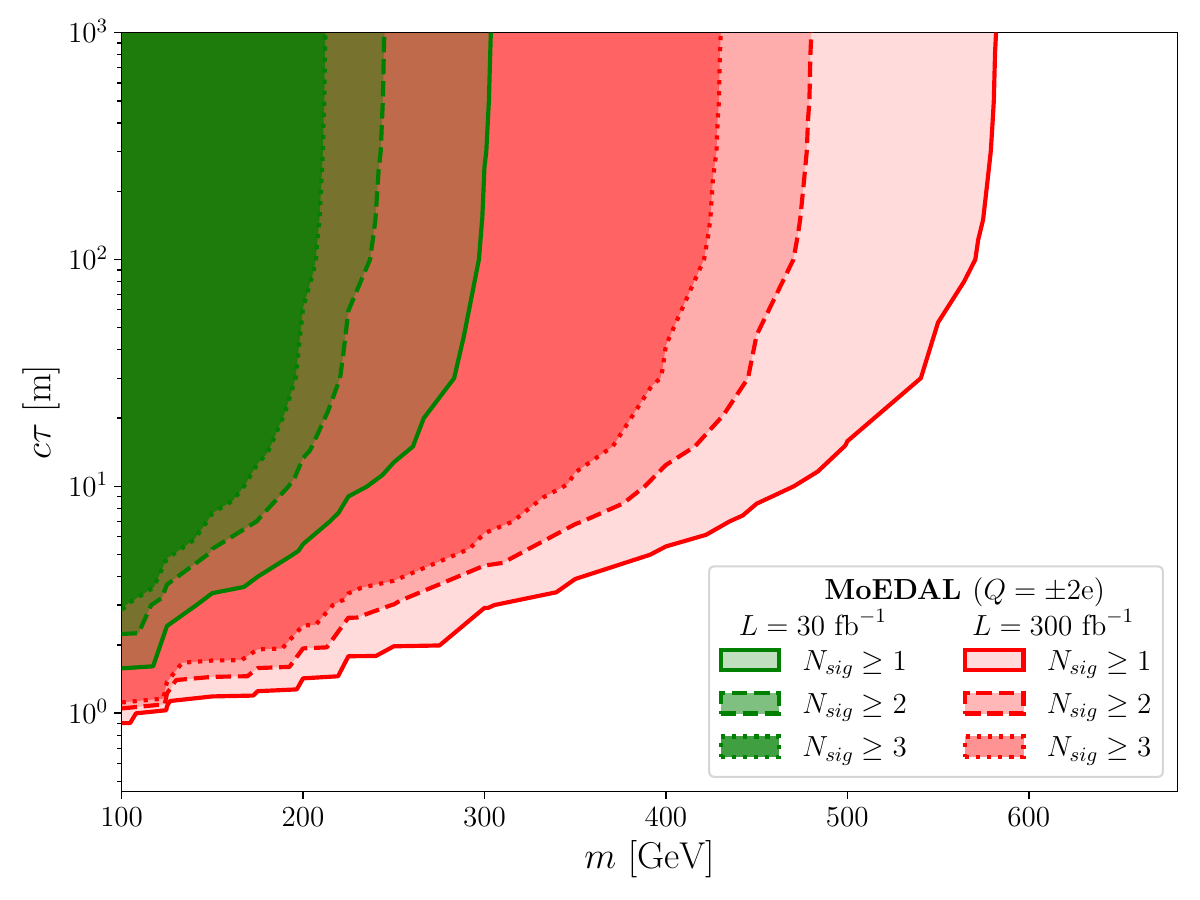}
      \includegraphics[width=0.4\textwidth]{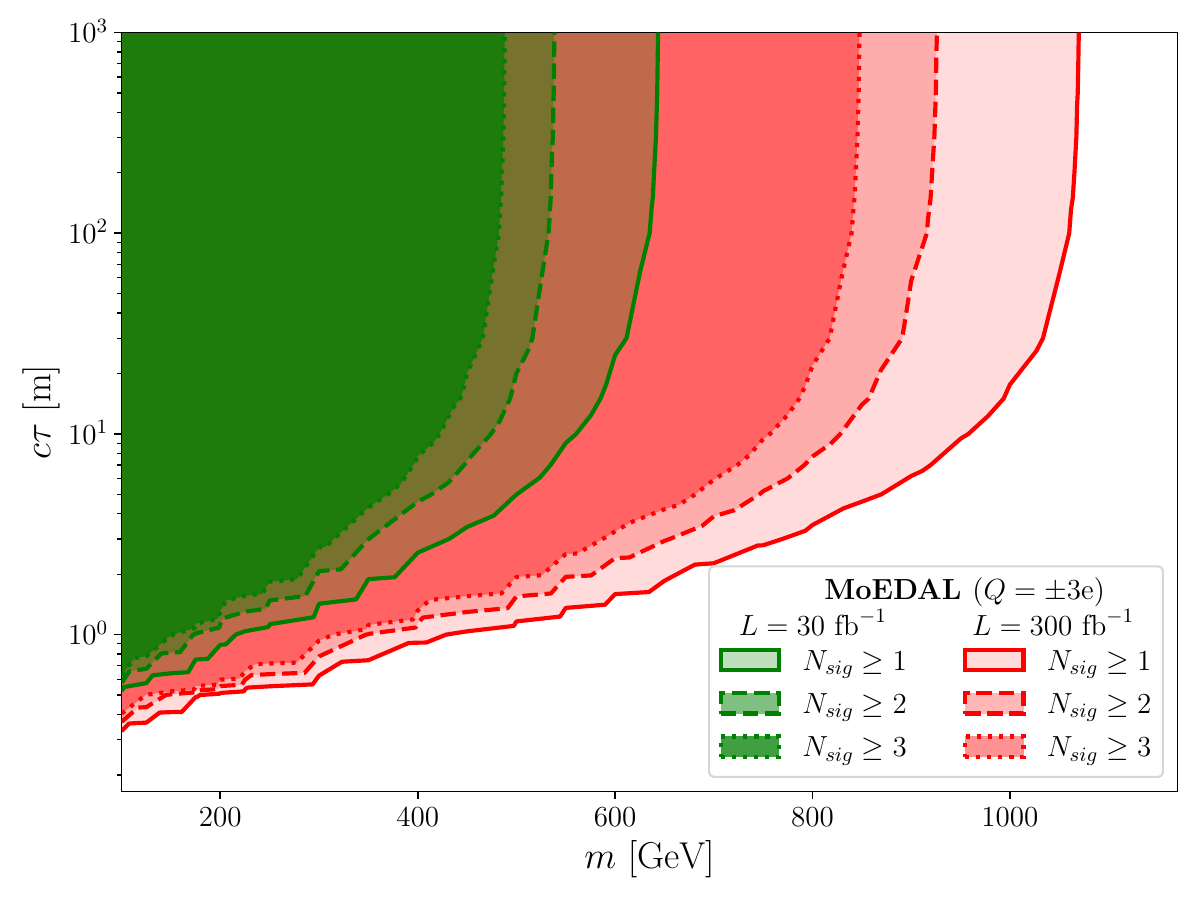}  \hspace{5mm}
      \includegraphics[width=0.4\textwidth]{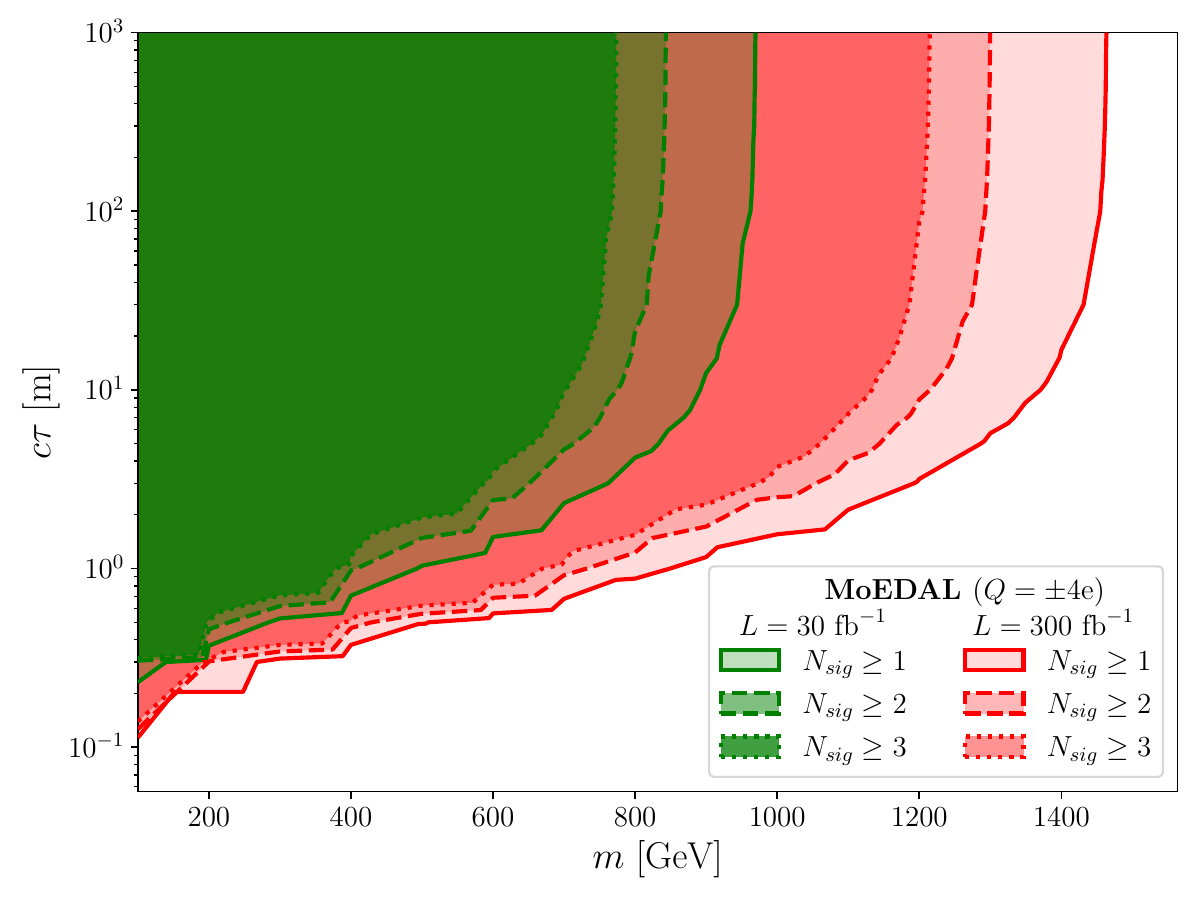}
      \includegraphics[width=0.4\textwidth]{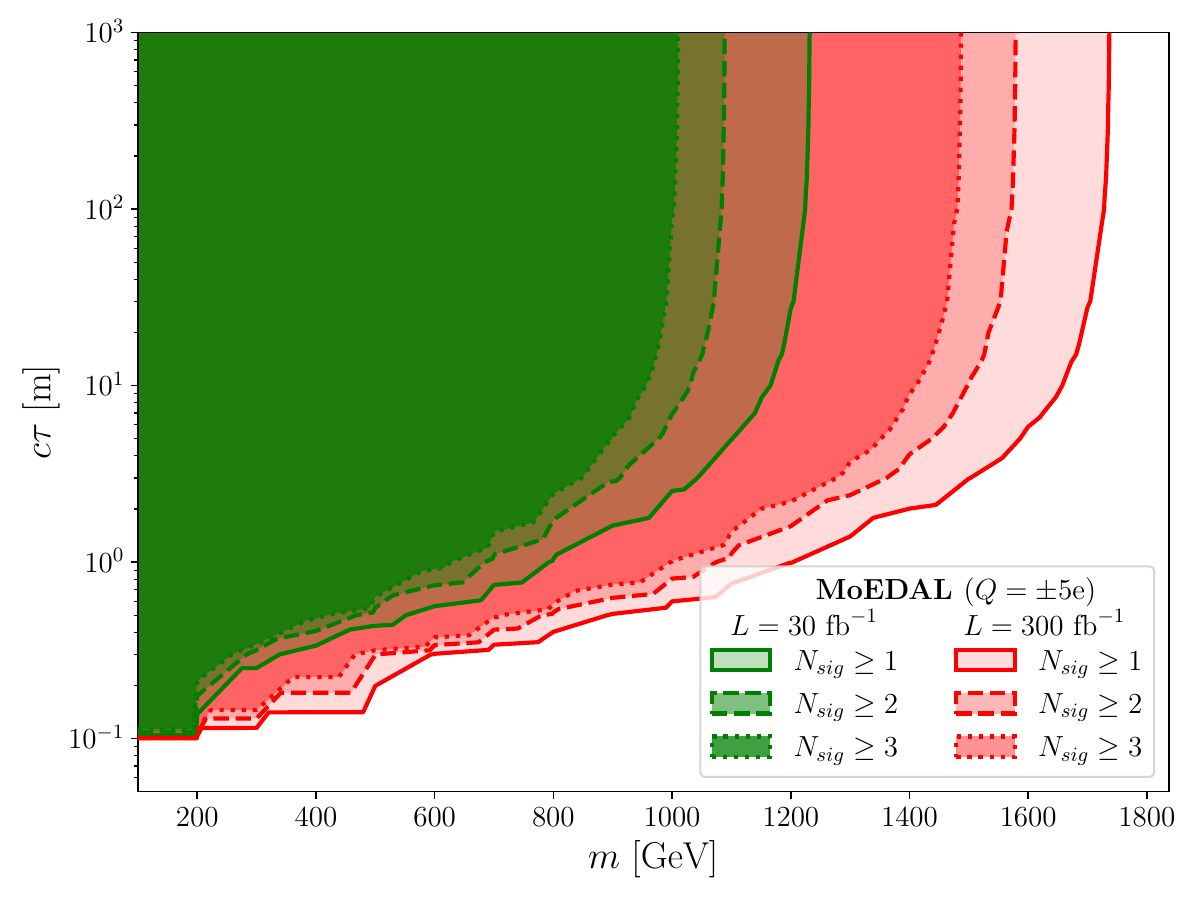}  \hspace{5mm}
      \includegraphics[width=0.4\textwidth]{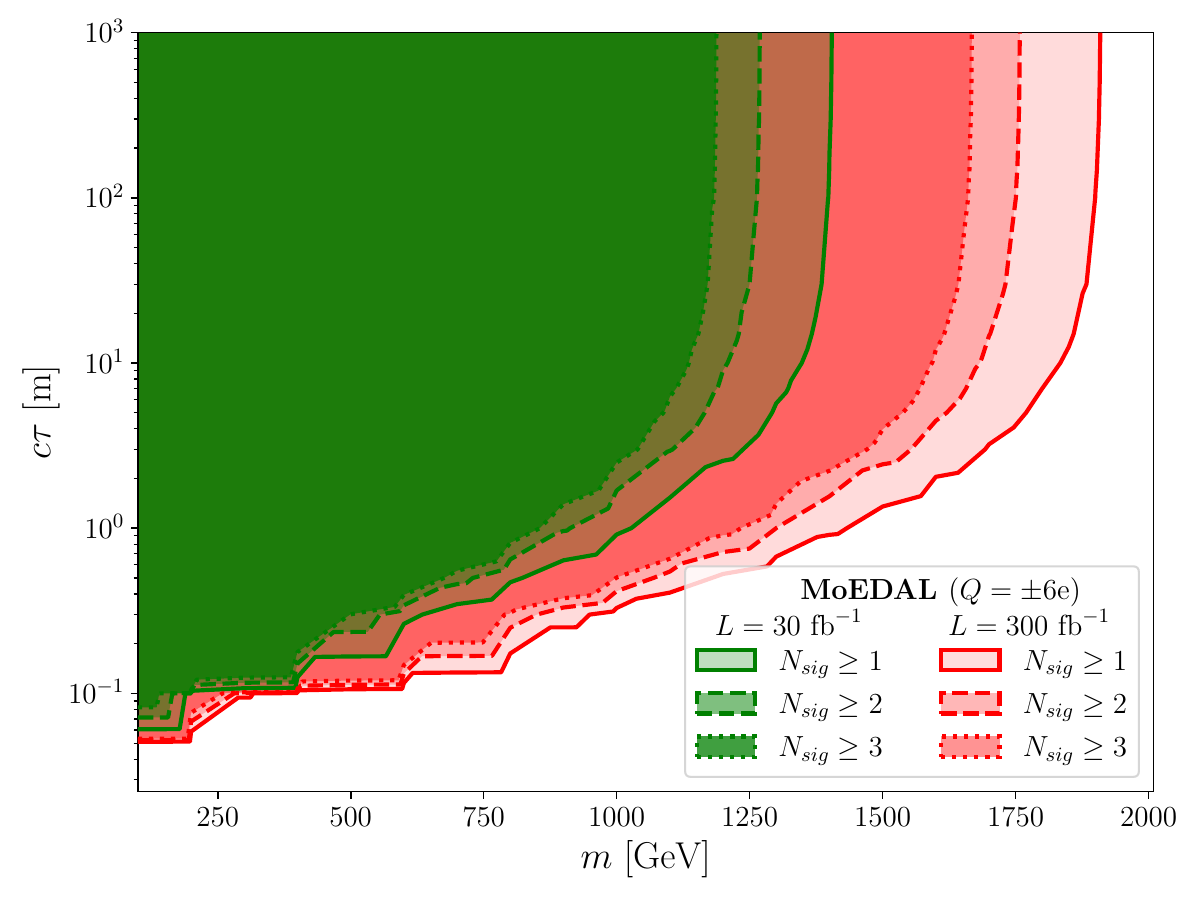}
      \includegraphics[width=0.4\textwidth]{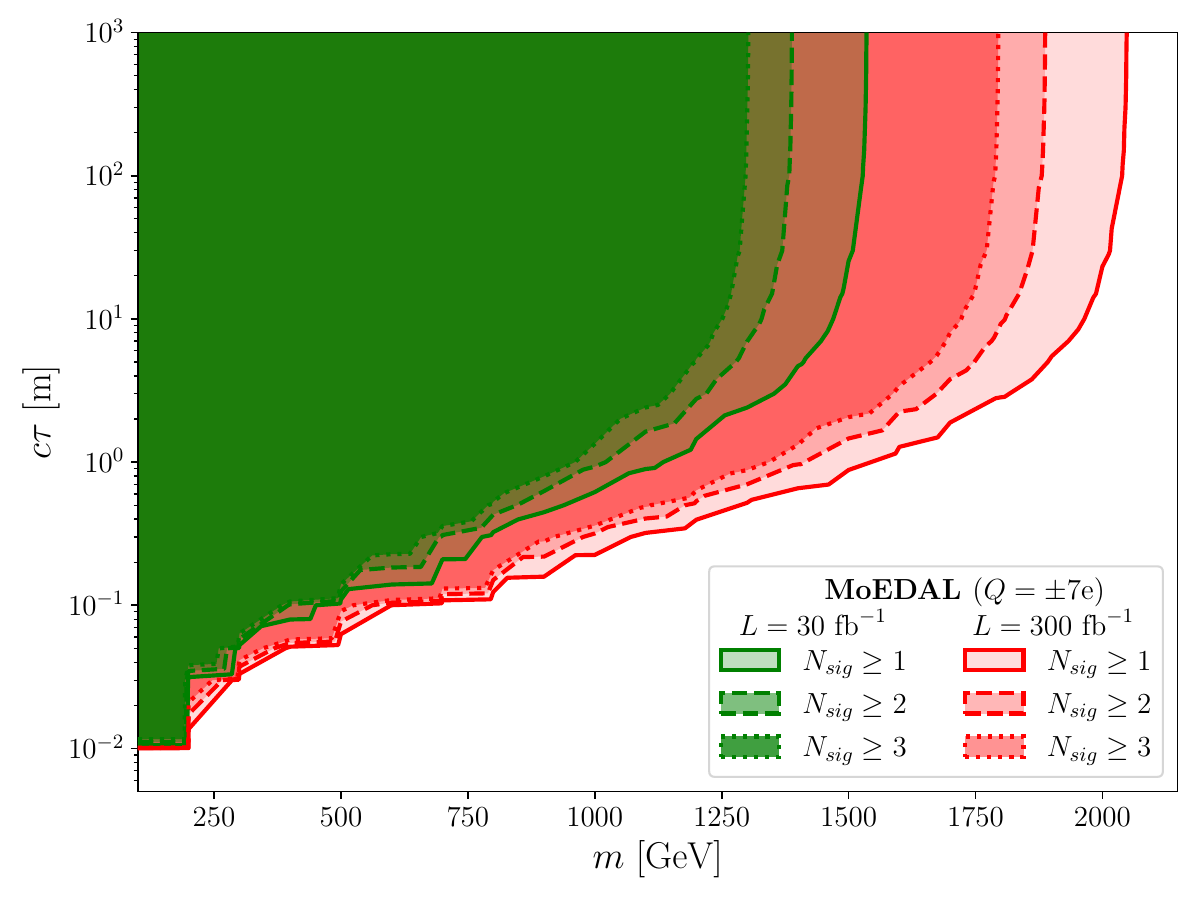}  \hspace{5mm}
      \includegraphics[width=0.4\textwidth]{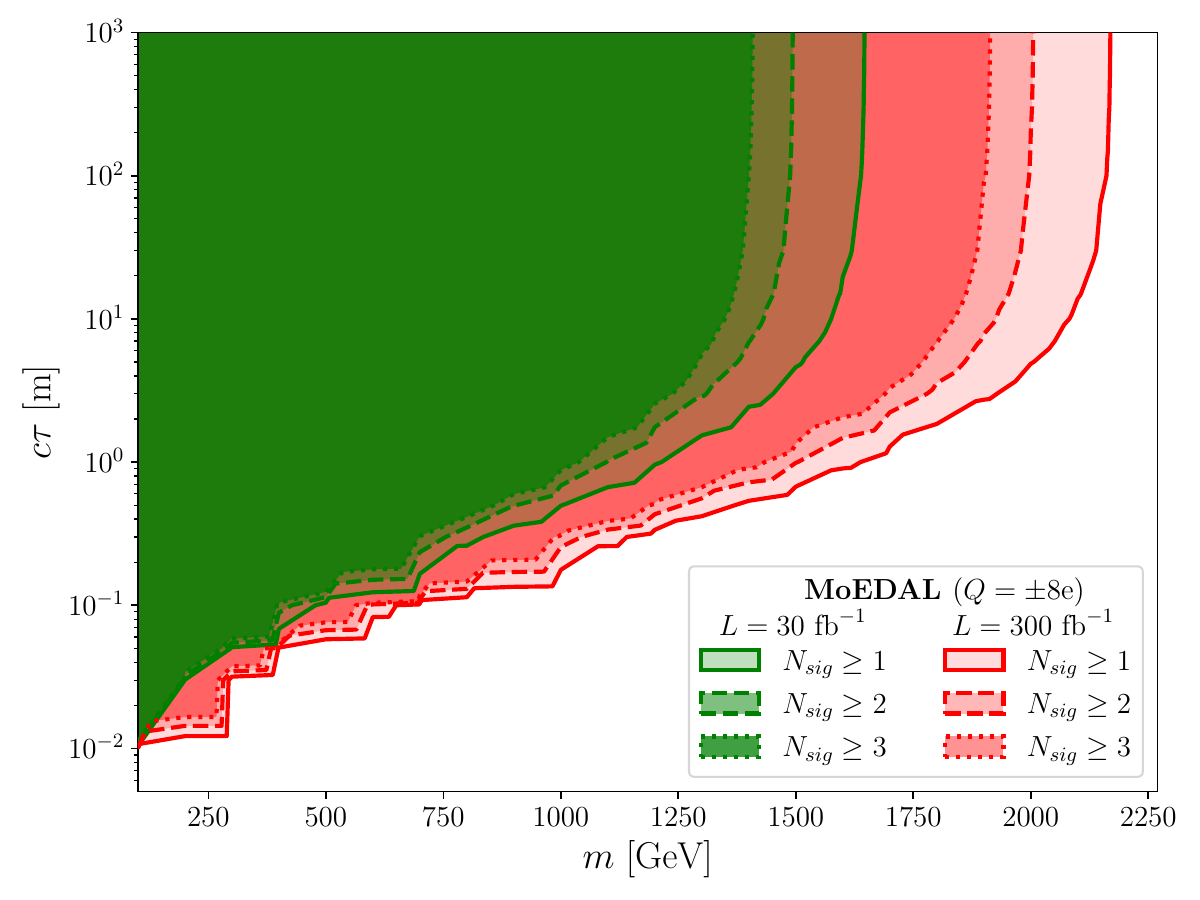}  
\caption{\small Model-independent detection reach of MoEDAL in the ($m$, $c
  \tau$) parameter plane for $SU(3)_C$-singlet scalars. Solid, dashed, and dotted contour lines correspond to $N_{\rm sig} = 1$, 2 and 3, respectively. Green and red colours represent results for Run 3 
  $(L=30$ $\mathrm{fb}{}^{-1})$ and HL-LHC $(L=300$ $\mathrm{fb}^{-1})$ data taking phases, respectively.
  }
\label{fig:lim_sHighQ}
\end{figure}

\begin{figure}[t!]
\centering
      \includegraphics[width=0.4\textwidth]{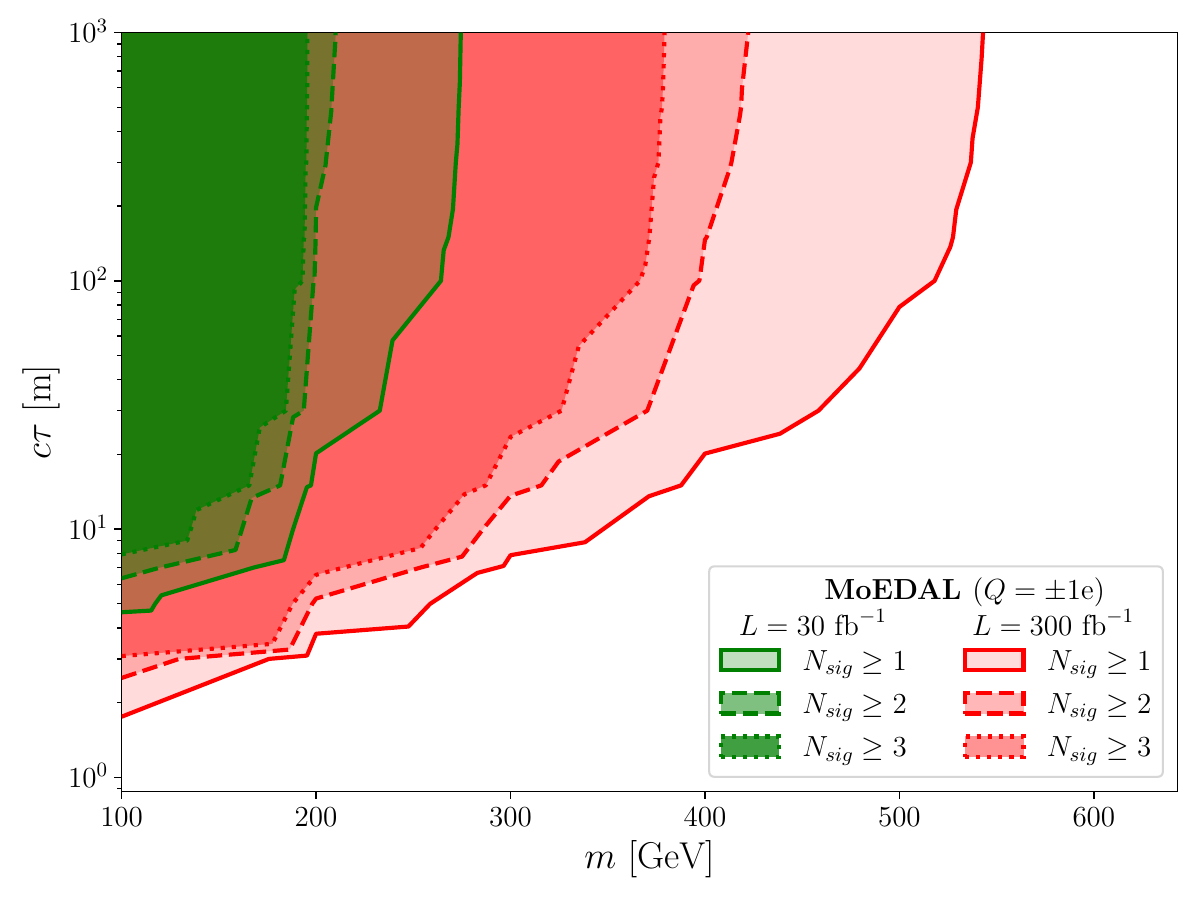} \hspace{5mm}
      \includegraphics[width=0.4\textwidth]{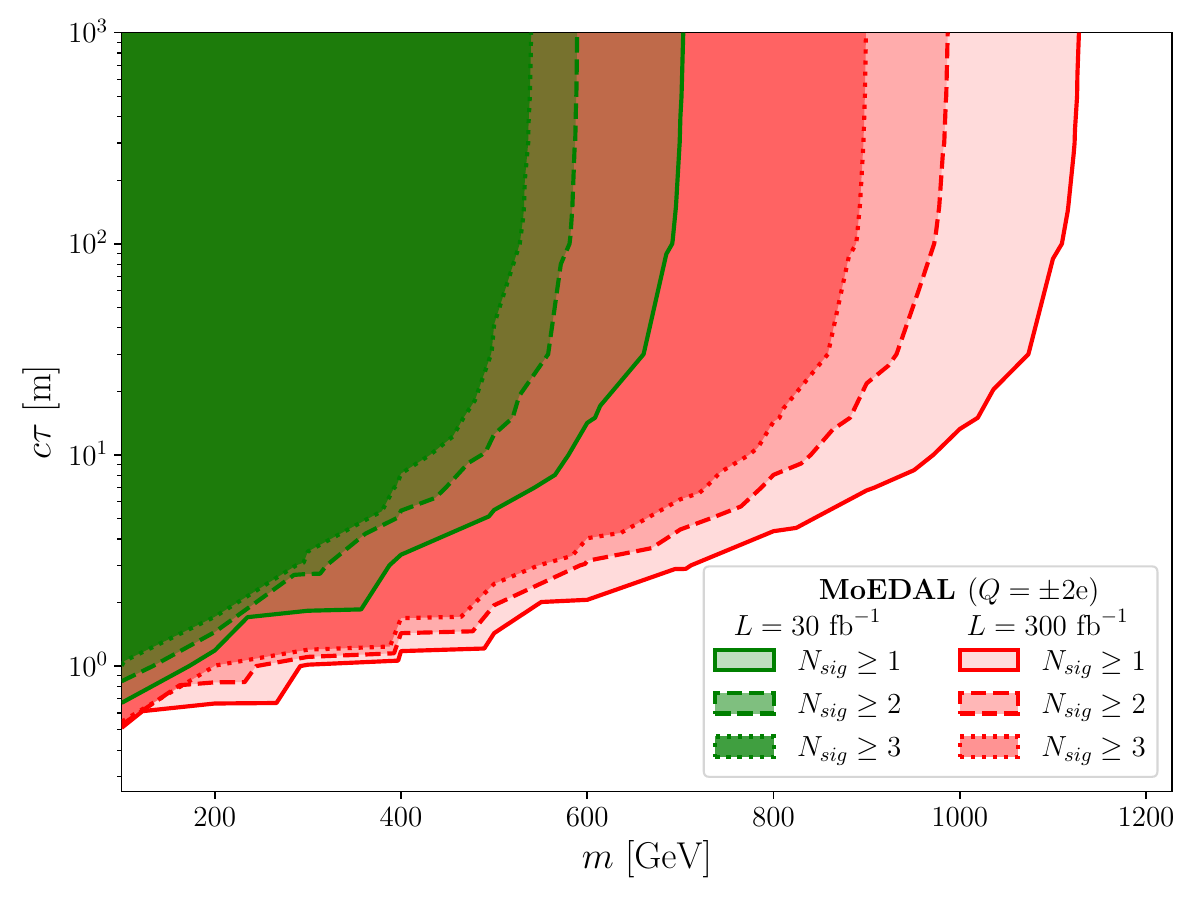}
      \includegraphics[width=0.4\textwidth]{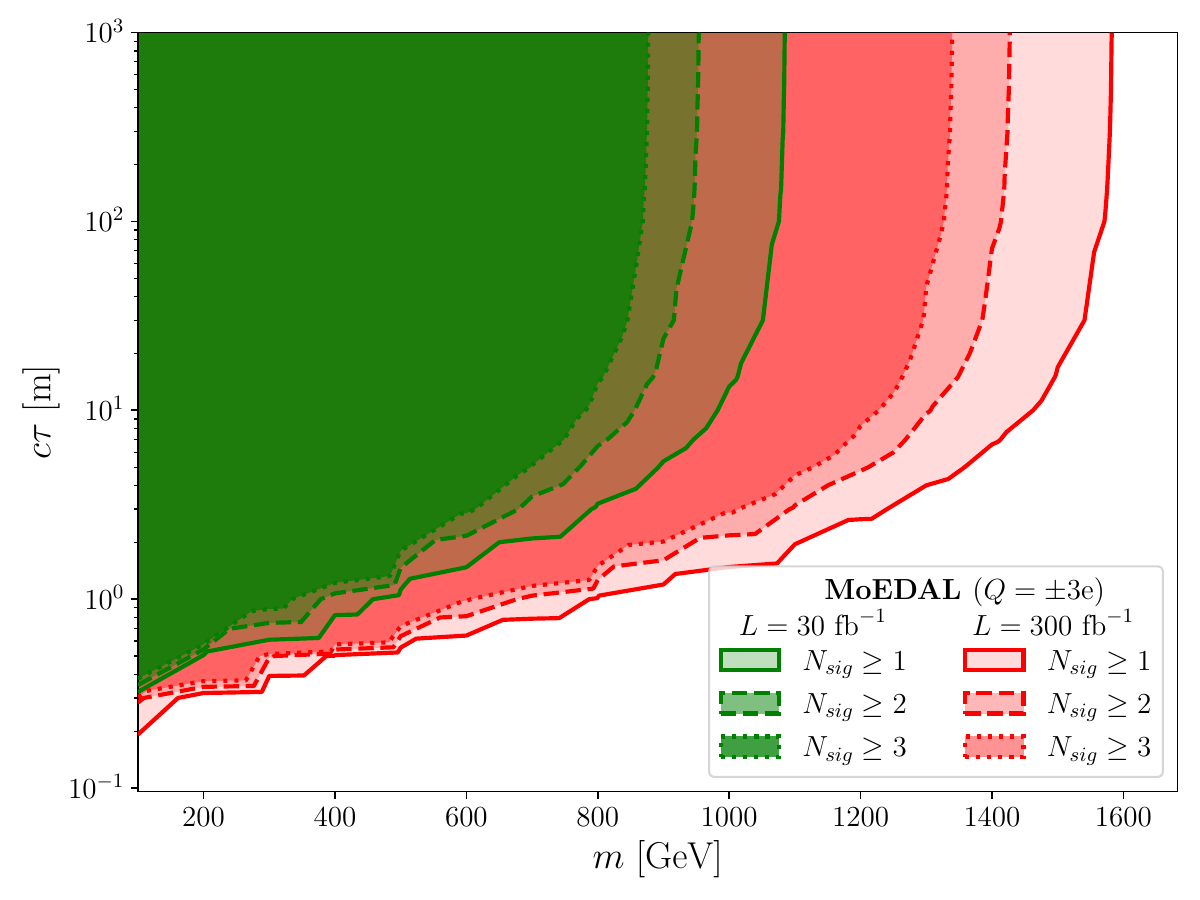}  \hspace{5mm}
      \includegraphics[width=0.4\textwidth]{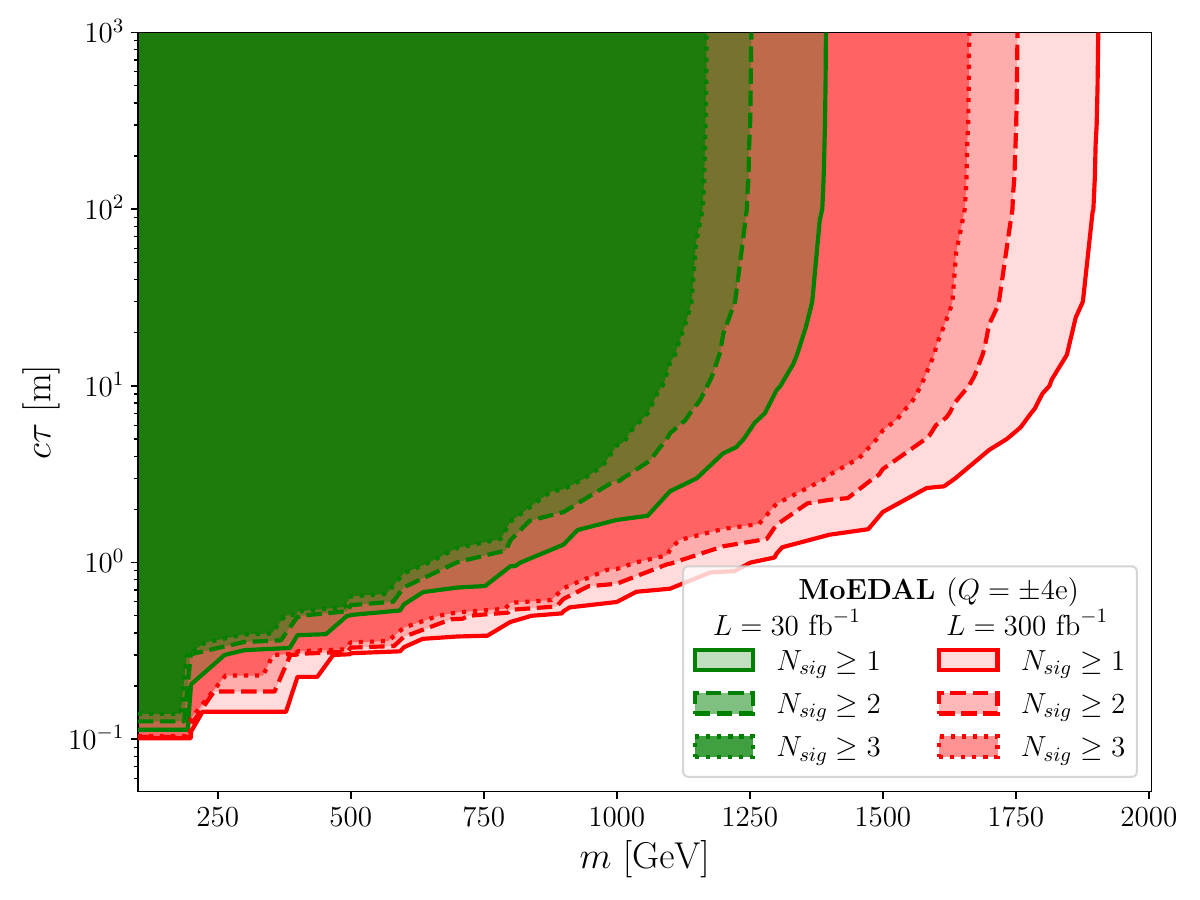}
      \includegraphics[width=0.4\textwidth]{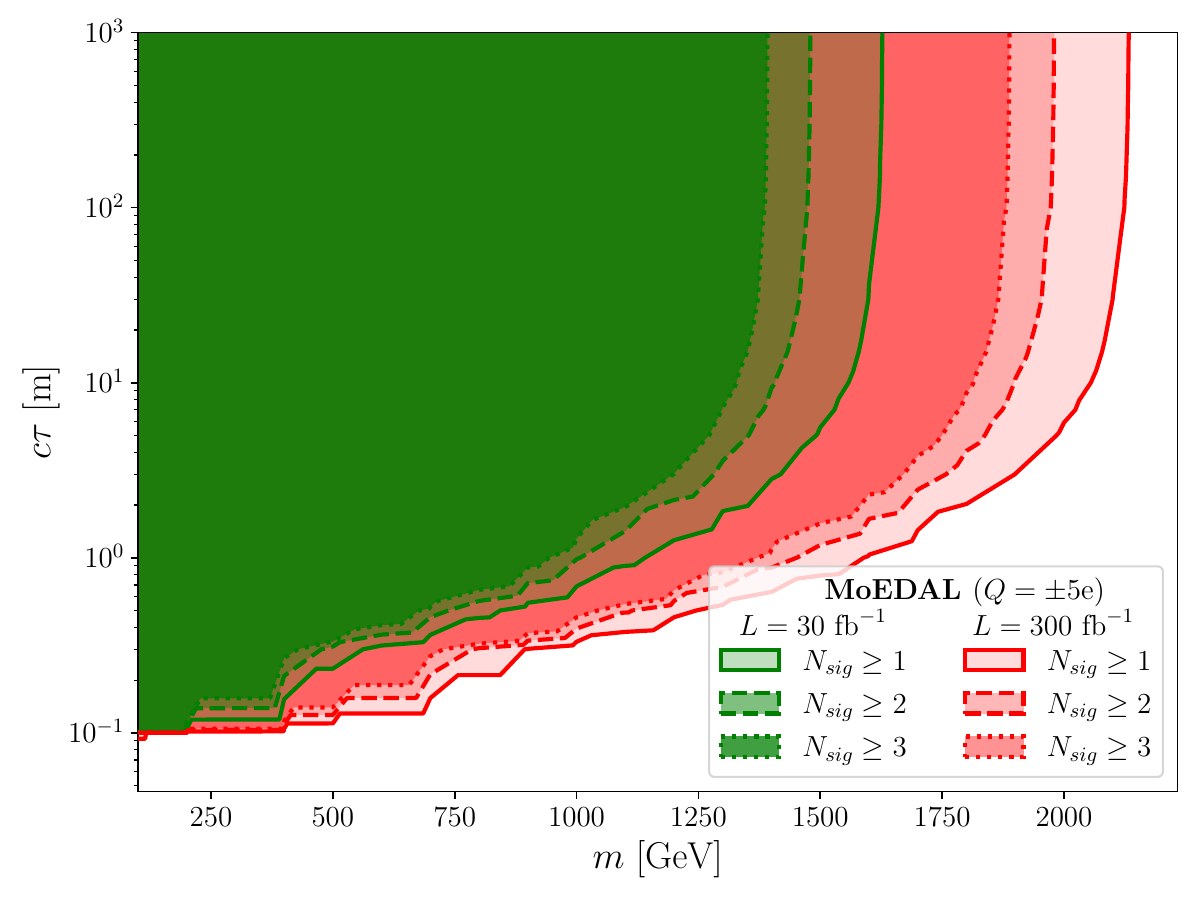}  \hspace{5mm}
      \includegraphics[width=0.4\textwidth]{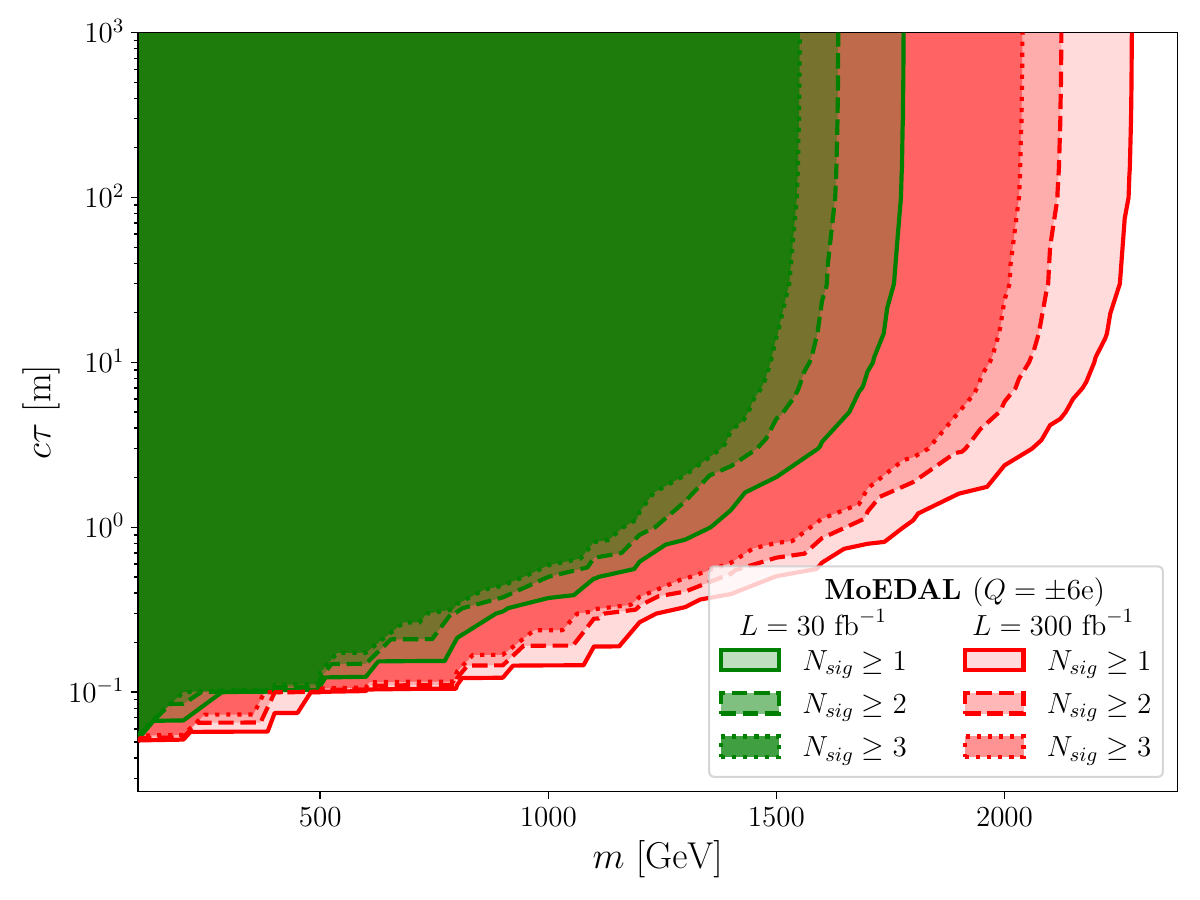}
      \includegraphics[width=0.4\textwidth]{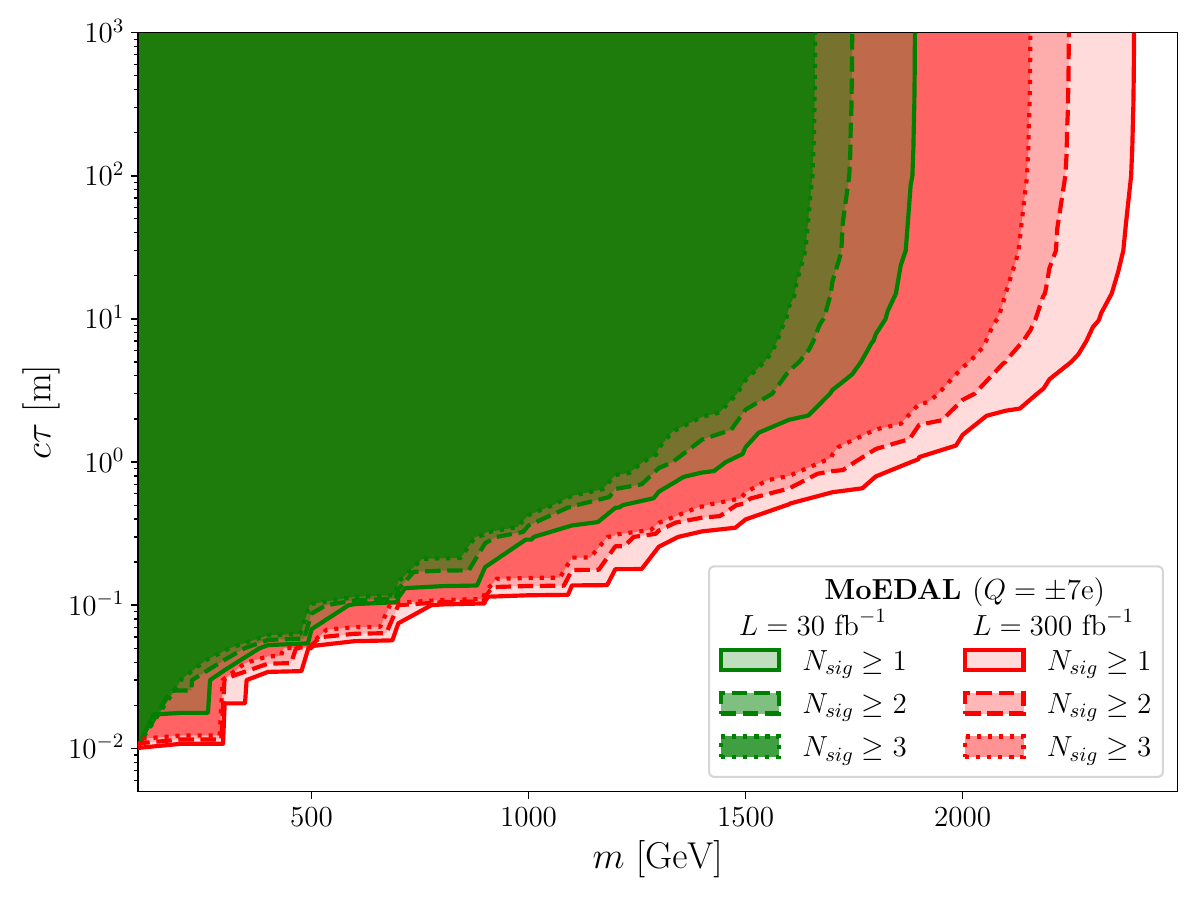}  \hspace{5mm}
      \includegraphics[width=0.4\textwidth]{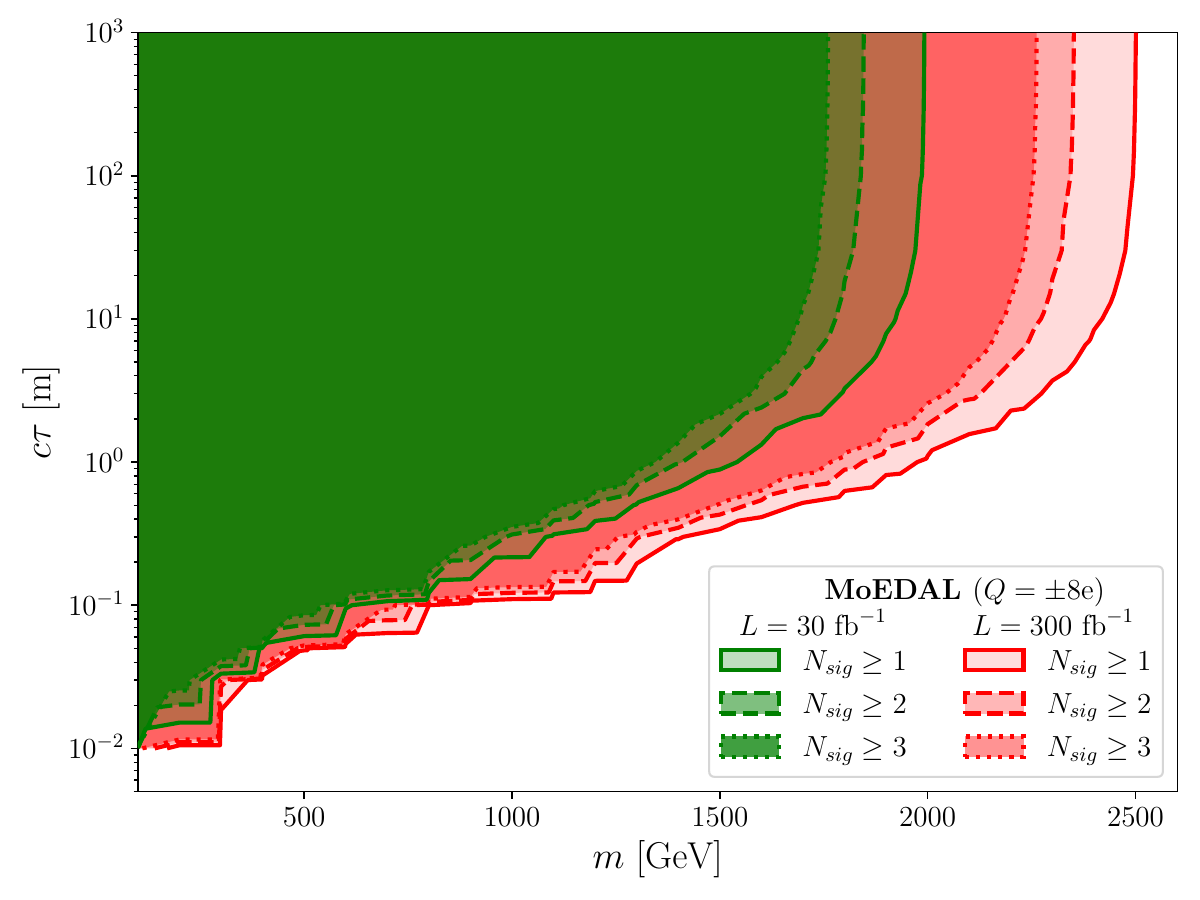}  
\caption{\small Model-independent detection reach of MoEDAL in the ($m$, $c
  \tau$) parameter plane for $SU(3)_C$-singlet fermions. Solid, dashed, and dotted contour lines correspond to $N_{\rm sig} = 1$, 2 and 3, respectively. Green and red colours represent results for Run 3 
  $(L=30$ $\mathrm{fb}{}^{-1})$ and HL-LHC $(L=300$ $\mathrm{fb}^{-1})$ data taking phases, respectively.
  }
\label{fig:lim_fHighQ}
\end{figure}

\begin{figure}[t!]
\centering
      \includegraphics[width=0.4\textwidth]{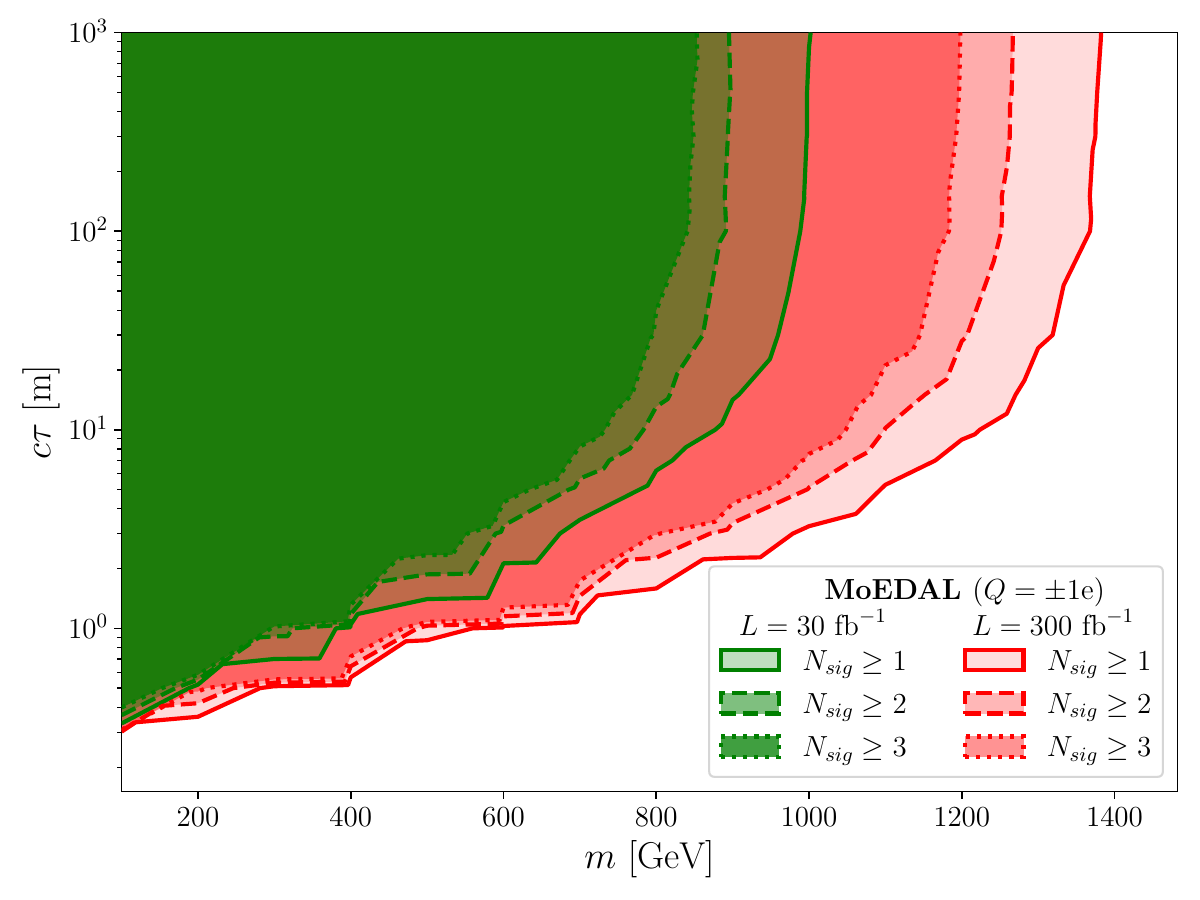} \hspace{5mm}
      \includegraphics[width=0.4\textwidth]{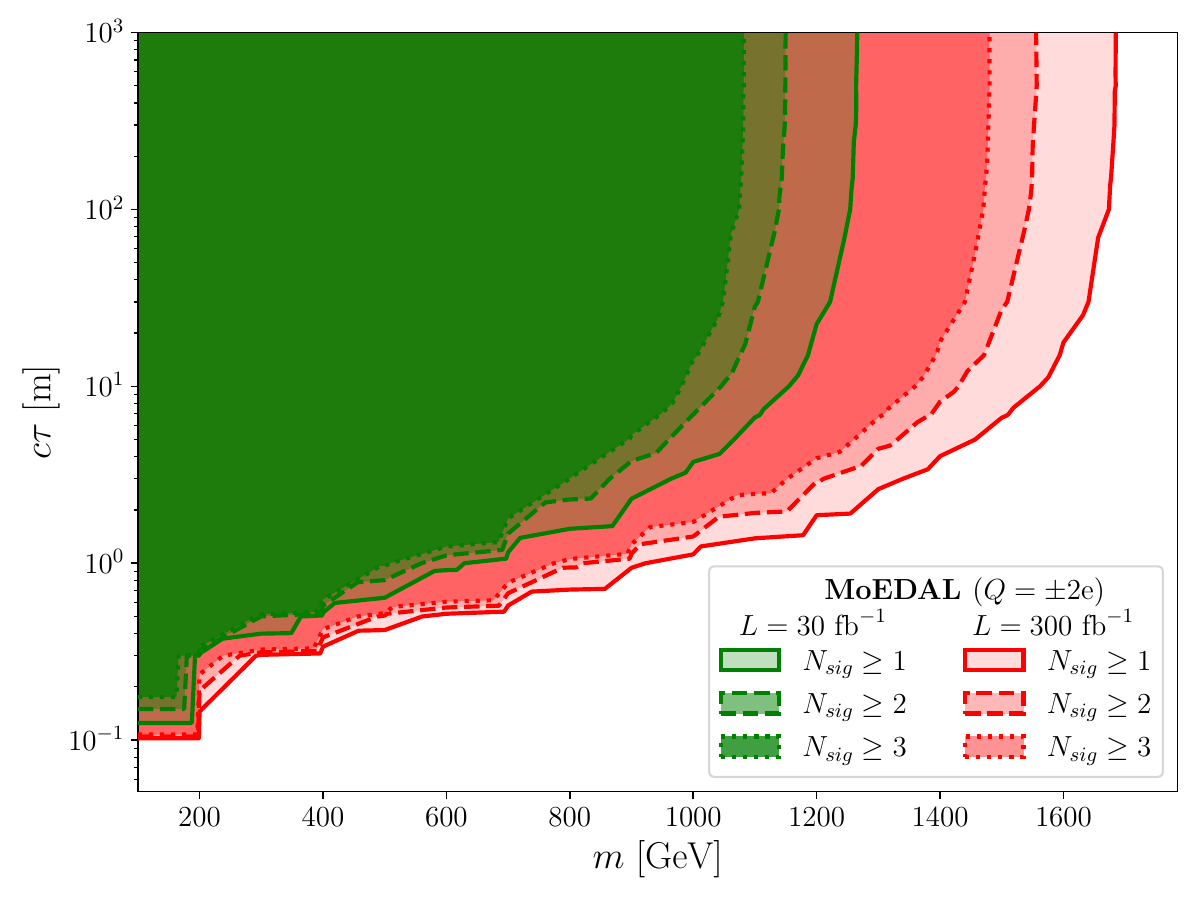}
      \includegraphics[width=0.4\textwidth]{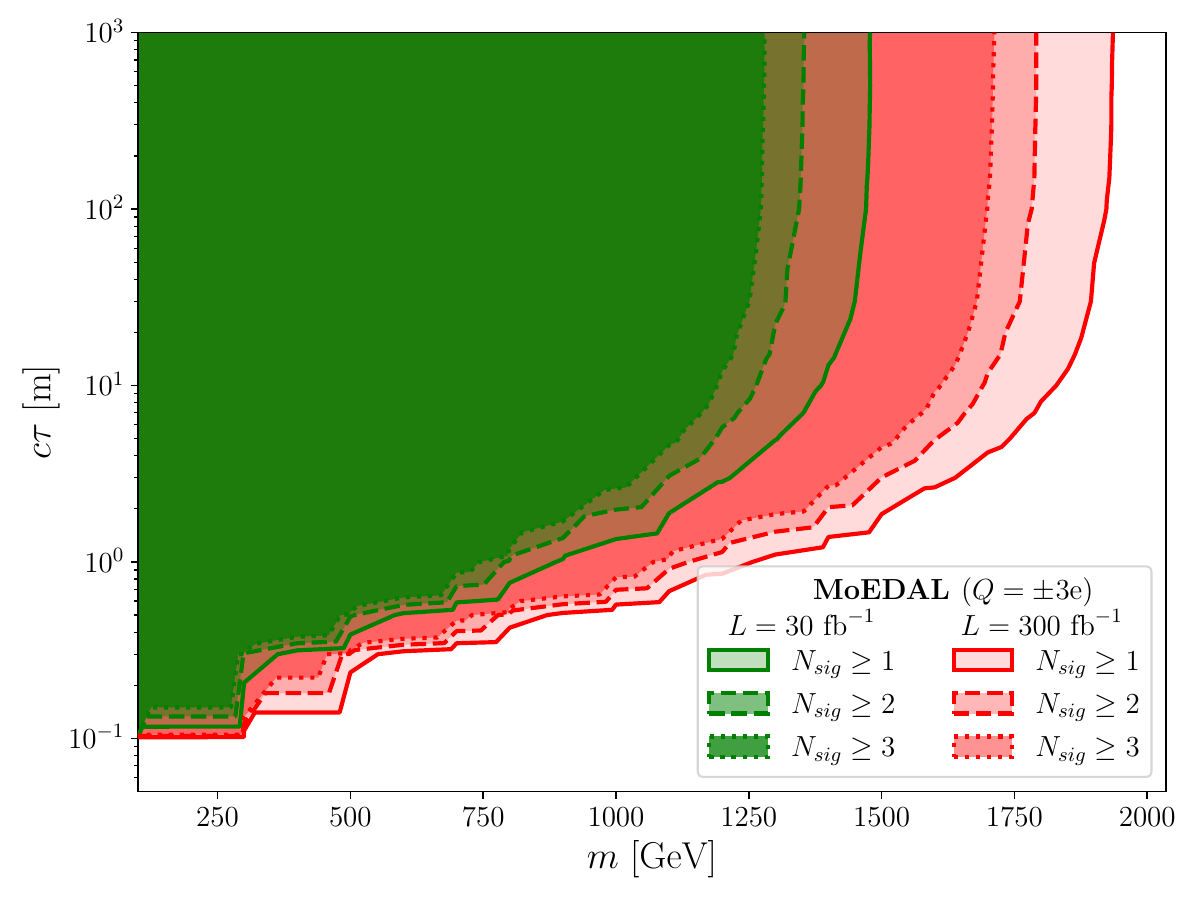}  \hspace{5mm}
      \includegraphics[width=0.4\textwidth]{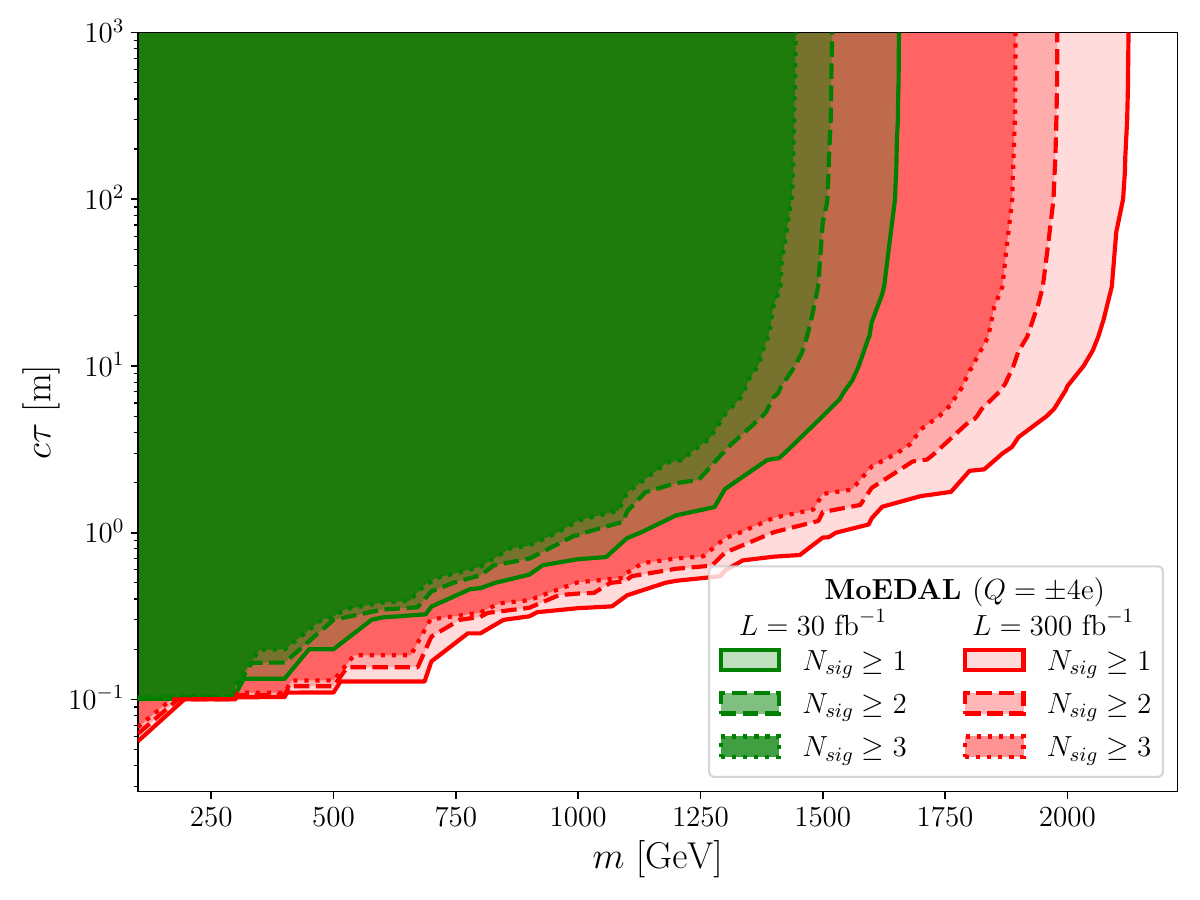}
      \includegraphics[width=0.4\textwidth]{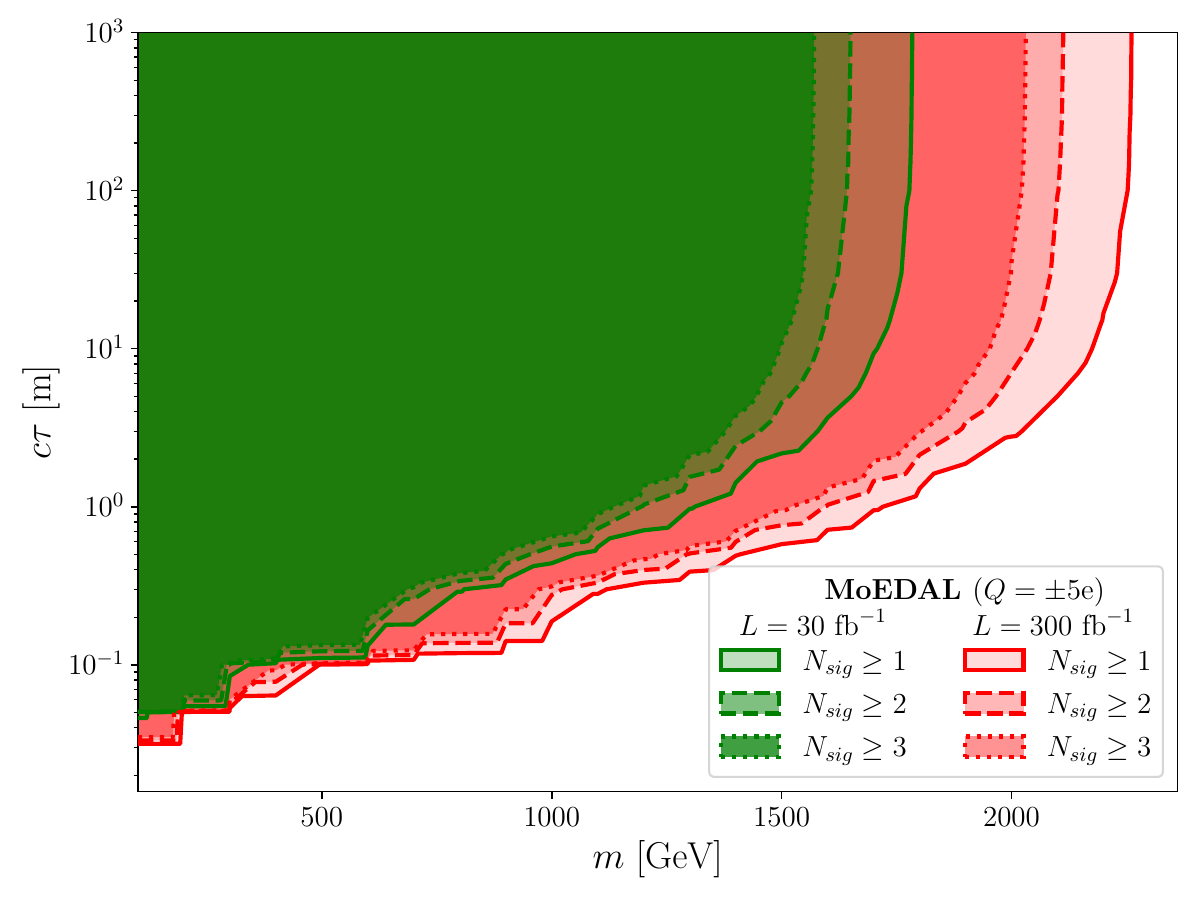}  \hspace{5mm}
      \includegraphics[width=0.4\textwidth]{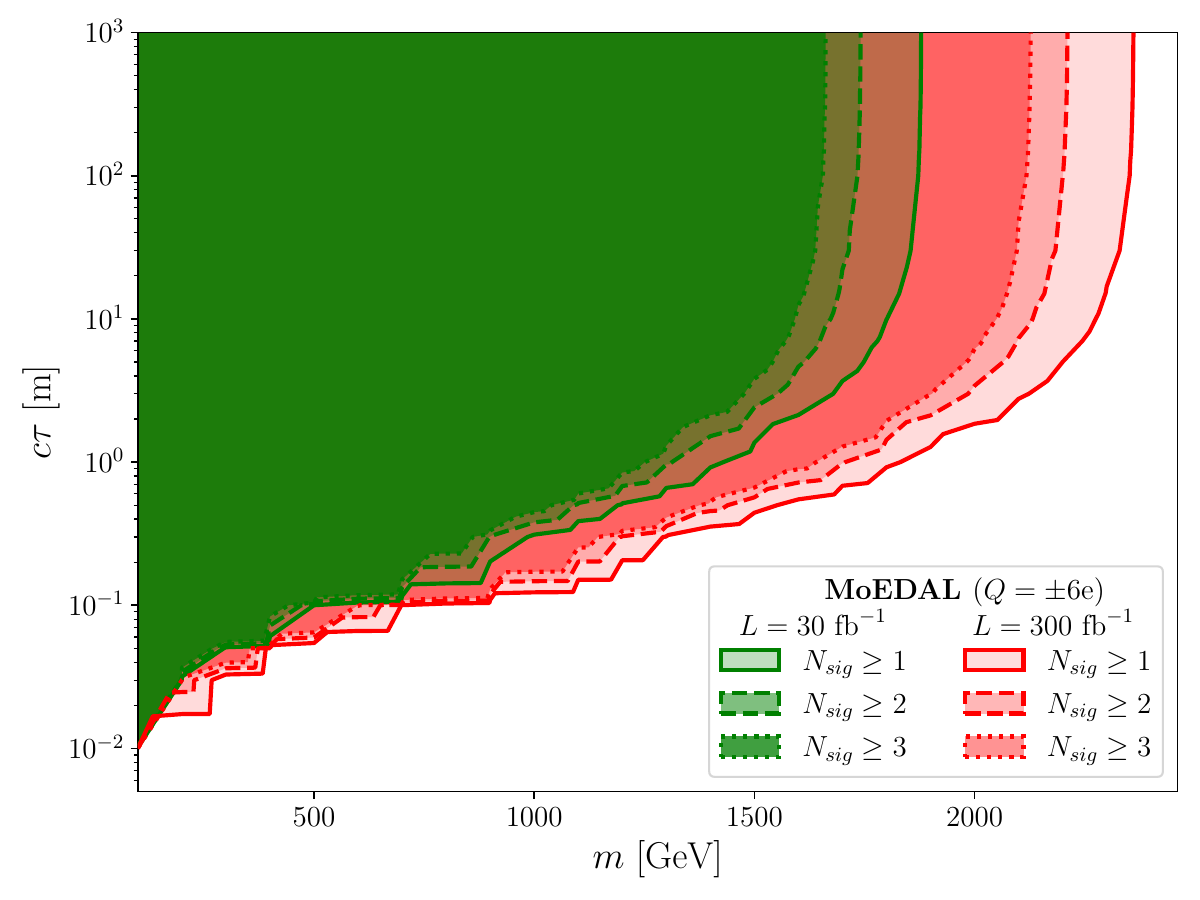}
      \includegraphics[width=0.4\textwidth]{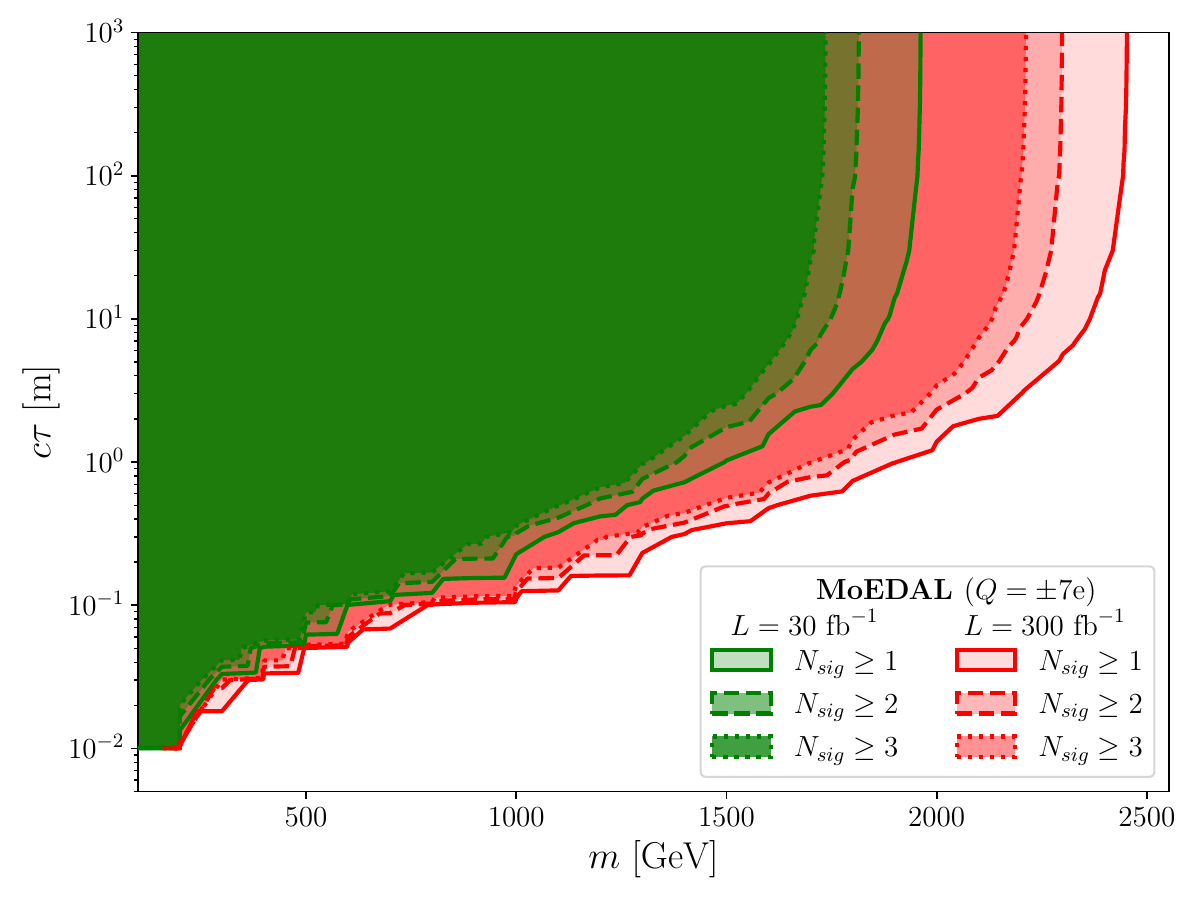}  \hspace{5mm}
      \includegraphics[width=0.4\textwidth]{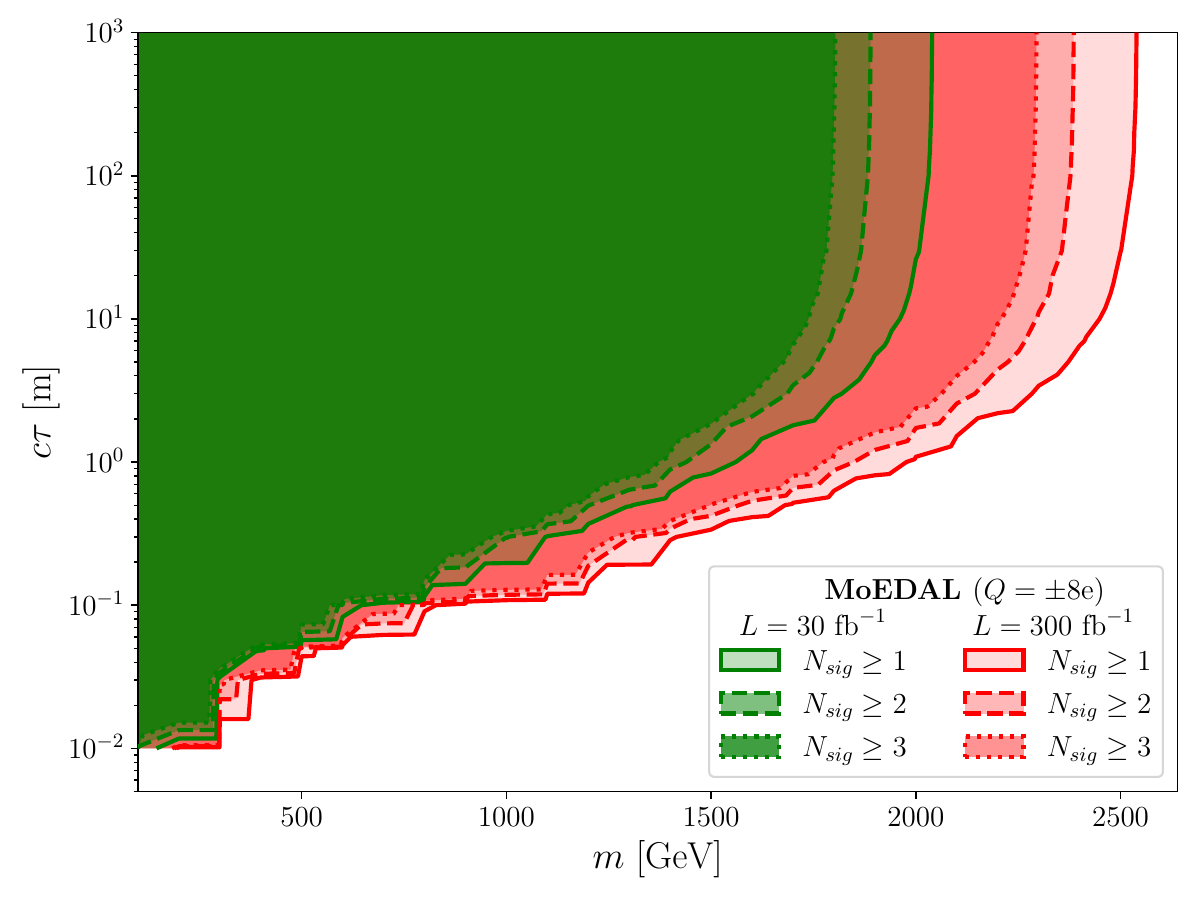}  
\caption{\small Model-independent detection reach of MoEDAL in the ($m$, $c
  \tau$) parameter plane for $SU(3)_C$-triplet scalars. Solid, dashed, and dotted contour lines correspond to $N_{\rm sig} = 1$, 2 and 3, respectively. Green and red colours represent results for Run 3 
  $(L=30$ $\mathrm{fb}{}^{-1})$ and HL-LHC $(L=300$ $\mathrm{fb}^{-1})$ data taking phases, respectively.
  }
\label{fig:lim_csHighQ}
\end{figure}

\begin{figure}[t!]
\centering
      \includegraphics[width=0.4\textwidth]{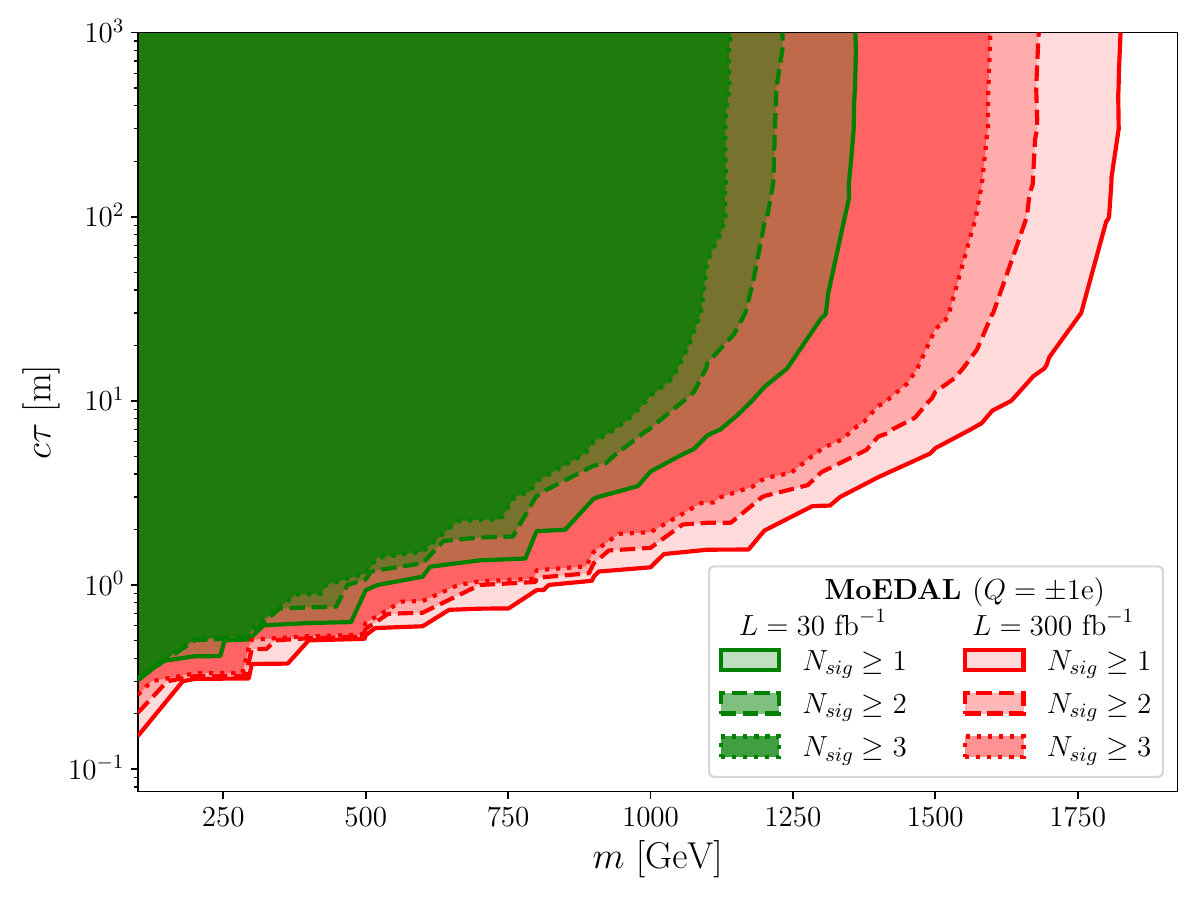} \hspace{5mm}
      \includegraphics[width=0.4\textwidth]{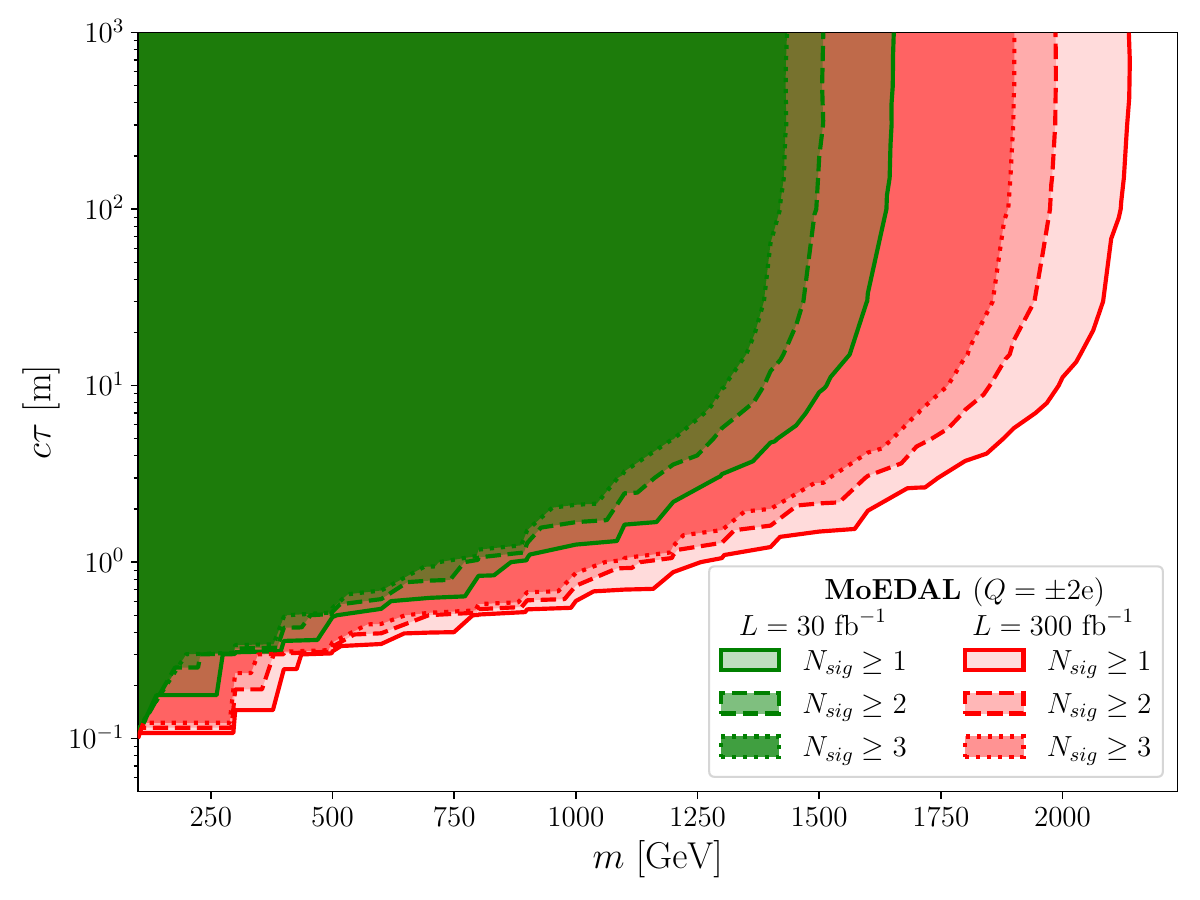}
      \includegraphics[width=0.4\textwidth]{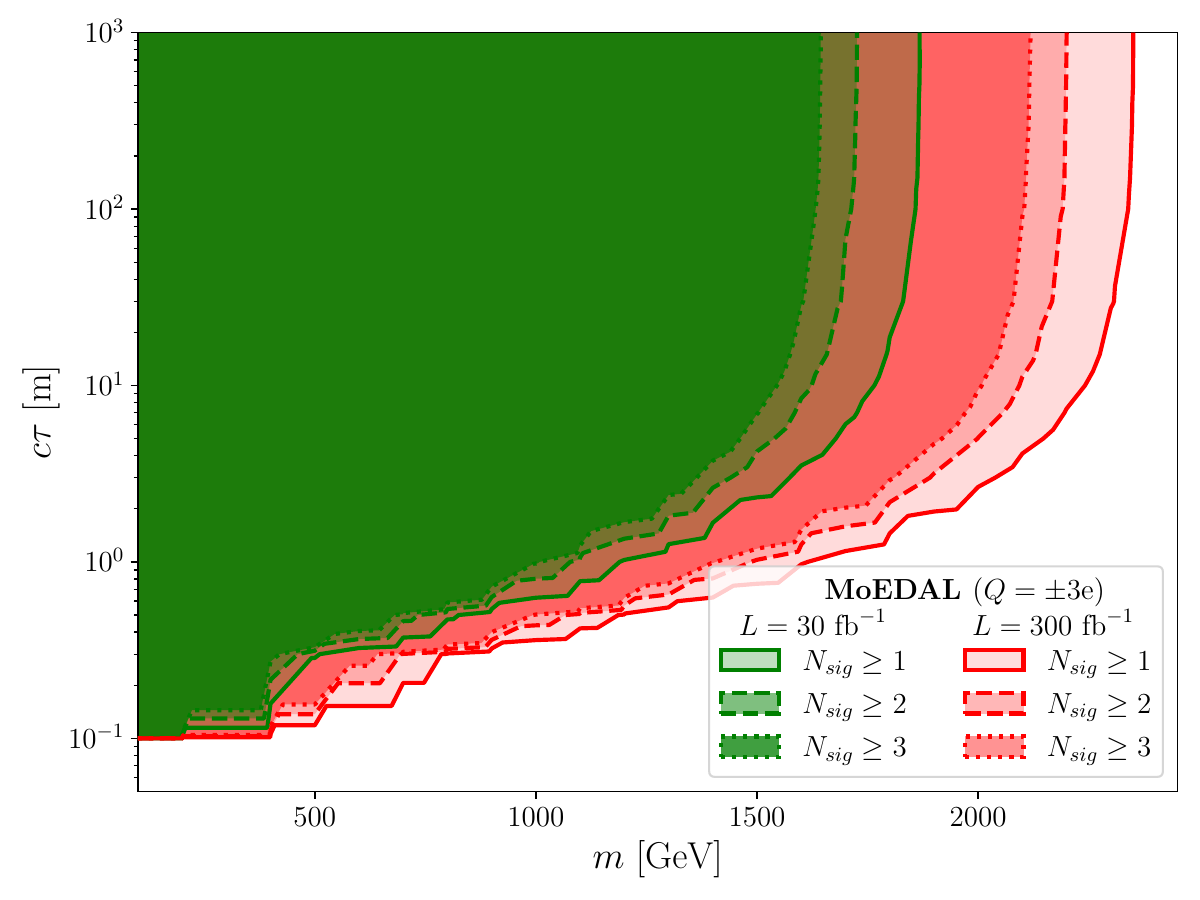}  \hspace{5mm}
      \includegraphics[width=0.4\textwidth]{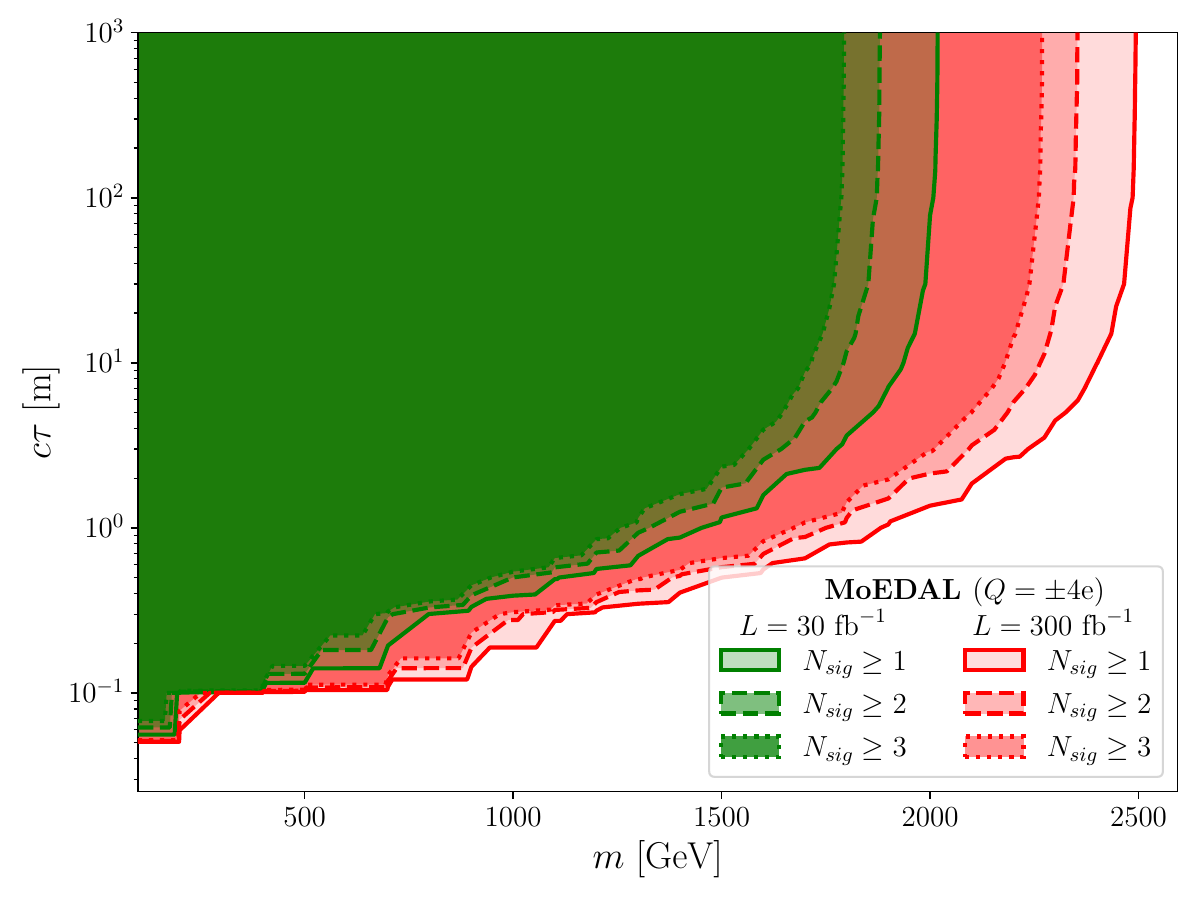}
      \includegraphics[width=0.4\textwidth]{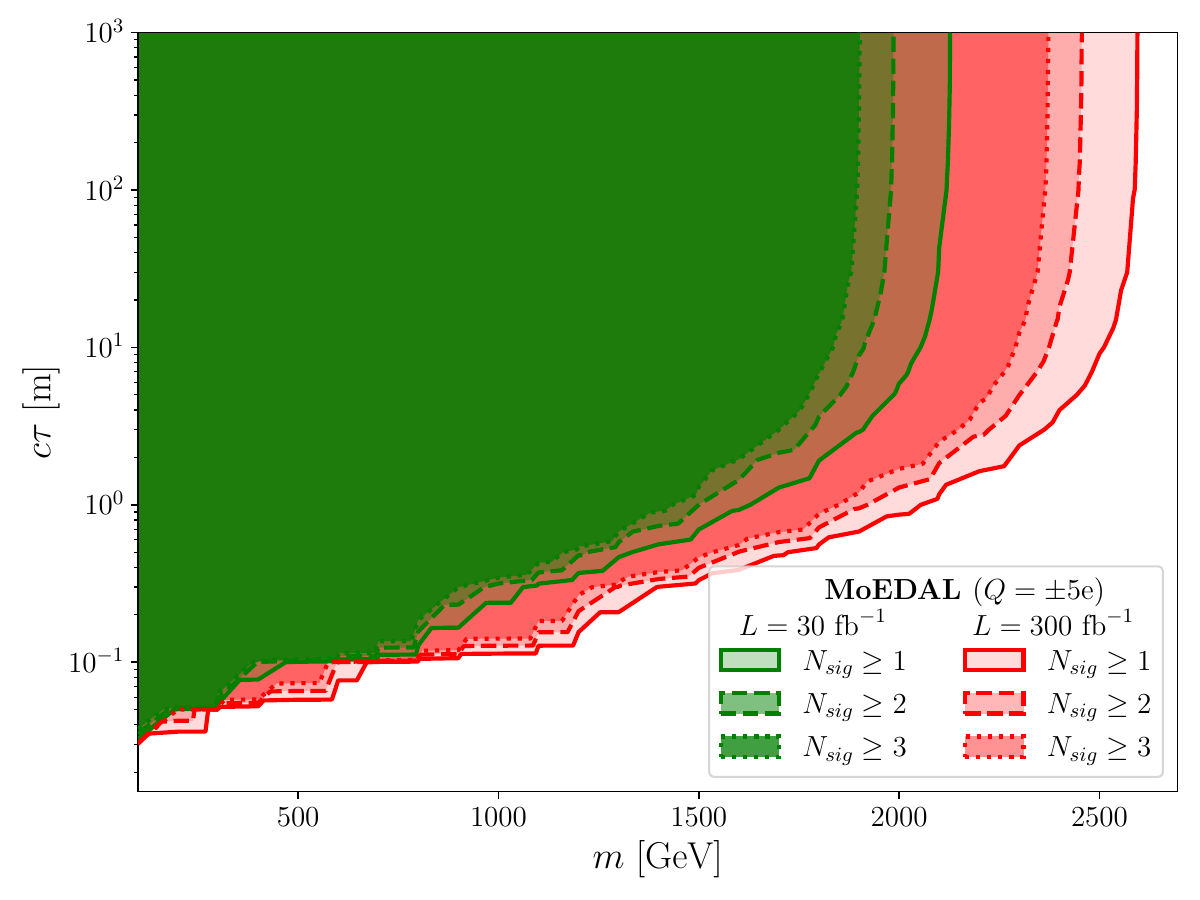}  \hspace{5mm}
      \includegraphics[width=0.4\textwidth]{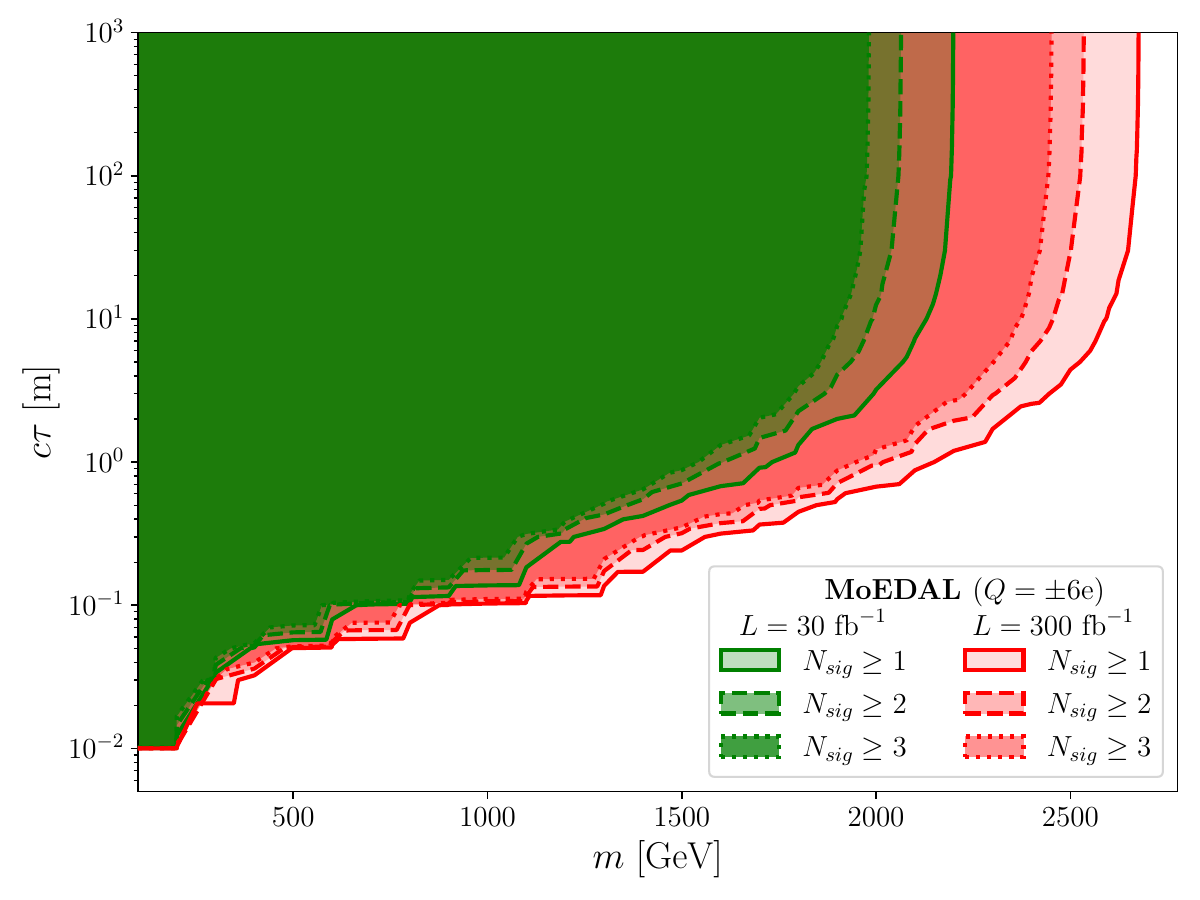}
      \includegraphics[width=0.4\textwidth]{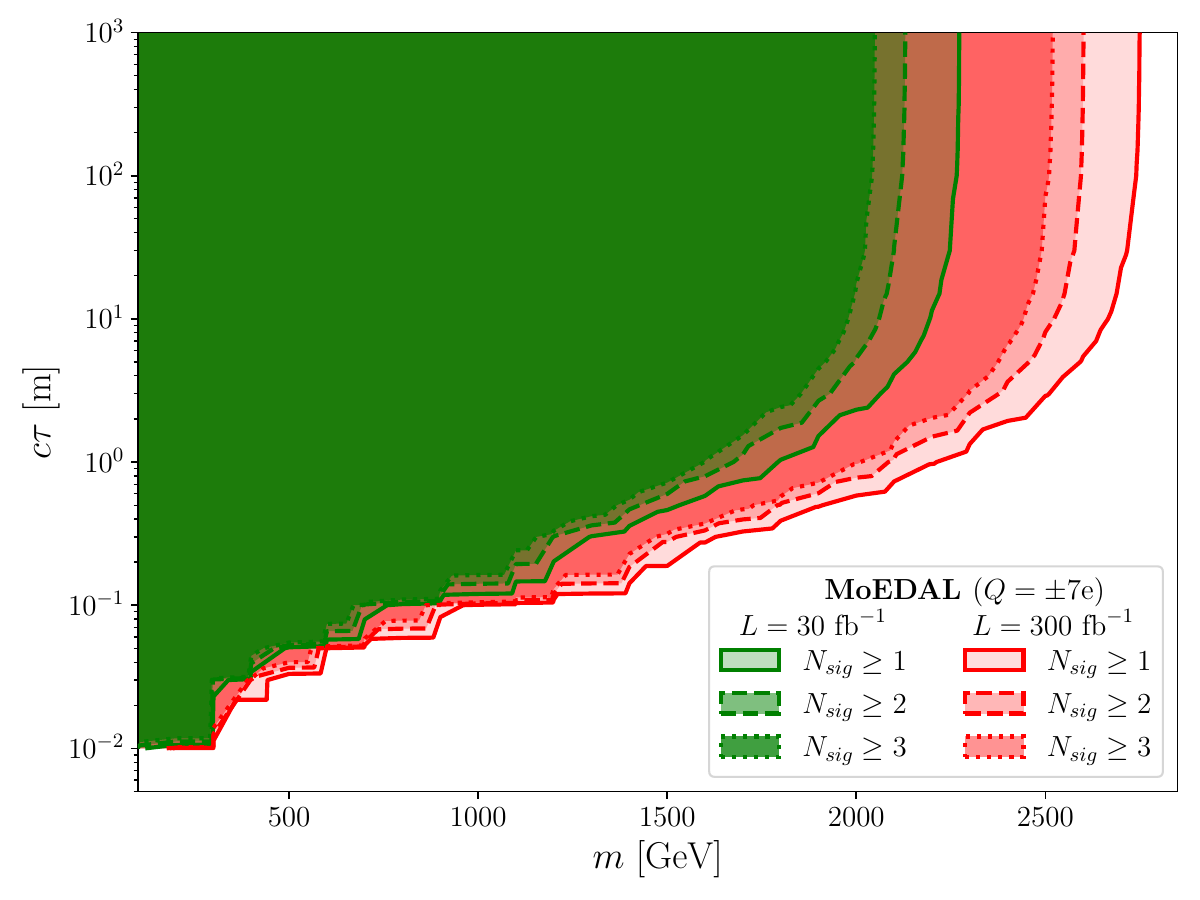}  \hspace{5mm}
      \includegraphics[width=0.4\textwidth]{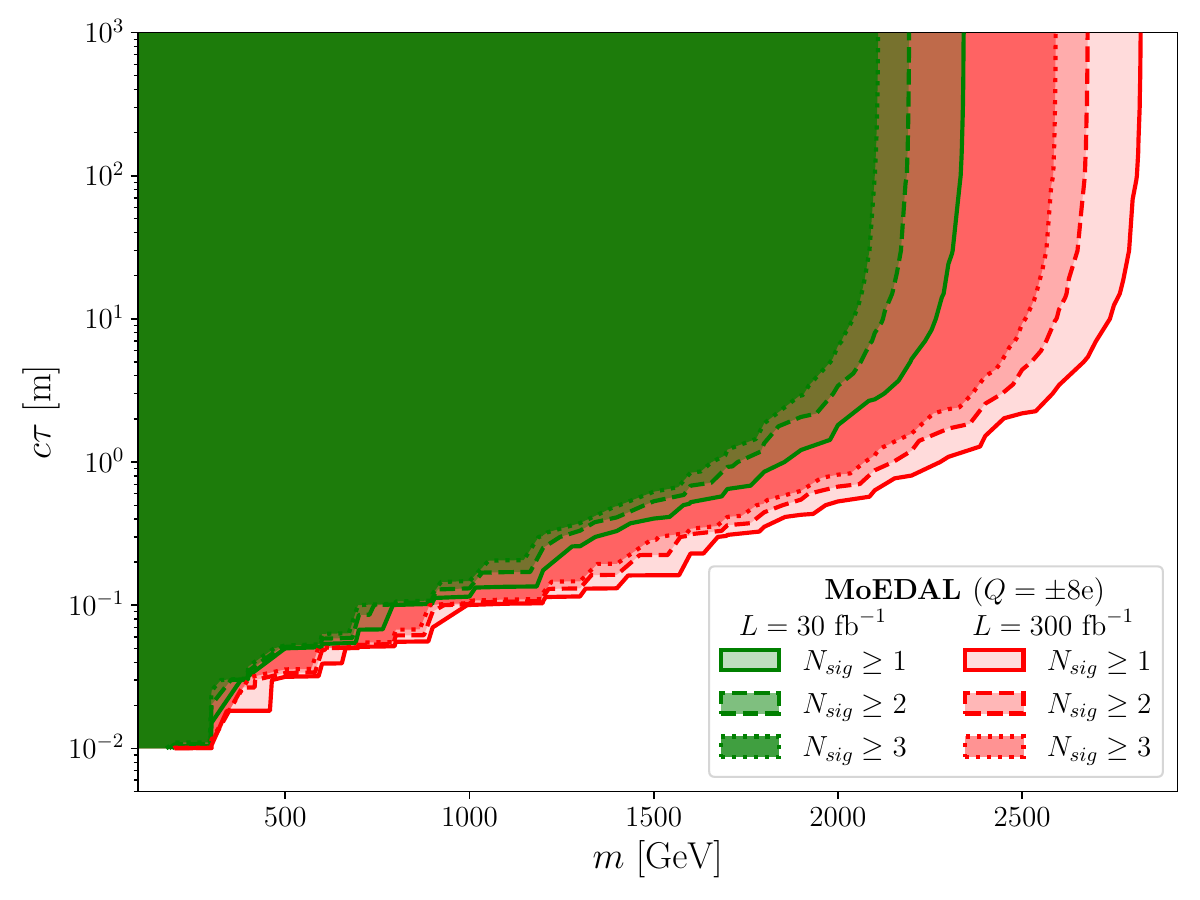}  
\caption{\small Model-independent detection reach of MoEDAL in the ($m$, $c
  \tau$) parameter plane for $SU(3)_C$-triplet fermions. Solid, dashed, and dotted contour lines correspond to $N_{\rm sig} = 1$, 2 and 3, respectively. Green and red colours represent results for Run 3 
  $(L=30$ $\mathrm{fb}{}^{-1})$ and HL-LHC $(L=300$ $\mathrm{fb}^{-1})$ data taking phases, respectively.
  }
\label{fig:lim_cfHighQ}
\end{figure}

    \chapter{Summary}\label{chap:five}

This thesis summarises four years of research in Particle Physics Phenomenology and concetrates on the prospects 
for detection of long-lived particles at the \textit{Large Hadron Collider (LHC)}. The main goal of the conducted research was to provide a comprehensive and 
an updated overview of the field and estimate the progress that is expected to happen during Run 3 and High Luminosity LHC data 
taking phases.

The first chapter of the thesis embeded the conducted research in a broader perspective of Particle Physics. It began by recalling the 
historical evolution of our understanding of the microscopic physical reality, which eventually led to the formulation of Quantum 
Mechanics and the birth of Particle Physics. Next, the current paradigm of the field, the \textit{Standard Model of Particle Physics (SM)}, was 
introduced. The basic ingredients of the theory and the most important phenomena were discussed. Issues and shortcomings of the 
Standard Model were listed to support the claim about the existence of yet unseen New Physics, the so-called \textit{Beyond the 
Standard Model (BSM)} Physics. A comprehensive overview of experimental searches for BSM Physics at the LHC was provided. So 
far none of these searches had resulted in undisputed evidence for New Physics, therefore, the necessity to focus on more exotic and 
unexplored theoretical scenarios was argued. Searches for BSM particles with large lifetimes, referred to as \textit{long-lived particles 
(LLPs)}, were presented as an interesting and promising alternative to ordinary studies.

The second chapter of the thesis was devoted to a description of some of the most popular New Physics scenarios, which predicted 
the existence of long-lived particles. The discussion started with supersymmetry, which is one of the most favoured BSM scenarios capable 
of resolving the hierarchy problem and providing a viable Dark Matter candidate. In the beginning, a general concept of supersymmetry, being 
a symmetry relating bosons and fermions, was introduced. Next, a mathematical framework of the theory was presented and applied to 
construct the \textit{Minimal Supersymmetric (extension of the) Standard Model (MSSM)}. Particle content and phenomenological 
consequences of the MSSM were discussed. Finally, theoretic scenarios resulting in supersymmetric LLPs were reviewed.
The second half of the chapter was devoted to neutrino mass models. First, the main idea of seesaw models was introduced. The seesaw 
mechanism provides a simple explanation of the origin and scale of the neutrino masses, by introducing new fields on top of the 
Standard Model. Some of these scenarios predict the existence of doubly or triply charged particles, which may be long-lived due to 
phase space suppression, mass degeneracy or small couplings. The other class of neutrino models are radiative neutrino models, 
in which neutrino masses arise through loop diagrams. 
One of such models, particularly interesting due to the presence of multiply charged $|Q| > 1 e$ long-lived scalars, was discussed in detail.

Chapter 3 provided an overview of searches for long-lived particles at the LHC. 
First, major general-purpose experiments, ATLAS 
and CMS, were discussed. These experiments have a similar design, they both consist of three main subdetector systems deployed in 
concentric cylindrical layers around the beam. The inner system is a tracker capable of measuring the momentum of produced particles, 
thanks to a powerful magnetic field. The intermediate layer is a calorimetric system, consisting of electromagnetic and hadronic 
calorimeters. The outermost detector layer is a muon system.

In the successive section, different kinds of 
searches for long-lived particles were reviewed. Searches for LLPs rely on a plethora of exotic signatures, e.g.: displaced vertices, 
disappearing tracks, anomalous jets, enlarged ionisation loss, and many more.
The second part of Chapter 3 was devoted to the MoEDAL detector, designed primarily to search for magnetic monopoles, 
however, as proven in the thesis, MoEDAL is also capable of searching for charged particles with large decay length $c\tau \gtrsim 1~\rm{m}$. The design and principle of operation of MoEDAL are completely different from ATLAS and CMS detectors. MoEDAL is a mostly 
passive detector, sensitive only to slowly moving particles with velocities $\beta \lesssim 0.15 \cdot |Q/e|$, where $Q$ is the 
electric charge of a particle. ATLAS and CMS, on the other hand, have active read out prone to pile-up, and they are sensitive only 
to particles with relativistic velocities $\beta \gtrsim 0.5$. 
The biggest advantage of MoEDAL is 
that the Standard Model background is negligibly small.
Its worst shortcoming is much fewer data available, due to placement next to the LHCb experiment, which opts for 
smaller instantaneous luminosity. After the description of the MoEDAL detector and its principle of operation, a simulation 
framework developed in the course of the PhD was described. The framework allowed for a simplified simulation of the detector and 
estimation of the number of signal events that MoEDAL is expected to observe in Run 3 and HL-LHC.

Chapter 4 contained research results and constituted the essence of the thesis. The results were grouped into four projects which 
corresponded to four studies conducted and published in the course of the PhD. The projects were described in chronological order, 
and not only did they provide a detailed description of the conducted research but also demonstrated the development of the research 
methodology and understanding of the underlying physics.

The first project was titled ``Prospects for discovering supersymmetric long-lived particles with MoEDAL''. The goal of the project 
was to investigate the MoEDAL detector, which at the time of conducting the study was known only for magnetic monopole 
searches, in the context of searches for long-lived supersymmetric particles. In particular, a specific supersymmetric scenario was considered, 
in which a pair of gluinos was produced ($pp \to \tilde g \tilde g$), each of them promptly decayed to SM jets and a long-lived neutralino ($\tilde g \to j j \tilde \chi_1^0$), which then decayed to metastable 
stau and off-shell tau ($\tilde \chi_1^0 \to \tilde \tau_1 \tau^*$), schematically:
$
pp \to \tilde g \tilde g \to \left( \tilde \chi_1^0 j j  \right) \left( \tilde \chi_1^0 j j  \right) \to 
\left( \tilde \tau_{1}  \tau^* jj \right)
\left( \tilde \tau_{1}  \tau^* jj \right)$.
A small mass difference between gluino and neutralino, $m_{\tilde g} - m_{\tilde \chi_1^0}=30~\rm{GeV}$, was assumed, and approximate mass degeneracy between neutralino and metastable stau, $m_{\tilde \chi_1^0} - m_{\tilde \tau_1}=1~\rm{GeV}$, was imposed. This theoretical scenario had some peculiar properties leading to enhanced sensitivity in the MoEDAL detector and 
deteriorated detection capability in ATLAS and CMS. Firstly, the production of gluinos was characterised by a large cross section, 
hence a higher chance for detection at the LHC. Secondly, gluino decayed promptly to long-lived neutralino, which could escape the 
ATLAS and CMS detectors unnoticed, if only its lifetime was long enough. Finally, the decay of the long-lived neutralino to charged 
metastable stau allowed MoEDAL to test this scenario because MoEDAL is sensitive only to highly ionising particles. The obtained 
result is very interesting. It was revealed that Run 3 MoEDAL can probe the supersymmetric model in a region of mass ($m$) vs. 
decay length ($c\tau$) parameter space, $m \in (1000,1350)$ GeV and $c\tau \in (10^3, 10^4)$ cm, where the most recent (at the time of realising the project)
ATLAS search for heavy stable charged particles \cite{ATLAS:2019gqq} (13 TeV 36.1 fb$^{-1}$) is not sensitive. Moreover, it was 
shown that by rearranging the deployment of the MoEDAL detector such that it would directly face the interaction point, MoEDAL could 
significantly increase its detection reach and compete with the projected Run 3 sensitivity of ATLAS. This result influenced the MoEDAL 
collaboration, and the necessary changes to the detector placement were applied for Run 3. The optimistic conclusion of the first 
study inspired three follow-up projects, which are all described in this thesis.

The second project was titled ``Prospects of searches for long-lived charged particles with MoEDAL'' and it is a direct continuation 
of the first study. In this project, a direct detection of various pair-produced supersymmetric LLPs at MoEDAL was studied. The 
candidates for long-lived particles were: gluinos ($\tilde g$), stops ($\tilde t$), five light-flavour squarks ($\tilde q = (\tilde u, \tilde d, \tilde c, \tilde s, \tilde b)$), charginos ($\tilde \chi_1^\pm$) and sleptons ($\tilde l$). 
A model-independent approach was adopted, in which masses and lifetimes of supersymmetric particles were varied as free parameters.
In the case of coloured 
supersymmetric particles, a simple hadronisation model was proposed in order to describe formation of R-hadrons in the final 
state. 
Re-interpretation for the Run 3 of the most recent ATLAS (13 TeV, $L=36.1~\rm{fb}^{-1}$) \cite{ATLAS:2019gqq} and CMS (13 TeV, $L=2.5~\rm{fb}^{-1}$) \cite{CMS:2016kce}
searches for heavy stable charged particles allowed us to compare MoEDAL's sensitivity to that of general-purpose experiments. 
The highest masses that MoEDAL could probe during Run 3 ($L=30~\rm{fb}^{-1}$), requiring $N_{\rm sig}=1$, were:
1600 GeV ($\tilde g$), 1920 GeV ($\tilde q$), 920 GeV ($\tilde t$), 670 ($\widetilde W$), 530 GeV ($\tilde h$), and 61 GeV ($\tilde \tau$). This was compared with the recast limits by ATLAS (CMS): 2000 (1500) GeV, 2310 GeV, 1350 (1000) GeV, 1090 GeV, 1170 GeV, and 430 (230) GeV, respectively.
We 
found that MoEDAL cannot compete with ATLAS nor CMS in the considered supersymmetric scenarios. Nonetheless, we would like 
to stress that constraints provided by MoEDAL are valuable because they are obtained using totally different experimental 
techniques, which might be sensitive to some regions of the model space overlooked by ATLAS and CMS, e.g. scenarios with small RPV couplings in which there is no missing transverse energy required by ATLAS/CMS data selection.

Another types of BSM particles considered were doubly charged scalars and 
spin-1/2 fermions, inspired by type II and type III seesaw models. Particles were 
assumed to be colourless, and transforming under a singlet or triplet $SU(2)_L$ 
representation. Their masses and lifetimes were treated as free parameters in 
order to obtain model-independent results. MoEDAL's sensitivity at the end 
of Run 3 was estimated, with $N_{\rm sig} = 1$, to be:
160 GeV, 650 GeV, 340 GeV, and 1130 GeV, for scalar $SU(2)_L$-singlet, fermion $SU(2)_L$-singlet, scalar $SU(2)_L$-triplet, fermion $SU(2)_L$-triplet, 
respectively. The corresponding 95\% CL limits set by the (recast) CMS analysis \cite{CMS:2016kce} were: 320 GeV, 680 GeV, 590 GeV, and 900 GeV, respectively. The 
conclusion was that MoEDAL, although always providing a valuable cross-check 
results, could compete with CMS only in the case of $SU(2)_L$-triplet fermion. The 
reason why MoEDAL is sensitive to this type of particle is twofold. Firstly, $SU(2)_L$-triplet production cross section is enhanced with respect to the singlet, 
but more importantly, the production of scalar particles suffers from the p-wave 
suppression, i.e. $\sigma \to 0$ for $\beta \to 0$ due to the angular 
momentum conservation in an s-channel production with $\gamma^*/Z^*$ 
exchange. Since MoEDAL is sensitive only to slowly moving particles, i.e. $\beta < 0.15 \cdot |Q/e|$, the p-wave suppression results in an unfavourable velocity 
distribution for produced BSM particles. This effect is absent for fermions, 
which typically move with lower velocities and are more likely to be detected in 
MoEDAL.

The third project, titled ``Detecting long-lived multi-charged particles in 
neutrino mass models with MoEDAL'', concentrated on particular 1-loop 
radiative neutrino mass generation model \cite{R:2020odv}, which predicted the existence of 
long-lived scalar particles: $S^{\pm 2}$, $S^{\pm 3}$, $S^{\pm 4}$, with 
electric charges $\pm 2e$, $\pm 3e$, and $\pm 4e$, respectively. These particles originated from a scalar $SU(2)_L$-triplet ($S_3$). The model contained also a triply charged fermion, $F^{\pm 3}$, which was expected to be short-lived. Particles in 
the basic (uncoloured) version of the model were colour-singlets, however, a 
coloured version of the scenario was obtained by promoting the BSM fields to $SU(3)_C$-(anti)triplets and adjusting the Lagrangian density appropriately. 
Long-lived particles in the coloured version of the model were: $\tilde S^{\pm 4/3}$, $\tilde S^{\pm 7/3}$, and $\tilde S^{\pm 10/3}$, with electric charges $\frac{4}{3} e$, $\frac{7}{3} e$, and $\frac{10}{3} e$, respectively. These particles originated from a coloured scalar $SU(2)_L$-triplet ($\tilde S_3$)

In this project we have pointed out the relevance of the photon fusion 
production process to searches for multiply charged particles at the LHC. 
Up to that point,
ATLAS and CMS analyses targeting multiply charged particles had been considering 
only the s-channel Drell-Yan production, while completely ignoring photon 
fusion, which is negligible for singly charged particles. However, when the 
electric charge of the particle is large, photon-induced production processes become significant. 
Fortunately for MoEDAL, including photon fusion not only enhances the total 
production cross section for the BSM particles but also alters their velocity 
distributions such that the MoEDAL's acceptance increases.

The analysis presented in the project was divided into two 
parts. First, model-independent limits on the multiply charged particles were 
estimated for Run 3 ($L=30~{fb}^{-1}$) and HL-LHC ($L=300~{fb}^{-1}$) 
MoEDAL. Limits on these particles were obtained by treating their mass and lifetime as free parameters. 

In the case of the uncoloured model, the estimated model-independent detection reach for Run 3 (HL-LHC) MoEDAL with $N_{\rm sig}=1$ was: 290 (600) GeV, 610 (1100) GeV, 960 (1430) GeV, and 1030 (1550) GeV, for $S^{\pm 2}$, $S^{\pm 3}$, $S^{\pm 4}$, and $F^{\pm 3}$, respectively. This has to be compared with current ATLAS limits (13 TeV 36.1 fb$^{-1}$) \cite{ATLAS:2018imb}: 650 GeV ($S^{\pm 2}$), 780 GeV $S^{\pm 3}$, 920 GeV$S^{\pm 4}$, 1130 GeV ($F^{\pm 3}$). For the Run 3 phase ($L=300~\rm{fb}^{-1})$, the only available ATLAS projection \cite{Jager:2018ecz} was 1500 GeV for $F^{\pm 3}$.

When it comes to the coloured LLPs, a model-independent MoEDAL detection reach for Run 3 (HL-LHC) with $N_{\rm sig}=1$ was estimated to: 1050 (1400) GeV, 1250 (1650) GeV, and 1400 (1800) GeV, for $\tilde S^{\pm 4/3}$, $\tilde S^{\pm 7/3}$, and $\tilde S^{\pm 10/3}$, respectively. The current (future) exclusion (sensitivity) limits from ATLAS/CMS were: 1450 (1700) GeV, 1480 (1730) GeV and 1510 (1790) GeV, for $\tilde S^{\pm 4/3}$, $\tilde S^{\pm 7/3}$, and $\tilde S^{\pm 10/3}$, respectively.

The second part of the analysis concerned the possibility to contrain 
parameters of the two considered models with the MoEDAL detector. Three 
phenomenologically interesting quantities are: lepton number violating 
coupling $\lambda_5$, the product of Yukawa couplings $h_F h_{\bar F}$, and 
coupling of the neutrino mass model to the SM $h_{ee}$. In our analysis we 
varied masses of the BSM fields and $\lambda_5$ parameter, keeping $h_{ee}$ 
($h_{ed}$ in the coloured version) fixed, and fitted neutrino data to get 
phenomenologically viable $h_F h_{\bar F}$. We found that the highest 
sensitivity of MoEDAL was for $\lambda_5\sim 10^{-5}$ in both versions of the 
model. This value of $\lambda_5$ corresponded to the longest lifetime of 
$S^{\pm 4}$ ($\tilde S^{\pm 10/3}$) in the uncoloured (coloured) model.
In addition, we investigated how different values of the $h_{ee}$ ($h_{ed}$) 
coupling influenced the sensitivity of the MoEDAL, and we found that while the 
maximum testable mass of the scalar $SU(2)_L$-triplet BSM field did not 
change much, it was accessible for a wider range of $\lambda_5$ values.

Our conclusion from the third project was that most of the parameter space 
accessible to Run 3 MoEDAL had been constrained by the large general-purpose LHC 
experiments, however, the prospects for BSM detection during the HL-LHC phase 
were more optimistic.

The last of the studied projects, named ``Discovery prospects for long-lived 
multiply charged particles at the LHC '', continued the research direction 
established in the previous projects. In the fourth study, a comprehensive 
overview of prospects for the detection of multiply charged LLPs at the LHC was 
provided. Four types of $SU(2)_L$-singlet particles were considered: colour-singlet and colour-triplet, scalars and spin-1/2 fermions. Projected sensitivities 
of direct detection searches at MoEDAL and ATLAS/CMS were compared, together 
with possible limits from the formation of new bound states decaying to $\gamma \gamma$. We studied model-independent limits in the $(m, c\tau)$ parameter plane for particles with electric charges in the range $1e \leq |Q| \leq 8e$.

In the last project we investigated in detail the impact of different production processes on the prospects for detection of multiply 
charged BSM particles, and we confirmed the relevance of photon-induced processes to the total production cross section and velocity 
distributions. In the case of colourless particles, for $|Q|\sim (3-4)e$ the photon fusion production process became comparable 
to Drell-Yan, while for larger charges it dominated. This effect also altered the velocity distributions, especially for scalars, for 
which the Drell-Yan production led to velocity suppression, i.e. $\sigma \to 0$ when $\beta \to 0$. The prevalence of 
photon fusion resulted in typically slower particles, for which it was easier to satisfy the MoEDAL sensitivity criterium, $\beta < 0.15 \cdot |Q/e|$. In the case of the coloured particles the effect was much milder because other production processes were contributing, i.e. gluon-gluon and gluon-photon fusion, which depended on the electric charge as $\propto Q^0$ and $\propto Q^2$, respectively.

In order to estimate the current limits on direct detection of multiply charged LLPs, the CMS search for large $dE/dx$ \cite{CMS:2016kce}, 
based on 13 TeV 2.5 fb$^{-1}$ dataset, was re-interpreted. In addition, for the $SU(3)_C$-singlet fermions
constraints from the ATLAS search \cite{ATLAS:2018imb} for large $dE/dx$, based on 13 TeV 36.1 fb$^{-1}$ data set, were available.
In this part of the analysis, we treated BSM particles as collider stable, i.e. traversing the whole volume of the CMS detector without decaying.
In the recast procedure two important experimental issues were 
investigated in order to provide more precise limits. The first effect was the underestimation of the transverse momentum of a multiply 
charged particle, due to implicit assumption $Q=\pm 1 e$ of the reconstruction algorithms. The other concern was the loss of 
signal efficiency due to slowly moving LLPs that reach the muon system already after the succeeding bunch crossing. In our procedure, we corrected for these two effects and included photon-induced production processes.
With the help of the Monte Carlo simulation, the effective production cross section for the considered BSM particles was obtained and compared with the experimental limits provided by the CMS. Moreover, projection for the Run 3 ($L=300~\rm{fb}^{-1}$) and HL-LHC ($L=3~\rm{ab}^{-1}$) luminosities was made. 

Charged LLPs could also be directly detected in the MoEDAL experiment. In order to assess the detector's sensitivity, a similar 
approach as in the previous three projects was used. Masses and lifetimes of particles were treated as free parameters, which
allowed to derive MoEDAL's sensitivity reach and the expected number of signal events $N_{\rm sig}$ for Run 3 ($L=30~\rm{fb}^{-1}$) and HL-LHC ($L=300~\rm{fb}^{-1}$). For very large decay lengths, $c \tau \gtrsim 100~\rm{m}$, MoEDAL's sensitivity saturated.

Finally, we considered a creation of bound states formed by a pair of new BSM particles, in analogy to positronium and quarkonium 
systems. We used a narrow-width approximation to calculate the cross section, $\sigma_{pp\to \cal B \to \gamma \gamma}$, corresponding to the production of a bound state in a 13 TeV $pp$ collision and a subsequent decay to two photons. Searches for two photons, referred to as \textit{diphoton channel}, are the most sensitive and constitute a discovery channel. We recast the most recent diphoton search by ATLAS \cite{ATLAS:2021uiz}, which was based on 139 fb$^{-1}$ data set, and compared it with the theoretically calculated cross section $\sigma_{pp\to \cal B \to \gamma \gamma}$ in order to derive lower mass limits, including the projections for the Run 3 and HL-LHC.

All three types of searches were compared and the following conclusions were drawn. For Run 3, large $dE/dx$ searches by ATLAS and 
CMS were by far the most sensitive for lower magnitudes of the electric charge, $|Q| \lesssim (3-4)e$. This is because for low 
charges MoEDAL had poor sensitivity due to the requirement $\beta < 0.15 \cdot |Q/e|$, while bound state formation was very 
unlikely. For larger charges, $|Q| \gtrsim (4-5)e$, the probability of forming a bound state increased rapidly, as $Q^{10}$, which 
made this kind of search the most sensitive. MoEDAL typically provided an intermediate sensitivity. However, the situation 
became more interesting for the HL-LHC phase. ATLAS and CMS analyses, both large $dE/dx$ and diphoton resonance searches, 
suffered from the increased background, which significantly reduced the benefit of increased luminosity. MoEDAL, on the other hand, was 
effectively background free, hence the signal rate increased dramatically and MoEDAL's sensitivity surpassed the projected results 
of major general-purpose LHC experiments for intermediate electric charges, i.e. $3 e \lesssim |Q| \lesssim 7e$. However, this projection is to be taken with a pinch of salt, because it does not include detector changes nor improvements to analyses.

All four projects provided a detailed and comprehensive overview of prospects for the detection of long-lived charged BSM particles in 
Run 3 and HL-LHC. Various New Physics scenarios were tested, e.g. supersymmetry, seesaw, radiative neutrino mass generation, 
and simplified models. For the first time the expected sensitivity of MoEDAL was estimated and compared to
major general-purpose experiments. While typically ATLAS and CMS have better sensitivity than MoEDAL, in some particular 
scenarios, e.g. for multiply charged LLPs with $3 e \lesssim |Q| \lesssim 7e$, MoEDAL might successfully compete with the large experiments, especially for the HL-LHC phase. The presented work resulted not only in qualitative research, but also influenced the field. First of all, it led to 
the rearrangement of MoEDAL subdetectors in order to increase the overall sensitivity and allow the experiment to effectively compete 
with ATLAS and CMS. Moreover, the content of this thesis was presented at several international conferences, where the relevance 
of the photon-induced production processes on the correct interpretation of experimental results for multiply charged LLPs was 
presented to the community. The very recent ATLAS and CMS studies, e.g. \cite{ATLAS:2022cob}, take photon-fusion into account.

   \begingroup
 	\renewcommand{\section}[2]{}%
      \bibliographystyle{utphys}
 \bibliography{bibliografia.bib}
    \endgroup

\end{document}